\documentclass[11pt]{report}
\usepackage{upennstyle} 

\usepackage{hyzsyntax}
\usepackage{thesis_tikz}
\usepackage{thesis_misc}

\usepackage[round]{natbib}
\title{GENERALIZED SYMMETRIES IN SUPERGRAVITIES AND \\SUPERCONFORMAL FIELD THEORIES VIA STRING THEORY}
\author{Hao Zhang}
\gradgroup{Physics and Astronomy} 
\date{2023} 
\supervisor{Mirjam Cveti\v{c}}
\supervisortitle{Professor of Physics and Astronomy}
\cosupervisor{Jonathan J. Heckman} 
\cosupervisortitle{Associate Professor of Physics and Astronomy} 
\gradchair{Vijay Balasubramanian, Professor of Physics and Astronomy}
\committee{Vijay Balasubramanian, Professor of Physics and Astronomy}
\committee{Ron Donagi, Professor of Mathematics}
\committee{Elliot Lipeles, Associate Professor of Physics and Astronomy} 

\authorlegal{Hao Zhang} 

\dedication{This dissertation is dedicated to my parents and my wife.} 

\acknowledgement{The completion of my doctoral study cannot be possible without the help, support, and encouragement of many people.

First of all, I am extremely grateful to my advisor, Mirjam Cveti\v{c}. Her strong support and warm encouragements are crucial for me to finish my Ph.D. and proceed in academia. I benefited hugely not only from the strict standard that she set up for our results, but also from her earnest expectations that I make my unique contribution to our collaborations. Her research style emphasizing a balance of novelty, rigor, and feasibility, will continue to be my treasure.

I am also enormously indebted to my co-advisor, Jonathan Heckman, for his lasting patience and inspiration in discussing with me whenever I show up. Starting from the first project he assigned to me, he has been very open to involving me in collaborations and brainstorming new ideas, giving me many experiences of how different types of research projects evolve and proceed. His rapid and sharp intuition fundamentally influenced my approach to physics in the future.

In addition, I am particularly thankful to both Mirjam Cveti\v{c} and Jonathan Heckman for encouraging me to work closely with more than one faculty member but still actively influencing me with their research style and taste with full dedication. 

Throughout my Ph.D., I am very fortunate to get involved in multiple research collaborations and to interact with many fascinating senior and juniors scholars: Ling Lin, Markus Dierigl, Craig Lawrie, Max H\"{u}bner, Gianluca Zoccarato, Tom Rudelius, Sandipan Kundu, Michele Del Zotto, Fabio Apruzzi, Falk Hassler, and Marco Fazzi. I am indebted to all of them so much that I have to express my gratitude in this excessively long and detailed manner. 

I am particularly grateful to Ling Lin for his strict training in research discussions and presentations during collaborations. I greatly benefit from his standard of seeking insightful and unambiguous statements and organizing them toward influential objectives. I am also very thankful to Markus Dierigl for his patience in introducing me to many novel and sophisticated concepts in our field, and for the influence of his pedagogical presentation style. I would like to sincerely thank Craig Lawrie for impressing me with the capability of extensive studies, and for his patience in correcting my mistakes. I greatly appreciate Max H\"{u}bner for his sharp understanding of geometry, for insisting on having a clear physical understanding of each step, and for being very open about explaining the basics of physics and math. I very much appreciate Gianluca Zoccarato for his solid control of various aspects of our collaborations, as shown in his real-time derivations. It's also my great pleasure to thank Tom Rudelius for his careful guidance in aspects of 6D SCFTs, and for bearing with my mistakes and sloppiness at an introductory stage. Many thanks to Sandipan Kundu for his patience in explaining aspects of effective field theories, and for his willingness to have prolonged discussions with me on general aspects of academia. I would like to give special thanks to Michele Del Zotto for stimulating collaborations, and for introducing me to the topic of higher-group symmetries. In the end, I would like to thank Fabio Apruzzi, Marco Fazzi, and Falk Hassler for inspiring collaborations and instructions. 

In addition, I benefited enormously from interacting with excellent fellow graduate students, including Ethan Torres, Muyang Liu, Thomas Rochais, Xingyang Yu, Shani Nadir Meynet, and Robert Moscrop. It is my great pleasure to have many entertaining collaborations and discussions with all of them. I want to first express my gratitude to my office mate, Ethan Torres, for continuing to impress me with his broad range of knowledge, and for many interesting collaborations and discussions. I appreciate Thomas Rochais for introducing me to many details in my first project, and for continuing to have long and patient discussions in our later collaborations. I am grateful to Xingyang Yu for his clear explanation during collaborations and for generously sharing his general view of academia with me. I am thankful to Muyang Liu for her detailed help in algebraic geometry and toric geometry, both formally and computationally, and for many interesting discussions. I thank Shani Nadir Maynet for his patience in explaining the details of BPS quivers to me. I also thank Robert Moscrop for his strong ability to consolidate several aspects of our collaboration regarding mathematical rigor. 

I thank many faculties with whom I had general research discussions: Vijay Balasubramanian, Mark Trodden, and Justin Khoury. Moreover, I thank various faculty members who gave me lectures in physics and mathematics: Jonathan Heckman, Ron Donagi, Andrea Liu, Antonella Grassi, and Burt Ovrut. I thank Mirjam Cveti\v{c}, Jonathan Heckman, and Ron Donagi for sitting on my qualification exam committee. I also thank Vijay Balasubramanian and Elliot Lipeles for joining my dissertation committee. I would like to thank all the administrative members in the department of physics and astronomy, Millicent Minnick and Michele Last, for their extensive help.

During the application season for postdoctoral positions, I am very grateful to Mirjam Cveti\v{c}, Jonathan Heckman, Ling Lin, Craig Lawrie, and Max H\"{u}bner for extensive discussions, for helping me to write and revise my application materials, and for their great patience in relieving my anxiety. I thank Mirjam Cveti\v{c}, Jonathan Heckman, Michele Del Zotto, and Ron Donagi for their recommendation letters supporting my applications.

It is also a great pleasure to spend my Ph.D. time with many friends at Penn: Qiuyue Liang, Pok Man Tam, Andrew Turner, Florent Baume, Martin Bies, Javiar Magan, Edgar Shagoulian, Tianyue Liu, Marielle Ong, Michio Tanaka, Jicheng Jin, Mark Giovinazzi, Ian Graham, Cathy Li, Meng Luo, Zhuoliang Ni, Jianrong Tan, Riley Xu, Valerie Yoshioka, Meng-Xiang Lin, and many more.

I would like to thank my parents and my extended family for supporting my pursuit of an academic career. Throughout my growth, they have been very good at stimulating my curiosity and encouraging me to pursue a career of my interest, which became crucial for me to enter graduate school in the first place. I am also grateful that they left nothing for me to worry about during these years so that I could entirely focus on my research.

In the end, I would like to thank my wife, Yuxi Ai, for being together with me throughout these five years. Our shared dream of making long-lasting scientific contributions brought us to Penn as graduate students, motivating me to always aim higher. Her experience in biology constantly brings me insights into academia from an entirely different perspective. Her encouragements and praises are my constant sources of confidence. Her everlasting love and accompany have been substantial in driving me toward completing my Ph.D. study.

This work is supported by Department of Physics and Astronomy and by Department of Mathematics, School of Physics, University of Pennsylvania, and also by the Simons Foundation Collaboration grant \#724069 on “Special Holonomy in Geometry, Analysis and Physics”.} 

\abstract{In this dissertation, we study the generalized symmetries in supergravities and superconformal field theories from the string theory perspective. 

Part one is devoted to the study of string universality in high spacetime dimensions. Answering this question requires us to combine the following two approaches. In the "top-down" approach, We focus on supergravity theories in 7, 8, and 9 dimensional spacetime with 16 supercharges. We emphasize two discrete aspects of these theories: generalized global symmetries and frozen singularities. We give an exhaustive classification of IIB supergravity theory in 8D, particularly emphasizing these two discrete aspects. In the "bottom-up" approach, we present a consistency condition of general 8D supergravity theories involving their higher-form symmetries use it to rule out many global structures of the gauge groups in 8D supergravity theories that do not admit string theory constructions. 

Part two studies the generalized global symmetries of geometrically-engineered quantum field theories via string theory. We examined branes wrapping on relative topological cycles that give heavy defects that are charged under generalized global symmetries, which can then be used to construct new lower-dimensional theories. By investigating the string theory origin of the topological operators, we provide a general construction of these topological operators in the context of geometric engineering as branes wrapped on the homological cycles in the asymptotic boundary of the internal geometry. We illustrate this proposal by determining non-invertible 2-form symmetries in 6D superconformal field theories. Furthermore, by wrapping type IIB 7-brane on the entire asymptotic boundary of the internal manifold, we explicitly give a unified string-theoretic construction of two different types of field-theoretic non-invertible duality defects.} 

\begin{document}
\maketitle 
\setcounter{page}{2}

\makecopyright 

\makededication 

\makeacknowledgement 

\makeabstract
\tableofcontents

\clearpage \phantomsection \addcontentsline{toc}{chapter}{LIST OF TABLES} \begin{singlespacing} \listoftables \end{singlespacing}

\clearpage \phantomsection \addcontentsline{toc}{chapter}{LIST OF ILLUSTRATIONS} \begin{singlespacing} \listoffigures \end{singlespacing}

\clearpage \phantomsection \addcontentsline{toc}{chapter}{INTRODUCTION}

\textbf{\huge INTRODUCTION}

\textbf{\Large \section{Quantum Field Theories}}

Quantum Field Theory (QFT) is a comprehensive framework that describes our current understanding of microscopic physics. QFT takes a spacetime and assigns a collection of quantum fields to it. (See \cite{Peskin:1995ev,schwartz_2013,weinberg_1995,Weinberg:1996kr,Zee:2003mt} for standard textbooks).

To explain the meaning of fields, we step back and start with classical fields. A crucial property of a field is spin, which is a phase dependence of the field on spacetime rotations. In four spacetime dimensions and above, spin can only be integers or half integers. Integer spin fields are called Bosonic fields which are classically commutative. For example, the classical Maxwell theory of Electromagnetism was phrased using classical bosonic fields. In contrast, half-integer spin fields are called Fermionic fields which are anti-commutative - one needs to express them in terms of anti-commuting Grassmanian variables. 

The procedure of uplifting a classical field to a quantum field is called second quantization. Concretely, one decomposes the classical field into momentum eigenstates and promotes the coefficient of each coefficient into creation and annihilation operators. Thus we can view the ``vacuum" of a QFT as a state without any excitation of quantum fields, on top of which a creation operator acting on the vacuum state would create a ``particle".

Perturbatively, Feynman diagrams are very powerful at performing scattering amplitude computations of different asymptotic states. For a specific combination of incoming and outgoing particles, their scattering amplitude can be viewed as a summation of an infinite number of intermediate processes. This summation can be organized by the number of loops in the Feynman diagrams, where each loop results in an extra factor of the coupling constant. Therefore, the tree diagram itself can describe the scattering amplitude under the limit that the coupling constant goes to zero. However, one must honestly sum over this infinite series of Feynman diagrams if the coupling constant is set to a finite value. Even evaluating every single diagram requires integration over the space of all momenta ``running over the loop", which would sometimes result in an infinity. Ultimately, we are summing over an infinite number of infinities. To resolve this issue, physicists came up with the renormalization approach: to remove such an infinite number of infinities at the expense of introducing a finite number of infinities. The physical interpretation of renormalization is to view all the masses and coupling constants as parameters that depend on the energy scale and then introduce suitable counterterms to compensate for their divergence at the high energy limit, such that only a finite difference remains. Whenever such a prescription works, we define the QFT as a \textit{renormalizable} QFT - physicists can generate accurate predictions for expected experimental outcomes as long as the perturbative approximation remains valid.

However, at a sufficiently large coupling constant (beyond the radius of convergence), the series of loop diagram summation does not converge, and the perturbative approach breaks down. Such a case is called the strong coupling scenario of a quantum field theory. We will soon see that a strong coupling scenario is important even in real-world physics.
 
Within the framework of QFT, the Standard Model (SM) of particle physics has been constructed. It successfully unifies three out of four fundamental interactions: electromagnetism, strong interaction, and weak interaction. The standard model of particle physics has sharp predictive power in that it can be verified down to a very high precision against collider experiments. However, there are situations where an accurate theoretical prediction is not possible. Concretely, Quantum ChromoDynamics (QCD), as a component of the SM describing the strong force, exhibits confinement at low energies - it takes a huge amount of energy to separate pairs of quarks farther away than the length scale of a baryon. So macroscopically, one can only see color-neutral states. Thus, QCD is a strongly coupled theory at low energies, and Feynman diagram calculation breaks down. Physically relevant quantities in QCD need to be computed numerically via lattice QCD. With these difficulties in mind, it would be desirable to attempt full control of non-perturbative behaviors of QCD, which could potentially benefit from understanding non-perturbative QFTs in general.

In addition to understanding the microscopic world, quantum field theory also has successful applications in other branches of physics. This includes condensed matter physics at the mesoscopic scale and cosmology at the macroscopic scale.

\subsubsection{Condensed Matter Physics} 

At a mesoscopic scale, QFT plays a fundamental role in understanding condensed matter physics, especially in phase transitions and strongly coupled matter systems (see \cite{Wen:2004ym,fradkin_2013} for standard textbooks). To begin with, Landau's paradigm of second-order phase transition using field theory applies to many different systems. Later, Wilson and other physicists managed to compute some universal classes of scaling coefficients in phase transitions using renormalization group flows. QFT also underlies our theoretical understanding of the quantum hall effect - we can use the system of 3D Chern-Simons theory with a gapped 2D boundary theory to explain the integer quantization phenomenon observed in quantum hall systems (\cite{Tong:2016kpv} for a pedagogical review). On top of that, incorporating emergent degrees of freedom can also keep track of the more intriguing behavior of fractional quantum hall systems with fractional quantization levels. More recently, QFT has had more formal applications in studying topological phases that are non-local systems with long-range entanglement. There, an important concept called Symmetry Protected Topological (SPT) phases describes a set of topological quantum field theories. SPT provides a general understanding of almost trivial phases of matter, except for having non-trivial boundary conditions as another QFT living in one dimension lower.

\subsubsection{Cosmic Inflation} 

At a macroscopic scale, QFT also finds many applications in cosmology. A prominent example is the theory of inflation (see \cite{2005pfc..book.....M,Weinberg:2008zzc} for reviews), which is modeled by a fluctuating scalar field (``the inflaton") with a potential (``the inflaton potential") that controls the rate of inflation.

The conception of inflation stems from the observation that the Cosmic Microwave Background (CMB) has a surprising homogeneity, namely a small temperature fluctuation (at order $10^{-5}$) from all directions. Such homogeneity cannot be explained by causal correlation. Therefore, the most natural explanation of such homogeneity turns out to be cosmic inflation, a rapid expansion from $10^{-36}$ second to $10^{-32}$ second after the Big Bang. However, the microscopic mechanism of cosmic inflation is unknown. Physicists only have some phenomenological models as opposed to first-principle understandings. Such phenomenological models are constructed under the framework of ``effective field theory" (viewing the theory at hand as an IR limit of an unspecified higher-energy theory, where some unspecified massive fields have already been integrated out to land on this effective theory) in order to explain cosmic inflation.

Most of such phenomenological models introduce a scalar field called the \textit{inflaton}. Inflation is affected by an inflaton potential, and different inflaton potential results in different behaviors of inflation. Within the framework of effective field theory, the power spectrum and many other observable can be determined. By comparing these observables with cosmological observations, we can, in turn, constrain various parameters in the models. In addition, QFT also plays a fundamental role in other cosmological scenarios such as early-universe particle physics, cosmic re-ionization, and many more.

\textbf{\Large \section{Generalized Global Symmetries}}

Having introduced many phenomenological applications of QFTs, we move on to their formal properties. Of primary interest are those properties that are universal to any QFT, which do not depend on any particular description (such as an explicit Lagrangian) of a QFT.

Global symmetry is a prominent property of a QFT. As we learned from Noether's theorem, every continuous global symmetry gives rise to a conserved quantity. Discrete symmetries are equally relevant in that they give rise to selection rules based on certain quantum numbers in scattering processes. In general, continuous and discrete global symmetries are both intrinsic properties of a QFT, which should remain identical even when one uses a different description of the same QFT.

Regular global symmetries are assumed to act on point-like operators and excitations. However, the concept of symmetries has recently been rephrased in the language of symmetry operators, such that the symmetry operator for a (regular) global symmetry is thus a codimension-1 topological operator. Locally it can be deformed without any cost, but once crossing a charged point-like operator, the topological operator imposes a group action onto the charged operator.  \cite{Gaiotto:2014kfa}

Under this rephrasing, there is a natural generalization of global symmetries by considering topological operators of higher codimensions. One can see such new topological operators acting on \textit{extended} charged operators/excitations, whose spacetime support has non-trivially linking with that of the topological operator. These higher codimension operators are avatars of \textit{global categorical symmetries/generalized global symmetries}. Almost all properties of ordinary global symmetries have their counterparts in generalized global symmetries - gauging, spontaneous breaking, Hooft anomalies, and so on. Sometimes one uses the term \textit{higher symmetries}, especially when we want to refer to not only generalized global symmetry but also their gauged version in a single term. (See \cite{Cordova:2022ruw} for reviews of generalized global symmetries).

Generalized global symmetries have many elegant applications. For example, the confinement-deconfinement phase transition in QCD can be rephrased in terms of (the existence and potential spontaneous breaking of) center 1-form symmetries associated with $SU(N)/Z_k$ gauge group (where $k$ can be any factor of $N$). The Green-Schwarz anomaly cancellation can also be rephrased in the language of 2-group symmetries, the simplest non-trivial examples of higher-categorical symmetries. The perspective of higher symmetries often requires us to include extra data to specify this theory fully, and different specifications of generalized global symmetries are often related by various gauging operations and implementing duality transformations. Said differently, what we naively thought of as a single QFT may now be a family of their QFTs after specifying their higher symmetries. Higher symmetries can also have 't Hooft anomalies, which obstruct their gauging. Moreover, one can use generalized global symmetries preserved along RG flows to constrain properties of the IR theory (such as their anomalies). 

\textbf{{\Large \section{Gravitational Physics}}}

Returning to our original goal of unifying all fundamental interactions, we have seen up till now that three out of four fundamental interactions can be incorporated by QFT. However, the fourth fundamental interaction, gravity, has many distinctive behaviors compared to the other three. First of all, only gravity would always \textit{accumulate} on a macroscopic scale (as opposed to averaging out as in electromagnetism), and thus eventually becoming the most prominent interaction on astrophysical and cosmological length scales. The contemporary understanding of gravitation is Einstein's general theories of relativity, which view gravity as the interplay between matter and geometry of spacetime. See \cite{misner2017gravitation,wald2010general,carroll2019spacetime} for standard textbooks.
 
More concretely, general relativity re-examined the conception of (infinitesimal) distance in spacetime. Previously in special relativity, spacetime is considered to be static, with the distance $ds$ defined such that $ds^2 = -dt^2 + dx^2 + dy^2 + dz^2$. Equivalently, one could view it as $ds^2 = \eta^{\mu\nu} dx_\mu dx_\nu$ by introducing the metric $\eta^{\mu\nu}$ that is a constant tensor. However, Einstein revolutionized such conception on top of his own work of special relativity by promoting the constant metric $\eta^{\mu\nu}$ to a dynamical tensor field $g^{\mu\nu}$. This brings us to General Relativity (GR).
 
Within GR, the presence of matter field assigns a certain curvature to the spacetime according to Einstein's field equation,
 \begin{equation}
     G_{\mu\nu} = \frac{8\pi G}{c^4} T^{\mu\nu}.
\end{equation}
Namely, the Einstein tensor $G_{\mu\nu}$ is proportional to the energy-momentum tensor, with the numerical coefficient further proportional to Newton's constant $G$ - the coupling constant of gravity. Here, the Einstein tensor is a second-order tensor that solely depends (in a sophisticated way) on the metric tensor $g_{\mu\nu}$ together with its derivative with the spacetime coordinates of first and second order. Very roughly speaking, the Einstein tensor is the ``acceleration" of the metric, whereas the energy-momentum tensor can be viewed as ``force exerted by matter onto the spacetime metric." 

After having understood the dynamical behavior of the metric tensor, the next step is to consider the behavior of probe particles. Such probe particles, whose energy and momentum are, by definition, too small to affect the spacetime metric, would exactly move along the geodesics (which should be viewed as the analog of a ``straight line" in curved spacetime). Taking a weak field limit in GR would recover for us the (previously understood) special theory of relativity.

Early observational verification of General Relativity includes the Mercury precession and deflection of starlight via solar gravitation, both of which took place soon after the discovery of general relativity. Later on, all subsequent experimental and observational pieces of evidence confirm general relativity by increasingly high precision. The most mysterious prediction of general relativity is the existence of black holes - which feature a singularity of spacetime surrounded by an ``event horizon". These objects gained indirect observational evidence in later 20th-century astronomy. A piece of more direct evidence is the recent detection of gravitational waves. Gravitational waves are propagating fluctuations of spacetime metric that can be created when a pair of black holes merge into one. Their detection is made by LIGO (Laser Interferometric Gravitational Observatory) and VIRGO (a European gravitational wave observatory). \footnote{Determination of theoretical predictions for Gravitational Wave signals of black hole merger events involves a strong coupling phenomenon of gravitational fields, where numerical computations are necessary.}

\textbf{\Large \section{String Theory as a Unification}}

After having explained two theoretical frameworks of fundamental interactions: QFT and GR, it is the utmost challenge and desire of theoretical high-energy physicists to consistently combine these two frameworks. Since, after all, there are places in our Universe, such as the event horizon of a black hole, where both quantum physics and strong-field gravitational physics become simultaneously relevant. 

If one attempts to \textit{incorporate} GR into QFT, then one inevitably runs into a large number of infinities which, furthermore, cannot be removed by renormalization. Said differently, it is impossible to construct a quantum field theory that unifies all four fundamental interactions. The most widely-accepted resolution of this issue until now is string theory (see \cite{Green:1987sp,Green:1987mn,Polchinski:1998rq,Polchinski:1998rr,becker_becker_schwarz_2006} for standard textbooks). String theory is a theory that involves spatially 1-dimensional dynamical objects, the fundamental strings. Under a limit where one recovers quantum field theory, different fields should come from different vibration modes of the fundamental strings. In particular, the dynamics on the string worldsheet can be described by the following sigma-model action:
\begin{equation}
    S_\sigma = -\frac{1}{2}T \int d^2 \sigma \sqrt{-h} h^{\alpha\beta} \p_\alpha X \cdot \p_\beta X.
\end{equation}
Here the dynamical metric tensor $h^{\alpha\beta}$ can be pulled back to the metric tensor in the spacetime, which is exactly the main object in general relativity. In addition, the coefficients of different vibration modes of the fundamental string would correspond to fields with different spin values. In this sense, string theory successfully unifies general relativity and quantum field theory into different aspects of a single theory of fundamental strings.

There are many consistency conditions of string theory. To begin with, if we demand having stable string theory vacuum, then the common prescription is to restrict ourselves to supersymmetric string theories (i.e., superstring theories), whose worldsheet contains pair of Bosonic and Fermionic degrees of freedom that are related by supersymmetric transformations.

We notice that consistency (namely conformal anomaly) of the string worldsheet requires that superstring theory be defined in 10 spacetime dimensions. However, superstring theories are not unique, and indeed multiple versions of superstring theories in 10D have been discovered: type I string, type IIA and type IIB string, $E_8 \times E_8$ heterotic string, and $SO(32)$ heterotic string. Several dualities among these string theories are later discovered, each stating that one string theory on a certain type of spacetime background (possibly containing some compact dimensions) is equivalent to another string theory on another type of spacetime background (also possibly containing some compact dimensions). 

\begin{figure}
    \centering
    \includegraphics{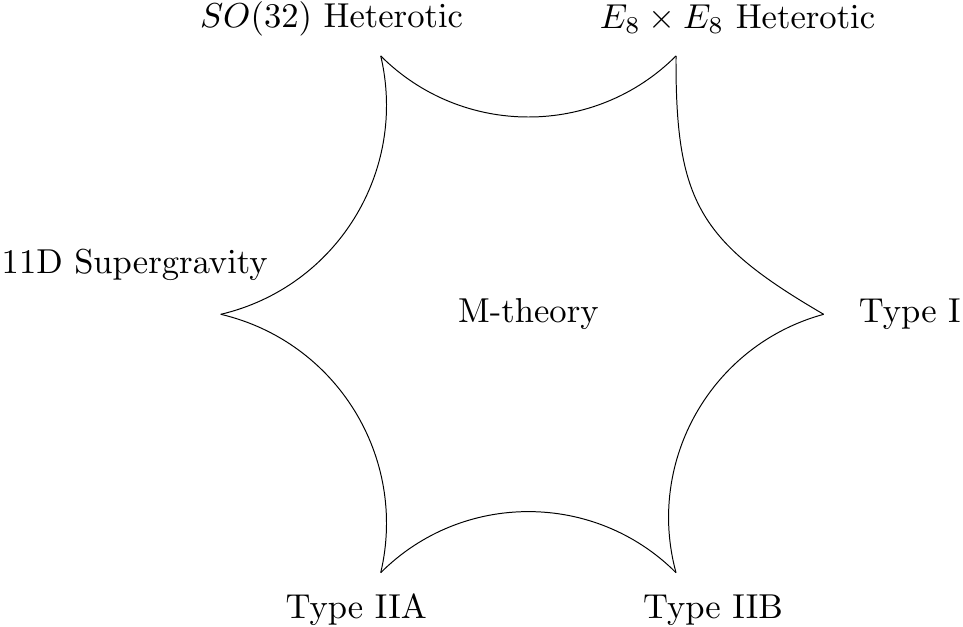}
    \caption{Depiction of all five types of 10D superstring theories and the 11D supergravity as arising from different limits of the (11D) M-theory}
    \label{fig:hexagon}
\end{figure}

Afterward, Witten proposed in 1995 (\cite{Witten:1995ex}) that all of these five string theories (together with a sixth theory of 11-dimensional supergravity) are various limits of a single 11-dimensional strongly-coupled quantum theory, known as M-theory. M-theory is no longer a theory of strings but instead a theory of membranes. The relation of M-theory and all known string theories is often illustrated via the ``hexagon diagram" of string theory, with M-theory living in the bulk, whose different limits to a corner result in either one of the five 10D string theories or the 11D supergravity. See Figure \ref{fig:hexagon} as an illustration.

In string theory, a dimensionful parameter is the string tension $T$, which is sometimes expressed in terms of the Regge slope $\alpha' = \frac{1}{2\pi T}$, or string length $l_s$ such that $l_s^2 = 2\alpha' = \frac{1}{\pi T}$. The only \textit{dimensionless} parameter is $g_s$, the string coupling constant. Therefore, as a quantum theory that incorporates GR and QFT, string theory is highly constrained.

In a small string coupling regime, one can also compute the scattering amplitudes of string theory by Feynman diagrams. In this case, the worldsheet of the interpolating string worldsheet would become a Riemann surface without any singular loci. The genus (number of holes) is now the analog of the number of loops back in Feynman diagrams of quantum field theory! The Feynman diagram of a string theory has a nice property that they are again renormalizable! This property of renormalizability can be intuitively explained by noticing the absence of sharp turning on the smooth Riemann surface. Such a property allows us to claim that string theory is a consistent quantum theory of gravity.

Another feature in string theory is that the spacetime coordinates can be reinterpreted as scalar fields on the worldsheet. For an open string, one can further examine the behavior of such Bosonic fields at the boundary of its worldsheet. For example, one could associate fixed (Dirichlet) or free (Neumann) boundary conditions to each such spacetime coordinate. These boundary conditions are well-understood, at least in a weak coupling regime. Nevertheless, it was later discovered that, in the strong coupling regime, the boundary conditions would themselves (very surprisingly) turn into dynamical objects. Such objects are called D-branes (\cite{Dai:1989ua}, see \cite{Johnson:2003glb} for a standard textbook) since their spacetime spans come from the loci that are associated with Dirichlet boundary condition of open strings. Looking backward, one notices that D-brane are infinitely heavy at weak couplings, which therefore appears to be ``merely a boundary condition". In contrast, at strong couplings, D-branes carry finite mass and thus become dynamical. 

Perturbatively, we know the worldvolume action of the D-brane, but the full non-perturbative description of string theory incorporating D-brane physics remains unknown. Instead, M-theory (at least the controllable aspects of it) is capable of describing several strong coupling phenomena which cannot be accessed via perturbative string theories. In fact, to obtain M-theory, one has to take the strong coupling limit of IIA string theory by sending the only free parameter $g_s$ to infinity, and then the spacetime would grow an extra dimension as a circle. Therefore, M-theory is intrinsically strongly coupled, and it does not have any free parameters.

On the formal side of theoretical high-energy physics, looking for the correct physical principles and their mathematical formulations of M-theory remains a long-term objective.

\textbf{\Large \section{Geometric Engineering of QFT and Quantum Gravity via String Theory}}

After many formal treatments of various string theories and extra spacetime dimensions, we remind the readers that string theory was originally discovered as the theory of quantum gravity, which was, in turn, motivated by the objective of describing Nature. Unfortunately, string theories have to live in 10 dimensions (and M-theory in 11 dimensions), but the physical world is 4-dimensional instead. Therefore, one might wonder how string theory can say anything about reality. 

Nonetheless, it turns out that we can still realize our 4-dimensional physics from 10-dimensional string theory, using \textit{compactification} \cite{Candelas:1985en}. Namely, we consider string theory on $\mathbb{R}^{1, 3} \times M_6$, where $M_6$ is a compact internal manifold - we require it to be a Calabi-Yau manifold to preserve supersymmetry \cite{Argyres:2022mnu}. In this general setup, one can, in principle, find suitable choices of this $M_6$ so that the 4D theory gives rise to a Minimally Supersymmetric Standard Model (MSSM), reproducing the matter spectrum of the standard model of particle physics together with their undiscovered ``superpartners". In this sense, we see that string theory is indeed a framework that is conceptually capable of unifying all four types of fundamental interactions in our 4-dimensional spacetime. \footnote{One can also get supersymmetric theory by putting M-theory on a 7-dimensional manifold with $G_2$ holonomy, see \cite{Atiyah:2001qf}.}

However, at the energy level currently measurable at particle colliders (TeV Scale), string theory does not give rise to any measurable deviation from the combination of general relativity and particle physics. Therefore, one needs to expect that experiments will not be capable of detecting signatures that could prove or disprove string theory until the far future (such as a few centuries later).

Despite the absence of observational evidence, part of formal studies of string theory gradually decouples from the objective of making predictions of real-world phenomena. Instead, such formal studies of string theory are becoming increasingly intriguing and profound research objectives for their own sake. A deep and clear understanding of string theory (or its certain limits that give rise to QFTs) on different spacetime geometries can help us better understand string theory as a whole, especially its strong-coupling behaviors. 

In addition, some aspects of string theory motivate mathematical research in an inspiring way. Some ``physically proven" statements are reformulated into mathematically surprising conjectures, some of which can later be proven at the level of mathematical rigor. Among them, mirror symmetry of Calabi-Yau manifolds is a well-known example. Namely, there are physical pieces of evidence that type IIA string on one Calabi-Yau threefold is equivalent to type IIB string on a ``mirror dual" Calabi-Yau threefold. This condition is later reformulated as ``homological mirror symmetry" by Kontsevich in \cite{Kontsevich:1994dn}, or alternatively via relation to T-duality as in \cite{Strominger:1996it}, and later proven physically in \cite{Hori:2000kt} by applying T-duality to the linear Sigma model (see \cite{Hori:2003ic} for a comprehensive review).

\textbf{Mapping out the String Theory Landscape}


String compactifications sit as the intersection between formal (down to any dimensions) and phenomenological studies (down to 4 dimensions without extended supersymmetry) of string theory. Specifically, dimensionally reducing string theory on a compact $M$ gives rise to a quantum theory of gravity, whereas reducing string theory on a non-compact $M$ gives rise to a quantum field theory. More restricted holonomy on $M$ would lead to more preserved supersymmetry in the lower-dimensional theory. Compactifications of string theory on lower-dimensional internal manifold folds will produce higher-dimensional supergravity theories with a large number of supercharges. Notice that one has fewer choices for a lower-dimensional internal manifold, corresponding to the fact that string-constructed higher-dimensional quantum theories of gravity are tightly constrained. With all possible choices of $M$, the collection of low-energy effective theories is called the String Landscape.

For QFTs and quantum gravities in relatively high dimensions with a sufficient amount of supercharges, we can classify all such theories constructed from string theory \cite{Argyres:2022mnu}. The main idea is to translate the physical constraints in spacetime into geometric or topological constraints of the internal manifold. One important remark is that this type of classification often needs to be supplemented by torsional fluxes of higher-form fields that exist in string theory, producing the so-called frozen singularities. One must consider such a case with particular care to ensure that one eventually reaches a complete classification of all string compactifications in a given dimension and number of supercharges. For example, there is a geometrical classification of 6D superconformal field theories by putting F-theory on non-trivial elliptic fibration over $\mathbb{C}^2/\Gamma_{U(2)}$ (\cite{Heckman:2013pva,Heckman:2015bfa}, see \cite{Heckman:2018jxk} for a review). 
We remark that worldvolume theories of extended objects in string theory on a certain background geometry also lead to geometrically engineered quantum field theories \cite{Katz:1996fh}. For example, in M-theory on $\mathbb{R}^{5, 1} \times \mathbb{R}_{\perp} \times \mathbb{C}^2/\Gamma$, putting some M5 branes along the $\mathbb{R}^{5, 1}$ directions also results in 6D superconformal field theories, which constitute a subset of those that can be engineered in directly compactifying F-theory as mentioned above.

\textbf{\Large \section{String Universality and the Swampland Program}}

In the end, the study of string theory is driven by the theoretical objective that not only is string theory the most studied theory of quantum gravity, but it might also be the only theory that underlies all supergravity theories. Therefore comes the question of whether string theory can be \textit{proven} to be the only consistent quantum theory of gravity and thus the unique underlying physical theory of Nature.

One approach to answering this question is to determine the complete set of consistent QG theories in our 4-dimensional spacetime and to see if they all come from string theory. Quite surprisingly, studying the consistency of QG in more than four dimensions turns out to be an excellent starting point since, to construct them in string theory, the internal manifold has fewer dimensions than 6 (=10-4). In 10 dimensions, it has already been shown that consistent QG all come from string theory. So a natural way to proceed is to gradually go down in dimensionality and then attempt the same set of strategies. Such an endeavor goes under the name of string universality.

String universality (see \cite{Adams:2010zy} and \cite{Kumar:2009us} for early works) precisely fits into the broad paradigm of the Swampland program (see \cite{Palti:2019pca,vanBeest:2022fss} for reviews), which is the quest for the complete set of consistency conditions of quantum theories of gravity. We now have a collection of conjectures called ``Swampland conjectures" with a varying level of rigor and predictive power. 

Among them, a fundamental and widely accepted conjecture is the \textit{no global symmetry conjecture} \cite{Banks:1988yz,Banks:2010zn}, stating that there should not be any global symmetries in a consistent quantum theory of gravity. The following thought experiment can help explain the no global symmetry conjecture. If there were such a global symmetry, one could take an arbitrarily large number of particles charged under this global symmetry and throw it into a black hole. Then, one waits for a long time for the Hawking radiation of this black hole to take place, and one will eventually end up with a remnant with an arbitrarily large charge of this global symmetry, signifying the sickness of this theory. If one appears to have identified a global symmetry in a theory of quantum gravity, then it has to be gauged or broken in a complete description of this theory. We remark that such a global symmetry could be either an ordinary or generalized global symmetry. 

Furthermore, one can take no global symmetry conjecture and examine the particle spectrum of any gauge symmetry. It turns out that for any possible gauge charge of gauge symmetry, there has to be a state carrying that charge, where we allow the state to be either fundamental or compound. Otherwise, there will be a global symmetry which is, in turn, forbidden by the no global symmetry conjecture.

We remark that the no global symmetry conjecture implies that there should be \textit{neither ordinary nor generalized} global symmetries in consistent quantum theories of gravities. From this, one can also derive the cobordism conjecture \cite{McNamara:2019rup}, stating that all quantum theory of gravity must be connected to the ``bubble of nothing" via a domain wall, also called an end-of-the-world brane. More formally, a quantum theory of gravity needs a trivial ``QG bordism group" $\Omega^{\text{QG}} = 0$, where $\Omega^{\text{QG}}$ is defined such that it reduces to the geometric cobordism class of the internal manifold for geometric compactifications of string theory.

String universality (and the Swampland program in general) remains an active area of research in theoretical high-energy physics. Some primary goals include identifying more connections among different Swampland conjectures and, ultimately, looking for fundamental principles that underlie all known Swampland conjectures.

\begin{mainf} 

\part{Generalized Symmetries in Supergravities}

\chapter{STRING UNIVERSALITY AND NON-SIMPLY-CONNECTED GAUGE GROUPS IN 8D}


\section{Introduction}

One of the important lessons from string theory is that consistency conditions of quantum gravity are highly restrictive.
In the low-energy limit, they result in a small and possibly finite subset of effective descriptions, leaving behind a vast ``Swampland'' of seemingly consistent quantum field theories coupled to gravity \cite{Vafa:2005ui}.
Recent attempts to specify the Swampland's boundary (cf.~\cite{Brennan:2017rbf,Palti:2019pca} for reviews) have reinforced the idea of String Universality: every consistent quantum gravity theory is in the string landscape.

Prototypical examples of String Universality appear in eleven and ten dimensions, where low energy limits of M- and string theory give rise to the only consistent supergravity theories.
In ten dimensions (10d), this requires more subtle field theoretic arguments \cite{Adams:2010zy}, or the incorporation of extended dynamical objects in the theory \cite{Kim:2019vuc}, to ``drain'' the 10d supergravity Swampland.

In lower dimensions, one observes a broader spectrum of string-derived supergravity theories, but these nevertheless show some intricate structures not naively expected from field theory considerations.
For example, the rank $r_G$ of the gauge group in known string compactifications is bounded by $r_G \leq 26 - d$ in $d$ dimensions, and satisfies $r_G \equiv 1 \mod 8$ and $r_G \equiv 2 \mod 8$ in $d=9$ and $d=8$, respectively.
Likewise, not all gauge algebras have string realizations.
In particular, there are no string compactifications to 8d with $\mathfrak{so}(2n+1) \, (n\geq 3)$, $\mathfrak{f}_4$ and $\mathfrak{g}_2$.
Again, novel Swampland constraints \cite{Kim:2019ths,Montero:2020icj} and refined anomaly arguments \cite{Garcia-Etxebarria:2017crf} reproduce these restrictions, thus downsizing the 9d and 8d Swampland considerably.\footnote{As $\mathfrak{g}_2$ does not suffer similar anomalies, it remains an open question if it truly belongs to the 8d Swampland.}

The goal of this work is to provide similar constraints for the \emph{global structure} of the gauge group of 8d ${\cal N}=1$ theories, by deriving a field theoretic consistency condition for the gauge group to take the form $G/Z$, with $Z \subset Z(G)$ a discrete subgroup of the center of $G$.
Taking inspiration from F-theory \cite{Vafa:1996xn,Morrison:1996na,Morrison:1996pp}, where the gauge group structure is encoded in the Mordell--Weil group of the elliptically-fibered compactification space \cite{Aspinwall:1998xj,Mayrhofer:2014opa,Cvetic:2017epq}, it appears that the allowed gauge groups $G/Z$ are heavily restricted. 
For example, there are no 8d string compactifications, including constructions beyond F-theory, that have gauge group $SU(n)/\bbZ_n$, whereas $SU(n)$ groups are ubiquitous.

These restrictions are mathematically well known from the classification of elliptic K3 surfaces \cite{MirandaPersson,MirandaPersson_extremal,shimada_k3} (see also \cite{Hajouji:2019vxs}).
Focusing on $G$ a simply-connected non-Abelian Lie group\footnote{
More precisely, the most general gauge group is $\frac{G \times U(1)^r}{Z \times Z_\text{f}}$, with $Z \subset Z(G)$, i.e., $Z \cap U(1)^r = \{ 1 \}$.
In this chapter we consider constraints for $Z$ exclusively, leaving a more detailed study including $Z_\text{f} \subset Z(G \times U(1)^r) \cong Z(G) \times U(1)^r$, based on \cite{Cvetic:2017epq}, for future work. \label{footnote:1}
}, the geometry restricts $Z$; e.g., when $Z \cong \bbZ_\ell$, then $\ell\leq 8$.
Moreover, for each of the cases $\ell=7,8$, there is exactly one elliptic K3 on which F-theory compactifies to an 8d theory with $G = SU(7)^3/\bbZ_7$ and $[SU(8)^2 \times SU(4) \times SU(2)]/\bbZ_8$, respectively.
Analogous restrictions on gauge groups also appear in heterotic compactifications \cite{Font:2020rsk}.

A natural question is, whether these restrictions reflect limitations of string theory, or previously unknown consistency conditions of quantum gravity in 8d.

In this chapter, we show that the latter is the case.
The key is to realize a non-simply-connected group $G/Z$ by gauging the $Z$ 1-form center symmetry \cite{Kapustin:2014gua, Gaiotto:2014kfa}.
Thus, charting the Swampland of gauge groups $G/Z$ (in any dimension) can be equivalently tackled by studying consistency conditions for gauging $Z$ in gravitational theories.
As we will discuss below, in 8d ${\cal N}=1$ theories, one such condition is the absence of a mixed anomaly between the center 1-form symmetries and gauge transformations of higher-form supergravity fields, which would obstruct the gauging of $Z$.
This rules out a vast set of seemingly acceptable 8d ${\cal N}=1$ theories without known string constructions, and, in particular, reproduce the geometric restrictions in models with F-theory realization, thus providing further evidence for String Universality in 8d.

The anomaly originates from a generalization of the familiar $\theta$-term, $\theta \, \text{Tr}(F^2)$, in 4d.
There, the fractional shift of the instanton density $\text{Tr}(F^2)$, due to the presence of a background field for the $Z$ 1-form symmetry, breaks the $2\pi$-periodicity of $\theta$ \cite{Kapustin:2014gua,Gaiotto:2014kfa,Gaiotto:2017yup,Gaiotto:2017tne,Cordova:2019jnf,Cordova:2019uob,Cordova:2019bsd}.
In higher dimensions, $\text{Tr}(F^2)$ couples to higher-form fields (e.g., to vector fields in 5d and tensors in 6d), which themselves possess gauge symmetries.
These can lead to mixed anomalies with the $Z$ 1-form center symmetry \cite{Apruzzi:2020zot, BenettiGenolini:2020doj}.\footnote{See also \cite{Morrison:2020ool, Albertini:2020mdx,Closset:2020scj,DelZotto:2020esg,Bhardwaj:2020phs} for recent treatments of higher-form symmetries in higher-dimensional setups and \cite{Dierigl:2020myk} for an analysis of the global gauge group in 6d SCFTs.}

The analogous coupling in 8d involves a 4-form $B_4$.
Crucially, while such a term is absent in a pure 8d supersymmetric gauge theory (as there are no appropriate fields $B_4$ in the ${\cal N}=1$ vector multiplet), the coupling $\sum_i m_i B_4 \wedge \text{Tr}(F_i^2)$ necessarily exists if one includes a gravity multiplet, which contains a unique tensor $B_2$ that is dual to $B_4$ \cite{Awada:1985ag}.
Supersymmetry further demands that $m_i \neq 0$, \cite{Salam:1985ns}.
A mixed anomaly involving the symmetries of $B_4$, which must be gauged, and the center 1-form symmetry $Z$ can, therefore, obstruct the gauging of the latter.
The vanishing of this anomaly is, hence, a \emph{necessary} condition to obtain a non-simply-connected gauge group $G/Z$.
Remarkably, in models with $m_i=1$ this condition turns out to reproduce geometric properties of elliptic K3 manifolds!
In combination with other Swampland criteria that constrain the coefficients $m_i$, this anomaly restricts possible combinations of simply-connected $G = \prod_i G_i$ and $Z \subset Z(G)$ in 8d. 
With this, we can consequently ``drain'' large portions of the 8d Swampland, and make predictions in corners of theory space where the global gauge group structure in corresponding string models is yet to be explored.

\section{Mixed anomaly for center symmetries in 8d supergravity}
\label{sec:anomaly}

Let $G = \prod_i G_i$ be a non-Abelian group, where $G_i$ are simple simply-connected Lie groups with algebra $\mathfrak{g}_i$.
In 8d ${\cal N}=1$, the gauge potential $A_i$, with field strength $F_i$, of the $\mathfrak{g}_i$ gauge symmetry comes in a vector multiplet with adjoint fermions.
There are no other massless charged matter states, so at low energies one expects a discrete $Z(G) = \prod_i Z(G_i)$ 1-form symmetry \cite{Gaiotto:2014kfa}.
Moreover, since the only massless fermions transform in a real representation, there are no pure gauge anomalies \cite{Taylor:2011wt}.

Besides the vector multiplets, 8d ${\cal N}=1$ supergravity contains the gravity multiplet with a 2-form gauge field $B_2$ as one of its component fields \cite{Salam:1985ns}. 
The field strength $H_3$ of this 2-form field obeys a modified Bianchi identity involving the gauge fields of the theory,
\begin{align}
H_3 = d B_2 + \sum_{i} m_i \, \text{CS}(A_i) \,.
\end{align}
Here, $\text{CS}(A_i)$ are the Chern--Simons functionals for the gauge factor $G_i$.

The positive integers $m_i$ associated with each gauge factor $G_i$, which we will refer to as the ``level'' of $G_i$, are a priori free parameters of the supergravity theory.
They can be interpreted as the magnetic charge of gauge instantons under $B_2$ --- more apparent in the dual formulation, with $B_2$ replaced by its magnetic-dual 4-form $B_4$. The most general Lagrangian contains the coupling \cite{Awada:1985ag}
\begin{align}
	\int_{M_8} \sum_i B_4 \wedge m_i \, \text{Tr}(F_i \wedge F_i) =: \int_{M_8} \sum_i B_4 \wedge m_i \, I_4(G_i)\, ,
	\label{eq:anomcoup}
\end{align}
where the trace is normalized such that the instanton density $I_4(G_i) = 1$ for a one-instanton configuration of a $G_i$-bundle on a 4-manifold $M_4$. 

The center 1-form symmetry of $G_i$ can be coupled to a 2-form background field $C_2^{(i)}$ which takes values in $Z(G_i)$.
When $C_2^{(i)}$ is non-trivial, it twists the $G_i$-bundle into a $G_i/Z(G_i)$-bundle with second Stiefel--Whitney class $w_2(G_i/Z(G_i)) = C_2^{(i)}$ \cite{Kapustin:2014gua,Gaiotto:2014kfa} that contributes to \eqref{eq:anomcoup},
\begin{align}
I_4(G_i/Z(G_i)) \equiv \alpha_{G_i} \mathfrak{P}\big(C^{(i)}_2\big) \mod \bbZ \, ,
\end{align}
with $\mathfrak{P}$ the Pontryagin square.
This contribution is, in general, fractional due to the coefficients $\alpha_{G_i}$ derived in \cite{Cordova:2019uob}, which we reproduce here:
\begin{center}
\renewcommand*{\arraystretch}{1.4}
\begin{tabular}{| c | c | c |}
\hline
$G_i$ & $Z(G_i)$ & $\alpha_{G_i}$ \\ \hline \hline
$SU(n)$ & $\mathbb{Z}_n$ & $\tfrac{n-1}{2n}$ \\ \hline
$Sp(n)$ & $\mathbb{Z}_2$ & $\tfrac{n}{4}$ \\ \hline
$Spin(2n+1)$ & $\mathbb{Z}_2$ & $\tfrac{1}{2}$ \\ \hline
$Spin(4n+2)$ & $\mathbb{Z}_4$ & $\tfrac{2n+1}{8}$ \\ \hline
$Spin(4n)$ & $\mathbb{Z}_2^{(L)} \times \mathbb{Z}_2^{(R)}$ & $\big( \tfrac{n}{4}, \tfrac{1}{2}\big)$ \\ \hline
$E_6$ & $\mathbb{Z}_3$ & $\tfrac{2}{3}$ \\ \hline
$E_7$ & $\mathbb{Z}_2$ & $\tfrac{3}{4}$ \\ \hline
\end{tabular}
\end{center}

Analogous to the situation in 6d \cite{Apruzzi:2020zot}, the coupling \eqref{eq:anomcoup} combines the fractional instanton configuration with a large gauge transformation $B_4 \rightarrow B_4 + b_4$, with $b_4$ a closed 4-form with integer periods, into a phase $2\pi i {\cal A}(b_4, C^{(i)}_2)$ for the partition function
\begin{align}\label{eq:anomaly}
	{\cal A}(b_4, C^{(i)}_2) = \sum_i m_i \alpha_{G_i} \int_{M_8} b_4 \cup \mathfrak{P}(C^{(i)}_2) \, .
\end{align}
While $\int_{M_8} b_4 \cup \mathfrak{P}(C^{(i)}_2) \in \bbZ$ for arbitrary $b_4$, the whole expression is, in general, fractional due to $\alpha_{G_i}$.
By generalizing the arguments presented in \cite{Hsieh:2020jpj,Apruzzi:2020zot}, the electrically charged objects for $B_4$ would acquire a fractional charge if this anomalous phase is non-trivial.
Since this violates charge quantization, the fractional shift \eqref{eq:anomaly} cannot be compensated and can be understood as an anomaly between the large gauge transformations of $B_4$ and the center 1-form symmetries.
As the former symmetry is gauged, one cannot allow for background fields $C_2^{(i)}$ where \eqref{eq:anomaly} is non-trivial.
Similar to the 6d setting \cite{Apruzzi:2020zot}, we expect that the violation of charge quantization is tied to the lack of counterterms that could absorb this anomaly. 
Moreover, we expect that arguments developed in \cite{Cordova:2019bsd} suggest that there cannot be a topological Green--Schwarz mechanism that cancels the above anomaly.\footnote{
Note that \cite{Cordova:2019bsd} discusses precisely the 4d analog of the anomaly \eqref{eq:anomaly} involving the $\theta$-angle instead of $B_4$.}

In general, while the individual centers $Z(G_i)$ are anomalous, there can be a non-trivial subgroup $Z \subset \prod_i Z(G_i)$ that is anomaly-free.
Assuming that there are no other obstructions to switch on a background for this subgroup $Z$ of the center, or other breaking mechanisms, this combination should be gauged, in line with common lore that in quantum theories of gravity no global symmetries (including discrete and higher-form symmetries) are allowed \cite{Brennan:2017rbf,Palti:2019pca,Harlow:2018tng}.
In turn, this leads to the gauge group $G/Z$.

\subsection{Condition for Anomaly-Free Center Symmetries}

In the following, we will discuss how to determine subgroups $\bbZ_\ell \cong Z \subset Z(G)$, for which a 1-form symmetry background has no fractional contribution \eqref{eq:anomaly} --- a necessary condition to gauge $Z$.

Let $Z(G) = \prod_{i=1}^s \bbZ_{n_i}$, and $(k_1,...,k_s) \in \prod_{i=1}^s \bbZ_{n_i}$ be the generator for $Z \cong \bbZ_\ell$.
This means that $\ell$ is the smallest integer such that $k_i \ell \equiv 0 \mod n_i$ for all $i$.
The generic background for the $Z(G)$ 1-form symmetry consists of fields $C_2^{(i)}$ for each $\bbZ_{n_i}$ factor of $Z(G)$.
Specifying a specific background for a subgroup then amounts to correlating the a priori independent $C_2^{(i)}$'s \cite{Cordova:2019uob}.
In particular, the background $C_2$ for $Z \cong \bbZ_\ell$ corresponds to setting $C_2^{(i)} = k_i C_2$.

For concreteness, let $G = \prod_{i=1}^s SU(n_i)$.
Then, the anomalous phase \eqref{eq:anomaly} in a non-trivial $C_2$ background of the subgroup $Z \subset Z(G)$ is
\begin{align}
\mathcal{A}(b_4, C_{2}^{(i)}) 
	= \,  \left( \sum_{i=1}^s \frac{n_i - 1}{2n_i} k_i^2 m_i \right) \int_{M_8} b_4 \cup \mathfrak{P}(C_2) \, ,
\end{align}
where we used $\mathfrak{P}(k C) = k^2 \mathfrak{P}(C)$.
Thus, the anomaly vanishes if the coefficient is integral.

Note that the anomaly contribution of non-$SU$ groups can be written as a sum of contributions from $SU(n)$-subgroups \cite{Cordova:2019uob}.
Therefore, by further restricting ourselves to rank$(G)\leq 18$ (which is the 8d bound for the total gauge rank \cite{Montero:2020icj}), we can exhaustively scan for all possible groups $G$ that have an anomaly-free $\bbZ_\ell \subset Z(G)$ with given $\ell$, by finding $s$ triples of integers $(n_i, k_i, m_i)$ such that
\begin{align}\label{eq:anomaly_product_of_su}
	\sum_{i=1}^s \frac{n_i - 1}{2n_i} \, k_i^2 m_i \in \bbZ \, , \quad \text{with} \quad k_i \cdot \ell \equiv 0 \mod n_i \, .
\end{align}

Clearly, the levels $m_i$ play an important role.
From an effective field theory perspective, these are free parameters that define the theory. However, these parameters themselves are constrained by Swampland criteria.
By the Completeness Hypothesis \cite{Polchinski:2003bq,Banks:2010zn}, the 2-form field $B_2$ couples to strings which carry localized degrees of freedom sensitive to the gauge group. 
These left-moving, charged excitations on the string have to cancel the worldvolume anomalies arising due to anomaly inflow \cite{Kim:2019vuc, Katz:2020ewz}. However, in $d$ dimension the left-moving central charge for such a string is bounded by $c_L \leq 26 - d$. 
While each $U(1)$ gauge factor contributes to $c_L$ with $c_{U(1)}=1$, each non-Abelian simple gauge factor $G_i$ with level $m_i$ contributes $c_i = \tfrac{m_i \, \text{dim}(G_i)}{m_i + h_i}$, with $h_i$ the dual Coxeter number of $G_i$.
Hence, we have
\begin{align}
\sum_{i} \frac{m_i \, \text{dim}(G_i)}{m_i + h_i} + n_{U(1)} \leq 18 \, .
\label{eq:anominfl}
\end{align}
Combined with the constraint that the rank of the total gauge group of the 8d supergravity theory can be only 2, 10, or 18 \cite{Montero:2020icj}, the $m_i$ are considerably restricted.
In particular, it is easily shown that in the rank-18 case, all $m_i$ must be 1 and all non-Abelian factors must have simply-laced algebras (see appendix \ref{app:m} for more details).
This is well-known in string compactifications, where $m_i$ are the levels of the worldsheet current algebra realizations of spacetime gauge groups, and are all $m_i=1$ on the rank-18 branch of the ${\cal N}=1$ moduli space.
As we will see now, the anomaly matches known geometric limitations in the F-theory realization of 8d rank-18 theories, which restricts the possible global gauge group structures. 
In the lower-rank cases these conditions can constrain gauge groups, whose algebras and levels fit in constructions such as the CHL string \cite{Chaudhuri:1995fk}, but whose global structure is yet to be explored.

\subsection{Anomaly-Free Centers in Theories of Rank 18}
\label{sec:singlefac}

All rank-18 ${\cal N}=1$ supergravity theories with a known string origin have a construction via F-theory \cite{Vafa:1996xn}, where physical features, including the global gauge group structure, are encoded in the geometry of elliptically-fibered K3 surfaces \cite{Taylor:2011wt,Weigand:2018rez,Cvetic:2018bni}.
In particular, there are beautiful arithmetic results \cite{MirandaPersson,MirandaPersson_extremal} which asserts that F-theory compactifications with non-Abelian gauge group $G/Z$, where $G$ consists only of $SU(n_i)$ factors, must satisfy 
\begin{align}\label{eq:anomaly_cancellation_rank18}
	\sum_{i=1}^s \frac{n_i -1}{2n_i} k_i^2 \equiv 0 \mod \bbZ \,.
\end{align}
with $(k_1,...,k_s) \in \prod_i Z(SU(n_i))$ the generator of any $\bbZ_\ell \subset Z$ subgroup.

While we defer a more detailed explanation of the geometric origin to this formula to appendix \ref{secapdx:geometry}, it is obvious that it fully agrees with the cancellation condition for every $\bbZ_\ell$ subgroup of the center 1-form symmetry \eqref{eq:anomaly_product_of_su}, as for rank-18 theories all levels are fixed to $m_i = 1$. 
We therefore find a deep connection between the mixed anomaly of the supergravity theory and the geometrical properties of F-theory compactifications.

The constraint is particularly powerful when the order $\ell$ of the gauged center subgroup is the power of a prime number.
For such $\ell \geq 9$, one can show that there are no possible sets $\{(n_i,k_i)\}$ for which the anomaly vanishes with gauge groups of rank $\leq 18$.
For $\ell=7$, there is exactly one configuration with three simple non-Abelian factors, $n_1 = n_2 = n_3 = 7$ and $(k_1,k_2,k_3) = (1,2,3)$, corresponding to an $SU(7)^3/ \bbZ_7$ theory.
This agrees with the classifications of K3 surfaces \cite{MirandaPersson_extremal} for F-theory constructions as well as possible heterotic realizations \cite{Font:2020rsk}.
Likewise, in the case $\ell = 8 =2^3$, the 1-form anomaly \eqref{eq:anomaly_cancellation_rank18} allows for only $G = SU(8)_1 \times SU(8)_2 \times SU(4) \times SU(2)$, into which the $\bbZ_8$ sub-center embeds as $(k_1, k_2, k_{SU(4)}, k_{SU(2)}) =(1,3,1,1)$.
Furthermore, if we also take inspiration from geometric properties of K3 surfaces --- there always is one $SU(n_i)$ factor with $\ell$ dividing $n_j$ --- we can show that there are \emph{no} possible configurations $(n_i,k_i)$ for all $\ell \geq 10$.
This also matches the dual heterotic constructions \cite{Font:2020rsk}.

\subsection{Predictions for Simple Groups}

To further showcase the constraining power of the field-theoretic anomaly argument, we use \eqref{eq:anomaly} to rule out 8d ${\cal N}=1$ theories with gauge group $G/Z$, where $G$ is a simple Lie group and $Z\subset Z(G)$ a non-trivial subgroup.
For $G$ with $m=1$ and rank$(G) \leq 18$, any $G/Z$ is inconsistent except 
\begin{align}
		& \frac{SU(16)}{\bbZ_2} \, , \quad \frac{SU(18)}{\bbZ_3} \, , \quad  \frac{Spin(32)}{\bbZ_2} \, , \label{eq:allowed_groups1}  \\
		& \frac{Sp(4)}{\bbZ_2} \, , \quad \frac{Sp(8)}{\bbZ_2} \, , \quad \frac{SU(8)}{\bbZ_2} \, , \quad \frac{SU(9)}{\bbZ_3} \, , \label{eq:allowed_groups2}  \\
		& \frac{Spin(16)}{\bbZ_2} \, , \quad \frac{Sp(12)}{\bbZ_2} \, , \quad \frac{Sp(16)}{\bbZ_2} \, . \label{eq:allowed_groups3}
\end{align}
The groups \eqref{eq:allowed_groups1} indeed correspond to the only cases with simple $G$ realizable via F-theory on elliptic K3's.
The groups in \eqref{eq:allowed_groups2} are subgroups of $Sp(10)$, which at $m=1$ can be constructed from the CHL string \cite{Chaudhuri:1995fk}.
Note that this rules out all other $Sp(k)/\bbZ_2 (k<10)$ theories, which seemed consistent based on the perturbative CHL spectrum \cite{Mikhailov:1998si}.
As we are not aware of any systematic study of the global gauge group structure in CHL compactifications, we view this as a prediction based on the 1-form anomaly \eqref{eq:anomaly}, which is also consistent with other Swampland arguments \cite{Montero:2020icj}.
Groups in \eqref{eq:allowed_groups3} have no known 8d string realization at $m=1$.
However, while $Sp(12)/\bbZ_2$ and $Sp(16)/\bbZ_2$ are excluded at any $m$ due to the bound \eqref{eq:anominfl}\footnote{In particular, \eqref{eq:anominfl} provides a physical explanation to the limitation $k \leq 10$ for $\mathfrak{sp}(k)$ gauge algebras known in 8d string constructions.}, $Spin(16)/\bbZ_2$ does arise at $m=2$ as a Wilson line reduction of the $E_8$ CHL string.

More generally, at level $m=2$, the center anomaly in conjunction with the bound \eqref{eq:anominfl} can rule out all $G/Z$ theories with simple $G$ except for
\begin{align}
	\begin{split}
		& \frac{SU(4)}{\bbZ_2} \, , \quad \frac{SU(8)}{\bbZ_2} \, , \quad \frac{SU(9)}{\bbZ_3} \, ,\quad \frac{Sp(2)}{\bbZ_2} \, , \quad \frac{Sp(4)}{\bbZ_2} \, ,  \\
		& \frac{Spin(8)}{\bbZ_2} \, , \quad \frac{Spin(16)}{\bbZ_2} \, , \quad SO(2n) \quad \text{with} \quad 2 \leq n \leq 9 \, ,
	\end{split}
\end{align}
all of which could, in principle, arise in CHL compactifications \cite{Mikhailov:1998si}.
We will leave an explicit verification and analysis of the global gauge group in these types of 8d string models for future works.
Note that $SO(2n)$ ($n$ odd) and $Sp(2)/\bbZ_2$ seem to be ruled out in 8d by independent Swampland arguments \cite{Montero:2020icj}, indicating mechanisms beyond the anomaly \eqref{eq:anomaly} that break the 1-form center symmetry.
It would be interesting to find an explicit description for these breaking mechanisms.

\section{Discussion and Outlook}
\label{sec:discussion}

Using a mixed anomaly \eqref{eq:anomaly}, we have presented a necessary condition for an 8d ${\cal N}=1$ theory with given non-Abelian gauge algebras $\mathfrak{g}_i$ at level $m_i$ to have a non-simply-connected gauge group $[\prod_i G_i]/Z$.
In combination with a set of Swampland criteria that restrict the gauge rank and the levels $m_i$, this condition rules out a vast set of possible gauge groups for 8d theories.
The constraints are especially powerful for theories of rank 18, where they reduce to known geometric properties of elliptic K3 surfaces.
As these properties control the global gauge group structure in F-theory compactifications, the anomaly provides a purely physical explanation for the intricate patterns of realizable gauge groups in F-theory.
The anomaly can further make predictions for inconsistent models in lower-rank cases, where the global gauge group structure in the corresponding string compactifications is yet to be explored systematically.

We stress that the absence of the anomaly \eqref{eq:anomaly} is only a necessary, but not sufficient condition for the gauge group to be $G/Z$.
Indeed, for F-theory constructions of the non-simply-connected gauge groups \eqref{eq:allowed_groups1}, there also exist K3 surfaces that realize the simply-connected versions in F-theory \cite{shimada_k3}.
There are also other instances where both $G$ and $G/Z$ are realized in different compactifications; this is also confirmed in the heterotic picture \cite{Font:2020rsk}.
As the center $Z$ in all these cases is non-anomalous, this is consistent with our findings.
At the same time, it is pointing toward additional breaking mechanisms, e.g., in terms of massive states charged under $Z$.
It would be interesting to investigate if these mechanisms are captured by an effective description involving the 1-form center symmetry.

There are also non-anomalous cases that have no realization in known classes of 8d string models.
A particular set of such cases are products $\frac{G_1}{Z_1} \times \frac{G_2}{Z_2}$ of anomaly-free factors, which would again be anomaly-free.
For example, the anomaly-free gauge group $[SU(5)^2/\bbZ_5] \times [SU(2)^4/\bbZ_2] = [SU(5)^2 \times SU(2)^4]/\bbZ_{10}$ as the non-Abelian part of a rank-18 theory (and, thus, all $m_i=1$) has no string realization.
As we have mentioned above, the F-theory geometry would forbid this case, because there is no $SU(n)$ factor with 10 dividing $n$.
Currently, we do not know an adequate physical argument providing the same restriction.
In terms of identifying gauged $\bbZ_\ell$ center symmetries, one plausible possibility is the existence of some mechanism that forces the presence of a $U(1)$ gauge factor into which, similarly to the hypercharge in the Standard Model, that $\bbZ_\ell$ embeds.
Such a theory would not be in contradiction to F-theory models, as center symmetries embedded in $U(1)$s have a different geometric origin \cite{Cvetic:2017epq} (see \cite{Cvetic:2015txa,Cvetic:2018ryq,Cvetic:2019gnh,Cvetic:2020fkd} for direct implications for 4d particle physics models) not subject to the restriction \eqref{eq:anomaly_cancellation_rank18}.
Moreover, in 8d F-theory, there are additional sources for $U(1)$ factors (harmonic $(1,1)$-forms on K3's that are not algebraic), whose center-mixing with non-Abelian gauge factors needs further investigation.
To complete the geometric picture from the field-theoretic side, one must also extend the discussion of anomalies to include $U(1)$ gauge sectors, which we defer to future studies.

We further suspect that other discrete symmetries of the theory can interact non-trivially with 1-form center symmetries, leading to further constraints on the gauge group structure.
For example, it has been pointed out \cite{deBoer:2001wca} that the gauge symmetry of the $E_8 \times E_8$ heterotic string should be augmented by an outer automorphism $\bbZ_2$ exchanging the $E_8$ factors, so that the gauge group is $(E_8 \times E_8) \rtimes \bbZ_2$.
In fact, the 9d CHL string arises as the $S^1$-reduction with holonomies in this $\bbZ_2$.
Such an identification would also be possible for, e.g., $[SU(2)^4 / \bbZ_2] \times [SU(2)^4 / \bbZ_2]$, all at $m_i = 1$, which in 8d is free of the anomaly \eqref{eq:anomaly}, but not realized in terms of a string compactification.
If one could establish other field theory / Swampland arguments for why the $\bbZ_2$ outer automorphism must be gauged in this case, there could be other mixed anomalies involving the 1-form symmetries such that only a diagonal $\bbZ_2$ center survives, leading to the realizable $[SU(2)^8]/\bbZ_2$ theory.

Finally, to fully classify the global gauge group structure in 8d ${\cal N}=1$ theories based on the 1-form anomaly \eqref{eq:anomaly}, it will be important to have more stringent constraints on the possible levels $m_i$ for given simple gauge factors $G_i$.
While for rank-18 theories, \eqref{eq:anominfl} fixes all $m_i=1$, they cannot be fully determined by this method alone for rank-10 or -2 theories, and will require new tools and concepts to predict these independently from concrete string realizations.
Perhaps, new ideas can arise by establishing a connection between higher-form anomalies and the Swampland ideas \cite{Montero:2020icj} that also rule out certain non-simply-connected gauge groups.
These insights can hopefully lead to a complete understanding of the global gauge group structure, and prove full String Universality for non-simply-connected groups in 8d.


\chapter{HIGHER-FORM SYMMETRIES AND THEIR ANOMALIES IN M-/F-THEORY DUALITY}

\section{Introduction and Summary}

Geometric engineering provides a powerful framework to study quantum field theories and their non-perturbative aspects.
Exploiting known features of a higher-dimensional theory on spacetime ${\cal M}$, one can uncover details of lower dimensional field and gravitational theories on $M$ by ``engineering'' a compactification ${\cal M} = M_D \times Y_d$ on an internal space $Y_d$.
In this approach, physical data of the $D$-dimensional theory on spacetime $M_D$ are mapped, using a ``dictionary'' specific to the theory on ${\cal M}$, onto properties of the internal space $Y_d$.\footnote{Throughout this chapter, we denote by $d$ the dimension over $\mathbb{R}$ of the internal space $Y_d$.}
$Y_d$ can then be studied using geometric tools which are not necessarily bound by limitations such as perturbative control.
The success of this process clearly hinges on the ``completeness'' of this dictionary, i.e., our ability to identify the relevant geometric structures associated to a particular physical aspect.

One such aspect is the set of generalized, or higher-form symmetries \cite{Gaiotto:2014kfa} of a quantum field theory.
Formulating their corresponding ``dictionary entries'' in various compactification scenarios in string theory has attracted a lot of recent attention \cite{DelZotto:2015isa,Garcia-Etxebarria:2019cnb,Dierigl:2020myk,Morrison:2020ool,Albertini:2020mdx,Bah:2020uev,Closset:2020scj,DelZotto:2020esg,Bhardwaj:2020phs,DelZotto:2020sop,Bhardwaj:2021pfz,Apruzzi:2021phx,Apruzzi:2021vcu, Hosseini:2021ged}.
In this chapter, we extend the discussion to compactifications of F- and M-theory on elliptically-fibered Calabi--Yau two- and three-folds $Y_d \stackrel{\pi}{\rightarrow} \mathfrak{B}_{d-2}$ \cite{Vafa:1996xn}.

We will focus on discrete 1-form symmetries $\Gamma$ that arise as the center symmetry of non-Abelian gauge dynamics, and whose gauging enforces the non-trivial gauge group topology $G = G_\text{sc} / \Gamma$, where $G_\text{sc}$ is the simply-connected gauge group with center $Z(G_\text{sc}) \supset \Gamma$ \cite{Gaiotto:2014kfa}.
For $\mathfrak{B}$ a compact base (and hence with gravity dynamical), the latter has a characterization in terms of the Mordell--Weil group of $\pi: Y \rightarrow \mathfrak{B}$ \cite{Aspinwall:1998xj,Mayrhofer:2014opa,Cvetic:2017epq}, or string junctions on $\mathfrak{B}$ \cite{Fukae:1999zs,Guralnik:2001jh}.
Using M-/F-duality, we will establish the explicit connection of these descriptions to the asymptotic $G_4$-fluxes that encode the 1-form symmetries in M-theory compactified on $Y$ \cite{Morrison:2020ool,Albertini:2020mdx}.

We will approach this by inspecting the local geometry defining a non-Abelian gauge algebra $\mathfrak{g}$ associated to a simply-connected group $G_\text{sc}$.
That is, $Y \rightarrow \mathfrak{B}$ is a non-compact fibration, with singular fibers realizing $\mathfrak{g}$ in F-theory.
Reducing the theory on an $S^1$ yields M-theory compactified on $Y$. 
The topology of the asymptotic boundary $\partial Y$ --- which encodes the asymptotic fluxes, and thus the 1-form symmetry in M-theory on $Y$ --- is determined by the $SL(2,\bbZ)$ monodromy around the singular fibers in the bulk of $Y$, which in turn is related to the (local) Mordell--Weil group as well as asymptotic string junctions.
Physically, this identifies the possible line operators charged under the 1-form symmetry from M2-branes wrapping non-compact 2-cycles in M-theory, with asymptotic $(p,q)$-string states in F-theory.
A valid compact model can be understood as ``gluing together'' several local patches along their boundaries.
In general, only a subgroup of the 1-form symmetry in each patch will survive, since new massive states break part of the 1-form symmetry explicitly.
The ``compatible'' boundary fluxes are then exactly captured by the global sections of the glued geometry.
Moreover, the cycles associated to the global sections of the geometry become compact, and one has to sum over the distinct flux configurations representing the modified gauge backgrounds of non-simply connected gauge groups. At the same time, the now dynamical magnetic states break the dual $(D-3)$-form symmetry explicitly.

An interesting aspect of 1-form symmetries is their 't Hooft anomalies.
Specifically, for center 1-form symmetries of gauge theories in spacetime dimension $D > 4$, there is a potential mixed anomaly involving the $(D-5)$-form instanton $U(1)$ symmetry \cite{Apruzzi:2020zot,Cvetic:2020kuw,BenettiGenolini:2020doj}.
This anomaly is a consequence of the fractionalization of the instanton number in the presence of a non-trivial background field for the 1-form center symmetry \cite{Gaiotto:2014kfa,Kapustin:2014gua,Gaiotto:2017yup,Cordova:2019uob}.
In ${\cal N}=1$ compactifications of M-theory to $D=7$ and $D=5$, we show that this anomaly arises from the reduction of the eleven-dimensional (11d) Chern--Simons term in the presence of asymptotic fluxes for $G_4$:
By expressing the boundary contributions as a \emph{fractional} linear combination of compactly supported fluxes, we derive the fractional instanton shift on the Coulomb branch of the effective gauge theory.
We demonstrate that, in case the compactification space $Y$ is elliptically fibered, this anomaly matches that of the six- or eight-dimensional (6d/8d) F-theory compactification \cite{Apruzzi:2020zot,Cvetic:2020kuw}.
Intriguingly, the M-theory computation reveals a mixed anomaly between two 1-form symmetries in 5d, which uplifts to a mixed anomaly between the 6d center symmetry, and a discrete 2-form symmetry of instanton strings \cite{DelZotto:2015isa}.
Moreover, we find for 5d gauge theories with a genuine 5d UV fixed point, that the geometrically determined instanton shift deviates from the value naively expected from the effective gauge description.
This indicates a non-perturbative correction to the 't Hooft anomaly from the superconformal dynamics at the UV fixed point, which would be interesting to scrutinize in the future.
It is important to point out that there can potentially be counterterms, e.g., from topological sectors, cancelling these anomalies field-theoretically, which in M-theory compactifications are not arising from the 11d Chern--Simons term.
We refer to a recent work \cite{Apruzzi:2021vcu} where an example of such a topological sector is discussed in the context of M-theory engineering of 5d SCFTs .

The rest of the chapter is organized as follows.
In Section \ref{sec:3app}, we study the higher-form symmetries of M-theory on elliptically fibered Calabi--Yaus in the framework put forward in \cite{Morrison:2020ool,Albertini:2020mdx}.
In Section \ref{sec:geometry}, we then compare the results with known characterizations of the gauge group in F-theory via the Mordell--Weil group \cite{Aspinwall:1998xj,Mayrhofer:2014opa} and string junctions \cite{Fukae:1999zs,Guralnik:2001jh}.
For simplicity, we focus mostly on F-/M-theory compactifications to 8d/7d, where the correspondences between these different approaches can be made concrete.
In Section \ref{sec:anomalies}, we analyze the dimensional reduction of the M-theory Chern--Simons term with boundary fluxes that parametrize the center 1-form symmetry of gauge theories, and derive their 't Hooft anomalies associated with instanton fractionalization for M-/F-theory compactifications to 7d/8d as well as 5d/6d.
Some computational details are collected in the appendices.

\section{Center Symmetries of M-theory on Elliptic Fibrations}
\label{sec:3app}

Yang--Mills theories with a fixed non-Abelian gauge algebra $\mathfrak{g}$ can have different topologies for its gauge group $G$, which generally takes the form
\begin{align}
G = \frac{G_{\text{sc}}}{\mathcal{Z}} \,.
\label{eq:modgauge}
\end{align}
Here, $\mathcal{Z}$ is a subgroup of the center $Z(G_{\text{sc}})$ of the simply-connected group $G_{\text{sc}}$ associated to $\mathfrak{g}$.
In the context of generalized global symmetries \cite{Gaiotto:2014kfa}, the non-trivial global structure \eqref{eq:modgauge} arises from gauging the subgroup ${\cal Z}$ of the global $Z(G_\text{sc})$ 1-form symmetry, which act on electric (Wilson) line charges of $G_\text{sc}$.
The presence of dynamical charged particles in representations ${\bf R}_i$, which in general do not need to be massless, explicitly breaks the center 1-form symmetry to the subgroup of $Z(G_\text{sc})$ that leaves all ${\bf R}_i$ invariant.
This happens due to the fact that the objects charged under the electric 1-form symmetries, i.e., Wilson line operators, can end on these charged particle states, and cease to define 1-form charges.

In $D$ spacetime dimensions, a $\mathfrak{g}$ gauge algebra also has a dual magnetic $Z(G_\text{sc})$ $(D-3)$-form symmetry, which acts on magnetically charged objects.
There is a mixed 't Hooft anomaly between the electric and magnetic higher-form symmetries, which form a so-called ``defect group structure'' \cite{Freed:2006ya,Freed:2006yc,DelZotto:2015isa}, and which is a generalization of Dirac's quantization condition for electric and magnetic charges.
In terms of the defect group, the global form \eqref{eq:modgauge} can also be understood as a choice of ${\cal Z} = \pi_1(G)$ magnetic higher-form symmetry, together with a ``mutually-local'' electric 1-form symmetry.
That is, the electric flux operators present in the $G = G_\text{sc}/{\cal Z}$ theory have integral pairing under the defect group pairing with the magnetic flux operators of the $(D-3)$-form ${\cal Z}$ symmetry.

In string theory realizations of quantum field theories, the charged objects of higher-form symmetries generally arise from branes wrapping asymptotic cycles (more precisely, relative cycles with respect to the asymptotic boundary) of appropriate dimensions in the non-compact internal space $Y$ \cite{DelZotto:2015isa,Garcia-Etxebarria:2019cnb,Morrison:2020ool,Albertini:2020mdx,Bah:2020uev,Closset:2020scj,DelZotto:2020esg,Bhardwaj:2020phs,DelZotto:2020sop,Bhardwaj:2021pfz,Apruzzi:2021phx,Apruzzi:2021vcu}.
These wrapped branes generate flux quanta, whose spacetime part represents the flux operators associated to the charged objects, and whose internal pieces are characterized by cohomology classes on the asymptotic boundary $\partial Y$.
In gauging the electric 1-form ${\cal Z}$ symmetry, the $G_\text{sc}/{\cal Z}$ theories include the magnetically charged objects from branes wrapping the corresponding relative cycles, which transform under the $(D-3)$-form ${\cal Z}$ symmetry.

In gravitational theories, global symmetries, including higher-form symmetries, are believed to be inconsistent \cite{Banks:1988yz,Kallosh:1995hi,Banks:2010zn,Harlow:2018tng}.
Since, in a theory with gauge group $G_\text{sc}/{\cal Z}$, the defect group structure forbids the simultaneous gauging of both ${\cal Z}$ 1-form and $(D-3)$-form symmetries, the magnetic one has to be broken.
Because the compactification space $Y$ is compact in gravitational models, this breaking happens explicitly due to the magnetically charged objects becoming dynamical (as the relative cycles become compact themselves).
As we will highlight below in the M-theory framework (and, by duality, also in F-theory), one can think of the compact model arising from gluing together local (non-compact) patches along their boundaries.

\subsection{Higher-Form Symmetries in M-Theory}
\label{subsec:Mapp}

In M-theory compactifications on local (non-compact) Calabi--Yau manifolds $Y_d$, the information about the 1-form symmetries is encoded in terms of geometric data \cite{Morrison:2020ool, Albertini:2020mdx}:
The electrically charged objects (Wilson lines) $\Gamma_{\text{el.}}$ are associated to M2-branes that stretch from the asymptotic boundary to the interior of $Y_d$, and are classified by classes in the relative homology $H_2 (Y_d, \partial Y_d)$.\footnote{Note that unless otherwise specified, all (co)homology groups $H_n(Y; R) \equiv H_n(Y)$ have coefficients $R = \bbZ$, which is suppressed in the notation.}
However, by an analog of 't Hooft's screening argument, their 1-form symmetry charges are subject to an equivalence relation induced by the addition of M2-branes wrapped over compact 2-cycles.
Mathematically, this is encoded in the quotient
\begin{align}\label{eq:1-form-charges_general}
	\Gamma_\text{el.} \equiv \Gamma = \frac{H_2(Y_d, \partial Y_d)}{\text{im}(\jmath_2)} \cong \frac{H_2(Y_d, \partial Y_d)}{\text{ker}(\partial_2)} \cong \text{im}(\partial_2) \subset H_1(\partial Y_d) \, ,
\end{align}
extracted from the long exact sequence of relative homology,\footnote{The map $\imath$ is induced by the inclusion $\partial Y \hookrightarrow Y$, and $\jmath_n$ is induced by the quotient map onto relative $n$-chains. As usual $\partial_n$ denotes the boundary map.}
\begin{align}\label{eq:long_seq_rel_hom}
\ldots \rightarrow H_n (\partial Y_d) \overset{\imath_n}{\rightarrow} H_n (Y_d) \overset{\jmath_n}{\rightarrow} H_n (Y_d, \partial Y_d) \overset{\partial_n}{\rightarrow} H_{n-1} (\partial Y_d) \rightarrow \ldots \, .
\end{align}

Similarly, by wrapping M5-branes over the relative $(d-2)$-cycles one obtains extended magnetically charged objects in the effective theory:
\begin{align}\label{eq:mag-charges_general}
	\Lambda_\text{mag.} \equiv \Lambda = \frac{H_{d-2}(Y_d, \partial Y_d)}{\text{im}(\jmath_{d-2})} \cong \frac{H_{d-2}(Y_d, \partial Y_d)}{\text{ker}(\partial_{d-2})} \cong \text{im}(\partial_{d-2}) \subset H_{d-3}(\partial Y_d) \, ,
\end{align}
The ``defect group'' pairing between the electric and magnetic charges is then given by the torsion linking pairing
\begin{align}\label{eq:linking_pairing_homology}
	L: \text{Tors}(H_1(\partial Y_d)) \times \text{Tors}(H_{d-3}(\partial Y_d)) \rightarrow \mathbb{Q}/\bbZ \, .
\end{align}
As a generalization of the requirement of mutual locality, imposed by Dirac quantization condition, in 4d, a choice of physical electric charges $\{ \Omega \} \subset \Gamma$ enforces the restriction $L(\Omega, \tilde{\Omega}) = 0$ for allowed magnetic charges $\{\tilde\Omega\} \subset \Lambda$, and vice versa. 

To compute \eqref{eq:1-form-charges_general} and \eqref{eq:mag-charges_general}, we assume that $H_n(Y_d)$ is torsion-free for any $n$ (which holds for all cases relevant to the present discussion).
Then Poincar\'e--Lefschetz duality and the Universal Coefficient theorem provide the identification
\begin{align}
	H_n(Y_d, \partial Y_d) \cong \text{Hom}(H_{d-n}(Y_d),\mathbb{Z}) \, .
\end{align}
Furthermore, $Y_d$ comes with an intersection pairing,
\begin{align}
	\langle \cdot , \cdot \rangle_n : H_{n}(Y_d) \times H_{d-n}(Y_d) \rightarrow \mathbb{Z} \, .
\end{align}
Using the above isomorphism, the maps $\jmath_n$ in \eqref{eq:long_seq_rel_hom} are then given by 
\begin{align}\label{eq:j_map_general}
  \jmath_n(\upsilon) = \langle \upsilon, \cdot \rangle_{n} = \langle \cdot, \upsilon \rangle_{d-n} \in \text{Hom}(H_{d-n}(Y_d), \bbZ) \, .
\end{align}
Picking a basis $\sigma_a$ for $H_{d-2}(Y_d)$ and a basis $\gamma_i$ for $H_2(Y_d)$ defines the $r_{d-2} \times r_2$ intersection matrix $M_{ai} = \langle \sigma_a, \gamma_i \rangle_{d-2} \equiv \langle \sigma_a, \gamma_i \rangle$, where $r_k = \text{rank}(H_k(Y_d))$, with $r_2 - r_{d-2} \equiv f \geq 0$.
Then $\jmath_{d-2}(\sigma_a) = \sum_i M_{ai} \eta_i$, where $\eta_i \in \text{Hom}(H_2(Y_d), \bbZ)$ is the dual basis of $\gamma_i$, i.e., $\eta_i(\gamma_j) = \delta_{ij}$.
Similarly, $\jmath_2(\gamma_i) = \sum_a (M^T)_{ia} \nu_a$, with $\nu_a \in \text{Hom}(H_{d-2}(Y_d),\bbZ)$ dual to $\sigma_a$.
Through a Smith decomposition,
\begin{align}\label{eq:smith_decomp}
	 M_{ai} = \sum_{b, j} \, S_{ab} D_{b j} \, T_{ji} \,,
\end{align}
with $S \, \,(r_{d-2} \times r_{d-2})$, $T \, \, (r_2 \times r_2)$ invertible integer matrices, and
\begin{align}
	  D_{b j} = \begin{pmatrix}
		N_1 & 0 & \ldots & 0 & 0 & \ldots \\
		0 & N_2 & \ldots & 0 & 0 & \ldots \\
		\vdots & \vdots & \ddots & \vdots & \vdots \\
		0 & 0 & \ldots & N_{r_{d-2}} & 0 & \ldots
	\end{pmatrix}_{bj} \, ,
\end{align}
we have
\begin{align}\label{eq:explicit_formula_Gamma}
	\begin{split}
		\Gamma & = \frac{H_2(Y_d, \partial Y_d)}{\text{im}(\jmath_2)} = \frac{\text{Hom}(H_{d-2}(Y_d), \mathbb{Z})}{\text{im}(D^T)} \cong \bigoplus_{k=1}^{r_{d-2}} \mathbb{Z}_{N_k} \, , \\
		\Lambda & = \frac{H_{d-2}(Y_d, \partial Y_d)}{\text{im}(\jmath_{d-2})} = \frac{\text{Hom}(H_{2}(Y_d), \mathbb{Z})}{\text{im}(D)} \cong \Gamma \oplus \mathbb{Z}^f \, .
	\end{split}
\end{align}
More precisely, the Smith decomposition tells us that we can define new bases $\xi_a = \sum_b (S^{-1})_{ab} \, \sigma_b$ for $H_{d-2}(Y_d)$ and $\epsilon_i = \sum_j T_{ij} \, \eta_j$ for $\text{Hom}(H_2(Y_d),\bbZ)$ such that $\jmath_{d-2}(\xi_a) = N_a \, \epsilon_a$ (no sum over $a = 1,...,r_{d-2}$), and similarly for $H_2(Y_d)$ and $\text{Hom}(H_{d-2}(Y_d), \bbZ)$.
For an $N_t$-torsional generator $T \in \Lambda_\text{mag.}$, we thus have a representative $\tilde\sigma_t \in H_{d-2}(Y_d, \partial Y_d)$, satisfying
\begin{align}\label{eq:lin_rel_homology}
  N_t \, \tilde\sigma_t & = \sum_a \lambda_a \, \jmath_{d-2}(\sigma_a) \, , \\
  \text{with} \quad \lambda_a & = (S^{-1})_{ta} \, . \label{eq:lambda_via_smith_decomp}
\end{align}

There is a dual description of the higher-form symmetries in terms of boundary conditions for background fluxes.
In this framework, the defect group structure arises from the non-commutativity of general flux operators at the asymptotic boundary of the compactification space, measured precisely by (the cohomological version of) the linking pairing \eqref{eq:linking_pairing_homology} \cite{Freed:2006ya, Freed:2006yc,DelZotto:2015isa,Garcia-Etxebarria:2019cnb}. In the effective field theory, the boundary conditions for background fluxes in the higher-dimensional theory parametrize allowed background gauge fields $B$ of higher-form symmetries. 
These restrict the possibilities to wrap M2- and M5-branes over elements in relative homology and in turn constrain the spectrum of extended operators, see \cite{Morrison:2020ool}. 
In Section \ref{sec:anomalies}, we will discuss how the background gauge fields enter the dimensional reduction of the M-theory Chern--Simons term.

\subsection{Higher-Form Symmetries on Elliptic Fibrations}
\label{subsec:Mell}

F-theory describes non-perturbative vacua of type IIB string theory, whose space\-time-de\-pen\-dent axio-dilaton field is captured by the complex structure of an auxiliary torus  (see \cite{Taylor:2011wt, Weigand:2018rez, Cvetic:2018bni} for reviews). 
Therefore, F-theory geometries are described by elliptically-fibered $Y_d$, whose base $\mathfrak{B}_{d-2}$ is part of the physical type IIB spacetime:
\begin{equation}
\begin{split}
  T^2 \enspace \hookrightarrow \enspace & Y_d \\
  & \, \downarrow \\
  & \mathfrak{B}_{d-2}
\end{split}
\end{equation}
By duality, F-theory compactified on $Y_d \times S^1$ is equivalent to M-theory compactified on $Y_d$, whose higher-form symmetries we now examine.

For non-compact backgrounds, i.e., where $\mathfrak{B}_{d-2}$ is non-compact, this induces a fibration structure on the asymptotic boundary as well:
\begin{equation}
\begin{split}
  T^2 \enspace \hookrightarrow \enspace & \partial Y_d \\
  & \, \downarrow \\
  & \partial \mathfrak{B}_{d-2}
\end{split}
\end{equation}
In general, the fibration on $\mathfrak{B}_{d-2}$ has singular fibers over (complex) codimension-one loci, which themselves extend to the asymptotic boundary $\partial \mathfrak{B}_{d-2}$.
Their effect on the boundary homology depends very much on the precise type of singular fibers.
It would be important to study such examples in more detail, e.g., in the context of F-/M-theory realizations of 6d/5d SCFTs.

Aiming for a more intuitive understanding in this chapter, we avoid these complications, and instead focus on $\dim_\mathbb{R} (Y_d) \equiv d = 4$, i.e., F-/M-theory compactified to 8d/7d.
In this case, the base part of the asymptotic boundary is a circle $\partial \mathfrak{B}_2 \simeq S^1$. Moreover, for situations relevant for supersymmetric F-theory backgrounds,\footnote{That is, backgrounds leading to a local Calabi--Yau 2-fold, i.e., a local patch of an elliptically-fibered K3, for which the base is $\mathbb{P}^1$.} $\mathfrak{B}_2$ itself can be identified with a disc $D_2$. 
We further demand that there is non-trivial gauge dynamics in the effective theory. 
This requires the presence of a singularity in $Y_4$ that can be interpreted as a singular fiber of the elliptic fibration. 
It is the topology of the asymptotic boundary to this fiber singularity that will determine the allowed flux backgrounds and, consequently, the gauge group in the M-theory setup. 
The internal fiber singularity induces a non-trivial fibration on the boundary circle which is associated to a non-trivial $SL(2,\mathbb{Z})$ monodromy, see Figure \ref{fig:singgeom}.\footnote{
The monodromy is not affected by local modifications in the interior such as a resolution of the fiber singularity that corresponds to a Coulomb branch deformation of the effective action derived from M-theory on $Y_4$.}
\begin{figure}[ht]
\centering
\includegraphics[width = 0.5 \textwidth]{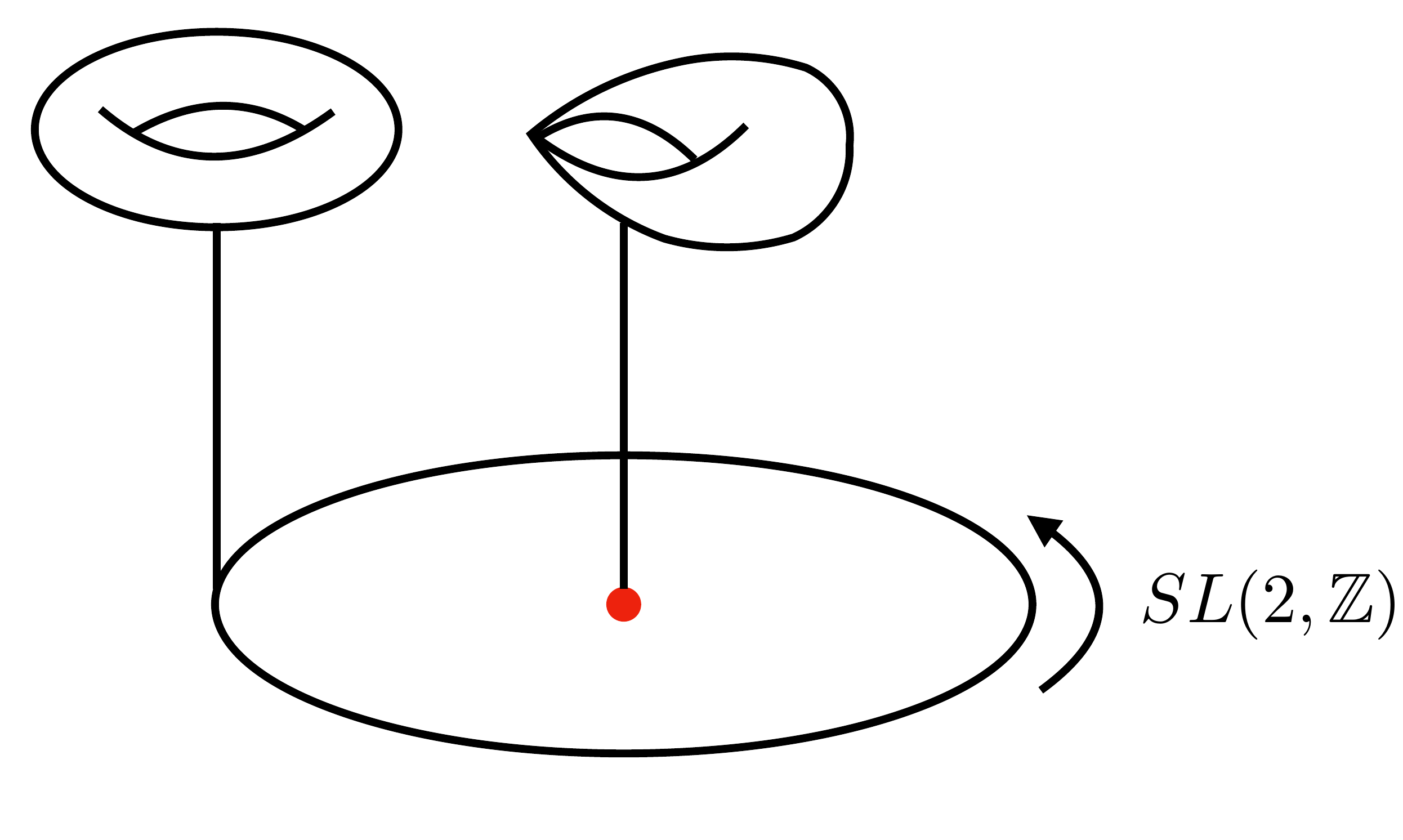}
\caption{Singular geometry with fiber singularity in the interior that induces an $SL(2,\mathbb{Z})$ monodromy around the boundary circle.}
\label{fig:singgeom}
\end{figure}

\subsection*{Boundary Geometry}

The boundary\footnote{In the remainder of this section we restrict to the case $d = 4$ and will not denote the dimension of the various spaces explicitly.} $\partial Y$ has the structure of a mapping torus
\begin{align}
\partial Y = T^2 \times [0,1] / \sim \,, \quad \text{with identification} \enspace (x,0) \sim (K(x),1) \,,
\end{align}
where $K: T^2 \rightarrow T^2$ is the overall $SL(2,\bbZ)$ monodromy around all singular fibers in the interior of $\mathfrak{B}_2$. This description allows for an application of a generalized Mayer--Vietoris sequence (see, e.g., \cite{hatcher2002algebraic}), which yields a long exact sequence for the homology groups of $\partial Y$. For our interest the relevant section of this long exact sequence is given by
\begin{equation}\label{eq:mapping_torus_seq}
\begin{split}
0 & \rightarrow H_3 (\partial Y) \rightarrow \underbrace{H_2 (T^2)}_{\cong \bbZ} \stackrel{\alpha}{\rightarrow} \underbrace{H_2 (T^2)}_{\cong \bbZ} \rightarrow H_2 (\partial Y) \rightarrow \underbrace{H_{1} (T^2)}_{\cong \bbZ \oplus \bbZ} \stackrel{\kappa}{\rightarrow} \underbrace{H_{1} (T^2)}_{\cong \bbZ \oplus \bbZ} \rightarrow \\ 
& \rightarrow H_{1} (\partial Y) \rightarrow \underbrace{H_{0} (T^2)}_{\cong \bbZ}  \stackrel{\beta}{\rightarrow} \underbrace{H_{0} (T^2)}_{\cong \bbZ} \rightarrow H_{0} (\partial Y) \rightarrow 0 \,.
\end{split}
\end{equation}
The maps from $H_n (T^2)$ (regarded as a $\bbZ$-module) to itself, denoted by $\alpha$, $\beta$, and $\kappa$, are given by $({\bf 1} - K_{\ast})$, where $K_*$ is the action on $H_n(T^2)$ induced by the $SL(2,\mathbb{Z})$ monodromy $K$.
Since any $SL(2,\bbZ)$ monodromy induces the identify action on $H_2(T^2)$ and $H_0(T^2)$ (it maps the full $T^2$, the generator of $H_2$, onto itself, and one point onto another, both being homologous, i.e., identical in $H_0$), the maps $\alpha$ and $\beta$ are 0.
This fixes $H_3 (\partial Y) \cong \mathbb{Z}$, $H_0 (\partial Y) \cong \mathbb{Z}$, and leaves
\begin{align}\label{eq:split_mapping_torus_seq}
  \begin{split}
    0 \rightarrow \bbZ \rightarrow H_2(\partial Y) \rightarrow \bbZ \oplus \bbZ \stackrel{\kappa}{\rightarrow} \bbZ \oplus \bbZ \rightarrow H_1(\partial Y) \rightarrow \bbZ \rightarrow 0 & \, .
  \end{split}
\end{align}
The remaining homology groups are determined by the monodromy induced map $\kappa$ on $H_1(T^2) \cong \bbZ \oplus \bbZ$, whose generators $(1,0)$ and $(0,1)$ are the usual $\mathcal{A}$- and $\mathcal{B}$-cycle, respectively. The \mbox{(co)kernel} of $\kappa$ splits \eqref{eq:split_mapping_torus_seq} to
\begin{align}
\begin{split}
  0 \rightarrow \bbZ \rightarrow H_2 (\partial Y) \rightarrow \text{ker}(\kappa) \rightarrow 0 \, , \quad \text{and} \quad  0 \rightarrow \text{coker}(\kappa) \rightarrow H_1(\partial Y) \rightarrow \mathbb{Z} \rightarrow 0 \, .
\end{split}
\end{align}
The last term in both of these sequences are free ($\bbZ$ is trivially free, and $\text{ker}(\kappa)$ is a subgroup of a free group $\bbZ \oplus \bbZ$), so both sequences split:
\begin{align}
  H_2(\partial Y) = \bbZ \oplus \text{ker}(\kappa) \, , \quad H_1(\partial Y) = \text{coker}(\kappa) \oplus \bbZ \, .
  \label{eq:boundhom}
\end{align}
From \eqref{eq:mapping_torus_seq}, we see that the $\mathbb{Z}$ factor in $H_2(\partial Y)$ originates from $H_2(T^2)$, i.e., is generated by the class $\mathfrak{f}$ of the torus fiber.
Meanwhile, the $\mathbb{Z}$ factor in $H_1 (\partial Y)$ comes from $H_0(T^2)$, hence corresponds to a marked point on the fiber, i.e., a section of the torus bundle which is a copy of the base circle. We will identify this class with the restriction of the zero-section to the boundary $S_0|_{\partial Y}$.

With a list of fiber singularities provided by the Kodaira classification, see e.g. \cite{Fukae:1999zs}, and their induced $SL(2,\mathbb{Z})$ monodromy up to an overall conjugation, we can determine the respective torsion groups, cf.~Table \ref{tab:fibers_and_kappa}.
\begin{table}[ht]
\renewcommand{\arraystretch}{1.25}
\begin{align*}
\begin{array}{| c | c | c | c | c |}
\hline
\text{fiber type} & \text{brane content} & \mathfrak{g} & \kappa & H_1 (\partial Y, \bbZ)_{\text{tors}} = \text{coker}(\kappa)_\text{tors} \\ \hline \hline
I_N & A^N & \mathfrak{su}(N) & \begin{pmatrix} 0 &  N \\ 0 & 0 \end{pmatrix} & \mathbb{Z}_N \\ \hline
II & A C & - & \begin{pmatrix} 0 & 1 \\ -1 & 1 \end{pmatrix} & - \\ \hline
III & A^2 C & \mathfrak{su}(2) & \begin{pmatrix} 1 & 1 \\ - 1 & 1 \end{pmatrix} & \mathbb{Z}_2 \\ \hline
IV & A^3 C & \mathfrak{su}(3) & \begin{pmatrix} 2 & 1 \\ -1 & 1 \end{pmatrix} & \mathbb{Z}_3 \\ \hline
I_{2n}^* & A^{4 + 2n} B C & \mathfrak{so}(4n + 8) & \begin{pmatrix} 2 & -2n \\ 0 & 2 \end{pmatrix} & \mathbb{Z}_2 \oplus \mathbb{Z}_2 \\ \hline
I_{2n+1}^* & A^{5 + 2n} BC & \mathfrak{so}(4n + 10) & \begin{pmatrix} 2 & - (2n+1) \\ 0 & 2  \end{pmatrix} & \mathbb{Z}_4 \\ \hline
IV^* & A^5 B C^2 & \mathfrak{e}_6 & \begin{pmatrix} 2 & -1 \\ 1 & 1 \end{pmatrix} & \mathbb{Z}_3 \\ \hline
III^* & A^6 B C^2 & \mathfrak{e}_7 & \begin{pmatrix} 1 & -1 \\ 1 & 1 \end{pmatrix} & \mathbb{Z}_2 \\ \hline
II^* & A^7 B C^2 & \mathfrak{e}_8 & \begin{pmatrix} 0 & -1 \\ 1 & 1 \end{pmatrix} & - \\ \hline
\end{array}
\end{align*}
\caption{Simple algebras $\mathfrak{g}$ realized via Kodaira fibers / $[p,q]$-7-branes, and the corresponding homology map $\kappa$, as well as $\text{coker}(\kappa)_{\text{tors}} \cong H_1(\partial Y, \bbZ)_\text{tors}$. \label{tab:fibers_and_kappa}}
\end{table}
This coincides with the table given in \cite{Garcia-Etxebarria:2019cnb, Albertini:2020mdx} for M-theory on lens spaces, which realizes all ADE algebras $\mathfrak{g}$, albeit not in an elliptic fibration (and thus have no F-theory uplift).
Moreover, we also find agreement between the linking pairing on boundary torsion cycles, and the defect group of 7d gauge theories with gauge algebra $\mathfrak{g}$, see Appendix \ref{app:defect_group_elliptic}.
This shows that the 7d theories that descend from an 8d F-theory compactification with simple gauge algebra $\mathfrak{g}$ has the expected electric and magnetic higher-form symmetries.
The only difference is that these 7d theories further contain $U(1)$ global symmetry, the Kaluza--Klein $U(1)$, whose background fluxes / asymptotic charges are captured by $\bbZ \subset H_1(\partial Y)$.

\subsection*{Bulk Geometry}

Equipped with the boundary homology groups $H_n(\partial Y)$, we can now examine the long exact sequence \eqref{eq:long_seq_rel_hom}, which encodes the extended charged objects under the higher-form symmetries.
For our investigation the relevant part of the long exact sequence above is given by
\begin{align}
\ldots \rightarrow H_2 (\partial Y) \stackrel{\imath_2}{\rightarrow} H_2 (Y) \stackrel{\jmath_2}{\rightarrow} H_2 (Y, \partial Y) \stackrel{\partial_2}{\rightarrow} H_1 (\partial Y) \stackrel{\imath_1}{\rightarrow} H_1 (Y) \rightarrow \ldots \, ,
\end{align}
Let us focus on the case with a single fiber in the interior of $Y$, corresponding to a simple gauge algebra $\mathfrak{g}$.
Then the resolution of this singularity introduces $\text{rank}(\mathfrak{g})$ compact 2-cycles (divisors in $Y$) $\sigma_a$, which intersect according to the Dynkin diagram of $\mathfrak{g}$.
Together with the generic fiber $\mathfrak{f}$, these form a basis for $H_2 (Y)$.
Since $\mathfrak{f}$ is homologous also to the torus fiber on the boundary, i.e., the factor $\bbZ$ in \eqref{eq:boundhom}, we see that $\mathfrak{f} \in \text{im} (\imath_2)= \text{ker} (\jmath_2)$.
This agrees with \eqref{eq:j_map_general}: the generic fiber $\mathfrak{f}$ on elliptic surfaces satisfies the intersection properties $\langle \mathfrak{f}, \mathfrak{f} \rangle = \langle \mathfrak{f}, \sigma_a \rangle = 0$.
On the other hand, two different resolution divisors cannot have the same intersection numbers with all 2-cycles, so $\jmath_2(\sigma_a) = \jmath_2(\sigma_b)$ if and only if $a = b$.
Therefore, the long exact sequence splits into the piece
\begin{align}
0 \rightarrow \langle \sigma_a \rangle \stackrel{\jmath_2}{\rightarrow} H_2 (Y,\partial Y) \stackrel{\partial_2}{\rightarrow} \underbrace{\mathbb{Z} \oplus \text{coker}(\kappa)_{\text{tors}} \oplus  \text{coker}(\kappa)_{\text{free}}}_{H_1(\partial Y)} \stackrel{\imath_1}{\rightarrow} H_1 (Y) \rightarrow \dots \,.
\end{align}

We have already explained above that the factor $\bbZ \equiv \bbZ_\text{KK} \subset H_1(\partial Y)$ is in ker$(\imath_1)$, as it encodes the background data for the KK $U(1)$ that is universal in M-theory on elliptic fibrations.
Since $H_1 (Y)$ is torsion-free\footnote{This assumption holds also for the elliptically fibered geometries we consider in this chapter. It would be interesting to study the physics of higher-form symmetries in models with non-trivial torsion in $H_1(Y)$.}, the torsion part $\text{coker}(\kappa)_{\text{tors}}$ cannot be mapped non-trivially into it.
Hence, $\text{coker}(\kappa)_{\text{tors}} \subset \text{ker}(\imath_1) = \text{im}(\partial_2) \subset \Gamma$ in \eqref{eq:1-form-charges_general}.
Therefore, for any $N_t$-torsional element $T \in \text{coker}(\kappa)_{\text{tors}}$ there is a $\tilde{\sigma} \in H_2(Y, \partial Y)$ with $\partial_2(\tilde{\sigma}) = T$.
Then, $N_t \tilde{\sigma} \in \text{ker}(\partial_2) = \text{im}(\jmath_2)$. 
Since $H_2 (Y, \partial Y) \cong \text{Hom}(H_2(Y), \bbZ)$ is also torsion-free, this means there are non-zero integers $\lambda_a$ such that
\begin{align}
\tilde{\sigma} = \frac{1}{N_t} \sum_a \lambda_a \jmath_2(\sigma_a) \in H_2(Y, \partial Y) \,,
\label{eq:reltors}
\end{align}
where the coefficients $\lambda_a$ can be understood modulo $N_t$ since one can always add an integer linear combination of $\jmath(\sigma_a)$ which is in $\text{ker}(\partial_2)$. 
Of course, these are the same coefficients as in \eqref{eq:lambda_via_smith_decomp}, determined via Smith decomposition of the intersection pairing on $Y$.
For example, as we will compute in Section \ref{subsec:7d_anomalies}, the generator of $N$-torsional boundary 1-cycles for $Y$ containing an $I_N$ fiber is represented by
\begin{align}\label{eq:boundary_torsion_rep_I_N}
\begin{split}
  \tilde\sigma & = \tfrac{1}{N} \sum_{a=1}^{N-1} a\, \jmath_2(\sigma_a) \\
  & = \tfrac{1}{N} \sum_{a=1}^{N-1} \underbrace{\tfrac{a-N}{N}}_{(-C^{-1})_{1,a}} \jmath_2(\sigma_a) \mod \text{ker}(\partial_2) \, ,
\end{split}
\end{align}
where $C$ is the Cartan matrix of $SU(N)$.
In Appendix \ref{app:defect_group_elliptic}, where we compute coker$(\kappa)$ for all Kodaira singularities, we see that only $I_N$ fibers have non-trivial coker$(\kappa)_\text{free} \cong \bbZ$, and that it further maps non-trivially under $\imath_1$.
Therefore, coker$(\kappa)_\text{free}$ never contributes to the higher-form symmetries.

In summary, we have seen that the higher-form symmetries \eqref{eq:explicit_formula_Gamma} of M-theory compactified on a local elliptically fibered four-manifold $Y_4 \equiv Y$ are entirely encoded in terms of the monodromy $K \equiv {\bf 1} + \kappa$ around the singular fiber in $Y_4$:
\begin{align}
  \Gamma \cong \text{coker}(\kappa)_\text{tors} \oplus \bbZ_\text{KK} \, .
\end{align}
In Section \ref{sec:geometry}, we will connect this result with ``established'' methods to describe gauge groups with different center symmetries, namely via Mordell--Weil torsion, and string junctions, and show that coker$(\kappa)_\text{tors}$ indeed describes the higher-form symmetries of the F-theory model in one higher dimension.

\subsection{Semi-simple Algebras, Adjoint Higgsing, and Compact Models}

So far, we have only discussed explicit examples with a single Kodaira fiber in $Y$ with monodromy $K$, corresponding to a simple ADE Lie algebra $\mathfrak{g}$.
In these cases, the boundary homology coker$(\kappa) \subset H_1(\partial Y)$ perfectly captures the higher-form symmetry expected from field theory.
However, since $\kappa = {\bf 1} - K$ is an endomorphism on $H_1(T^2) \cong \bbZ^2$, coker$(\kappa)$ can have at most two torsion factors.
This begs the question how this can capture the center symmetries of a semi-simple algebra like, e.g.,  $\mathfrak{g} = \mathfrak{su}(N)^3$, which can be realized by three $I_N$ fibers in $Y$.

The key missing component is dynamical $\mathfrak{u}(1)$ factors which generically arise in the presence of multiple singular fibers.
To see their importance, consider a model $\tilde{Y}$ with one $I_{N-1}$ and one $I_1$ fiber which are mutually local.
That is, in an $SL(2,\bbZ)$ frame where the monodromy of the $I_{N-1}$ fiber is $K_{N-1} = \left(\begin{smallmatrix} 1 & 1-N \\ 0 & 1 \end{smallmatrix} \right)$, the $I_1$ fiber induces $K_1 = \left(\begin{smallmatrix} 1 & -1 \\ 0 & 1 \end{smallmatrix} \right)$.
Therefore, the overall monodromy is $K = K_{N-1} K_1 = \left(\begin{smallmatrix} 1 & -N \\ 0 & 1 \end{smallmatrix} \right)$, with $\text{coker}(\kappa)_\text{tors} = \text{coker}({\bf 1} - K)_\text{tors} = \bbZ_N$.
But F-theory on $\tilde{Y}$ naively has only an $\mathfrak{su}(N-1)$ algebra, whose center cannot possibly accommodate a $\bbZ_N$ 1-form symmetry.
Moreover, $\bbZ_N$ does not have a $\bbZ_{N-1}$ subgroup, so the boundary homology seems to not capture the higher-form symmetries of $\mathfrak{su}(N-1)$ at all.

However, there is also the additional $I_1$ fiber.
Because it is mutually local with the $I_{N-1}$ fiber, they share the same the vanishing cycle, in this case the ${\cal A}$-cycle.
By fibering this 1-cycle between the two singular fibers (see left of Figure \ref{fig:1formHiggs}), we obtain a compact 2-cycle $\sigma_N$, in addition to the $N-1$ resolution divisors of the $I_N$ fiber, which gives rise to a dynamical $\mathfrak{u}(1)$.
This $\mathfrak{u}(1)$ can be viewed as the one arising in the adjoint Higgsing $\mathfrak{su}(N) \rightarrow \mathfrak{su}(N-1) \times \mathfrak{u}(1)$, which geometrically precisely corresponds to the deformation of an $I_N$ fiber into an $I_{N-1}$ and a mutually local $I_1$ fiber.

\begin{figure}[ht]
\centering
\includegraphics[width = 0.8 \textwidth]{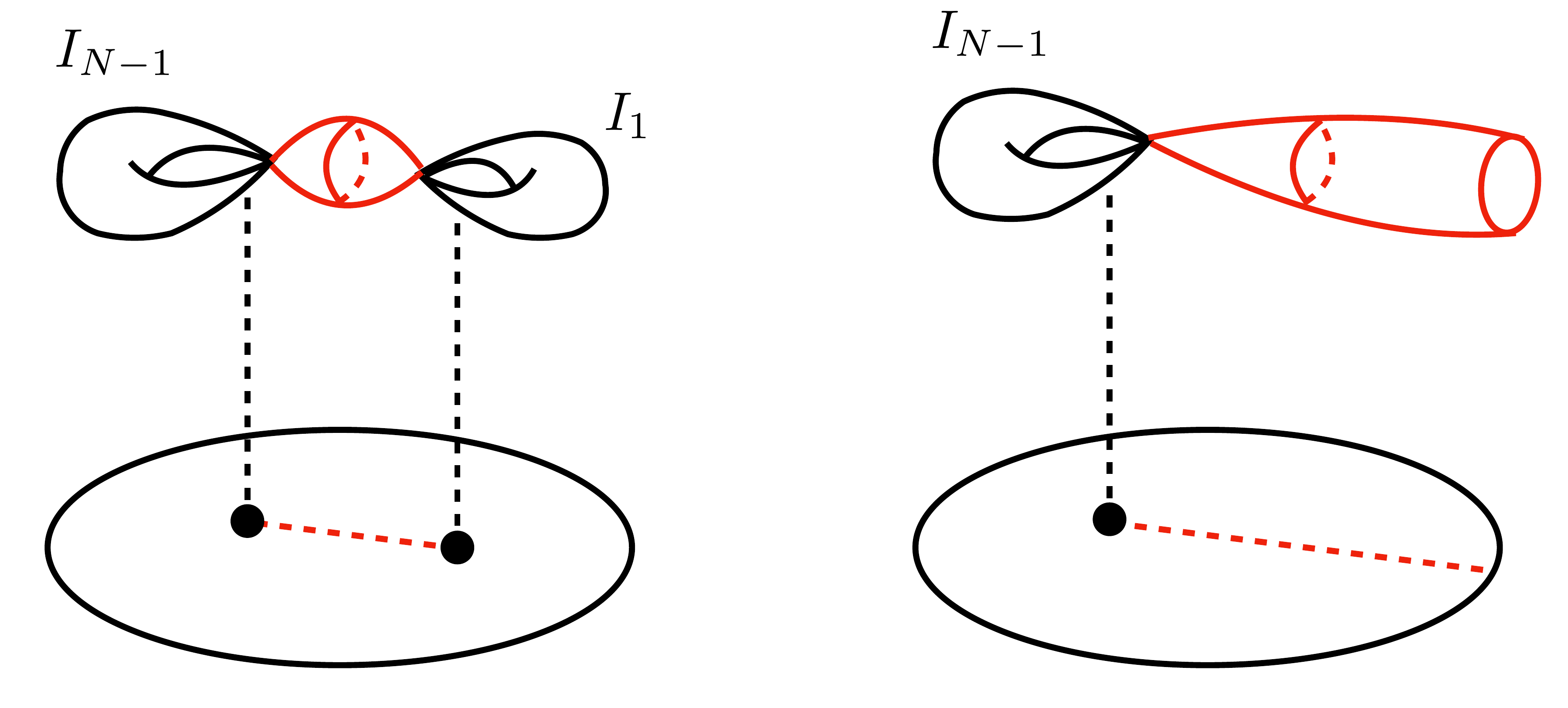}
\caption{Cartoon of a Higgsing transition, where one of the singular fibers is moved outside of the disk, thereby changing the monodromy. \label{fig:1formHiggs}}
\end{figure}

In this Higgsing transition, the fundamental and adjoint representations of the original $\mathfrak{su}(N)$ decompose as
\begin{align}
 {\bf N} \rightarrow ({\bf N-1})_{1} \oplus {\bf 1}_{1-N} \, , \quad  {\bf adj}(N) \rightarrow {\bf adj}(N-1)_0 + ({\bf N-1})_{N} + (\overline{\bf N-1})_{-N} + {\bf 1}_0 \, ,
\end{align}
where the subscripts denote the $\mathfrak{u}(1)$ charge, normalized such that every state has integer charge.
Therefore, the Wilson lines in the fundamental representation of $\mathfrak{su}(N)$, which are the charged objects under the $\mathbb{Z}_N$ 1-form symmetry in the theory prior to Higgsing, gives rise to line operators with $\mathfrak{u}(1)$ charges $1 \mod N$.
However, the bifundamental states from the decomposition of the adjoint representation, which correspond to M2-branes wrapping the red 2-cycle in the left of Figure \ref{fig:1formHiggs}, carry $\mathfrak{u}(1)$ charge $N$.
These screen the $\mathfrak{u}(1)$ charges of line operators in the Higgsed phase, and therefore break the $U(1)$ 1-form symmetry of the $\mathfrak{u}(1)$ gauge factor to $\bbZ_N$.
It is this $\bbZ_N$ 1-form symmetry (and its magnetically dual 3-form symmetry) which is captured by the boundary homology.
The bifundamental states $({\bf N - 1})_N$ also break the $\bbZ_{N-1}$ 1-form symmetry of the $\mathfrak{su}(N-1)$ factor explicitly.
Their presence further forbids the charged objects of the $\bbZ_{N-1}$ magnetic symmetry, but does allow for a linear combination between the $\mathfrak{su}(N-1)$ and the $\mathfrak{u}(1)$ magnetic charges which correspond to M5-branes wrapping the 2-cycle $\sigma_N$.
These have charge $N$ with respect to the magnetic $U(1)$ 3-form symmetry of the $\mathfrak{u}(1)$ gauge factor, so that this is also broken to a $\bbZ_N$.
All this agrees with the fact that the simply-connected group $SU(N)$ actually has $[SU(N-1) \times U(1)]/\bbZ_{N-1}$ as a subgroup, which can be interpreted as the $U(1)$ gauging the center of $SU(N-1)$.

The logic applies to any deformation of Kodaira fibers of ADE type $\mathfrak{g}$ into multiple fibers of type $\mathfrak{g}_i$, corresponding to an adjoint Higgsing which also produces additional $\mathfrak{u}(1)$ gauge factors.
The boundary homology is not affected by such a deformation, and thus the higher-form symmetries of the full system are still those of an $\mathfrak{g}$ gauge theory, albeit embedded as a subgroup of the $U(1)$ higher-form symmetry of the Abelian factors.

To recover the higher-form symmetries of the individual $\mathfrak{g}_i$ factors, one has to decouple the $\mathfrak{u}(1)s$.
Geometrically, this can be easily achieved by pushing all other singular fibers to infinity, or, equivalently, restricting to the local neighborhood of the $\mathfrak{g}_i$ fiber with its monodromy at the boundary.
As an example, consider again the model with an $I_{N-1}$ and an $I_1$ fiber.
As depicted schematically in the right panel of Figure \ref{fig:1formHiggs}, decoupling the $\mathfrak{u}(1)$ sends the $I_1$ fiber to infinity, which turns the previously compact 2-cycle into a relative cycle.
Physically, this turns the dynamical bifundamental states into infinitely massive probe particles in the fundamental representation of $\mathfrak{su}(N-1)$, whose world-lines then constitute the correct charged objects of the $\bbZ_{N-1}$ 1-form symmetry.

Note that not every configuration with multiple fibers allows for an interpretation in terms of a deformation / Higgsing of a single fiber / simple ADE algebra.
In such cases, one has to study more carefully the set of compact 2-cycles stretched between different singular fibers.
For example, if we add an $I_1$ fiber to a mutually non-local $I_{N-1}$ fiber, there is no compact 2-cycle that one can form by fibering a 1-cycle between them, because they have linearly independent vanishing cycles.
Such a configuration would not have an additional $\mathfrak{u}(1)$ factor, and consequently, no way to modify the center symmetries as above.
This can also be seen from the boundary homology.
For concreteness, consider, in an $SL(2,\bbZ)$ frame with $K_{N-1} = \left(\begin{smallmatrix} 1 & N-1 \\ 0 & 1 \end{smallmatrix} \right)$, an $I_1$ fiber with monodromy $K_1 = \left(\begin{smallmatrix} 1 & 0 \\ 1 & 1 \end{smallmatrix} \right)$, such that the overall monodromy is $K = K_{N-1} K_1 = \left(\begin{smallmatrix} N & N-1 \\ 1 & 1 \end{smallmatrix} \right)$.\footnote{The reversed ordering is related to this one by an $SL(2,\bbZ)$ conjugation, so it is equivalent.}
Then, it is straightforward to compute $\text{coker}(\kappa) = \text{coker}\left(\left(\begin{smallmatrix} N-1 & N-1 \\ 1 & 0 \end{smallmatrix} \right) \right) = \bbZ_{N-1}$.

If we have three or more singular fibers, then there can be at most two linearly independent vanishing cycles, simply because any vanishing cycle can be represented as a vector in $\bbZ^2$.
That is, given, e.g., $k$ $I_N$ fibers, each of which has one vanishing cycle ${\cal C}_i$, $i=1,...,k$, there are $(k-2)$ linear relations $\sum_{i=1}^k n^{(\ell)}_i {\cal C}_i = 0$, $\ell = 1,...,k-2$.
If we fiber from the $i$-th singular fiber the vanishing cycle $n^{(\ell)}_i {\cal C}_i$ to a marked smooth fiber $\mathfrak{f}_p$, we have a 2-chain with boundary $n^{(\ell)}_i {\cal C}_i$ in $\mathfrak{f}_p$, which cancel out the boundaries of the corresponding 2-chains from the other singular fibers due to the $\ell$-th relation.
This gives rise to $(k-2)$ compact 2-cycles.
The resulting $\mathfrak{u}(1)$ gauge factors then gauge parts of the overall $(\bbZ_N)^k$ center symmetry, leaving a subgroup with at most two discrete factors.

\subsection*{Gluing Patches to Compact Models}

We can use the above insights to describe the process of passing from local to global models.
Effectively, this is done by ``gluing'' the local patches $Y_i$ of individual singular fibers along the boundaries.
In each pairwise gluing, relative 2-cycles in $Y_{i_1}$, whose boundary 1-cycle is a vanishing cycle in $Y_{i_2}$, can be screened by additional compact 2-cycles that are formed between the singular fibers; this corresponds to the situation in Figure \ref{fig:1formHiggs}, read from right to left.
This modifies the higher-form symmetry charges, as seen from the boundary homology in terms of (torsional) 1-cycles on $\partial Y$.

To obtain a compact geometry, we must demand that any 1-cycle in the torus fiber can shrink (possibly after decomposing) on singular fibers, in accordance with the fact that there is no (non-trivial) boundary.
Equivalently, this means that the overall monodromy around all singular fibers must be trivial.
Additionally, for a valid F-theory geometry, $Y_4$ must be a K3-surface, which further limits the possible combination of singular fibers.
A more subtle effect that becomes relevant in compact models is that there might be certain linear relations between 2-cycles, such that the physically distinct number of $\mathfrak{u}(1)$s can be reduced. 
Regardless, compact 2-cycles stretched between several singular fibers gauges a diagonal subgroup of the corresponding center 1-form symmetry, as in the $\mathfrak{su}(N-1) \times \mathfrak{u}(1)$ example above. 
Phrased in the language of string junctions, whose local picture we will discuss momentarily, these phenomena have been discussed in \cite{Guralnik:2001jh}.

For example, a valid K3 can be obtained by gluing together four local patches with an $I_0^*$ fiber, each with monodromy $K = \left( \begin{smallmatrix} -1 & 0 \\ 0 & -1 \end{smallmatrix} \right)$.
Since each $I_0^*$ fiber has two independent vanishing cycles, which also generate the two $\bbZ_2$ factors of $Z(Spin(8))$, one would find $8-2 = 6$ linear relations between them, corresponding to six compact 2-cycles stretching between the four singular fibers.
However, from the compactness condition there are two additional relations among these, such that there are only four independent $\mathfrak{u}(1)$ gauge factors, which in total gives a rank 20 gauge group.\footnote{Note that two of these $\mathfrak{u}(1)$s come from the 8d ${\cal N}=1$ gravity multiplet.
These are always present, though the embedding of the center from the non-Abelian gauge symmetry into them is model specific.
}
Nevertheless the six compact 2-cycles lead to the gauging of six independent $\bbZ_2$ subgroups of the full $(\bbZ_2)^8$ 1-form symmetry \cite{Guralnik:2001jh}.
Two are orthogonal to the $\mathfrak{u}(1)$s, so that the non-Abelian part of the gauge group is $G = Spin(8)^4 / [\bbZ_2 \times \bbZ_2]$, where the denominator is the ``diagonal'' $\bbZ_2 \times \bbZ_2$ subgroup of $(\bbZ_2)^8$.
As for the Abelian factors, one can choose an appropriate basis for them, such that the $\bbZ^{\text{diagonal}}_2 \subset \bbZ_2 \times \bbZ_2$ subgroup of each $Spin(8)$ factor is embedded into one of the $U(1)$s.
The full gauge group is therefore
\begin{align}
  \frac{(Spin(8)^4 / [\bbZ_2 \times \bbZ_2]) \times U(1)^4}{(\bbZ_2)^4} \cong \frac{Spin(8)^4 \times U(1)^4}{(\bbZ_2)^6} \, .
\end{align}

\section{Center Symmetries in M-/F-theory Duality}
\label{sec:geometry}

In global, compact F-theory models, there are two equivalent ways to characterize the gauge group topology: either via the Mordell--Weil group of rational sections \cite{Aspinwall:1998xj,Mayrhofer:2014opa,Cvetic:2017epq} (see also \cite{Grimm:2015wda}), or through so-called (fractional) null junctions \cite{Fukae:1999zs,Guralnik:2001jh}.
The purpose of this section is to relate these ideas to the characterization of the gauge group via higher-form symmetries presented above.
We begin with the string junctions, since these have direct visualizations in terms of relative cycles in the local setting.

\subsection{String Junctions}
\label{subsec:StrJapp}

We begin with a brief review of F-theory / type IIB in terms of string junctions.
In this picture, the gauge dynamics are associated to the world-volume of non-perturbative 7-branes, which are classified by their $[p,q]$-type. 
The gauge degrees of freedom on such branes are described by $(p,q)$-strings, i.e., bound states of $p$ fundamental strings and $q$ D1-strings, which can end on a 7-brane of type $[p,q]$ \cite{Gaberdiel:1997ud, Gaberdiel:1998mv}. 
Much like the geometrization of the axio-dilaton in terms of an auxiliary torus, the $(p,q)$-labels encode the transformation properties, or charges, under the $SL(2,\bbZ)$ duality group of 7-branes and strings in type IIB string theory.
The induced $SL(2,\mathbb{Z})$ monodromy around a general $[p,q]$ 7-brane is given by
\begin{align}
K_{[p,q]} = \begin{pmatrix} 1 + p q & -p^2 \\ q^2 & 1 - p q \end{pmatrix} \,.
\label{eq:branemon}
\end{align}
This monodromy transformation is implemented by a branch-cut that emanates from the brane stack and stretches to infinity. When an $(r,s)$-string stretches across this branch-cut its charges change as indicated in Figure \ref{fig:junctiondeformation}.
\begin{figure}[ht]
\centering
\includegraphics[width = .8 \textwidth]{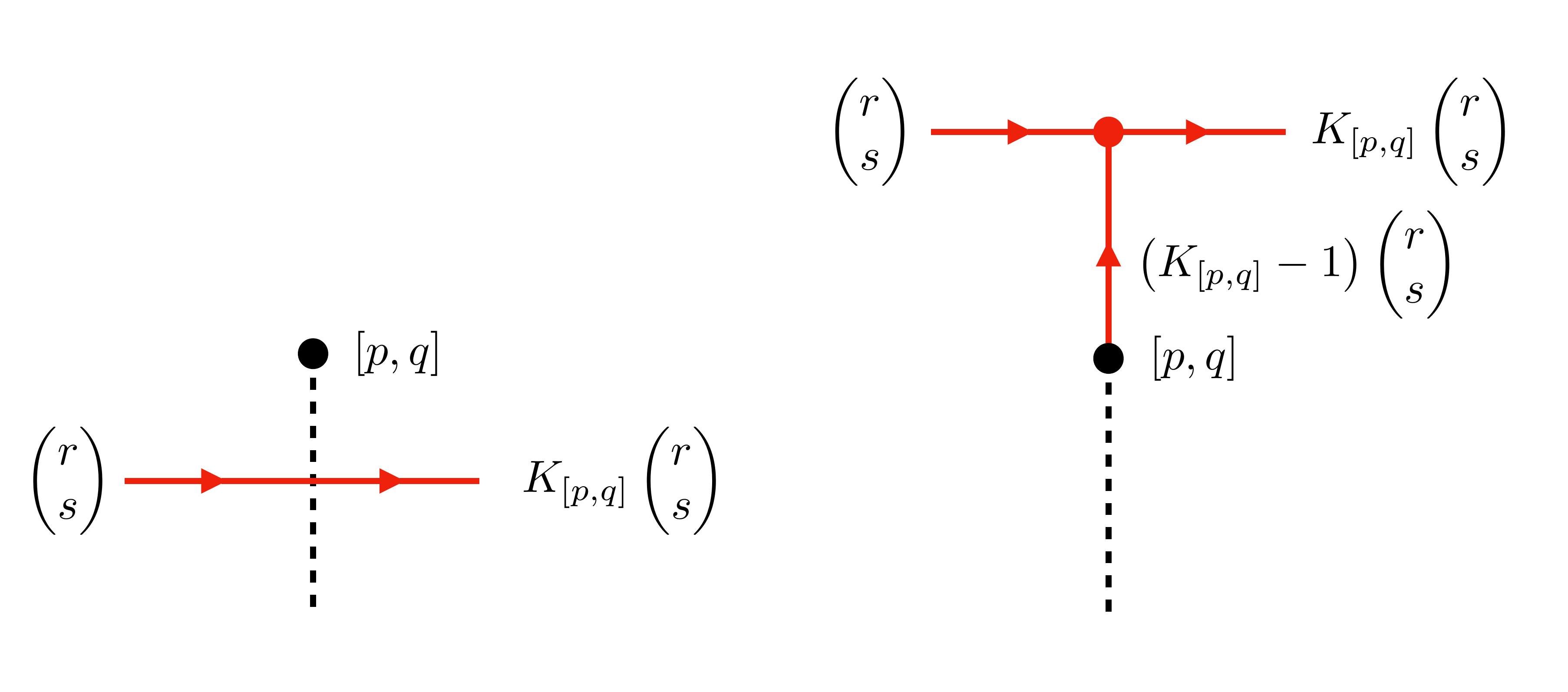}
\caption{Transformation properties of a general $(r,s)$-string passing through the branch-cut of a $[p,q]$-brane.}
\label{fig:junctiondeformation}
\end{figure}
This configuration can be deformed across the brane. As in Hanany--Witten transitions, a new string connected to the brane stack appears, see Figure \ref{fig:junctiondeformation}.

In F-theory a single $[p,q]$-brane corresponds to an $I_1$ fiber singularity. More general fiber degenerations can then be understood by stacking several 7-branes on top of each other. In the process some of the strings stretching between the individual branes become massless and constitute the gauge theory degrees of freedom on the eight-dimensional brane worldvolume. The overall $SL(2,\bbZ)$ monodromy is given by the product of the constituents. In this way one can reconstruct the full list of Kodaira singularities \cite{Gaberdiel:1997ud,DeWolfe:1998zf}. This determines the gauge algebra of the system, but is not enough to provide information about the gauge group. 
In the following we will focus on the analysis of compactifications to eight dimensions, where 7-branes are parallel, and in which case global tadpole cancellation requires 24 7-branes with overall trivial monodromy, whose transverse positions is parametrized by a $\bbP^1 \cong \mathfrak{B}_2$ that is the base of an elliptically fibered K3-surface.

To gain access to the information of the gauge group one has to analyze the full lattice of string junctions. 
Since Gauss' law forbids the presence of asymptotic charges on \emph{compact} spaces, all junctions are either closed, or have prongs ending on 7-brane stacks.
In other words, no junction can be allowed to have a free prong, whose $(p,q)$ label would be so-called asymptotic charges of the junction.
In determining the gauge group, a special role is played by the so-called null junctions. 
They are constructed by encircling all singularities of the compact model, and thus experience no net monodromy.
These junctions have vanishing pairing\footnote{A precise definition of the junction pairing is given in \cite{DeWolfe:1998zf}. For the present discussion, it suffices to note that this pairing can be identified with the intersection pairing in homology in the dual M-theory frame.} with all other junctions, and can be viewed as a ``trivial'' physical state.
More precisely, since the null junction encircles all 7-branes, one can close the loop ``on the other side'' of the $\bbP^1$, thus removing the string completely.
On the other hand, using the Hanany--Witten transition discussed above one can pull the string through all of the 7-brane stacks, leading to a configuration of a multi-pronged strings of vanishing asymptotic charge that ends on the individual stacks. 
It is obvious that a global null junction can be thought of as the sum of ``local'' null junctions, i.e., junctions that encircle one brane stack, and have a prong that emanates from the circle to connect with other local null junctions.
See Figure \ref{fig:nulljunc} for a schematic depiction of local null junctions.
\begin{figure}[ht]
\centering
\includegraphics[width = 0.9 \textwidth]{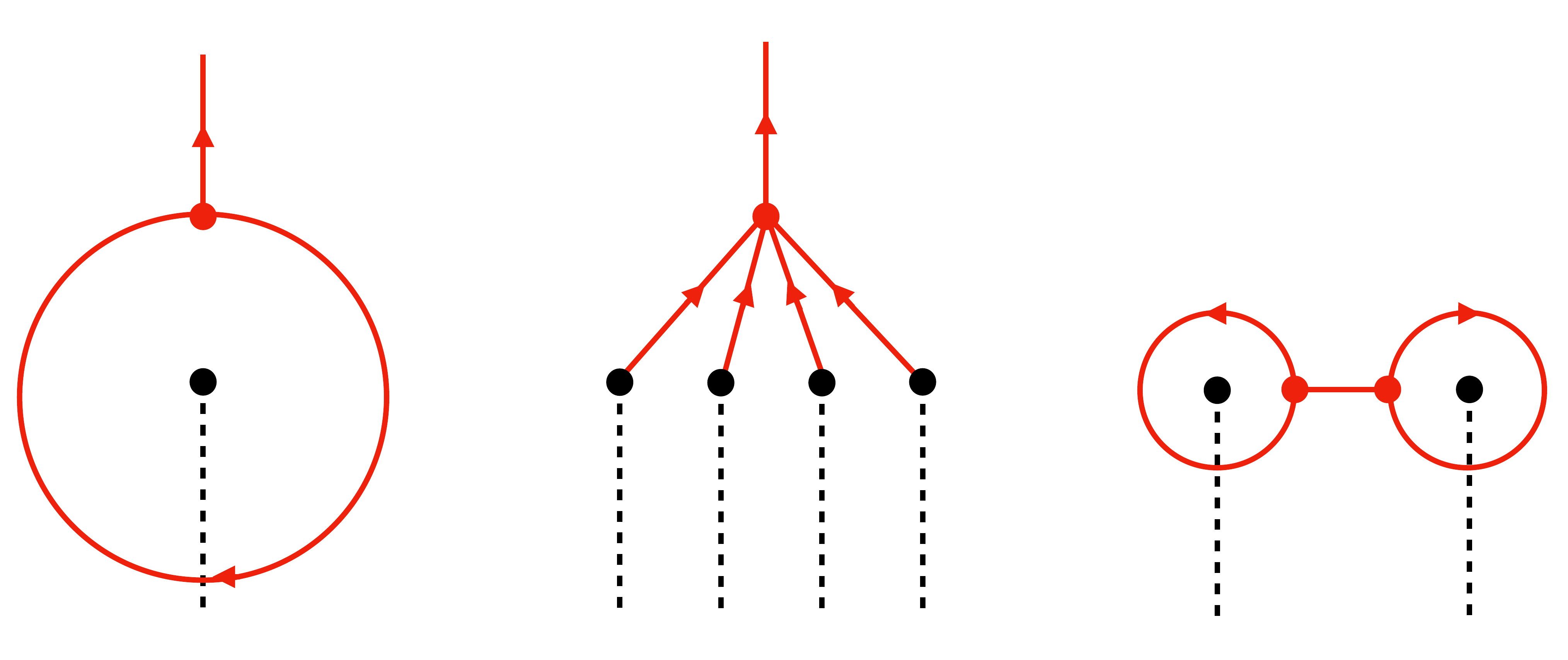}
\caption{A local null junction (left) obtained from encircling a brane stack with a string. In general, it carries an asymptotic charge.
Via a Hanany--Witten transition, the null junction can also be presented as joining prongs from the constituent branes of the stack (middle).
One can connect local null junctions via their asymptotic charges; the global null junctions on a compact $\bbP^1$ have no net asymptotic charge (schematically on the right).
}
\label{fig:nulljunc}
\end{figure}

Zooming onto the local patch around a single brane stack, realizing the algebra $\mathfrak{g}$ with simply-connected cover $G_\text{sc}$, such local null junctions generally carry asymptotic $(p,q)_\text{asymp}$ charge, represented by the prong going off to infinity on the left of Figure \ref{fig:nulljunc}.
If the stack is encircled by a $(r,s)$-string, this charge is
\begin{align}\label{eq:asymp_charge_null_junc}
\begin{pmatrix} p \\ q \end{pmatrix}_{\text{asymp}} = (K - \mathbf{1}) \begin{pmatrix} r \\ s \end{pmatrix} \equiv -\kappa \begin{pmatrix} r \\ s \end{pmatrix} \,,
\end{align}
where $K$ is the monodromy of the stack.
It turns out to be useful to consider \emph{all} possible charges $(r,s)$ such that the asymptotic charges $(p,q)_\text{asymp}$ are integers.
For integral $(r,s)$, the resulting null junction is called a proper, or integer null junction\footnote{Such a junction can be represented by the left panel of Figure \ref{fig:nulljunc}; by making the circle infinitely large, it is obvious that integral null junctions decouple from the local gauge dynamics of the brane stack.
}, whose asymptotic charges are
\begin{align}\label{eq:integer_null_junctions}
  \text{integer null junctions} = \left\{ \left. \kappa \begin{pmatrix} r \\ s \end{pmatrix} \, \right| \, r,s \in \bbZ \right\} = \text{im}(\kappa: \bbZ^2 \rightarrow \bbZ^2) \, .
\end{align}
However, with suitably fractional $(r,s)$, one can generate all integer asymptotic charges $(p,q)_\text{asymp}$, if $\mathfrak{g} \neq \mathfrak{su}(N)$; if $\mathfrak{g} = \mathfrak{su}(N)$, then $\eqref{eq:asymp_charge_null_junc}$ generates all integer charges of the form $(p,0)_\text{asymp}$, up to $SL(2,\bbZ)$ conjugacy.
If one performs a Hanany--Witten transition for these so-called fractional null junctions, then the individual prongs on the constituent branes of the stack (as depicted schematically on the right in Figure \ref{fig:nulljunc}) are in general fractional as well.
Clearly, there is always an integer which multiplies a fractional null junction into an integer null junction.
Then, considering the quotient, we find (fractional null junctions)/(integer null junctions) $\cong \text{coker}(\kappa)_\text{tors}$.

For the various brane configurations that realize Kodaira fibers, we list the generators of local fractional null junctions (also known as \emph{extended weight junctions} \cite{DeWolfe:1998zf}), as well as their fractional prongs ending on the brane constituents of the central brane stack, in \eqref{eq:frac_null_junc_list} \cite{DeWolfe:1998zf,Guralnik:2001jh}.\footnote{We denote the fractional prongs ending on the individual constituents according to their ordering in the second column.}
As is evident from this table, we can identify (fractional null junctions)/(integer null junctions) $\cong Z(G_\text{sc})$.
\begin{table}[ht]
\begin{align}\label{eq:frac_null_junc_list}
\renewcommand{\arraystretch}{1.25}
\begin{array}{| c | c | c |}
\hline
\text{Kodaira / $\mathfrak{g}$} & \text{branes} & \text{generating fractional junction}_{(p,q)_{\text{asymp}}} \\ \hline \hline
I_N \, /\, \mathfrak{su}(N) & A^N & \big\{ \tfrac{1}{N}, \dots, \tfrac{1}{N} \big\}_{(1,0)} \\ \hline
II \, / \, -& A C & - \\ \hline
III \, / \, \mathfrak{su}(2) & A^2 C & \big\{ \tfrac{1}{2}, \tfrac{1}{2}, 0 \big\}_{(1,0)} \, , \quad \big\{\tfrac12, \tfrac12, -1 \big\}_{(0,1)} \\ \hline
IV \, / \, \mathfrak{su}(3) & A^3 C & \big\{ \tfrac{1}{3}, \tfrac{1}{3}, \tfrac{1}{3}, 0 \big\}_{(1,0)} \, , \quad \big\{ \tfrac13,\tfrac13,\tfrac13,-1 \big\}_{(0,1)} \\ \hline
\multirow{2}{*}{$I^*_{n-4} \, / \, \mathfrak{so}(2n)$} & \multirow{2}{*}{$A^{4 + 2n} B C$} & \{ 0,...,0, \tfrac12, \tfrac12 \}_{(1,0)} \\ 
& & \{ \tfrac12,...,\tfrac12, \tfrac{-n-1}{2}, \tfrac{1-n}{2} \}_{(0,1)} \\\hline
\multirow{2}{*}{$IV^* \, / \, \mathfrak{e}_6$} & \multirow{2}{*}{$A^5 B C^2$} & \{ -\tfrac13, ..., -\tfrac13, \tfrac43 , \tfrac23 ,\tfrac23\}_{(1,0)} \\ 
& &   \{ 1, ..., 1, -3,-1,-1\}_{(0,1)} \\ \hline
\multirow{2}{*}{$III^* \, / \, \mathfrak{e}_7$} & \multirow{2}{*}{$A^6 B C^2$} & \{ -\tfrac12, ..., -\tfrac12, 2,1,1\}_{(1,0)} \\ 
& &   \{ \tfrac32, ..., \tfrac32, -5,-2,-2\}_{(0,1)} \\ \hline
\multirow{2}{*}{$II^* \, / \, \mathfrak{e}_8$} & \multirow{2}{*}{$A^7 B C^2$} & \{ -1,...,-1,4,2,2\}_{(1,0)}\\
& & \{ 3,...,3 -11, -5, -5\}_{(0,1)} \\ \hline
\end{array}
\end{align}
\end{table}

In a global model, the requirement of vanishing asymptotic charge ``selects'' a subgroup of the center $Z(G^{(i)}_\text{sc})$ of the $i$-th brane stack, represented by a local fractional null junction, which is then combined with the fractional junctions of other stacks into a global fractional null junction (cf.~right panel of Figure \ref{fig:nulljunc}).
The gauge group $G_\text{global}$ of the full theory then satisfies $\pi_1(G_\text{global}) \cong$(global fractional null junctions)/(global integer null junctions) \cite{Guralnik:2001jh}.

It is important to point out that the null junctions are not actually physical junctions \cite{DeWolfe:1998zf}, as, by construction, they have vanishing pairing with all other junctions.\footnote{Note that the local fractional null junctions have vanishing intersection with all root junctions $\{\sigma_a\}$ in the interior, hence, they can be thought of as trivial linear maps, $\{\sigma_a\} \rightarrow \bbZ$, induced by the junction pairing.}
On the other hand, the asymptotic charge \eqref{eq:asymp_charge_null_junc} is a physical quantity.
Hence, the triviality of the fractional null junction implies that this asymptotic charge, which is carried by the prong stretching to infinity in Figure \ref{fig:nulljunc}, must be generated by the prongs emanating from the individual branes.
The latter can be expressed in terms of a fractional linear combination of the root junctions, as we will demonstrate now.

Let us consider a concrete example of a model with $\mathfrak{su}(N)$ gauge algebra.
This is realized by a stack of $N$ mutually local branes, whose $N-1$ simple roots $\sigma_a$ are single-prong strings which stretch between two consecutive branes.
In the notation of \eqref{eq:frac_null_junc_list}, these are
\begin{align}
  \sigma_a = \{0,..., 0, \underbrace{1}_{a\text{-th}}, -1, 0, ..., 0\} \, .
\end{align}
Then, the fractional junction with asymptotic charge $(p,q)_\text{asymp} = (1,0)$ is
\begin{align}
\begin{split}
\big\{ \tfrac{1}{N}, \dots, \tfrac{1}{N} \big\} & = \{1,0, ..., 0\} + \tfrac{1-N}{N} \{1,-1,0,...\} + \tfrac{2-N}{N} \{0,1,-1,0,...\} + ... + \tfrac{1}{N} \{0,...,1,-1\} \\
& = \{1,0,...,0\} + \sum_{a=1}^{N-1} \tfrac{a-N}{N} \sigma_a = \{1,0,...,0\} + \sum_{a=1}^{N-1} (-C^{-1})_{1,a} \, \sigma_a \, ,
\end{split}
\end{align}
where $C$ is the Cartan matrix of $SU(N)$.
This shows the equivalence between different presentations of the asymptotic junction $\big\{ \tfrac{1}{N}, \dots, \tfrac{1}{N} \big\}$, as depicted in Figure \ref{fig:nullroots} for $N=4$.
\begin{figure}[ht]
\centering
\includegraphics[width=0.6\textwidth]{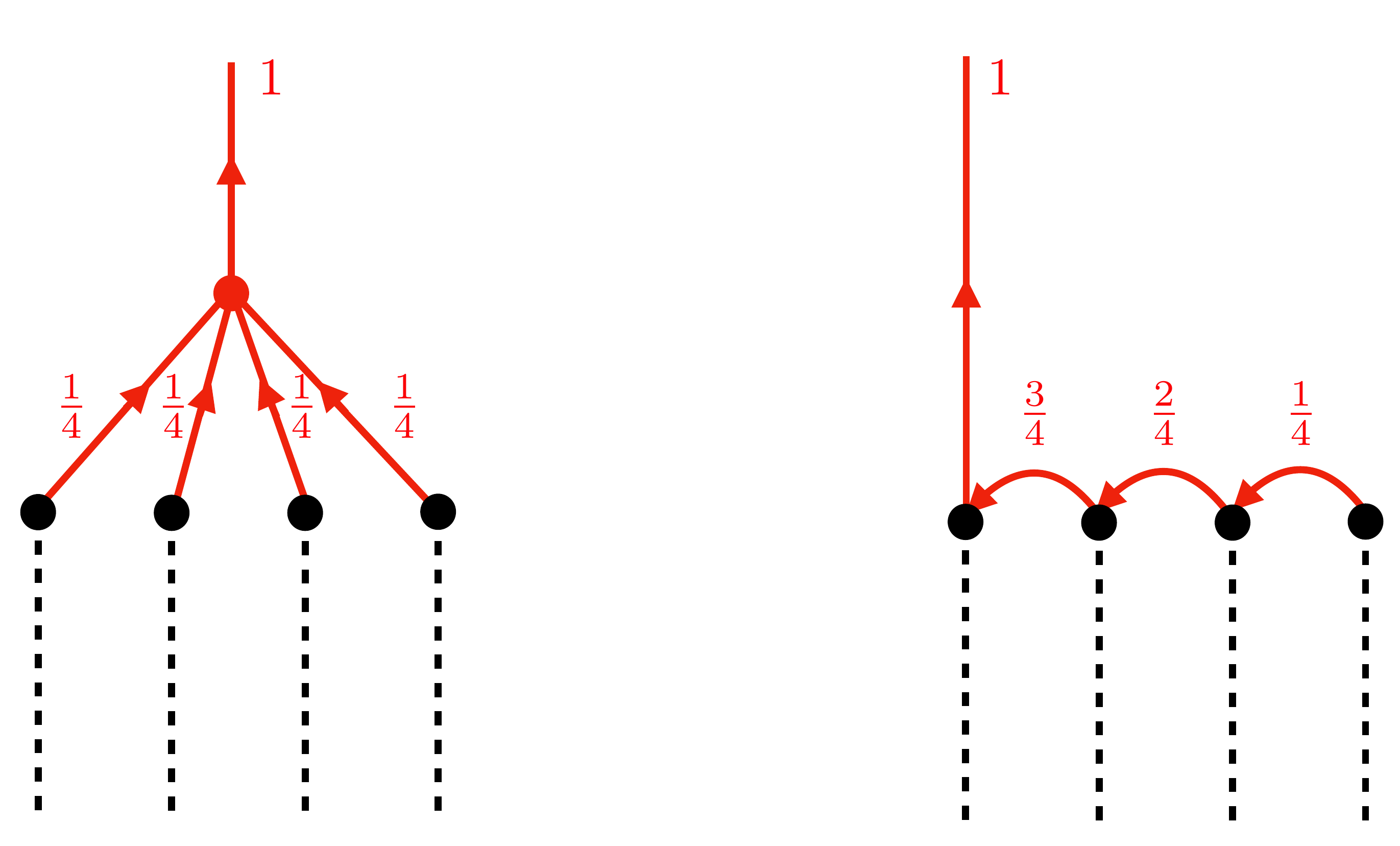}
\caption{Schematic depiction of a local contribution to a fractional null junction in terms of a physical asymptotic junction and a fractional combination of root junctions for $A^4$ stack (which we depicted as separated for convenience).}
\label{fig:nullroots}
\end{figure}

Because the null junction itself is trivial, we see that the integral prongs stretching to infinity, and which carry the asymptotic charge, are a fractional linear combination of the root junctions:
\begin{align}\label{eq:asymp_junction_suN_generator}
  \{1,0,...,0\} \simeq \sum_{a=1}^{N-1} (C^{-1})_{1,a} \sigma_a \, ,
\end{align}
which, up to a sign, takes the same form as the representation \eqref{eq:boundary_torsion_rep_I_N} in terms of the relative homology of the elliptic fibration.
Moreover, we claimed that the fractional null junctions, modulo integer null junctions, represent the center $\bbZ_N$.
That is, $k \in \bbZ_N$ is represented by the fractional null junction $\big\{ \tfrac{k}{N}, ..., \tfrac{k}{N} \big\}$.
To rewrite this in a non-trivial manner, note that the inverse Cartan matrix of $SU(N)$ is \cite{2017arXiv171101294W}
\begin{align}
  (C^{-1})_{ab} = \text{min}(a,b) - \frac{a\,b}{N} \, ,
\end{align}
which satisfies
\begin{align}
\begin{split}
  k \, (C^{-1})_{1,b} & = k \, \frac{b}{N} \mod \bbZ \\
  & = (C^{-1})_{k b} \mod \bbZ \, .
\end{split}
\end{align}
Then, we have
\begin{align}\label{eq:k_frac_junc}
\begin{split}
  \{k,0,...,0\} & \simeq  \sum_{a=1}^{N-1} k (C^{-1})_{1,a} \, \sigma_a \\
  & = \sum_{a=1}^{N-1} (C^{-1})_{ka} \sigma_a + (\text{root junctions}) \, .
\end{split}
\end{align}

\subsection*{1-form Symmetry Charges from String Junctions}

The correspondence between the junction picture and the M-theory description of Section \ref{sec:3app} is on the nose, if we identify string junctions with wrapped M2-branes \cite{Grassi:2013kha}:
a $(p,q)$-prong of a string junction corresponds, in the dual M-theory frame, to M2-branes wrapping a 2-cycle that is the fibration of the 1-cycle
\begin{align}
\mathcal{C} = p \mathcal{A} + q \mathcal{B}
\end{align} 
in the torus fiber over a curve in the base $\mathfrak{B}$.
The splitting of this prong into other prongs $(p_i, q_i)$ corresponds to a linear relation ${\cal C} = \sum_i p_i {\cal A} + q_i {\cal B}$ in $H_1(T^2)$.
The prong can also end on a 7-brane, in which case the cycle ${\cal C}$ is a vanishing cycle in the singular fiber corresponding to the 7-brane.

If a junction has only prongs ending on 7-branes, then these correspond to M2-branes wrapping compact 2-cycles that stretch between fiber singularities in $Y$, i.e., these define elements in $H_2 (Y)$.
In a local model, we have the additional option for a prong to extend to infinity, and thus leaving behind an asymptotic charge $(p,q)_\text{asymp}$.
In the geometry, this then corresponds to a relative cycles in $H_2(Y,\partial Y)$ whose boundary is given by $\mathcal{C} = p{\cal A} + q{\cal B} \subset \partial Y$. 
\begin{figure}
\centering
\includegraphics[width = 0.8 \textwidth]{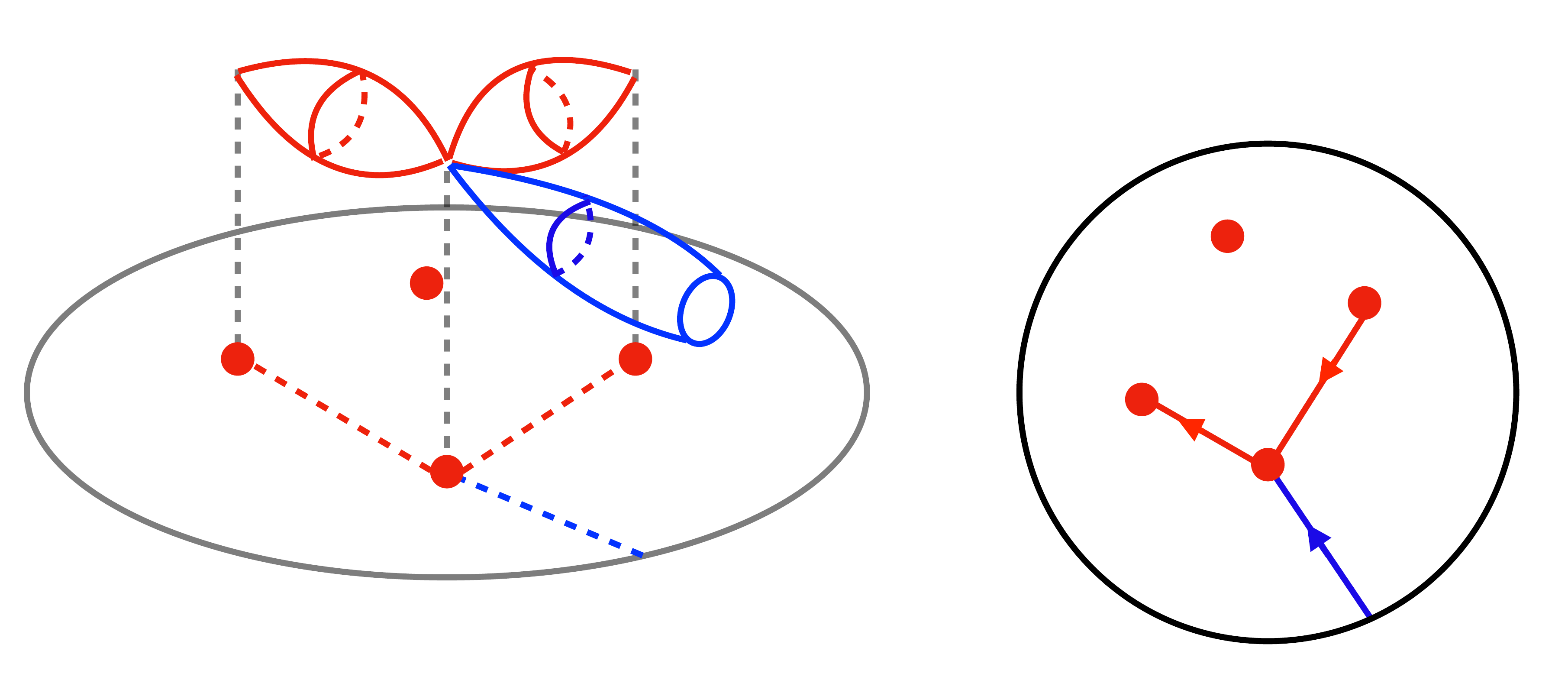}
\caption{The schematic relation, for an $I_4$ singularity, between string junctions and M2-brane states that correspond to roots (red) and asymptotic charges (blue).}
\label{fig:geomjunc}
\end{figure}
Similarly, $\sigma_a \in H_2(Y)$ would have vanishing asymptotic $(p,q)$ charge, and correspond to the root junctions.
See Figure \ref{fig:geomjunc} for a schematic depiction of both types in case of an $I_4$ singularity.\footnote{Note that we are making use here of the equivalence on elliptic surfaces between blow-up resolution, where one introduces 2-cycles into the fiber over a point, and deformation smoothing, where the new 2-cycles arise as fibrations of 1-cycles between the $I_1$ fibers into which a general Kodaira singularity has been deformed into.}

We can now easily identify the geometric counterpart of the local null junctions.
The key is that the asymptotic charges \eqref{eq:asymp_charge_null_junc} of null junction are defined by the same monodromy-induced map $\kappa$ that determines the boundary homology via \eqref{eq:split_mapping_torus_seq}.
Since the set of possible asymptotic $(p,q)$ charges correspond to 1-cycles ${\cal C} = p{\cal A} + q{\cal B}$ on the boundary torus fiber, we immediately see that $\text{coker}(\kappa)_\text{tors} \cong \text{(fractional null junctions)/(integer null junctions)}$.
These boundary 1-cycles are represented in the bulk by the relative 2-cycles \eqref{eq:reltors}, which mirrors the representation of the asymptotic junctions in terms of fractional root junctions in \eqref{eq:asymp_junction_suN_generator} and \eqref{eq:k_frac_junc}.

Without specifying the exact brane content that is encircled by the null junctions, imposing the existence of specific fractional null junctions can restrict the possible $SL(2,\mathbb{Z})$ monodromy induced by the stack. 
Suppose that one demands the existence of fractional null junctions of the form
\begin{align}
\begin{pmatrix} r \\ s \end{pmatrix} = \begin{pmatrix} \tfrac{1}{N} \\ 0 \end{pmatrix} \,.
\label{eq:specnnull}
\end{align}
The requirement of integer asymptotic charge around a stack with monodromy $K = \left(\begin{smallmatrix} a & b \\ c & d \end{smallmatrix} \right) \in SL(2,\mathbb{Z})$ then reads
\begin{align}
(K - \mathbf{1}) \begin{pmatrix} \tfrac{1}{N} \\ 0 \end{pmatrix} = \frac{1}{N} \begin{pmatrix} a-1 \\ c\end{pmatrix} \,,
\end{align}
leading to the constraints $a = 1 \, \text{mod} \, N$, $c = 0 \, \text{mod} \, N$. 
Similarly, one can consider the fractional null junction with \eqref{eq:specnnull}, which encircles the brane stack, and passes its the branch-cut in the opposite direction. 
This corresponds to turning the arrow on the left-hand side of Figure \ref{fig:nulljunc} around. The associated monodromy is generated by $K^{-1}$ and one has
\begin{align}
(K^{-1} - \mathbf{1}) \begin{pmatrix} \tfrac{1}{N} \\ 0 \end{pmatrix} = \frac{1}{N} \begin{pmatrix} d-1 \\ -c \end{pmatrix} \,,
\end{align}
which yields $d = 1 \, \text{mod} \, N$. Together with the constraints above this can be summarized as
\begin{align}
\begin{pmatrix} a & b \\ c & d \end{pmatrix} = \begin{pmatrix} 1 & \ast \\ 0 & 1 \end{pmatrix} \text{ mod } N \,.
\end{align}
This means that the allowed monodromies $K$ are in the congruence subgroup $\Gamma_1 (N)$ of $SL(2,\mathbb{Z})$.\footnote{
In compact models, there are interesting implications for the allowed congruence subgroups \cite{Dierigl:2020lai} imposed by the cobordism conjecture \cite{McNamara:2019rup,Montero:2020icj}.}
An elliptic fibration $Y \stackrel{\pi}{\rightarrow} \mathfrak{B}$ with such a restricted monodromy is known to preserve $N$-torsional points in the fiber \cite{diamond2006first}, which form torsional sections of $\pi$ that are also known to characterize the gauge group topology in F-theory.

\subsection{F-theory and Torsional Sections}
\label{subsec:Fapp}

Sections of an elliptic fibration $\pi: Y_d \rightarrow \mathfrak{B}$ form the so-called Mordell--Weil group, with the zero-section $S_0$ being the neutral element.
It is a finitely generated Abelian group,
\begin{align}
  \text{MW}(\pi) = \mathbb{Z}^s \times \mathbb{Z}_{N_1} \times \mathbb{Z}_{N_2} \, ,
\end{align}
whose torsional part can only contain up to two independent generators, whose orders $N_t$ are bounded by $8$ on compact elliptic fibrations suitable for F-theory models \cite{Hajouji:2019vxs} (see \cite{Park:2011wv,Lee:2019skh,Grassi:2021wii} for discussions on bounds for $s$ in this context).
In compact models, the gauge group is shown to be \cite{Aspinwall:1998xj,Mayrhofer:2014opa,Cvetic:2017epq}
\begin{align}
G = \Big( \prod_i G_{\text{sc},i} \times U(1)^{s} \Big) / (\bbZ_{n_1} \times \dots \times \bbZ_{n_{s}} \times \bbZ_{N_1} \times \bbZ_{N_2}) \, ,
\label{eq:gaugemod}
\end{align}
where $G_{\text{sc},i}$ are the simply-connected non-Abelian groups associated with the gauge algebras $\mathfrak{g}_i$ from 7-branes / singular fibers over (complex) codimension-one loci in $\mathfrak{B}$.
The factors $\mathbb{Z}_{n_i}$, associated to one of the $s$ free generators of the Mordell--Weil group, are always embedded in one of the $U(1)$ factors \cite{Cvetic:2017epq}.
We will ignore these factors, and focus on the finite factors $\mathbb{Z}_{N_t}$ generated by the torsional sections, which are embedded entirely in the non-Abelian factors $G_{\text{sc},i}$ \cite{Aspinwall:1998xj,Mayrhofer:2014opa}.

The divisors $\hat{S}^{(N)} \in H_{d-2}(Y_d)$ associated to an $N$-torsional section $S^{(N)}$ satisfy
\begin{align}
N \big( \hat{S}^{(N)} - \hat{S}_0 - \pi^{-1} (\delta) \big) = \sum_{a} \lambda_a \sigma_a \equiv \sum_{i} \sum_{b=1}^{\text{rank}(\mathfrak{g}_i)} \lambda_{i,b} \sigma_{i,b} \,, \quad \text{with} \enspace \lambda_{i,b} \in \mathbb{Z} \,,
\label{eq:Ftors}
\end{align}
where we have re-grouped the resolution divisors $\sigma_a$ on the right-hand side into their corresponding simple non-Abelian algebra $\mathfrak{g}_i$.
The term $\pi^{-1} (\delta)$ denotes a vertical divisor (i.e., pull-back of a base divisor $\delta$) that depends on the intersection properties between $\hat{S}^{(N)}$ and $\hat{S}_0$, which will not affect the discussion below.
This shows that the elements $\big(\hat{S}^{(N)} - \hat{S}_0 - \pi^{-1} (\delta) \big)$ are torsional up to the contribution of the resolution divisors $\sigma_{i,b}$, which in general dimensions are $\bbP^1$-fibered, with fiber class $\gamma_{i,b}$, over a divisor in $\mathfrak{B}$.
The coefficients $\lambda_{i,b}$ are determined by the so-called Shioda map \cite{10.3792/pjaa.65.268,2001math.....12259W,Park:2011ji,Morrison:2012ei} as follows.
The section $\hat{S}^{(N)}$ intersects at most \emph{one} of the rational fibers $\gamma_{i,b}$ of the divisors $\sigma_{i,b}$, say, $\gamma_{i,k}$, once, i.e., $\langle \hat{S}^{(N)} , \gamma_{i,b} \rangle = \delta_{k,b}$.
Then, we have
\begin{align}\label{eq:coefficients_shioda}
  \frac{\lambda_{i,b}}{N} = \sum_{c = 1}^{\text{rank}(\mathfrak{g}_i)} \langle \hat{S}^{(N)} , \gamma_{i,c} \rangle \, (C_{(i)})^{-1}_{cb} = (C_{(i)})^{-1}_{kb} \, , \quad \text{where } (C_{(i)})_{bc} = -\langle \sigma_{i,b} , \gamma_{i,c} \rangle
\end{align}

It is the existence of the element $\big(\hat{S}^{(N)} - \hat{S}_0 - \pi^{-1} (\delta) \big) \in H_{d-2} (Y_d)$ that restricts the global realization of the gauge group and accordingly the allowed spectrum of charged dynamical fields. 
By M-/F-theory duality, matter states in F-theory descend to M2-branes wrapping 2-cycles, which must have integer intersection pairing with elements in $H_{d-2}(Y_d)$, the existence of the divisor
\begin{align}\label{eq:shioda_map_relation}
\big(\hat{S}^{(N)} - \hat{S}_0 - \pi^{-1} (\delta) \big) = \tfrac{1}{N} \sum_a \lambda_a \sigma_a \,,
\end{align}
imposes, due to the fractional pre-factors $\tfrac{\lambda_a}{N}$, non-trivial constraints on the intersection numbers of 2-cycles with the divisors $\sigma_a$, which in turn determine the $\mathfrak{g}$-representation in which the matter transforms in.
Hence, \eqref{eq:shioda_map_relation} can be interpreted as an element in the cocharacter lattice, which enforces the non-trivial global structure $\pi_1(G) \cong \text{cocharacters} / \text{coroots}$ \cite{Mayrhofer:2014opa}.

While the above results are derived in compact models, the relationship between the monodromy reduction and the invariant torsion points on the generic fiber exist also in local models $Y$. 
If every singular fiber induces a monodromy in a congruence subgroup of $SL(2,\bbZ)$, then, as explained above, $Y$ has some torsional sections.
In a local geometry, sections are non-compact divisors, i.e., sit in $H_{d-2}(Y, \partial Y)$.
A relationship of the form \eqref{eq:shioda_map_relation} then implies that $\hat{S}^{(N)} - \hat{S}_0$ represents a torsional element in $H_{d-2}(Y, \partial Y)/\{\jmath_{d-2}(\sigma_a)\} \subset \Lambda$ in \eqref{eq:mag-charges_general}.\footnote{We have suppressed the vertical part $\pi^{-1}(\delta)$ here to reduce cluttering.
In general, this can be also decomposed into a compact $\pi^{-1}(\delta_c)$ and non-compact piece $\pi^{-1}(\delta_{nc})$.
The torsional generator for $H_{d-2}(Y, \partial Y)/\text{im}(\jmath_{d-2})$ is then $\hat{S}^{(N)} - \hat{S}_0 - \pi^{-1}(\delta_{nc})$.
}

This can be most easily seen for $d=4$, i.e., F-theory compactifications to eight dimensions.
Consider, for concreteness, a single $I_N$ fiber, corresponding to $\mathfrak{g} = \mathfrak{su}(N)$.
In this case, $\sigma_a = \gamma_a$, and the generic fiber $\mathfrak{f}$, which satisfies $\langle \mathfrak{f} , \mathfrak{f} \rangle = \langle \mathfrak{f} ,\sigma_a \rangle = 0$, form a basis of $H_2(Y_4)$.
The monodromy around the fiber preserves $N$-torsional points, which in the (resolved) $I_N$ fiber are situated on one of the $N$ fiber components ($\sigma_a$ and the affine component, $\sigma_0 := \mathfrak{f} - \sum_{a=1}^{N-1} \sigma_a$) each.
By ``fibering'' each point over the non-compact base $\mathfrak{B}$, we obtain a non-compact 2-cycle $\hat{S}^{(N)}_k$.
That is, they each define a class in $H_2(Y_4, \partial Y_4) \cong \text{Hom}(H_2(Y_4), \bbZ)$, characterized by the ``intersection'' with $\sigma_a$, $\hat{S}^{(N)}_k : \sigma_a \mapsto \delta_{a,k}$ for $0 \leq a,k \leq n-1$, which also implies $\hat{S}_k^{(N)}(\mathfrak{f}) = 1$ for any $k$.
Note that the zero-section is the one meeting the affine node, i.e., $\hat{S}_0 = \hat{S}^{(N)}_0$. 
With $\langle \sigma_a, \sigma_b \rangle = -C_{ab}$, it is straightforward to check that, for $k \neq 0$,
\begin{align}
\begin{split}
  & \sum_{b=1}^{N-1} (C^{-1})_{kb} \, \langle \sigma_b, \sigma_a \rangle = \delta_{k,a} = \left(\hat{S}_k^{(N)} - \hat{S}_0 \right)(\sigma_a)\, , \qquad a=1,...,N-1 \, , \\
  & \sum_{b=1}^{N-1} (C^{-1})_{kb} \, \langle \sigma_b, \mathfrak{f} \rangle = 0 = \left(\hat{S}_k^{(N)} - \hat{S}_0 \right)(\mathfrak{f}) \, ,
\end{split}
\end{align}
showing that
\begin{align}\label{eq:shioda_as_relative_cycle}
  \hat{S}_k^{(N)} - \hat{S}_0= \sum_{b=1}^{N-1} \underbrace{(C^{-1})_{kb}}_{\equiv \lambda_b / n} \, \langle \sigma_b , \cdot \rangle \in \text{Hom}(H_2(Y_4),\bbZ) \cong H_2(Y_4, \partial Y_4) \, .
\end{align}
The coefficients $\lambda_b$ are precisely as defined in \eqref{eq:coefficients_shioda};
since $(-C)^{-1}$ is the inverse Cartan matrix of $SU(N)$, $N$-times any of its entries is integral, thus showing \eqref{eq:Ftors}.\footnote{For non-compact elliptic surfaces, the vertical part $\pi^{-1}(\delta) \cong m \times \mathfrak{f}$ for some $m \in \bbZ$ always defines a trivial map, $\langle \mathfrak{f}, \cdot \rangle = 0 \in \text{Hom}(H_2(Y_4), \bbZ)$.}
Since $\langle \sigma_c, \cdot \rangle = \jmath_{d-2}(\sigma_c)$ in \eqref{eq:j_map_general}, relations of the sort \eqref{eq:Ftors} directly identify the torsional section (more precisely, the linear combination $\hat{S}^{(N)}_k - \hat{S}_0$) as a representative higher-form symmetry charges \eqref{eq:boundary_torsion_rep_I_N}.
Note that this expression also agrees with the relationship between the asymptotic and root junctions \eqref{eq:k_frac_junc}, serving as further proof that the two concepts are equivalent.

\subsection*{Torsional Sections in the Boundary Homology}

A relationship of the form \eqref{eq:shioda_as_relative_cycle} implies that the 2-cycle $\hat{S}^{(N)}_k - \hat{S}_0$ maps to a torsion element in $H_2(Y_4, \partial Y_4)/\text{im}(\jmath_2) \cong \text{im}(\partial_2) \subset H_1(\partial Y)$.
To see this explicitly, consider the points, $z_\text{tors}$ and $z_0$, marked by the two sections $S^{(N)}_k$ and $S_0$, respectively, on a reference $T^2$ fiber $\mathfrak{f}_p$ of the boundary fibration $\partial Y \rightarrow \partial \mathfrak{B} \cong S^1$.
Then, $z_\text{tors}$ traces out the 1-cycle $\partial_2(\hat{S}^{(N)}_k) = \hat{S}^{(N)}_k|_{\partial Y}$ on $\partial Y$, as we move it through the family of fibers over the base $S^1$; the same applies to $z_0$ tracing out $\partial_2(\hat{S}_0) = \hat{S}_0|_{\partial Y}$.
In any individual fiber, the two points are homologous.
But, by encircling the base $S^1$ once, the monodromy $K$ ``twists'' the section $S^{(N)}_k$ around the zero-section.
This twist corresponds to the 1-cycle $\big( \hat{S}^{(N)}_k - \hat{S}_0 \big)\big|_{\partial Y} = {\cal C} \in H_1(\partial Y)$.

To quantify this twist, we can use the standard presentation of the torus as $\bbC / (m \, \tau_p + n )$, with $z_0 \mapsto 0$, where $\tau_p$ is the complex structure of the torus $\mathfrak{f}_p$.
Then, every point on the torus can be represented as $\left( \begin{smallmatrix} x \\ y \end{smallmatrix} \right) \equiv x \, \tau_p+ y + (m \, \tau_p + n)$ with $0 \leq x,y < 1$.
The monodromy map $K$ acts via matrix multiplication, $\left( \begin{smallmatrix} x \\ y \end{smallmatrix} \right) \mapsto K \left( \begin{smallmatrix} x \\ y \end{smallmatrix} \right)$, which fixes $z_0 = \left( \begin{smallmatrix} 0 \\ 0 \end{smallmatrix} \right)$.
In the covering space $\bbC$ of the torus, this defines a translation by $(K-{\bf 1}) \left( \begin{smallmatrix} x \\ y \end{smallmatrix} \right)$, which on the quotient $\bbC / (m \,  \tau_p+ n)$ corresponds to a 1-chain ${\cal C}$.

Being preserved under the monodromy now precisely means that $z_\text{tors} \equiv \left( \begin{smallmatrix} x_t \\ y_t \end{smallmatrix} \right)$ maps onto itself in $\bbC / (m \, \tau_p+ n)$.
That is, the chain ${\cal C} = a {\cal A} + b {\cal B}$ is a 1-\emph{cycle} on $\mathfrak{f}_p$, expressed in terms of the ${\cal A}$ and ${\cal B}$ cycles, with coefficients given by $(K - {\bf 1}) \left( \begin{smallmatrix} x_t \\ y_t \end{smallmatrix} \right) = \left( \begin{smallmatrix} a \\ b \end{smallmatrix} \right)$ with $a,b \in \bbZ$.
Finally, the fact that $z_\text{tors}$ is an $N$-torsional point means that $(x_t, y_t) = (\chi/N, \upsilon/N)$ for some $\chi, \upsilon \in \{0,...,N-1\}$ (see, e.g., \cite{diamond2006first}).\footnote{By definition, the Mordell--Weil group law for sections in elliptic fibration is just the fiberwise addition of points, the latter of which can be represented as the ``usual'' addition in $\bbC / (m\tau_p + n)$. This makes the given presentation of the torsion points apparent.}
Therefore, we see from $N \left( \begin{smallmatrix} a \\ b \end{smallmatrix} \right) = (K - {\bf 1}) \left( \begin{smallmatrix} N x_t \\ N y_t \end{smallmatrix} \right) = (K - {\bf 1}) \left( \begin{smallmatrix} \chi \\ \upsilon \end{smallmatrix} \right)$ that $N {\cal C} \in \text{im}(K - {\bf 1}) = \text{im}(\kappa)$.
From \eqref{eq:boundhom}, we see that ${\cal C} = \big( \hat{S}^{(N)}_k - \hat{S}_0 \big)\big|_{\partial Y}$ indeed represents an $N$-torsional element in $\text{coker}(\kappa) \subset H_1(\partial Y)$.

\section{Anomalies of 1-Form Center Symmetries in M-theory}
\label{sec:anomalies}

The defect group structure represents an 't Hooft anomaly between the electric 1-form and magnetic $(D - 3)$-form symmetry \cite{Freed:2006ya,Freed:2006yc,Albertini:2020mdx}.
Another such potential anomaly involving the 1-form center symmetry arises in spacetime dimension $D \geq 5$ \cite{Apruzzi:2020zot,Cvetic:2020kuw,BenettiGenolini:2020doj}, as a generalization of the ``anomaly in the coupling-space'' \cite{Cordova:2019jnf,Cordova:2019uob} in $D=4$, where a non-trivial background field for the 1-form center symmetry affects the periodicity of the theta angle \cite{Aharony:2013hda}.
In $D\geq 5$ dimensions, this turns into a genuine mixed 't Hooft anomaly between 1-form center symmetries and $(D-5)$-form $U(1)_I$ instanton symmetries.\footnote{
The anomaly restricts possible gaugings of center symmetries --- i.e., it affects physically allowed global gauge groups --- whenever $U(1)_I$ must be gauged for consistency \cite{Apruzzi:2020zot,Cvetic:2020kuw}.}

In this section, we discuss the origin of the anomaly in gauge theories from M-theory compactifications on Calabi--Yau spaces $Y_d$ ($\dim_\mathbb{R} Y_d \equiv d = 4,6$).
As we will see, one way to derive the anomaly is to reduce the 11d Chern--Simons term in the presence of boundary fluxes that parametrize the 1-form symmetry background.
Similar to the discussion in Section \ref{sec:3app}, the computation is performed in the Abelian phase, i.e., on the Coulomb branch of the ${\cal N}=1$ gauge theory in 7d or 5d, which corresponds to a desingularized internal space $Y_d$.
We then interpret the result in the singular / non-Abelian limit, as well as in cases that admit an 8d / 6d F-theory description.
Note that there could be counterterms / topological sectors which (partly) cancel this anomaly field-theoretically.
These will not be captured by our analysis of the Chern--Simons term in the presence of boundary fluxes, but could manifest in other aspects of the M-theory geometry, see \cite{Apruzzi:2021vcu} for a recent discussion.

\subsection{Background Fields for 1-Form Symmetries in M-theory}

Consider M-theory on a spacetime ${\cal M} = M_{11-d} \times Y_d$, where the $d$-dimensional ``internal'' space $Y_d$ is non-compact with asymptotic boundary $\partial Y_d$.
Assuming that $M_{11-d}$ has a topologically trivial boundary, boundary fluxes of the M-theory 3-form potential $C_3$ are then encoded in fluxes on $\partial Y_d$.
More precisely, dual to \eqref{eq:long_seq_rel_hom}, there is a long exact sequence,
\begin{align}\label{eq:long_seq_cohomology}
\ldots \rightarrow H^n({\cal M}, \partial {\cal M}) \stackrel{\hat\jmath_n^*}{\rightarrow} H^n({\cal M}) \stackrel{\hat\imath^*_n}{\rightarrow} H^n(\partial {\cal M}) \rightarrow H^{n+1}({\cal M}, \partial {\cal M}) \rightarrow \ldots
\end{align}
involving the relative cohomology $H^n({\cal M}, \partial {\cal M})$.
A non-trivial boundary flux of $C_3$ corresponds to an element in $\text{im}(\hat\imath^*_4) \subset H^4(\partial {\cal M})$, where $\hat\imath^*_n$ is the map on $n$-forms induced by the natural inclusion $\hat\imath: \partial {\cal M} \rightarrow {\cal M}$, and is the cohomological version of $\imath_n$ in \eqref{eq:long_seq_rel_hom}.
With ${\cal M} = M_{11-d} \times Y_d$ and the assumption that $M_{11-d}$ is closed, i.e., $\partial {\cal M} = M_{11-d} \times \partial Y_d$, the boundary fluxes are encoded in the map
\begin{align}
\begin{aligned}
	H^4({\cal M}) \cong \bigoplus_{p+q=4} H^p(M_{11-d}) & \otimes H^q(Y_d) &&\stackrel{\imath^*_4}{\longrightarrow} && \bigoplus_{p+q=4} H^p(M_{11-d}) \otimes H^q(\partial Y_d) \cong H^4(\partial {\cal M})\, , \\
	F & \otimes \omega && \mapsto && F \otimes \imath_q^*(\omega) \, ,
\end{aligned}
\end{align}
with $\imath^*_q: H^q(Y_d) \rightarrow H^q(\partial Y_d)$ the analogous map in the long exact sequence \eqref{eq:long_seq_cohomology} associated to the relative cohomology for $\partial Y_d \subset Y_d$.

In the following, we focus on $p=q=2$, as gauge fields $A_a$ in $M_{11-d}$ arise from the Kaluza--Klein decomposition of the M-theory 3-form potential,
\begin{align}\label{eq:C3_KK_decomp}
	C_3 = C^{(M)}_3 + \sum_{\mathrm{w}_a \in H^2(Y_d)} A_a \wedge \mathrm{w}_a \, ,
\end{align}
where $C^{(M)}_3$ is a 3-form in $M_{11-d}$.
If we only include $\mathrm{w}_a = \jmath_2^*(\omega_a) \in H^2_{\text{cpt}}(Y_d) \equiv \text{im}(\jmath_2^*) = \text{ker}(\imath^*_2)$ with compact support, the associated $A_a$'s correspond to the Cartan $U(1)$s of dynamical gauge symmetries.
The flux, or the field strength, of such a configuration is then
\begin{align}
	G_4 = G_4^{(M)} + \sum_{\omega_a \in H^2(Y_d, \partial Y_d)} F_a \otimes \jmath_2^*(\omega_a) \, ,
\end{align}
where $F_a$ represents the first Chern-class of a line bundle in $H^2(M_{11-d})$, and corresponds to the field strength of the gauge field $A_a$.
In the following, we will assume $G_4^{(M)}=0$, as we are only interested in contributions to 2-form backgrounds in $M_{11-d}$.

Non-trivial ``boundary'' fluxes are labelled by elements in $ H^2(Y_d) / \text{im}(\jmath_2^*) = H^2(Y_d) / \text{ker}(\imath^*_2)$ $\cong \text{im}(\imath^*_2) \subset H^2(\partial Y_d)$ \cite{Morrison:2020ool}.
They can be represented by classes $\tilde\omega_k \in H^2(Y_d)$ with $\imath^*_2(\tilde\omega_k) \neq 0$.
``Turning on'' these boundary fluxes means that we include additional terms
\begin{align}\label{eq:G_4_with_boundary_fluxes}
	G_4 = \sum_{a} F_a \otimes \jmath_2^*(\omega_a) + \sum_k B_k \otimes \tilde\omega_k\, .
\end{align}
These additional terms can be related to the electrically and magnetically charged objects, \eqref{eq:1-form-charges_general} and \eqref{eq:mag-charges_general}, of the higher-form symmetry.
By virtue of the commutative diagram via Poincar\'e--Lefschetz duality (see, e.g., \cite{hatcher2002algebraic}),
\begin{equation}\label{eq:comm_diag}
	\begin{tikzcd}[row sep = normal, column sep= normal  ]
		\ldots \arrow[r] & H^1(\partial Y_d) \arrow{r}{}{d_1} \arrow{d}{}{\cong} & H^2(Y_d, \partial Y_d) \arrow{r}{}{\jmath_2^*} \arrow{d}{}{\cong} & H^2(Y_d) \arrow{r}{}{\imath_2^*} \arrow{d}{}{\cong} & H^2(\partial Y_d) \arrow{d}{}{\cong} \arrow[r] & \ldots \\
		\ldots \arrow[r] & H_{d-2}(\partial Y_d) \arrow{r}{}{\imath_{d-2}} & H_{d-2}(Y_d) \arrow{r}{}{\jmath_{d-2}} & H_{d-2}(Y_d, \partial Y_d) \arrow{r}{}{\partial_{d-k}} & H_{d-3}(\partial Y_d) \arrow[r] & \ldots
	\end{tikzcd}
\end{equation}
we have $H^2(Y_d) / \text{im}(\jmath^*_2) \cong H_{d-2}(Y_d, \partial Y_d) / \text{im}(\jmath_{d-2}) = \Lambda_\text{mag.} \cong \bbZ^f \oplus \Gamma$ from \eqref{eq:explicit_formula_Gamma}.

Thus, the additional terms $B_k \otimes \tilde{\omega_k}$ in $G_4$ arrange into
\begin{align}\label{eq:free_and_torsion_fluxes}
	 H^2(M_{11-d}) \otimes (\mathbb{Z}^f \oplus \Gamma) \cong H^2(M_{11-d})^{\otimes f} \oplus H^2(M_{11-d}; \Gamma) \, .
\end{align}
Contributions in $H^2(M_{11-d})^{\otimes f}$ correspond to background gauge fields of flavor symmetries in $M_{11-d}$.
As they are in the free part of $H^2(\partial Y_d) \cong H_{d-3}(\partial Y_d)$, they have trivial linking pairing\footnote{This is dual to the pairing \eqref{eq:linking_pairing_homology} in the boundary homology.} with any other boundary flux, and hence commutes with any other flux background.
We will return to these backgrounds later, and first focus on the torsional part $H^2(M_{11-d}; \Gamma)$, which physically correspond to background fields for global 1-form $\Gamma$ symmetries in $M_{11-d}$.

Turning on a background flux in $H^2(M_{11-d}; \Gamma) \ni b \equiv B \otimes \tilde\omega$ corresponds to a \emph{torsional} internal flux $\tilde\omega_t \in \Gamma \subset H^2(Y_d) / \text{ker}(\imath^*_2)$, i.e., $\imath_2^*(\tilde\omega_t) \neq 0 \in H^2(\partial Y_d)$, but $N \, \tilde\omega_t \in \text{ker}(\imath_2^*) = \text{im}(\jmath_2^*)$ for some $N \in \mathbb{N}_0$.
This means that there is an integer linear combination
\begin{align}\label{eq:linear_relation_boundary_flux}
	N \, \tilde\omega_t = \sum_a (S^{-1})_{ta} \, \jmath_2^*(\omega_a) \equiv \sum_a \lambda_a \, \jmath_2^*(\omega_a) \, ,
\end{align}
which is Poincar\'e-dual to the homology relation \eqref{eq:lin_rel_homology}.
Since $\tilde\omega_t$ is only defined modulo $\text{ker}(\imath^*_2)= \text{im}(\jmath_2^*)$, we can restrict $\lambda_a \in \{0,...,N-1\}$.
Thus, we can formally write
\begin{align}\label{eq:G4_shifted_U1s}
	G_4 = \sum_a F_a \otimes \jmath_2^*(\omega_a) + B \otimes \tilde\omega_t = \sum_a \left( F_a + \frac{\lambda_a}{N} B \right) \otimes \jmath_2^*(\omega_a) \, ,
\end{align}
which can be interpreted as a $N$-fractional shift of the Cartan fluxes by the 1-form symmetry.
This interpretation agrees with the field theoretic description of 1-form symmetry transformation in the Abelian phase of the gauge theory \cite{Hsin:2018vcg,Cordova:2019uob}.
Note that only the $N$-fractional part of the shift to the Cartan fluxes in \eqref{eq:G4_shifted_U1s} is well-defined, since the boundary flux $\tilde\omega_t$ is only defined modulo $H^2_\text{cpt}(Y_d)$.

The presentation \eqref{eq:G4_shifted_U1s} has the advantage that we can ``straightforwardly'' perform the usual KK-reduction of the M-theory Chern--Simons term 
\begin{align}\label{eq:11d-CS-term}
	\frac16 \int_{\cal M} C_3 \wedge G_4 \wedge G_4 \, .
\end{align}
By that, we mean that the integral is strictly speaking only defined for compactly supported cohomology forms on non-compact spaces.
Presumably, a mathematically more rigorous definition of this coupling in terms of differential cohomology classes \cite{Witten:1996hc,Witten:1999vg,Belov:2006jd,Belov:2006xj,Fiorenza:2012mr,Fiorenza:2012ec}, which we will not attempt to utilize here, can encompass contributions from both compactly supported and boundary fluxes.
For the $N$-torsional fluxes that parametrize the 1-form symmetry backgrounds, \eqref{eq:G4_shifted_U1s} allows us to circumvent this process and evaluate the integrals of products of the compactly supported 2-forms $\jmath_2^{*} (\omega_a)$, albeit with the fractional coefficients.
As we will see, this approach is sufficient to derive the \emph{fractionalization}, i.e., a fractional shift of the instanton density of gauge theories in the presence of a 1-form symmetry background that matches field theory results.

\subsection{Compactification to 7d}
\label{subsec:7d_anomalies}

Let us apply the above results to $\dim Y_d \equiv d = 4$, i.e., M-theory compactified to seven dimensions.
In the case the ansatz \eqref{eq:C3_KK_decomp} for $C_3$ includes only compactly supported fluxes in $Y_4$, the reduction of the 11d Chern--Simons term \eqref{eq:11d-CS-term} produces the term
\begin{align}\label{eq:CS-term-7d}
	& \frac{1}{6} \int_{{\cal M}_{11}} C_3 \wedge G_4 \wedge G_4 = \frac12 \sum_{a,b} \int_{M_{7}} C_3^{(M)} \wedge  F_a \wedge F_b \times \int_{Y_4} \jmath_2^*(\omega_a) \wedge \jmath_2^*(\omega_b) \, ,
\end{align}
where the factor of 3 comes from the assumption that the boundary of the 7d spacetime $M_7$ is trivial, allowing for Stoke's theorem on $G_4^{(M)} = dC_3^{(M)}$.
If $F_a$ are the Cartan $\mathfrak{u}(1)$s of a non-Abelian gauge symmetry $\mathfrak{g}$ in $M_7$, then $\int_{Y_4}\jmath_2^*(\omega_a) \wedge \jmath_2^*(\omega_b) \equiv -C_{ab}$ is the (negative) Cartan matrix of $G$.
Moreover, it coincides with the matrix $M_{ai}$ in \eqref{eq:smith_decomp}, where the basis $\sigma_a \in H_{d-2}(Y_4) = H_2(Y_4)$ and $\gamma_i \in H_2(Y_4)$ is formed by the 2-cycles dual to $\omega_a \in H^2(Y_4)$ in both cases.
In the non-Abelian limit (i.e., when we blow-down the compact curves dual to $\omega_a$ in $Y_4$), this term produces the 7d-coupling
\begin{align}\label{eq:7d_instanton_coupling}
	\sum_{a,b} \int_{M_{7}} \frac{(-C_{ab})}{2} \, C_3^{(M)} \wedge  F_a \wedge F_b \rightarrow \int_{M_7} C_3^{(M)} \wedge \text{Tr}(F^2)
\end{align}
between the instanton density $\text{Tr}(F^2)$\footnote{The trace is normalized such that a 1-instanton configuration integrates to an integer over any integer 4-cycle in $M_7$.} of $G$ and the 3-form $C_3^{(M)}$.

Including an $N$-torsional boundary flux $\tilde\omega_t$, and its fractional shift \eqref{eq:G4_shifted_U1s} it induces on the Cartan $\mathfrak{u}(1)$s, the coupling becomes\footnote{We have implicitly used the ``continuum description'' \cite{Kapustin:2014gua,Gaiotto:2014kfa,Gaiotto:2017yup} for the 1-form background gauge field $B$ as an ordinary differential form, for which the wedge product makes sense.
Regarding $B \in H_2(M_D; \Gamma)$ as a differential cohomology class, one should replace $B \wedge B$ by the Pontryagin square operation $\mathfrak{P}(B)$.
}
\begin{align}\label{eq:CS-term_reduced_to_7d}
\begin{split}
	& \sum_{a,b} \frac{(-C_{ab})}{2} \int_{M_7} C_3^{(M)} \wedge \left( F_a + \frac{\lambda_a}{N} B \right) \wedge \left( F_b + \frac{\lambda_b}{N} B \right) \\
	= & \sum_{a,b} \int_{M_7} C_{ab} \, C_3^{(M)} \wedge \left(\frac12 F_a \wedge F_b + \frac{\lambda_a}{N} F_b \wedge B + \frac{\lambda_a \lambda_b}{2N^2} B \wedge B \right) .
\end{split}
\end{align}
With $-C_{ab} \equiv M_{ab}$ in \eqref{eq:smith_decomp}, we see from \eqref{eq:lambda_via_smith_decomp} that 
\begin{align}
	\sum_a C_{ab} \frac{\lambda_a}{N} = \sum_{a,c,j} S_{ac} \, D_{cj} \, T_{jb} \frac{(S^{-1})_{ta}}{N} = \sum_j \frac{D_{tj} \, T_{jb}}{N} = \frac{n_t}{N} T_{tb} = T_{tb} \in \bbZ \, ,
\end{align}
because $N \equiv n_t$ is the torsion order of the boundary flux $\tilde\omega_t$ that we turned on.
This means that the cross terms $C_3^{(M)} \wedge F_b \wedge B$ in \eqref{eq:CS-term_reduced_to_7d} actually have integer coefficients.

Since $F_a$ and $B$ are all integer 2-forms (more precisely, 2-cocycles) in $M_7$, we see that a non-trivial 1-form symmetry background corresponding to $\tilde\omega_t$ leads to a shift
\begin{align}
	\sum_{a,b} \frac{-C_{ab}}{2} \, F_a \wedge F_b + \sum_{a,b} (-C_{ab}) \frac{\lambda_a \, \lambda_b}{2N^2} B \wedge B + \text{integer contributions} \, .
\end{align}
In the non-Abelian limit, the instanton coupling \eqref{eq:7d_instanton_coupling} thus becomes
\begin{align}\label{eq:shift_instanton_general}
	\int_{M_7} C_3^{(M)} \wedge \text{Tr}(F^2) \rightarrow \int_{M_7} C_3^{(M)} \wedge \Big( \text{Tr}(F^2) + \frac{1}{2N} \underbrace{\sum_{a,b} (-C_{ab}) \frac{\lambda_a \lambda_b}{N}}_{= -\sum_b T_{tb} (S^{-1})_{tb} \in \bbZ} B \wedge B + \text{integer 4-form} \Big) .
\end{align}

This fractional shift leads to an 't Hooft anomaly between the 1-form center symmetry, and the large gauge transformations of the $U(1)$ symmetry $C_3^{(M)} \rightarrow C_3^{(M)} + \Lambda^{(M)}_3$, where $\Lambda^{(M)}_3$ is a closed 3-form \cite{Apruzzi:2020zot,Cvetic:2020kuw,BenettiGenolini:2020doj}.
While $C_3^{(M)}$ is a background field in 7d when gravity is decoupled, in supergravity, it becomes the dynamical field dual of the anti-symmetric 2-tensor in the gravity multiplet, and as such must enjoy an unbroken $U(1)$ symmetry.
A mixed anomaly with a 1-form symmetry thus prevents the gauging of this 1-form symmetry, and thus restricts possible $\pi_1(G)$, if the 1-form symmetry corresponds to a center symmetry.
It is straightforward to uplift this to term and the anomaly to 8d, where $C_3^{(M)}$ now becomes a 4-form gauge potential $B_4$ coupling to the instanton density, with analogous implications for global gauge group structures in 8d supergravity \cite{Cvetic:2020kuw}.

\paragraph{Example} Consider M-theory on $Y_4 = \mathbb{C}^2/\bbZ_N$, which gives rise to a 7d theory with $G = SU(N)$.
The corresponding exact sequence in relative homology \eqref{eq:comm_diag} collapses in this case to a short exact sequence (see, e.g., \cite{Garcia-Etxebarria:2019cnb}),
\begin{align}\label{eq:relative_hom_sequence_C2/Gamma}
\begin{split}
	& 0 \longrightarrow H_2(Y) \stackrel{\jmath_2}{\longrightarrow} H_2(Y, \partial Y) \stackrel{\partial_2}{\longrightarrow} H_1(\partial Y) \rightarrow 0 \, , \\
	\text{with} \quad &  H_2(Y, \partial Y) \cong \text{Hom}(H_2(Y) , \bbZ) \, , \quad H_1(\partial Y) \cong \bbZ_N \, .
\end{split}
\end{align}
For $Y_4 = \mathbb{C}^2/\bbZ_N$, it is well-known that $H_2(Y)$ is spanned by $N-1$ $\bbP^1$'s ($\bbP^1_a$, $a=1,...,N-1$), which intersect each other in the form of an $SU(N)$ Dynkin diagram, that is, 
\begin{align}\label{eq:suN_cartan_matrix_7d}
	C_{ab} = \langle \bbP^1_a , \bbP^1_b \rangle = 
	\begin{pmatrix}
		-2 & 1 & 0 & \ldots & 0 \\
		1 & -2 & 1 & \ddots & 0 \\
		0 & 1 & -2 & \ddots & \vdots \\
		\vdots & \ddots & \ddots & \ddots \\
		0 & 0 & \ldots & 1 & -2
	\end{pmatrix} \, .
\end{align}
A Smith decomposition $C = S D T$ yields
\begin{align}\label{eq:smith_decomp_suN_cartan}
\begin{split}
	& S = 
	\begin{pmatrix}
		1 & \ldots & 1 & \ldots & \ldots & 1 \\
		\vdots & 2 & 2 & \ldots & \ldots & 2 \\
		1 & 2 & 3 & \ldots & \ldots & 3 \\
		\vdots & \vdots & 3 & \ddots & & \vdots \\
		\vdots & \vdots & \vdots & & N-2 & N-2\\
		1 & 2 & 3 & \ldots & N-2 & N-1
	\end{pmatrix}^{-1} \, , \quad 
	T = 
	\begin{pmatrix}
		1 & 0 & \ldots & 0 & 1 \\
		0 & 1 & \ddots & \vdots & 2  \\
		\vdots & \ddots & \ddots &  & \vdots \\
		0 & 0 & \ldots & 1 & N-2 \\
		0 & 0 & \ldots & & 1
	\end{pmatrix} \, ,\\
	& D = \text{diag}[\underbrace{-1, -1, \ldots,  -1, -N}_{N-1}] \, ,
\end{split}
\end{align}
confirming $\text{Tor}(H_2(Y, \partial Y)/\text{im}(\jmath_2)) \cong \bbZ_N = H_1(\partial Y)$ in \eqref{eq:relative_hom_sequence_C2/Gamma}, corresponding to the $(N-1)$-th entry in the diagonal matrix $D$.
Hence, the coefficients in \eqref{eq:shift_instanton_general} are
\begin{align}
	\lambda_a = (S^{-1})_{N-1, a} = a \, .
\end{align}
As a cross check, we find the standard identity for $SU(N)$ Cartan matrices,
\begin{align}\label{eq:sum_cartan_a}
	\sum_{a=1}^{N-1} (-C_{ab}) \frac{\lambda_a}{N} = \sum_{a=1}^{N-1} (-C_{ab}) \frac{a}{N}= \begin{cases}
		0 \in \bbZ \, , \quad b = 1,..., N-2 \, ,\\
		1 \in \bbZ \, , \quad b = N - 1 \, .
	\end{cases}
\end{align}

Thus, a background flux $B$ for the $\bbZ_N$ 1-form symmetry induces shift \eqref{eq:G4_shifted_U1s} of the Cartan fluxes given by
\begin{align}
	F_a \rightarrow F_a + \frac{a}{N} B
\end{align}
which agrees with the action of the 1-form symmetry in the maximally Abelian phase of the gauge theory \cite{Hsin:2018vcg,Cordova:2019uob}.
Moreover, it also leads to the fractional shift
\begin{align}\label{eq:instanton_shift_suN_7d}
	\frac{1}{2N} \sum_{a,b=1}^{N-1} (-C_{ab}) \frac{\lambda_a \lambda_b}{N} B \wedge B = \frac{N-1}{2N} B \wedge B \, ,
\end{align}
to the instanton density \eqref{eq:shift_instanton_general}.
This agrees with the field theoretic results about the fractionality of $SU(N)$ instantons in the presence of a 1-form center background field \cite{Kapustin:2014gua,Gaiotto:2014kfa,Gaiotto:2017yup,Hsin:2018vcg,Cordova:2019uob}.

\paragraph{Models with F-theory uplift}
The result \eqref{eq:shift_instanton_general} applies, mutatis mutandis, to M-theory on elliptically fibered $Y_4$.
These models can be interpreted as an $S^1$-reduction of F-theory compactified on $Y_4$.
If this 8d theory has gauge symmetry $\mathfrak{g}$, then, in 7d, there are $\text{rank}(\mathfrak{g}) + 1$ compact 2-cycles $\sigma_a \in H_2(Y_4)$.
The additional 2-cycle is the generic fiber $\mathfrak{f}$ of $\pi: Y_4 \rightarrow B_2$, and gives rise to the vector multiplet in 7d obtained by integrating the Ramond-Ramond 2-form field in 8d\footnote{In a type IIB description, this 2-form field is the reduction of the 10d RR-field $C_4$ on the base $B_2$.} over the $S^1$.
Because $\mathfrak{f}$ has intersection number 0 with any compact 2-cycle in $Y_4$, it does not contribute to the Chern--Simons term \eqref{eq:CS-term-7d}.
In the presence of a boundary flux, the shifted 7d Chern--Simons terms \eqref{eq:shift_instanton_general} thus are equivalent to a fractional shift of the $G$-instantons inherited from an 8d 1-form background field.

As a concrete example, consider $Y_4$ the neighborhood of an $I_N$ fiber, which realizes an $\mathfrak{su}(N)$ gauge symmetry in 8d F-theory.
The set of compact curves, $\{\sigma_a\}$, $a=1,...,N$, intersect in the affine $SU(N)$ Dynkin diagram:
\begingroup\makeatletter\def\f@size{10}\check@mathfonts
\begin{align}
\begin{split}
	&\langle \sigma_a , \sigma_b \rangle
	= \left(\begin{array}{ccccc|c}
		& & & & & 1 \\
		& & & & & 0 \\
		& & C & & & \vdots\\
		& & & & & 0 \\
		& & & & & 1 \\ \hline
		1 & 0 & \hdots & 0 & 1 & -2
	\end{array}
	\right) = \left( \begin{array}{cc|c}
		& &  0 \\
		\multicolumn{2}{c|}{S^{-1}}  & \vdots \\
		& & 0 \\ \hline
		-1 & \hdots & -1
	\end{array} \right)
	\left( \begin{array}{c|c}
		D & 0\\ \hline
		0 & 0
	\end{array} \right)
	\left( \begin{array}{ccc|c}
		& & & -2 \\
		& & & -3 \\
		& T & & \vdots \\
		& & &  1-N \\ 
		& & & -1 \\ \hline
		0 & \hdots & 0 & 1
	\end{array} \right),
\end{split}
\end{align}
\endgroup
with the $(N-1) \times (N-1)$ matrices $(S, D, T)$ given in \eqref{eq:smith_decomp_suN_cartan}.
As expected, $\jmath_2((S^{-1})_{N,b} \,  \sigma_b) = -\sum_{a=1}^N \sigma_a = -\mathfrak{f}$ has trivial intersection with any compact 2-cycle $\sigma_c$.
The remaining $N-1$ 2-cycles gives rise to the same structure as the $\mathfrak{su}(N)$ example from $Y_4 = \mathbb{C}^2 / \bbZ_N$ above, with the $N$-torsional boundary flux given by $\tilde\sigma_t = \tfrac{1}{N} (S^{-1})_{N-1,c} \, \sigma_c = \sum_{c=1}^{N-1} \tfrac{c}{N} \sigma_c$.
Analogously, the fractional shift of the instanton density of the $SU(N)$ symmetry is \eqref{eq:instanton_shift_suN_7d}, which is the same as the shift in 8d \cite{Cvetic:2020kuw}.

\subsection{Compactification to 5d}

In compactifications on $Y_6$ to five dimensions, a reduction analogous to \eqref{eq:CS-term_reduced_to_7d} of the M-theory Chern--Simons term with the ansatz \eqref{eq:C3_KK_decomp} give rise to the 5d Chern--Simons terms \cite{Cadavid:1995bk,Ferrara:1996hh,Ferrara:1996wv}
\begin{align}\label{eq:CS-term_reduced_to_5d}
	\frac16 \sum_{\alpha,\beta,\gamma} \int_{M_5} K_{\alpha \beta \gamma} \, A_\alpha \wedge F_\beta \wedge F_\gamma \, .
\end{align}
When we only consider dynamical gauge fields in 5d spacetime (for which we use Latin indices $(\alpha,\beta,\gamma) \rightarrow (a,b,c)$), the internal pieces $\mathrm{w}_{\alpha,\beta,\gamma}$ in \eqref{eq:C3_KK_decomp} are all compactly supported 2-forms $\jmath_2^*(\omega_a)$.
In this case, the coefficients $K_{abc} = \int_{Y_6} \jmath_2^*(\omega_a) \wedge \jmath_2^*(\omega_b) \wedge \jmath_2^*(\omega_c)$ have a natural interpretation as the integral of products of compactly supported 2-forms, or, dually, as intersection number of 4-cycles $\sigma_a \in H_4(Y_6)$.
Physically, $K_{abc}$ encode the Coulomb branch dynamics of the effective 5d gauge theory \cite{Morrison:1996xf,Intriligator:1997pq}.

In the following, we are interested in the terms with $\omega_\alpha \equiv \tilde\omega_I \in H^2(Y_6)$ fixed, such that $\imath^*_2(\tilde\omega_I) \in \mathbb{Z}^f \subset H^2(\partial Y_6)$ (cf.~formula \eqref{eq:free_and_torsion_fluxes}), and let the indices $(\beta,\gamma) \rightarrow (b,c)$ run over compactly supported 2-forms that span the Cartan subgroup of the gauge group $G$.
As stated above, $\tilde{F}_I \otimes \tilde\omega_I$ corresponds to a background Cartan $U(1)_I$ flux of the global 0-form symmetries.
Since the Poincar\'e--Lefschetz-dual 4-cycle is a relative homology class, $\text{PD}(\tilde\omega_I) = \epsilon_I \in H_4(Y_6, \partial Y_6)$, that is \emph{not} in the image of $\jmath_4$, it may be regarded as a non-compact 4-cycle in $Y_6$.
In general, the corresponding Chern--Simons coefficients $K_{Ibc}$ encode mass parameters of the $G$ gauge theory \cite{Jefferson:2018irk}.
Crucially, the global 0-form symmetry include $U(1)$ factors that charge the instanton particles of the $G$ gauge theory, which can enhance the flavor symmetry (the part of the global symmetry charging hypermultiplets of the effective gauge theory) at the UV fixed point \cite{Seiberg:1996bd}.
For our discussion, we focus on the effective gauge theory phase, in which $\tilde\omega_I$ corresponds to an instantonic $U(1)_I$ global symmetry, rather than a Cartan $U(1)$ of the (classical) flavor symmetry.

Passing, for convenience, to the Poincar\'e--Lefschetz-dual homology description, we have $\text{PD}(\omega_{b,c}) = \sigma_{b,c} \in H_4(Y_6)$, and we can form the intersection product $\sigma_b \cdot \sigma_c \equiv \gamma_{bc} \in H_2(Y_6)$, which yields a 2-cycle in $Y_6$.
On the other hand, $\tilde\omega_I \in H^2(Y_6)/\text{im}(\jmath_2^*)$ is represented by an element in $H^2(Y_6)$, which we abusively also denote by $\tilde\omega_I$.
Then $\text{PD}(\tilde\omega_I) = \epsilon_I \in H_4(Y_6, \partial Y_6) \cong \text{Hom}(H_2(Y_6), \bbZ)$.
This now gives a straightforward way to ``define'' $K_{Ibc} = \epsilon_I (\gamma_{bc})$.

In the 5d Chern--Simons terms \eqref{eq:CS-term_reduced_to_5d}, turning on the 1-form symmetry background \eqref{eq:G4_shifted_U1s} leads to
\begin{align}\label{eq:fractional_shift_CS-terms_5d}
\begin{split}
	& \frac16 \sum_{\alpha,\beta,\gamma} \int_{M_5} K_{\alpha \beta \gamma} \, A_\alpha \wedge F_\beta \wedge F_\gamma \supset \frac12 \sum_{b,c} \int_{M_5} K_{Ibc} \, A_I \wedge F_b \wedge F_c \\
	\longrightarrow \quad & \frac12 \sum_{b,c} \int_{M_5}  K_{Ibc} \, A_I \wedge \left( F_b \wedge F_c + \frac{2\lambda_b}{N} F_c \wedge B + \frac{\lambda_b \lambda_c}{N^2} B \wedge B \right) \, .
\end{split}
\end{align}
To match this with the field theory results, one needs to show that the cross terms $F_c \wedge B$ again have integer coefficients.
Arguing for the integrality analogously to the 7d case requires the intersection pairing between the basis of $(d-2)$- and 2-cycles.
However, the spaces $H_4(Y_6)$ and $H_2(Y_6)$ are in general very different, and the intersection product $H_4(Y_6) \times H_4(Y_6) \rightarrow H_2(Y_6)$ depends on details of $Y_6$.
As such, it is difficult to make a general argument that applies to all geometries.
Instead, we will look at concrete examples, where we can calculate \eqref{eq:fractional_shift_CS-terms_5d} explicitly.

\subsubsection{Examples with 5d UV Fixed Point}

Let us illustrate this first in a simple example, namely, for the rank one 5d ${\cal N}=1$ theory with gauge symmetry $G = SU(2)$, theta angle $\theta = 0$, and no matter.
The latter two conditions ensure that the 1-form $\mathbb{Z}_2$ center symmetry is not explicitly broken by any matter or instanton particles \cite{Morrison:2020ool,Albertini:2020mdx}.
This gauge theory has as (continuous) global 0-form $U(1)$ instanton symmetry, which enhances to an $SU(2)$ at the UV fixed point \cite{Seiberg:1996bd,Morrison:1996xf,Intriligator:1997pq}.
The (non-compact) Calabi--Yau threefold $Y_6$ that describes this theory via M-theory is local neighborhood of an $\mathbb{F}_0 \cong \bbP^1 \times \bbP^1 \equiv \sigma$ surface, which generates $H_4(Y_6) \cong \bbZ$.
Furthermore, $H_2(Y_6) \cong \bbZ^2$ is generated by two $\bbP^1$s, $\gamma_1$ and $\gamma_2$, inside $\sigma$, with intersection pairing $\langle \sigma, \gamma_1 \rangle = \langle \sigma, \gamma_2 \rangle = -2$.
The corresponding Smith decomposition \eqref{eq:smith_decomp} is simple,
\begin{align}
	(-2, -2) = (1) (-2, 0) \begin{pmatrix} 1 & 1 \\ 0 & 1 \end{pmatrix} \, ,
\end{align}
implying that $\epsilon_1 = \eta_1 + \eta_2 = \tfrac12 \jmath_2^*(\sigma)$ is the generator of the $\Gamma \cong \mathbb{Z}_2$ 1-form symmetry backgrounds (and so $\lambda_b \equiv \lambda_\sigma = 1$ and $N=2$ in \eqref{eq:fractional_shift_CS-terms_5d}).
Meanwhile, the ``non-compact divisor'' $\epsilon_I \in H_4(Y_6, \partial Y_6) \cong \text{Hom}(H_2(Y_6),\bbZ)$ corresponding to the $U(1)_I$ global symmetry is given by
\begin{align}
	\epsilon_I = \eta_2: H_2(Y_6) \rightarrow \bbZ \, , \quad a_1\gamma_1 + a_2\gamma_2 \mapsto a_2 \, .
\end{align}
Additionally, we need $\sigma \cdot \sigma = -2(\gamma_1 + \gamma_2)$.
This means that $K_{Ibc} \equiv K_{I, \sigma, \sigma} = \epsilon_I(\sigma \cdot \sigma) = -2$.

In the presence of the $\bbZ_2$ 1-form symmetry background, the shifted Chern--Simons term \eqref{eq:fractional_shift_CS-terms_5d} then becomes
\begin{align}
\begin{split}
	& \frac12 \int_{M_5} \sum_{b,c} K_{Ibc} \, A_I \wedge \left( F_b \wedge F_c + \frac{2\lambda_b}{N} F_c \wedge B + \frac{\lambda_b \lambda_c}{N^2} B \wedge B \right) \\
	= & - \int_{M_5} A_I \wedge \left( F_\sigma \wedge F_\sigma + F_\sigma \wedge B + \frac14 B \wedge B \right) \\
	\stackrel{\text{non-ab.~limit}}{\longrightarrow} \quad & - \int_{M_5} A_I \wedge \left( \text{Tr}(F_{SU(2)}^2) + \frac{1}{4} B \wedge B + \text{integer terms} \right)\, .
\end{split}
\end{align}
The mixed 't Hooft anomaly between $U(1)_I$ and the center 1-form symmetry of $SU(2)$ resulting from this fractional shift of the instanton density indeed agrees with expectations from field theory \cite{BenettiGenolini:2020doj}.

Note that this geometric computation is strictly speaking only valid on the Coulomb branch of the 5d theory.
While the extrapolation to the sublocus, where we have an effective non-Abelian gauge theory, agrees with previous work, our approach cannot preclude a cancellation of this anomaly through a topological sector which is hidden on the Coulomb branch.
Indeed, recent work \cite{Apruzzi:2021vcu} suggests the existence of such sectors on the Higgs branch, which geometricly can only be accessed by passing through the strongly-coupled SCFT point via deformation.
Because of this, we do not make any claims about how anomaly lifts to the UV theory.

In general, there is no reason to expect that modifications from the effective field theory expectations can only arise from the Higgs branch.
Indeed, certain UV-effects can also be found on the Coulomb branch, which indicates a more subtle effect of turning on 1-form symmetry backgrounds in the SCFT.
For that, we consider setups realizing pure $SU(N \geq 3)_k$ gauge theories with Chern--Simons level $2-N < k < N-2$.
These theories have an $\bbZ_{\text{gcd}(N,k)}$ 1-form symmetry, and rank $f=1$ global symmetry given by the instantonic $U(1)_I$ \cite{Morrison:2020ool,Albertini:2020mdx}.
A possible M-theory geometry is a local Calabi--Yau neighborhood $Y_6$ of $N-1$ Hirzebruch surfaces $\sigma_a \cong \bbF_{n_a}$ that intersect transversely in a chain.
Leaving the detailed computation to appendix \ref{app:smith_decomp_suN_5d}, we present here the shift of the Chern--Simons terms by the $\bbZ_{\text{gcd}(N,k)}$-torsional boundary flux:
\begin{align}\label{eq:instanton_shift_5d_suN}
\begin{split}
	& \frac12 \sum_{a,b} \int_{M_5}  K_{Iab} \, A_I \wedge \left( F_a \wedge F_b  + \frac{\lambda_a \lambda_b}{\text{gcd}(N,k)^2} B \wedge B \right) \\
	= & \int_{M_5} \left( \sum_{a,b}  \frac{Q_{22}}{2} (-C_{ab}) A_I \wedge F_a \wedge F_b - \frac{Q_{21}(N+k)}{2} A_I \wedge F_{N-1} \wedge F_{N-1} \right) \\
	+ & \int_{M_5} \frac{N-1}{2N} \ell^2 \left( Q_{22}  - Q_{21} (N+k) \frac{N-1}{N} \right) A_I \wedge B \wedge B \, ,
\end{split}
\end{align}
where $\ell = N/\text{gcd}(N,k) \mod N$ is the generator of the $\bbZ_{\text{gcd}(N,k)}\subset \bbZ_{N}$ subgroup of the center of $SU(N)$.
The coefficients $Q_{22}$ and $Q_{21}$ are integers fixed by a Euclidean Algorithm on $(N, k-N)$:
\begin{align}\label{eq:relation_Qs}
  \text{gcd}(N, k-N) = \text{gcd}(N,k) = Q_{22} N - Q_{21} (k-N)\, .
\end{align}

This result does not immediately agree with the expectations from the effective $SU(N)_k$ gauge description.
Neglecting the contribution proportional to $Q_{21} A_I \wedge F_{N-1} \wedge F_{N-1}$ above, one would interpret the first term, $\sum_{a,b}  \frac{Q_{22}}{2} (-C_{ab}) A_I \wedge F_a \wedge F_b$, as the Coulomb branch expression of $A_I \wedge (Q_{22} \text{Tr}(F^2))$.
That is, instanton density of $SU(N)$ is coupled to $U(1)_I$ with charge $Q_{22}$.
Then --- again neglecting the term proportional to $Q_{21}$ --- the last line in \eqref{eq:fractional_shift_CS-terms_5d} would precisely correspond to the fractional shift of $SU(N)/\bbZ_{\text{gcd}(N,k)}$ instantons.
Thus, the $Q_{21}$-terms are expected to be non-perturbative corrections to the effective $SU(N)_k$ gauge description.
Furthermore, we have not fully explored the invariance of \eqref{eq:relation_Qs} under $\big( Q_{12}, Q_{22} \big) \rightarrow \big(Q_{21} + m \tfrac{N}{\text{gcd}(N,k)} , Q_{22} + m \tfrac{k-N}{\text{gcd}(N,k)} \big)$, though this seems to be related to a redefinition of the generator for $U(1)_I$, cf.~\eqref{eq:def_top_U1_suN_5d_example}.
To gain a better understanding of these terms, it would be instructive to find a field theoretic description of such non-perturbative corrections, and/or verify the geometric result from another construction of the 5d SCFT that is the UV-completion of the $SU(N)_k$ gauge theory.

\subsubsection{5d KK-Theories and 6d Anomalies}

If $Y_6$ is elliptically fibered (over a K\"ahler manifold $\mathfrak{B}_4$), then M-theory on $Y_6$ gives rise to a so-called 5d KK theory.
The UV-completion of such a gauge theory is not a genuine 5d SCFT, but rather a 6d SCFT on an $S^1$ (hence the name).
In this reduction, the 5d 1-form symmetry receives contributions from both 1-form and 2-form symmetries in 6d \cite{Morrison:2020ool}.
Moreover, the instanton density of the 6d gauge symmetry couples via a Green--Schwarz mechanism to dynamical tensor fields, which on an $S^1$ reduce to vector multiplets associated to additional $U(1)$ (0-form) gauge symmetries in 5d.
Therefore, in the 5d KK-theory, the 6d mixed anomaly between the 1-form symmetry and the large gauge transformations of tensor fields \cite{Apruzzi:2020zot} are encoded in the Chern--Simons terms $K_{abc}$ involving three compact divisors.

For an example, consider the 6d non-Higgsable $SU(3)$ theory.
The corresponding 5d KK-theory is M-theory compactified on an elliptically fibered $\pi: Y_6 \rightarrow \mathfrak{B}_4$, given by the Calabi--Yau neighborhood of three intersecting $\bbF_1$ surfaces \cite{DelZotto:2017pti}:\footnote{
The basis of 2-cycles on each $\bbF_1$ surface is $\{f,e\}$, which on $\bbF_1$ intersect as $e \cdot e = -1, e \cdot f =1, f \cdot f = 0$.
}
\begin{equation}
	\begin{tikzcd}[row sep = normal, column sep= normal  ]
		\sigma_1 \cong \bbF_1 \arrow[dash]{r}{e} \arrow[dash]{d}{e} & \bbF_1 \cong \sigma_2 \\
		\sigma_3 \cong \bbF_1 \arrow[dash]{ur}{e} &
	\end{tikzcd} \, .
\end{equation}
The intersection $\sigma_1 \cdot \sigma_2 = \sigma_2 \cdot \sigma_3 = \sigma_1 \cdot \sigma_3 = e$ is the $(-1)$-curve in each $\sigma_a$, which is the section of the $\bbP^1$-fibration on $\sigma_a$ with fiber $f_a$.
The generic elliptic fiber is $\mathfrak{f} = f_1 + f_2 + f_3$ in homology.
Thus, all $\sigma_a$ are fibered over the same genus-0 curve $C \subset B_4$, which has self-intersection number $-3$ inside $\mathfrak{B}_4$.
Furthermore, $\sigma_a \cdot \sigma_a = -2(e + 3f_a)$.

A basis of $H_2(Y_6)$ is given by $\gamma_i \in \{f_1, f_2, f_3, e\}$.
From the intersection matrix,
\begingroup\makeatletter\def\f@size{10}\check@mathfonts
\begin{align}\label{eq:smith_decomp_su3_NHC}
	M_{ai} = \langle \sigma_a , \gamma_j \rangle = \begin{pmatrix}
		-2 & 1 & 1 & -1 \\
		1 & -2 & 1 & -1 \\
		1 & 1 & -2 & -1
	\end{pmatrix}
	=
	\underbrace{\begin{pmatrix}
		1 & 1 & 0\\
		1 & 2 & 0\\
		1 & 1 & 1
	\end{pmatrix}^{-1}}_{S^{-1}}
	\begin{pmatrix}
		- 1 & 0 & 0 & 0\\
		0 & -3 & 0 & 0\\
		0 & 0 & -3 & 0
	\end{pmatrix}
	\underbrace{\begin{pmatrix}
		1 & 1 & -2 & 2 \\
		0 & 1 & -1 & 1 \\
		0 & 0 & 0 & 1 \\
		0 & 0 & 1 & -1
	\end{pmatrix}}_{T},
\end{align}
\endgroup
we see that the surfaces $\sigma_{1,2}$ form the Cartan $U(1)$s of the $SU(3)$ gauge symmetry, whose $\bbZ_3$ 1-form center symmetry is encoded in the boundary flux (cf.~\eqref{eq:lin_rel_homology})
\begin{align}\label{eq:boundary_flux_su3_NHC}
	\tilde\sigma_t = \frac13 \sum_a \lambda_a \jmath_4(\sigma_a) \equiv \frac{1}{3} \sum_a (S^{-1})_{2,a} \jmath_4(\sigma_a) = \frac{1}{3} (\jmath_4(\sigma_1) + 2 \jmath_4(\sigma_2)) \, .
\end{align}
The additional dynamical $U(1)_\text{d}$ gauge field is dual to $\xi_3 = (S^{-1})_{3,a} \sigma_a = \sigma_1 + \sigma_2 + \sigma_3 = \pi^{-1}(C)$, which is a vertical divisor in $Y_6$, and hence corresponds to the $S^1$-reduction of the 6d tensor field.

Thus, the coefficients of the relevant Chern--Simons terms,
\begin{align}\label{eq:CS-term-SU3-NHC}
	\frac12 \sum_{a,b =1}^2 \int_{M_5} K_{\text{d}ab} \, A_\text{d} \wedge F^{(SU(3))}_a \wedge F^{(SU(3))}_b \, ,
\end{align}
are $K_{\text{d} ab} = \langle \xi_3 , \sigma_a \cdot \sigma_b \rangle = 3 (-C^{(SU(3))})_{ab}$, with $(-C^{(SU(3))})$ the Cartan matrix of $SU(3)$.
In the limit where we collapse the fibers $f_{1,2}$ in $Y_6$, and thereby enhancing the gauge symmetry to $SU(3) \times U(1)_\text{d}$, these Chern--Simons terms become
\begin{align}
	-3 \int_{M_5} A_\text{d} \wedge \text{Tr}(F^{(SU(3))} \wedge F^{(SU(3))}) \, ,
\end{align}
indicating that $U(1)_\text{d}$ gauges the instanton symmetry of $SU(3)$ ``with charge 3'' \cite{Morrison:2020ool}.
This also agrees with the reduction of the corresponding 6d Green--Schwarz coupling with tensor charge $3$.
Furthermore, this charge prefactor also ensures that the 1-form center symmetry has no mixed anomaly with the large gauge transformations of $U(1)_\text{d}$, since the fractional shift induced by the boundary flux \eqref{eq:boundary_flux_su3_NHC} is
\begin{align}
	\sum_{a,b=1}^2 K_{\text{d} ab} \, \frac{\lambda_a \lambda_b}{2 \cdot 3^2} \, B \wedge B = \frac{1}{6} \sum_{a,b=1}^{2}  a \, b \,(-C^{(SU(3))})_{ab} \, B \wedge B = B \wedge B \, .
\end{align}
This matches the absence of the corresponding 1-form anomaly in 6d \cite{Apruzzi:2020zot}.

Let us further consider $\tilde\rho := \eta_4 \in \text{Hom}(H_2(Y_6), \bbZ) \cong H_4(Y_6, \partial Y_6)$, which from \eqref{eq:smith_decomp} satisfies $-\jmath_4(\xi_3) = 3 \eta_4$, i.e., is the generator of the second $\bbZ_3$ factor in the 5d 1-form symmetry.
Since $\tilde\rho(f_a)=0$, $\tilde\rho(e)=1$, $\tilde\rho$ can be interpreted as a non-compact \emph{vertical} divisor in $Y_6$, whose projection $\pi(\tilde\rho)$ onto $\mathfrak{B}_4$ intersects the compact $(-3)$-curve $C$ once.
In the 6d F-theory setting, wrapping D3-branes on $\pi(\tilde\rho)$ gives rise to string-like surface defects that are charged under the 6d $\bbZ_3$ 2-form symmetry \cite{DelZotto:2015isa,Morrison:2020ool}.

In 5d, we can now study the effects of turning on background fields $B$ for the $\bbZ_3$ 1-form center symmetry of $SU(3)$, as well as $B_\text{d}$ for the $\bbZ_3$ 1-form symmetry that descends from the 6d 2-form symmetry.
While the first corresponds to the shift $F_a \rightarrow F_a + \frac{a}{3} B$ for the $SU(3)$ Cartan fluxes, the second shifts the field strength of the $U(1)_\text{d}$ as $F_\text{d} \rightarrow F_\text{d} + \frac13 B_\text{d}$.
The latter can be viewed as a transformation $A_\text{d} \rightarrow A_\text{d} + \frac13 \epsilon$, where $\epsilon$ is a flat $\bbZ_3$ connection \cite{Cordova:2019jnf,Cordova:2019uob,BenettiGenolini:2020doj}.
From \eqref{eq:CS-term-SU3-NHC}, we would then obtain the term
\begin{align}
	\frac12 \sum_{a,b=1}^2 K_{\text{d}ab} \, \frac13 \epsilon \wedge \frac{a}{3} B \wedge \frac{b}{3} B = \frac13 \epsilon \wedge B \wedge B \, .
\end{align}
This constitutes a mixed 't Hooft anomaly between the two $\bbZ_3$ 1-form symmetries, which can be written in terms of a 6d anomaly theory,
\begin{align}
  {\cal A}[B_\text{d}, B] = \int_{X_6} \frac13 B_\text{d} \wedge B^2 \, ,
\end{align}
where $\partial X_6 = M_5$ is an auxiliary manifold whose boundary is the 5d spacetime.

It appears natural to uplift this anomaly to the 6d gauge theory that corresponds to F-theory compactified on $Y_6$.
Here, this would be a mixed 't Hooft anomaly between the 6d 2-form $\bbZ_3$ symmetry for the instanton strings, and the 1-form $\bbZ_3$ center symmetry of the non-Higgsable $SU(3)$ gauge sector.
One intuitive explanation for this anomaly is that, by turning on a background field for the 1-form center symmetry, the instanton number fractionalizes, being now instantons of an $SU(3)/\bbZ_3$ bundle.
Compared to the instanton strings of $SU(3)$, which have charge $0 \mod 3$ under the $\bbZ_3$ 2-form symmetry, the $SU(3)/\bbZ_3$ instantons have charge $1 \mod 3$, and hence screen all asymptotic charges of the 2-form symmetry.
It would be interesting to investigate potential field theoretic counterterms for this anomaly, and, if present, their imprints in geometry.



\chapter{GAUGE GROUP TOPOLOGY OF 8D CHL VACUA}


\section{Introduction}

Supersymmetric string compactifications on low-dimensional internal manifolds have seen a resurgence of interest within the Swampland program \cite{Vafa:2005ui,Ooguri:2006in}.
One of the main reasons is that, thanks to the large amount of supersymmetry, one can essentially classify all supergravity models that arise as the low-energy description of such compactifications.
Therefore, they provide an excellent ``laboratory'' to test our understanding of the physical principles that separate the Landscape from the Swampland.

Given the profound role of gauge symmetries in our mathematical formulation of effective theories, principles that delineate the boundary between consistent and inconsistent gauge groups of supergravity models are of particular interest.
In the context of 8d ${\cal N}=1$ supergravity theories, significant progress in this direction has been made recently, which not only explains the absence of specific gauge \emph{algebras} \cite{Garcia-Etxebarria:2017crf,Montero:2020icj,Hamada:2021bbz} in the 8d string landscape, but also some of the intricate patterns of the possible global structures, i.e., topology, of the gauge \emph{group} \cite{Cvetic:2020kuw,Montero:2020icj}.
In particular, the ideas pertaining to the gauge group topology have been mostly tested and confirmed for 8d theories with total gauge rank\footnote{
Within the known 8d ${\cal N}=1$ string landscape, the total gauge rank can be either 4, 12, or 20; this limitation can be understood as a quantum-gravitational consistency condition, by invoking Swampland arguments \cite{Montero:2020icj}.
Different from the rank counting in that work, which organizes the theories into having rank 2, 10, or 18, we include the contributions of the ${\cal N}=1$ gravity multiplet which always contains two graviphotons, because the associated $U(1)$ factors are generally involved in the overall gauge group topology.
} 20 in their F-theory realization \cite{Vafa:1996xn}, where the relevant geometric features \cite{Aspinwall:1998xj,Mayrhofer:2014opa,Cvetic:2017epq} have been classified \cite{shimada_k3}.

To lend further credence, but, more importantly, to collect additional ``data'' to eventually sharpen these arguments,\footnote{The arguments of \cite{Cvetic:2020kuw,Montero:2020icj} provide necessary, but not sufficient criteria for a non-trivial global gauge group structure, see \cite{Cvetic:2020kuw} for a detailed discussion.
} it would be desirable to also study other branches of the 8d moduli space.
Unfortunately, there does not exist a classification of gauge group topologies in rank 12 or rank 4 theories as comprehensive as in the case of rank 20 theories \cite{shimada_k3,Font:2020rsk}.
With this motivation in mind, the purpose of this work is to provide the general framework to determine the gauge group topology in 8d ${\cal N}=1$ string models, with a focus on rank 12 theories.

Rank 12 theories arise as $S^1$-reductions of the CHL string \cite{Chaudhuri:1995fk,Chaudhuri:1995bf}.
The physical states, which are characterized by the winding numbers and momenta of the CHL string, live in an even lattice $\Lambda_M$ of rank 12, the so-called Mikhailov lattice \cite{Mikhailov:1998si}.
Then, any non-Abelian gauge \emph{algebra} $\mathfrak{g}$ that can arise in an 8d CHL vacuum must have a root lattice $\Lambda_\text{r}^\mathfrak{g}$ that embeds in a specific way into $\Lambda_M$.
Such lattice embeddings can be classified \cite{Font:2021uyw} in an analogous fashion as for rank 20 theories based on their heterotic realization \cite{Font:2020rsk}, where the corresponding string momentum lattice is the rank 20 Narain lattice $\Lambda_N$ \cite{Narain:1985jj,Narain:1986am}.

On the other hand, as we will elaborate in Section \ref{sec:2}, the information about the global structure of the gauge \emph{group} $G = \widetilde{G} / {\cal Z}$, with $\widetilde{G}$ the simply-connected group with algebra $\mathfrak{g}$, is encoded in the lattice \emph{dual} to the string momentum lattice $\Lambda_S$ with $\Lambda_S = \Lambda_N$ or $\Lambda_M$.
Roughly speaking, the definition of the dual lattice $\Lambda_S^* \subset \Lambda_S \otimes \mathbb{R}$ as having integer pairing with all vectors in $\Lambda_S$ can be regarded as a constraint on the representations of the physical states in $\Lambda_S$.
More precisely, the fundamental group,
\begin{align}
    {\cal Z} = \pi_1(G) = \Lambda_\text{cc}^G / \Lambda_\text{cr}^\mathfrak{g} \, ,
\end{align}
depends on the \emph{cocharacter} lattice $\Lambda_\text{cc}^G$, which is a sublattice of the \emph{coweight} lattice $\Lambda_\text{cw}^\mathfrak{g} = \Lambda_\text{r}^*$.
This is the dual of the character lattice $\Lambda_\text{c}^G$, which corresponds to the charge lattice occupied by physical states,\footnote{Here, we adapt the notation from \cite{Gukov:2006jk}.
It is also common (see, e.g. \cite{bump2004lie,Hall2015}) to refer to $\Lambda_\text{c}^G$ ($\Lambda_\text{cc}^G$) as the (co-)weight lattice \emph{of the group} $G$.
}
which clearly is the momentum lattice $\Lambda_S$.
From this perspective, the self-duality of the Narain lattice  (imposed by modularity of the heterotic worldsheet), together with the fact that rank 20 theories only have ADE-algebras (whose (co-)root lattices $\Lambda_\text{r}^\mathfrak{g} = \Lambda_\text{cr}^\mathfrak{g}$ agree), appear as a coincidence that makes it straightforward to compute the fundamental group ${\cal Z} = \pi_1(G)$ as (the torsional piece\footnote{The free part corresponds to $U(1)$ symmetries, which in fact can also have a non-trivial global structure with the non-Abelian group; we will elaborate on this in detail below.} of) $\Lambda_N / \Lambda_\text{r}^\mathfrak{g}$, as done in \cite{Font:2020rsk}.
This is confirmed via duality by F-theory geometries \cite{shimada_k3}, where the corresponding data are encoded in the Mordell--Weil group \cite{Aspinwall:1998xj,Mayrhofer:2014opa,Cvetic:2017epq}.
In the rank 12 case, this quotient is no longer the correct object to compute, due to $\Lambda_M \neq \Lambda_M^*$, as well as the appearance of non-simply laced $\mathfrak{sp}$ algebras with $\Lambda_\text{r}^\mathfrak{sp} \neq \Lambda_\text{cr}^\mathfrak{sp}$.
Instead, as we shall demonstrate explicitly in Section \ref{sec:2}, the correct prescription for $\pi_1(G)$ of CHL vacua is captured by the ``mismatch'' between $\Lambda_\text{cr}^\mathfrak{g}$ and $\Lambda_M^*$.

Moreover, our approach naturally computes the global gauge group structure including the $U(1)$ gauge factors.
That is, given the embedding data $\Lambda_\text{r}^\mathfrak{g} \subset \Lambda_S$ of the non-Abelian root lattice into the momentum lattice, we can determine the entire gauge group topology, which takes the generic form
\begin{align}
    \frac{[\widetilde{G}/{\cal Z}] \times U(1)^{r_F}}{{\cal Z}'} \, ,
\end{align}
with $r_F = \text{rank}(\Lambda_S) - \text{rank}(\mathfrak{g})$.
As we will explain, the quotient ${\cal Z}'$, which may be interpreted as a constraint on the $U(1)$ charges of states in certain representations of $\mathfrak{g}$, arises due to lattice generators of $\Lambda_S^*$ that are not in the plane containing $\Lambda_\text{r}^\mathfrak{g}$.
For rank 20 theories, our approach is equivalent to methods based on string junctions that describe the dual F-theory model \cite{Guralnik:2001jh,Cvetic:2021sjm}, and we will demonstrate its efficacy also in a concrete CHL model below.

An important consequence, which we prove in Section \ref{subsec:1-form_anomalies}, is that the non-Abelian gauge group topology $\widetilde{G}/{\cal Z}$ is consistent with a gauged 1-form ${\cal Z}$ symmetry \cite{Gaiotto:2014kfa}, in both heterotic and CHL vacua.
That is, there is no mixed anomaly that would obstruct such a gauging, consistent with the findings in \cite{Cvetic:2020kuw}.
We verify this explicitly by computing ${\cal Z} = \pi_1(G)$ for all maximally enhanced CHL models (i.e., those with rank$(G)=10$), which is presented in Appendix \ref{app:big_table}. 
We also find a consistent cross-check for two of these models, which are subject to constraints posed in \cite{Montero:2020icj}.
To facilitate the computation of ${\cal Z}$, we show, in Section \ref{sec:3}, that for any CHL model, specified by an embedding $\Lambda_\text{r}^\mathfrak{g} \subset \Lambda_M$, the corresponding gauge group topology can be directly inferred from that of a ``parent'' rank 20 heterotic model with $G_\text{het} = \widetilde{G}_\text{het} / {\cal Z}_\text{het}$.\footnote{Via string dualities, the CHL model corresponds to IIB with an O7$_+$ plane, or, equivalently, F-theory on a K3-surface with a (partly) ``frozen'' singularity \cite{Witten:1997bs,Tachikawa:2015wka,Bhardwaj:2018jgp}.
The same K3, when interpreted without the frozen singularity, defines a rank 20 F-theory model that is dual to the ``parent'' heterotic model.
}
This then allows for an easy extraction of ${\cal Z}$ via ${\cal Z}_\text{het}$, the latter of which can be obtained from the heterotic classification \cite{Font:2020rsk,Cvetic:2021sjm}.
We conclude in Section \ref{sec:conclusions} with some outlook to related topics.

\section{Gauge Groups from Momentum Lattices}
\label{sec:2}

We begin this section by reviewing the group-theoretic definition of the global gauge group structure in terms of the various lattices.
We then discuss how these structures emerge in root lattice embeddings into the momentum lattice $\Lambda_S$ of string states.
We will highlight the key differences between rank 20 heterotic theories with $\Lambda_S = \Lambda_N$ the Narain lattice, and rank 12 CHL theories with $\Lambda_S = \Lambda_M$ the Mikhailov lattice.

\subsection{Lattices and Gauge Group Topology}

Any non-Abelian gauge algebra $\mathfrak{g}$ of rank $r$ is specified by a root system $\Phi_\mathfrak{g}$, which is a finite subset of a Euclidean vector space $E \cong \mathbb{R}^r$ satisfying certain axioms (see, e.g., \cite{bump2004lie,Hall2015} for a broader introduction; we follow the conventions of \cite{Gukov:2006jk}). 
Relevant to us in the following will be that the root lattice $\Lambda^\mathfrak{g}_\text{r} \supset \Phi_\mathfrak{g}$ --- spanned by integer linear combinations of simple roots $\boldsymbol{\mu} \in \Phi_\mathfrak{g}$ --- is a rank $r$ lattice inside $E$.
The space $E$ comes equipped with a bilinear pairing $( \cdot , \cdot ) : E \times E \rightarrow \mathbb{R}$ which induces a pairing on $\Lambda^\mathfrak{g}_\text{r}$.
The normalization is such that $(\boldsymbol{\nu}, \boldsymbol\nu)= 2$ for $\boldsymbol\nu \in \Phi$ a short root, and $(\boldsymbol\nu, \boldsymbol\nu)= 4$ for the long root of $\mathfrak{sp}(n)$.
The axioms also assert that $2(\boldsymbol\nu_1, \boldsymbol\nu_2)/(\boldsymbol\nu_2, \boldsymbol\nu_2) \in \bbZ$ for any two roots $\boldsymbol\nu_1, \boldsymbol\nu_2 \in \Phi_\mathfrak{g}$, ensuring that the coroots,
\begin{align}\label{eq:coroots_def}
    \Phi_\mathfrak{g}^\vee =\left\{ \left. \boldsymbol\nu^\vee := \frac{2\boldsymbol\nu}{(\boldsymbol\nu, \boldsymbol\nu)} \, \right| \, \boldsymbol\nu \in \Phi_\mathfrak{g} \right\} \subset E \, ,
\end{align}
and their integer span $\Lambda^\mathfrak{g}_\text{cr}$, the coroot lattice, have integer pairings with roots.
For $\mathfrak{g}$ an ADE algebra, we have $\Lambda^\mathfrak{g}_\text{r} = \Lambda^\mathfrak{g}_\text{cr}$, because all ADE roots have length squared 2.
One then defines the weight and coweight lattices, $\Lambda^\mathfrak{g}_\text{w}$ and $\Lambda^\mathfrak{g}_\text{cw}$, as their respective dual lattices:\footnote{Given a lattice $\Lambda$ with pairing $(\cdot, \cdot)$, the dual lattice is defined to be $\Lambda^* = \{\overline{\bf a} \in \Lambda \otimes \mathbb{R} \, | \, (\overline{\bf a}, {\bf v}) \in \bbZ \, \text{ for all } \, {\bf v} \in \Lambda \}$.
$\Lambda^*$ has the same rank as $\Lambda$.
}
\begin{align}\label{eq:weight_coweight_lattice_def}
    \begin{split}
        \Lambda^\mathfrak{g}_\text{w} & := (\Lambda^\mathfrak{g}_\text{cr})^* = \{ {\bf w} \in E \, | \, ({\bf w}, \boldsymbol\alpha^\vee) \in \bbZ \, \text{ for all } \, \boldsymbol\alpha^\vee \in \Lambda^\mathfrak{g}_\text{cr} \} \supset \Lambda^\mathfrak{g}_\text{r} \, ,\\
        \Lambda^\mathfrak{g}_\text{cw} & := (\Lambda^\mathfrak{g}_\text{r})^* = \{ \overline{\bf w} \in E \, | \, (\overline{\bf w}, \boldsymbol\alpha) \in \bbZ \, \text{ for all }\, \boldsymbol\alpha \in \Lambda^\mathfrak{g}_\text{r} \} \supset \Lambda^\mathfrak{g}_\text{cr} \, .
    \end{split}
\end{align}
Note that all these lattices are of rank $r$, i.e., they span $E$ over $\mathbb{R}$.
If $\mathfrak{g} = \oplus_j \mathfrak{g}_j$ is a sum of simple factors, there is an orthogonal decomposition $E = \oplus_j E_j$, where $E_j$ are spanned by the roots $\Phi_{\mathfrak{g}_j}$ and their associated lattices of the corresponding simple factor $\mathfrak{g}_j$.

So far, all data are defined by the gauge algebra $\mathfrak{g}$ with roots $\Phi_\mathfrak{g}$.
The actual gauge group $G$ is specified by a third pair of lattices, the character lattice $\Lambda^G_\text{c}$ and the cocharacter lattice $\Lambda^G_\text{cc}$, which are intermediate lattices,
\begin{align}\label{eq:character_cochar_lattice_def}
    \begin{split}
        \Lambda^\mathfrak{g}_\text{r} \subset\, & \Lambda^G_\text{c} \subset \Lambda^\mathfrak{g}_\text{w} \, , \\
        \Lambda^\mathfrak{g}_\text{cr} \subset\, & \Lambda^G_\text{cc} \subset \Lambda^\mathfrak{g}_\text{cw} \, ,
    \end{split}
\end{align}
that are dual to each other, $(\Lambda^G_\text{c})^* = \Lambda^G_\text{cc}$, with respect to the pairing $( \cdot, \cdot)$.
A gauge theory with group $G$ can only have dynamical states whose weight vectors lie in $\Lambda_\text{c}^G$, which is also often called the weight lattice \emph{of the group} $G$.\footnote{One can show, see, e.g., \cite{bump2004lie}, that $\Lambda^G_\text{c}$ is isomorphic to character group Hom$(T, \mathbb{C}^\times)$ of the maximal torus $T \subset G$ of the \emph{group}.
}
In terms of the (co-)character lattices, the center and the fundamental group of $G$ are:
\begin{align}\label{eq:group_topology_via_lattices_general}
\begin{split}
    Z(G) & = \Lambda^\mathfrak{g}_\text{cw} / \Lambda^G_\text{cc} \cong \Lambda^G_\text{c} / \Lambda^\mathfrak{g}_\text{r}\, , \\
    \pi_1(G) & = \Lambda^G_\text{cc} / \Lambda^\mathfrak{g}_\text{cr} \cong \Lambda^\mathfrak{g}_\text{w} / \Lambda^G_\text{c}\, .
\end{split}
\end{align}

If $G = \widetilde{G}$ is the simply-connected group with algebra $\mathfrak{g}$, then $\Lambda^{\widetilde{G}}_\text{c} = \Lambda^\mathfrak{g}_\text{w}$ and $\Lambda^{\widetilde{G}}_\text{cc} = \Lambda^\mathfrak{g}_\text{cr}$.
Elements $c$ in the center $Z(\widetilde{G}) = \Lambda^\mathfrak{g}_\text{cw} / \Lambda^\mathfrak{g}_\text{cr}$, represented by a coweight $\overline{\bf v}_c \in \Lambda^\mathfrak{g}_\text{cw}$, act on a weight by a phase $\exp(2\pi i c({\bf w}))$, where the fractional number
\begin{align}\label{eq:center_charge_weight}
    c({\bf w}) = ({\bf w}, \overline{\bf v}_c) \equiv ({\bf w}, \overline{\bf v}_c + \boldsymbol{\alpha}^\vee) \mod \bbZ \quad \text{ for any } \quad \boldsymbol{\alpha}^\vee \in \Lambda_\text{cr}^\mathfrak{g} \, ,
\end{align}
can be interpreted as the charge of ${\bf w}$ under the center element $c$ represented by $\overline{\bf v}_c \mod \Lambda^\mathfrak{g}_\text{cr}$.
Note that this center charge is invariant for all weights of an irreducible representation ${\bf R}$ of $\mathfrak{g}$, because $c({\bf w} + \boldsymbol{\alpha}) = ({\bf w} + \boldsymbol{\alpha}, \overline{\bf v}_c) = ({\bf w}, \overline{\bf v}_c) \mod \bbZ$ for roots $\boldsymbol{\alpha} \in \Lambda_\text{r}^\mathfrak{g}$.

Since $\Lambda_\text{c}^G \subset \Lambda_\text{w}^\mathfrak{g} \equiv \Lambda_\text{c}^{\widetilde{G}}$, we can regard a character ${\bf w} \in \Lambda_\text{c}^G$ of a non-simply connected group $G$ as weights of $\widetilde{G}$.
Then we see that they are acted on trivially by $\pi_1(G) = \Lambda_\text{cc}^G / \Lambda_\text{cr}^\mathfrak{g} \subset \Lambda_\text{cw}^\mathfrak{g} / \Lambda_\text{cr}^\mathfrak{g} = Z(\widetilde{G})$, because they have center charges $c({\bf w}) = ({\bf w}, \overline{\bf v}_c) = 0 \mod \bbZ$ for $\overline{\bf v}_c \in \Lambda_\text{cc}^G$.
Hence, we can also view the ``non-trivial global structure'' $G = \widetilde{G} / \pi_1(G)$ of a gauge group as imposed by requiring a subgroup ${\cal Z} \equiv \pi_1(G) \subset Z(\widetilde{G})$ to act trivially on all dynamical representations.

\subsection{Gauge Group Topology from Lattice Embeddings}\label{subsec:gauge_group_from_lattice}

Compactifications of the heterotic or CHL string to 8d are characterized by a lattice $\Lambda_S$ with a symmetric non-degenerate bilinear pairing $\langle \cdot, \cdot \rangle_S: \Lambda_S \times \Lambda_S \rightarrow \bbZ$ of signature $(2,R)$.
For the heterotic string, $\Lambda_S$ is the rank 20 Narain lattice $\Lambda_N$ with $R=18$ \cite{Narain:1985jj,Narain:1986am}.
For the CHL string, $\Lambda_S$ is the rank 12 Mikhailov lattice $\Lambda_M$ with $R=10$ \cite{Mikhailov:1998si}.
In either case, we can linearly extend $\Lambda_S$ to vector space $V$ with a symmetric non-degenerate bilinear pairing $\langle \cdot , \cdot \rangle$:
\begin{align}
    V_S := \Lambda_S \otimes \mathbb{R} \, , \quad \langle \lambda_1 {\bf v}_1, \lambda_2 {\bf v}_2\rangle = \lambda_1 \lambda_2 \langle {\bf v}_1, {\bf v}_2\rangle_S \quad \text{for} \quad {\bf v}_1, {\bf v}_2 \in \Lambda_S, \, \, \lambda_1, \lambda_2 \in \mathbb{R} .
\end{align}
Since $\Lambda_S \subset V_S$, we will identify the lattice pairing $\langle \cdot , \cdot \rangle_S$ with the vector space pairing $\langle \cdot , \cdot \rangle$ in the following.
Then there is a dual lattice $\Lambda_S^* \subset V_S$ defined with respect to $\langle \cdot , \cdot \rangle$.
The Narain lattice is self-dual, $\Lambda_N^* = \Lambda_N$, but for the Mikhailov lattice, $\Lambda_M^* \neq \Lambda_M$.

By tuning the compactification moduli, the gauge symmetry of the effective theory in 8d can change.
Roughly speaking, this tuning amounts to setting the masses of certain states to 0, which can furnish the W-bosons of non-Abelian gauge symmetries.
The question of which non-Abelian gauge algebras $\mathfrak{g}$ are realizable in this way can be answered by cataloging all embeddings of the root lattices $\Lambda_\text{r}^\mathfrak{g}$ into $\Lambda_S$, whose roots $\Phi_\mathfrak{g}$ satisfy the worldsheet conditions which guarantee their masslessness.
This process has been recently carried out in detail \cite{Fraiman:2018ebo,Font:2020rsk,Font:2021uyw}, which in particular resulted in the full list of realizable gauge algebras with maximal rank (i.e., rank$(\mathfrak{g}) = 18$ for $\Lambda_\text{r}^\mathfrak{g} \hookrightarrow \Lambda_N$, and rank$(\mathfrak{g})=10$ for $\Lambda_\text{r}^\mathfrak{g} \hookrightarrow \Lambda_M$).

The purpose of this chapter is not to reiterate the necessary and sufficient criteria to find such embeddings, but to focus on the extraction of the \emph{global form} of the gauge group from the embedding data.
To this end, our working assumption will be that any root lattice embedding $\Lambda_\text{r}^\mathfrak{g} \stackrel{\imath}{\hookrightarrow} \Lambda_S$ we consider in the following satisfies these criteria, which guarantees that the corresponding 8d compactification (be it heterotic or CHL) has a non-Abelian symmetry algebra $\mathfrak{g}$. 
From this starting point, let us now distill the properties pertaining to the gauge group topology.

At the level of vector spaces we introduced above, an embedding $\Lambda_\text{r}^\mathfrak{g} \stackrel{\imath}{\hookrightarrow} \Lambda_S$ extends to an injective homomorphism $\imath: E \hookrightarrow V_S$, with $E = \Lambda_\text{r}^{\mathfrak{g}} \otimes \mathbb{R}$, such that
\begin{enumerate}
    \item \label{crit1} $\langle (\imath({\bf v}), \imath({\bf w}) \rangle = ({\bf v}, {\bf w})$ for any ${\bf v}, {\bf w} \in E$;
    \item \label{crit2} $\imath(\Lambda^\mathfrak{g}_\text{r})$ is a sublattice of $\Lambda_S \subset V_S$;
    \item \label{crit3} $\imath(\Lambda^\mathfrak{g}_\text{cr})$ is a sublattice of $\Lambda_S^* \subset V_S$.
\end{enumerate}
The first and second points are just a careful restatement of ``$\Lambda_\text{r}^\mathfrak{g} \stackrel{\imath}{\hookrightarrow} \Lambda_S$ is a lattice embedding''.
For heterotic vacua, the third point is equivalent to the second, since $\Lambda_N = \Lambda_N^*$, and $\Lambda_\text{r}^\mathfrak{g} = \Lambda_\text{cr}^\mathfrak{g}$ for an ADE algebra $\mathfrak{g}$.
For the CHL string this is a non-trivial criterion, which however is satisfied in valid embeddings \cite{Mikhailov:1998si}, as we will discuss below.
From criterion \ref{crit1}, it is straightforward to show that $\imath(\Lambda^*) = \imath(\Lambda)^*$ for any lattice $\Lambda \subset E$.
Then, the second and third conditions imply $\imath(\Lambda_\text{cw}^\mathfrak{g}) = \imath((\Lambda_\text{r}^\mathfrak{g})^*) \supset \Lambda_S^* \cap \imath(E)$, and $\imath(\Lambda^\mathfrak{g}_\text{w}) = \imath((\Lambda_\text{cr}^\mathfrak{g})^*) \supset \Lambda_S \cap \imath(E)$.

Given such an embedding $\imath: E \hookrightarrow V_S$, we naturally have an orthogonal decomposition 
\begin{align}\label{eq:decomp_V_into_E+F}
    V_S = \imath(E) \oplus F \, , \quad \text{where} \quad F = \{{\bf v} \in V \, | \, \langle {\bf v}, \imath({\bf w}) \rangle = 0 \, \text{ for all } \, {\bf w} \in E \} \, ,
\end{align}
because the restriction of $\langle \cdot , \cdot \rangle$ to $\imath(E)$ is the pairing $(\cdot, \cdot)$ which is non-degenerate.
For later convenience, we define the projections 
\begin{align}
\pi_F: \imath(E) \oplus F &\rightarrow F , \\
\pi_E: \imath(E) \oplus F &\rightarrow \imath(E).
\end{align} This decomposition determines the number of independent $\mathfrak{u}(1)$ gauge factors to be $\dim_\mathbb{R} (F) \equiv r_F= 2+R - \text{rank}(\mathfrak{g})$.

The lattice points of $\Lambda_S \subset V_S$ define physical states, and lattice points of $\Lambda_S^*$ impose constraints on the $\mathfrak{g}$-representations and $\mathfrak{u}(1)$ charges of these states, because they have to pair integrally with points in $\Lambda_S$.
These constraints can be interpreted as a non-trivial global structure of the gauge group of the form
\begin{align}\label{eq:global_gauge_group_general}
    \frac{[\widetilde{G}/{\cal Z}] \times U(1)^{r_F}}{{\cal Z}'} \, ,
\end{align}
where $\widetilde{G}$ is the simply-conncted version of the non-Abelian group with algebra $\mathfrak{g}$, ${\cal Z} \subset Z(\widetilde{G})$ a subgroup of the center, and ${\cal Z}'$ embeds into both $Z(\widetilde{G})$ and $U(1)^{r_F}$.

Let us first understand the ``purely non-Abelian'' constraints, i.e., those that specify the non-Abelian group $G = \widetilde{G} / {\cal Z}$.
These are restrictions on the physically realized weights that form the character lattice $\Lambda^G_\text{c} \subset \Lambda_\text{w}^\mathfrak{g} \subset E$.
In the string realization, any physical state corresponds to a lattice point ${\bf s} \in \Lambda_S$, which can be decomposed orthogonally as ${\bf s} = {\bf s}_E + {\bf s}_F \in \imath(E) \oplus F$.
The weight ${\bf w} \in \Lambda_\text{w} \subset E$ of such a state ${\bf s}$ under the non-Abelian part $G = \widetilde{G} / {\cal Z}$ is then the orthogonal projection of $\bf s$ onto $\imath(E)$, i.e., ${\bf s}_E  = \pi_E ({\bf s})$.\footnote{
More precisely, we identify ${\bf s}_E = \imath({\bf w})$.
Recalling that any weight is specified by its Dynkin labels ${\bf w}_i = ({\bf w}, \boldsymbol\mu_i^\vee)$, where $\boldsymbol\mu_i^\vee \in \Lambda_\text{cr} \subset E$ are the simple coroots, we have $\langle {\bf s}, \imath(\boldsymbol\mu_i^\vee) \rangle = \langle {\bf s}_E, \imath(\boldsymbol\mu_i^\vee) \rangle = ({\bf w}, \boldsymbol\mu_i^\vee)$.}

In other words, \textit{the character lattice of $G$ is the orthogonal projection of $\Lambda_S$ onto $\imath(E)$}: 
\begin{align}\label{eq:char_lat_non-ab}
    \Lambda^G_\text{c} \cong \pi_E (\Lambda_S) \subset \imath(E) \, .
\end{align}
The vectors ${\bf s} \in \Lambda_S$ are subject to the constraint that they pair integrally with all points in $\Lambda_S^*$.
Consider in particular a constraint associated with a point ${\bf c} \in \Lambda_S^* \cap \imath(E) \subset \imath(\Lambda_\text{cw})$, and let $\overline{\bf v} \in \Lambda_\text{cw}$ be such that $\imath(\overline{\bf v}) = {\bf c}$.
By orthogonality, we have 
\begin{align}
    \langle {\bf s}, {\bf c} \rangle = \langle\pi_E({\bf s}) , {\bf c}\rangle = \langle \imath({\bf w}), \imath(\overline{\bf v})\rangle = ({\bf w}, \overline{\bf v}) \in \bbZ \, .
\end{align}
This shows that $\Lambda_S^* \cap \imath(E)$ can be identified with the cocharacter lattice $\Lambda_\text{cc}^G$ of $G$.
So, from \eqref{eq:group_topology_via_lattices_general}, the non-Abelian gauge \emph{group} $G$ satisfies
\begin{align}\label{eq:non_ab_group_from_string_lat}
    \begin{split}
        Z(G) &= \frac{\Lambda^G_\text{c}}{\Lambda^\mathfrak{g}_\text{r}} = \frac{\pi_E(\Lambda_S)}{\imath(\Lambda^\mathfrak{g}_\text{r})} \, , \qquad \pi_1(G) = \frac{\Lambda^G_\text{cc}}{\Lambda^\mathfrak{g}_\text{cr}} = \frac{\Lambda_S^* \cap \imath(E)}{\imath(\Lambda^\mathfrak{g}_\text{cr})} \, . 
    \end{split}
\end{align}

Equivalently to \eqref{eq:char_lat_non-ab}, the projection of the lattice $\Lambda_S$ of physical states onto $F$ gives the ``characters'' of the $U(1)$s, i.e., the possible $U(1)$ charges.
Just as how the non-Abelian weight ${\bf w}({\bf s})$ of a state is specified by the Dynkin labels ${\bf w}_i = ({\bf w}, \boldsymbol\mu_i^\vee) = \langle {\bf s}_E, \imath(\boldsymbol\mu_i^\vee) \rangle$, where the simple coroots $\boldsymbol\mu_i^\vee$ span $E$ (over $\mathbb{R}$), the $U(1)$ charges are defined by the pairing with basis vectors of $F$.
To fix the normalization of the $U(1)$s, we use a lattice basis $\boldsymbol\xi_{\ell}$, $\ell=1,..., r_F$, for $\Lambda^*_S \cap F$, i.e., the orthogonal complement of $\imath(\Lambda^\mathfrak{g}_\text{cr})$ inside $\Lambda_S^*$ (we will see momentarily that $\Lambda^*_S \cap F \neq \emptyset$):
\begin{align}
    q_\ell({\bf s}) := \langle {\bf s}, \boldsymbol\xi_\ell \rangle \, .
\end{align}
In this normalization, states ${\bf s} \in \Lambda_S$ that are singlets under the non-Abelian gauge algebra $\mathfrak{g}$, i.e., $\pi_E({\bf s}) = 0 \Leftrightarrow {\bf s} \in \Lambda_S \cap F$, clearly have integer $U(1)$ charges $q_\ell$. 
The lattice points of $\Lambda_S^*$ that are not inside $\imath(E)$ now constrain the $U(1)$-charges $q_i({\bf s})$ and the non-Abelian weights ${\bf w}({\bf s})$ of a physical state corresponding to ${\bf s} \in \Lambda_S$.
To see this, we orthogonally decompose $\Lambda_S^* \ni {\bf c} = {\bf c}_E + {\bf c}_F$.
Note that, in general, neither ${\bf c}_E$ nor ${\bf c}_F$ are lattice points of $\Lambda_S^*$!
But, because for a root $\imath(\boldsymbol\alpha) \in \imath(\Lambda^\mathfrak{g}_\text{r}) \subset \Lambda_S \cap \imath(E)$, we have $\bbZ \ni \langle {\bf c}, \imath(\boldsymbol\mu) \rangle = \langle {\bf c}_E, \imath(\boldsymbol\mu) \rangle$, this guarantees that ${\bf c}_E = \imath(\overline{\bf v}) \in \imath(\Lambda^\mathfrak{g}_\text{cw})$ for some coweight $\overline{\bf v}$ of $\mathfrak{g}$.

Then, since $\Lambda^\mathfrak{g}_\text{cr} \subset \Lambda^\mathfrak{g}_\text{cw}$ are lattices of the same rank, we know that for any $\overline{\bf v} \in \Lambda^\mathfrak{g}_\text{cw}$ there is a smallest positive integer $k$ such that $\imath(k \overline{\bf v}) = k {\bf c}_E \in \imath(\Lambda^\mathfrak{g}_\text{cr}) \subset \Lambda_S^*$, so $k {\bf c}_F = k {\bf c} - k{\bf c}_E \in \Lambda_S^* \cap F$ is an \textit{integer} linear combination of $\boldsymbol\xi_{\ell}$. 
This means that $\langle {\bf c}_F, {\bf s} \rangle = \sum_\ell \lambda_\ell q_\ell({\bf s})$ is a $k$-fractional linear combination of the $U(1)$ charges $q_\ell({\bf s})$ of ${\bf s}$.
Therefore, the vector ${\bf c} \in \Lambda_S^*$ of the dual lattice imposes that 
\begin{align}
    \sum_\ell \lambda_\ell q_\ell({\bf s}) + ({\bf w}({\bf s}), \overline{\bf v}) \in \bbZ \, .
    \label{eq:cocharacter_constraint}
\end{align}
Moreover, from the above considerations it is clear that $k\lambda_\ell \in \bbZ$ and $k({\bf w}({\bf s}), \overline{\bf v})=({\bf w}({\bf s}), k\overline{\bf v}) \in \bbZ$.
Hence, the constraint is a $\bbZ_k$ constraint, in that it becomes trivial when it is multiplied by $k$.
It can be interpreted as identifying a $\bbZ_k \subset Z(\widetilde{G})$ with a subgroup of $U(1)^{r_F}$, i.e., it defines a $\bbZ_k$ subgroup of ${\cal Z}'$ in \eqref{eq:global_gauge_group_general}.
Just by counting dimension of $\Lambda_S^* / (\Lambda_S^* \cap \imath(E)$), there are at most $r_F$ linearly independent such constraints that are also independent of the ``non-Abelian constraints'' in ${\cal Z}$, i.e., ${\cal Z}' \cong \prod_{\ell=1}^{r_F} \bbZ_{k_\ell}$.
Then, analogously to \eqref{eq:non_ab_group_from_string_lat}, we have
\begin{align}\label{eq:ab_group_from_string_lat}
    {\cal Z}' \cong \frac{\Lambda_\text{cc}'}{\imath(\Lambda^\mathfrak{g}_\text{cr})} \quad \text{with} \quad \Lambda_\text{cc}' := \pi_E(\Lambda_S^*) \subset \imath(\Lambda^\mathfrak{g}_\text{cw}) \, .
\end{align}

In general, $\mathfrak{g} = \oplus \mathfrak{g}_i$ will be a sum of simple algebras, with a orthogonal decomposition of the lattice $\Lambda_\text{cw}^\mathfrak{g} = \oplus_i \Lambda_\text{cw}^{\mathfrak{g}_i}$.
Then, any lattice vector ${\bf c} \in \Lambda_\text{cc}'$ or ${\bf c} \in \Lambda_\text{cc}^G$ has a unique decomposition ${\bf c} = \sum_i \imath(\overline{\bf w}_i)$, where $\overline{\bf w}_i \in \Lambda_\text{cw}^{\mathfrak{g}_i}$ defines an equivalence class $[\overline{\bf w}_i] \equiv k_i \in \Lambda_\text{cw}^{\mathfrak{g}_i} / \Lambda_\text{cr}^{\mathfrak{g}_i} = Z(\widetilde{G}_i)$.
Then, the equivalence class of ${\bf c}$ in $\Lambda_\text{cc}' / \imath(\Lambda_\text{cw}^\mathfrak{g})$ (or $\Lambda_\text{cc}^G / \imath(\Lambda_\text{cw}^\mathfrak{g})$) $\subset \imath(\Lambda_\text{cw}^\mathfrak{g}) / \imath(\Lambda_\text{cr}^\mathfrak{g}) \cong Z(\widetilde{G})$ can be represented by the tuple $(k_i) \in \prod_i Z(\widetilde{G}_i) = Z(\widetilde{G})$.

In the following, we will exemplify the above structures in 8d heterotic and CHL vacua.
To ease the notation, we will from now on drop the explicit embedding map $\imath$, and regard all occurring lattices and subspaces as embedded into $V_S := \Lambda_S \otimes \mathbb{R} = \Lambda_S^* \otimes \mathbb{R}$, with all pairings inherited from $\langle \cdot, \cdot \rangle$ on $V_S$.

\textbf{Global gauge group structure of 8d heterotic vacua}

For 8d heterotic vacua with gauge rank 20, the topology of the non-Abelian gauge symmetry $G = \widetilde{G} / {\cal Z}$ has been recently studied through lattice embeddings in $\Lambda_N$ in \cite{Font:2020rsk}.
There, the crucial data were an overlattice $M$ of the root lattice $\Lambda_\text{r}^\mathfrak{g}$, whose length-squared 2 lattice points coincide with $\Lambda_\text{r}^\mathfrak{g}$, that embeds primitively inside $\Lambda_N$.
Then, the identification ${\cal Z} = \pi_1(G) = M / \Lambda_\text{r}^\mathfrak{g}$ was cross-checked with the classification of Mordell--Weil torsion of elliptic K3-surfaces in \cite{shimada_k3}, which is known to provide an equivalent characterization of the non-Abelian gauge group topology of 8d heterotic vacua via F-theory \cite{Aspinwall:1998xj}.

Comparing with the general formula \eqref{eq:group_topology_via_lattices_general} for $\pi_1(G)$, this identification seems to be at odds at first, since it is the \textit{coroot lattice} $\Lambda^\mathfrak{g}_\text{cr}$ rather than root lattice $\Lambda^\mathfrak{g}_\text{r}$ that appears in the quotients characterizing the fundamental group.
Of course, this is remedied by the fact that, in 8d heterotic vacua, only ADE algebras $\mathfrak{g}$ can be realized, which have $\Lambda^\mathfrak{g}_\text{r} = \Lambda^\mathfrak{g}_\text{cr}$.
Then, to be consistent with \eqref{eq:group_topology_via_lattices_general}, the overlattice $M$ should be identified with the cocharacter lattice $\Lambda_\text{cc}^G$.
Indeed, because $M$ contains $\Lambda^\mathfrak{g}_\text{r} = \Lambda^\mathfrak{g}_\text{cr}$, the requirement that it embeds primitively into $\Lambda_N$ means that it contains all points of $\Lambda_N \cap E$.
Furthermore, as $M/\Lambda_\text{r}^\mathfrak{g}$ is of finite order, $M$ has the same rank as $\Lambda_\text{r}^\mathfrak{g}$, which is the same as the dimension of $E$, so it cannot contain more points than $\Lambda_N \cap E$.
Therefore, we indeed find $M = \Lambda_N \cap E = \Lambda_N^* \cap E$ to be the cocharacter lattice as in \eqref{eq:non_ab_group_from_string_lat}. 

Our proposal can further determine the non-trivial constraints ${\cal Z}'$ between the non-Abelian group and the $U(1)$s of the heterotic compactification.
Note that, through duality to F-theory, there is an independent method to determine this structure via string junctions \cite{Guralnik:2001jh}.
While a full proof of the equivalence between these two methods has been handled in a companion work \cite{Cvetic:2021sjm}, we remark here that we indeed find identical results for 8d heterotic string vacua.
We will present, for completeness, an example of a heterotic model with $\mathfrak{g} = \mathfrak{su}(2)^2 \oplus \mathfrak{su}(4)^2 \oplus \mathfrak{so}(20)$ in Appendix \ref{app:heterotic_example}, where we show that the global gauge group is
\begin{align}\label{eq:heterotic_example_gauge_group}
    \frac{[SU(2)^2 \times SU(4)^2 \times Spin(20)]/[\bbZ_2 \times \bbZ_2] \times U(1)^2}{\bbZ_4 \times \bbZ_4} \, .
\end{align}

\subsection{Gauge Group Topology of 8d CHL Vacua}
\label{subsec:CHL_example}

Our main focus is to derive the gauge group topology of 8d CHL vacua.
The important difference from heterotic vacua is the fact that the string lattice $\Lambda_S$ is no longer self-dual in this case.
As found in \cite{Mikhailov:1998si}, the rank 12 momentum lattice is the Mikhailov lattice
\begin{align}
    \Lambda_M = U(2) \oplus U \oplus \text{E}_8 \cong U \oplus U \oplus \text{D}_8 \, .
\end{align}
Here, $\text{E}_8$ (D$_8$) denotes the root lattice of the Lie group $E_8$ ($Spin(16)$).
The rank 2 lattice $U = \{ l {\bf e} + n {\bf f} \, | \, (n,l) \in \bbZ^2 \}$ is defined by the Gram matrix
\begin{align}\label{eq:gram_matrix_U_lattice}
    \begin{pmatrix} \langle {\bf e}, {\bf e}\rangle & \langle {\bf e}, {\bf f}\rangle \\ \langle {\bf f}, {\bf e}\rangle & \langle {\bf f}, {\bf f} \rangle \end{pmatrix}  = \begin{pmatrix} 0 & 1 \\ 1 & 0 \end{pmatrix} \,.
\end{align} 
In this basis, $U(2) = \{ l{\bf e} + n{\bf f} \, | \, l \in 2\bbZ, n \in \bbZ \}$ . 
The dual Mikhailov lattice is then
\begin{align}
    \Lambda_M^* = \overline{U}(\tfrac{1}{2}) \oplus U \oplus \text{E}_8 \cong U \oplus U \oplus \text{D}_8^* \, ,
\end{align}
with $\overline{U}(\tfrac{1}{2})= \{ l{\bf e} + n{\bf f} \, | \, l \in \bbZ, n \in \tfrac12 \bbZ \}$.

The criteria for embeddings of root lattices into $\Lambda_M$ have also been studied in \cite{Mikhailov:1998si}.
One key novelty, compared to heterotic vacua, is that one can realize non-simply-laced $\mathfrak{sp}(n)$ gauge algebras in 8d CHL vacua.
Importantly, one criterion of the associated root lattice embedding $\Lambda_\text{r}^{\mathfrak{sp}(n)} \stackrel{\imath}{\hookrightarrow} \Lambda_M$ is that a long root $\boldsymbol\nu_L$ (with $( \boldsymbol\nu_L, \boldsymbol\nu_L) = 4$) of $\mathfrak{sp}(n)$ must embed such that it has \emph{even} pairing with all points in $\Lambda_M$:
\begin{align}
    \langle \imath(\boldsymbol{\nu}_L) , {\bf v} \rangle \in 2\bbZ \quad \text{for all} \quad {\bf v} \in \Lambda_M \, .
\end{align}
This in turn means that the short coroots, $\boldsymbol\nu_L^\vee = 2\boldsymbol\nu_L / (\boldsymbol\nu_L , \boldsymbol\nu_L) = \frac12 \boldsymbol\nu_L$, pair integrally with $\Lambda_M$.
In particular, this means $\boldsymbol\nu_L^\vee \in \Lambda_M^*$.
Since all other roots have length 2, and thus map to themselves as coroots, the coroot lattice $\Lambda_\text{cr}^{\mathfrak{sp}(n)}$ is guaranteed to embed into $\Lambda_M^*$, which is our condition \ref{crit3} for the embedding map $\imath: E \hookrightarrow V$.
As a result, the methods outlined in Section \ref{subsec:gauge_group_from_lattice} carry through.

To highlight the difference from the process for heterotic vacua, note that, in general, the ``overlattice'' $\Lambda_M \cap E$ of the root lattice $\Lambda_\text{r}^\mathfrak{g} \subset E$ neither contains all points of actual cocharacter lattice $\Lambda_\text{cc}^G = \Lambda_M^* \cap E$, nor those of the character lattice $\Lambda_\text{c}^G = \pi_E (\Lambda_M)$. 
For example, this discrepancy means that the quotient $(\Lambda_M \cap E) / \Lambda_\text{r}^\mathfrak{g}$ generally gives only a \emph{subgroup} of the center $Z(G) = \pi_E (\Lambda_M) / \Lambda_\text{r}^\mathfrak{g}$, and is not directly related to the fundamental group $\pi_1(G)$.

\textbf{Example}

To explicitly demonstrate our approach, we will consider a CHL model with $\mathfrak{g} = \mathfrak{su}(2)^2 \oplus \mathfrak{su}(4)^2 \oplus \mathfrak{sp}(2)$.
To this end, we represent $\mathbf{v}^{(\ell)} \in V_M := \Lambda_M \otimes \mathbb{R}$
\begin{align}
    \mathbf{v}^{(\ell)} = (l^{(\ell)}_1, l^{(\ell)}_2, n^{(\ell)}_1, n^{(\ell)}_2; \sigma^{(\ell)}_1, \dots, \sigma^{(\ell)}_8 ) \, ,
\end{align}
with pairing
\begin{equation}
    \langle \mathbf{v}^{(1)}, \mathbf{v}^{(2)} \rangle = l^{(1)}_1 n^{(2)}_1 + l^{(2)}_1 n^{(1)}_1 + l^{(1)}_2 n^{(2)}_2 + l^{(2)}_2 n^{(1)}_2 + \sum_{j = 1}^{8} \sigma_j^{(1)}  \sigma_j^{(2)} \, .
\end{equation}
Then, in the presentation $\Lambda_M = U \oplus U(2) \oplus \text{E}_8$ of the Mikhailov lattice, the $U$ lattice is spanned by $(l_1, 0, n_1, 0; 0, ...)$ with $l_1, n_1 \in \bbZ$, while the $U(2)$ part is spanned by $(0,l_2,0,n_2;0,...)$ with $l_2 \in 2\bbZ, n_2 \in \bbZ$ (see also \eqref{eq:gram_matrix_U_lattice}).
The $\text{E}_8$ lattice is then generated by $(0,0,0,0;\boldsymbol\sigma)$ with
\begin{align}
    \text{E}_8 \cong \left\{ \boldsymbol\sigma = (\sigma_1 , ..., \sigma_8) \in \left(\frac{1}{2}\bbZ\right)^8  \, \middle| \,  \sum_{i=1}^8 \sigma_i \in 2 \bbZ \ \ \text{and} \ \ \sigma_i - \sigma_j \in \bbZ \ \ \forall i, j \right\} \, .
\end{align}
For $\overline{\bf v} \in \Lambda_M^* = U \oplus \overline{U}(\tfrac12) \oplus \text{E}_8 \subset V_M$, the only difference for the conditions on the coefficients is that $l_2 \in \bbZ$ and $n_2 \in \tfrac12 \bbZ$.

The root lattice embedding which realizes the gauge algebra $\mathfrak{g} = \mathfrak{su}(2)^2 \oplus  \mathfrak{su}(4)^2 \oplus \mathfrak{sp}(2)$ has been computed in \cite{Font:2021uyw}.
$\Lambda_\text{r}^\mathfrak{g}$ is specified by the embedding of the simple roots $\boldsymbol\mu$ into $\Lambda_M$ in the above representation: 
\begin{equation}
\left[ \begin{array}{c}
\boldsymbol\mu_1 \\ \hline
\boldsymbol\mu_2 \\ \hline
\boldsymbol\mu_3 \\
\boldsymbol\mu_4 \\
\boldsymbol\mu_5 \\ \hline
\boldsymbol\mu_6 \\
\boldsymbol\mu_7 \\
\boldsymbol\mu_8 \\ \hline
\boldsymbol\mu_9 \\ 
\boldsymbol\mu_{10}
\end{array}\right]
=
\left[\begin{array}{cccc|cccccccc}
   1   & 2 &-1 &-1 & 0 & 0 & 0 & 1 & 1 & 1 & 1 &-2 \\\hline
   1   & 0 &-1 &-1 & 0 & 0 & 0 & 0 & 0 & 0 & 0 &-2 \\\hline
\ 0\ \ & 0 & 0 & 0 & 1 &-1 & 0 & 0 & 0 & 0 & 0 & 0 \\
   0   & 0 & 0 & 0 & 0 & 1 &-1 & 0 & 0 & 0 & 0 & 0 \\
   0   & 0 & 0 & 0 &-1 &-1 & 0 & 0 & 0 & 0 & 0 & 0 \\\hline
   0   & 0 & 0 & 0 & 0 & 0 & 0 & 0 & 0 & 1 &-1 & 0 \\
   0   & 0 & 0 & 0 & 0 & 0 & 0 & 0 & 1 &-1 & 0 & 0 \\
   0   & 0 & 0 & 1 & 0 & 0 & 0 &-1 &-1 & 0 & 0 & 0 \\\hline
   1   & 0 & 1 & 0 & 0 & 0 & 0 & 0 & 0 & 0 & 0 & 0 \\
   0   & 2 &-2 &-3 & 0 & 0 & 0 & 0 & 2 & 2 & 2 &-2
\end{array}\right].
\end{equation}
Here, the first two rows ${\boldsymbol\mu}_1$, ${\boldsymbol\mu}_2$ are the simple roots of $\mathfrak{su}(2)^2$, the next groups of three are the simple roots of the two $\mathfrak{su}(4)$'s, and the last two rows are simple roots of $\mathfrak{sp}(2)$, with ${\boldsymbol\mu}_{10}$ being the long root.
The corresponding coroot lattice is spanned by ${\boldsymbol\mu}^\vee_i = {\boldsymbol\mu}_i$ for $i<9$, and ${\boldsymbol\mu}_{10}^\vee = \frac12 {\boldsymbol\mu}_{10}$.
The coweight lattice is then spanned by $\overline{\bf w}_i = (C^{-1})_{ij} {\boldsymbol\mu}_j$, with $C_{ij} = \langle {\boldsymbol\mu}_i, {\boldsymbol\mu}_j \rangle$, which we re-express in terms of the coroots:
\begin{align}\label{eq:CHL_example_coweights_as_coroots}
    \begin{split}
        & \mathfrak{su}(2)^2 : \quad \overline{\bf w}_i = \tfrac12 {\boldsymbol\mu}_i^\vee \quad (i = 1, 2) \, , \\
        & \mathfrak{su}(4)^2: \quad \overline{\bf w}_{m+i} = \begin{pmatrix}
            \nicefrac34 & \nicefrac12 & \nicefrac14 \\
            \nicefrac12 & 1 & \nicefrac12 \\
            \nicefrac14 & \nicefrac12 & \nicefrac34
        \end{pmatrix}_{ij} 
            {\boldsymbol\mu}^\vee_{m+j} \quad (m=2,5) \, ,\\
        & \mathfrak{sp}(2): \quad \overline{\bf w}_9 = {\boldsymbol\mu}_9^\vee + {\boldsymbol\mu}_{10}^\vee \, , \quad \overline{\boldsymbol\mu}_{10} = \tfrac12 {\boldsymbol\mu}_9^\vee + {\boldsymbol\mu}_{10}^\vee \, .
    \end{split}
\end{align}
The orthogonal complement $F$ of $\Lambda_\text{r}^\mathfrak{g}$ in $\Lambda_M \otimes \mathbb{R}$ is spanned by
\begin{align}
    \begin{split}
    \boldsymbol\xi_1 &= (-2, 0, 2, 0; 0, 0, 0, 0, 0, 0, 0, 2), \\
    \boldsymbol\xi_2 &= (2, 4, -2, -5; 0, 0, 0, 1,  3, 3, 3, -4) \\
    \langle \boldsymbol\xi_1, \boldsymbol\xi_1 \rangle &= \langle \boldsymbol\xi_2, \boldsymbol\xi_2 \rangle = -4, \ \ \langle \boldsymbol\xi_1, \boldsymbol\xi_2 \rangle = 0 \, .
    \end{split}
\end{align}
These give the generators of the two independent $U(1)$s.

As explained above, any vector of $\overline{\bf v} \in \Lambda_{M}^*$ can be written as an integer linear combination of coweight and the $U(1)$ generators:
\begin{equation}
    \overline{\bf v} = (l_1, l_2, n_1, n_1; \sigma_1, \dots, \sigma_8) = \sum_{j = 1}^{10} k_j \overline{\bf w}_{j} + m_1 \boldsymbol\xi_1 + m_2 \boldsymbol\xi_2,\ \ \ k_j \in \bbZ \, .
\end{equation}
To determine the gauge group data \eqref{eq:non_ab_group_from_string_lat} and \eqref{eq:ab_group_from_string_lat}, we then need to express the basis of $\Lambda_M^*$ in this fashion.
This is a straightforward, but rather cumbersome exercise in linear algebra.
Sparing the details, the key step is to find the generators of $\Lambda_M^*$ that are linearly independent modulo the coroots ${\boldsymbol\mu}_i^\vee$.
For $\Lambda_M^* \cap E$, where $E = \Lambda_\text{r}^\mathfrak{g} \otimes \mathbb{R}$, there are two such generators:
\begin{align}
    \begin{split}
    {\bf c}_1 &= (1, 1, -1, -2; 0, 0, -1, 0, 1, 1, 1, -2)= \overline{\bf w}_2 + \overline{\bf w}_4 + \overline{\bf w}_{10}  \, ,\\
    {\bf c}_2 &= (1, 1, 0, 0; 0, 0, 0, 0, 1, 0, 0, -1) = \overline{\bf w}_1  + \overline{\bf w}_7 + (\overline{\bf w}_9 - \overline{\bf w}_{10})  \, .
    \end{split}
\end{align}
These generate $\pi_1(G) = (\Lambda_M^*\cap E) / \Lambda_\text{cr}^\mathfrak{g}$ as follows.
From \eqref{eq:CHL_example_coweights_as_coroots}, we see that ${\bf c}_1$ projects onto $\overline{\bf w}_2 = \frac12 {\boldsymbol\mu}^\vee_2 \in \Lambda_\text{cw}^{\mathfrak{su}(2)}$ of the second $\mathfrak{su}(2)$ factor, which is an order two element in $\Lambda_\text{cw}^{\mathfrak{su}(2)} / \Lambda_\text{cr}^{\mathfrak{su}(2)} \cong \bbZ_2$.
Likewise, the component $\overline{\bf w}_4 = \tfrac12 {\boldsymbol\mu}_3^\vee + {\boldsymbol\mu}_4^\vee + \tfrac12 {\boldsymbol\mu}_5^\vee$ projects onto the order two element in $\Lambda_\text{cw}^{\mathfrak{su}(4)} / \Lambda_\text{cr}^{\mathfrak{su}(4)} \cong \bbZ_4$ of the first $\mathfrak{su}(4)$ factor.
Finally, the component $\overline{\bf w}_{10} = \tfrac12 {\boldsymbol\mu}_9^\vee + {\boldsymbol\mu}_{10}^\vee$ projects onto the order 2 element in $\Lambda_\text{cw}^{\mathfrak{sp}(2)}/\Lambda_\text{cr}^{\mathfrak{sp}(2)} \cong \bbZ_2$.
Therefore, ${\bf c}_1$ itself projects onto an order 2 element in $\Lambda_\text{cw}^\mathfrak{g} / \Lambda_\text{cr}^\mathfrak{g} = Z(SU(2) \times SU(2) \times SU(4) \times SU(4) \times Sp(2)) = \bbZ_2 \times \bbZ_2 \times \bbZ_4 \times \bbZ_4 \times \bbZ_2$.
Moreover, this analysis shows that this element must be
\begin{align}
    z({\bf c}_1) = (0,1,2,0,1) \in \bbZ_2 \times \bbZ_2 \times \bbZ_4 \times \bbZ_4 \times \bbZ_2 \, .
\end{align}
An analogous argument shows that ${\bf c}_2$ also projects onto an order 2 element in $Z(\widetilde{G})$, given by
\begin{align}
    z({\bf c}_2) = (1,0,0,2,1) \in \bbZ_2 \times \bbZ_2 \times \bbZ_4 \times \bbZ_4 \times \bbZ_2 \, .
\end{align}
So the global structure of the non-Abelian gauge group $G$ is:
\begin{equation}
    G = \frac{SU(2)^2 \times SU(4)^2 \times Sp(2)}{\bbZ^{(1)}_2 \times \bbZ^{(2)}_2} \, ,
\end{equation}
where the embedding of each $\bbZ_2^{(i)}$ into $Z(\widetilde{G})$ is given by $z({\bf c}_i)$.
Once more, notice the importance of the dual momentum lattice in determining the global gauge group.
Neither ${\bf c}_1$ nor ${\bf c}_2$ are elements of $\Lambda_M$, since $l_2 = 1 \notin 2\bbZ$, so just inspecting points in $\Lambda_M$ would not have yielded this result.
However, ${\bf c}_1 + {\bf c}_2 \in \Lambda_M$, from which one might be tempted to deduce that $\pi_1(G) = \bbZ_2$, which is the diagonal $\bbZ_2 \subset \bbZ^{(1)}_2 \times \bbZ^{(2)}_2$.
Note that this $\bbZ_2$ embeds trivially into $Z(Sp(2))$.

We can explicitly verify, from the generators $z({\bf c}_i)$, that the ${\cal Z} =\bbZ^{(1)}_2 \times \bbZ^{(2)}_2$ 1-form symmetry is free of the anomaly \cite{Cvetic:2020kuw}.
Indeed, we will prove momentarily that this is guaranteed for the non-Abelian gauge group topology of any 8d CHL vacua.

From the lattice embedding, we can also determine the gauge group structure involving the $U(1)$s.
Two generators of $\Lambda_M^*$ that are not contained in $\Lambda_M^* \cap E$ are
\begin{align}
    \begin{split}
    {\bf c}_3 &= (0, 0, 0, 0; \tfrac{1}{2}, -\tfrac{1}{2}, -\tfrac{1}{2}, -\tfrac{1}{2}, \tfrac{1}{2}, -\tfrac{1}{2}, -\tfrac{1}{2}, -\tfrac{1}{2})= \tfrac{1}{4}\boldsymbol\xi_1 + \overline{\bf w}_2 + \overline{\bf w}_3 + \overline{\bf w}_{7} \, , \\
    {\bf c}_4 &= (1, 2, -1, -1; 0, 0, -1, 0, 1, 1, 1, -2)= \tfrac{1}{4}\boldsymbol\xi_2 + \overline{\bf w}_1 + \overline{\bf w}_4 + \overline{\bf w}_8 \, .
    \end{split}
\end{align}
In $\Lambda_\text{cc}' = \pi_E (\Lambda_M^*) \subset \Lambda_\text{cw}^\mathfrak{g}$, we then have $\pi_E ({\bf c}_3) = \overline{\bf w}_2 + \overline{\bf w}_3 + \overline{\bf w}_{7}$ and $\pi_E({\bf c}_4) = \overline{\bf w}_1 + \overline{\bf w}_4 + \overline{\bf w}_8$, whose equivalence class in $Z(\widetilde{G}) = \bbZ_2 \times \bbZ_2 \times \bbZ_4 \times \bbZ_4 \times \bbZ_2$ are
\begin{equation}
    z({\bf c}_3) = (0, 1, 1, 2, 0)  \, ,\quad z({\bf c}_4) = (1, 0, 2, 1, 0) \, ,
\end{equation}
which each generate a $\bbZ_4$ subgroup.
The first $\bbZ_4$, generated by ${\bf c}_3$, is a subgroup of the $U(1)$ generated by $\boldsymbol\xi_1$, whereas the second $\bbZ_4$ generated by ${\bf c}_4$ is in the $U(1)$ associated with $\boldsymbol\xi_2$.
So, in summary, the global form of the full gauge group is
\begin{equation}
    \frac{[(SU(2)^2 \times SU(4)^2 \times Sp(2))/(\bbZ_2 \times \bbZ_2)] \times U(1)^2}{\bbZ_4 \times \bbZ_4} \, .
\end{equation}

\subsection{Absence of 1-Form Anomalies}
\label{subsec:1-form_anomalies}

A non-trivial global structure $G = \widetilde{G} / {\cal Z}$ for the non-Abelian gauge group can be interpreted as having gauged the subgroup ${\cal Z}$ of the $Z(\widetilde{G})$ 1-form center symmetry of the simply-connected group $\widetilde{G}$ \cite{Gaiotto:2014kfa}.
In supergravity theories of dimension five or higher, such a gauging may be obstructed due to a mixed anomaly involving the large gauge transformations of the tensor field in the supergravity multiplet \cite{Apruzzi:2020zot,Cvetic:2020kuw,BenettiGenolini:2020doj}.
For 8d ${\cal N}=1$ theories, this obstruction can be quantified as follows \cite{Cvetic:2020kuw}.
Let $\widetilde{G} = \prod_i \widetilde{G}_i$, where $\widetilde{G}_i$ are simple factors, with $Z(\widetilde{G}_i) \cong \bbZ_{n_i}$, or $Z(\widetilde{G}_i) \cong \bbZ_2 \times \bbZ_2$ for $\widetilde{G}_i = Spin(4N_i)$.
Then a generator $z$ of ${\cal Z} \subset \prod_i Z(\widetilde{G}_i)$ is specified by a tuple $(k_i)$, where $k_i \mod n_i \in \bbZ_{n_i}$.\footnote{For $\widetilde{G}_i = Spin(4N_i)$ with $Z(\widetilde{G}_i) \cong \bbZ_2 \times \bbZ_2$, we would have two integers $k_i^{(1)}$ and $k_i^{(2)}$ modulo 2 specifying the embedding of $z$ into $Z(\widetilde{G}_i)$.}
The absence of the anomaly that would obstruct the gauging of ${\cal Z}$ requires that for any generator $z \simeq (k_i)$, we have
\begin{align}\label{eq:1-form_anomaly_condition_general}
    \sum_i m_i \, \alpha_{\widetilde{G}_i} \, k_i^2 = 0 \mod \bbZ \, ,
\end{align}
where $m_i$ is the Kac-Moody level of the worldsheet current algebra realization of $\widetilde{G}_i$.
The non-triviality of this condition is due to the fractional numbers $\alpha_{\widetilde{G}_i}$; for $\widetilde{G}$ with non-trivial $Z(\widetilde{G})$ that can appear in 8d supergravity, these are \cite{Cordova:2019uob}:\footnote{For $\widetilde{G} = Spin(4N)$ with $Z(\widetilde{G}) = \bbZ_2 \times \bbZ_2$, there are two inequivalent anomaly coefficients, $(\alpha^{(1)}, \alpha^{(2)})$.
The first is the same for both generators $(1,0)$ and $(0,1)$ of each $\bbZ_2$ factor; the second coefficient is associated with the generator $(1,1)$ of the diagonal $\bbZ_2$ subgroup.
In this identification, the (co-)spinor representation is charged under $(1,0)$ and $(0,1)$, respectively; hence, both are charged under $(1,1)$.
The vector representation is charged under both $(1,0)$ and $(0,1)$, but invariant under $(1,1)$.
\label{footnote:Spin_1}}
\begin{align}\label{eq:list_alpha}
    \renewcommand{\arraystretch}{1.2}
    \begin{array}{c||c|c|c|c|c|c}
        \widetilde{G} & SU(N) & Spin(4N+2) & Spin(4N) & E_6 & E_7 & Sp(N)\\ \hline 
        \alpha_{\widetilde{G}} & \frac{N-1}{2N} & \frac{2N+1}{8} & \left(\frac{N}{4}, \frac12\right) & \frac{2}{3} & \frac34 & \frac{N}{4}
    \end{array}
\end{align}
In the following, we show that for any non-Abelian gauge group $G = \widetilde{G}/{\cal Z}$ realized via lattice embeddings into the Narain or the Mikhailov lattice, as described above, \eqref{eq:1-form_anomaly_condition_general} is satisfied.

To do so, we first recall that any generator $z \simeq (k_1, k_2, ...) \in {\cal Z} = \pi_1(G)$ may be represented by a cocharacter vector ${\bf c} = \imath(\overline{\bf v}_c) \in \Lambda_\text{cc}^G = \Lambda_S^* \cap \imath(E)$.
If $\widetilde{G} = \prod_i \widetilde{G}_i$ with simple factors $\widetilde{G}_i$, then $E = \oplus_i E_i$, where $E_i = \Lambda^{\mathfrak{g}_i}_\text{r} \otimes \mathbb{R}$, is an orthogonal decomposition of $E$.
So 
\begin{align}\label{eq:splitting_of_cochar_into_coweights}
\begin{split}
    \overline{\bf v}_c = \sum_i \overline{\bf v}^{(i)}_c \, , \quad & \text{with } \, \overline{\bf v}_c^{(i)} \in \Lambda_\text{cw}^{\mathfrak{g}_i} \, \text{ representing } \, k_i \in \frac{\Lambda_\text{cw}^{\mathfrak{g}_i}}{\Lambda_\text{cr}^\mathfrak{g}} = Z(\widetilde{G}_i) \, , \\
    & \text{and } \, (\overline{\bf v}_c^{(i)}, \overline{\bf v}_c^{(j)}) = 0 \, \text{ for } \, i \neq j \, .
\end{split}
\end{align}

The key feature to prove \eqref{eq:1-form_anomaly_condition_general} is that
\begin{align}
\begin{split}
    \langle {\bf c}, {\bf c} \rangle = (\overline{\bf v}_c, \overline{\bf v}_c) = \sum_i (\overline{\bf v}_c^{(i)}, \overline{\bf v}_c^{(i)}) \in \begin{cases}
        2\bbZ & \text{for} \quad {\bf c} \in \Lambda_N^* = \Lambda_N \, , \\
        \bbZ & \text{for} \quad {\bf c} \in \Lambda_M^* \, .
    \end{cases}
\end{split}
\end{align}

Then, to prove \eqref{eq:1-form_anomaly_condition_general} for heterotic vacua (i.e., ${\bf c} \in \Lambda_N$), which only allows ADE-type groups $\widetilde{G}_i$ with $m_i = 1$, we need to show that, for any $\overline{\bf v}_c^{(i)} \in \Lambda_\text{cw}^{\mathfrak{g}_i} \subset E_i$ which represents $k_i \in Z(\widetilde{G}_i)$, its length square satisfies $(\overline{\bf v}_c^{(i)}, \overline{\bf v}_c^{(i)}) = 2 \alpha_{\widetilde{G}_i} k_i^2 \mod \bbZ$.
For CHL vacua with ${\bf c} \in \Lambda_M^*$, we need $(\overline{\bf v}_c^{(i)}, \overline{\bf v}_c^{(i)}) = 2 \alpha_{\widetilde{G}_i} k_i^2 \mod \bbZ$ for ADE-type $\widetilde{G}_i$ at level 2, and $(\overline{\bf v}_c^{(i)}, \overline{\bf v}_c^{(i)}) = \alpha_{\widetilde{G}_i} k_i^2 \mod \bbZ$ for $\widetilde{G}_i = Sp(N_i)$ at level 1.

For ADE-groups, this simplifies due to $\Lambda_\text{r}^\mathfrak{g} = \Lambda_\text{cr}^\mathfrak{g}$ being an even self-dual lattice.
In this case, $Z(\widetilde{G}) = \Lambda_\text{cw}^\mathfrak{g} / \Lambda_\text{cr}^\mathfrak{g} = (\Lambda_\text{r}^\mathfrak{g})^* / \Lambda_\text{cr}^\mathfrak{g} = (\Lambda_\text{r}^\mathfrak{g})^* / \Lambda_\text{r}^\mathfrak{g}$ is the so-called discriminant group of $\Lambda_\text{r}^\mathfrak{g}$ (see \cite{Nikulin_1980} for more details).
Via the pairing on $\Lambda_\text{r}^\mathfrak{g} \otimes \mathbb{R}$, one can use
\begin{align}
\begin{split}
    & \tfrac12 (\overline{\bf w} + \boldsymbol{\alpha}, \overline{\bf w} + \boldsymbol{\alpha}) = \tfrac12 (\overline{\bf w}, \overline{\bf w}) + (\overline{\bf w}, \boldsymbol{\alpha}) + \tfrac12 (\boldsymbol{\alpha},\boldsymbol{\alpha}) = \tfrac12 (\overline{\bf w}, \overline{\bf w})  \mod \bbZ \, , \\
    & \text{for} \quad \boldsymbol{\alpha} \in \Lambda_\text{r}^\mathfrak{g} \quad \text{and} \quad \overline{\bf w} \in \Lambda_\text{cw}^\mathfrak{g} = (\Lambda_\text{r}^\mathfrak{g})^* \, ,
\end{split}
\end{align}
to define a quadratic form $q: Z(\widetilde{G}) \rightarrow \mathbb{Q}/\bbZ$, which is a quadratic refinement of the so-called discriminant pairing on $Z(\widetilde{G})$.
Then, if the vector $\overline{\bf v}_c^{(i)} \in \Lambda_\text{cw}^{\mathfrak{g}_i}$ projects onto $k_i \in (\Lambda_\text{r}^{\mathfrak{g}_i})^* / \Lambda_\text{r}^{\mathfrak{g}_i} = Z(\widetilde{G}_i)$, we evidently have $(\overline{\bf v}_c^{(i)}, \overline{\bf v}_c^{(i)}) = 2 q(k_i) \mod \bbZ$.
The upshot of this detour is that the discriminant form of ADE root lattices and its quadratic refinements are well-known (see, e.g., \cite{shimada_k3}), and given by
\begin{align}
    \begin{aligned}
        \mathfrak{su}(N) & : \quad Z(\widetilde{G}) = \bbZ_N\, , \quad && q(k) = k^2 \cdot \tfrac{N-1}{2N} = k^2\, \alpha_{SU(N)}  \, , \\
        \mathfrak{so}(4N+2) & : \quad Z(\widetilde{G}) = \bbZ_4 \, , \quad && q(k) = k^2 \cdot \tfrac{2N+1}{8} = k^2 \, \alpha_{Spin(4N+2)} \, , \\
        \mathfrak{so}(4N) & : \quad Z(\widetilde{G}) = \bbZ_2 \times \bbZ_2 \, , \quad && q( k_1, k_2) = \tfrac{N}{4} (k_1^2 + k_2^2) + \tfrac{N-1}{2} k_1 k_2 \, , \\ 
        \mathfrak{e}_6 &: \quad Z(\widetilde{G}) = \bbZ_3 , \quad && q(k) = k^2 \cdot \tfrac23 = k^2 \, \alpha_{E_6}  \, , \\
        \mathfrak{e}_7 &: \quad Z(\widetilde{G}) = \bbZ_2 , \quad && q(a,b) = k^2 \cdot \tfrac34 = k^2 \, \alpha_{E_7} \, .
    \end{aligned}
\end{align}
Hence, for all simple ADE-type $\widetilde{G}_i$, the quadratic form gives $(\overline{\bf v}_c, \overline{\bf v}_c) = 2q(k_i) = 2k_i^2 \alpha_{\widetilde{G}_i}$, as required to show \eqref{eq:1-form_anomaly_condition_general} for both heterotic and CHL vacua.\footnote{For $\mathfrak{so}(4N)$, the generators $\vec{k} = (1,0), (0,1) \in \bbZ_2 \times \bbZ_2$ both satisfy $q(\vec{k}) = \tfrac{N}{4} = \alpha_{Spin(4N)}^{(1)} \mod \bbZ$, and $\vec{k}=(1,1)$ satisfies $q(k) = N - \tfrac12 = \tfrac12 \mod \bbZ = \alpha^{(2)}_{Spin(4N)}$.
This agrees with the mixed 1-form anomalies with the individual $\bbZ_2$ subgroups, see footnote \ref{footnote:Spin_1}.
}

For $\widetilde{G}_i = Sp(N_i)$, $Z(\widetilde{G}_i) = \bbZ_2$ is no longer the discriminant group of the root lattice, since $(\Lambda_\text{r}^\mathfrak{sp})^* \neq \Lambda_\text{cr}^\mathfrak{sp}$.
So we need to find an explicit representation of $k=1 \in \bbZ_2 = \Lambda_\text{cw}^{\mathfrak{sp}} / \Lambda_\text{cr}^{\mathfrak{sp}}$ in terms of a coweight $\overline{\bf v}$, and compute its length squared.
One way to represent the simple roots of $Sp(N)$ inside $E \cong \mathbb{R}^{N}$, with standard basis $\{\boldsymbol{e}_m\}$, is $\boldsymbol\mu_m = \boldsymbol{e}_m - \boldsymbol{e}_{m+1}$ for $m<N$, and $\boldsymbol\mu_{N} = 2 \boldsymbol{e}_{N}$ (see, e.g., \cite{Hall2015}).
Then, a basis $\overline{\bf w}_l$ for $\Lambda_\text{cw}^{\mathfrak{sp}(N)} = (\Lambda_\text{r}^{\mathfrak{sp}(N)})^*$, which is the dual basis of $\{\boldsymbol{\mu}_m\}$, i.e., $(\overline{\bf w}_l, \boldsymbol{\mu}_m) = \delta_{lm}$, is given by
\begin{align}
    \overline{\bf w}_l = \sum_m (M^{-1})_{l m} \, \boldsymbol{\mu}_m \quad \text{with} \quad M_{l m} = (\boldsymbol{\mu}_l, \boldsymbol{\mu}_m) = \begin{pmatrix}
        2 & -1 & 0 & \ldots & 0 \\
        -1 & 2 & -1 & \ddots & 0 \\
         & \ddots & \ddots & \ddots \\
        0 & \ldots & -1 & 2 & -2 \\
        0 & \ldots & 0 & -2 & 4
    \end{pmatrix}.
\end{align}
The inverse is
\begin{align}
\begin{split}
    (M^{-1})_{k m} & = \min(k, m) \, , \quad k,m < N \, , \\
    (M^{-1})_{Nm} & = (M^{-1})_{mN} = \tfrac{m}{2} \, , \quad m < N \, , \quad (M^{-1})_{NN} = \tfrac{N}{4} \, .
\end{split}
\end{align}
Now, because the coroot lattice $\Lambda_\text{cr}^{\mathfrak{sp}(N)}$ is spanned by $\boldsymbol{\mu}^\vee_m = \boldsymbol{\mu}_m$ for $m<N$, and $\boldsymbol{\mu}^\vee_N = \tfrac12 \boldsymbol{\mu}_N = \boldsymbol{e}_N$, the coweight basis vectors $\overline{\bf w}_k = \sum_m (M^{-1})_{km} \boldsymbol{\mu}_m$ with $k<N$ are actually integer vectors in $\Lambda_\text{cr}^{\mathfrak{sp}(N)}$, and hence represent $0 \in Z(Sp(N)) = \Lambda_\text{cw}^{\mathfrak{sp}(N)} / \Lambda_\text{cr}^{\mathfrak{sp}(N)}$.
The non-trivial element $1 \in \bbZ_2 \cong Z(Sp(N))$ must therefore be the equivalence class of $\overline{\bf w}_N = \sum_{m=1}^{N-1} \tfrac{m}{2} \boldsymbol{\mu}^\vee_m + \tfrac{N}{2} \boldsymbol{\mu}^\vee_N$.
Then, one can explicitly compute that
\begin{align}
\begin{split}
    (\overline{\bf w}_N, \overline{\bf w}_N) = \tfrac{N^2}{2}-\tfrac{N}{4} - 2N+2 &= \tfrac{N^2}{2} - \tfrac{N}{4} - \underbrace{\tfrac{N(N-1)}{2}}_{\in \bbZ \, \, \forall N} \! \! \mod \bbZ  \\
    &= \tfrac{N}{4} \! \! \mod \bbZ  \\
    &= \alpha_{Sp(N)} \mod \bbZ\, ,
\end{split}
\end{align}
which indeed is the form needed to prove \eqref{eq:1-form_anomaly_condition_general} for CHL vacua with $\mathfrak{sp}$ gauge factors.

\section{CHL Gauge Groups from Heterotic Models}
\label{sec:3}

In this section, we show how we can recover the data $({\cal Z}, {\cal Z}')$ about the gauge group topology (cf.~\eqref{eq:global_gauge_group_general}) of any 8d CHL vacua from the corresponding data of an 8d heterotic configuration.
Physically, this is based on the duality between CHL vacua and heterotic compactifications ``without vector structure'' \cite{Witten:1997bs}, or, equivalently F-theory with O7$_+$-planes encoded in ``frozen'' singularities \cite{Tachikawa:2015wka,Bhardwaj:2018jgp}.
In either of these duality frames, an 8d CHL vacuum with non-Abelian gauge algebra $\mathfrak{g} = \mathfrak{sp}(n) \oplus \mathfrak{h}$, with $\mathfrak{h}$ of ADE-type, arises from a heterotic or F-theory model with gauge algebra $\mathfrak{g}_\text{het} = \mathfrak{so}(16+2n) \oplus \mathfrak{h}$ (see also \cite{Hamada:2021bbz}).
Indeed, our CHL example in Section \ref{subsec:CHL_example} with $\mathfrak{g} = \mathfrak{sp}(2) \oplus \mathfrak{su}(4)^2 \oplus \mathfrak{su}(2)^2$ can be obtained from the heterotic example with $\mathfrak{g}_\text{het} = \mathfrak{so}(20) \oplus \mathfrak{su}(4)^2 \oplus \mathfrak{su}(2)^2$, whose global structure we compute in Appendix \ref{app:heterotic_example}.
A direct comparison shows that both examples have the same data, $({\cal Z}, {\cal Z}') = ({\cal Z}_\text{het}, {\cal Z}'_\text{het})$, which specifies the global structure of the gauge group.
Though, in general, these two pairs need not be identical, the identification of the CHL gauge group data $({\cal Z}, {\cal Z}')$ is straightforward to obtain, given the corresponding information about the heterotic/F-theory model.
The information about the latter can be extracted from various sources, e.g., from K3-data \cite{shimada_k3} specifying the F-theory setting, or the lattice embeddings of heterotic models \cite{Font:2020rsk}.
An alternative way is to use string junctions \cite{Guralnik:2001jh}, which will be explored in full detail in an upcoming work \cite{Cvetic:2021sjm}.

To describe the procedure, let us assume that we have the explicit embedding of the subgroup $({\cal Z}_\text{het}, {\cal Z}_\text{het}')$ into $Z(\widetilde{G}_\text{het}) = Z(Spin(16+2n)) \times Z(\widetilde{H})$, where $\widetilde{G}_\text{het}$ and $\widetilde{H}$ are the simply-connected groups with algebra $\mathfrak{g}_\text{het}$ and $\mathfrak{h}$, respectively.
Then, any generator $z = (z_\mathfrak{sp} , z_\mathfrak{h}) \in Z(Sp(n)) \times Z(\widetilde{H}) = \bbZ_2 \times Z(\widetilde{H})$ of the group ${\cal Z}$ (or ${\cal Z}'$, respectively) specifying the CHL gauge topology arises from a generator $\hat{z} = (\hat{z}_\mathfrak{so} , \hat{z}_\mathfrak{h})  \in Z(Spin(16+2n))\times Z(\widetilde{H})$ of ${\cal Z}_\text{het}$ (or ${\cal Z}'_\text{het}$, respectively), via the map
\begin{align}\label{eq:map_CHL_center_from_het_center}
\begin{split}
    z_\mathfrak{h} & = \hat{z}_\mathfrak{h} \in Z(\widetilde{H}) \, , \\
    z_\mathfrak{sp} & = \begin{cases}
        \hat{z}_\mathfrak{so} \mod 2 \, , & \hat{z}_\mathfrak{so} \in \bbZ_4 = Z(Spin(16+2n)) \quad (n \, \text{ odd}) \, , \\
        \hat{z}_\mathfrak{so}^{(1)} + \hat{z}_\mathfrak{so}^{(2)} \mod 2 \, , & \hat{z}_\mathfrak{so} = (\hat{z}_\mathfrak{so}^{(1)}, \hat{z}_\mathfrak{so}^{(2)}) \in \bbZ_2 \times \bbZ_2 = Z(Spin(16+2n)) \quad (n \, \text{ even}) \, .
    \end{cases}
\end{split}
\end{align}

While we will provide a proof of the validity of this map momentarily, it allows us to readily determine the gauge groups of all CHL vacua, given their heterotic ``parent''.
We illustrate this for all maximally enhanced cases (i.e., where the non-Abelian algebra $\mathfrak{g}$ has the maximally allowed rank of 10) in Table \ref{tab:big_table} in Appendix \ref{app:big_table}.
We find not only instances with ${\cal Z} = \bbZ_2$, but also examples with ${\cal Z} = \bbZ_2 \times \bbZ_2$.
Moreover, as an explicit check of the claims of Section \ref{subsec:1-form_anomalies}, all these center embeddings correspond to anomaly-free 1-form center symmetries.

\subsection*{Proving the Validity of the Map}

The proof of the validity of \eqref{eq:map_CHL_center_from_het_center} proceeds in three steps.
First, we review the embedding of the Mikhailov lattice and its dual into the Narain lattice \cite{Mikhailov:1998si}, and highlight the role of the $\mathfrak{so}(16) \subset \mathfrak{so}(16+2n)$ subalgebra.
Second, we construct the roots and coroots of the CHL gauge algebra $\mathfrak{g}$ from those of the parent heterotic algebra $\mathfrak{g}_\text{het}$.
Because $\mathfrak{g} \supset \mathfrak{sp}(n)$, the coroot lattice of $\mathfrak{g}$ will no longer be a sublattice of the heterotic configuration.
In the third step, we show that the cocharacter lattice of the CHL configuration is obtained from a suitable projection of the cocharacter lattice of the heterotic model.
Analogously to Section \ref{subsec:1-form_anomalies}, each cocharacter $\hat{\bf c}$ projects onto coweights of each gauge factor, thereby specifying a generator of the fundamental group as embedded into the center of the simply-connected cover.
Then, by identifying the generators of $\Lambda_\text{cw}^\mathfrak{g} / \Lambda_\text{cr}^\mathfrak{g}$ in terms of those of $\Lambda_\text{cw}^{\mathfrak{g}_\text{het}} / \Lambda_\text{cr}^{\mathfrak{g}_\text{het}}$, we will establish the map \eqref{eq:map_CHL_center_from_het_center}.

\textbf{Finding Mikhailov inside Narain}

As argued in \cite{Mikhailov:1998si}, there is, up to isomorphisms, a unique embedding
\begin{align}\label{eq:unique_D8_embedding}
    \Lambda_\text{r}^{\mathfrak{so}(16)} \equiv \text{D}_8 \hookrightarrow \Gamma_{16} \subset \Gamma_{16} \oplus U \oplus U = \Lambda_N
\end{align}
of the root lattice of $\mathfrak{so}(16)$ into the Narain lattice.
Denoting by $V_M$ the subspace of $V_N := \Lambda_N \otimes \mathbb{R}$ that is orthogonal to $\text{D}_8$, with projection $P_M: V_N \rightarrow V_M$, the Mikhailov lattice $\Lambda_M$ and its dual $\Lambda_M^*$ are found as
\begin{align}\label{eq:mikhailov_inside_narain}
\begin{split}
    P_M(\Lambda_N) & \cong \text{D}_8^* \oplus U \oplus U \cong \Lambda_M^* \, , \\
    \Lambda_N \cap V_M & \cong \text{D}_8 \oplus U \oplus U \cong \Lambda_M \, .
\end{split}
\end{align}

To give an ``intuitive'' argument for this, note that the lattice $\Gamma_{16}$ can be identified with the character lattice of $Spin(32)/\bbZ_2$, i.e., it is generated by the root lattice of $\mathfrak{so}(32)$ together with the weights of the spinor representation ${\bf S}_{\mathfrak{so}(32)}$.
The embedding $\text{D}_8 \hookrightarrow \Gamma_{16}$ then corresponds to the embedding of the roots of an $\mathfrak{so}(16) \subset \mathfrak{so}(32)$ subalgebra.
Conversely, the branching $\mathfrak{so}(32) \supset \mathfrak{so}(16) \oplus \mathfrak{so}(16)$ corresponds to an orthogonal decomposition $\Gamma_{16} \otimes \mathbb{R} = (\text{D}_8 \otimes \mathbb{R}) \oplus (\text{D}_8 \otimes \mathbb{R}) \equiv V_{D_8} \oplus V_{D_8}'$.
We can extend this decomposition to
\begin{align}\label{eq:unique_D8_embedding_vector_space}
\begin{split}
    & V_N = \Lambda_N \otimes \mathbb{R} = V_{D_8} \oplus \underbrace{V_{D_8}' \oplus (U \otimes \mathbb{R}) \oplus (U \otimes \mathbb{R})}_{=:V_M} \, , \\
    & P_M: V_N \rightarrow V_M \, , \quad {\bf s} = {\bf s}_D + {\bf s}_M \mapsto {\bf s}_M \, .
\end{split}
\end{align}
From this, the nature of the two lattices $P_M(V_N)$ and $\Lambda_N \cap V_M$ can be inferred from the group-theoretic decomposition
\begin{align}
\begin{split}
    \mathfrak{so}(32) & \supset \mathfrak{so}(16) \oplus \mathfrak{so}(16) \, ,\\
    {\bf adj}_{\mathfrak{so}(32)} & \rightarrow ({\bf adj}_{\mathfrak{so}(16)}, {\bf 1}) \oplus ({\bf 1}, {\bf adj}_{\text{so}(16)}) \oplus ({\bf V}_{\mathfrak{so}(16)}, {\bf V}_{\mathfrak{so}(16)}) \, ,\\
    {\bf S}_{\mathfrak{so}(32)} & \rightarrow ({\bf S}_{\mathfrak{so}(16)}, {\bf S}_{\mathfrak{so}(16)}) \oplus ({\bf C}_{\mathfrak{so}(16)} , {\bf C}_{\mathfrak{so}(16)}) \, ,
\end{split}
\end{align}
where ${\bf adj}$, ${\bf V}$ and ${\bf C}$ denote the adjoint, vector, and co-spinor representations, respectively.
At the level of lattices, the lack of any non-adjoint representations that are charged under just one of the $\mathfrak{so}(16)$ factors means that the only lattice points in the hyperplane $V'_{D_8}$ correspond to adjoint weights, i.e., $\Gamma_{16} \cap V'_{D_8} \cong \text{D}_8$.\footnote{The symmetry between the two $\text{D}_8$'s is an isomorphism of $\Lambda_N$. The results below would be the same if we swapped their roles in the subsequent discussion.}
However, since the bi-charged representations project onto the (co-)spinors and vectors of each $\mathfrak{so}(16)$, the projection of $\Gamma_{16}$ onto $V'_{D_8}$ is $\Lambda_\text{w}^{\mathfrak{so}(16)} = \left(\Lambda_\text{cr}^{\mathfrak{so}(16)} \right)^* = \left(\Lambda_\text{r}^{\mathfrak{so}(16)} \right)^* = \text{D}_8^*$.
Since the copies of $U$ lattices in \eqref{eq:unique_D8_embedding} and \eqref{eq:unique_D8_embedding_vector_space} are merely spectators in this argument, we find the Mikhailov lattice $\Lambda_M$ and its dual as given in \eqref{eq:mikhailov_inside_narain}.

\textbf{Constructing the CHL (co-)roots}

Since the heterotic gauge algebras $\mathfrak{g}_\text{het} = \mathfrak{h} \oplus \mathfrak{so}(16+2n)$, which are of interest to us, contain an $\mathfrak{so}(16+2n)$ algebra, we can identify an $\mathfrak{so}(16) \subset \mathfrak{so}(16+2n)$ subalgebra, whose root lattice may be identified with $\text{D}_8$ in \eqref{eq:unique_D8_embedding}.
By orthogonality \eqref{eq:unique_D8_embedding_vector_space}, the root lattice $\Lambda_\text{r}^\mathfrak{h} \subset \Lambda_N$ of the ADE-algebra $\mathfrak{h}$ must then lie in the plane $V_M$, and hence, by \eqref{eq:mikhailov_inside_narain}, $\Lambda_\text{r}^\mathfrak{h} = \Lambda_\text{cr}^\mathfrak{h}  \subset \Lambda_M$.

In order to obtain the $\mathfrak{sp}(n)$, first consider a basis for $\Lambda_\text{r}^{\mathfrak{so}(16+2n)}$ formed by the simple roots $\hat{\boldsymbol{\mu}}_i$, $i=1,...,8+n$, of $\mathfrak{so}(16+2n)$, with $\hat{\boldsymbol{\mu}}_{n+7}$ and $\hat{\boldsymbol{\mu}}_{n+8}$ forming the ``branched nodes'' in the $\mathfrak{so}(16+2n)$ Dynkin diagram:
\begin{equation}
\begin{split}
\begin{tikzpicture}
        \node [style=A] (13) at (1, 0) {$\ \hat{\mu}_{1}\ $};
        \node [style=none] (10) at (2, 0) {};
        \node [style=none] (18) at (3, 0) {$\cdots \cdots$};
        \node [style=none] (17) at (4, 0) {};
		\node [style=A] (4) at (5, 0) {$\hat{\mu}_{n + 5}$};
		\node [style=A] (5) at (7, 0) {$\hat{\mu}_{n + 6}$};
		\node [style=A] (6) at (8.6, 1) {$\hat{\mathbf{\mu}}_{n + 7}$};
		\node [style=A] (7) at (8.6, -1) {$\hat{\mu}_{n + 8}$};
		\draw [in=180, out=0] (4) to (5);
		\draw (5) to (6);
		\draw (5) to (7);
		\draw (13) to (10.center);
		\draw (17.center) to (4);
\end{tikzpicture}
\end{split}
\end{equation}
Then, associated with the branching $\mathfrak{so}(16+2n) \supset \mathfrak{so}(16) \oplus \mathfrak{so}(2n)$, the subspace $V_{D_8}$ in \eqref{eq:unique_D8_embedding_vector_space} is spanned by the $\mathfrak{so}(16)$ roots $\{ \hat{\boldsymbol{\mu}}_{n+1}, ..., \hat{\boldsymbol{\mu}}_{n+8}\}$.
Orthogonal to that will be the root lattice $\Lambda_\text{r}^{\mathfrak{so}(2n)}$ of $\mathfrak{so}(2n)$ inside $\text{D}_8 \oplus U \oplus U \cong \Lambda_M$, with simple roots
\begin{align}\label{eq:simple_roots_remainig_so}
    \hat{\boldsymbol{\rho}}_1 = \hat{\boldsymbol{\mu}}_{n-1} \, , \quad \hat{\boldsymbol{\rho}}_2 = \hat{\boldsymbol{\mu}}_{n-2} , \quad ... \quad  , \, \hat{\boldsymbol{\rho}}_{n-1} = \hat{\boldsymbol{\mu}}_{1} \, , \quad \hat{\boldsymbol{\rho}}_n = \hat{\boldsymbol{\mu}}_1 + \sum_{i=2}^{n+6} 2 \hat{\boldsymbol{\mu}}_{i} + \hat{\boldsymbol{\mu}}_{n+7} + \hat{\boldsymbol{\mu}}_{n+8} \, .
\end{align}
At the level of lattices, we have $\Lambda_\text{r}^{\mathfrak{so}(2n)} = \Lambda_\text{r}^{\mathfrak{sp}(n)}$, but the simple roots differ.
In terms of the $\mathfrak{so}(2n)$ roots $\hat{\boldsymbol{\rho}}$, the simple roots $\boldsymbol{\rho}$ of $\mathfrak{sp}(n)$ are \cite{Hall2015}
\begin{align}\label{eq:simple_roots_sp_from_so}
    {\boldsymbol{\rho}}_1 = \hat{\boldsymbol{\rho}}_1 \, , \quad {\boldsymbol{\rho}}_2 = \hat{\boldsymbol{\rho}}_2 \, , \quad ... \quad , \quad {\boldsymbol{\rho}}_{n-1} = \hat{\boldsymbol{\rho}}_{n-1} \, , \quad {\boldsymbol{\rho}}_n = -(\hat{\boldsymbol{\rho}}_{n-1} + \hat{\boldsymbol{\rho}}_{n}) \, ,
\end{align}
where $\boldsymbol{\rho}_n$ is the long root of $\mathfrak{sp}(n)$.
This modifies to coroot lattice $\Lambda_\text{cr}^\mathfrak{so(2n)} \neq \Lambda_\text{cr}^\mathfrak{sp(n)} \subset \Lambda_M^*$, with basis $\boldsymbol{\rho}_i^\vee = \boldsymbol{\rho}_i$ for $i=1,...,n-1$, and $\boldsymbol{\rho}_n^\vee = \tfrac12 \boldsymbol{\rho}_n$.
Under the projection $P_M: V_N \rightarrow V_M$, we have
\begin{align}\label{eq:projection_of_so_roots_as_sp}
\begin{split}
    P_M(\hat{\boldsymbol{\mu}}_i) & = 0 \quad \text{for} \quad i = n+1,...,n+8 \, , \\
    P_M(\hat{\boldsymbol{\mu}}_i) & = \hat{\boldsymbol{\mu}}_i = \boldsymbol{\rho}_{n-i} = \boldsymbol{\rho}_{n-i}^\vee \quad \text{for} \quad i=1,...,n-1 \, , \\
    P_M(\hat{\boldsymbol{\mu}}_n) & = \frac12 P_M \left(\hat{\boldsymbol{\rho}}_n - \hat{\boldsymbol{\mu}}_1 - \sum_{i=2}^{n-1} 2\hat{\boldsymbol{\mu}}_i - \sum_{i=n+1}^{n+6} 2\hat{\boldsymbol{\mu}}_i - \hat{\boldsymbol{\mu}}_{n+7} - \hat{\boldsymbol{\mu}}_{n+8} \right) \\ 
    & = \frac12 \left( \hat{\boldsymbol{\rho}}_n - \hat{\boldsymbol{\rho}}_{n-1} - 2\sum_{i=1}^{n-2} \hat{\boldsymbol{\rho}}_i \right) = - \frac{\boldsymbol{\rho}_n}{2} - \sum_{i=1}^{n-1} \boldsymbol{\rho}_i = - \boldsymbol{\rho}_n^\vee - \sum_{i=1}^{n-1} \boldsymbol{\rho}_i^\vee \, .
\end{split}
\end{align}

Note that, by our working assumption, the $\mathfrak{so}$-roots \eqref{eq:simple_roots_remainig_so} satisfy the masslessness condition for the heterotic string. 
A legitimate question is, then, if the $\mathfrak{sp}$-roots \eqref{eq:simple_roots_sp_from_so} satisfy the analogous conditions of the CHL string, i.e., whether the corresponding CHL vacuum indeed has an $\mathfrak{sp}(n) \oplus \mathfrak{h}$ gauge symmetry. This is indeed the case, since the embedding above is directly related to the realization of $\mathfrak{sp}$ gauge algebras given in \cite{Mikhailov:1998si}.

\textbf{CHL cocharacters from heterotic cocharacters}

Having identified the (co-)root lattices, we now need to show that every cocharacter of the CHL configuration arises from a cocharacter in the heterotic model.
By defining $\widehat{E} = \Lambda_\text{r}^{\mathfrak{g}_\text{het}} \otimes \mathbb{R} \subset V_N$ and $E = \Lambda_\text{r}^\mathfrak{g} \otimes \mathbb{R} \subset V_M$, we first want to show that
\begin{align}\label{eq:projection_commutes_w_cap}
    P_M(\Lambda_N) \cap E = P_M( \Lambda_N \cap \widehat{E})\,.
\end{align}
For this, we use the fact that the branching $\mathfrak{g}_\text{het} = \mathfrak{so}(16+2n) \oplus \mathfrak{h}  \supset \mathfrak{so}(16)  \oplus \mathfrak{so}(2n) \oplus \mathfrak{h} $ induces the orthogonal decomposition
\begin{align}
    \widehat{E} = V_{D_8} \oplus ( \underbrace{\Lambda_\text{r}^{\mathfrak{so}(2n)} \oplus \Lambda_\text{r}^{\mathfrak{h}}}_{= \Lambda_\text{r}^\mathfrak{g}}) \otimes \mathbb{R} = V_{D_8} \oplus E \subset V_{D_8} \oplus V_M = V_N\, .
\end{align}
Combining this with the general decomposition \eqref{eq:decomp_V_into_E+F} of $V_N = \widehat{E} \oplus F$, where $F$ is the hyperplane containing the $U(1)$s, \eqref{eq:unique_D8_embedding_vector_space} implies that
\begin{align}\label{eq:V_N_complete_decomp}
    V_M = E \oplus F \subset \underbrace{V_{D_8} \oplus E}_{\widehat{E}} \oplus F = V_N \, .
\end{align}
Notice that, in particular, the number of independent $U(1)$ gauge factors, $r_F = \dim F$, is the same for the CHL and the heterotic vacuum.
So, any ${\bf s} \in V_N$ can be written as ${\bf s} = {\bf s}_D + {\bf s}_E + {\bf s}_F$ with ${\bf s}_D \in V_{D_8}$, ${\bf s}_E \in E$, and ${\bf s}_F \in F$.
Then,
\begin{align}
\begin{split}
    {\bf s} \in P_M(\Lambda_N) \cap E & \quad \Leftrightarrow \quad {\bf s} = {\bf s}_E \in E \ \ \text{and} \ \ \exists \ {\bf s}_D \in V_{D_8} : {\bf s}_E + {\bf s}_D \in \Lambda_N \\
    & \quad \Leftrightarrow \quad \exists \ {\bf s}' = {\bf s}_E + {\bf s}_D \in \Lambda_N \cap \widehat{E} : {\bf s} := {\bf s}_E = P_M({\bf s}') \\
    & \quad \Leftrightarrow \quad {\bf s} \in P_M(\Lambda_N \cap \widehat{E}) \, .
\end{split}
\end{align}
The significance of \eqref{eq:projection_commutes_w_cap} is that we can identify the cocharacter lattice $\Lambda^G_\text{cc}$ of the CHL vacuum as the projection of the heterotic cocharacter lattice $\Lambda^{G_\text{het}}_\text{cc}$ under $P_M$.\footnote{We can also determine the CHL group structure including the $U(1)$s, following \eqref{eq:ab_group_from_string_lat}, from the parent heterotic theory. We will focus on the non-Abelian part, because the relevant data for its group topology are encoded in known K3-data \cite{shimada_k3}.
}
Namely, from the general prescription \eqref{eq:non_ab_group_from_string_lat}, we have
\begin{align}
    \Lambda^G_\text{cc} = \Lambda_M^* \cap E \stackrel{\eqref{eq:mikhailov_inside_narain}}{=} P_M(\Lambda_N) \cap E \stackrel{\eqref{eq:projection_commutes_w_cap}}{=} P_M(\Lambda_N \cap \widehat{E}) = P_M (\Lambda^{G_\text{het}}_\text{cc}) \, .
\end{align}

Analogously, we can infer the CHL cocharacters $\Lambda_\text{cc}'$, that encode the constraints involving the $U(1)$ charges, from the corresponding ones of the heterotic model, $\widehat{\Lambda}_\text{cc}^{'}$.
Namely, at the level of vector spaces, \eqref{eq:V_N_complete_decomp} implies that the projections $\pi_{\widehat{E}}: V_N \rightarrow \widehat{E}$ and $P_M: V_N \rightarrow V_M$ commute, and in fact compose to the projection $\pi_E: V_N \rightarrow E$.
Then, from \eqref{eq:ab_group_from_string_lat}, we have 
\begin{align}
    \Lambda'_{\text{cc}} = \pi_E (\Lambda_M^*) = \pi_E (P_M(\Lambda_N)) = P_M(\pi_E(\Lambda_N)) = P_M (\widehat{\Lambda}'_{\text{cc}}) \, .
\end{align}

In summary, we see that any cocharacter ${\bf c}$ of the CHL vacuum arises as the projection of a heterotic cocharacter $\hat{\bf c}$ under $P_M$.
Any such cocharacter $\hat{\bf c} \in \Lambda_N$ can be written as $\hat{\bf c} = \hat{\bf c}_{\mathfrak{so}(16+2n)} + \hat{\bf c}_{\mathfrak{h}} + \hat{\bf c}_F$.
If $\hat{\bf c}_F \in F$ is 0, then $\hat{\bf c} \in \Lambda_N \cap \widehat{E} = \Lambda_\text{cc}^{G_\text{het}}$ specifies an element $\hat{z}$ of $\pi_1(G_\text{het}) = {\cal Z}_\text{het} = \Lambda_\text{cc}^{G_\text{het}} / \Lambda_\text{cr}^{\mathfrak{g}_\text{het}} \subset \Lambda_\text{cw}^{\mathfrak{g}_\text{het}} / \Lambda_\text{cr}^{\mathfrak{g}_\text{het}} = Z(\widetilde{G}_\text{het})$.
If $\hat{\bf c}_F \neq 0$, then $\hat{\bf c}$ defines an element $\hat{z}$ of ${\cal Z}'_\text{het} \subset Z(\widetilde{G}_\text{het}) \times U(1)^{r_F}$.
In particular, each component $\hat{\bf c}_\mathfrak{h} \in \Lambda_\text{cw}^\mathfrak{h}$ and $\hat{\bf c}_{\mathfrak{so}(16+2n)} \in \Lambda_\text{cw}^{\mathfrak{so}(16+2n)}$ specifies a center element $\hat{z}_{\mathfrak{h}} \in Z(\widetilde{H})$ and $\hat{z}_\mathfrak{so} \in Z(Spin(16+2n))$, respectively, which is the restriction of $\hat{z}$ to the corresponding center subgroup.

To establish \eqref{eq:map_CHL_center_from_het_center}, all we need to determine is, given the generators of ${\cal Z}_\text{het}$ and ${\cal Z}'_\text{het}$ in terms of $(\hat{z}_\mathfrak{h}, \hat{z}_\mathfrak{so}) \in Z(\widetilde{H}) \times Z(Spin(16+2n)) = Z(\widetilde{G}_\text{het})$, what the corresponding center element $z = (z_\mathfrak{h}, z_{\mathfrak{sp}}) \in Z(\widetilde{H}) \times Z(Sp(n)) = Z(\widetilde{G})$ is.
For $z_\mathfrak{h}$, this is easy to answer.
Since the coroot lattice $\Lambda_\text{cr}^\mathfrak{h}$ remains invariant when passing from the heterotic to the CHL model, the component $P_M(\hat{\bf z}_\mathfrak{h}) = \hat{\bf z}_\mathfrak{h}$ defines the same element $z_{\mathfrak{h}} = \hat{z}_\mathfrak{h} \in Z(\widetilde{H}) = \Lambda_\text{cw}^\mathfrak{h} / \Lambda_\text{cr}^\mathfrak{h}$.

However, the same does not hold for $z_\mathfrak{sp}$, since in the CHL vacuum, we have to compare $P_M(\hat{\bf c}_{\mathfrak{so}(16+2n)})$ to the coroots of $\mathfrak{sp}(n)$, rather than those of $\mathfrak{so}(2n) \subset \mathfrak{so}(16+2n)$.
To this end, we first construct the coweights $\widehat{\overline{\bf w}}$ of $\mathfrak{so}(16+2n)$ which represent $Z(Spin(16+2n)) = \Lambda_\text{cw}^{\mathfrak{so}(16+2n)} / \Lambda_\text{cr}^{\mathfrak{so}(16+2n)}$.
As duals of the roots $\hat{\boldsymbol{\mu}}_i$, $i=1,...,8+n$, a basis for these are given by
\begin{align}
    \widehat{\overline{\bf w}}_l = (C^{-1})_{l i} \hat{\boldsymbol{\mu}}_i \quad \text{with} \quad C_{il} = (\hat{\boldsymbol{\mu}}_i, \hat{\boldsymbol{\mu}}_l) = \begin{pmatrix}
        2 & -1 & 0 & \ldots & \ldots & 0 \\
        -1 & 2 & \ddots & \ddots & & \\
        & \ddots & \ddots & \ddots & \\
        0 & \ldots & -1 & 2 & -1 & -1 \\
        0 & \ldots & 0 & -1 & 2 & 0 \\
        0 & \ldots & 0 & -1 & 0 & 2
    \end{pmatrix}
\end{align}
The inverse of the Cartan matrix $C$ of $\mathfrak{so}(16+2n)$ is (see, e.g., \cite{2017arXiv171101294W})
\begin{align}\label{eq:inverse_so_cartan_matrix}
\begin{split}
    (C^{-1})_{ij} & = (C^{-1})_{ji} = \min(i,j) \quad \text{for} \quad i,j < n+6 \, , \\
    (C^{-1})_{n+7,j} & = (C^{-1})_{j,n+7} = (C^{-1})_{n+8,j} = (C^{-1})_{j,n+8} = \tfrac{j}{2} \quad \text{for} \quad j < n+6 \, , \\
    (C^{-1})_{n+7,n+8} & = (C^{-1})_{n+8,n+7} = \tfrac{n+6}{4} \, , \quad (C^{-1})_{n+7,n+7} = (C^{-1})_{n+8,n+8} = \tfrac{n+8}{4} \, .
\end{split}
\end{align}
Since we are ultimately interested in the equivalence classes of $\mathfrak{sp}(n)$-coweights ${\bf c}_{\mathfrak{sp}}$ in $\bbZ_2 \cong \Lambda_\text{cw}^{\mathfrak{sp}(n)} / \Lambda_\text{cr}^{\mathfrak{sp}(n)}$, we use \eqref{eq:simple_roots_sp_from_so} to compute, for later convenience,
\begin{align}\label{eq:projection_so_coweights}
\begin{split}
    P_M(\widehat{\overline{\bf w}}_{n+7}) & = \sum_{j=1}^{n+8} (C^{-1})_{n+7, j} \, P_M(\hat{\boldsymbol{\mu}}_j) \\
    & = \sum_{j=1}^{n-1} (C^{-1})_{n+7,j} \, \boldsymbol{\rho}_{n-j}^\vee - (C^{-1})_{n+7,n} \left( \boldsymbol{\rho}_n^\vee + \sum_{j=1}^{n-1} \boldsymbol{\rho}_j^\vee \right) \\
    & = \sum_{j=1}^{n-1} (C^{-1})_{n+8,j} \, \boldsymbol{\rho}_{n-j}^\vee - (C^{-1})_{n+8,n} \left( \boldsymbol{\rho}_n^\vee + \sum_{j=1}^{n-1} \boldsymbol{\rho}_j^\vee \right) = P_M(\widehat{\overline{\bf w}}_{n+8}) \\
    & = \sum_{j=1}^{n-1} \frac{j-n}{2} \boldsymbol{\rho}_j^\vee - \frac{n}{2} \boldsymbol{\rho}_n^\vee \, .
\end{split}
\end{align}
Clearly, for any $n\geq 1$, at least one of the summands has a fractional coefficient.
And since $2P_M(\widehat{\overline{\bf w}}_{n+7}) = 2P_M(\widehat{\overline{\bf w}}_{n+8}) \in \Lambda_\text{cr}^{\mathfrak{sp}(n)}$, this means that $P_M(\widehat{\overline{\bf w}}_{n+7}) = P_M(\widehat{\overline{\bf w}}_{n+8})$ map to $1 \in \bbZ_2 \cong \Lambda_\text{cw}^{\mathfrak{sp}(n)} / \Lambda_\text{cr}^{\mathfrak{sp}(n)}$.
Moreover, since $\hat{\boldsymbol{\mu}}^\vee_i = \hat{\boldsymbol{\mu}}_i$, we can easily verify that
\begin{align}
\begin{split}\label{eq:relation_coweights_so}
    & \widehat{\overline{\bf w}}_{n+7} + \widehat{\overline{\bf w}}_{n+8} = \sum_{j=1}^{n+8} ((C^{-1})_{n+7,j} + (C^{-1})_{n+8,j}) \hat{\boldsymbol{\mu}}_j \\
    = & \sum_{j=1}^{n+6} j \, \hat{\boldsymbol{\mu}}_j + \frac{2n+14}{4} ( \hat{\boldsymbol{\mu}}_{n+7} + \hat{\boldsymbol{\mu}}_{n+8}) \\
    = & \frac{n+1}{2} ( \hat{\boldsymbol{\mu}}_{n+7}^\vee + \hat{\boldsymbol{\mu}}_{n+8}^\vee) \mod \Lambda_\text{cr}^{\mathfrak{so}(16+2n)} \, .
\end{split}
\end{align}
Now it is instructive to differentiate between even and odd $n$.

For odd $n$, for which we know $\Lambda_\text{cw}^{\mathfrak{so}(16+2n)} / \Lambda_\text{cr}^{\mathfrak{so}(16+2n)} \cong \bbZ_4$, the above equation is $\widehat{\overline{\bf w}}_{n+7} + \widehat{\overline{\bf w}}_{n+8} = 0 \mod \Lambda_\text{cr}^{\mathfrak{so}(16+2n)}$.
At the same time, since $2(n+6)$ and $2(n+8)$ cannot be divisible by 4 with odd $n$, both $\widehat{\overline{\bf w}}_{n+7}$ and $\widehat{\overline{\bf w}}_{n+8}$ are order 4 elements modulo $\Lambda_\text{cr}^{\mathfrak{so}(16+2n)}$.
The order 2 element in $\Lambda_\text{cw}^{\mathfrak{so}(16+2n)} / \Lambda_\text{cr}^{\mathfrak{so}(16+2n)}$ is then represented by $2 \widehat{\overline{\bf w}}_{n+7} = 2 \widehat{\overline{\bf w}}_{n+8} \mod \Lambda_\text{cr}^{\mathfrak{so}(16+2n)} = \widehat{\overline{\bf w}}_{2j-1} \mod \Lambda_\text{cr}^{\mathfrak{so}(16+2n)}$ for $1\leq j \leq n+6$ (as also evident from \eqref{eq:inverse_so_cartan_matrix}).
Then, if $\hat{\bf c}_{\mathfrak{so}(16+2n)}$ projects onto an order 4 element in $\Lambda_\text{cw}^{\mathfrak{so}(16+2n)} / \Lambda_\text{cr}^{\mathfrak{so}(16+2n)} \cong \bbZ_4$, it must be in the same equivalence class as either $\widehat{\overline{\bf w}}_{n+7}$ or $\widehat{\overline{\bf w}}_{n+8}$, which by \eqref{eq:projection_of_so_roots_as_sp} both map onto the order 2 element in $\Lambda_\text{cw}^{\mathfrak{sp}(n)} / \Lambda_\text{cr}^{\mathfrak{sp}(n)} \cong \bbZ_2$, confirming \eqref{eq:map_CHL_center_from_het_center} for odd $n$.

For even $n$, we see from \eqref{eq:inverse_so_cartan_matrix} and \eqref{eq:relation_coweights_so} that, in the quotient $\Lambda_\text{cw}^{\mathfrak{so}(16+2n)} / \Lambda_\text{cr}^{\mathfrak{so}(16+2n)} \cong \bbZ_2 \times \bbZ_2$, the equivalence classes of $\widehat{\overline{\bf w}}_{n+7}$, $\widehat{\overline{\bf w}}_{n+8}$, and $\widehat{\overline{\bf w}}_{n+7} + \widehat{\overline{\bf w}}_{n+8}$ all define order 2 elements.
This means that $\widehat{\overline{\bf w}}_{n+7}$ and $\widehat{\overline{\bf w}}_{n+8}$ represent the generators $(1,0)$ and $(0,1)$, respectively, while $\widehat{\overline{\bf w}}_{n+7} + \widehat{\overline{\bf w}}_{n+8}$ represents $(1,1)$.
Again, since $P_M(\widehat{\overline{\bf w}}_{n+7}) = P_M(\widehat{\overline{\bf w}}_{n+8})$ map to $1 \in \bbZ_2 = Z(Sp(n))$, this confirms \eqref{eq:map_CHL_center_from_het_center} for even $n$.

\section{Summary and Outlook}
\label{sec:conclusions}

In this chapter, we have presented an explicit identification of the gauge group topology
\begin{align}
    \frac{[\widetilde{G}/{\cal Z}] \times U(1)^{r_F}}{{\cal Z}'}
\end{align}
of 8d ${\cal N}=1$ compactifications of heterotic and CHL string theories, based on the embedding of the root lattice $\Lambda_\text{r}^\mathfrak{g}$ of the non-Abelian gauge algebra $\mathfrak{g}$ (with simply-connected cover $\widetilde{G}$) into the momentum lattice $\Lambda_S$ of string states.
For rank 20 theories, this agrees with known results from the heterotic \cite{Font:2020rsk} or the F-theory duality frame \cite{Aspinwall:1998xj,Guralnik:2001jh,shimada_k3,Cvetic:2021sjm}.
For CHL vacua, we have highlighted the necessity to distinguish between $\Lambda_S$ and its dual, as well as between the root $\Lambda_\text{r}^\mathfrak{g}$ and coroot lattice $\Lambda_\text{cr}^\mathfrak{g}$.
If this is taken into account, the resulting non-Abelian gauge group topology $G = \widetilde{G}/{\cal Z}$ is guaranteed to have no anomalies for the corresponding ${\cal Z} \subset Z(\widetilde{G})$ 1-form symmetry \cite{Cvetic:2020kuw}.
This can be verified explicitly for all 61 maximally enhanced CHL vacua, for which we have compiled the non-Abelian gauge group topology ${\cal Z}$ in Appendix \ref{app:big_table}.

We have also demonstrated in an explicit example how to compute the subgroup ${\cal Z}' \subset Z(\widetilde{G})$, which is identified with a subgroup of the Abelian gauge factor $U(1)^{r_F}$.
As we have argued for in Section \ref{sec:3}, the global gauge group structure $({\cal Z}, {\cal Z}')$ of any CHL vacuum can be in principle inferred from the corresponding data $({\cal Z}_\text{het}, {\cal Z}'_\text{het})$ of a parent heterotic model.
While ${\cal Z}_\text{het}$ can be readily obtained from existing data (e.g., from \cite{Font:2020rsk}), a comprehensive list of the part ${\cal Z}_\text{het}'$ involving the $U(1)$s has been covered in a companion work \cite{Cvetic:2021sjm}, from which we can then also classify ${\cal Z}'$ for CHL vacua.
There, we will also extend the analysis to include 8d ${\cal N}=1$ theories with gauge rank 4 \cite{Dabholkar:1996pc,Witten:1997bs,Aharony:2007du}.
Additionally, it should be straightforward to apply the machinery to 7d heterotic compactifications \cite{Fraiman:2021soq}.

Another interesting direction would be to understand the results about the global gauge group structures involving geometrically engineered $\mathfrak{sp}$ gauge symmetries in the language of higher-form symmetries \cite{Gaiotto:2014kfa}.
This would require a refinement of the framework of \cite{Morrison:2020ool,Albertini:2020mdx} to M-theory compactifications with frozen singularities \cite{Witten:1997bs,Tachikawa:2015wka,Bhardwaj:2018jgp}.
Furthermore, it would be interesting to reproduce the $\mathfrak{sp}(n)$-contribution to the mixed 1-form anomalies from a dimensional reduction of the M-theory Chern--Simons term in the presence of boundary fluxes which encode the 1-form symmetry background \cite{Cvetic:2021sxm}.
Lastly, having a complete catalog of gauge group topology including the $U(1)$s could provide a guideline to formulate field-theoretic constraints on allowed topologies ${\cal Z}'$, in similar fashion to \cite{Cvetic:2020kuw,Montero:2020icj}.



\chapter{ALL EIGHT- AND NINE-DIMENSIONAL STRING VACUA FROM JUNCTIONS}

\section{Introduction}

Supergravity theories in a large number of dimensions form an ideal laboratory to investigate the manifestation of quantum gravitational consistency conditions in the low-energy limit of string theory.
In particular, they provide a concrete class of models which corroborates the conjecture of ``string universality'', or ``string lamppost principle'', stating that all consistent (super)gravity theories arise from string theory.

Formulating and sharpening the relevant conditions on consistent effective theories of quantum gravity are at the heart of the Swampland Program \cite{Vafa:2005ui,Ooguri:2006in}.
Arguably, among the best motivated of these constraints is that quantum gravity theories should have no exact global symmetries. This statement can also be applied to higher-form global symmetries, such as 1-form center symmetries of non-Abelian gauge sectors. The condition demands that these symmetries are either gauged or broken. However, if they are to be gauged, one needs to demand the absence of obstructions/anomalies to turning on the gauge fields of these generalized symmetries.
This absence of anomalies of center 1-form symmetries can lead to severe restrictions on the global topology of the allowed gauge groups in supersymmetric theories \cite{Apruzzi:2020zot,Cvetic:2020kuw}. Similarly, the absence of global symmetries requires certain topological invariants called bordism groups to be trivial \cite{McNamara:2019rup}, once more leading to powerful constraints, in particular on the total rank of the gauge symmetry, of consistent supergravity theories in more than six dimensions \cite{Montero:2020icj,Bedroya:2021fbu}.

Being confronted with the set of supergravity models that pass the above consistency tests, the remaining question is whether all of these can be realized in string theory. To answer this we therefore need good control of the realization of the global form of the gauge groups, i.e., the fate of the center 1-form symmetries in string theory constructions. In the chapter we focus on compactifications to eight and nine dimensions (8d and 9d) with 16 supercharges (i.e., ${\cal N}=1$).

A powerful approach that successfully utilizes the machinery of geometry is F-theory \cite{Vafa:1996xn}, which ties the algebraic and arithmetic properties of elliptic K3-surfaces to 8d gauge theories with ADE gauge algebras of total rank $(2,18)$.\footnote{Throughout this chapter, we collect the number $a$ of independent gravi-photons, and the maximal non-Abelian gauge rank $r$ into a pair $(a, r)$, which we often refer to as the (total) gauge rank.
At generic values of moduli, the gauge algebra is hence $\mathfrak{u}(1)^{a+r}$.}
In this context, the global gauge group structure is encoded in the Mordell--Weil group of rational sections of the elliptic fibration \cite{Aspinwall:1998xj,Mayrhofer:2014opa,Cvetic:2017epq}. 
Under M-/F-theory duality, this can be phrased in terms of gauging and breaking higher-form symmetries \cite{Gaiotto:2014kfa}, that is reflected geometrically in the gluing of torsional homology cycles in local patches containing the non-Abelian gauge dynamics \cite{Cvetic:2021sxm}.\footnote{The investigation of generalized symmetries within the geometric engineering framework has received broad attention in recent literature \cite{DelZotto:2015isa,Morrison:2020ool,Albertini:2020mdx,Dierigl:2020myk,Apruzzi:2020zot,Cvetic:2020kuw,Bhardwaj:2020phs,DelZotto:2020sop,Bhardwaj:2021pfz,Hosseini:2021ged,Cvetic:2021sxm,Bhardwaj:2021wif,Apruzzi:2021mlh,Apruzzi:2021nmk,Tian:2021cif,DelZotto:2022fnw}.}
Moreover, through suitable deformations that correspond to infinite distance points in the 8d moduli space, the F-theory geometry also classifies 9d ${\cal N}=1$ string vacua with gauge rank $(1,17)$ \cite{Lee:2021qkx,Lee:2021usk}.
This is consistent with the dual heterotic description, where the 9d moduli space --- described via the rank $(1,17)$ Narain lattice --- is contained in that of the 8d moduli space, with a rank $(2,18)$ Narain lattice description (see \cite{Font:2020rsk} for a recent comprehensive study).

However, the dictionary between geometry and physics is less understood in the presence of so-called frozen singularities \cite{Witten:1997bs,deBoer:2001wca,Tachikawa:2015wka,Bhardwaj:2018jgp}.
While these are known to be the necessary ingredient for an F-theory description of $\mathfrak{sp}$ gauge algebras on the 8d ${\cal N}=1$ moduli branches of gauge ranks $(2,10)$ and $(2,2)$, the characterization of, e.g., the gauge group topology is no longer purely geometric (i.e., given by the Mordell--Weil group) \cite{Cvetic:2021sjm}.
Likewise, it is not immediately clear how to identify decompactification limits on these moduli spaces.
On the other hand, advances in the Swampland program \cite{Hamada:2021bbz} strongly suggest, that all 8d ${\cal N}=1$ vacua should have a characterization in terms of an elliptically-fibered K3.

As we will demonstrate in this chapter, string junctions provide a \emph{unified framework} that encompasses all these features.
In this description, the underlying elliptic K3 is encoded in the configuration of $[p,q]$-7-branes of type IIB string theory, whose $[p,q]$-type are in one-to-one correspondence to elliptic singularities characterized by an $SL(2,\bbZ)$-monodromy $M_{[p,q]}$.
The junctions are then $(p,q)$-strings or -5-branes stretched between the 7-branes.
In their original formulation \cite{Gaberdiel:1997ud,Gaberdiel:1998mv,DeWolfe:1998zf} that is equivalent to F-theory without frozen singularities, junctions describe the 8d gauge dynamics with ADE gauge algebras,\footnote{String junctions have been also used to construct lower-dimensional theories, see \cite{Bonora:2010bu,Cvetic:2011gp,Garcia-Etxebarria:2013tba,Grassi:2014ffa,Agarwal:2016rvx,Grassi:2018wfy,Hassler:2019eso,Heckman:2020svr,Grassi:2021ptc}.} as well as their higher-form symmetries \cite{Cvetic:2021sxm}.
To account for a junction description of all 8d ${\cal N}=1$ vacua, we extend the discussion to include O7$^+$-planes, which are the IIB avatars of frozen singularities, and have the same monodromy as an elliptic $\text{D}_8$ singularity \cite{Witten:1997bs,deBoer:2001wca,Tachikawa:2015wka}.

A concept that will be key to this chapter are so-called fractional null junctions, which are certain fractional (and hence, unphysical) $\colpq{p}{q}$-charges encircling all 7-branes, i.e., \emph{loop junctions}.
In the absence of O7$^+$-planes, these are known to be equivalent to Mordell--Weil torsion of the underlying elliptic K3 \cite{Fukae:1999zs,Guralnik:2001jh}.
To correctly account for the electric and magnetic center symmetries and the gauge group topology for the $\mathfrak{sp}$ gauge symmetries that arise in the presence of O7$^+$, it turns out to be instrumental to study separately $(p,q)$-strings and -5-branes.
The key difference is, while any integer number of 5-branes can end on an O7$^+$, the number of \emph{string}-prongs there must be even.
Indeed, with this modification, we find that the junction description of 8d vacua with one O7$^+$ is equivalent to so-called CHL vacua \cite{Chaudhuri:1995fk,Chaudhuri:1995bf} of rank $(2,10)$, including the characterization of the global gauge group structure \cite{Cvetic:2021sjm,Font:2021uyw}.
Moreover, it is straightforward to include two O7$^+$-planes, thereby giving a junction-esque classification of 8d string vacua with gauge rank $(2,2)$ including their gauge group topology, for which there is no known heterotic or CHL string description.

In addition, we also propose a junction description for decompactification limits to 9d including O7$^+$-planes.
Parallel to the 9d uplifts of the rank $(2,18)$ setting \cite{Lee:2021qkx,Lee:2021usk} (see also \cite{Collazuol:2022jiy} for a related discussion of 10d uplifts of 9d heterotic vacua), we identify the corresponding infinite distance limits with O7$^+$-planes by the emergence of loop junctions that affinize the 8d gauge algebra.
Again, the subtle differences from having modified boundary conditions for strings and 5-branes can be cross-checked with the momentum lattice description of the CHL string for uplifting 8d rank $(2,10)$ theories to 9d rank $(1,9)$ theories.
Like in 8d, the junction description naturally encodes the gauge group topology of 9d vacua.
For the rank $(2,2)$ theories without a momentum lattice analog, the 9d theories with rank $(1,1)$ that result from the junction description live on two disconnected moduli branches that are only connected through an $S^1$-reduction to 8d, which matches other stringy constructions \cite{Aharony:2007du}.
This further establishes junctions as a complimentary framework to sharpen aspects of the Swampland Distance conjecture \cite{Ooguri:2006in} in string compactifications.

The rest of the chapter is organized as follows.
After reviewing the junction framework with ordinary $[p,q]$-7-branes in Section \ref{sec:local_analysis}, we discuss, in Section \ref{subsec:junctions_on_O7}, the modified boundary conditions for strings and 5-branes on an O7$^+$-plane that give rise to the correct higher-form symmetries of $\mathfrak{sp}$ gauge algebras in 8d.
In Section \ref{sec:global}, we then describe global 8d models by ``gluing'' together local patches with 7-brane stacks involving O7$^+$-planes.
A particular focus will be on the gauge group topology that is encoded in the fractional null junctions.
We then examine, in Section \ref{sec:9d}, the infinite distance limits described via 7-branes and junctions that correspond to 9d ${\cal N}=1$ vacua, for which we will also determine the global gauge group structure.
The appendices contain some technical aspects, as well as the full list of gauge group topologies for all 8d vacua with maximally-enhanced non-Abelian symmetries in Appendix \ref{app:results}.

\section{String and 5-brane junctions}
\label{sec:local_analysis}

String junctions provide an efficient way to classify electrically charged states with respect to gauge symmetries localized on 7-brane stacks in type IIB string theory. Therefore, they also contain information about the electric 1-form center symmetries and the global realization of the 8d gauge group \cite{Fukae:1999zs,Guralnik:2001jh,Cvetic:2021sxm}. 
The magnetically dual perspective is provided by 5-brane webs, which can also be described by junctions \cite{Aharony:1997bh,Leung:1997tw,Kol:1998cf,DeWolfe:1999hj}.

In this section we recall some properties of string and 5-brane junctions in the presence of a general $[p,q]$-7-brane stack. This provides a local construction of the charged states. Importantly, the charge under the center symmetry is related to the appearance of certain \emph{fractional junctions}, the extended weight junctions, that determine the global properties of the model \cite{Cvetic:2021sxm}. We then generalize the discussion of string and 5-brane junctions to backgrounds containing O7$^+$-planes, whose geometric interpretation in F-theory is more challenging \cite{Tachikawa:2015wka,Bhardwaj:2018jgp}.
With the help of string junctions we can successfully extract the correct properties of these configurations and identify the electric center symmetries also for symplectic gauge groups.
This analysis is repeated with 5-brane junctions, which, as opposed to the ADE algebras realized without O7$^+$'s, have a subtle distinction from string junctions that is precisely needed to correctly account for the magnetic center symmetry.

\subsubsection[Basics of \texorpdfstring{$[p,q]$}{[p,q])}-7-branes and string junctions]{Basics of \boldmath{$[p,q]$}-7-branes and junctions}

In this section we will recall the basics of the junction description for 8d ${\cal N}=1$ dynamics from type IIB compactifications \cite{Gaberdiel:1997ud,DeWolfe:1998zf}.
The key players are spacetime filling $[p,q]$-7-branes $\mathbf{X}_{[p,q]}$, which we will denote with square brackets.
A single 7-brane must have coprime $p$ and $q$.
In the plane perpendicular to its worldvolume, $\mathbf{X}_{[p,q]}$ induces a singularity in the axio-dilaton profile $\tau = C_0 + i e^{- \phi}$, composed of the RR 0-form $C_0$ and the dilaton field $\phi$, which is characterized by an $SL(2,\bbZ)$ monodromy
\begin{align}
M_{[p,q]} = \begin{pmatrix} 1 + p q & - p^2 \\ q^2 & 1 - p q \end{pmatrix} \in SL(2,\mathbb{Z}) \, ,
\end{align}
which acts on $\tau$ by a M{\" o}bius transformation, and in the doublet representation,
\begin{align}
    \colpq{B_2}{C_2} \mapsto M_{[p,q]} \colpq{B_2}{C_2} \, ,
\end{align}
on the NSNS- and RR-2-form fields $(B_2, C_2)$.
This monodromy can be captured by a branch cut in the perpendicular plane that emanates form the 7-brane.
In the following it will prove useful to introduce conventions for some special 7-branes, that will later appear in the construction of non-Abelian gauge algebras
\begin{equation}
\begin{split}
&\mathbf{A} = \bX_{[1,0]}: \quad M_{[1,0]} = \begin{pmatrix} 1 & -1 \\ 0 & 1 \end{pmatrix} \,, \\
&\mathbf{B} = \bX_{[1,-1]}: \quad M_{[1,-1]} = \begin{pmatrix} 0 & -1 \\ 1 & 2 \end{pmatrix} \,, \\
& \mathbf{C} = \bX_{[1,1]}: \quad M_{[1,1]} = \begin{pmatrix} 2 & -1 \\ 1 & 0 \end{pmatrix} \,, \\
& \mathbf{N} = \bX_{[0,1]}: \quad M_{[0,1]} = \begin{pmatrix} 1 & 0 \\ 1 & 1 \end{pmatrix} \,.
\end{split}
\label{eq:branespecies}
\end{equation}

In a local model, where the perpendicular plane is non-compact (i.e., is $\mathbb{R}^2 = \mathbb{C} \ni z$), it is customary to extend the branch cuts all downwards (meeting at $z = -i \infty$, without crossing each other before).
Starting from a configuration describing a certain 8d vacuum, we can obtain another one on the same 8d ${\cal N}=1$ moduli space by moving the 7-branes.
When $\bX_{[p_1, q_1]}$ crosses the branch cut of $\bX_{[p_2,q_2]}$ from the left to right, the $[p,q]$-type changes according to:
\begin{align}\label{eq:brane_move_left-to-right}
    \begin{split}
        \underrightarrow{{\bf X}_{[p_1, q_1]}} {\bf X}_{[p_2, q_2]} \rightarrow {\bf X}_{[p_2, q_2]} {\bf X}_{[p_1 + D \cdot p_2, \, q_1 + D \cdot q_2]} \, ,
    \end{split}
\end{align}
where $D \equiv \det\left( \begin{smallmatrix} p_1 & p_2 \\ q_1 & q_2 \end{smallmatrix} \right)$, and the arrow indicating the branch-cut-crossing 7-brane.
Likewise, when it crosses from right to left, one has
\begin{align}\label{eq:brane_move_right-to-left}
    {\bf X}_{[p_2, q_2]} \underleftarrow{{\bf X}_{[p_1, q_1]}} \rightarrow & {\bf X}_{[p_1 + D \cdot p_2, \, q_1 + D \cdot q_2]} {\bf X}_{[p_2, q_2]} \, .
\end{align}
For any concrete configuration, we can arrange the 7-branes along a horizontal axis, and denote them as $\bX_{[p_1, q_1]} \bX_{[p_2,q_2]} \, ...$ by labelling from left to right.
The $SL(2,\bbZ)$ monodromy around any (connected) part of this chain is the product of the individual monodromies of the encircled branes \emph{from right to left}.
To obtain a valid global configuration (i.e., where the perpendicular plane is $\mathbb{P}^1$) describing an 8d ${\cal N}=1$ supergravity model, tadpole cancellation requires exactly 24 $[p,q]$-7-branes, whose overall monodromy must be the identity.
Note that, while their \emph{relative} $[p,q]$-types are pivotal for distinguishing different physical configurations, an \emph{overall} $SL(2,\bbZ) \ni g$ transformation,
\begin{align}
    \left[ \begin{smallmatrix} p_i \\ q_i \end{smallmatrix} \right] \mapsto g \left[ \begin{smallmatrix} p_i \\ q_i \end{smallmatrix} \right] \, , \quad M_{[p_i, q_i]} \mapsto g M_{[p_i, q_i]} g^{-1} \quad \text{for all } \, i \, ,
\end{align}
does not matter physically for a model (local or global) described by a collection of 7-branes $\bX_{[p_i, q_i]}$.

BPS-particles in 8d arise from $(p,q)$-strings --- a bound state of $p$ fundamental and $q$ D-strings with electric charge $\colpq{p}{q}$ under $\colpq{B_2}{C_2}$ --- anchored on the 7-branes of the same $[p,q]$-type, and extending as a directed line into the perpendicular plane.
Their magnetically dual objects, which are four-dimensional in 8d, are given by $(p,q)$-5-branes --- a bound state of $p$ NS5- and $q$ D5-branes with magnetic charge $\colpq{p}{q}$ under $\colpq{B_2}{C_2}$ --- that share four common spatial directions as the 7-brane and also project to lines in the perpendicular plane.
Since strings and 5-branes can fuse and split, so long as the overall $\colpq{p}{q}$-charge is conserved at every vertex, they form junctions, see left of Figure \ref{fig:Juncrules}.
Note that, by flipping the direction on any prong, its $\colpq{p}{q}$-charge acquires a minus sign.\footnote{To make contact with the F-theory description of type IIB, note that (directed) junctions can be interpreted as (oriented) 2-cycles in an elliptic K3, on which M2- and M5-branes can be wrapped, which are the objects dual to strings and 5-branes under M-/F-theory duality. See, e.g., \cite{Cvetic:2021sxm} for details of this correspondence.}

As charged objects of the 2-form fields $(B_2, C_2)$, strings and 5-branes also experience $SL(2,\bbZ)$ monodromies as they are transported around 7-branes.
This action can be represented in the perpendicular plane, after choosing the branch cuts, by an analogous transformation
\begin{align}\label{eq:monodromy_on_prongs}
\colpq{r}{s} \rightarrow \colpq{r'}{s'} =  M_{[p,q]} \colpq{r}{s} =  \colpq{r}{s} + (q r - p s) \colpq{p}{q} \,
\end{align}
on the $\colpq{r}{s}$-charges of a junction-prong as it crosses the branch cut of a $[p,q]$-7-brane, see middle of Figure \ref{fig:Juncrules}.
Finally, in analogy to Hanany--Witten transitions \cite{Hanany:1996ie}, the same junction can be expressed, by moving the branch-cut-crossing prong across the 7-brane, as a junction with an additional prong on the 7-brane, see right of Figure \ref{fig:Juncrules}.
\begin{figure}[ht]
    \centering
    \includegraphics[width = \textwidth]{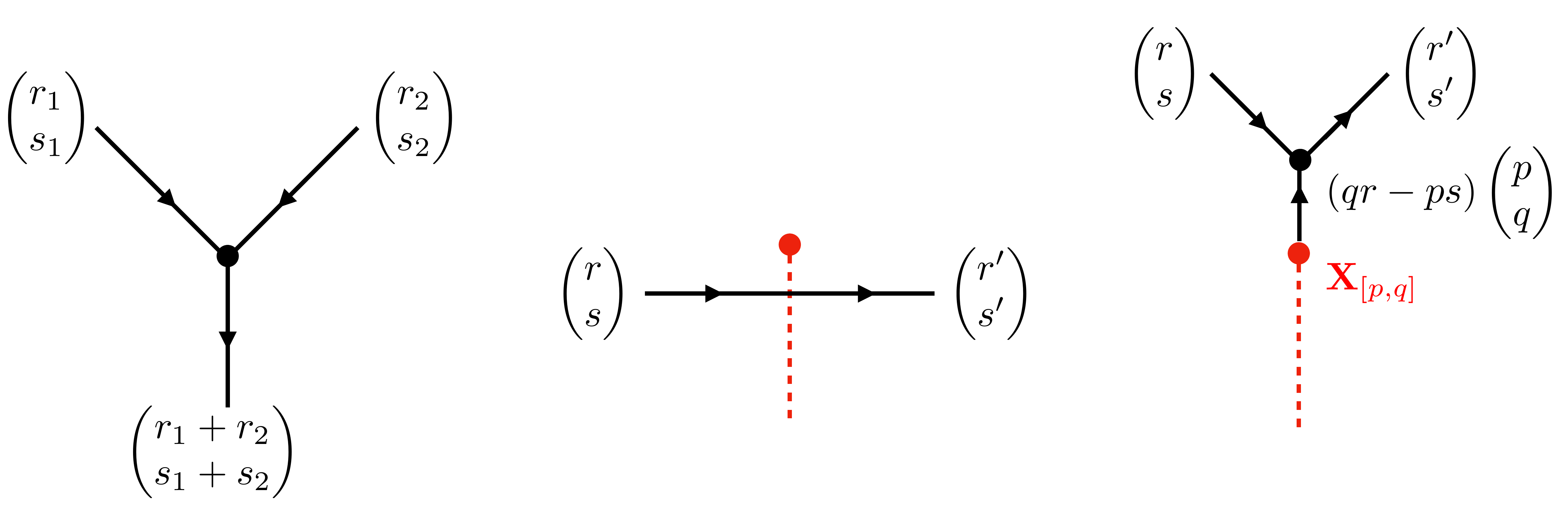}
    \caption[matrix in text]{Strings and 5-branes, which are represented as lines in the perpendicular plane, form junctions, where the $\colpq{p}{q}$-charge at each vertex is conserved (left). In the presence of 7-branes, they undergo monodromy transformations \eqref{eq:monodromy_on_prongs} when they cross a branch cut (middle). By a Hanany--Witten transition, the same junction can be represented as having a prong on the 7-brane (right).}
    \label{fig:Juncrules}
\end{figure}

To have non-Abelian gauge dynamics in 8d, we have to collide 7-branes to form stacks.
Strings that stretch between different constituents of a stack then become light and form massless W-bosons of the enhanced gauge symmetry.
In terms of the 7-brane types \eqref{eq:branespecies}, ADE-gauge algebras are realized when the following stacks form\footnote{We have chosen a particular $SL(2,\bbZ)$-frame that is common in the literature, but any $SL(2,\bbZ)$-conjugated configuration would obviously give the same gauge algebra.}:
\begin{align}\label{eq:ADE_brane_stacks}
\begin{array}{| c | c | c |}
\hline
\text{Lie algebra} & \text{brane constituents} & \text{monodromy} \\ \hline \hline
\mathfrak{su}_{n} & \mathbf{A}^{n} & \begin{pmatrix} 1 & - n \\ 0 & 1 \end{pmatrix} \\ \hline
\mathfrak{so}_{2n} & \mathbf{A}^n \mathbf{B} \mathbf{C} & \begin{pmatrix} -1 & n-4 \\ 0 & -1 \end{pmatrix} \\ \hline
\mathfrak{e}_{n \geq 1} & {\bf A}^{n-1} {\bf B} {\bf C}^2 & \begin{pmatrix} -2 & 2n-9 \\ -1 & n-5 \end{pmatrix} \\ \hline
\tilde{\mathfrak{e}}_{n \geq 0} & {\bf A}^n {\bf X}_{[2,-1]} {\bf C} & \begin{pmatrix} -3 & 3n-11 \\ -1 & n-4 \end{pmatrix} \\ \hline
\end{array}
\end{align}
where we have used exponents to group the same type of branes that are appear consecutively.
The overall monodromy of a 7-brane stack is the product of the individual branes from right to left; e.g., $M_{\mathfrak{so}_{2n}} = M_{[1,1]} M_{[1,-1]} M_{[1,0]}^n$.
The realizations of the exceptional algebras are physically equivalent, i.e., equal up to 7-brane moves inside the stack and $SL(2,\bbZ)$ conjugations, for $n\geq 2$,\footnote{We use the standard identifications $\mathfrak{e}_2 \cong \mathfrak{su}_2 \oplus \mathfrak{u}(1)$, $\mathfrak{e}_3 \cong \mathfrak{su}_3 \oplus \mathfrak{su}_2$, $\mathfrak{e}_4 \cong \mathfrak{su}_5$, $\mathfrak{e}_5 \cong \mathfrak{so}_{10}$.} while $\mathfrak{e}_1 \cong \mathfrak{su}_2$ and $\tilde{\mathfrak{e}}_1 \cong \mathfrak{u}(1)$; finally, the $\tilde{\mathfrak{e}}_0$ configuration corresponds to a trivial gauge algebra.
There are additional strongly coupled versions of the Lie algebra $\mathfrak{su}_n$ with $n \in \{2,3\}$ of the form $\mathbf{A}^{n+1} \mathbf{C}$.
In the remaining part of this section, we will focus mainly on the ``standard'' cases $\mathfrak{su}_n$, $\mathfrak{so}_{2n}$ and $\mathfrak{e}_{n\geq 6}$, while $\tilde{\mathfrak{e}}_{n}$ will be relevant in Section \ref{sec:9d}.
Of course there is a beautiful relation between the 7-branes stacks above with their induced $SL(2,\mathbb{Z})$ monodromies, and the classification of singularities in elliptic fibrations by Kodaira, which is central in F-theory (see \cite{Weigand:2018rez,Cvetic:2018bni} for recent reviews and additional references). 
In the following, we will focus solely on the junction perspective.

\subsubsection{The junction lattice}\label{subsec:junction_lattice}

In the following, we give an abstract definition of junctions as lines in the plane perpendicular to the 7-branes satisfying the axioms above.
In principle, one has to specify if they represent $(p,q)$-strings or 5-brane webs to attach physical meaning to them.

Consider the junctions formed by a single prong extending from one 7-brane $\bX_{[p,q]}$, which we denote with a lower case letter as $\bx_{[p,q]}$, and sometimes call a unit junction.
In analogy to the different types defined in \eqref{eq:branespecies}, there are then also junctions
\begin{align}
\mathbf{a}\,, \enspace \mathbf{b} \,, \enspace \mathbf{c} \,, \enspace \mathbf{n} \,.
\label{eq:prongs}
\end{align}
Since a general string or 5-brane junction takes the form of a linear combination of the individual prongs, the set of all physical junctions (strings or 5-branes) on a 7-brane configuration, $\bX_{[p_1, q_1]} \bX_{[p_2,q_2]} \cdots \bX_{[p_i,q_i]} \cdots$, form a $\bbZ$-module,
\begin{align}
J_{\text{phys}} =  \big\{ {\bf j} = \sum_i a^i \mathbf{x}_{[p_i,q_i]} \, | \, a^i \in \mathbb{Z} \big\} \, .
\label{eq:Jphys}
\end{align}
One important physical invariant is the net, or \emph{asymptotic} $\colpq{p}{q}$-charge of a junction ${\bf j}$, given by $\colpq{p}{q}_\text{asymp} = \sum_i a^{i} \colpq{p_i}{q_i}$.

One further defines a symmetric bi-linear pairing $(.,.)$ on this module as follows.
For the basis junction $\bx_{[p_i, q_i]}$ (note that the ordering of the 7-branes is important), one defines\footnote{Here, we simply present the rules as stated in \cite{DeWolfe:1998zf}. It can be shown that they agree with the geometric intersection pairing for the elliptic K3 of the dual F-theory description.}
\begin{align}
\begin{split}
& \big( \mathbf{x}_{[p_i,q_i]}, \mathbf{x}_{[p_j,q_j]} \big) = \big( \mathbf{x}_{[p_j,q_j]}, \mathbf{x}_{[p_i,q_i]} \big) = \begin{cases}
    -1 \, , & \text{if } \, i=j \\
    \tfrac12 \det \left( \begin{smallmatrix} p_i & p_j \\ q_i & q_j \end{smallmatrix} \right) \, , & \text{if } \, \bX_{[p_i,q_i]} \, \text{ is on the left of } \, \bX_{[p_j, q_j]} \, .
\end{cases}
\end{split}
\label{eq:selfintersec}
\end{align}
By linearly extending to the module $J_\text{phys}$, we endow it with a lattice structure, which will be called the (physical) junction lattice.
For example, consider an arrangement of only $\bA$, $\bB$ and $\bC$ branes which are ordered ``alphabetically'',
\begin{align}
    \bA_1 \, \cdots \, \bA_\alpha \, \cdots \, \bB_1 \, \cdots \, \bB_\beta \, \cdots \, \bC_1 \, \cdots \, \bC_\gamma \, .
\end{align}
For this 7-brane configuration, we have
\begin{align}
\begin{split}
    & (\ba_\alpha, \ba_{\alpha'}) = -\delta_{\alpha, \alpha'} \, , \quad (\bb_\beta, \bb_{\beta'}) = - \delta_{\beta,\beta'} \, , \quad (\bc_{\gamma}, \bc_{\gamma'}) = -\delta_{\gamma,\gamma'} \, , \\
    & (\ba_\alpha, \bb_\beta) = -\tfrac12 \, , \quad (\ba_\alpha, \bc_\gamma) = \tfrac12 \, , \quad (\bb_\beta, \bc_\gamma) = 1 \, .
\end{split}
\end{align}

An important property of the pairing \eqref{eq:selfintersec} is that it is invariant under 7-brane motions.
That is, given a fixed set of 7-branes $\bX_{[p_i,q_i]}$, the lattice $(J_\text{phys}, ( \cdot, \cdot))$ changes only up to a unimodular transformation (i.e., change of basis) when we move the 7-branes.
To see this it suffices to consider a two-branes configuration $\bX_{[p_1, q_1]} \bX_{[p_2,q_2]}$ with $J_\text{phys} = \{a^1 \bx_{[p_1,q_1]} + a^2 \bx_{[p_2,q_2]} \}$, for which the pairing matrix is $\left( \begin{smallmatrix} -1 & D/2 \\ D/2 & -1 \end{smallmatrix} \right)$, with $D = \det\left( \begin{smallmatrix} p_1 & p_2 \\ q_1 & q_2 \end{smallmatrix} \right)$.
After moving $\bX_{[p_1,q_1]}$ across the branch cut to the right, as in \eqref{eq:brane_move_left-to-right} (the other direction, \eqref{eq:brane_move_right-to-left}, works analogously), the configuration $\bX_{[p_2,q_2]} \bX_{[p_1 + D p_2, q_1 + D q_2]} \equiv \bX_{l} \bX_{r}$ has the lattice
\begin{align}
    J_\text{phys} = \left\{ a_l \bx_{l} + a_r \bx_{r}  \right\} \, , \ \text{with} \ (\bx_{i} ,\bx_{j}) = \left( \begin{smallmatrix}
        -1 & -\tfrac{D}{2} \\
        -\tfrac{D}{2} & -1
    \end{smallmatrix} \right) = \left( \left( \begin{smallmatrix}
        -D & 1 \\
        1 & 0
    \end{smallmatrix} \right)^{-1} \right)^T \left( \begin{smallmatrix}
        -1 & \tfrac{D}{2} \\
        \tfrac{D}{2} & -1
    \end{smallmatrix} \right) \left( \begin{smallmatrix}
        -D & 1 \\
        1 & 0
    \end{smallmatrix} \right)^{-1} .
\end{align}
The unimodular transformation $\left( \begin{smallmatrix} -D & 1 \\ 1 & 0 \end{smallmatrix} \right)$ precisely traces how the original unit prongs $\{\bx_{[p_i,q_i]}\}$ are expressed in terms of the new basis $\{\bx_l, \bx_r\}$ after the 7-brane transition \eqref{eq:brane_move_left-to-right},
\begin{align}
    \bx_{[p_1,q_1]} \rightarrow -D \bx_{l} + \bx_{r} \, , \quad \bx_{[p_2,q_2]} \rightarrow \bx_{l} \, .
\end{align}

\subsubsection*{Loop junctions and their self-pairings}

A junction type that will be particularly important to our discussions are \emph{loop junctions}.
These are formed by encircling a collection of 7-branes with an $\colpq{r}{s}$-charge, that undergoes $SL(2,\bbZ)$ transformations as it crosses their branch cuts.
As a convention for nomenclature, we use the $\colpq{r}{s}$-charge it starts out with to label the loop junction $\boldsymbol\ell_{(r,s)}$, even if its $(p,q)$-type changes after it comes back, see Figure \ref{fig:loop_junction}.
If the overall monodromy of the encircled stack is $M$, then such a loop has asymptotic charge $\colpq{p}{q} = (M - \mathbb{1}) \colpq{r}{s}$.
For two loops, $\boldsymbol\ell_{(r,s)}$ and $\boldsymbol\ell_{(u,v)}$, encircling the same 7-branes, one clearly has $\boldsymbol\ell_{(r,s)} + \boldsymbol\ell_{(u,v)} = \boldsymbol\ell_{(r+u,s+v)}$.
In principle, any such loop can be turned into the standard basis \ref{eq:Jphys} with prongs on 7-branes by pulling the loop across the encircled 7-branes via a Hanany--Witten transition, which allows to compute pairings involving loop junctions.
However, since the loop does not touch the encircled 7-branes, but only sees their overall monodromy, the self-pairing of a loop should be computable just with this data.
\begin{figure}[ht]
    \centering
    \includegraphics[width = .35 \textwidth]{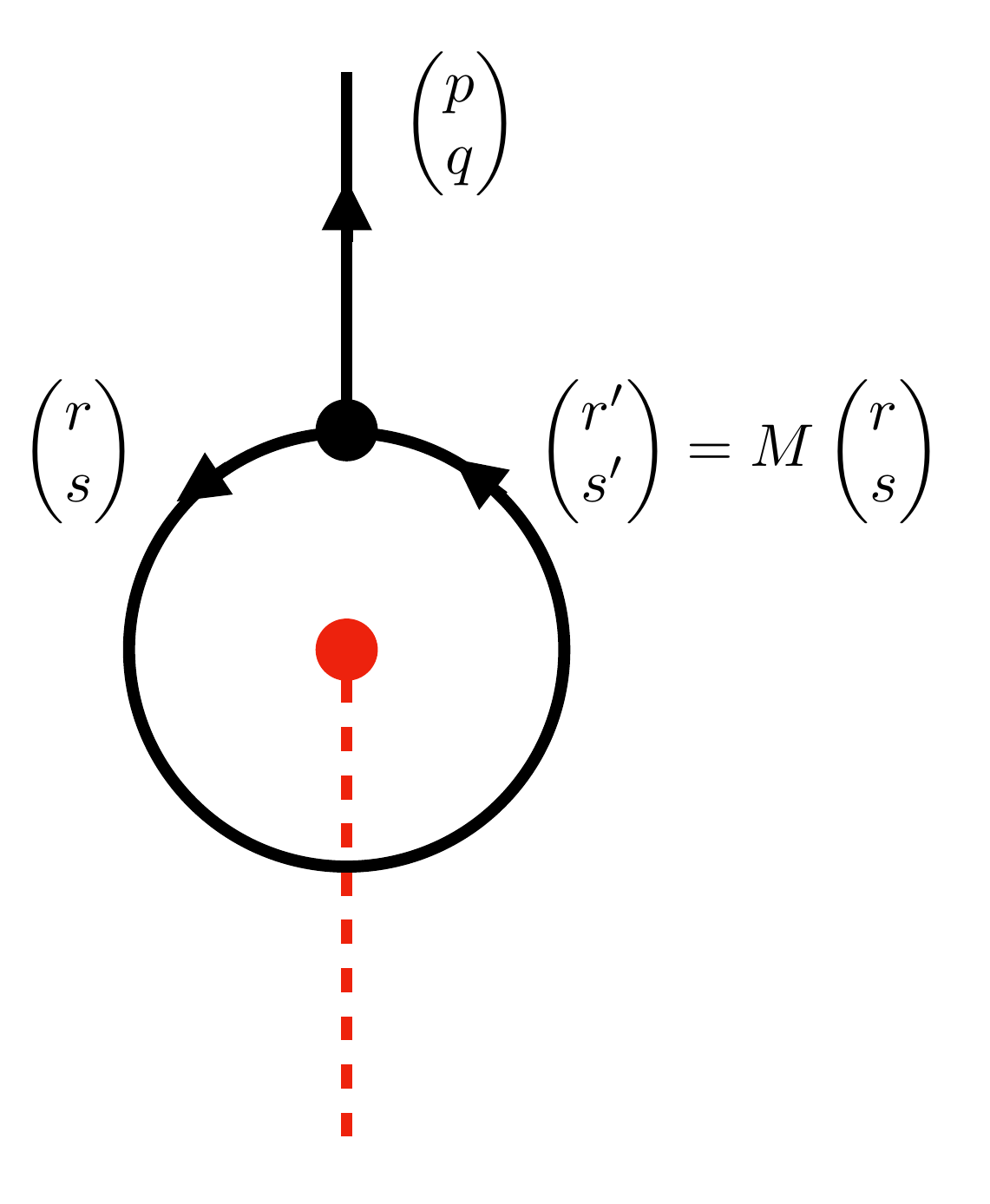}
    \caption[includes a matrix]{A loop junction $\boldsymbol\ell_{(r,s)}$ around a collection of 7-branes with overall monodromy $M$.
    The asymptotic charge $\colpq{p}{q} = \colpq{r'}{s'} - \colpq{r}{s} = (M - \mathbb{1}) \colpq{r}{s}$ is in general non-zero.
    }
    \label{fig:loop_junction}
\end{figure}

To do so, first consider the junction ${\bf j} = \mathbf{x}_{[p,q]} + \mathbf{x}_{[r,s]}$ as depicted on the left of Figure \ref{fig:Juncexa}.
According to \eqref{eq:selfintersec}, we have
\begin{align}\label{eq:self_pairing_3-pronged}
    ({\bf j}, {\bf j}) = ({\bf x}_{[p,q]}, { \bf x}_{[p,q]}) + ({\bf x}_{[r,s]}, {\bf x}_{[r,s]}) + 2({\bf x}_{[p,q]},{\bf x}_{[r,s]}) = -2 + \det \begin{pmatrix} p & r \\ q& s \end{pmatrix} \, .
\end{align}
As pointed out in \cite{DeWolfe:1998zf}, this result can also be interpreted as the sum of the contributions from the two end points of the 7-branes (each contribution $-1$), and the contribution of the 3-pronged vertex.
The latter must therefore be
\begin{align}
    \det \begin{pmatrix} p & r \\ q& s \end{pmatrix} = ps - rq = \det \begin{pmatrix} r & -(p+r) \\ s& -(q+s) \end{pmatrix} = \det \begin{pmatrix} -(p+r) & p \\ -(q+s) & q \end{pmatrix} \, ,
\end{align}
i.e., the determinant of two of the three $\colpq{p}{q}$-charge vectors, arranged in their counter-clockwise ordering (and all prongs either ingoing or outgoing).

\begin{figure}[ht]
    \centering
    \includegraphics[width = 0.7 \textwidth]{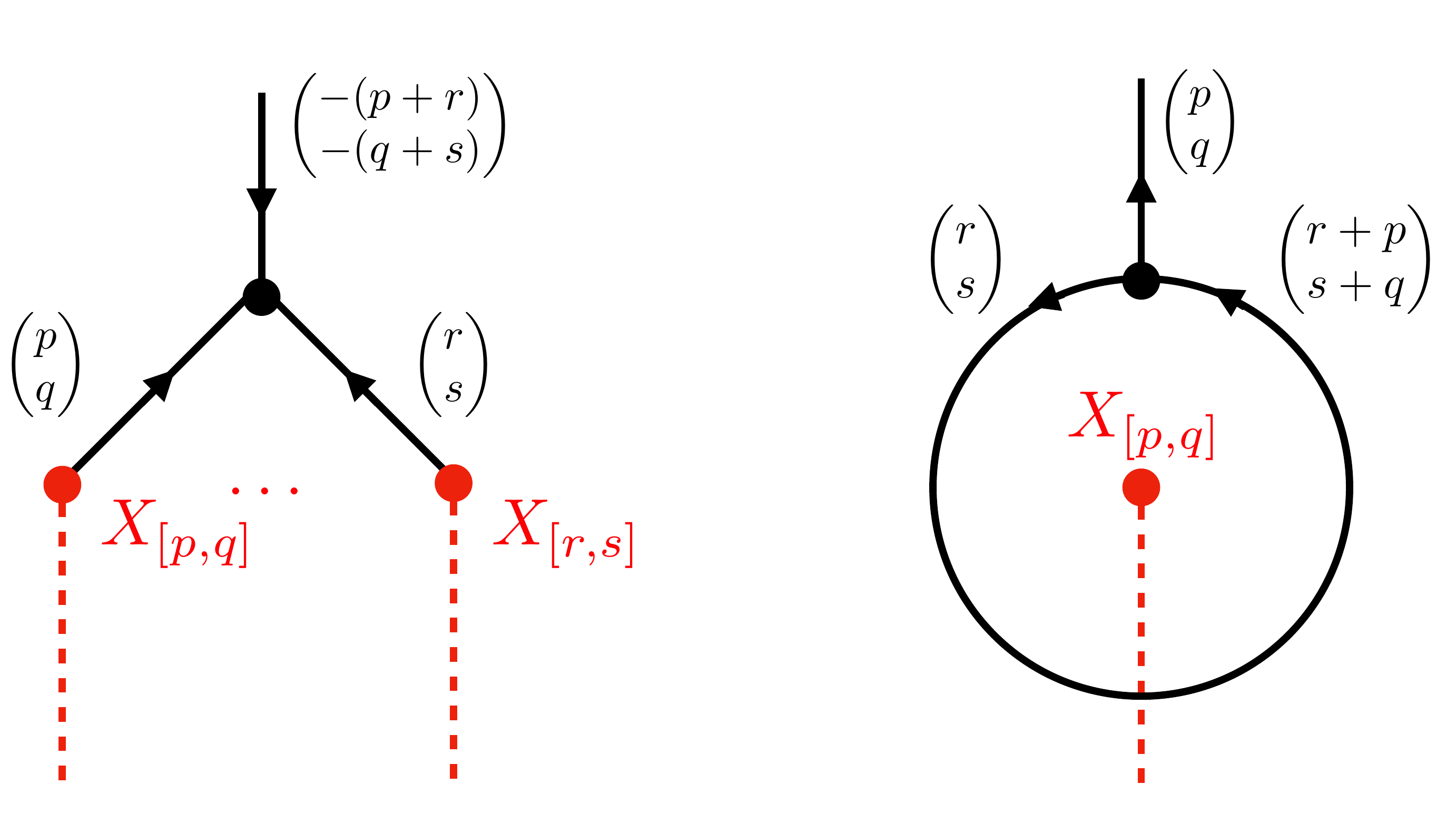}
    \caption{The self-pairing of a 3-pronged junction (left) can be separated into contributions from the ends on 7-branes and the vertex, see \eqref{eq:self_pairing_3-pronged}.
    When there are no prongs ending on 7-branes, such as for loop juncions (right), the only contribution is that of the vertex.
    }
    \label{fig:Juncexa}
\end{figure}

This logic can now be easily applied to compute self-pairings of loop junctions.
Since such a junction has no endpoints on 7-branes, the only contribution to the self-pairing must come from the 3-pronged vertex.
For the junction $\boldsymbol\ell_{(r,s)}$ in Figure \ref{fig:loop_junction}, this contribution evaluates to (after accounting for the signs necessary to have all prongs in- or outgoing)
\begin{align}
    (\boldsymbol\ell_{(r,s)}, \boldsymbol\ell_{(r,s)}) = \det \begin{pmatrix} p & r \\ q & s \end{pmatrix} = - \det \begin{pmatrix} r & r' \\ s & s' \end{pmatrix} \, .
\end{align}
As a consistency check, consider a loop junction $\boldsymbol{\ell}_{(r,s)}$ around a single $[p,q]$-7-brane such that the asymptotic charge is $\colpq{p}{q}$ (see right of Figure \ref{fig:Juncexa}), i.e.,
\begin{align}
    (M_{[p,q]} - \mathbb{1}) \colpq{r}{s} = (qr - ps) \colpq{p}{s} \stackrel{!}{=} \colpq{p}{q} \quad \Leftrightarrow \quad (qr - ps) \stackrel{!}{=} 1 \, ,
\end{align}
which always has a solution for $(r,s)$ since the labels of a single $\bX_{[p,q]}$ must be coprime.
Then, the self-pairing is $(\boldsymbol\ell_{(r,s)}, \boldsymbol\ell_{(r,s)}) = \det \left( \begin{smallmatrix} p & r \\ q & s \end{smallmatrix} \right) = ps - qr = -1 = (\bx_{[p,q]}, \bx_{[p,q]})$.
This was expected, since by construction, this loop is equivalent, by a Hanany--Witten transition, to the unit junction $\bx_{[p,q]}$.

\subsubsection*{(Co-)weight lattices from junctions}

For a single brane stack of ADE type \eqref{eq:ADE_brane_stacks}, the physical junctions \emph{without} asymptotic charges are generated by
\begin{equation}
\begin{split}
\mathfrak{su}_n:& \quad \boldsymbol\alpha_i = \mathbf{a}_i - \mathbf{a}_{i + 1} \,, \enspace i \in \{1, \dots, n-1\} \,, \\
\mathfrak{so}_{2n}:& \quad \boldsymbol\alpha_i = \mathbf{a}_i - \mathbf{a}_{i + 1} \,, \enspace i \in \{1, \dots, n-1 \} \,, \enspace \boldsymbol\alpha_n = \mathbf{a}_{n-1} + \mathbf{a}_n - \mathbf{b} - \mathbf{c} \,, \\
\mathfrak{e}_n:& \quad \boldsymbol\alpha_i = \mathbf{a}_i - \mathbf{a}_{i + 1} \,, \enspace i \in \{1, \dots, n-2 \} \,, \enspace \boldsymbol\alpha_{n-1} = \mathbf{a}_{n-2} + \mathbf{a}_{n-1} - \mathbf{b} - \mathbf{c}_1 \,, \enspace \boldsymbol\alpha_n = \mathbf{c}_1 - \mathbf{c}_2 \,,
\end{split}\label{eq:roots_junctions_ADE}
\end{equation}
where we have indexed 7-branes and their associated unit junctions of the same $[p,q]$-type.
Computing their mutual bi-linear pairing of $\boldsymbol\alpha_i$ one finds
\begin{align}
(\boldsymbol\alpha_i, \boldsymbol\alpha_j) = A_{ij} \,,
\end{align}
with $A_{ij}$ the \textit{negative} Cartan matrix. 
Indeed, strings represented by the junctions above are associated to the W-bosons which lead to the enhanced gauge symmetry on the 7-brane stack.
We will call them \emph{root junctions} for obvious reasons, and they span the root junction lattice of the ADE algebra $\Lambda_\text{r} \subset J_\text{phys}$.

In complete analogy to representation theory (save for a minus sign for the pairing), the bi-linear pairing allows the definition of the \emph{coroot junctions}, whose span is the coroot junction lattice $\Lambda_\text{cr}$, as follows
\begin{align}
\boldsymbol\alpha^{\vee}_i = \frac{2}{-(\boldsymbol\alpha_i, \boldsymbol\alpha_i)} \boldsymbol\alpha_i \, .
\end{align}
Since for ADE algebras all roots have length-square $2$, these coincide with the root junctions.
However, physically, these should be thought of as the magnetically dual states, and hence arise from 5-brane webs represented by the junctions.
We can therefore also identify the pairing between two junctions, where one represents a string and the other a 5-brane, as the Dirac-pairing between electric and magnetic operators of the 8d gauge theory.

One further defines the \emph{weight junctions} ${\bf w}_i$, which are dual to the coroot junctions with respect to $(.,.)$ (or, more precisely, its $\mathbb{Q}$-linear extension),
\begin{align}
    ({\bf w}_i, \boldsymbol\alpha^{\vee}_j) = -\delta_{ij} \,.
\end{align}
They span the weight junction lattice $\Lambda_\text{w}$, and correspond to the electric states of the gauge symmetry if they represent a string. Similarly, one defines the \emph{coweight junctions} ${\bf w}_i^\vee$ and their lattice $\Lambda_\text{cw}$ via
\begin{align}
({\bf w}_i^{\vee}, \boldsymbol\alpha_j) = -\delta_{ij} \,,
\end{align}
which, when representing a 5-brane, is a magnetic state.

Note that the (co-)weights and (co-)roots are in a very real sense localized degrees of freedom.
For any additional 7-brane $\bX_{[r,s]}$ that is added to the system, we can explicitly compute from \eqref{eq:roots_junctions_ADE} that $(\bx_{[r,s]}, \boldsymbol\alpha^{(\vee)}_j) = 0$.
Therefore, any junction that has no prong on the 7-brane stack represents an uncharged state under the gauge symmetry on that stack.

For ADE algebras the coweights and weights again agree, and there the distinction between string and 5-brane junctions is only of formal nature.
However, it will become important once we include O7$^+$-planes.
Before that, we have to introduce the concept of so-called extended (co-)weight junctions \cite{DeWolfe:1998zf}.

\subsubsection{Extended (co-)weights and higher-form center symmetries}

In general, the (co-)weight junctions,
\begin{align}
    {\bf w}^\vee_i = \sum_j (-A^{-1})_{ij} \boldsymbol\alpha_j \, , \quad {\bf w}_i = \sum_j (-\tilde{A}^{-1})_{ij} \boldsymbol\alpha^\vee_j \, ,
\end{align}
with $\widetilde{A}_{ij} = (\boldsymbol\alpha_i^\vee, \boldsymbol\alpha_j^\vee)$, will have fractional coefficients in front of the unit prongs $\mathbf{x}_{[p_i,q_i]}$.
This implies that they are not physical junctions on their own.
However, they can be made physical by adding certain other fractional junctions with \emph{non-zero} asymptotic $\colpq{p}{q}$-charges, resulting in an integer (i.e., physical) junction with a prong that extends away from the 7-brane stack.
Equivalently, it formalizes the intuition that non-adjoint matter states (carrying weights that are not roots) on a 7-brane stack arise from open strings that have ends on other 7-branes (possibly at infinity).

As a simple example, consider $\fkg = \mathfrak{su}_2$, realized on an $\bA_1 \bA_2$-stack.
While the (co-)weight junction ${\bf w}^{(\vee)} = \tfrac12 (\ba_1 - \ba_2)$ without any asymptotic $(p,q)$-charge is non-physical, we can consider the unit string junctions $\ba_1$ or $\ba_2$, each of which carries an asymptotic $\colpq{p}{q} = \colpq{1}{0}$ charge.
From $(\ba_1 , \boldsymbol\alpha^\vee) = (\ba_1, \ba_1 - \ba_2) = -(\ba_2, \ba_1 - \ba_2) = -1$, we expect these (string) junctions to be fundamental matter of the $\mathfrak{su}_2$.
Note that we can formally write
\begin{align}
    \ba_1 = \tfrac12 (\ba_1 + \ba_2) + {\bf w} \, , \quad \ba_2 = \tfrac12 (\ba_1 + \ba_2) - {\bf w} \, .
\end{align}
Because $\tfrac12 (\ba_1 + \ba_2) \equiv \boldsymbol\omega$ has asymptotic charge $\colpq{1}{0}$, and satisfies $(\boldsymbol\omega, \boldsymbol\alpha) = 0$, we can interpret the above rewriting as separating the $\mathfrak{su}_2$ gauge charges of the unit junctions, captured by the summand proportional to $\bf w$, from the asymptotic $SL(2,\bbZ)$-charges, captured by $\boldsymbol\omega$.
By linearity, this separation can be done for any physical junction ${\bf j} = n_1 \ba_1 + n_2 \ba_2$.
For $\mathfrak{su}_2$, the state corresponding to ${\bf j} = s {\bf w} + k \boldsymbol\omega \in J_\text{phys}$ is a weight of an spin-$s/2$ representation, which has charge $s \mod 2$ under the $\bbZ_2$-center.
It is easy to see in this case, the physicality condition, i.e., for ${\bf j}$ to have integer number of prongs on the 7-branes, relates $s \equiv k \mod 2$.
Therefore, the coefficient of any physical junction in front of $\boldsymbol\omega$ provides an equivalent way to encode the center charge of that corresponding state.
This line of argument can be generalized to any ADE-stack \cite{DeWolfe:1998zf}.

\begin{figure}
    \centering
    \includegraphics[width = 0.5 \textwidth]{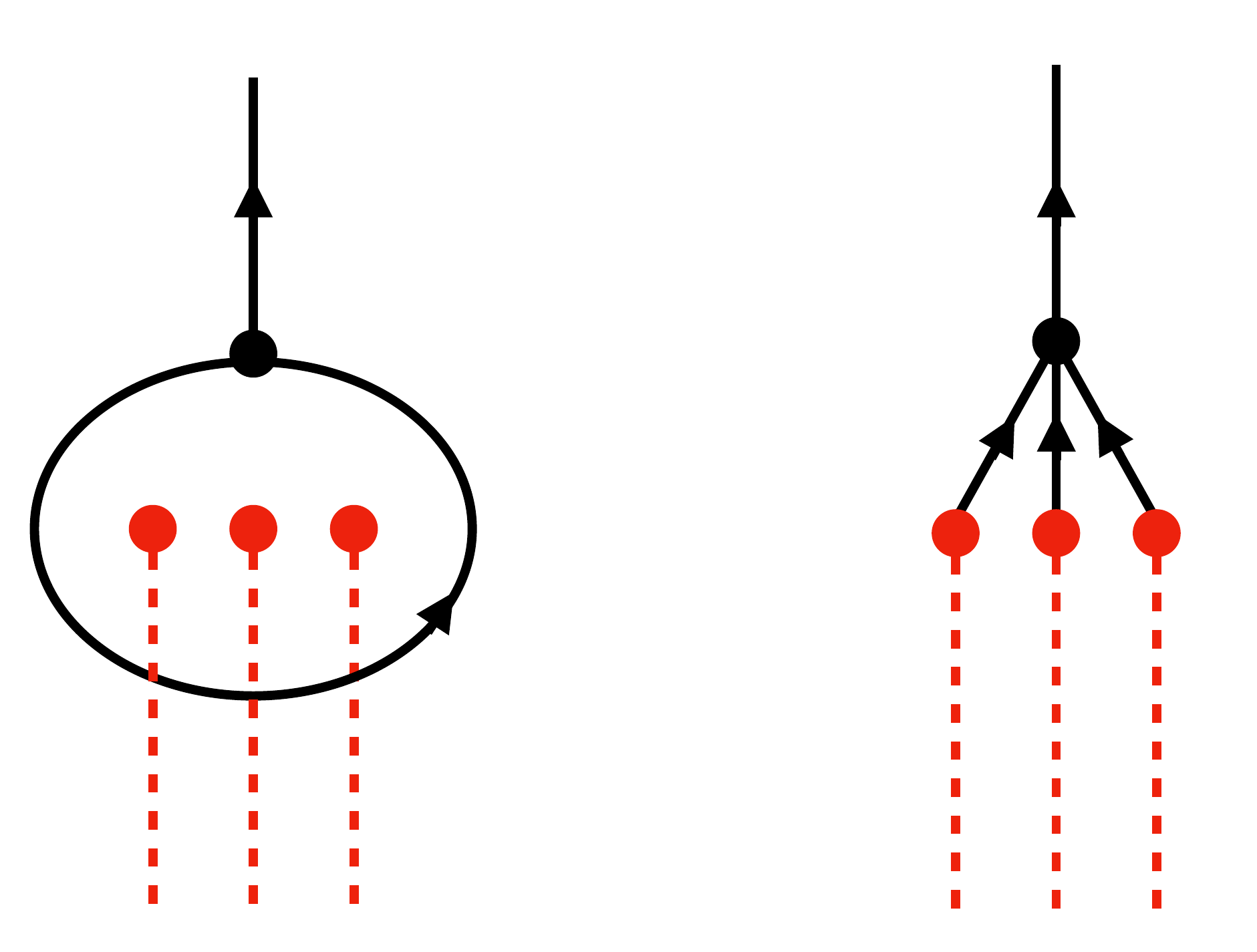}
    \caption[matrix]{Construction of extended weight junctions. Since the $\colpq{r}{s}$-charges that appear in the loop are in general fractional, the prongs ending on the 7-branes after pulling the loop across also have fractional coefficients.}
    \label{fig:ExtWeight}
\end{figure}

First, notice that, by a Hanany--Witten transition, $\boldsymbol\omega = \boldsymbol\ell_{(0,-\nicefrac{1}{2})}$ is a loop junction around the $\bA_1 \bA_2$ stack.
For a general stack with monodromy $M$, one defines $\boldsymbol\omega$, which are called extended weight junctions, as the generators of all loop junctions $\boldsymbol\ell_{(r,s)}$ (with possibly fractional $(r,s)$) encircling the stack that have integer asymptotic $\colpq{p}{q}$-charge, i.e.,
\begin{align}
    (M - \mathbb{1}) \colpq{r}{s} \in \mathbb{Z}^2 \, .
\end{align}
For the ADE algebras realized via the stacks as given in \eqref{eq:ADE_brane_stacks}, a standard basis for these are denoted $\boldsymbol\omega_{p,q}$ with asymptotic charges
\begin{align}
\boldsymbol\omega_p: \quad \colpq{p}{q}_{\text{asymp}} = \colpq{1}{0} \,, \quad \boldsymbol\omega_q: \quad \colpq{p}{q}_{\text{asymp}} = \colpq{0}{1} \,.
\label{eq:basisextweight}
\end{align}
For $\mathfrak{su}_n$ stacks, $(M-\mathbb{1})$ has only rank 1, so there is only one generator, $\boldsymbol\omega_p$ with asymptotic $\colpq{1}{0}$ charge.
Due to the generally fractional $(r,s)$-prong that crosses the branch cuts, the prongs that end on the constituent branes of the stack are also fractional after a Hanany--Witten transition, see Figure \ref{fig:ExtWeight}.
Explicitly, the extended weight junctions and their pairings are given by\footnote{Note that we can infer $\big(\boldsymbol{\omega}_p, \boldsymbol{\omega}_q\big)$ from the self-pairing of $\boldsymbol{\omega}_p + \boldsymbol{\omega}_q = \boldsymbol\ell_{(r_p, s_p)} + \boldsymbol\ell_{(r_q, s_q)} = \boldsymbol\ell_{(r_p+r_q, s_p+s_q)}$, which can be computed from the contribution of the single 3-pronged vertex, as in Figure \ref{fig:Juncexa}.}
\begin{align}
\begin{split}
    \mathfrak{su}_{n} \quad (\bA^n): & \quad \boldsymbol\omega_p = \boldsymbol\ell_{(0, -\nicefrac{1}{n})} = \tfrac{1}{n} \sum_{i = 1}^{n} \mathbf{a}_i \,, \quad (\boldsymbol\omega_p , \boldsymbol\omega_p) = -\tfrac{1}{n} \, ,  \\
    \mathfrak{so}_{2n} \quad (\bA^n \bB \bC): & \quad 
    \begin{cases}
        \boldsymbol\omega_p = \boldsymbol\ell_{(-\nicefrac{1}{2},0)} = \tfrac{1}{2}(\mathbf{b} + \mathbf{c}) \,, \\
        \boldsymbol\omega_q = \boldsymbol\ell_{(1-\nicefrac{n}{4}, -\nicefrac{1}{2})} = \tfrac{1}{2}(\sum_{i} \mathbf{a}_i - \mathbf{b} + \mathbf{c} - n \boldsymbol{\omega}_p) \,,
    \end{cases}  
    \ (\boldsymbol\omega_\alpha, \boldsymbol\omega_\beta) = 
    \begin{pmatrix}    0 & 0 \\ 0 & \tfrac{n}{4}-1    \end{pmatrix}_{\alpha\beta} ,
\\
    \mathfrak{e}_6 \quad (\bA^5 \bB \bC^2): & \quad
    \begin{cases}
        \boldsymbol\omega_{p} = \boldsymbol\ell_{(0,\nicefrac{1}{3})} = -\tfrac{1}{3} \sum_{i=1}^{5} \mathbf{a}_{i}+\tfrac{4}{3} \mathbf{b}+\tfrac{2}{3} \sum_{i=1}^{2} \mathbf{c}_{i} \,, \\ 
        \boldsymbol{\omega}_{q} = \boldsymbol\ell_{(-1,-1)} =\sum_{i=1}^{5} \mathbf{a}_{i}-3 \mathbf{b}-\sum_{i=1}^{2} \mathbf{c}_{i} \,,
    \end{cases}
    \ (\boldsymbol\omega_\alpha, \boldsymbol\omega_\beta) = 
    \begin{pmatrix}    \tfrac13 & -\tfrac12 \\ -\tfrac12 & 1 \end{pmatrix}_{\alpha\beta} ,
 \\
    \mathfrak{e}_7 \quad (\bA^6 \bB \bC^2): & \quad 
    \begin{cases}
        \boldsymbol{\omega}_{p} = \boldsymbol\ell_{(\nicefrac12, \nicefrac12)} =-\tfrac{1}{2} \sum_{i=1}^{6} \mathbf{a}_{i}+2 \mathbf{b}+\sum_{i=1}^{2} \mathbf{c}_{i} \,, \\
        \boldsymbol{\omega}_{q} = \boldsymbol\ell_{(-\nicefrac52, -\nicefrac32)} =\tfrac{3}{2} \sum_{i=1}^{6} \mathbf{a}_{i}-5 \mathbf{b}-2 \sum_{i=1}^{2} \mathbf{c}_{i} \,,
    \end{cases} 
    \ (\boldsymbol\omega_\alpha, \boldsymbol\omega_\beta) = 
    \begin{pmatrix}    \tfrac12 & -1 \\ -1 & \tfrac52 \end{pmatrix}_{\alpha\beta} ,
    \\
    \mathfrak{e}_8 \quad (\bA^7 \bB \bC^2): & \quad 
    \begin{cases}
        \boldsymbol{\omega}_{p}= \boldsymbol\ell_{(2,1)} =-\sum_{i=1}^{7} \mathbf{a}_{i}+4 \mathbf{b}+2 \sum_{i=1}^{2} \mathbf{c}_{i} \,, \\
        \boldsymbol{\omega}_{q}= \boldsymbol\ell_{(-7,-3)} =3 \sum_{i=1}^{7} \mathbf{a}_{i}-11 \mathbf{b}-5 \sum_{i=1}^{2} \mathbf{c}_{i} \,,
    \end{cases} 
    \ (\boldsymbol\omega_\alpha, \boldsymbol\omega_\beta) = 
    \begin{pmatrix}    1 & -\tfrac52 \\ -\tfrac52 & 7 \end{pmatrix}_{\alpha\beta} .
\end{split}
\label{eq:extADE}
\end{align}
All physical junctions associated to a 7-brane stack, i.e., junctions with prongs of only integer $\colpq{p}{q}$-charge, can be written uniquely in terms of a linear combination of weight and extended weight junctions 
\begin{align}
\mathbf{j} = \sum_i a^i {\bf w}_i + a^p \boldsymbol\omega_p + a^q \boldsymbol\omega_q \,, \quad a^i, a^p, a^q \in \mathbb{Z} \,.
\label{eq:nAstates}
\end{align}
Physically, this means that a physical $(p,q)$-string/-5-brane is fully characterized by its asymptotic electric/magnetic $\colpq{p}{q}$-charge under $(B_2, C_2)$, and the weight/coweight charges under the 7-brane gauge algebras.

In turn, it can be verified that every possible weight junction ${\bf w} = \sum_i a^i {\bf w}_i$ ($a^i \in \bbZ$) of the gauge algebra $\fkg$ can be completed into a physical junction by the addition of an integer linear combination ${\bf j}_e = a^p \boldsymbol\omega_p + a^q \boldsymbol\omega_q$ of extended weights \cite{DeWolfe:1998zf}.
Such integer linear combination is not unique and is determined only up to multiples $n^p \boldsymbol\omega_p + n^q \boldsymbol\omega_q$ which have integer charges for each prong.
This non-uniqueness can be understood as the fact that ${\bf j}_e$ is determined by the charge of ${\bf w}$ under the center $Z(\widetilde{G})$ of the simply-connected group $\widetilde{G}$ associated to $\mathfrak{g}$.
Intuitively, this is expected because the charge under the center of a specific state ${\bf w}$ is encoded in its prefactors of the weight basis ${\bf w}_i$, which in turn introduces fractional prongs that can only be cancelled by the extended weights.
Analogously to how weights can be screened by the W-bosons for observers ``at infinity'', there are multiples of specific asymptotic $\colpq{p}{q}$-charges that can be added and subtracted without affecting the local gauge dynamics on the 7-branes.\footnote{Describing the gauge dynamics by F-theory on a non-compact K3, this is reflected by the homology of the asymptotic boundary exhibiting discrete torsion, associated to the fact that $n^p \times \text{(A-cycle)} + n^q \times \text{(B-cycle)}$ on the generic torus fiber shrinks at the singularity \cite{Cvetic:2021sxm}.}

The precise connection between higher-form symmetries and extended weights have been described in \cite{Cvetic:2021sxm}.
Formally, we can define the lattice,
\begin{align}
J_{\text{ext}} = \{ \mathbf{j}_e = a^p \boldsymbol{\omega}_p + a^q \boldsymbol{\omega}_q \, | \, a^p, a^q \in \mathbb{Z} \} \, ,
\end{align}
whose elements are arbitrary integer linear combinations of extended weights that may be fractional.
Then, the screening arguments for the center symmetries, together with the junction characterization of gauge degrees of freedom, translates into:
\begin{align}
Z (\widetilde{G}_{\text{ADE}}) = \frac{\text{weights}}{\text{roots}} = \frac{\text{coweights}}{\text{coroots}} = \frac{J_{\text{ext}}}{J_{\text{phys}} \cap J_{\text{ext}}} \,,
\end{align}
where $(J_{\text{phys}} \cap J_{\text{ext}})$ denotes extended weight junctions that are themselves physical, i.e., do not contain fractional prongs.
Note that, since $\boldsymbol\omega_\circ = \boldsymbol\ell_{(r_\circ, s_\circ)}$ are loop junctions of the form depicted on the left of Figure \ref{fig:ExtWeight}, $(J_{\text{phys}} \cap J_{\text{ext}})$ are precisely the loops $\ell_{(r,s)} = n^p \ell_{(r_p, s_p)} + n^q \ell_{(r_q, s_q)}$ with integer $(r,s)$.
Concretely, in terms of the extended weights summarized in \eqref{eq:extADE}, one finds
\begin{align}\label{eq:center_charge_table_ADE}
    \renewcommand{\arraystretch}{1.8}
    \begin{array}{c|c}
        \mathfrak{g} & J_{\text{ext}} / ( J_\text{phys} \cap J_{\text{ext}} ) \\ \hline
        \mathfrak{su}_n & \displaystyle\frac{\{a^p \boldsymbol{\omega}_p\}}{(n \boldsymbol{\omega}_p)} \cong \big\{ a^p \ \text{mod } n \big\} \cong \bbZ_n \\
        \mathfrak{so}_{4n} & \displaystyle\frac{\{a^p \boldsymbol{\omega}_p + a^q \boldsymbol{\omega}_q\}}{(2 \boldsymbol{\omega}_p, 2\boldsymbol{\omega}_q ) } \cong \{ (a^p \ \text{mod } 2, a^q \ \text{mod } 2)\} \cong  \mathbb{Z}_2 \oplus \mathbb{Z}_2 \\
        \mathfrak{so}_{4n+2} & \displaystyle\frac{\{a^p \boldsymbol{\omega}_p + a^q \boldsymbol{\omega}_q\}}{(2 \boldsymbol{\omega}_p, 4 \boldsymbol{\omega}_q, \boldsymbol{\omega}_p + 2 \boldsymbol{\omega}_q )} \cong \big\{ 2a^p + a^q \mod 4 \big\} \cong \bbZ_4 \\
        \mathfrak{e}_6 & \displaystyle\frac{\{a^p \boldsymbol{\omega}_p + a^q \boldsymbol{\omega}_q\}}{(3 \boldsymbol{\omega}_p, \boldsymbol{\omega}_q)} \cong \big\{ a^p \ \text{mod } 3 \} \cong \bbZ_3 \\
        \mathfrak{e}_7 & \displaystyle\frac{\{a^p \boldsymbol{\omega}_p + a^q \boldsymbol{\omega}_q\}}{(2 \boldsymbol{\omega}_p, 2 \boldsymbol{\omega}_q, \boldsymbol{\omega}_p + \boldsymbol{\omega}_q )} \cong \big\{ a^p + a^q \mod 2 \big\} \cong \bbZ_2 
    \end{array}
\end{align}
In the language of higher-form symmetries (see also \cite{Cvetic:2021sxm}), a physical string/5-brane junction ${\bf j} = \sum_i a^i {\bf w}_i + a^p \boldsymbol\omega_p + a^q \boldsymbol\omega_q$ carries an electric/magnetic $Z(\widetilde{G})$ 1-form/5-form symmetry charge prescribed by \eqref{eq:center_charge_table_ADE}.

\section[Junctions on \texorpdfstring{\textbf{O7}$^+$}{O7+} and center symmetries of \texorpdfstring{$\mathfrak{sp}$}{sp} dynamics]{Junctions on O7\boldmath{$^+$} and center symmetries of \boldmath{$\mathfrak{sp}$} dynamcis}
\label{subsec:junctions_on_O7}

So far, we have reviewed the junction framework for ordinary $[p,q]$-7-branes, which succinctly encode the 8d ${\cal N}=1$ gauge dynamics with simply-laced gauge algebras.
However, field theoretically, one can also have $\mathfrak{sp}_n$ algebras. 
In the type IIB string constructions these are linked to the presence of O7$^+$-planes, which was not considered systematically within the junction framework previously.
Therefore, we need to generalize the above analysis.

First, we note that the O7$^+$-plane, unlike the the O7$^-$-plane, does not split, at finite string coupling, into constituents represented by ordinary $[p,q]$-7-branes.
Therefore, we will represent it by a single, albeit special, 7-brane.
The monodromy generated by one O7$^+$-plane is in the same $SL(2,\bbZ)$ conjugacy class as a 7-brane stack with $\mathfrak{g} = \mathfrak{so}_{16}$.
In the following local analysis, we use the same presentation as in \eqref{eq:ADE_brane_stacks},
\begin{align}\label{eq:O7_monodromy}
M_{\text{O7}^+} = \begin{pmatrix} -1 & 4 \\ 0 & -1 \end{pmatrix} \,.
\end{align}
There are multiple ways of arguing for this physically.
The prevalent interpretation of an O7$^+$ in recent literature \cite{Witten:1997bs, deBoer:2001wca} is as the remnant of ``freezing'' the $\mathfrak{so}_{16}$ gauge dynamics on an ordinary 7-brane stack, see also \cite{Tachikawa:2015wka, Bhardwaj:2018jgp}.

However, even after two decades, the freezing operation remains somewhat mysterious.
In particular, a geometric derivation of its effect on higher-form symmetries in the M-theory frame \cite{Morrison:2020ool,Albertini:2020mdx} appears to be challenging.
However, as we will argue now, one can obtain a complete picture, at least in the IIB duality frame, of the mechanism using junctions.
The key distinction to $[p,q]$-branes is that the physicality condition for prongs that end on an O7$^+$ \emph{differ} between strings and 5-branes.

From the perturbative IIB picture, only pairs of fundamental strings can end on an O7$^+$ \cite{Imamura:1999uf}, which one can see in a perturbative picture via Chan--Paton factors.
Via various dualities it can also be argued that only an even number of D-strings can end on the O7$^+$-plane, see \cite{Imamura:1999uf, Bergman:2001rp}. 
Thus, the physical string junctions emanating from the O7$^+$-plane have $\colpq{p}{q}$-charge restricted by $p, q \in 2 \mathbb{Z}$.
In contrast, $(p,q)$-5-branes can end with any integer number on O7$^+$.
This is relevant, e.g., in the construction of 5d SCFTs via 5-brane webs \cite{Bergman:2015dpa}.
Hence, a prong of a physical 5-brane junction can end with arbitrary integer $\colpq{p}{q}$-charge on an O7$^+$.
As we will see momentarily, these conditions naturally give rise to a consistent description of $\mathfrak{sp}_n$ gauge algebras, including their center symmetries, from the junction lattice.
Moreover, once we have set up the notation in Section \ref{sec:global}, we can derive these conditions independently in global models with one O7$^+$-plane that realize 8d rank $(2,10)$, by appealing to the dual description of these models via the CHL-string (see Appendix \ref{apdx:evenness}).

It is worthwhile to compare the boundary conditions for strings and 5-branes on O7$^+$ with the unfrozen $\mathfrak{so}_{16}$-stack.
First, since both generate the same monodromy, the loop junctions that generate all integer asymptotic $\colpq{p}{q}$-charges in the presence of a single O7$^+$ are the same as for an $\mathfrak{so}_{16}$ stack \eqref{eq:extADE},
\begin{align}
\boldsymbol{\omega}_p^{\text{O7}^+} = \boldsymbol{\ell}_{(- \nicefrac{1}{2},0)} \,, \quad \boldsymbol{\omega}_q^{\text{O7}^+} = \boldsymbol{\ell}_{(-1, - \nicefrac{1}{2})} \,.
\end{align}
We will call these the extended weight junctions of the O7$^+$, even though a single O7$^+$ (unlike its unfrozen cousin) has no gauge dynamics, and hence no root or weight lattice to begin with.
By collapsing the loop or, equivalently, performing a Hanany--Witten transition across the orientifold plane, these extended weight junctions have a $\colpq{1}{0}$ and a $\colpq{0}{1}$ prong, respectively, on the O7$^+$.
The set of junctions ending/emanating from one O7$^+$, which is entirely characterized by its total $\colpq{p}{q} = \colpq{a^p}{a^q}$-charge, can therefore be written as
\begin{align}
    {\bf j} = a^p \, \boldsymbol{\omega}_p^{\text{O7}^+} + a^q \, \boldsymbol\omega_q^{\text{O7}^+} \, , \quad a^p, a^q \in \bbZ \, .
\end{align}
Physical string junctions must then have $(a^p, a^q) \equiv (0 \ \text{mod 2}, 0\ \text{mod 2}) $, which agrees with the physicality condition for junctions on an $\mathfrak{so}_{16}$-stack that has no (unscreenable) gauge charge, see \eqref{eq:center_charge_table_ADE}.
Equivalently, any string loop $\boldsymbol\ell_{(r,s)}$ that encircles the O7$^+$ is physical if and only if $r$ and $s$ are both integer.
In contrast, a physical 5-brane junction with odd $(a^p, a^q)$, which can end on an O7$^+$, would not be admissible on an $\mathfrak{so}_{16}$-stack without picking up (unscreenable) gauge charge.
In particular, this means that \emph{physical} 5-brane loops $\boldsymbol\ell_{(r,s)}$ around the O7$^+$ could have \emph{half-integer} valued $r$ and $s$.

To fully incorporate O7$^+$'s in the junction framework, we also need to define the bi-linear pairing.
It is hard to come up with a rule for prongs ending on the O7$^+$ by appealing to any geometric counterpart in a dual M-/F-theory picture, because of the presence of frozen singularities there.
However, since the extended weights can be viewed as loops which is only sensitive to the induced $SL(2,\bbZ)$ monodromy $M$, but not the ``microscopics'' of a 7-brane stack, one would naturally expect that the pairing of such junctions is insensitive to whether $M$ is sourced by an O7$^+$, or an $\mathfrak{so}_{16}$ stack.
Following the discussion around Figure \ref{fig:Juncexa}, we can therefore directly compute (rather than define, which would require further justifications) from the loop junction representation:
\begin{align}
\big(\boldsymbol{\omega}_p^{\text{O7}^+}, \boldsymbol{\omega}_p^{\text{O7}^+}\big) = 0 \,, \quad \big(\boldsymbol{\omega}_q^{\text{O7}^+}, \boldsymbol{\omega}_q^{\text{O7}^+}\big) = 1 \,, \quad \big(\boldsymbol{\omega}_p^{\text{O7}^+}, \boldsymbol{\omega}_q^{\text{O7}^+}\big) = 0 \,.
\label{eq:bilinext}
\end{align}
Considering O7$^+$-planes together with general $[r,s]$-7-branes (which we assume to be on the left of the O7$^+$), one further finds
\begin{align}
(\boldsymbol{\omega}_p^{\text{O7}^+}, \mathbf{x}_{[r,s]}) = -\tfrac{s}{2} \,, \quad (\boldsymbol{\omega}_q^{\text{O7}^+}, \mathbf{x}_{[r,s]}) = \tfrac{r}{2} \,.
\end{align}

\subsubsection*{$\mathfrak{sp}$ gauge algebras and their higher-form symmetries from junctions}

From the perturbative IIB picture, it is well-known that we can generate 8d $\mathfrak{sp}_n$ gauge dynamics on a 7-brane stack formed by $n$ $\bA$-branes on top of one O7$^+$.
This 7-brane stack, of the form $\mathbf{A}^n {\bf O7}^+$, has the same monodromy as an $\mathfrak{so}_{16+2n}$ stack:
\begin{align}\label{eq:monodromy_sp}
M_{\mathfrak{sp}_n} = M_{\mathbf{A}^n {\bf O7}^+} = \begin{pmatrix} -1 & 4 + n \\ 0 & -1 \end{pmatrix} \,.
\end{align}
This allows us to straightforwardly define the extended weight junctions, as loops $\ell_{(r,s)}$ around the entire stack (including the O7$^+$) such that an asymptotic $\colpq{1}{0}$- (for $\boldsymbol\omega_p$) or $\colpq{0}{1}$-charge (for $\boldsymbol\omega_q$) remains.
Then, after performing the suitable Hanany--Witten transitions, we find
\begin{align}\label{eq:sp_extended_weights}
\begin{split}
    & \boldsymbol\omega_p^{\mathfrak{sp}_n} = \boldsymbol\ell_{(-\nicefrac12,0)} = \boldsymbol\omega_p^{\text{O7}^+} \,, \quad \boldsymbol\omega_q^{\mathfrak{sp}_n} = \boldsymbol\ell_{(-1 - \nicefrac{n}{4}, -\nicefrac12)} =\tfrac{1}{2} \sum_{i = 1}^n \mathbf{a}_i - \tfrac{n}{2} \boldsymbol\omega_p^{\text{O7}^+} + \boldsymbol\omega_q^{\text{O7}^+} \, .
\end{split}
\end{align}
Since $M_{\mathfrak{sp}_n} = M_{\mathfrak{so}_{16+2n}}$, the loop junctions look identical to those of $\mathfrak{so}_{16+2n}$.
Hence, with $J_\text{ext} = \{ a^p \boldsymbol\omega_p^{\mathfrak{sp}_n} + a^q \boldsymbol\omega_q^{\mathfrak{sp}_n} \ | \ a^p, a^q \in \bbZ \}$, the physicality condition on linear combinations of extended junctions is captured by
\begin{align}\label{eq:wrong_electric_center_sp}
    J_\text{ext} / \big( J_\text{ext} \cap J_\text{phys, strings}) = \begin{cases}
        \big\{ (a^p \ \text{mod 2} , a^q \ \text{mod 2}) \} \cong \bbZ_2 \times \bbZ_2 \, , & n \ \text{even} \, , \\
        \big\{ 2a^p + a^q \ \text{mod 4} \} \cong \bbZ_4 \, , & n \ \text{odd} \, .
    \end{cases}
\end{align}
On the other hand, for 5-brane junctions, we have
\begin{align}
    J_\text{ext} / \big( J_\text{ext} \cap J_\text{phys, 5-branes}) = 
        \big\{  a^q \ \text{mod 2} \} \cong \bbZ_2 \, .
\end{align}
Comparing to the discussion around \eqref{eq:center_charge_table_ADE}, one might be tempted to identify $Z(Sp(n))_\text{electric} = \bbZ_2 \times \bbZ_2$ or $\bbZ_4$, and $Z(Sp(n))_\text{magnetic} = \bbZ_2$, which is clearly not correct.

To rectify this, we must instead consider in detail the role of the extended weights as completing weights and coweights into physical strings and 5-branes, respectively.
In analogy to the ADE-stacks, we first construct the $\mathfrak{sp}_n$ roots $\boldsymbol\alpha_i$ as string junctions with no asymptotic $\colpq{p}{q}$-charge, that stretch between the constituents of this stack.
A basis for such junctions are
\begin{align}
\boldsymbol{\alpha}_i = \mathbf{a}_i - \mathbf{a}_{i+1} \,, \enspace i \in \{1, \dots, n-1\} \,, \enspace \boldsymbol{\alpha}_n = 2 \mathbf{a}_n - 2 \boldsymbol{\omega}_p^{\text{O7}^+} \,,
\end{align}
see also Figure \ref{fig:sproots}.
With this and the bi-linear pairings in \eqref{eq:bilinext}, one straightforwardly verifies
\begin{align}
(\boldsymbol{\alpha}_n , \boldsymbol{\alpha}_n) = -4 \, , \quad (\boldsymbol{\alpha}_i , \boldsymbol{\alpha}_n) = 2 \, \delta_{i, n-1} \, , \quad  (\boldsymbol{\alpha_i}, \boldsymbol{\alpha}_j) = 
    \begin{cases} -2 \, , & i = j \, , \\
        1 \, , & | i -j | = 1\, , \\
        0 \, , & \text{else} \, ,
    \end{cases}
     \quad 1 \leq i,j \leq n-1 \,,
\end{align}
which precisely reproduces the negative of the Cartan matrix of an $\mathfrak{sp}_n$ algebra.
In particular, we see that, while the short roots (those with length squared 2) arise from single-pronged strings between the $\bA$-branes, just as for ADE-algebras, the long root $\boldsymbol{\alpha}_n$ is only a generator due to the evenness condition for strings ending on the O7$^+$.

\begin{figure}[ht]
    \centering
    \includegraphics[width = 0.4 \textwidth]{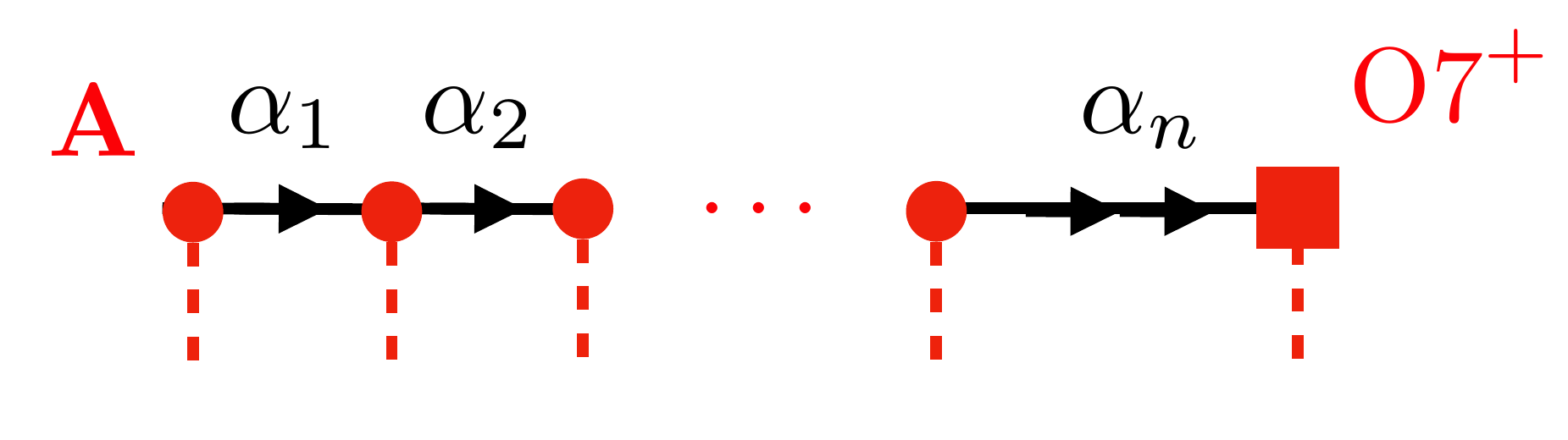}
    \caption{Root junctions of $\mathfrak{sp}$ algebra (the double arrow denotes the factor of 2 required by evenness on O7$^+$).}
    \label{fig:sproots}
\end{figure}

As a non-simply-laced algebra, $\mathfrak{sp}$ has different coroots than roots:
\begin{align}
\boldsymbol{\alpha}_i^\vee = \frac{2 \boldsymbol{\alpha}_i}{- (\boldsymbol{\alpha}_i, \boldsymbol\alpha_i)} = 
\begin{cases} 
    \boldsymbol\alpha_i \, , & 1 \leq i \leq n-1 \, ,\\
    \frac{1}{2} \boldsymbol\alpha_n = {\bf a}_n - \boldsymbol\omega^{O7{^+}}_p \, , & i = n \,.
\end{cases}
\end{align}
Since these are the magnetic objects in a gauge theory context, they arise from 5-brane junctions, which, consistent with the boundary conditions, can have a single prong on the O7$^+$ that is needed to form the short coroot $\boldsymbol\alpha^\vee_n$ with length squared 1.

The distinction between strings and 5-branes are of course also important for the weights and coweights.
The weight junctions are obtained as dual to the coroot junctions. Defining the matrix $\widetilde{A}_{ij} = \big( \boldsymbol\alpha^{\vee}_i, \boldsymbol\alpha_j^{\vee} \big)$, these can be written as
\begin{align}
\mathbf{w}_i = \big( - \widetilde{A}^{-1} \big)_{ij} \, \boldsymbol\alpha^{\vee}_j = \sum_j \text{min}\{i,j\} \, \boldsymbol\alpha_j^\vee = \sum_{j=1}^{n-1} \text{min}\{i,j\} \, \boldsymbol\alpha_j + \tfrac12 \text{min}\{i,n\} \, \boldsymbol\alpha_n  \,.
\label{eq:spweights}
\end{align}
Similarly, one obtains the coweight junctions as duals of the root junctions. With the negative Cartan matrix $A = (\alpha_i, \alpha_j)$ one has
\begin{align}
\mathbf{w}_i^\vee = \big( - A^{-1} \big)_{ij} \, \boldsymbol\alpha_j = 
\begin{cases}
    \sum_{j=1}^{n-1} \text{min}\{i,j\} \, \boldsymbol\alpha^\vee_j + i \, \boldsymbol\alpha^\vee_n \, , & i < n \, , \\
    \sum_{j=1}^{n} \frac{j}{2} \, \boldsymbol\alpha_j^\vee \, , & i = n \,,
\label{eq:spcoweights}
\end{cases}
\end{align}
expressed as fractional linear combinations of coroots.

In terms of the junctions, the electric center symmetry,
\begin{align}
    Z\big(Sp(n) \big) = \text{weights} / \text{roots} \cong \frac{ \{ \tfrac{k}{2} \boldsymbol\alpha_n \} }{(\boldsymbol\alpha_n)} \cong \bbZ_2,
\end{align}
is precisely generated by multiples of $\tfrac12 \boldsymbol\alpha_n = \ba_n - \boldsymbol\omega_p^{\text{O7}^+}$, which is unphysical as a string junction because of the odd prong on the O7$^+$.
To obtain a physical junction with the same $\mathfrak{sp}_n$ gauge charge, we must therefore add linear combinations of the extended weights \eqref{eq:sp_extended_weights} with integer prongs on $\bA$, and odd numbers of prong of $\colpq{p}{q} = \colpq{1}{0}$-charge on the O7$^+$.
For even $n$, this requirement is met by $a^p \boldsymbol\omega_p^{\mathfrak{sp}_n} + a^q \boldsymbol\omega_q^{\mathfrak{sp}_n}$ with $a^p \equiv 1 \ \text{mod } 2$ and $a^q \equiv 0 \ \text{mod } 2$, whereas for odd $n$, we need $2a^p + a^q \equiv 2 \ \text{mod } 4$.
Each of these generate a $\bbZ_2$-subgroup of the putative center \eqref{eq:wrong_electric_center_sp}, which is the correct presentation of $Z\big(Sp(n) \big)_\text{electric} \cong Z\big( Sp(n) \big)$.
The mismatch from \eqref{eq:wrong_electric_center_sp} is because not all integer linear combinations of extended weight junctions can be completed into a physical string junction with an $\mathfrak{sp}_n$ weight.
For the magnetic center symmetry,
\begin{align}
    \text{Hom}( Z\big(Sp(n) \big), \bbZ) \cong Z\big(Sp(n) \big) = \text{coweights} / \text{coroots} \cong \frac{ \{ k{\bf w}_n^\vee \} }{\text{coroots}} \, ,
\end{align}
the unphysical 5-brane coweight junctions come from the half-integer valued prongs in
\begin{align}
    {\bf w}^{\vee}_n = \sum_{i=1}^n \tfrac{i}{2} \boldsymbol\alpha_i^\vee = \tfrac{1}{2} \sum_{i = 1}^n \mathbf{a}_i - \tfrac{n}{2} \boldsymbol\omega_p^{\text{O7}^+} = \boldsymbol\omega_q^{\mathfrak{sp}_n} - \boldsymbol\omega_q^{\text{O7}^+}\, ,
\end{align}
which can be made physical by adding $a^p \boldsymbol\omega_p^{\mathfrak{sp}_n} + a^q \boldsymbol\omega_q^{\mathfrak{sp}_n}$ with $a^q \equiv 1 \ \text{mod 2}$.

In summary, for a physical string junction that ends on an $\mathfrak{sp}_n$ stack, which can be uniquely decomposed as
\begin{align}
    {\bf j} = \sum_i a^i {\bf w}_i + a^p \boldsymbol\omega_p^{\mathfrak{sp}_n} + a^q \boldsymbol\omega_q^{\mathfrak{sp}_n}
\end{align}
in terms of the extended weights \eqref{eq:sp_extended_weights} and weight junctions \eqref{eq:spweights}, the $\bbZ_2$ charge of the corresponding state under the electric center symmetry is
\begin{align}\label{eq:center_charge_string_sp}
\begin{cases}
    a^p \ \text{mod 2} \, , & \text{$n$ even,} \\
    a^p + \tfrac12 a^q \ \text{mod 2} \, , & \text{$n$ odd (then $a^q$ must be even).}
\end{cases}
\end{align}
For a physical 5-brane junction, decomposed into extended weights and coweights \eqref{eq:spcoweights},
\begin{align}
    {\bf j} = \sum_i a^i {\bf w}^\vee_i + a^p \boldsymbol\omega_p^{\mathfrak{sp}_n} + a^q \boldsymbol\omega_q^{\mathfrak{sp}_n} \, ,
\end{align}
its magnetic $\bbZ_2$-center charge is
\begin{align}\label{eq:center_charge_5-brane_sp}
    a^q \ \text{mod 2} \, .
\end{align}

This completes the list of local building blocks of simple gauge algebras that can be combined into a global model describing 8d supergravity.
As we will discuss now, the consistent combination of the individual brane stacks then determines the global structure of the gauge dynamics in these models.

\section{8d string vacua and their global structure from junctions}\label{sec:global}

In this section, we combine the above local descriptions of simple gauge algebras into a compact setting, to classify all 8d $\calN = 1$ string vacua using junctions.
These vacua fall into three moduli branches, which have gauge rank $(2,18)$, $(2,10)$, and $(2,2)$, respectively.
As shown in \cite{Hamada:2021bbz}, the gauge symmetries of the effective supergravity descriptions can be classified by the $SL(2,\bbZ)$ monodromies assoiated with each simple gauge factor, with a few additional consistency conditions.
In theories of rank $(2,18)$, which enjoy a description as F-theory on an elliptically-fibered K3 surface, these restrictions are met by 24 $[p,q]$-7-branes with trivial total monodromy \cite{Morrison:1996na,Morrison:1996pp,Douglas:2014ywa}.
Via the freezing procedure \cite{Witten:1997bs,Tachikawa:2015wka,Bhardwaj:2018jgp}, one can further obtain all rank $(2,10)$ or $(2,2)$ vacua, if one replaces any rank $(2,18)$ 7-brane configuration with one or two $\mathfrak{so}_{16+2n}$ stacks with the corresponding frozen $\mathfrak{sp}_{n}$ algebra that contains an O7$^+$ \cite{Hamada:2021bbz}.
Based on this, junctions provide a unified description of the gauge dynamics, in particular, the gauge group topology, for all these vacua.

Before we dive into the details, let us give a schematic description of this approach.
For a given 7-brane configuration (with or without O7$^+$) in a global model, the set of characters or cocharacters (i.e., a sublattice of the weight or coweight lattice that is occupied by dynamical states) correspond to physical (string or 5-brane) junctions that have zero asymptotic $\colpq{p}{q}$-charge.
Since the $\colpq{p}{q}$-charge of a prong ending on a stack is entirely captured by extended weight junctions, which in turn encodes the center charge of the (co-)characters represented by the junction, enumerating all linear combinations of extended weights from different stacks that add up to zero $\colpq{p}{q}_\text{asymp}$ also enumerates the center charge of all dynamical gauge charges.
In particular, computing the center charges of all string junctions that have no $\mathfrak{u}(1)$-charges determines $Z(G)$, and those of 5-brane junctions determine $\pi_1(G) \cong {\cal Z}$, where $G = \widetilde{G} / {\cal Z}$ is the physically realized non-Abelian gauge group, with simply-connected cover $\widetilde{G}$.

For the rank $(2,18)$ branch, the information about $\mathcal{Z}$ has been shown to be conveniently encoded in so-called fractional null junctions \cite{Guralnik:2001jh}, which are certain fractional multiples of physical loop junctions $\boldsymbol\ell^N_{(r,s)}$ around all 7-branes (with trivial total monodromy).
As we will see, this correspondence continues in realizations of rank $(2,10)$ with one O7$^+$-plane, and rank $(2,2)$ theories with two O7$^+$'s.
While for rank $(2,10)$, we can crosscheck the results with those obtained from a dual CHL string description \cite{Font:2021uyw,Cvetic:2021sjm}, the junction description provides a prediction for the gauge group topology of rank $(2,2)$ vacua which are inaccessible via the heterotic/CHL string.
To illustrate the procedure we will explicitly work out an example for each of the three branches of the 8d moduli space.

\subsubsection{Gauge group topology from global null junctions}
\label{subsec:global_junciton_lattice_null_junctions}

The construction of supersymmetric 8d theories with a dynamical gravity sector and rank $(2,18 - 8k)$ gauge sector requires the identification of a set of $(24 - 10k)$ $[p,q]$-7-branes and $k$ O7$^+$-planes with vanishing overall monodromy. These configurations can then be placed on a $\mathbb{P}^1$ which compactifies the underlying type IIB theory from ten to eight dimensions\footnote{The number of branes in the setup can also be understood as the demand that their cumulative gravitational backreaction in terms of the induced deficit angle adds up to $4 \pi$ as appropriate for the 2-sphere $S^2 \sim \mathbb{P}^1$.}.

Following the conventions laid out in the previous section, we arrange the 7-branes along a horizontal axis in the perpendicular plane (which is now a compact $\mathbb{P}^1$), and enumerate (from the left to right) the ordinary $[p,q]$-7-branes and the O7$^+$'s separately.
Within the vector space of all possible junctions (which carries a pairing given by simply linearly-extending the rules \eqref{eq:selfintersec} and \eqref{eq:bilinext}),
\begin{align}\label{eq:junction_lattice_compact}
    J = \left\{ \sum_{i=1}^{24-10k} a^i \bx_{[p_i,q_i]} + \sum_{j=1}^k \left( b^j \boldsymbol\omega_p^{\text{O7}^+_j} + \tilde{b}^j \boldsymbol\omega_q^{\text{O7}^+_j} \right) \ \big| \ a^i, b^j, \tilde{b}^j \in \mathbb{Q} \right\} \, ,
\end{align}
string junctions giving rise to the electrically charged states must have integral number of prongs on 7-branes, with ``integrality'' on the O7$^+$ being defined as having even number of prongs. 
Analogously, magnetically charged states are described by physical 5-branes with integral number of prongs on all 7-branes, including the O7$^+$'s.
However, since the 7-branes move on a compact $\mathbb{P}^1$, physical junctions must have vanishing asymptotic $\colpq{p}{q}$-charge, i.e., have no open ends.
This means that the \emph{physical} string / 5-brane junction lattice, corresponding to dynamical electric / magnetic states of the 8d supergravity theory, is
\begin{align}
\begin{aligned}
    J_\text{phys}^\text{el} = & \left\{
    \renewcommand{\arraystretch}{1.5}
    \begin{array}{c}
        \mathbf{j} = \sum_{i = 1}^{24 - 10k} a^i \, \mathbf{x}_{[p_i,q_i]} + \sum_{j = 1}^k \Big( 2 b^j \, \boldsymbol\omega^{\text{O7}^+_j}_p + 2 \tilde{b}^j \, \boldsymbol\omega_q^{\text{O7}^+_j} \Big) \, , \\
    \text{with} \ a^i, b^j, \tilde{b}^j \in \mathbb{Z} \, , \quad \sum_i a^i p_i + \sum_j 2b^j = 0 \, , \quad \sum_i a^i q_i + \sum_j 2\tilde{b}^j = 0 
    \end{array}
    \right\} \, , \\
    J_\text{phys}^\text{mag} = & \left\{
    \renewcommand{\arraystretch}{1.5}
    \begin{array}{c}
        \mathbf{j} = \sum_{i = 1}^{24 - 10k} a^i \, \mathbf{x}_{[p_i,q_i]} + \sum_{j = 1}^k \Big( b^j \, \boldsymbol\omega^{\text{O7}^+_j}_p + \tilde{b}^j \, \boldsymbol\omega_q^{\text{O7}^+_j} \Big) \, , \\
    \text{with} \ a^i, b^j, \tilde{b}^j \in \mathbb{Z} \, , \quad \sum_i a^i p_i + \sum_j b^j = 0 \, , \quad \sum_i a^i q_i + \sum_j \tilde{b}^j = 0 
    \end{array}
    \right\} \, .
\end{aligned}
\end{align}
Obviously, these lattices are of rank $24 - 10k -2 = 22-10k$.

Furthermore, by Hanany--Witten transitions, different elements in these lattices can represent the same physical junction.
Equivalently, we can add arbitrary multiples of so-called \emph{(global) integer null junctions}, 
\begin{align}
    J^{N,\text{el/mag}}_\text{int} = \left\{ \boldsymbol\delta_{(r,s)}^N \in J^\text{el/mag}_\text{phys} \ \big| \  \boldsymbol\delta_{(r,s)}^N = \boldsymbol\ell_{(r,s)} \ \text{loops around \emph{all} 7-branes} \right\} \, ,
\end{align}
where it is understood that, \textit{a priori}, there are different integral null junctions for strings and 5-branes.
Because of the compactness, such a loop can be shrunk to a point ``on the other side'' of the $\mathbb{P}^1$ without crossing any 7-branes, and thus are physically trivial.
However, they would appear as a non-trivial element in $J_\text{phys}$ after pulling them through the 7-branes, which must therefore be modded out before we can identify the junction lattice with the physical charge lattice.
Note that, by construction, $\boldsymbol\delta_{(r,s)}^N \in J^N_\text{int}$ has trivial pairing with any other $\boldsymbol\delta^N \in J^N_\text{int}$, as well as no asymptotic $\colpq{p}{q}$-charge (since they encircle a configuration with trivial overall monodromy).
Moreover, $(\boldsymbol\delta_{(r,s)}^N, {\bf j}) = 0$ for all $\boldsymbol\delta_{(r,s)}^N \in J_\text{int}^N$ if and only if ${\bf j}$ has zero asymptotic charge \cite{DeWolfe:1998zf}.
Hence, $\boldsymbol\delta_{(r,s)}^N \in J_\text{int}^N$ has trivial pairing with all physical junctions, explaining the prefix ``null''.
As a notational convention, we shall denote any loop junctions $\boldsymbol\ell_{(r,s)}$ with no asymptotic charge by $\boldsymbol\delta_{(r,s)}$.

This allows us now to identify the {\it (co-)character lattice} $\Lambda_\text{c}$ ($\Lambda_\text{cc}$), the lattice of all electrically (magnetically) charged states present in the supergravity theory, as
\begin{align}
    \Lambda_\text{c} \cong J^\text{el}_\text{phys} / J_\text{int}^{N,\text{el}} \, , \quad \Lambda_\text{cc} \cong J^\text{mag}_\text{phys} / J_\text{int}^{N,\text{mag}} \, ,
\end{align}
which are rank $20 - 8k$ lattices.
Since we mod out a sublattice which is null, the junction pairing on \eqref{eq:junction_lattice_compact} induces a non-degenerate pairing on these lattices, whose signature can be shown to be $(2, 18-8k)$.
For $[{\bf j}_e] \in \Lambda_\text{c}$ and $[{\bf j}_m] \in \Lambda_\text{cc}$ with representatives ${\bf j}_{e}, {\bf j}_m \in J^{\text{el/mag}}_{\text{phys}}$, integrality of the Dirac pairing requires $({\bf j}_e , {\bf j}_m) \in \bbZ$.
Moreover, the Completeness Hypothesis for quantum gravity implies that the two lattices are dual to each other, $\Lambda_\text{c} = (\Lambda_\text{cc})^*$, i.e., for any $[{\bf j}_e]$ there is a $[{\bf j}_m]$ such that $({\bf j}_e, {\bf j}_m) = 1$ and vice versa.
This can be explicitly checked, as we will discuss later.

Now suppose that the 7-branes give rise to the full 8d gauge algebra
\begin{align}
    \mathfrak{g} = \bigoplus_\sigma \mathfrak{g}_\sigma \oplus \mathfrak{u}(1)^{\oplus r_A} \,, \quad \text{with} \enspace r_A = 20 - 8k - \sum_{a} \text{rank} (\mathfrak{g}_a) \,.
\end{align}
Since for each 7-brane stack with gauge factor $\fkg_\sigma$, we have (co-)weight junctions ${\bf w}^{(\vee)}_{\sigma;i_\sigma}$, $i_\sigma=1,...,\text{rank}(\fkg_\sigma)$, we can uniquely (up to global null junctions) decompose\footnote{Note that each stack $\sigma$ can appear with a monodromy that is conjugated by $g_\sigma \in SL(2,\bbZ)$ compared to the ``standard frame'' \eqref{eq:ADE_brane_stacks} or \eqref{eq:monodromy_sp} chosen in the previous section, so that the extended weights have $(p,q)$-charges $g_\sigma\colpq{1}{0}$ for $\boldsymbol\omega_p^\sigma$ and $g_\sigma \colpq{0}{1}$ for $\boldsymbol\omega_q^\sigma$, respectively.}
\begin{align}\label{eq:decomp_j_em}
    {\bf j}_{e \, (m)} = \sum_{\sigma} \left( \sum_{i_\sigma} a^{i_\sigma}_\sigma {\bf w}^{(\vee)}_{\sigma;i_\sigma} + a^p_\sigma \boldsymbol\omega_{p}^{\sigma} + a^q_\sigma \boldsymbol\omega_{q}^{\sigma}  \right) + \sum_s b_s \bx_s \in J_\text{phys}^{\text{el (mag)}}\, ,
\end{align}
where $s$ labels the remaining 7-branes (including potential O7$^+$'s) that are not part of the non-Abelian stacks, on which the prongs must be integral, $b_s \in \bbZ$ (or satisfy the corresponding integrality condition on O7$^+$'s).
Since gauge charges under $\fkg_\sigma$ are carried by the (co-)weights ${\bf w}^{(\vee)}_{\sigma;i_\sigma}$, the states with \emph{only} Abelian charges live in the subspace orthogonal to the (co-)weights,
\begin{align}\label{eq:abelian_junctions}
    J^{\text{el/mag}}_A := \left\{ P_A({\bf j}_{e (m)}) \ | \ {\bf j}_{e (m)} \in J_\text{phys}^\text{el (mag)} \right\}_\text{phys} \equiv \left\{ \sum_{\sigma} \left( a^p_\sigma \boldsymbol\omega_{p}^{\sigma} + a^q_\sigma \boldsymbol\omega_{q}^{\sigma}  \right) + \sum_s b_s \bx_s \right\} \cap J_\text{phys}^{\text{el/mag}} \, ,
\end{align}
where $P_A$ is the projection onto the orthogonal complement of the non-Abelian (co-)weights.
In particular, since global null junctions have zero pairing with all physical junctions, we have $J_\text{int}^{N, \text{el/mag}} \subset J^{\text{el/mag}}_A$.

Because the overall $\colpq{p}{q}$ charge of any physical junction must be zero, only specific linear combinations of extended weights, and therefore, only (co-)weights of $\fkg_\sigma$ with specific center charges, can be completed into a physical junction in \eqref{eq:decomp_j_em} with the available singlet branes.
If the resulting string junctions give rise to representations that are all invariant under a subgroup ${\cal Z}$ of the center, then the gauge group has some non-trivial global structure.

More precisely, the most general global gauge group structure is 
\begin{align}\label{eq:general_gauge_group_topology}
G = \frac{[\prod_\sigma \widetilde{G}_\sigma / \mathcal{Z} ] \times U(1)^{r_A}}{\mathcal{Z}'} \,,
\end{align}
where $\widetilde{G}_\sigma$ is the simply-connected realization of the gauge algebra $\mathfrak{g}_\sigma$. The finite group $\mathcal{Z}$ embeds into the overall center $\prod_\sigma Z(\widetilde{G}_\sigma)$ of the non-Abelian factors with trivial map to the Abelian groups, whereas $\mathcal{Z}'$ does have a non-trivial map into the Abelian sector.
We will now explain how to extract these discrete groups from junctions.

We first focus on the factor $\mathcal{Z}$, which demands that electric states can appear only in certain irreducible representations under $\prod_\sigma \widetilde{G}_\sigma$ that are invariant under ${\cal Z}$.
Equivalently, this can be understood as the existence of magnetic states $[{\bf j}^\text{nA}_m] \in \Lambda_\text{cc}$ charged only under the non-Abelian gauge factors, which via the Dirac pairing condition $({\bf j}_e, {\bf j}^\text{nA}_m) \in \bbZ$ enforces the absence of electric states that are not invariant under ${\cal Z} \subset \prod_\sigma Z(\widetilde{G}_\sigma)$.
Decomposing such a junction,
\begin{align}
    {\bf j}^\text{nA}_{m} = \sum_{\sigma} \left( \sum_{i_\sigma} a^{i_\sigma}_\sigma {\bf w}^{\vee}_{\sigma;i_\sigma} + a^p_\sigma \boldsymbol\omega_{p}^{\sigma} + a^q_\sigma \boldsymbol\omega_{q}^{\sigma}  \right) + \sum_s b_s \bx_s \in J_\text{phys}^\text{mag} \, ,
\end{align}
the assumption that this junction is only charged under the non-Abelian factors implies that the Abelian part,
\begin{align}
    \sum_{\sigma} \left( a^p_\sigma \boldsymbol\omega_{p}^{\sigma} + a^q_\sigma \boldsymbol\omega_{q}^{\sigma}  \right) + \sum_s b_s \bx_s \in J_A^\text{mag} \otimes \mathbb{Q} \, ,
\end{align}
has also zero pairing with every junction in $J_A^\text{mag}$.
However, this is only possible if it is proportional to a linear combination of global integer null junction with rational coefficients, i.e.,
\begin{align}\label{eq:fracnull}
    P_A({\bf j}_m^\text{nA}) = \sum_{\sigma} \left( a^p_\sigma \boldsymbol\omega_{p}^{\sigma} + a^q_\sigma \boldsymbol\omega_{q}^{\sigma}  \right) + \sum_s b_s \bx_s = b_m \boldsymbol\delta_{(r_m,s_m)}^N \, , \quad \boldsymbol\delta_{(r_m,s_m)}^N \in J_\text{int}^{N, \text{mag}} \, , \ b_m \in \mathbb{Q} \, .
\end{align}

Considering such decompositions for all 5-brane junctions ${\bf j}_m^\text{nA} \in J_\text{phys}^\text{mag}$ with no Abelian charge, one obtains the lattice
\begin{align}
    J^{N, \text{mag}}_\text{frac} = \left\{ b_m \boldsymbol\delta_{(r_m,s_m)}^N \ \bigg| \ {\bf j}_m^\text{nA} = \sum_{\sigma} \sum_{i_\sigma} a^{i_\sigma}_\sigma {\bf w}^{\vee}_{\sigma;i_\sigma} + b_m \boldsymbol\delta_{(r_m,s_m)}^N \in J_\text{phys}^\text{mag} \right\} \supset J^{N,\text{mag}}_\text{int} \, ,
\end{align}
of what is called (global) fractional null junctions \cite{Fukae:1999zs,Guralnik:2001jh}.
Let us further denote the smallest positive integer $n_m$ such that $n_m b_m \boldsymbol\delta_{(r_m,s_m)}^N \in J_\text{int}^{N, \text{mag}}$.
Since the prongs on $\bX_s$ are already integral, due to the physicality of ${\bf j}_m^\text{nA}$, $n_m$ is also the smallest positive integer such that $n_m (a^p_\sigma \boldsymbol\omega_{p}^{\sigma} + a^q_\sigma \boldsymbol\omega_{q}^{\sigma}) \in J^{\sigma}_\text{phys, 5-branes}$ is physical on every non-Abelian stack $\sigma$.
At the same time, according to the discussions around \eqref{eq:center_charge_table_ADE} and \eqref{eq:center_charge_5-brane_sp}, the coefficients $(a^p_\sigma, a^q_\sigma)$ specify an element 
\begin{align}
    z_m = (z_\sigma) \in \prod_\sigma \frac{J^{\sigma}_\text{ext}}{J_\text{phys}^\sigma \cap J^\sigma_\text{ext}} \cong \prod_\sigma Z(\widetilde{G}_\sigma) \, .
\end{align}
Since the set of all such $z_m$ generate the discrete factor ${\cal Z}$ in \eqref{eq:general_gauge_group_topology}, we find
\begin{align}
    {\cal Z} \cong \frac{J^{N, \text{mag}}_{\text{frac}}}{J^{N, \text{mag}}_{\text{int}}} \,.
\label{eq:juncZ}
\end{align}

The main advantage of this formula is that we can conveniently compute $J_\text{frac}^{N, \text{mag}}$ from pulling the two generators $\boldsymbol\delta^N_a$ of $J_\text{int}^{N, \text{mag}}$ across all 7-brane stacks, which yields
\begin{align}
    \boldsymbol\delta^N_a = \sum_\sigma \big( c_{a;\sigma}^p \boldsymbol\omega_p^\sigma + c_{a;\sigma}^q \boldsymbol\omega_q^\sigma) + \sum_s c_{a;s} \bx_s \in J_\text{phys}^\text{mag} \, .
\end{align}
Then $J_\text{frac}^{N, \text{mag}}$ is generated by $\mathbb{Q}$-linear combinations,
\begin{align}
    \lambda_1 \boldsymbol\delta^N_1 + \lambda_2 \boldsymbol\delta^N_2 = \sum_\sigma \left( \big( \lambda_1 c_{1;\sigma}^p + \lambda_2 c_{2;\sigma}^p \big) \boldsymbol\omega_p^\sigma + \big( \lambda_1 c_{1;\sigma}^q + \lambda_2 c_{2;\sigma}^q \big) \boldsymbol\omega_q^\sigma \right) + \sum_s \big( \lambda_1 c_{1;s} + \lambda_2 c_{2;s} \big) \bx_s \, ,
\end{align}
such that $\big( \lambda_1 c_{1;\sigma}^p + \lambda_2 c_{2;\sigma}^p \big)$, $\big( \lambda_1 c_{1;\sigma}^q + \lambda_2 c_{2;\sigma}^q \big)$ are integer, and $\big( \lambda_1 c_{1;s} + \lambda_2 c_{2;s} \big)$ satisfies the physicality condition on $\bX_s$.
As advertised, this procedure applies indiscriminately to configurations with or without O7$^+$-planes, as long as the integrality conditions on O7$^+$'s and $\mathfrak{sp}_n$ stacks follow the prescription in Section \ref{subsec:junctions_on_O7}.

The second discrete factor ${\cal Z}' \subset \prod_\sigma Z(\widetilde{G}_\sigma) \times U(1)^{r_A}$ in \eqref{eq:general_gauge_group_topology} correlates the representations under the non-Abelian factors $\prod_\sigma \widetilde{G}_\sigma$ of electric states to their $\mathfrak{u}(1)$ charges, such that their transformation under $\prod_\sigma Z(\widetilde{G}_\sigma)$ is compensated by a ${\cal Z}'$ subgroup in $U(1)^{r_A}$.
Analogously as above, this subgroup can be viewed as being enforced by the presence of magnetic states, now with non-trivial $U(1)$-charges, and hence have a junction representation $[{\bf j}_m] \in \Lambda_\text{cc}$ with
\begin{align}
    J_\text{phys}^\text{mag} \ni {\bf j}_{m} = \sum_{\sigma} \left( \sum_{i_\sigma} a^{i_\sigma}_\sigma {\bf w}^{\vee}_{\sigma;i_\sigma} + a^p_\sigma \boldsymbol\omega_{p}^{\sigma} + a^q_\sigma \boldsymbol\omega_{q}^{\sigma}  \right) + \sum_s b_s \bx_s = \sum_{\sigma} \sum_{i_\sigma} a^{i_\sigma}_\sigma {\bf w}^{\vee}_{\sigma;i_\sigma} + {\bf j}_{A,m}\,,
    \label{eq:magdecompA}
\end{align}
where ${\bf j}_{A,m} = P_A({\bf j}_m) \in J_A \otimes \mathbb{Q}$ is no longer a null junction.
Nevertheless, as the mismatch of ${\bf j}_{A,m}$ from being a physical junction is still determined by the coefficients $(a_\sigma^p, a_\sigma^q)$, we have
\begin{align}\label{eq:abelian_quotient_from_junctions}
\mathcal{Z}' = \frac{P_A\left( J_\text{phys}^\text{mag} \right)}{J_A^\text{mag}}  \,.
\end{align}
Intuitively, this measures the ``fractionality'' of the $\mathfrak{u}(1)$-charges of all magnetic objects (living in $J_\text{phys}^\text{mag}$) with respect to the charges of those that are uncharged under any non-Abelian symmetry (and hence live in $J_A^\text{mag})$.
Note that, since $P_A(J_\text{int}^N) = J_\text{int}^N$, the null junctions do not affect these quotients.

\subsubsection{Duality to heterotic and CHL descriptions}

Theories of rank $(2,18)$ and $(2,10)$ have a dual construction in terms of the heterotic and CHL string, respectively, which provides a crosscheck for the junction description. 
In both cases, electrically charged states are identified as elements from a momentum lattice for perturbative string excitations.
For the heterotic string the momentum lattice is the so-called Narain lattice \cite{Narain:1985jj}:
\begin{align}
\Lambda_{\text{Narain}} = (- \text{E}_8 ) \oplus (- \text{E}_8 ) \oplus U \oplus U \,,
\label{eq:Narainlattice}
\end{align}
with $(- \text{E}_8)$ the negative of the E$_8$ root lattice and $U$ the hyperpolic lattice defined by the bi-linear form
\begin{align}
U: \quad \begin{pmatrix} 0 & 1 \\ 1 & 0 \end{pmatrix} \,.
\label{eq:bilinU}
\end{align}
The CHL string is determined by the Mikhailov lattice \cite{Mikhailov:1998si}
\begin{align}
\Lambda_{\text{Mikhailov}} = (- \text{D}_8) \oplus U \oplus U = (- \text{E}_8) \oplus U \oplus U(2) \,,
\label{eq:Mikh}
\end{align}
with the negative Spin$(16)$ root lattice $(- \text{D}_8)$. Here, $U(x)$ denotes a lattice of rank two with bi-linear form \eqref{eq:bilinU} multiplied by $x$.

Following the notation of \cite{Cvetic:2021sxm}, we will denote them collectively as $\Lambda_S$. 
Then, if the duality holds, we expect $\Lambda_S = \Lambda_\text{c}$.
Equivalently, points in the dual momentum lattice\footnote{Note that $\Lambda_{\text{Narain}}^* = \Lambda_{\text{Narain}}$ is self-dual while $\Lambda_{\text{Mikhailov}}^* \neq \Lambda_{\text{Mikhailov}}$ is not.} correspond to physical magnetically charged states and therefore must be associated to 5-brane junctions (modulo null junctions) in $\Lambda_\text{cc}$.
The non-Abelian gauge factors are then specified by an embedding of the corresponding (negative) root lattice into $\Lambda_S$; the coroot lattice then naturally embeds in the dual lattice $\Lambda^*_S$.
It is worth noting that the computation of the gauge group topologies from this data \cite{Font:2020rsk,Font:2021uyw,Cvetic:2021sxm} is in a sense complimentary to the junctions approach outlined above.
While in both scenarios, the setting is fully characterized by the non-Abelian gauge algebras (by specifying either the 7-brane stacks or the embedding of the (co-)root lattices), the gauge group is concisely encoded in the projection of the full physical lattice onto the Abelian junctions (see \eqref{eq:juncZ} and \eqref{eq:abelian_quotient_from_junctions}), the methods in \cite{Font:2020rsk,Font:2021uyw,Cvetic:2021sxm} extract the gauge group from the projection onto the (co-)root lattice.

To corroborate the equivalence of the two approaches, we describe in the following the precise identification of the momentum lattices with string junctions on 7-brane configurations with zero or one O7$^+$-plane.

\subsubsection*{Narain lattice from junctions}

To construct the Narain lattice \eqref{eq:Narainlattice} from junctions, it is easiest to find a 7-brane configuration in which the two $(-\text{E}_8)$ factors are manifest via the root junctions on two $\mathfrak{e}_8$ 7-brane stacks, and make use of the fact that the lattice structure does not change as we move 7-branes.
Each of these $\mathfrak{e}_8$ stacks contains ten 7-branes, leaving a remaining four branes to specify the compact type IIB background.
A convenient configuration of this sort has been presented in Section 7 of \cite{DeWolfe:1998pr}, and takes the form
\begin{align}
\bA (\bA^7 \bB \bC \bC) \bX_{[3,1]}\bA' (\bA^7 \bB \bC \bC)' \bX'_{[3,1]} \,,
\label{eq:parent_brane_config}
\end{align}
up to possible $SL(2,\bbZ)$ conjugations. 
In fact, the above configuration has two identical parts, consisting of $\bA (\bA^7 \bB \bC \bC) \bX_{[3,1]}$, each having a trivial SL$(2,\mathbb{Z})$ monodromy.\footnote{Note however that the Spin cover of the monodromy is non-trivial and given by $(-1)^F \in \text{Mp}(2,\mathbb{Z})$, see e.g. \cite{Pantev:2016nze, GarciaEtxebarria:2020xsr, Dierigl:2020lai}.}
In addition, by pushing the twelve branes onto a single stack, one enhances the symmetry to the so-called double loop algebra $\hat{\mathfrak{e}}_9$, whose significance we will explain further in Section \ref{sec:9d}.
In an F-theory description, one may interpret this configuration as a stable degeneration limit of the elliptically-fibered K3 into two $dP_9$ surfaces.

Note that for $\mathfrak{e}_8$, whose extended weights \eqref{eq:extADE} are physical, roots and weights junctions agree, so the span of all physical string prongs on each $\mathfrak{e}_8$ stack (with decomposition as in \eqref{eq:nAstates}) contains two copies of the $(-\text{E}_8)$-lattice.
Next we need to find the factor $U \oplus U$ in the orthogonal complement of the $\mathfrak{e}_8$ root lattices.
A convenient set of generators for these two hyperbolic lattices can be expressed as\footnote{This is the same result as in \cite{DeWolfe:1998pr} (see their Figure 8).}
\begin{align}\label{eq:U_lattices_gens_Narain}
\begin{split}
U:& \quad \big( \boldsymbol{\delta}_{(1,0)} \,, \boldsymbol\delta_{(1,0)} + \bx_{[3, 1]} - \bx'_{[3, 1]} \big) \,, \\
U:& \quad \big( \boldsymbol{\delta}_{(3,1)} \,, \boldsymbol\delta_{(3, 1)} - \ba + \ba' \big) \,.
\end{split}
\end{align}
Here $\boldsymbol{\delta}_{(r,s)} = \boldsymbol\ell_{(r,s)}$ denotes a $(r,s)$-string loop around one $\bA (\bA^7 \bB \bC \bC) \bX_{[3,1]}$ configuration, which has no asymptotic charge since this stack has no overall monodromy.
Its orthogonality to the $\mathfrak{e}_8$ root lattice is evident from the fact that this junction has no prongs on any of the two stacks.
Note that, as is evident from Figure \ref{fig:rank18}, such a loop automatically encircles the other $\bA (\bA^7 \bB \bC \bC) \bX_{[3,1]}$ configuration.
Using Hanany--Witten transition one can rewrite them in terms of integer strings ending on the brane constituents in the interior (say, on the left in Figure \ref{fig:rank18}), e.g., 
\begin{align}
\begin{split}
\boldsymbol{\delta}_{(1,0)} &= 3 \boldsymbol{\omega}^{\mathfrak{e}_8}_p + \boldsymbol{\omega}^{\mathfrak{e}_8}_q - \bx_{[3,1]} \,, \\
\boldsymbol{\delta}_{(3,1)} &= - \ba + \boldsymbol{\omega}_p^{\mathfrak{e}_8} \,,
\end{split}
\end{align}
or equivalently for the primed stack.

\begin{figure}[ht]
    \centering
    \includegraphics[width = 0.6 \textwidth]{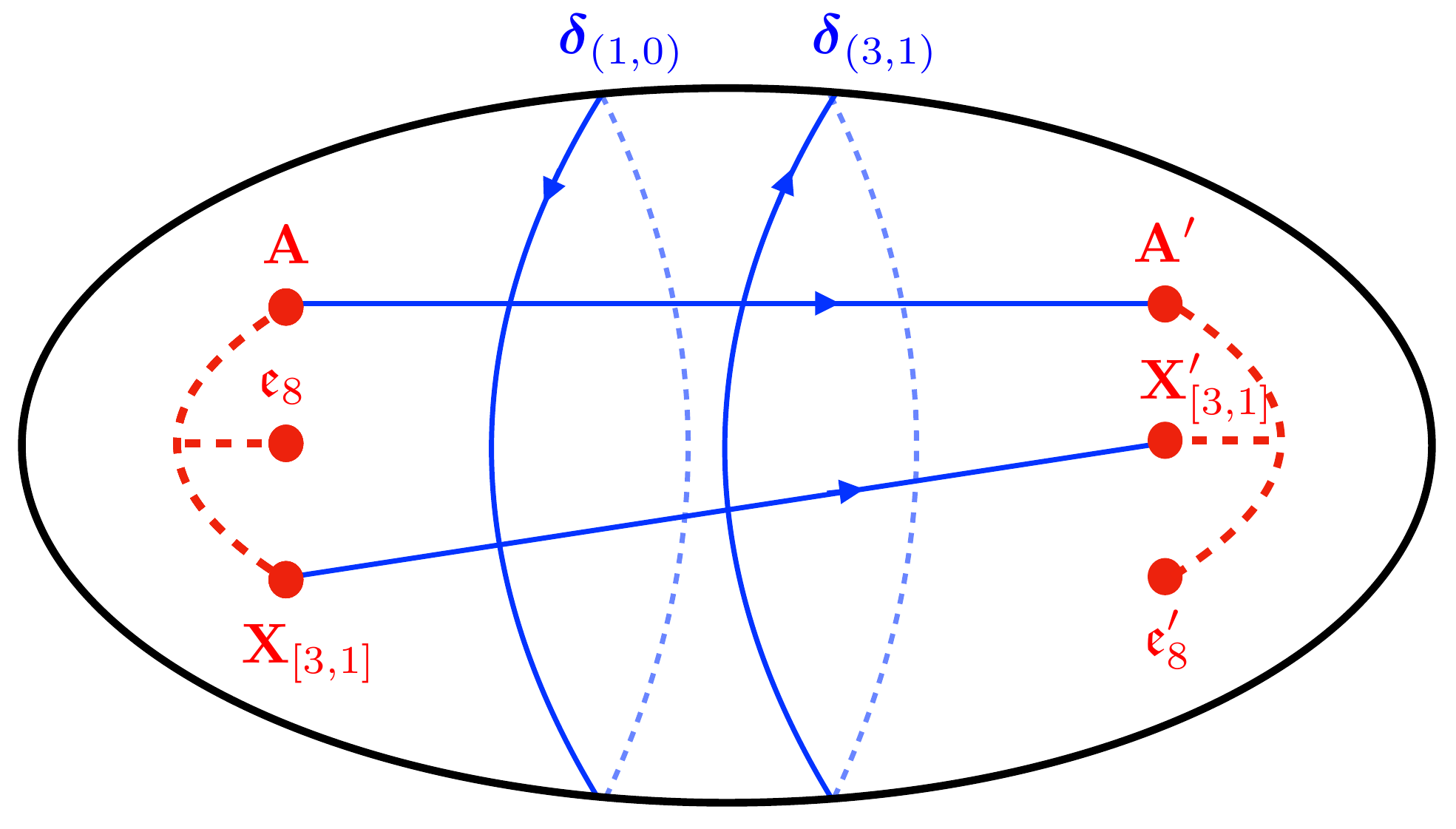}
    \caption{String junction lattice for rank $(2,18)$ theories.}
    \label{fig:rank18}
\end{figure}

This accounts for a rank 20 sublattice $(-\text{E}_8)^{\oplus 2} \oplus U^{\oplus 2}$ of the full string junction lattice $J_\text{phys}^\text{el}$.
The remaining two directions are spanned by the global null junctions $J_\text{int}^N$, for which the canonical basis is
\begin{align}
\begin{split}
    &\boldsymbol\delta_{(1,0)}^N = -(3 \boldsymbol\omega_p^{\mathfrak{e}_8} + \boldsymbol\omega_q^{\mathfrak{e}_8}) + \bx_{[3,1]} - (3 \boldsymbol\omega_p^{\mathfrak{e}'_8} + \boldsymbol\omega_q^{\mathfrak{e}'_8}) + \bx_{[3,1]}' \, , \\
    &\boldsymbol\delta_{(0,1)}^N = -\ba  + 10 \boldsymbol\omega_p^{\mathfrak{e}_8} + 3 \boldsymbol\omega_q^{\mathfrak{e}_8} - 3\bx_{[3,1]} -\ba'  + 10 \boldsymbol\omega_p^{\mathfrak{e}_8'} + 3 \boldsymbol\omega_q^{\mathfrak{e}_8'} - 3\bx_{[3,1]}' \, .
\end{split}
\end{align}
Since these have trivial pairing with all physical junctions, the lattice pairing of the above generators (including the $\mathfrak{e}_8$ roots and those of the $U$-lattices) desecend, without modification, to the quotient
\begin{align}
\Lambda_\text{c} = \frac{J_\text{phys}^\text{el}}{J_\text{int}^{N, \text{el}}} = \frac{(-\text{E}_8)^{\oplus 2} \oplus U^{\oplus 2} \oplus J_\text{int}^{N, \text{el}}}{J_\text{int}^{N, \text{el}}} \cong \Lambda_{\text{Narain}} \, .
\end{align}
Since there are no O7$^+$-planes, the 5-brane junction lattices are the same as their stringy counterparts, so we immediately find
\begin{align}
    \Lambda_\text{cc} = J_\text{phys}^\text{mag} / J_\text{int}^{N, \text{mag}} \cong J_\text{phys}^\text{el} / J_\text{int}^{N, \text{el}} \cong \Lambda_\text{Narain} = \Lambda_\text{Narain}^* \, .
\end{align}

\subsubsection*{Mikhailov lattice from junctions}

The Mikhailov lattice describing the momentum lattice for the 8d CHL string is obtained as follows. 
We keep one of the $\hat{\mathfrak{e}}_9$ configurations unchanged, which still leads to an $(- \text{E}_8)$ factor in the string junction lattice. 
On the other side, we remove a $\mathbf{C}$ brane from the $\mathfrak{e}_8$ stack, but add to it the singlet $\mathbf{A}$-brane, which leads to an $\mathfrak{so}_{16}$ brane stack, which we then ``freeze'' into an O7$^+$-plane:
\begin{align}
\bA {\bf E}_8 \bX_{[3,1]} = \bA (\bA^7 \bB \bC \bC) \bX_{[3,1]} \rightarrow (\bA^8 \bB \bC) \bC \bX_{[3,1]} \rightarrow {\bf O7}^+ \bC \bX_{[3,1]} \,.
\end{align}
The resulting complete 7-brane configuration,
\begin{align}
{\bf O7}^+ \bC \bX_{[3, 1]} \bA' (\bA^7 \bB \bC^2) \bX'_{[3, 1]} \,,
\end{align}
and is depicted in Figure \ref{fig:rank10}.
\begin{figure}[ht]
    \centering
    \includegraphics[width = 0.6 \textwidth]{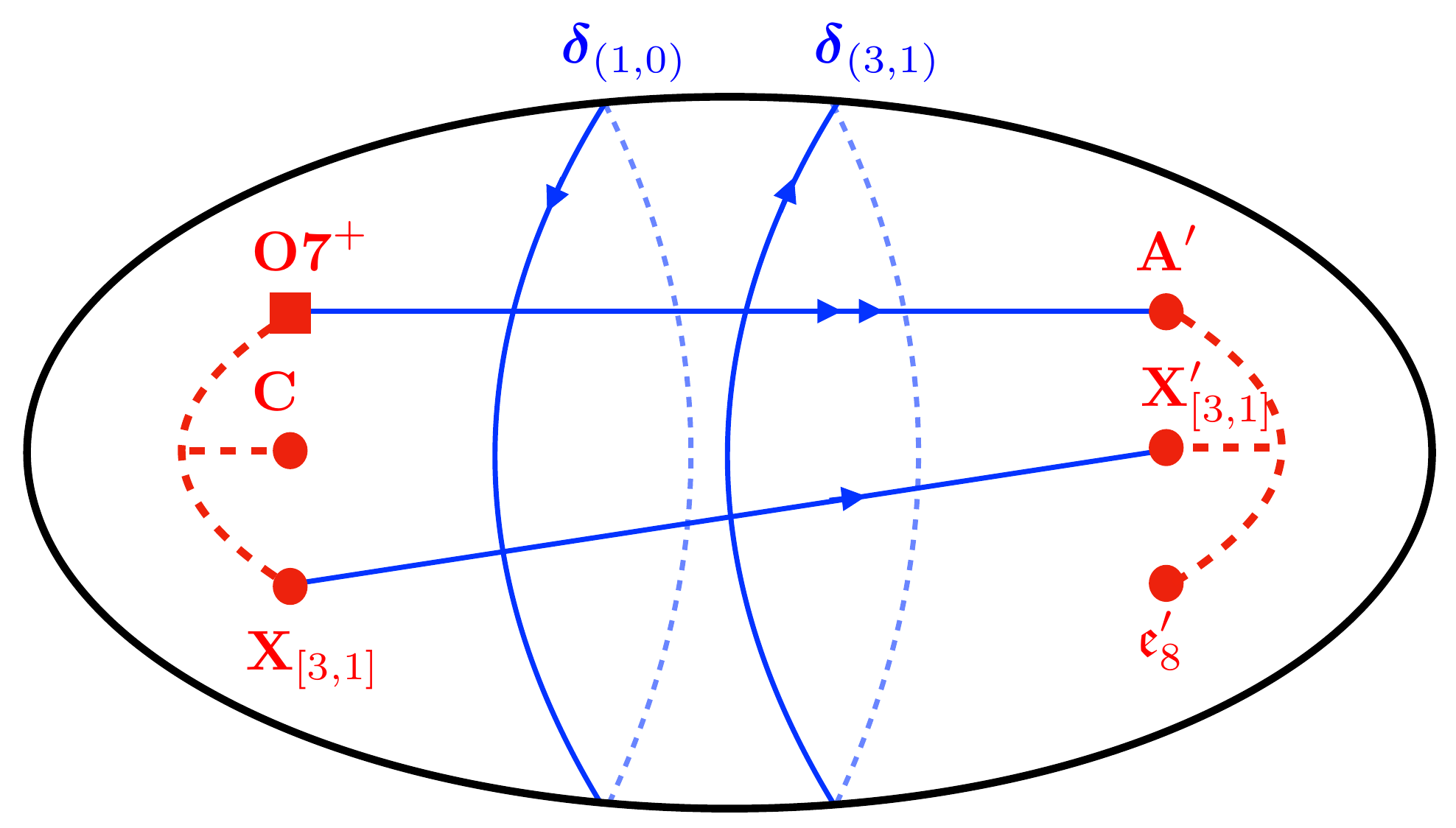}
    \caption{String junction lattice for rank $(2,10)$ theories.}
    \label{fig:rank10}
\end{figure}
The total rank of the junction lattice is now 14.

Inside the string junction lattice $J_\text{phys}^\text{el}$ of this configuration, we now need to identify $U \oplus U(2)$ orthogonal to the $\mathfrak{e}_8$ roots.
Given the similarities to the rank $(2,18)$ configuration, a natural choice for the generators would be a variation of \eqref{eq:U_lattices_gens_Narain}.
While the first set exists also for the frozen configuration, the second $U$-factor has a generator with a single $\ba$-prong, which would become part of the O7$^+$, and not be physical.
Therefore, the set of generators orthogonal to the E$_8$ root lattce are
\begin{align}
\begin{split}
U:& \quad \big( \boldsymbol{\delta}_{(1,0)} \,, \boldsymbol{\delta}_{(1,0)} + \bx_{[3,1]} - \bx_{[3,1]}' \big) \,, \\
U(2):& \quad \big( \boldsymbol{\delta}_{(3,1)} \,, 2 \boldsymbol{\delta}_{(3,1)} + 2 \boldsymbol{\omega}^{\text{O7}^+}_p - 2 \ba' \big) \,,
\end{split}
\label{eq:Mlattices}
\end{align}
where the second generator of $U(2)$ is primitive because of the evenness condition for strings ending on O7$^+$.
These have an equivalent representation with
\begin{align}
\begin{split}
\boldsymbol{\delta}_{(1,0)} &= 2 \boldsymbol{\omega}_p^{\text{O7}^+} + \bc - \bx_{[3,1]} \,, \\
\boldsymbol{\delta}_{(3,1)} &= - 2 \boldsymbol{\omega}^{\text{O7}^+}_p - 2 \boldsymbol{\omega}^{\text{O7}^+}_q + 2 \bc \,.
\end{split}
\label{eq:Mloops}
\end{align}
After quotienting out the global null junctions $J_\text{phys}^{N,\text{el}}$, with generators
\begin{align}\label{eq:null_junctions_rank_10}
    \begin{split}
    &\boldsymbol\delta_{(1,0)}^N = -2\boldsymbol\omega_p^{\text{O7}^+} - \bc + \bx_{[3,1]} - (3 \boldsymbol\omega_p^{\mathfrak{e}'_8} + \boldsymbol\omega_q^{\mathfrak{e}'_8}) + \bx_{[3,1]}' \, , \\
    &\boldsymbol\delta_{(0,1)}^N = 4\boldsymbol\omega_p^{\text{O7}^+} -2\boldsymbol\omega_q^{\text{O7}^+} + 5\bc - 3\bx_{[3,1]} -\ba'  + 10 \boldsymbol\omega_p^{\mathfrak{e}_8'} + 3 \boldsymbol\omega_q^{\mathfrak{e}_8'} - 3\bx_{[3,1]}' \, ,
\end{split}
\end{align}
one finds
\begin{align}
\Lambda_\text{c} \cong (-\text{E}_8) \oplus U \oplus U(2) = \Lambda_{\text{Mikhailov}} \,.
\end{align}

In this case it is interesting to also analyze the 5-brane junctions that correspond to the dual lattice. Here, it is important to remember that 5-branes have different physicality conditions, which allow for an arbitrary integer number of them to end on the O7$^+$-plane.
This does not affect the $\mathfrak{e}_8$ root junctions, and the $U$-factor in \eqref{eq:Mlattices}, but does imply that $\boldsymbol\delta_{(3,1)}$ in \eqref{eq:Mloops} is no longer a primitive 5-brane junction.
Instead, it is a multiple of
\begin{align}
\tfrac{1}{2} \boldsymbol{\delta}_{(3,1)} = \boldsymbol{\delta}_{(\nicefrac{3}{2},\nicefrac{1}{2})} = - \boldsymbol{\omega}^{\text{O7}^+}_p - \boldsymbol{\omega}^{\text{O7}^+}_q + \bc \,.
\end{align}
This implies that the there is an overlattice of $U(2)$ in \eqref{eq:Mlattices} inside the physical 5-brane junction lattice, given by
\begin{align}
U(\tfrac{1}{2}): \quad \big( \tfrac{1}{2} \boldsymbol{\delta}_{(3,1)} \,, \boldsymbol{\delta}_{(3,1)} + \boldsymbol{\omega}^{\text{O7}^+}_p - \ba' \big) \,.
\end{align}
Note that the null junction lattice spanned by \eqref{eq:null_junctions_rank_10} remain primitive as a sublattice of $J_\text{phys}^\text{mag}$.
Therefore one has for the magnetic 5-brane junction lattice
\begin{align}
\Lambda_{cc} = (- \text{E}_8) \oplus U \oplus U(\tfrac{1}{2}) = \Lambda_{\text{Mikhailov}}^* \,,
\end{align}
which precisely coincides with the dual of the Mikhailov lattice \eqref{eq:Mikh}.

\subsubsection[A rank \texorpdfstring{$(2,2)$}{(2,2)} momentum lattice]{A rank \boldmath{$(2,2)$} momentum lattice}

8d rank $(2,2)$ string vacua have no known constructions as $T^2$- or $S^1$-reductions of the heterotic or CHL string.
However, using the junctions, we propose an analogue of a momentum lattice description, which can be applied, in particular, to determine the gauge group topologies of these theories.
To this end, we start with a 7-brane configuration with two O7$^+$'s, that we obtain from further freezing an $\mathfrak{so}_{16}$ stack on the primed side of \eqref{eq:parent_brane_config}.
The overall brane configuration is then given by
\begin{align}
{\bf O7}^+ \bC \bX_{[3,1]} {{\bf O7}^+}' \bC' \bX'_{[3,1]} \,,
\end{align}
and is depicted in Figure \ref{fig:rank2}.
\begin{figure}[ht]
    \centering
    \includegraphics[width = 0.6 \textwidth]{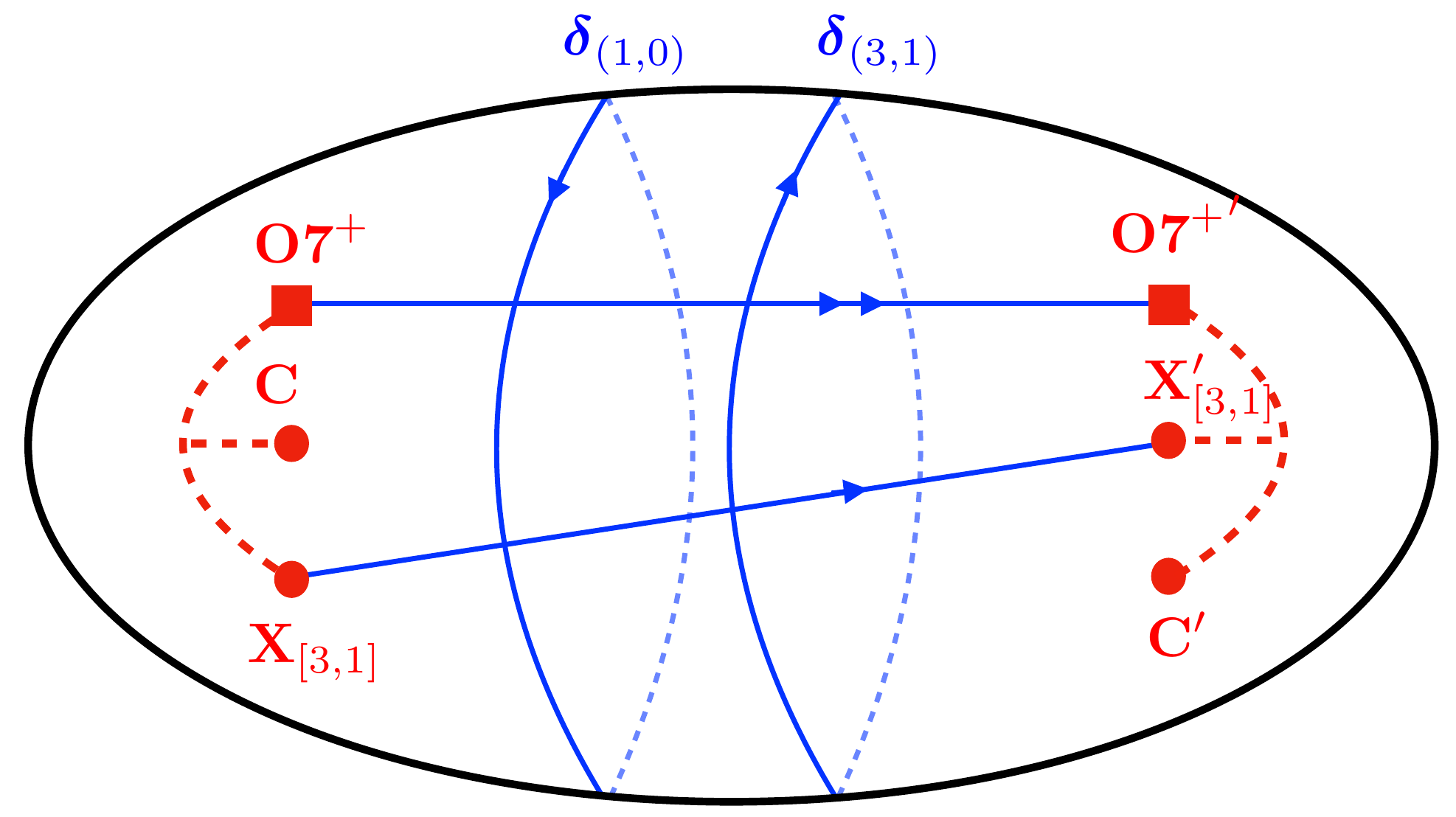}
    \caption{String junction lattice for rank $(2,2)$ theories.}
    \label{fig:rank2}
\end{figure}

The string junction lattice in this case has rank $6$, whose non-null directions are isomorphic to 
\begin{align}
    U \oplus U(2) \, .
\end{align}
The explicit generators in terms of physical string junctions are given by
\begin{align}
\begin{split}
U:& \quad \big( \boldsymbol{\delta}_{(1,0)} \,, \boldsymbol{\delta}_{(1,0)} + \bx_{[3,1]} - \bx_{[3,1]}' \big) \,, \\
U(2):& \quad \big( \boldsymbol{\delta}_{(3,1)} \,, 2 \boldsymbol{\delta}_{(3,1)} + 2 \boldsymbol{\omega}^{\text{O7}^+}_p - 2 \boldsymbol{\omega}^{{\text{O7}^+}'}_p \big) \,,
\end{split}
\end{align}
where the loop junctions satisfy the same relation as in \eqref{eq:Mloops}.
The string null junctions $J_\text{phys}^{N, \text{el}}$ have generators
\begin{align}
\begin{split}
    & \boldsymbol\delta_{(1,0)}^N = -2 \boldsymbol\omega_p^{\text{O7}^+} - \bc + \bx_{[3,1]} -2 \boldsymbol\omega_p^{{\text{O7}^+}'} - \bc' + \bx_{[3,1]}' \, , \\
    & \boldsymbol\delta_{(0,1)}^N = 4 \boldsymbol\omega_p^{\text{O7}^+} - 2 \boldsymbol\omega_q^{\text{O7}^+} + 5 \bc - 3 \bx_{[3,1]} + 4 \boldsymbol\omega_p^{{\text{O7}^+}'} - 2 \boldsymbol\omega_q^{{\text{O7}^+}'} + 5 \bc' - 3 \bx_{[3,1]}' \, .
\end{split}
\end{align}
The full physical string junction lattice is therefore $J_\text{phys}^\text{el} = U \oplus U(2) \oplus J_\text{phys}^{N, \text{el}}$.

As for the rank $(10,2)$ case above, we find that for 5-branes the $U(2)$ turns into a $U(\tfrac{1}{2})$, i.e., $J_\text{phys}^\text{mag} = U \oplus U(\tfrac12) \oplus J_\text{phys}^{N, \text{mag}}$.
A novel modification that will affect the gauge group computation is that also the null junctions are refined.
Namely, the generators of $J_\text{phys}^{N,\text{mag}}$ are
\begin{align}\label{eq:null_junction_gens_rank_2}
\begin{split}
    \tfrac12 (\boldsymbol\delta^N_{(1,0)} + \boldsymbol\delta^N_{(0,1)}) = \boldsymbol\delta^N_{(\nicefrac12,\nicefrac12)} & = \boldsymbol\omega_p^{\text{O7}^+} - \boldsymbol\omega_q^{\text{O7}^+} + 2\bc - \bx_{[3,1]} + \boldsymbol\omega_p^{{\text{O7}^+}'} - \boldsymbol\omega_q^{{\text{O7}^+}'} + 2\bc' - \bx_{[3,1]}' \, , \\
    \tfrac12 (\boldsymbol\delta^N_{(1,0)} - \boldsymbol\delta^N_{(0,1)}) = \boldsymbol\delta^N_{(\nicefrac12,-\nicefrac12)} & = -3 \boldsymbol\omega_p^{\text{O7}^+} + \boldsymbol\omega_q^{\text{O7}^+} - 3\bc - 2\bx_{[3,1]} - 3 \boldsymbol\omega_p^{{\text{O7}^+}'} + \boldsymbol\omega_q^{{\text{O7}^+}'} - 3\bc' + 2\bx_{[3,1]}' \, .
\end{split}
\end{align}
In summary, after modding out the null junctions, we have
\begin{align}
    \Lambda_\text{c} = U \oplus U(2) \, , \quad \Lambda_\text{cc} = U \oplus U(\tfrac{1}{2}) = \Lambda_\text{c}^*\,.
\end{align}

\subsubsection{Examples}

Using the techniques outlined in Section \ref{subsec:global_junciton_lattice_null_junctions}, we can determine the brane configurations and the resulting gauge group topologies for all 8d ${\cal N}=1$ string vacua.
This is done explicitly for all maximally-enhanced cases on each branch of the moduli space, as summarized in Appendix \ref{app:results}.
In the following we demonstrate the general procedure in specific examples.
For convenience, we focus 8d theories that were discussed in \cite{Font:2021uyw,Cvetic:2021sjm} from the perspective of the heterotic or CHL momentum lattice.
The generalization to rank $(2,2)$ theories is, to our knowledge, the first time in the literature the global gauge group topology has been computed for these string vacua.

\subsubsection{A rank \boldmath{$(2, 18)$} example}

The non-Abelian gauge algebra of the model is given by
\begin{align}
\mathfrak{g} = \mathfrak{so}_{20} \oplus \mathfrak{su}_4 \oplus \mathfrak{su}_4 \oplus \mathfrak{su}_2 \oplus \mathfrak{su}_2 \,,
\end{align}
which can be generated by the following brane configuration:
\begin{align}
(\bA^{10} \bB \bC) \bN^4 \bX_{[1,3]}^4 \bX_{[2,5]}^2 \bC^2 \,.
\end{align}
Note that for a consistent overall monodromy, the $\mathfrak{su}$ algebras are not associated to a stack of $\bA$-branes, but rather in some $SL(2,\mathbb{Z})$ conjugated frame. 
Accordingly, the associated extended weight junctions summarized in \eqref{eq:extADE} need to be conjugated, and are given by
\begin{align}
\begin{split}
\mathfrak{so}_{20} \quad (\bA^{10} \bB \bC):& \quad \boldsymbol\omega_{p} = \tfrac{1}{2} (\bb + \bc) \,, \quad \boldsymbol\omega_q = \tfrac{1}{2} \sum_{i = 1}^{10} \ba_i - 3\bb - 2\bc \,, \\
\mathfrak{su}_4 \quad (\bN^4):& \quad \boldsymbol\omega_{(0,1)} = \tfrac{1}{4}\sum_{i = 1}^4\bn_i \,, \\
\mathfrak{su}_4 \quad (\bX_{[1,3]}^4):& \quad \boldsymbol\omega_{(1, 3)} = \tfrac{1}{4}\sum_{i = 1}^4 \bx_{[1,3],i} \,, \\
\mathfrak{su}_2 \quad (\bX^2_{[2,5]}):& \quad \boldsymbol\omega_{(2, 5)} = \tfrac{1}{2}\sum_{i = 1}^2 \bx_{[2,5],i} \,, \\
\mathfrak{su}_2 \quad (\bC^2):& \quad \boldsymbol\omega_{(1,1)} = \tfrac{1}{2}\sum_{i = 1}^2 \bc_{i} \,,
\end{split}
\label{eq:exaext20}
\end{align}
where $\boldsymbol\omega_{(p,q)}$ is the extended weight of the corresponding $\mathfrak{su}$-stack with asymptotic $\colpq{p}{q}$-charge. 
In terms of these extended weights the two linearly independent integer null junctions are given by
\begin{align}
\begin{split}
\boldsymbol{\delta}^N_{(1,0)} &= - 2 \boldsymbol\omega_p - 4 \boldsymbol\omega_{(0,1)} + 4 \boldsymbol\omega_{(1, 3)} - 2 \boldsymbol\omega_{(2, 5)} + 2 \boldsymbol\omega_{(1,1)} \,, \\
\boldsymbol{\delta}^N_{(0,1)} &= 6 \boldsymbol\omega_p - 2 \boldsymbol\omega_q + 24 \boldsymbol\omega_{(0,1)} - 20 \boldsymbol\omega_{(1, 3)} + 8 \boldsymbol\omega_{(2, 5)} - 2 \boldsymbol\omega_{(1,1)} \,.
\end{split}
\label{eq:exa20null}
\end{align}
Notice how, in both junctions, the greatest common divisor of the coefficients is 2.
Therefore, the fractional null junctions are generated by
\begin{align}
J^{N, \text{mag}}_{\text{frac}} = \Big\{ \lambda^N_{(1,0)} \, \boldsymbol\delta^N_{(1,0)} + \lambda^N_{(0,1)} \, \boldsymbol\delta^N_{(0,1)} \ \Big| \ \lambda^N_{(1,0)}, \lambda^N_{(0,1)} \in \tfrac{1}{2} \mathbb{Z} \Big\} \,, 
\end{align}
and the global realization of the non-Abelian gauge group is determined by
\begin{align}
\mathcal{Z} = \frac{J^N_{\text{frac}}}{J^N_{\text{int}}} = \frac{ \{\tfrac{x}{2} \boldsymbol\delta_{(1,0)}^N \ | \ x \in \bbZ \} }{(\boldsymbol\delta_{(1,0)}^N)} \times \frac{ \{\tfrac{y}{2} \boldsymbol\delta_{(1,0)}^N \ | \ y \in \bbZ\} }{(\boldsymbol\delta_{(0,1)}^N)} \cong \mathbb{Z}_2 \times \mathbb{Z}_2 \,.
\end{align}
Moreover, the coefficients in front of the extended weights in $\tfrac12 \boldsymbol\delta^N$ determine, according to \eqref{eq:center_charge_table_ADE}, the generators of ${\cal Z} \subset Z(Spin(20)) \times Z(SU(4))^2 \times Z(SU(2))^2$ to be
\begin{align}
\begin{split}
    & \tfrac12 \boldsymbol\delta_{(1,0)}^N \simeq (1,0; 2; 2; 1; 1) \in (\bbZ_2 \times \bbZ_2) \times \bbZ_4 \times \bbZ_4 \times \bbZ_2 \times \bbZ_2 \, , \\
    & \tfrac12 \boldsymbol\delta_{(0,1)}^N \simeq (1,1; 0; 2 ; 0; 1) \in (\bbZ_2 \times \bbZ_2) \times \bbZ_4 \times \bbZ_4 \times \bbZ_2 \times \bbZ_2 \, .
\end{split}
\end{align}

Beyond the non-Abelian gauge factors, the theory contains two gravi-photons generating two $\mathfrak{u}(1)$ gauge factors.
These arise from Abelian junctions \eqref{eq:abelian_junctions} that are not null junctions, which for the present model can be easily determined, from \eqref{eq:exaext20}, to be linear combinations of
\begin{align}\label{eq:exaAbelian}
    \begin{split}
        & {\bf u}_1 = 4 (\boldsymbol\omega_{(0,1)} + \boldsymbol\omega_{(1,3)} - \boldsymbol\omega_{(2,5)} + \boldsymbol\omega_{(1,1)} ) \equiv 4 {\bf v}_1 \, , \\
        & {\bf u}_2 = 4 (\boldsymbol\omega_{(0,1)} - \boldsymbol\omega_q) \equiv 4 {\bf v}_2\, .
    \end{split}
\end{align}
${\bf v}_a$, while themselves non-physical, can be made physical by adding coweight junctions, so lie in $P_A(J_\text{phys}^\text{mag})$.
At the same time, there are no ``finer'' coweights to make fractions of ${\bf v}_a$ physical, so ${\bf v}_a$ generate $P_A(J_\text{phys}^\text{mag})$.
Then, the Abelian quotient ${\cal Z'}$, according to \eqref{eq:abelian_quotient_from_junctions}, is
\begin{align}
\mathcal{Z}' = \frac{P_A(J_\text{phys}^\text{mag})}{J_A^\text{mag}} = \frac{\{x{\bf v}_1 \ | \ x \in \bbZ\} }{({\bf u}_1)} \times \frac{\{ y{\bf v}_2 \ | \ y \in \bbZ\}}{({\bf u}_2)} = \mathbb{Z}_4 \times \mathbb{Z}_4 \,.
\end{align}
The generators of ${\cal Z}'$ are
\begin{align}\label{eq:abelian_quotient_generators_rk-20_example}
\begin{split}
    & {\bf v}_1 \simeq (0,0; 1; 1; 1; 1 \ | \ e^{i \pi /2}; 1) \in (\bbZ_2 \times \bbZ_2) \times \bbZ_4 \times \bbZ_4 \times \bbZ_2 \times \bbZ_2 \times U(1)_1 \times U(1)_2\, , \\
    & {\bf v}_2 \simeq (0,1; 1; 0; 0; 0 \ | \ 1; e^{i \pi /2}) \in (\bbZ_2 \times \bbZ_2) \times \bbZ_4 \times \bbZ_4 \times \bbZ_2 \times \bbZ_2 \times U(1)_1 \times U(1)_2 \, ,
\end{split}
\end{align}
where we have used a vertical line to separate the finite groups from the $U(1)$'s, whose trivial element is $1$.

Consequently, the global form of the gauge group is given by
\begin{equation}
    \frac{\left[\left(Spin(20) \times SU(2)^{2} \times SU(4)^{2}\right) /\left(\mathbb{Z}_{2} \times \mathbb{Z}_{2}\right)\right] \times U(1)^{2}}{\mathbb{Z}_{4} \times \mathbb{Z}_{4}} \,,
\end{equation}
in perfect agreement with the heterotic analysis in \cite{Cvetic:2021sjm}.

\subsubsection{A rank \boldmath{$(2,10)$} example}

Since the brane configuration above already containes an $\mathfrak{so}_{16}$ stack we can simply reinterpret this as an O7$^+$-plane, leading to the brane configuration
\begin{align}
(\bA^2 {\bf O7}^+) \bN^4 \bX_{[1,3]}^4 \bX_{[2,5]}^2 \bC^2 \,.
\end{align}
This setup has a non-Abelian gauge algebra given by
\begin{align}
\mathfrak{g} = \mathfrak{sp}_2 \oplus \mathfrak{su}_4 \oplus \mathfrak{su}_4 \oplus \mathfrak{su}_2 \oplus \mathfrak{su}_2 \,,
\end{align}
leading to a model with rank $(2,10)$ dual to a specific CHL background. 
Except for the $\mathfrak{sp}$ factor, the extended weights are the same as in \eqref{eq:exaext20}. 
For the $\mathfrak{sp}$ algebra one has
\begin{align}
    \mathfrak{sp}_2 \quad (\bA^2 {\bf O7}^+): \quad \boldsymbol\omega_p = \boldsymbol\omega_p^{\text{O7}^+} \, , \quad \boldsymbol\omega_q = \tfrac12 (\ba_1 + \ba_2) - \boldsymbol\omega_p^{\text{O7}^+} + \boldsymbol\omega_q^{\text{O7}^+} \, .
\end{align}
Formally, the global null junctions $\boldsymbol\delta^N_{(p,q)}$ are the same as in \eqref{eq:exa20null}, except that the $\boldsymbol\omega_{p,q}$ appearing there are now the extended weight junctions of $\mathfrak{sp}$.
Again, we can divide both by 2, obtaining the fractional 5-brane junctions
\begin{align}
    J^{N, \text{mag}}_{\text{frac}} = \Big\{ \lambda^N_{(1,0)} \, \boldsymbol\delta^N_{(1,0)} + \lambda^N_{(0,1)} \, \boldsymbol\delta^N_{(0,1)} \ \Big| \ \lambda^N_{(1,0)}, \lambda^N_{(0,1)} \in \tfrac{1}{2} \mathbb{Z} \Big\} \, ,
\end{align}
implying ${\cal Z} \cong \bbZ_2 \times \bbZ_2$, with generators
\begin{align}
\begin{split}
    & \tfrac12 \boldsymbol\delta_{(1,0)}^N \simeq (0; 2; 2; 1; 1) \in \bbZ_2 \times \bbZ_4 \times \bbZ_4 \times \bbZ_2 \times \bbZ_2 = Z(Sp(2) \times SU(4)^2 \times SU(2)^2) \, , \\
    & \tfrac12 \boldsymbol\delta_{(0,1)}^N \simeq (1; 0; 2; 0; 1) \in \bbZ_2 \times \bbZ_4 \times \bbZ_4 \times \bbZ_2 \times \bbZ_2 = Z(Sp(2) \times SU(4)^2 \times SU(2)^2) \, ,
\end{split}
\end{align}
where the entry for $\bbZ_2 = Z(Sp(2))$ is determined only by the coefficient in front of $\boldsymbol\omega_q$, see \eqref{eq:center_charge_5-brane_sp}.

In a similar way, there are two $\mathfrak{u}(1)$ generators that formally are the same as in \eqref{eq:exaAbelian}.
Since $({\bf u}_1, {\bf v}_1)$ have no prongs on the $\mathfrak{sp}$ stack, we get a $\bbZ_4$ factor in ${\cal Z}'$, with generator
\begin{align}
    {\bf v}_1 \simeq (0;1;1;1;1 \ | \ e^{i\pi /2} ; 1) \in \bbZ_2 \times \bbZ_4^2 \times \bbZ_2^2 \times U(1)_1 \times U(1)_2 \, .
\end{align}
Moreover, since ${\bf v}_2$ has an order-4 prong that is not on the O7$^+$, physicality conditions do not change the fact that we obtain another $\bbZ_4$ factor in ${\cal Z}'$, now with generator
\begin{align}
    {\bf v}_2 \simeq (1;1;0;0;0 \ | \ 1 ; e^{i\pi/2} ) \in \bbZ_2 \times \bbZ_4^2 \times \bbZ_2^2 \times U(1)_1 \times U(1)_2 \, .
\end{align}
To summarize, the global gauge group of this rank $(2,10)$ model is
\begin{equation}
    \frac{\left[\left(Sp(2) \times SU(2)^{2} \times SU(4)^{2}\right) /\left(\mathbb{Z}_{2} \times \mathbb{Z}_{2}\right)\right] \times U(1)^{2}}{\mathbb{Z}_{4} \times \mathbb{Z}_{4}} \, ,
\end{equation}
agreeing with the CHL result computed in \cite{Cvetic:2021sjm}.

\subsubsection{A rank \boldmath{$(2,2)$} example}\label{subsubsec:rank4}

The rank $(2,2)$ moduli branch has six special points with non-Abelian symmetry enhancements \cite{Hamada:2021bbz}.
We have enumerated the 7-brane configurations for all of these, as well as the resulting gauge group topologies in Appendix \ref{subapdx:rank4_catalog}. 

It turns out that there is only one whose non-Abelian gauge group is non-simply-connected.
For illustration, we consider this example in more detail.
The brane configuration is
\begin{equation}\label{eq:rank_2_example_config}
     {\bf O7}^+ {{\bf O7}^+}' \bB^2 \bC^2 \, ,
\end{equation}
where the monodromy of the ${{\bf O7}^+}'$ is $SL(2,\bbZ)$ conjugated to be $\left( \begin{smallmatrix} 7 & 16 \\ -4 & -9 \end{smallmatrix} \right)$.\footnote{This can be viewed as freezing both $\mathfrak{so}_{16}$ algebras of the rank $(2,18)$ configuration $(\bA^8 \bB \bC) (\bX_{[2, -1]}^8 \bB \bX_{[3, -1]}) \bB^2 \bC^2$.}
Notice that, since the extended weights of O7$^+$ generate all $\colpq{p}{q}$-charges, whose parity is invariant under $SL(2,\bbZ)$ conjugation, we use the canonical basis $\boldsymbol\omega_{p,q}^{{\text{O7}^+}'}$ for the ${{\bf O7}^+}'$, which have $\colpq{1}{0}$ and $\colpq{0}{1}$ prongs, respectively.

The non-Abelian gauge algebra of \eqref{eq:rank_2_example_config} is $\mathfrak{su}_2 \oplus \mathfrak{su}_2$, with extended weights
\begin{align}
    \begin{split}
        \mathfrak{su}_2 \quad (\bB^2) : & \quad \boldsymbol\omega_{(1,-1)} \equiv \boldsymbol\omega_b = \tfrac12 (\bb_1 + \bb_2) \, , \\
        \mathfrak{su}_2 \quad (\bC^2) : & \quad \boldsymbol\omega_{(1,1)} \equiv \boldsymbol\omega_c = \tfrac12 (\bc_1 + \bc_2) \, .
    \end{split}
\end{align}
The physical 5-brane null junctions $J_\text{phys}^{N,\text{mag}}$ (see \eqref{eq:null_junction_gens_rank_2}) are then
\begin{align}
    \begin{split}
        \boldsymbol\delta^N_{(\nicefrac12, \nicefrac12)} &= \boldsymbol\omega_p^{\text{O7}^+} - \boldsymbol\omega_q^{\text{O7}^+} + \boldsymbol\omega_p^{{\text{O7}^+}'} - \boldsymbol\omega_q^{{\text{O7}^+}'} - 2 \boldsymbol\omega_b \, , \\
        \boldsymbol\delta^N_{(\nicefrac12, -\nicefrac12)} &= -3\boldsymbol\omega_p^{\text{O7}^+} + \boldsymbol\omega_q^{\text{O7}^+} - 7\boldsymbol\omega_p^{{\text{O7}^+}'} + 5\boldsymbol\omega_q^{{\text{O7}^+}'} + 8\boldsymbol\omega_b + 2 \boldsymbol\omega_c\, ,
    \end{split}
\end{align}
from which we find that the fractional null junctions $J_\text{frac}^{N,\text{mag}}$ is generated by
\begin{align}
    \tfrac12 \big( \boldsymbol\delta^N_{(\nicefrac12, \nicefrac12)} + \boldsymbol\delta^N_{(\nicefrac12, -\nicefrac12)} \big) = \boldsymbol\delta^N_{(\nicefrac12, 0)} = -\boldsymbol\omega^{\text{O7}^+}_p - 3\boldsymbol\omega^{{\text{O7}^+}'}_p + 2\boldsymbol\omega^{{\text{O7}^+}'}_q + 3\boldsymbol\omega_b + \boldsymbol\omega_c \, .
\end{align}
It corresponds to the generator
\begin{align}
    (3 \ \text{mod 2}, 1 \ \text{mod 2}) = (1,1) \in \bbZ_2 \times \bbZ_2 = Z(SU(2) \times SU(2))
\end{align}
of ${\cal Z} = J_\text{frac}^{N,\text{mag}} / J_\text{phys}^{N,\text{mag}} = \bbZ_2$.

Together with the null junctions, the Abelian junction lattice $J_A^\text{mag}$ is generated by
\begin{align}
    {\bf u}_1 = 2{\bf v}_1 = 2 \big( -\boldsymbol\omega_p^{{\text{O7}^+}} + \boldsymbol\omega_q^{{\text{O7}^+}} + \boldsymbol\omega_b \big) \, , \quad {\bf u}_2 = 2{\bf v}_2 = 2 \big( -\boldsymbol\omega_p^{{\text{O7}^+}} - \boldsymbol\omega_q^{{\text{O7}^+}} + \boldsymbol\omega_c \big) \, .
\end{align}
Clearly, this leads to ${\cal Z}' = \bbZ_2 \times \bbZ_2$, with generators
\begin{align}
    \left. \begin{array}{c}
         {\bf v}_1 \simeq (1;0 \ | \ -1 ; 1) \\
         {\bf v}_2 \simeq (0;1 \ | \ 1 ; -1)
    \end{array} \right\}
    \in \bbZ_2 \times \bbZ_2 \times U(1)_1 \times U(1)_2 = Z\big(SU(2)_{\bB} \times SU(2)_{\bC} \times U(1)_1 \times U(1)_2  \big) \, .
\end{align}
The full gauge group is thus
\begin{align}
    \frac{ [SU(2) \times SU(2)]/\bbZ_2 \times U(1)^2}{\bbZ_2 \times \bbZ_2} \, .
\end{align}

\section{9d vacua via affine 7-brane stacks}\label{sec:9d}

Recently, it was argued that one can recover any 9d ${\cal N}=1$ string vacuum with gauge rank $(1,17)$ from F-theory on a suitably degenerated K3 geometry that lies at infinite distance in the complex structure moduli space \cite{Lee:2021qkx,Lee:2021usk}.
As shown in these works, such decompactification limits have a particularly convenient description in terms of $[p,q]$-7-branes and junctions that realize affine algebras.
In the following, we demonstrate how the methods from the previous sections naturally apply also to these limiting configurations, and compute the 9d gauge group topologies for rank $(1,17)$ vacua.

Since the affinization is characterized entirely by the $SL(2,\bbZ)$ monodromy, a natural proposition is that these configurations also describe 9d uplifts when we include O7$^+$-planes.
Indeed, (after resolving an ambiguity by string dualities) this straightforwardly reproduces the landscape of 9d rank $(1,9)$ vacua \cite{Font:2021uyw}, including their global gauge group structures.
Moreover, applying the same reasoning to configurations with two O7$^+$'s, we consistently find two branches of 9d rank $(1,1)$ vacua \cite{Aharony:2007du}, which are only connected through circle-reductions to 8d.

The key ingredient that enters the description for all ranks are 7-brane stacks realizing an \emph{affine} Lie algebra $\widehat{\mathfrak{e}}_k$, which we will now briefly recall.

As found in \cite{DeWolfe:1998yf,DeWolfe:1998eu,DeWolfe:1998pr}, the $\mathfrak{e}_n$ and $\tilde{\mathfrak{e}}_n$ algebras can be enhanced to their affine versions, by including a specific 7-brane on top:
\begin{align}\label{eq:affine_stacks}
\begin{split}
    \widehat{{\bf E}}_{n \geq 1} = & \underbrace{ {\bf A}^{n-1} {\bf B} {\bf C}^2}_{{\bf E}_n} {\bf X}_{[3,1]} = {\bf A}^{n-1} {\bf B} {\bf C} {\bf B} {\bf C} \, , \\
    \widehat{\tilde{{\bf E}}}_{n \geq 0} = & \underbrace{ {\bf A}^{n} {\bf X}_{[2,-1]} {\bf C}}_{\tilde{\bf E}_n} {\bf X}_{[4,1]} \, .
\end{split}
\end{align}
Note that for $n\geq 2$, these are equivalent up to 7-brane moves and $SL(2,\bbZ)$ conjugations \cite{DeWolfe:1998eu}.
It is straightforward to check that, in this $SL(2,\bbZ)$-frame, they have monodromy
\begin{align}
    M(\widehat{\bf E}_n) = M(\widehat{\tilde{\bf E}}_n) = \begin{pmatrix}
    1 & 9 - n \\ 0 & 1
    \end{pmatrix} \, .
\end{align}
The hallmark of these stacks is the existence of a special loop junction $\boldsymbol{\delta}_{(1,0)} \equiv \boldsymbol\delta$ around them (with no asymptotic charge), satisfying $(\boldsymbol{\delta}, \boldsymbol{\delta}) = (\boldsymbol{\delta}, \boldsymbol{\alpha}_i) = 0$, with $\boldsymbol\alpha_i$ the root junctions of ${\bf E}_n$ or $\tilde{\bf E}_n$.
By a Hanany--Witten transition, one finds the equivalent presentation \cite{DeWolfe:1998pr}
\begin{align}
    \boldsymbol\delta_{\widehat{\text{E}}} =  {\bf x}_{[3,1]} - {\bf b} - {\bf c}_1 - {\bf c}_2= {\bf b}_2 + {\bf c}_2 - {\bf b}_1 - {\bf c}_1 \, , \quad \boldsymbol\delta_{\widehat{\tilde{\text{E}}}} = {\bf x}_{[4,1]} - 2{\bf c} - {\bf x}_{[2,-1]} \, .
\end{align}

Representation theoretically, $\boldsymbol\delta$ plays the role of the imaginary root required for the affinization of $\mathfrak{e}_n$ or $\tilde{\mathfrak{e}}_n$.
They generate an infinite dimensional Kac--Moody algebra with roots $\{\boldsymbol\alpha + k \boldsymbol\delta \, | \, k \in \bbZ\}$, where $\boldsymbol\alpha$ is any root of $\mathfrak{e}_n$ or $\tilde{\mathfrak{e}}_n$.
When we seperate the affinizing $\bX$-branes in \eqref{eq:affine_stacks} from the $\mathfrak{e}_n$ or $\tilde{\mathfrak{e}}_n$ stacks, these junctions, as strings, give rise to BPS states with masses proportional to $k$.
In the affine limit, we thus obtain an infinite tower of massless BPS states.
Physically, string junctions of these type give rise to an infinite tower of massless BPS states.

A special extension exists for $n=8$.
Here, by adding an ${\bf A}$-brane from the left to the $\widehat{\bf E}_8$ or $\widehat{\tilde{\bf E}}_8$, the monodromy becomes $\mathbb{1} \equiv M(\widehat{\bf E}_9) = M(\widehat{\tilde{\bf E}}_9)$.
This would give rise to two independent towers of massless BPS states from loops of $\colpq{1}{0}$ and $\colpq{0}{1}$ string junctions, which lead to the double loop enhancement of $\mathfrak{e}_8$.
These special enhancements reflect a decompactification to 10d \cite{Lee:2021qkx,Lee:2021usk}.
For discussions of 9d vacua, we will not consider such double loop brane-stacks, but the necessary constituent branes form one half of the rank $(2,18)$ configuration \eqref{eq:parent_brane_config} that correspond geometrically to the singular fibers of a $dP_9$ surface.

Among the various types of infinite distance limits of F-theory compactified on K3 surfaces, those describing decompactification from 8d to 9d are captured by so-called Kulikov models of type III.a \cite{Lee:2021qkx,Lee:2021usk}.
In these geometries, the complex structure moduli have been tuned such that the K3 degenerates into a collection of intersecting elliptic and/or rational surfaces.
While we refer to those references for details, the relevant fact about these deformations is that they correspond to brane motions which generate one or two 7-brane stacks carrying an $\widehat{\mathfrak{e}}_n$ or $\widehat{\tilde{\mathfrak{e}}}_n$ algebra (with $n \leq 8$).
The tower of massless states from the imaginary root may then be identified with the momentum states of a Kaluza--Klein (KK) tower on a circle whose size becomes infinite at the infinite distance limit.
In the case with two affine stacks, the individual imaginary root junctions turn out to be identical in the global setting, consistent with having just one KK-tower \cite{Lee:2021usk}.

\subsubsection{Global structure of 9d vacua of rank 17}

As for the classification of 8d vacua, one can also categorize all brane configurations with such affine stacks.
Then, if the non-Abelian brane stacks correspond to the algebra $\mathfrak{h} \oplus \widehat{{\cal E}}_n$ or $\mathfrak{h} \oplus \widehat{{\cal E}}_n \oplus \widehat{{\cal E}}_m$ (where ${\cal E} = \mathfrak{e}$ or $\tilde{\mathfrak{e}}$) for some finite semi-simple, simply-laced algebra $\mathfrak{h}$, the associated non-Abelian gauge algebra in 9d is $\mathfrak{h} \oplus {\cal E}_n$ or $\mathfrak{h} \oplus {\cal E}_n \oplus {\cal E}_m$, respectively \cite{Lee:2021qkx,Lee:2021usk}.
This reproduces, e.g., all the maximally enhanced non-Abelian algebras (i.e., with rank 17) determined in the dual heterotic frame \cite{Font:2020rsk}.

To also analyze the gauge group topologies in this description, we need to examine the full junction lattice, including the branes away from the affine stack.
An important detail here is that the overall gauge rank reduces by 2 as we decompactify from 8d to 9d, corresponding to the re-interpretation of the KK-states (which become massless) and the decoupling of the winding states (which become infinitely heavy) as we increase the size of the compactification circle.
In the momentum lattice description of the 8d and 9d theories of rank $(2,18)$ and $(1,17)$, respective, we have
\begin{align}\label{eq:affinization_lattice}
    \Lambda^\text{het}_{\text{8d}} = \Lambda^\text{het}_{\text{9d}} \oplus U \quad \Rightarrow \quad \Lambda^\text{het}_{\text{9d}} \cong \Lambda^\text{het}_{\text{8d}} / U \, ,
\end{align}
with $U$ the rank 2 hyperbolic lattice that is spanned by the KK and winding states.
Since the momentum lattice is equivalently described by junctions $J_\text{phys}^{\text{el}} = J_\text{phys}^{\text{mag}} \cong \Lambda_\text{8d}^\text{het} \oplus J_\text{phys}^{N}$, with the KK-tower being generated by the junction $\boldsymbol\delta$, there must exist another non-null junction $\boldsymbol\epsilon$ that generates this $U$ factor with $\boldsymbol\delta$, i.e., satisfying
\begin{align}
    (\boldsymbol\delta, \boldsymbol\epsilon) = 1 \, , \quad (\boldsymbol\delta, \boldsymbol\delta) = (\boldsymbol\epsilon, \boldsymbol\epsilon) = (\boldsymbol\delta, {\bf j}) = (\boldsymbol\epsilon, {\bf j}) = 0 \, ,
\end{align}
for ${\bf j}$ any (co-)weight or (co-)root junction, or a non-null generator of the Abelian junctions $J_A$.
Such a $\boldsymbol\epsilon$-junction always exists, but the details depend on the specific configuration.

Since the junction lattice reproduces the 9d momentum lattice, it must also encode the global structure of the gauge group.
In particular, it allows us to use the intuition in terms of fractional null junctions to re-derive the results of \cite{Font:2020rsk}.
Let us demonstrate this for 9d models with maximally enhanced non-Abelian symmetries, for which there are two classes of 8d brane configurations \cite{Lee:2021usk}.

In the first class, the non-Abelian algebra (with the place holder ${\cal E} = \mathfrak{e}$ or $\tilde{\mathfrak{e}}$) is
\begin{equation}
\fkg_{\text{8d}, \infty} = \mathfrak{su}_{18 - m - n} \oplus \widehat{{\cal E}}_m \oplus \widehat{{\cal E}}_{n} \quad \Rightarrow \quad \fkg_\text{9d} = \mathfrak{su}_{18 - m - n} \oplus \mathcal{E}_m \oplus \mathcal{E}_n, \quad m, n \in \{0, 1, 3, \dots, 8\} \, ,
\end{equation}
whose brane configurations (together with the $U$-lattice generators) are
\begin{subequations}\label{eq:decomp_series_A}
\begin{align}
\begin{split}
    m \geq n \geq 1:& \quad \bA^{18 - m - n} (\overbrace{\bA^{m-1} \bX_{[n-10, 1]} \bX_{[n-8, 1]}^2 \bX_{[n-6, 1]}}^{\widehat{{\bf E}}_m}) (\overbrace{\bA^{n-1} \bB \bC^2 \bX_{[3, 1]}}^{\widehat{{\bf E}}_n}) \, , \\
    & \quad (\boldsymbol\delta, \boldsymbol\epsilon) = (\boldsymbol\delta^R_{(1,0)},\ \ (n-5)\boldsymbol\delta^R_{(1,0)} + {\boldsymbol\ell}^R_{(0, 1)} + \bx_{[n-6, 1]} - \bx_{[3, 1]} ) \, ,
\end{split} \label{eqn:2dP9_a}\\
\begin{split}
    m = 1, n = 0:& \quad \bA^{17}(\overbrace{\bX_{[10, -1]} \bX_{[8, -1]}^2 \bX_{[6, -1]}}^{\widehat{{\bf E}}_1}) (\overbrace{\bX_{[2, -1]} \bC \bX_{[4, 1]}}^{\widehat{\tilde{{\bf E}}}_0}) \, , \\
    & \quad (\boldsymbol\delta, \boldsymbol\epsilon) = (\boldsymbol\delta^R_{(1,0)},\ \ -17 \boldsymbol\omega_A + \bx_{[10, -1]} + 2\bx_{[4, 1]} - \bc) \, ,
\end{split} \label{eqn:2dP9_b} \\
\begin{split}
    m > n = 0;\ m \neq 1:& \quad \bA^{18-m} (\overbrace{ \bA^m \bX_{[11, -1]} \bX_{[8, -1]} \bX_{[5, -1]}}^{\widehat{\tilde{{\bf E}}}_{m \neq 1}}) (\overbrace{\bX_{[2, -1]} \bC \bX_{[4, 1]}}^{\widehat{\tilde{\bf E}}_0}) \, ,  \\
    & \quad (\boldsymbol\delta, \boldsymbol\epsilon) = (\boldsymbol\delta^R_{(1,0)},\ \ -\boldsymbol\delta^R_{(1,0)} + \bx_{[5, -1]} - 2\bx_{[2,-1]} - \bc) \, .
\end{split}\label{eqn:2dP9_c}
\end{align}
\end{subequations}
where $\boldsymbol\delta^R$ and $\boldsymbol\ell^R$ are loop junction that encircle counterclockwise around the second affine stack.
Note that, because the $\colpq{1}{0}$ loop is mutually-local with respect to the ${\bf A}$-branes, it is evident that, by pulling $\boldsymbol\delta_{(1,0)}$ across these, one obtains a loop junction around the other affine stack, showing explicitly that imaginary roots of each affine stack are identical.

The second class has non-Abelian gauge algebras
\begin{equation}
    \fkg_{\text{8d}, \infty} = \mathfrak{so}_{34-2k} \oplus \widehat{{\cal E}}_{k} \quad \Rightarrow \quad \fkg_\text{9d} = \mathfrak{so}_{34-2k} \oplus {\cal E}_k \, , \quad 0 \leq k \leq 8 , k\neq 2 \, , 
\end{equation}
whose brane configurations (and $U$-lattice generators) are
\begin{subequations}\label{eq:decomp_series_D}
\begin{align}
\begin{split}
    k = 1, 3, \dots, 8: & \quad ( \overbrace{\bA^{17 - k} \bX_{[k - 10, 1]} \bX_{[k - 8, 1]}}^{\textbf{D}_{17-k}}) \bX_{[k-8, 1]} (\overbrace{\bA^{k - 1} \bB \bC^2 \bX^{(1)}_{[3, 1]}}^{\widehat{\textbf{E}}_k}) \bX^{(2)}_{[3, 1]} \, , \\
    & \quad (\boldsymbol\delta, \boldsymbol\epsilon) = ( \boldsymbol\delta^R_{(1,0)},\ \ \boldsymbol\delta^R_{(1,0)} + \bx^{(1)}_{[3, 1]} - \bx^{(2)}_{[3, 1]}) \label{eqn:1dP9_a} 
\end{split}\\
\begin{split}
    k = 0 : & \quad (\overbrace{\bA^{17} \bX_{[10, -1]} \bX^{(1)}_{[8, -1]}}^{\textbf{D}_{17}}) \bX^{(2)}_{[8, -1]} (\overbrace{\bX_{[2, -1]} \bC \bX_{[4, 1]}}^{\widehat{\tilde{\textbf{E}}}_0}) \bX_{[3, 1]}  \label{eqn:1dP9_b} \, , \\
    & \quad (\boldsymbol\delta, \boldsymbol\epsilon) = (\boldsymbol\delta^R_{(1,0)},\ \ -2\boldsymbol\delta^R_{(1,0)} + \bx^{(2)}_{[8, -1]}  - 3\bx_{[2,-1]} - 2\bx_{[1,1]}) \, ,
\end{split}
\end{align}
\end{subequations}
again, with the imaginary root junction $\boldsymbol\delta^R_{(1,0)}$ being the $\colpq{1}{0}$-loop around the affine stack to the right.

By separating the ${\bf X}$-brane responsible for the affinization from each affine stack, we obtain a genuine 8d configuration.
For these configurations, we can apply the same procedure as in the previous section, and construct the global fractional null junctions that encode to the cocharacters of the 8d gauge symmetry.
Since the affinization \eqref{eq:affinization_lattice} mods out by physical junctions that are orthogonal to the root lattices of the 9d gauge factors, it does not affect the coefficients of global null junctions in front of the extended weights.
Therefore, the fractional null junctions are the same for the 8d configuration as for its affinized version.

As a concrete example, consider \eqref{eqn:1dP9_a} with $k=7$, which in 9d gives rise to $\fkg_{9d} = \mathfrak{so}_{20} \oplus \mathfrak{e}_7$.
Separating the singlet brane responsible for the affinization,
\begin{align}
    (\underbrace{\bA^{10} \bX_{[-3,1]} \bX_{[-1,1]}}_{\textbf{D}_{10}'}) \, \bX_{[-1,1]} \, (\underbrace{\bA^6 \bB \bC^2}_{\textbf{E}_7}) \, \bX^{(1)}_{[3,1]} \, \bX^{(2)}_{[3,1]} \ ,
\end{align}
we find the $\mathfrak{so}_{20}$-stack to have monodromy $\left( \begin{smallmatrix} -1 & 6 \\ 0 & -1 \end{smallmatrix} \right)$ in this $SL(2,\bbZ)$ frame, with extended weight junctions $\boldsymbol\omega'_{p,q}$ carrying asymptotic $\colpq{p}{q}$-charges as follows:
\begin{align}
    \boldsymbol\omega_p': \, \, \colpq{1}{0} \, , \qquad \boldsymbol\omega_q' : \, \, \colpq{-2}{1} \, .
\end{align}
The two global null junctions $\boldsymbol\delta^N_{(1,0)}$ and $\boldsymbol\delta^N_{(0,1)}$ can be then expressed in terms of the extended weights $\boldsymbol\omega'_{p,q}$ of $\mathfrak{so}_{20}$ and $\boldsymbol\omega_{p,q}$ of $\mathfrak{e}_7$ as
\begin{align}
    \begin{split}
        & \boldsymbol\delta^{N}_{(1,0)} = -2 \boldsymbol\omega_p' - \bx_{[-1,1]} - 5\boldsymbol\omega_p - \boldsymbol\omega_q + \bx^{(1)}_{[3,1]} + \bx^{(2)}_{[3,1]} \, , \\
        & \boldsymbol\delta^{N}_{(0,1)} = 2 \boldsymbol\omega_p' - 2 \boldsymbol\omega_q' + 5 \bx_{[-1,1]} + 17 \boldsymbol\omega_p + 3 \boldsymbol\omega_q + \bx^{(1)}_{[3,1]} + \bx^{(2)}_{[3,1]} \, .
    \end{split}
\end{align}
It is easy to see that the fractional null junctions are then multiples of
\begin{align}
    \tfrac12 (\boldsymbol\delta^N_{(1,0)} + \boldsymbol\delta^N_{(0,1)}) = - \boldsymbol\omega_q' + 2 \bx_{[-1,1]} + 6 \boldsymbol\omega_p + \boldsymbol\omega_q + \bx_{[3,1]}^{(1)} + \bx_{[3,1]}^{(2)} \, .
\end{align}
By \eqref{eq:center_charge_table_ADE}, this corresponds to the central element
\begin{align}
    (0,1;1) \in \bbZ_2^2 \times \bbZ_2 = Z(Spin(20) \times E_7),
\end{align}
which leads to the 9d non-Abelian gauge group $[Spin(20) \times E_7]/\bbZ_2$.

To determine the full gauge group, including the gravi-photon $U(1)$, we must first find the non-null generator of the Abelian junction lattice that is orthogonal to the $U$-lattice spanned by $(\boldsymbol\delta, \boldsymbol\epsilon)$.
In this example, it can be easily determined (by avoiding prongs on the $\mathfrak{e}_7$ stack or the $\bX_{[3,1]}$ branes),
\begin{align}
    {\bf u} = 2{\bf v} = 2( \boldsymbol\omega_p' + \boldsymbol\omega_q' - \bx_{[-1,1]})\, ,
\end{align}
which immediately gives ${\cal Z}' = \bbZ_2$, with generator
\begin{align}
    {\bf v} \simeq (1,1;0 \ | \ e^{i \pi} ) \in Z(Spin(20) \times E_7 \times U(1)) \, .
\end{align}
Therefore, the 9d gauge group is
\begin{align}
    \frac{[Spin(20) \times E_7]/\bbZ_2 \times U(1)}{\bbZ_2} \, .
\end{align}

By analogous computations, we compute the non-Abelian gauge groups of all models with maximally enhanced non-Abelian symmetry (summarized in Table \ref{table:9Drank17}), which agree with results from the heterotic picture \cite{Font:2020rsk}.

\begin{table}[ht!]
\renewcommand{\arraystretch}{1.3}
\centering
\begin{tabular}{|c|c||c|c|}
    \hline
    $(\fkg_\text{9d}, \pi_1(G_\text{9d}))$  & $\text{FNJ}$ & $(\fkg_\text{9d}, \pi_1(G_\text{9d}))$  & $\text{FNJ}$  \\ \hline \hline
       \multicolumn{4}{|c|}{$\fkg_{\text{8d}, \infty} = \mathfrak{su}_{18 - m - n} \oplus \widehat{\mathfrak{e}}_m \oplus \widehat{\mathfrak{e}}_n$}  \\ \hline
       $(\mathfrak{e}_8 \oplus \mathfrak{e}_8 \oplus \mathfrak{su}_2, -)$    & -  &        $(\mathfrak{e}_6 \oplus \mathfrak{su}_3 \oplus \mathfrak{su}_2 \oplus \mathfrak{su}_9 , \bbZ_3)$ & $\boldsymbol\delta^N_{(0,1)}/3$ \\
       $(\mathfrak{e}_8 \oplus \mathfrak{e}_7 \oplus \mathfrak{su}_3, -)$  & - &        $(\mathfrak{e}_6 \oplus \mathfrak{su}_2 \oplus \mathfrak{su}_{11}, -)$ & - \\
       $(\mathfrak{e}_8 \oplus \mathfrak{e}_6 \oplus \mathfrak{su}_4, -)$  & - &        $(\mathfrak{e}_6 \oplus \mathfrak{su}_{12}, \bbZ_{3})$ & $\boldsymbol\delta^N_{(0,1)}/3$ \\
       $(\mathfrak{e}_8 \oplus \mathfrak{so}_{10} \oplus \mathfrak{su}_5, -)$  & - &        $(\mathfrak{so}_{10} \oplus \mathfrak{so}_{10} \oplus \mathfrak{su}_8, \bbZ_4)$ & $\boldsymbol\delta^N_{(1,1)}/4$ \\
       $(\mathfrak{e}_8 \oplus \mathfrak{su}_5 \oplus \mathfrak{su}_6, -)$  & -  &        $(\mathfrak{so}_{10} \oplus \mathfrak{su}_5 \oplus \mathfrak{su}_9, -)$ & - \\
       $(\mathfrak{e}_8 \oplus \mathfrak{su}_3 \oplus \mathfrak{su}_2 \oplus \mathfrak{su}_7, -)$  & - &        $(\mathfrak{so}_{10} \oplus \mathfrak{su}_3 \oplus \mathfrak{su}_2 \oplus \mathfrak{su}_{10} , \bbZ_2)$ & $\boldsymbol\delta^N_{(1,-1)}/2$ \\
       $(\mathfrak{e}_8 \oplus \mathfrak{su}_9 \oplus \mathfrak{su}_2, -)$  & - &        $(\mathfrak{so}_{10} \oplus \mathfrak{su}_{2} \oplus \mathfrak{su}_{12}, \bbZ_4)$ & $\boldsymbol\delta^N_{(1,1)}/4$ \\
       $(\mathfrak{e}_8 \oplus \mathfrak{su}_{10}, -)$  & - &        $(\mathfrak{so}_{10} \oplus \mathfrak{su}_{13}, -)$ & - \\
       $(\mathfrak{e}_7 \oplus \mathfrak{e}_7 \oplus \mathfrak{su}_4, \bbZ_2)$   & $\boldsymbol\delta^N_{(1,1)}/2$ &        $(\mathfrak{su}_5 \oplus \mathfrak{su}_5 \oplus \mathfrak{su}_{10}, \bbZ_5)$ & $\boldsymbol\delta^N_{(1,2)}/5$ \\       
       $(\mathfrak{e}_7 \oplus \mathfrak{e}_6 \oplus \mathfrak{su}_5, -)$   & -  &        $(\mathfrak{su}_5 \oplus \mathfrak{su}_3 \oplus \mathfrak{su}_2 \oplus \mathfrak{su}_{11}, -)$ & - \\ 
       $(\mathfrak{e}_7 \oplus \mathfrak{so}_{10} \oplus \mathfrak{su}_6, \bbZ_2)$  & $\boldsymbol\delta^N_{(1,-1)}/2$ &        $(\mathfrak{su}_5 \oplus \mathfrak{su}_{2} \oplus \mathfrak{su}_{13}, -)$ & - \\      
       $(\mathfrak{e}_7 \oplus \mathfrak{su}_5 \oplus \mathfrak{su}_7, -)$ & - &        $(\mathfrak{su}_5 \oplus \mathfrak{su}_{14}, -)$ & - \\
       $(\mathfrak{e}_7 \oplus \mathfrak{su}_3 \oplus \mathfrak{su}_2 \oplus \mathfrak{su}_8, \bbZ_2)$  & $\boldsymbol\delta^N_{(1,1)}/2$ &        $((\mathfrak{su}_3 \oplus \mathfrak{su}_2) \oplus (\mathfrak{su}_3 \oplus \mathfrak{su}_2) \oplus \mathfrak{su}_{12}, \bbZ_6)$ & $\boldsymbol\delta^N_{(3,1)}/6$ \\
       $(\mathfrak{e}_7 \oplus \mathfrak{su}_2 \oplus \mathfrak{su}_{10}, \bbZ_2)$  & $\boldsymbol\delta^N_{(1,1)}/2$ &        $((\mathfrak{su}_3 \oplus \mathfrak{su}_2) \oplus \mathfrak{su}_{2} \oplus \mathfrak{su}_{14}, \bbZ_2)$ & $\boldsymbol\delta^N_{(1,-1)}/2$ \\
       $(\mathfrak{e}_7 \oplus \mathfrak{su}_{11}, -)$  & - &        $((\mathfrak{su}_3 \oplus \mathfrak{su}_2) \oplus \mathfrak{su}_{15}, \bbZ_3$ & $\boldsymbol\delta^N_{(0,1)}/3$ \\
       $(\mathfrak{e}_6 \oplus \mathfrak{e}_6 \oplus \mathfrak{su}_6, \bbZ_3)$ & $\boldsymbol\delta^N_{(0,1)}/3$ &        $(\mathfrak{su}_2 \oplus \mathfrak{su}_{2} \oplus \mathfrak{su}_{16}, \bbZ_4)$ & $\boldsymbol\delta^N_{(1,-1)}/4$  \\        
       $(\mathfrak{e}_6 \oplus \mathfrak{so}_{10} \oplus \mathfrak{su}_7, -)$ & - &        $(\mathfrak{su}_2 \oplus \mathfrak{su}_{17}, -)$ & - \\
       $(\mathfrak{e}_6 \oplus \mathfrak{su}_5 \oplus \mathfrak{su}_8, -)$ & -  &        $(\mathfrak{su}_{18}, \bbZ_3)$ & $\boldsymbol\delta^N_{(1,1)}/3$ \\
     \hline\hline
       \multicolumn{4}{|c|}{$\fkg_{\text{8d}, \infty} = \mathfrak{so}_{34 - 2k} \oplus \widehat{\mathfrak{e}}_k$}  \\ \hline
       $(\mathfrak{e}_8 \oplus \mathfrak{so}_{18}, -)$   & - & $(\mathfrak{su}_5 \oplus \mathfrak{so}_{26}, -)$ & - \\
       $(\mathfrak{e}_7 \oplus \mathfrak{so}_{20}, \bbZ_2)$  & $\boldsymbol\delta^N_{(1,1)}/2$ & $((\mathfrak{su}_3 \oplus \mathfrak{su}_2) \oplus \mathfrak{so}_{28}, \bbZ_2)$ & $\boldsymbol\delta^N_{(1,1)}/2$ \\
       $(\mathfrak{e}_6 \oplus \mathfrak{so}_{22}, -)$ & - & $(\mathfrak{su}_2 \oplus \mathfrak{so}_{32}, \bbZ_2)$ & $\boldsymbol\delta^N_{(1,1)}/2$ \\ 
       $(\mathfrak{so}_{10} \oplus \mathfrak{so}_{24}, \bbZ_2)$ & $\boldsymbol\delta^N_{(1,1)}/2$ & $(\mathfrak{so}_{34}, -)$ & - \\ \hline
\end{tabular}
\caption{Non-Abelian gauge group $G_\text{9d}$ of all maximally-enhanced 9d rank $(1,17)$ string vacua, seen as dimensional uplifts of 8d string junction vacua.
The generator of $\pi_1(G_\text{9d}) \cong \bbZ_{\ell}$ is represented as a fractional null junction (FNJ) $\boldsymbol\delta^N_{(p,q)} / \ell = \boldsymbol\delta^N_{(\nicefrac{p}{\ell}, \nicefrac{q}{\ell})}$.
}
\label{table:9Drank17}
\end{table}

\subsubsection{9d CHL vacua via junctions}

Having reproduced the maximal rank branch of the 9d moduli space, we would like to extend the junction method also to rank-reduced theories.
We start by matching the known circle compactification of the 9d CHL string in terms of junctions in the presence of a single O7$^+$-plane, focusing again on the cases with maximal non-Abelian gauge rank.

A key assumption here is that the decompactification limit of 8d vacua, even in the presence of O7$^+$-planes, is characterized by the appearance of singularities in the axio-dilaton profile that induce $SL(2,\bbZ)$ monodromy of affine type.
Though we do not have a proof for this, we expect the identification of the resulting loop junctions as the only BPS-tower compatible with decompactification to be valid also with O7$^+$-planes, given that the loop can be thought of as a $(p,q)$-string that is only sensitive to the monodromy, but not the details of the 7-branes.
Moreover, as we will see below, the results following this assumption agree with the momentum lattice description for the 8d and 9d CHL string \cite{Mikhailov:1998si,Font:2021uyw}.

Analogous to the procedure in previous sections of describing the O7$^+$ as freezing a $\mathfrak{so}_{16}$ stack in an ``ordinary'' rank $(2,18)$ setting, we therefore focus on those brane configurations in \eqref{eq:decomp_series_A} and \eqref{eq:decomp_series_D}, whose non-Abelian stack can host a $\mathfrak{so}_{16}$.
This is only possible if the configuration includes a ${\bf D}_{n \geq 8}$ or $\widehat{\textbf{E}}_8$ brane stack.
While for \eqref{eq:decomp_series_A}, there is only one rank $(2,18)$ configuration, with $\fkg_{\text{8d},\infty} = \mathfrak{su}_2 \oplus \widehat{\mathfrak{e}}_8 \oplus \widehat{\mathfrak{e}}_8$, there is an ambiguity for the class \eqref{eq:decomp_series_D}, in that we can naively embed the O7$^+$ inside the $\widehat{\mathfrak{e}}_8$ or the $\mathfrak{so}$ stack.
However, inspecting the set of allowed string and 5-brane junctions reveals a striking difference between the two options.

If we embed the O7$^+$ inside the $\mathfrak{so}$-stack, the freezing of the $\mathfrak{so}_{16}$ subalgebra and the modified boundary conditions for the junctions do not affect the $U$ lattice.
This is made explicit in \eqref{eq:decomp_series_D}, since $\boldsymbol\delta$ and $\boldsymbol\epsilon$ junctions only have prongs on the affine stack, which remains unmodified.
On the other hand, if we would embed the O7$^+$ inside an $\widehat{\textbf{E}}_8$, then the freezing procedure restricts the set of allowed string junctions to be orthogonal to the $\mathfrak{so}_{16}$ roots, and have even prongs on the orientifold plane.
As we will explain in detail in Appendix \ref{app:O7_into_E8}, the result is that we can no longer consistently define a $U$-lattice from the allowed junctions.
Instead, the evenness condition can at most accommodate a stretched hyperbolic lattice $U(2)$.

Based on the dual CHL string description, we propose that only the embeddings with a \textit{modified} $U$ lattice gives a consistent 9d uplift.
Namely, unlike the maximal rank case, the momentum lattice $\Lambda^\text{CHL}_{\text{8d}} \cong (-\text{E}_8) \oplus U \oplus U(2)$ of 8d CHL vacua is no longer self-dual, whereas the corresponding 9d lattice $\Lambda^\text{CHL}_{\text{9d}} \cong (-\text{E}_8) \oplus U$ is \cite{Mikhailov:1998si}.
The additional $U(2)$ in 8d arises from the winding and KK-states of the CHL string, and must therefore be represented in terms of the imaginary root junction around the affine stack, and another string junction that emanates from it.
If we embed the O7$^+$ inside the $\textbf{D}_{17-k}$ stack of \eqref{eq:decomp_series_D} instead, we would have an unstretched $U$-lattice for winding and KK-states.
Moreover, if we would naively identify the would-be 9d gauge algebra with that of 8d (replacing the affine symmetry with its non-affine version), this kind of embedding would lead to an $\mathfrak{sp}$ algebra in 9d, which again is not compatible with the CHL string.
While these arguments provide strong evidence in favor of the proposal, we leave a rigorous proof for future works, and discuss the resulting characterization of 9d CHL vacua in terms of string junctions.

Let us start from the 8d rank $(2,18)$ configuration \eqref{eqn:2dP9_a} with $n=8$, which, if we moved $\bX_{[2,1]}$ from $\widehat{\bf E}_m$ across the branch cut of $\widehat{\bf E}_{n=8}$, becomes \eqref{eqn:1dP9_a} with $k=8$.
Using the brane moves described in Appendix \ref{app:O7_into_E8}, we can turn the $\widehat{\textbf{E}}_8$ into an $SL(2,\bbZ)$-conjugated $\textbf{E}_9$ stack:
\begin{align}
    {\bf A}^{10-m} (\overbrace{\bA^{m-1} \bX_{[-2, 1]} \bX_{[0, 1]}^2 \bX_{[2, 1]}}^{\widehat{\textbf{E}}_m}) \underbrace{\overbrace{ {\bf X}^8_{[0,1]} {\bf X}_{[1,4]} {\bf X}_{[1,2]} }^{\textbf{D}_8'} {\bf X}_{[1,2]}}_{\textbf{E}_9' \cong \widehat{\textbf{E}}_8} \, .
\end{align}
The $\widehat{\bf E}_m$ stack can be conjugated by $g = \left( \begin{smallmatrix} 1 & 1 \\ 0 & 1 \end{smallmatrix} \right)$ to obtain the standard form from Section \ref{sec:local_analysis}.
In particular, this means that the standard extended weight junctions now carry asymptotic $\colpq{p}{q}$ charge given as
\begin{align}
    \boldsymbol\omega^{\mathfrak{e}_m}_p : \ g^{-1}\colpq{1}{0} = \colpq{1}{0} \, , \quad \boldsymbol\omega^{\mathfrak{e}_m}_q : \ g^{-1}\colpq{0}{1} = \colpq{-1}{1} \, .
\end{align}
The $\textbf{D}_8'$ stack can be conjugated by $g' = \left( \begin{smallmatrix} 3 & -1 \\ 1 & 0 \end{smallmatrix} \right)$ to the standard representation, $g' M(\textbf{D}_8') {g'}^{-1} = M({\bf A}^8 {\bf B} {\bf C})$.
Introducing the O7$^+$, i.e., ${\bf X}^8_{[0,1]} {\bf X}_{[1,4]} {\bf X}_{[1,2]} \rightarrow {{\bf O7}^+}'$ (where we use the prime to denote the non-standard $SL(2,\bbZ)$-frame), we obtain
\begin{align}\label{eq:9d_CHL_brane_example}
    {\bf A}^{10-m} (\overbrace{ \underbrace{\bA^{m-1} \bX_{[-2, 1]} \bX_{[0, 1]}^2}_{{\bf E}_m} \bX_{[2, 1]}}^{\widehat{\textbf{E}}_m}) \underbrace{ {\bf O7}^{+'} {\bf X}_{[1,2]}}_{\sim \textbf{E}_9' \cong \widehat{\textbf{E}}_8} \, .
\end{align}
Compared to the standard presentations discussed in Section \ref{sec:local_analysis} (i.e., where the monodromy of O7$^+$ is $M({\bf O7}^+) = \left( \begin{smallmatrix} -1 & 4 \\ 0 & -1 \end{smallmatrix} \right)$), the O7$^+$ monodromy in this $SL(2,\bbZ)$ frame is
\begin{align}
    M({{\bf O7}^+}') = {g'}^{-1} M({\bf O7}^+) g' = \left( \begin{smallmatrix} -1 & 0 \\ -4 & -1 \end{smallmatrix} \right) ,
\end{align}
and the standard extended weights $\boldsymbol\omega_{p,q}^{\text{O7}^+}$ have asymptotic $\colpq{p}{q}$-charges
\begin{align}
    \boldsymbol\omega_p' : {g'}^{-1}\colpq{1}{0} = \colpq{0}{-1} \, , \quad \boldsymbol\omega_q' : {g'}^{-1}\colpq{0}{1} = \colpq{1}{3} \, ,
\end{align}
for which the pairing relations \eqref{eq:bilinext} hold.
We can pull the imaginary root junction $\boldsymbol\delta \equiv \boldsymbol\delta^R_{(1,0)}$ across the branch-cuts, and obtain the equivalence 
\begin{align}
    \boldsymbol\delta = 2(-\boldsymbol\omega'_p - \boldsymbol\omega_q' + \bx_{[1,2]}) \, ,
\end{align}
which consistently has only even number of prongs on ${{\bf O7}^+}'$.
Moreover, the pairings are $(\boldsymbol\omega_p', \bx_{[1,2]}) = -(\boldsymbol\omega_q^{\text{O7}^+}, \bx_{[1,2]}) = \tfrac12$, and assert, together with \eqref{eq:bilinext}, that $(\boldsymbol\delta, \boldsymbol\delta) = 0$.
The $\boldsymbol\epsilon$-junction from \eqref{eqn:2dP9_a} cannot be realized as a string junction in the presence of the O7$^+$, because it requires a net $\colpq{p}{q} = \colpq{3}{1}$ charge to end on ${{\bf O7}^+}' {\bf X}_{[1,2]}$ (see Appendix \ref{app:O7_into_E8} for details).
Instead, the prongs of any physical string junction on the ${{\bf O7}^+}' {\bf X}_{[1,2]}$ stack must be
\begin{align}
    2 \lambda_p \boldsymbol\omega_p' + 2 \lambda_q \boldsymbol\omega_q' + \lambda {\bf x}_{[1,2]} \, , \quad \lambda_{p,q}, \lambda \in \bbZ \, ,
\end{align}
which necessarily has even $q$-charge, as well as even pairing with $\boldsymbol\delta$.
This means that, orthogonal to the $\mathfrak{su}_{10-m} \oplus \mathfrak{e}_m$ weight junctions in \eqref{eq:9d_CHL_brane_example}, we must have a $U(2)$ lattice, spanned by string junctions $\boldsymbol\delta$ and $\boldsymbol\epsilon' = - \boldsymbol\delta + 6 \boldsymbol\omega_p' + 4 \bx_{[1,2]} - 2\bx_{[2,1]}$.

In the magnetically dual picture, any integer number of 5-brane prongs can end on ${{\bf O7}^+}'$.
In particular, 5-brane junctions corresponding to $\tfrac12 \boldsymbol\delta$ and $\tfrac12 \boldsymbol\epsilon'$ are then physical, and would span a squeezed hyperbolic lattice $U(\tfrac12)$.
This is consistent with the fact that in the 9d uplift of CHL vacua, the momentum lattice ``loses'' a $U(\tfrac12)$ factor \cite{Mikhailov:1998si}:
\begin{align}
    \left(\Lambda^\text{CHL}_{\text{8d}}\right)^* \cong (-\text{E}_8) \oplus U \oplus U(\tfrac12) \, , \quad \left(\Lambda^\text{CHL}_{\text{9d}}\right)^* \cong \Lambda^\text{CHL}_{\text{9d}} \cong (-\text{E}_8) \oplus U \, .
\end{align}
The remaining moduli available in 9d are then the deformations that move the 7-branes outside the ${{\bf O7}^+}' \bX_{[1,2]}$ stack.
The resulting maximal non-Abelian enhancements can be equally characterized by an 8d configuration of type \eqref{eqn:2dP9_a} (with $n=8$) or \eqref{eqn:1dP9_a} (with $k=8$), but with $\widehat{\textbf{E}}_8$ frozen via the embedding of an O7$^+$ described above (as summarized in Table \ref{tab:9d_CHL_maximal}).

The null junctions for \eqref{eq:9d_CHL_brane_example} are
\begin{align}
    \begin{split}
        & \boldsymbol\delta^N_{(1,0)} = -3 \boldsymbol\omega_p^{\mathfrak{e}_m} - \boldsymbol\omega_q^{\mathfrak{e}_m} + \bx_{[2,1]} -2\boldsymbol\omega'_p - 2\boldsymbol\omega_q' + 2\bx_{[1,2]} \, , \\
        & \boldsymbol\delta^N_{(0,1)} = (m-10) \boldsymbol\omega_{\mathfrak{su}} + (18-m) \boldsymbol\omega^{\mathfrak{e}_m}_p + 3 \boldsymbol\omega^{\mathfrak{e}_m}_q - 3\bx_{[2,1]} + 4\boldsymbol\omega'_p + 2\boldsymbol\omega'_q - x_{[1,2]} \, ,
    \end{split}
\end{align}
from which one can straightforwardly determine the non-Abelian gauge group structure for specific $m$.
It so happens that they are all trivial in the maximally enhanced cases, which agrees with the CHL-string computations \cite{Font:2021uyw}.

\begin{table}[ht]
\renewcommand{\arraystretch}{1.3}
\begin{equation*}
\begin{array}{|c|c|c|c|}
\hline \fkg^\text{CHL}_\text{9d} & \pi_1(G_\text{9d}) & \fkg_{\text{8d}, \infty} & \text{8d brane config.}  \\
\hline 
\hline \mathfrak{su}_{10} & 0 & \mathfrak{su}_{10} + \widehat{\mathfrak{e}}_8 & (\ref{eqn:2dP9_c}),\ n=8 \\
\hline \mathfrak{su}_{9}\oplus \mathfrak{su}_{2} & 0 & \mathfrak{su}_9 \oplus \widehat{\mathfrak{e}}_1 \oplus \widehat{\mathfrak{e}}_8  &  (\ref{eqn:2dP9_c}),\ m=1,\ n=8  \\
\hline \mathfrak{su}_7\oplus \mathfrak{su}_{2} \oplus \mathfrak{su}_{3} & 0 &  \mathfrak{su}_7 \oplus \widehat{\mathfrak{e}}_3 \oplus \widehat{\mathfrak{e}}_8 &  (\ref{eqn:2dP9_a}),\ m=3,\ n=8  \\
\hline  \mathfrak{su}_{6} \oplus \mathfrak{su}_{5} & 0 & \mathfrak{su}_6 \oplus \widehat{\mathfrak{e}}_4 \oplus \widehat{\mathfrak{e}}_8  &  (\ref{eqn:2dP9_a}),\ m=4,\ n=8  \\
\hline \mathfrak{su}_{5} \oplus \mathfrak{so}_{10} & 0 & \mathfrak{su}_5 \oplus \widehat{\mathfrak{e}}_5 \oplus \widehat{\mathfrak{e}}_8  &  (\ref{eqn:2dP9_a}),\ m=5,\ n=8  \\
\hline  \mathfrak{su}_{4} \oplus \mathfrak{e}_{6} & 0 & \mathfrak{su}_4 \oplus \widehat{\mathfrak{e}}_6 \oplus \widehat{\mathfrak{e}}_8  &  (\ref{eqn:2dP9_a}),\ m=6,\ n=8  \\
\hline \mathfrak{su}_{3} \oplus \mathfrak{e}_{7} & 0 & \mathfrak{su}_3 \oplus \widehat{\mathfrak{e}}_7 \oplus \widehat{\mathfrak{e}}_8  &  (\ref{eqn:2dP9_a}),\ m=7,\ n=8  \\
\hline \mathfrak{su}_2 \oplus \mathfrak{e}_8  & 0 & \mathfrak{su}_2 \oplus \widehat{\mathfrak{e}}_8 \oplus \widehat{\mathfrak{e}}_8  & (\ref{eqn:2dP9_a}),\ m=8,\ n=8  \\
\hline \mathfrak{so}_{18} & 0 & \mathfrak{so}_{18} \oplus \widehat{\mathfrak{e}}_8 &  (\ref{eqn:1dP9_a}),\ k=8  \\	

\hline
\end{array}
\end{equation*}
\caption{Maximal non-Abelian enhancements on the 9d rank $(1,9)$ moduli space that has a dual description in terms of the CHL string, obtained from an affine 8d realization in which an $\widehat{\mathfrak{e}}_8$ is frozen.
Note that all cases have trivial non-Abelian gauge group topology $\pi_1(G_\text{9d})$.
\label{tab:9d_CHL_maximal}}
\end{table}

\subsubsection[Disconnected moduli branches for 9d rank \texorpdfstring{$(1,1)$}{(1,1)} vacua]{Disconnected moduli branches for 9d rank \boldmath{$(1,1)$} vacua}

The description of 9d rank $(1,9)$ theories presented above has a clear interpretation in terms of ``freezing'', i.e., introducing an O7$^+$-plane into the 7-brane system that describes a rank $(1,17)$ theory.
In parallel to the construction of 8d vacua discussed in Section \ref{sec:global}, it then is natural to propose that 9d rank $(1,1)$ theories arise by a further freezing.
Moreover, the duality to the CHL string strongly suggests that, in 9d, the freezing process requires an $\widehat{\mathfrak{e}}_8$ affine algebra, in which the $\mathfrak{e}_8$ root junctions, as well as odd multiples of the winding-state-junction (i.e., $\boldsymbol\epsilon$) are projected out.
Therefore, from the maximally-enhanced cases in Table \ref{tab:9d_CHL_maximal}, only the second to last (with brane configuration \eqref{eq:9d_CHL_brane_example}), but not the last entry, can undergo a further freezing.

After repeating the brane motions discussed in Appendix \ref{app:O7_into_E8}, now for the first affine stack in \eqref{eq:9d_CHL_brane_example}, the corresponding (doubly) frozen configuration looks like
\begin{align}\label{eq:9d_rank_1_example_config}
    \bA \bA \, (\widetilde{{\bf O7}^+} \bX_{[1,-2]}) \, ({{\bf O7}^+}' \bX_{[1,2]}) \, ,
\end{align}
where the $M(\widetilde{{\bf O7}^+}) = \left( \begin{smallmatrix} 3 & 4 \\ -4 & -5 \end{smallmatrix} \right)$ is the monodromy of the left O7$^+$-plane in this $SL(2,\bbZ)$-frame.
We obtain an enhanced $\fkg = \mathfrak{su}_2$ non-Abelian symmetry when the two $\bA$-branes are moved on top of each other, which is the maximal enhancement we can have in 9d.
In fact, if one could separate the two $\bX$-branes from their corresponding O7$^+$ (making the KK modes massive), and move them next to each other, they would be locally-mutual, thus allowing for another $\mathfrak{su}_2$ enhancement --- this would be nothing but the 8d rank $(2,2)$ example \eqref{eq:rank_2_example_config} studied in the previous section, which had an $SU(2)^2 /\bbZ_2$ non-Abelian gauge group.
However, since for the 9d uplift, one of them must be broken, the fractional null junction that generated this $\bbZ_2$ quotient no longer exists for the configuration \eqref{eq:9d_rank_1_example_config}.
Hence, the 9d non-Abelian gauge group must be $SU(2)$.

It is suggestive that this doubly frozen, rank $(1,1)$ moduli branch corresponds to M-theory on a Klein-bottle \cite{Aharony:2007du}.
Namely, starting from the rank $(1,17)$ theories with heterotic description, which is dual to M-theory on a cylinder, the first freezing led to CHL vacua in 9d, which are equivalent to M-theory on a M{\"o}bius strip, or a cylinder with one cross-cap.
Freezing once more, i.e., adding another cross-cap on the other side, then produces a Klein-bottle.

However, as pointed out in \cite{Aharony:2007du}, there is a second branch of 9d rank $(1,1)$ moduli space that is disconnected from M-theory on a Klein-bottle.
That is, it cannot be realized as freezing 9d rank $(1,9)$ models.
However, since after an $S^1$-reduction, the 8d rank $(2,2)$ moduli space \emph{is} connected, there should exist a junction description for this 9d branch, as a suitable infinite distance limit in which KK-states become light.

In fact, starting from the general 8d configuration with two O7$^+$'s, depicted in Figure \ref{fig:rank2}, it is not hard to identify such potential limits.
Starting from ${\bf O7}^+ \bC \bX_{[3,1]} \, {{\bf O7}^+} \bC \bX_{[3,1]}$, where both O7's now have the standard monodromy, we can either push the $\bC$-branes from the left on top of the orientifolds,
\begin{align}\label{eqn:rank2_a}
    (\underbrace{{\bf O7}^+ \bC}_{\text{affine}}) \bX_{[3,1]} \, (\underbrace{{{\bf O7}^+} \bC}_{\text{affine}}) \bX_{[3,1]} \, ,
\end{align}
which is just a slightly rearranged version of \eqref{eq:9d_rank_1_example_config}, or we can generate a $\widehat{\bf E}_1 = \bB \bC \bC \bX_{[3,1]}$ stack, by moving 7-branes as
\begin{equation}
\begin{split}
    & {\bf O7}^+ \, \bC \, \underrightarrow{\bX_{[3, 1]}} \, {\bf O7}^+ \bC \bX_{[3, 1]} \ \longrightarrow \ {\bf O7}^+ \, \underrightarrow{\bC} \, {\bf O7}^+\,  \bB  \bC \bX_{[3, 1]} \\
    \longrightarrow \ & {\bf O7}^+ \, {\bf O7}^+ \, \underrightarrow{\bX_{[3,-1]}} \bB \bC \bX_{[3,1]} \ \longrightarrow \ {\bf O7}^+ \, {\bf O7}^+  \, (\underbrace{\bB \bC \bC \bX_{[3,1]}}_{=\widehat{\bf E}_1}) \, .
\end{split}\label{eqn:rank2_b}
\end{equation}

First, notice that one cannot transition between \eqref{eqn:rank2_a} and \eqref{eqn:rank2_b} without separating branes making up the affine stack.
In other words, these configurations are connected only via the 8d moduli space.
Second, by the brane move $\bC \underleftarrow{\bX_{[3,1]}} \rightarrow \bB \bC$ inside the affine stack, we find that $\widehat{\bf E}_1 = \bB \bC \bC \bX_{[3,1]} \simeq (\bB \bC) (\bB \bC)$ is the strong-coupling version of two O7$^-$-planes on top of each other.
Therefore, \eqref{eqn:rank2_b} is T-dual to IIA on an interval with O8$^\pm$'s at each end, which further dualizes to the 9d Dabholkar--Park background in type IIB \cite{Dabholkar:1996zi,Witten:1997bs}.
This is indeed the branch of 9d rank $(1,1)$ moduli space that is disconnected from M-theory on a Klein-bottle \cite{Aharony:2007du}.

\section{Conclusions and outlook}

In this chapter, we have extended the framework of string junctions on $[p,q]$-7-branes \cite{Gaberdiel:1997ud,Gaberdiel:1998mv,DeWolfe:1998zf} to include O7$^+$-planes.
The key difference is the distinction between physical $(p,q)$-strings and 5-branes that can end on the O7$^+$: while the latter can end with arbitrary integer $\colpq{p}{q}$-charges on the O7$^+$, only even numbers of integer $(p,q)$-strings may do so.
When applied to the construction of 8d ${\cal N}=1$ gauge theories on stacks including both ordinary $[p,q]$-7-branes and O7$^+$'s, this modification consistently reproduces the root and coroot lattices of non-simply-laced $\mathfrak{sp}$-algebras, as well as their electric 1-form- and magnetic 5-form center symmetries.
Furthermore, this provides a junction description for all 8d rank $(2,10)$ string compactifications with a dual CHL-string description \cite{Font:2021uyw,Cvetic:2021sxm}, including their gauge group topologies that are succinctly characterized by loop junctions encircling all 7-branes.
In addition, using junctions, we find a previously unknown lattice description for 8d string vacua of rank $(2,2)$, that is analogous to the Narain lattice characterization of 8d and 9d heterotic/CHL vacua.
This establishes junctions as a unifying framework to describe gauge enhancements (including the global gauge group structure) of \emph{all} 8d string vacua.

Moreover, in synergy with Swampland ideas \cite{Lee:2021qkx,Lee:2021usk}, we have discovered a full classification of 9d ${\cal N}=1$ string vacua, including their global gauge group structures, by 7-brane configurations with affine stacks characterized by loop junctions for their imaginary roots.
Again, the consistent inclusion of O7$^+$-planes in the analysis of potential infinite distance limits on the 8d moduli space turns out to be vital to capture subtleties, such as the two components of the 9d rank $(1,1)$ moduli space that are connected only through an $S^1$-reduction to 8d \cite{Aharony:2007du}.

The 9d results motivate a string-independent classification of the 9d ${\cal N}=1$ supergravity landscape in a similar fashion to \cite{Hamada:2021bbz}, where the 8d landscape was classified based on a Swampland ``translation'' of the $SL(2,\bbZ)$ characterization of 7-branes and O7$^+$-planes.
While perhaps unexpected from their direct constructions, this chapter shows that 9d string compactifications also admit a completely analogous characterization.
Hence, it is suggestive that there should also be a parallel story for the moduli space of 9d instantons that can be studied by $SL(2,\bbZ)$ monodromies.
In particular, such a bottom-up analysis could provide an explanation independent of the CHL-string, for why the 9d analog of the freezing mechanism can only be performed with an $\widehat{\bf E}_8$, but not an ${\bf D}_{n \geq 8}$ stack.

Another useful insight from the junction perspective is on the stringy origin of center symmetries in 8d gauge theories with non-simply-laced algebra.
Via dualities, it would be interesting if one can use this insight to generalize the geometric engineering framework for higher-form symmetries in M- and F-theory \cite{Morrison:2020ool,Albertini:2020mdx,Cvetic:2021sjm} to include frozen singularities.
This may have promising applications to the study of 6d SCFTs constructed on such singularities \cite{Bhardwaj:2018jgp} as well as lower dimensional SCFTs, obtained either from dimensionally reducing 6d theories, or directly engineering them with junction techniques \cite{Garcia-Etxebarria:2013tba,Agarwal:2016rvx,Hassler:2019eso,Heckman:2020svr}.


\part{Generalized Symmetries in Superconformal Field Theories}

\chapter{HIGHER SYMMETRIES OF 5D ORBIFOLD SCFTS}

\section{Introduction}

Higher-form symmetries \cite{Gaiotto:2014kfa} provide a powerful way to constrain the non-perturbative data of a quantum field theory \cite{Kapustin:2013uxa,Sharpe:2015mja,Gaiotto:2017yup,Gaiotto:2017tne,Cordova:2018cvg,Cordova:2020tij,Brennan:2020ehu,Kaidi:2021gbs,Lee:2021crt,Choi:2021kmx,Kaidi:2021xfk}.
This is especially valuable in the case of $d > 4$ superconformal field theories since all known examples are intrinsically strongly
coupled. Indeed, the main method to construct such examples proceeds by taking a singular limit of a string / M-theory / F-theory compactification. With this in mind, it is important to extract the corresponding data of higher-form symmetries for
such systems directly from the singular geometry of a string compactification \cite{DelZotto:2015isa,Heckman:2017uxe,Eckhard:2019jgg,GarciaEtxebarria:2019caf,Albertini:2020mdx,Morrison:2020ool,Dierigl:2020myk,Closset:2020scj,DelZotto:2020esg,Apruzzi:2020zot,Bhardwaj:2020phs,Closset:2020afy,Heidenreich:2020pkc,DelZotto:2020sop,Gukov:2020btk,Bah:2020uev,Bhardwaj:2021pfz,Apruzzi:2021mlh,Apruzzi:2021phx,Hosseini:2021ged,Apruzzi:2021vcu,Bhardwaj:2021wif,Bhardwaj:2021zrt,Closset:2021lwy,Heidenreich:2021xpr,Buican:2021xhs,Cvetic:2021maf, Debray:2021vob,Apruzzi:2021nmk,Braun:2021sex,Bah:2021brs,Bhardwaj:2021mzl,Cvetic:2020kuw,Cvetic:2021sxm}.

In this chapter we determine the higher-form symmetries for 5d superconformal field theories (SCFTs) which originate from an orbifold singularity $\orb = \mathbb{C}^3 / \Gamma$ for $\Gamma$ a finite subgroup of $SU(3)$. We denote the resulting 5d SCFTs as $\TM$.
There is a full classification of finite subgroups of $\Gamma$ (including their group actions) which result in
Gorenstein Calabi-Yau threefold singularities \cite{yau1993gorenstein} (see also \cite{watanabe1982invariant}).
It also gives rise to a large class of well-known 5d SCFTs. For example, the trinion theory $T_N$ with flavor symmetry algebra
$\mathfrak{su}(N)^3$ arises from the singularity $\mathbb{C}^{3} / \mathbb{Z}_N \times \mathbb{Z}_N$ (see \cite{Benini:2009gi}).
Recently the physics and geometry of many such singularities were studied in references \cite{Tian:2021cif, Acharya:2021jsp}. For
further discussion of higher-form symmetries in 5d SCFTs, see in particular \cite{Albertini:2020mdx,Morrison:2020ool,BenettiGenolini:2020doj,Apruzzi:2021vcu,Genolini:2022mpi}. For additional
background on geometric engineering and 5d SCFTs, see \cite{Seiberg:1996bd, Katz:1996xe, Witten:1996qb, Morrison:1996xf, Douglas:1996xp, Katz:1996fh, Intriligator:1997pq, Aharony:1997ju, Aharony:1997bh, Diaconescu:1998cn, Bergman:2012kr, Bergman:2013koa} as well as \cite{DelZotto:2017pti, Jefferson:2018irk,Closset:2018bjz, Bhardwaj:2018yhy, Bhardwaj:2018vuu, Bhardwaj:2019fzv, Apruzzi:2019vpe, Apruzzi:2019opn, Apruzzi:2019enx}.

The higher symmetries of 5d SCFTs that have gauge theory phases can be determined directly from the corresponding Lagrangian description, exploiting standard techniques \cite{Gaiotto:2014kfa} --- with the subtlety that it can happen that 5d instantons are charged with respect to the center symmetry in the presence of a non-zero CS level. There are, however, many 5d SCFTs which do not have a gauge theory phase, and instead are defined purely by singular geometry. A pivotal example of this type is the famous $E_0$ theory \cite{Seiberg:1996bd}, which is realized as the
singular limit of the local Calabi-Yau threefold $O(-3) \rightarrow \mathbb{P}^2$, namely the orbifold $\mathbb C^3 / \mathbb Z_3$ \cite{Morrison:1996xf}. For these theories an alternative route to compute the corresponding higher form symmetries is given by exploiting the defect group of M-theory on the corresponding singularity \cite{DelZotto:2015isa,Albertini:2020mdx,Morrison:2020ool}. For instance proceeding in this way one can show that for the $E_0$ theory, the defect group is:
\begin{equation}
    \mathbb D(E_0) \supset (\mathbb Z_3)^{(1)}_e \oplus (\mathbb Z_3)^{(2)}_m\,,
\end{equation}
where the subscripts and
the superscripts refer to the fact that we have an electric 1-form symmetry and a magnetic 2-form symmetry. The one-form electric symmetry arises from M2-branes wrapped on two-cycles, and the two-form magnetic symmetry similarly arises from wrapped M5-branes on four-cycles. The two are related to different choices of global structures for the $E_0$ theory \cite{Albertini:2020mdx}.

So long as the singularity is isolated, it is straightforward to read off the corresponding electric one-form symmetry via the abelianization $\mathrm{Ab}[\pi_{1}(\partial \mathbb{C}^3 / \Gamma)]$, much as was done in the case of the 6d defect group in \cite{DelZotto:2015isa}. If, however, the group action $\Gamma$ results in a non-isolated singularity, then the boundary $\partial \mathbb{C}^{3} / \Gamma$ will also have singularities. For toric singularities, this problem was resolved in \cite{Albertini:2020mdx}.
For more general orbifold singularities, however, it is still an open question as to how to read off the resulting higher-form symmetries
directly from the singularity.

Our aim in this chapter will be to present two complementary solutions to the computation of higher-form symmetries for
such 5d orbifold SCFTs. First of all, there is a well-defined notion of the fundamental group $\pi_{1}(\partial \mathbb{C}^3 / \Gamma) = \pi_{1}(S^5 / \Gamma)$ even when the group action by $\Gamma$ has fixed points. We use this to directly extract the electric one-form symmetry of such theories.

Second of all, we can directly exploit the fact that the higher-form symmetries are closely related to extended defects of the 5d SCFT and that the 1-form and 2-form symmetries above, upon circle reduction, give both rise to 1-form symmetries for the corresponding 4d KK theory. The defect group of the 4d KK theory is then captured by the screening of the latter by BPS particles, which is in turn  specified by the BPS quiver of the 5d SCFT \cite{Closset:2019juk}, the supersymmetric quantum mechanics (SQM) which encodes the dynamics on the worldline of the BPS particles of the 4d KK theory. Indeed, since compactification on a further circle takes us to type IIA on the same singularity, the resulting quiver is just the one obtained from a D0-brane probing $\mathbb{C}^3 / \Gamma$. From the 5d BPS quiver analysis, we expect that the one-form symmetry part of the defect group of the 4d KK theory $D_{S^1}\TM$ has the form
\begin{equation}\label{eq:4defecto}
   \mathbb D(D_{S^1} \TM)^{(1)} = \mathbb G^{(1)} \oplus \mathbb G^{(1)}
\end{equation}
where
\begin{equation}
    \mathbb G \simeq \bigoplus_{\ell = 1}^r \mathbb Z_{n_\ell},.
\end{equation}
In equation \eqref{eq:4defecto} there are two identical factors of $\mathbb G$ that denote respectively the possible electric and magnetic 1-form symmetries that are controlled by a choice of global structure for the 4d KK theory. The positive integers $n_\ell$ can be completely determined by via a standard 't Hooft screening argument \cite{tHooft:1977nqb}  --- see e.g. \cite{Caorsi:2017bnp}. Moreover, the quiver also captures the Weyl pairing \cite{Caorsi:2017bnp} (or linking pairing \cite{GarciaEtxebarria:2019caf}) from which the resulting Heisenberg algebra of non-commuting fluxes \cite{ Freed:2006ya,Freed:2006yc} that governs the global structure of the theories \cite{Aharony:1998qu,Witten:2009at} can be reconstructed \cite{DelZotto:2022ras}. Knowing the 1-form defect group of the 4d KK theory, it is easy to recover the corresponding factors of the defect group of the associated 5d SCFT:
\begin{equation}
    \mathbb D (\TM) \supseteq \mathbb G^{(1)}_e \oplus \mathbb G^{(2)}_m\,.
\end{equation}

Whenever the 5d SCFT has a global structure which allows for a 1-form symmetry as well as a 0-form symmetry, the two can mix, and this can result in a non-trivial global 2-group symmetry -- see e.g. \cite{Kapustin:2013uxa,Cordova:2018cvg,Cordova:2020tij}.\footnote{See also  \cite{Sati:2008eg,Baez:2005sn,Fiorenza:2012tb,Fiorenza:2010mh,Sati:2009ic} for foundational work on higher group gauge symmetries.} As a further result in this chapter we begin exploring the 2-group symmetries of some orbifold 5d SCFTs with a Lagangian description \cite{Apruzzi:2021vcu}, reproducing the known features of such systems in terms of the abelianization of the orbifolding group $\Gamma$. Our result indicates that the 2-group structure is indeed a feature of the 5d SCFT rather than an emergent IR artifact.

The result of this chapter is organized as follows. In section \ref{sec:PRESCRIPTION}, after a brief review of the defect group
and its use in determining the higher-form symmetries of a 5d SCFT, we give a general prescription
for computing the higher-form symmetries of the 5d SCFT, both via a direct analysis of $\pi_1(S^5 / \Gamma)$,
and via the corresponding 5d BPS quiver. In section \ref{sec:EXAMP} we turn to a collection of examples, illustrating how our method works in
practice. In section \ref{sec:2group} we turn to a preliminary analysis of 2-group structures in such theories, and in particular its (conjectural) relation to the abelianization of $\Gamma$. We present our conclusions and potential future directions in section \ref{sec:CONC1}. The appendices contain some additional technical details as well as instructions for reproducing the relevant quiver and group theory computations.

\section{Defect Groups and Higher Symmetries in 5d} \label{sec:PRESCRIPTION}

In this section we discuss the interplay between the defect group and higher-form symmetries, with a particular emphasis on 5d theories.
Recall that the defect group is a general way to capture the spectrum of defects with charges which cannot be screened by dynamical states of the theory. This notion was first introduced in reference \cite{DelZotto:2015isa} in the context of 6d SCFTs, but it has far wider applicability, especially when combined with flux non-commutativity \cite{Freed:2006ya,Freed:2006yc}, as exploited, for example in references \cite{GarciaEtxebarria:2019caf,Albertini:2020mdx, Morrison:2020ool}. It is especially helpful in the context of higher-dimensional quantum field theories specified by a compactification of string theory, and we will mainly focus on this case in what follows.

In the context of string compactification, we obtain supersymmetric defects by wrapping branes on non-compact cycles of a local geometry. Branes of the same codimension which are wrapped on compact cycles amount to dynamical degrees of freedom which can screen the charges associated with these defects. Indeed, in many quantum field theories, the corresponding collection of defects needs to be supplemented by a choice of global structure which restricts the spectrum of extended objects \cite{Aharony:2013hda, Kapustin:2014gua, Gaiotto:2014kfa, DelZotto:2015isa}. This can happen whenever the corresponding torsional fluxes do not commute \cite{Aharony:1998qu, Freed:2006ya,Freed:2006yc, Tachikawa:2013hya, Monnier:2014txa, DelZotto:2015isa, Heckman:2017uxe, Monnier:2017klz}. Our conventions and treatment will follow that presented in \cite{Albertini:2020mdx}, to which we refer the interested reader for further details.

In any geometric engineering setup, the BPS spectrum of the resulting quantum field theory is captured by branes of various dimensions that are wrapping on shrinking cycles of a non-compact geometry $\mathbf{X}$, which in our case is a Calabi-Yau threefold singularity CY$_3$. When a $p$-brane wraps a compact $k$-cycle, it describes a $p-k+1$ dimensional BPS excitation. While, $p$-branes wrapped on non-compact $k$-cycles describe $p-k+1$ dimensional defects operators. Branes are charged with respect to flux operators, which can be used to construct the  corresponding quasi-topological symmetry defects that describe the charges of the extended objects. Of course, we can have a generalized 't Hooft screening, due to the possibility of defects to end on dynamical BPS objects, which breaks the associated higher-form symmetry. The remaining symmetry is captured by the defect group:
\begin{equation}
    \bbD := \bigoplus_n \bbD^{(n)} \quad \text{where}\ \  \bbD^{(n)} =  \bigoplus_{p\text{-branes}}\left(\bigoplus_{k \text{ s.t. } \newline p-k+1=n} \left(\frac{H_k(\mathbf{X}, \partial \mathbf{X})}{H_{k}(\mathbf{X})}\right)\right)
\end{equation}
In other words, the defect group $\mathbb{D}$ is the group of charges of higher symmetries acting on defects modulo screening. Moreover, together with the corresponding Heisenberg algebra of non-commuting fluxes, it encodes the quantum data of the Hilbert space at the boundary of the non-compact geometry. This construction captures all possible global structures realized by the geometry of the string compactification. We note that in principle, there could be additional emergent higher-form symmetries in the deep infrared of such a system, which would in turn signal the existence of additional defects. Our operating assumption---which is well-supported in practice--- is that such subtleties will not arise in the analysis to follow.

In this chapter our focus is on the 1-form and the 2-form symmetry parts of the defect group in the context of a geometric engineering of M-theory on a Calabi-Yau singularity $\orb$
\begin{equation}
\mathbb D(M/\orb) \supset \mathbb D^{(1)}_{M2} \oplus \bbD^{(2)}_{M5}
\end{equation}
where the subscripts denote the associated branes, and the superscript indicates that the M2-branes are associated with a one-form and the M5-branes with a two-form generalized symmetry.

Since M2s and M5s are mutually non-local and in general the singularity $\orb$ might have some torsional flux, one naturally expects to find examples with non-trivial higher symmetries and a non-trivial global structure. To see this, consider a geometry $\mathcal{M}_{11}=\mathcal{M}_5 \times \orb$ where for ease of exposition we take $\mathcal M_5$ compact and torsion free. The geometry $\mathcal{M}_{11}$ has a boundary at infinity given by $\partial \mathcal{M}_{11} = \mathcal M_5 \times \partial \orb$, to which we associate an Hilbert space $\mathcal{H}(\partial \mathcal{M}_{11})$. The resulting Hilbert space has selection sectors that can be thought of as states in a quantum mechanics where the role of operators is played by torsional fluxes, organized by a generalized cohomology group, $\mathbb{E}(\mathcal{M}_{11})$. The presence of a non-trivial torsion for the generalized cohomology,
\begin{gather}
    \mathrm{Tor}\,\mathbb{E}(\mathcal{M}_{11})=\bigoplus_i H^{i+1}(\mathcal{M}_5) \otimes \mathrm{Tor }\, \mathbb{D}^i,
\end{gather}
might cause the flux operators to form a non-commutative algebra \cite{Freed:2006ya,Freed:2006yc}. Indeed, fluxes corresponding to the M2 and M5-branes that contribute to the 1-form symmetry and the 2-form symmetry part of the defect groups satisfy the following relation
\begin{gather}
    \Psi_2 \Phi_5=\text{exp}\left( 2 \pi i  L(l_1,l_2) \int_{\mathcal{M}_5} \omega_1 \wedge \omega_2 \right)\Phi_5 \Psi_2,
\end{gather}
where the term in the exponential is a pairing of cocycles in $\rm{Tor}\,\mathbb{E}(\mathcal{M}_{11})$, $\omega_{1,2}$ represents the forms dual to the cycle where the extended objects have supports, $l_{1,2}$ are elements of $\mathbb{D}^i$ and $L(\cdot,\cdot)$ is the linking form on $\partial \orb$. To fully specify the quantum system, we need to select a maximal set of mutually commuting fluxes as a base for our Hilbert space.

This construction can be made more rigorous and general \cite{Albertini:2020mdx}. In particular, it is known how to compute the defect group from exact sequences in homology \cite{DelZotto:2015isa,Albertini:2020mdx,Morrison:2020ool}. In the next section we will review this result and use it to compute the defect group of orbifold singularities. Moreover, we will confirm the same result exploiting the corresponding 5d BPS quivers, building on \cite{Hosseini:2021ged,DelZotto:2022ras}.

The rest of this section is organized as follows. Again specializing to the case of 5d SCFTs obtained from M-theory on an orbifold singularity $\orb = \mathbb{C}^3 / \Gamma$, we show how to extract the defect group directly from the fundamental group of $S^5 / \Gamma$, which we refer to as the ``algebraic topology approach''. After this, we turn to a physical realization of the same data in terms of the 5d BPS quiver defined by the 5d SCFT. The physical interpretation of the quiver in terms of the Dirac pairing for BPS particles of the 4d KK theory provides a complementary method for extracting the same data on higher-form symmetries. We turn to examples later in section \ref{sec:EXAMP}.

\subsection{Algebraic Topology Approach}

Let us now turn to a computation of the defect group directly via the corresponding singular geometry specified by the orbifold group $\mathbb{C}^{3} / \Gamma$. We start by considering M-theory on $\mathcal M_5 \times \orb$. In order to capture the defect group one has to consider the long exact sequence of relative homology of $(\orb, S^5/\Gamma)$:
\begin{equation}
    \dots \rightarrow H_2(S^5/\Gamma) \overset{\imath_2}{\rightarrow} H_2(\orb) \overset{\jmath_2}{\rightarrow} H_2(\orb, S^5/\Gamma) \overset{\partial_2}{\rightarrow} H_1(S^5/\Gamma) \overset{\imath_1}{\rightarrow} \underbrace{H_1(\orb)}_{ = 0} \rightarrow \dots.
\end{equation}
Strictly speaking, some of the quantities in the above exact sequence may not involve smooth spaces, for example if $\Gamma$ has fixed points.
In the present context, we can always assume the existence of a crepant resolution, and work in terms of the resolved geometry. Since, however, our answer will be independent of a given choice of a resolution, there is a precise sense in which these objects ought to make sense even without an explicit blowup, consistently with the remark of \cite{Morrison:2020ool} that the higher form symmetries are independent from flop transitions in the resolved geometry. Indeed, we will shortly give a precise definition of $H_{1}(S^5/\Gamma)$ as the abelianization of $\pi_{1}(S^5 / \Gamma)$,
even when $\Gamma$ has a fixed point locus on the $S^5$.

Now, we expect $H_1(\orb)$ to vanish due to the fact that $\orb$ is a Calabi-Yau space, and, moreover, by definition \cite{Albertini:2020mdx,Morrison:2020ool}:
\begin{equation}
\begin{aligned}
    \mathbb D^{(1)} &\equiv \frac{H_2(\orb, S^5/\Gamma)}{\jmath_2(H_{2}(\orb))} \\
    &\simeq H_1(S^5/\Gamma)\\
    &\simeq \text{Ab}[\pi_1(S^5/\Gamma)]
\end{aligned}
\end{equation}
Using both Poincaré duality and the Universal Coefficient Theorem it can be shown that $\mathbb D^{(1)}_{M2} \simeq \mathbb D^{(2)}_{M5}$ whenever $H_3(\orb)$ vanishes, which is indeed the case for the orbifold singularities we are considering.

The defect group is then fully captured by $\pi_1(S^5/\Gamma)$, since the abelianization of this group is just $H_1(S^5 / \Gamma)$. An important subtlety here is that in general, $\Gamma$ may have a fixed point locus which complicates the analysis. In the special case where there are no fixed points, we have $\pi_{1}(S^5 / \Gamma_{\mathrm{no-fixed}}) = \Gamma_{\mathrm{no-fixed}}$. To extend this to the more general case which can include fixed points, we use a result proved by Armstrong in 1967 \cite{armstrong1968fundamental}:

\textit{Let $\Gamma$ be a discontinuous group of homeomorphisms of a path connected, simply connected, locally compact metric space $X$, and let $H$ be the normal subgroup of $\Gamma$ generated by those elements which have fixed points. Then the fundamental group of the orbit space $X/\Gamma$ is isomorphic to the factor group $\Gamma/H$.}

In other words, to compute $\pi_{1}(S^5 / \Gamma)$, we just need to enumerate the generators of $\Gamma$ which might have a fixed point locus on $S^5$. Specifying the particular group action induced via $f_\Gamma: \Gamma \rightarrow \text{Homeo}(\bbC^3)$, we denote by $H_{\Gamma, f_{\Gamma}} \trianglelefteq \Gamma$ the resulting normal subgroup of $\Gamma$. The end result is that $\pi_1(S^5/\Gamma) = \Gamma/H_{\Gamma, f_\Gamma}$, so the one-form symmetry part of the defect group is just:
\begin{equation}
\mathbb{D}^{(1)} = \mathrm{Ab}[\Gamma / H_{\Gamma, f_{\Gamma}}].
\end{equation}
This also shows that the higher-form symmetry is independent of a choice of resolution, and moreover,
provides a systematic way to determine this data without specifying a blowup.
We give explicit examples of this procedure in section \ref{sec:EXAMP}.

\subsection{Quiver Approach}

A complementary way to extract the same information on the higher-form symmetry is to determine the corresponding quiver associated with a given orbifold singularity. In physical terms, this arises from the 5d BPS quiver of the theory \cite{Closset:2019juk}. 5d BPS quivers are the quiver of supersymmetric quantum mechanics that capture the BPS spectrum of particles of the 4d KK theory $D_{S^1}\TM$. Exploiting the Kaluza-Klein (KK) circle as an M-theory circle, it is clear that the 4d KK theory associated to $\orb$ is obtained from type IIA on the same Calabi-Yau threefold \cite{Lawrence:1997jr}. The 5d BPS quivers are therefore captured from the BPS quivers of IIA on $\orb$ \cite{Closset:2019juk} (see also \cite{Duan:2020qjy}). For the case at hand, the structure of the quivers can be reproduced from the D0-brane probe of this singularity. The resulting supersymmetric quiver quantum mechanics follows from the general prescription of Douglas and Moore \cite{Douglas:1996sw}. The resulting 3d McKay quivers were obtained in \cite{Hanany:1998sd,Lawrence:1998ja}. For additional details on how to implement this procedure, see Appendix \ref{app:3dmckay}. In many of our quiver figures, we present an explicit indexing of the nodes, and in particular this does not refer to the rank of each gauge group.

The crucial part needed for our analysis is the physical interpretation of the BPS quivers (see e.g. \cite{Fiol:2000wx,Denef:2002ru,Cecotti:2010fi,Cecotti:2011rv,Cecotti:2011gu,Alim:2011ae,Alim:2011kw}). The nodes of the BPS quivers are in one-to-one correspondence with a basis of generators of the charge lattice of the theory. Let us denote the corresponding charges $\gamma_1,...,\gamma_N$. The states corresponding to the charges of the generators are viewed as a collection of elementary constituents, out of the bound states of which the whole spectrum of the theory can be reconstructed. Since all the states in the spectrum are formed by these bound states, their charges are integer multiples of the charges of the elementary constituents, and we can completely determine the 't Hooft screening in terms of the latter \cite{Caorsi:2017bnp}. The charges of the line defects are in turn valued in a dual lattice of charges, where the duality is determined by the Dirac pairing \cite{Gaiotto:2010be,Aharony:2013hda}. For Lagrangian theories, the relevant Dirac pairing is determined by a simple computation in the Coulomb phase of the theory, for non-Lagrangian theories, however, geometry is needed. Here another aspect of the structure of the BPS quiver quantum mechanics is crucial, namely that the adjacency matrix which determines the structure of the BPS quiver quantum mechanics, is indeed captured by the Dirac pairing among the charges of the elementary constituents
\begin{equation}
\langle \gamma_i,\gamma_j\rangle_D = B_{ij}\,.
\end{equation}
For this reason the relevant quotient, which captures the defect group from the IR \cite{DelZotto:2022ras}, is also reproduced by the cokernel of the Dirac pairing. The 4d KK theory is obtained from type IIA on the same orbifold singularity $\orb$ and the torsional generators of the defect group are:
\begin{equation}\label{eq:5dBPSoneform}
\text{Tor }\mathbb D^{(1)}(\mathrm{IIA}/\orb) = \text{Tor}( \text{coker} (B)) = \mathbb G^{(1)}_e \oplus \mathbb G^{(1)}_m
\end{equation}
where
\begin{equation}
\mathbb G^{(1)}_e \simeq  \mathbb G^{(1)}_m \simeq \bigoplus_{\ell=1}^r \mathbb Z_{n_{\ell}} \,.
\end{equation}
By construction the integers $n_\ell$ can be recovered by the Smith normal form of the matrix $B$  \cite{Caorsi:2017bnp}. The fact that $B$ is antisymmetric entails that one gets the two identical electric and magnetic factors in equation \eqref{eq:5dBPSoneform}.

Whenever we choose a global form for the 5d theory $\TM$ with a magnetic 2-form symmetry, the latter gives rise to a magnetic 1-form symmetry for the 4d KK theory, by wrapping the corresponding surface defects on the KK circle. For this reason we identify
\begin{equation}
\mathbb D^{(2)}(M/\orb) \simeq \mathbb G^{(1)}_m
\end{equation}
above. This strategy gives rise to interesting consistency checks for the entire construction.

\section{Examples} \label{sec:EXAMP}

In section \ref{sec:PRESCRIPTION} we presented a general prescription for how to extract the higher-form symmetries from 5d SCFTs defined by M-theory on the background $\mathbb{C}^3 / \Gamma$. Our plan in this section will be to show how this works in practice, illustrating with a number of examples that both methods produce the same result, and agree with previously established results available in the literature, including v1 and v2 of \cite{Tian:2021cif}.

To frame the discussion to follow, we first divide the subgroups of $SU(3)$ into three main families:
\begin{itemize}
\item Family 1: The abelian subgroups;
\item Family 2: The subgroups of $SU(3)$ induced from finite non-abelian subgroups of $U(2)$;
\item Family 3: The complement of families 1 and 2.
\end{itemize}
We proceed by way of example, illustrating how our method works in each of these cases. In the case of finite abelian subgroups of $SU(3)$, we find agreement with the results of \cite{Albertini:2020mdx, Morrison:2020ool} as well as those of \cite{Tian:2021cif} which  involved a direct analysis of the resolved geometry. The case of family 2 does not appear to have been treated in the existing literature, but again we find examples which contain non-trivial higher-form symmetries. In the case of family 3, all examples we considered have too
many elements in $\Gamma$ whose action on $\mathbb{C}^3$ contains a fixed point locus.
The resulting normal subgroup generated by such elements is so large that $\mathrm{Ab}[\pi_{1}(S^5 / \Gamma)]$ is trivial, and as such they all produce a trivial higher-form symmetry. This is also in agreement with the results of v1 and v2 of \cite{Tian:2021cif}.

\subsection{Abelian Subgroups of \texorpdfstring{$SU(3)$}{SU(3)}}
We now turn to the case of $\Gamma$ a finite abelian subgroup of $SU(3)$. In this case, we have a
diagonal group action on the holomorphic coordinates $(x,y,z)$ of $\mathbb{C}^3$.
Since the maximal torus of $SU(3)$ is just $U(1)^2$, the most general orbifold action is given by
\begin{equation}
(x,y,z) \mapsto (\omega^{a_1} \xi^{b_1} x, \omega^{a_2} \xi^{b_2}y , \omega^{a_3} \xi^{b_3} z),
\end{equation}
with $\omega$ and $\xi$ primitive $m^{\mathrm{th}}$ and $n^{\mathrm{th}}$ roots of unity, and integers $a_i$ and $b_i$ with
$\sum_i a_i=0 \, \mathrm{mod} \, m$ and $\sum_i b_i=0 \, \mathrm{mod} \, n$.

We now consider two specific set of actions that highlight the main properties of these orbifolds. More general actions can be considered, but a complete analysis is left for future work.


The first case we consider is when $n=1$, thus $\Gamma = \mathbb{Z}_m$. Here, the generators are of the form $\frac{1}{m}(1,a_2,a_3)$, with $1+a_2+a_3=0 \mod m$.
In this case the choice of group action dictates the fixed point locus. For these examples  we can state a general rule:
\begin{align}
    H_{\Gamma, f_\Gamma} = \bbZ_{gcd(m,a_2)} \times \bbZ_{gcd(m,a_3)}
\end{align}
Let us check this formula on few examples:
\begin{itemize}
\item One generator for all fixed points.

The subgroup of fixed points is generated by some unique power of the generator of the full group, i.e. $g^k$ has fixed points, with $g$ the generator of $\Gamma = \bbZ_m$. In these cases, $H_{\Gamma, f_\Gamma} = \bbZ_{m/k}$, and thus
\begin{equation}
\mathbb D^{(1)}_{M2} = \text{Ab}[\pi_1(S^5/\Gamma)] = \bbZ_{m}/\bbZ_{m/k} \cong \bbZ_k
\end{equation}

\textbf{Example 1}: For $g = \frac{1}{10}(1, 1, 8)$, only $k = 5$ gives $g^k = \frac{1}{2}(1, 1, 0)$ with fixed loci $|z| = 1$. So we have $\mathbb D^{(1)}_{M2}  = \bbZ_5$.\footnote{In particular, $g = \frac{1}{10}(1, 2, 7)$ is also such that $\mathbb D^{(1)}_{M2} = \bbZ_5$. We have confirmed by quiver computation that the statement of ``$\Lambda_{\text{el.}} = \bbZ_2$" in v1,v2 and v3 of \cite{Tian:2021cif} is a typo.}

\item More generators for all fixed points.

If there are more powers of the generator of the orbifold that lead to a fixed point, the normal subgroup is just the direct product of those factors.

\medskip

\textbf{Example 2}: For $\Gamma = \mathbb{Z}_6$ generated by $g = \tfrac{1}{6}(1, 2, 3)$, both $k = 2$ and $k=3$ have fixed points. $k=2$ leads to a $\bbZ_3$ subgroup, while $k=3$ to a $\bbZ_2$. So $H_{\Gamma, f_\Gamma} = \bbZ_3 \times \bbZ_2 = \Gamma$, and thus $\mathbb D^{(1)}_{M2} = 0$. We have also directly verified this by constructing the corresponding 5d BPS quiver --- see Figure \ref{fig:z6_123} (left).

\medskip

\textbf{Example 3}: For $\Gamma = \mathbb{Z}_{12}$ generated by $g=\tfrac{1}{12}(1, 2, 9)$, $g^4$ generates a $\bbZ_3$ subgroup and $g^6$ a $\bbZ_2$ one. So $H_{\Gamma, f_\Gamma} = \bbZ_2 \times \bbZ_3 \subset \bbZ_{12} = \Gamma$, and thus $\mathbb D^{(1)}_{M2}  = \bbZ_2$. The corresponding 5d BPS quiver can be found in Figure \ref{fig:z6_123} (right).
\end{itemize}

The second case we consider is $\Gamma = \mathbb{Z}_m \times \mathbb{Z}_n$, where we mainly consider the family of generators $\tfrac{1}{m}(1, m-1, 0)$ for $\bbZ_m$ and $\tfrac{1}{n}(0, 1, n-1)$ for $\bbZ_n$ that has been treated in \cite{Tian:2021cif}. In this case, we can rule out any non-trivial higher symmetry using the algebraic approach. Each of these manifestly have a fixed circle ($|z| = 1$ and resp., $|x| = 1$). So they both have $H_{\Gamma, f_\Gamma} = \Gamma$ and so Armstrong's theorem tells us that $\pi_1(S^5/\Gamma) = 0$. This result can also be established using BPS quivers. For example, for $n=m$, the 5d BPS quiver corresponding to models in this class all have a box-product form 
 \cite{Alim:2011kw}\footnote{\, Given two acyclic quivers $Q_1$ and $Q_2$ with adjacency matrices $B_i = S_i^t - S_i$ where $S_i$ are upper triangular 2d/4d $S$-matrices and $i=1,2$, the quiver $Q_1 \boxtimes Q_2$ is a quiver with adjacency matrix $B_\boxtimes = (S_1 \otimes S_2)^t - (S_1 \otimes S_2)$. If the quivers are not acyclic, which is the case in equation \eqref{eqn:AnBoxAn}, one can still define a $\boxtimes$ operation at the level of the corresponding path algebras: the path algebra of the quiver $Q_1\boxtimes Q_2$ is the tensor product of $\mathbb C Q_1$ and $\mathbb C Q_2$ with extra lagrange multipliers which implements the commutativity relation on the corresponding squares --- the quiver $Q_1\boxtimes Q_2$ also comes with a superpotential, which to a first approximation is a superposition of the superpotentials of $Q_1$ and $Q_2$ together with extra Lagrange multipliers which implement the commutativity of the tensor product operation on the path algebra --- see e.g. \cite{Cecotti:2010fi} for a review of the $\boxtimes$ operation for $Q_1$ and $Q_2$ acyclic.}
\begin{equation}\label{eqn:AnBoxAn}
\widehat A(n,0) \boxtimes \widehat A(n,0)
\end{equation}
where $\widehat A(n,0)$ is the quiver corresponding to a loop with $n$ arrows oriented clockwise and superpotential given by $\text{tr}(\prod_i \psi_i)$. A direct case by case analysis for $0\leq n \leq 30$ reveals that the cokernel is trivial in all these cases. Similar considerations hold for $n\neq m$, though in this case we do not have a single concise expression as in line (\ref{eqn:AnBoxAn}), and the quivers have to be extracted from the general procedure summarized in Appendix \ref{app:3dmckay}.

In addition, we want to point out that there are more possible actions of $\bbZ_m \times \bbZ_n$ than those considered above. Here we give an example of such action with a non-trivial 1-form symmetry.

\textbf{Example 4}: For $\omega = \frac{1}{9}(1, 1, 7)$ and $\xi = \frac{1}{3}(1, 2, 0)$.  The fixed loci are generated by $3\omega$ and $3\omega + \xi$, spanning a group of $H_{\Gamma, f_\Gamma} = \bbZ_3 \times \bbZ_3$. So we have $D_{M2}^{(1)} = \bbZ_3$.

\begin{figure}
    \centering
    \begin{tabular}{cc}
    \includegraphics[scale=0.2]{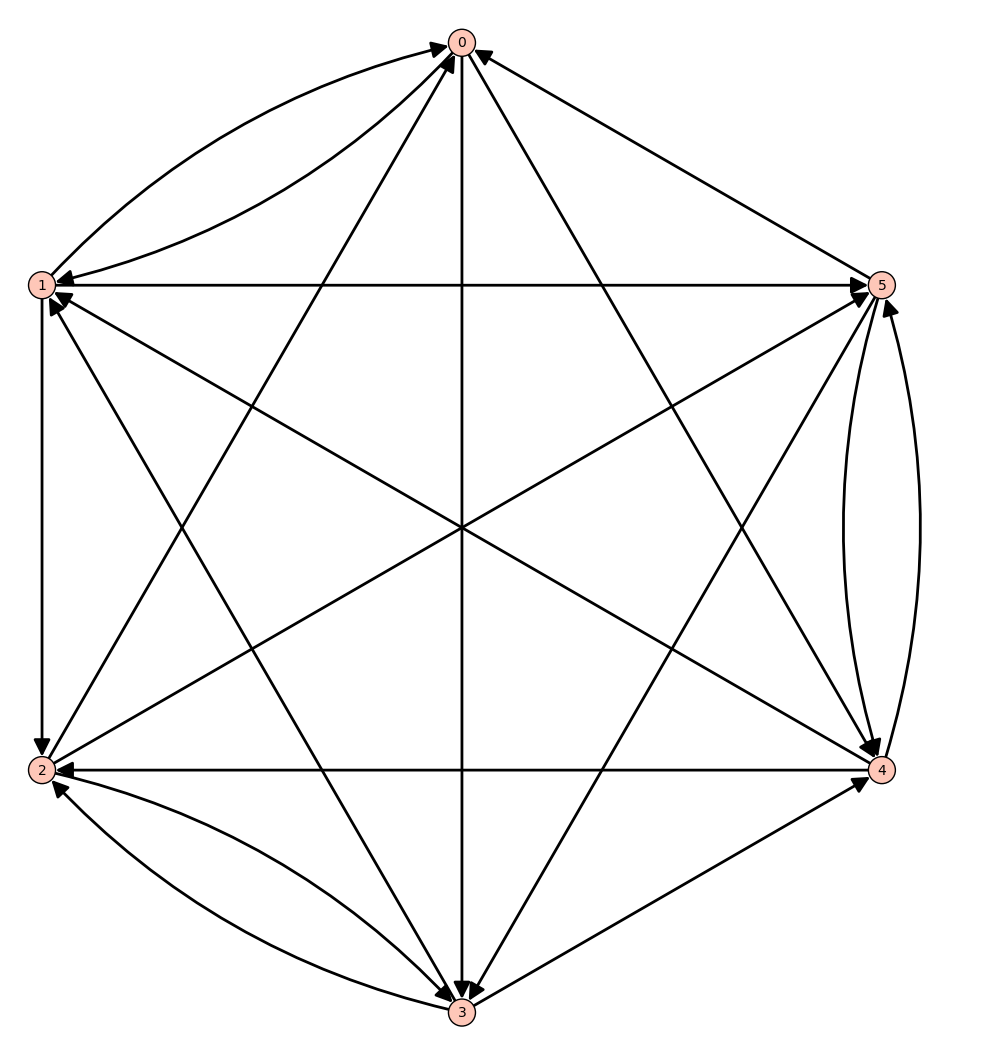}&\includegraphics[scale=0.23]{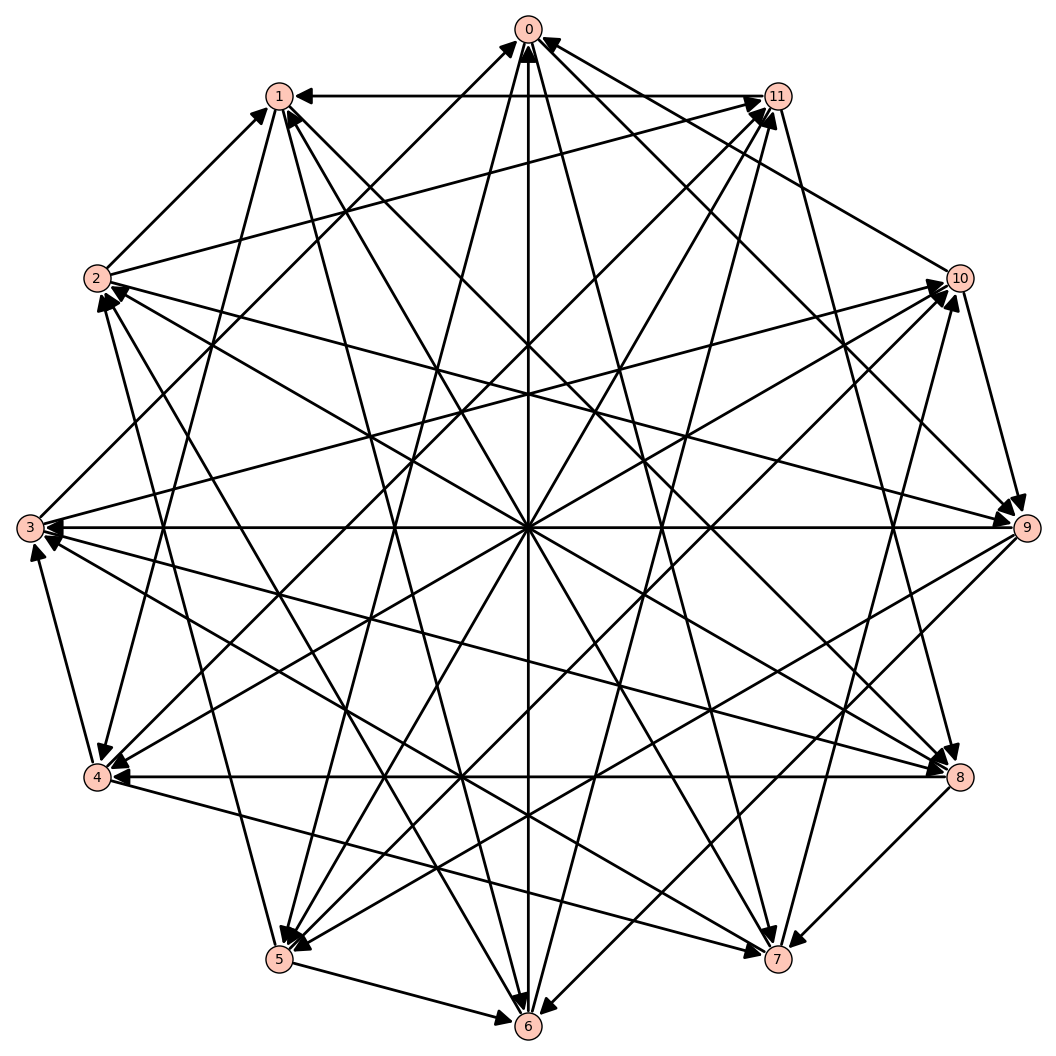}\\
    \end{tabular}
    \caption{\textsc{Left:} the orbifold quiver for the $\bbC^3/\mathbb{Z}_6$ theory generated by $g = \frac{1}{6}(1,2,3)$. \textsc{Right:} The orbifold quiver for the $\bbC^3/\mathbb{Z}_{12}$ theory generated by $g = \frac{1}{12}(1,2,9)$.}
    \label{fig:z6_123}
\end{figure}

\subsection{Examples Induced from Subgroups of \texorpdfstring{$U(2)$}{U(2)}}

Let us now turn to subgroups of $SU(3)$ induced from finite non-abelian subgroups of $U(2)$. These groups are obtained by taking some finite $\hat{\Gamma} \subset U(2)$ and mapping the elements $\hat{g}\in\hat{\Gamma}$ to
\begin{equation}
    g = \begin{pmatrix}
        \hat{g} & 0 \\
        0 & (\det \hat{g})^{-1}
    \end{pmatrix}.
\end{equation}
The finite subgroups of $U(2)$ are well known (see \cite{cohen_U2}, for example) and given by certain cyclic extensions of finite $SU(2)$ subgroups. In principle, this gives us an exhaustive method for generating such finite $SU(3)$ subgroups.

The first set of $U(2)$ derived groups we consider are those found in \cite{Tian:2021cif, watanabe1982invariant}. These groups form a special subclass of $SU(3)$ groups in the sense that their invariant subrings are complete intersection rings. Using the generators provided in \cite{Tian:2021cif}, we easily see that many elements have fixed points regardless of the group. In fact, in all of these cases we get that $H \cong \Gamma$ and hence
\begin{equation}
    \pi_1 (S^5/\Gamma) = 0.
\end{equation}
We list our findings in \ref{tbl:u2}.

\begin{table}[]
\centering
$\begin{array}{|c|c|c|c|}
     \hline\hline
    \Gamma & |\Gamma| & \mathbb D^{(1)}_{M2} & \text{Ab}[\Gamma] \\ \hline
    G_m &  8m      &  0 &  \bbZ_2 \times \bbZ_2 \times \bbZ_2 (2|m);\ \ \bbZ_2 \times \bbZ_2 (2 \!\!\not | m)  \\ \hline
    G_{p, q} & 8p q^2     & 0 & \bbZ_2^2 \times \bbZ_{2q} (2|p);\ \ \bbZ_2 \times \bbZ_{2q} (2\!\!\not | p) \\ \hline
    G'_{m}  & 8m      &  0  &  \bbZ_2 \times \bbZ_2 (2|m);\ \ \bbZ_4 \times \bbZ_2 (2 \!\!\not| m)\\ \hline\hline
    E^{(1)} & 72 & 0 & \bbZ_3 \times \bbZ_3\\ \hline
    E^{(2)} & 24 & 0 & \bbZ_3 \\ \hline
    E^{(3)} & 96 & 0 & \bbZ_2\times\bbZ_2 \\ \hline
    E^{(4)} & 48 & 0 & \bbZ_2 \\ \hline
    E^{(5)} & 96 & 0 & \bbZ_4 \\ \hline
    E^{(6)} & 48 & 0 & \bbZ_2 \times \bbZ_3 \\ \hline
    E^{(7)} & 144 & 0 & \bbZ_2\times\bbZ_3 \\ \hline
    E^{(8)} & 192 & 0 & \bbZ_2\times\bbZ_4 \\ \hline
    E^{(9)} & 240 & 0 & \bbZ_2 \\ \hline
    E^{(10)} & 360 & 0 & \bbZ_3 \\ \hline
    E^{(11)} & 600 & 0 & \bbZ_5 \\ \hline\hline
\end{array}$\caption{Data for orbifold theories derived from subgroups of $U(2)$ which have complete intersection invariant subrings \cite{watanabe1982invariant}.}\label{tbl:u2}
\end{table}

However, our approach means we can consider more general subgroups of $SU(3)$ derived from $U(2)$. An important class which we can consider are those derived from {\it small}\,\footnote{A group $G \subset GL(n,\mathbb{C})$ is small if there are no elements $g\in G$ with exactly $n-1$ many eigenvalues equal to $1$. In other words, $G$ contains no reflections.} subgroups of $U(2)$ \cite{yau1993gorenstein}. The ``small-ness'' condition restricts the number of elements with fixed points in $\Gamma$, and as such can potentially have a larger defect group when compared with a ``larger'' subgroup of $SU(3)$.

\begin{table}[t!]
\centering
$\begin{array}{|c|c|c|c|} \hline\hline
    \Gamma \phantom{\Big|}& |\Gamma| & \mathbb D^{(1)}_{M2} & \text{Ab}[\Gamma] \\ \hline
    T_m  \phantom{\Big|}& 24m &  \bbZ_{3m}\ (3|m)\ \ \bbZ_{m}\ (3\!\!\not| m) &  \bbZ_{3m} \\ \hline
    O_m  \phantom{\Big|}& 48m & \bbZ_m & \bbZ_2 \times \bbZ_m \\ \hline
    I_m  \phantom{\Big|}& 120m & \bbZ_m & \bbZ_m  \\ \hline
    D_{n, q} \phantom{\Big|}&   4qm\ (m = n-q)  & \bbZ_{2m}\ (2|m), \ \ \bbZ_{m}\ (2\!\!\not| m) &  \bbZ_{4m} (2 \!\!\not| q);\ \ \bbZ_{2m} \times \bbZ_2 (2 | q)  \\ \hline\hline
\end{array}$
\caption{Data for orbifold theories derived from small subgroups of $U(2)$. Note that the entries in this table depend are quite sensitive to the divisibility properties of $m,n$ and $q$, as discussed in \ref{app:conv}. To get a sense of the size of the normal subgroup with a fixed point locus, we have also listed the abelianization of $\Gamma$.}\label{tbl:smallu2}
\end{table}

Our findings in Table \ref{tbl:smallu2} can be confirmed again exploiting 5d BPS quivers. A more systematic study of the properties of the BPS categories of 5d orbifold SCFTs will appear elsewhere. For all these cases, we have a non-trivial defect group owing to the fact that they are built from small $U(2)$ subgroups. The analysis of these cases is uniform, and we carried out many consistency checks in this family. For the sake of brevity we report here only few salient examples. We refer to appendix \ref{app:conv} for our conventions about the representations we exploit in the analysis.

\medskip

\noindent\textbf{The $D_{5,3}$ orbifold SCFT}. Let us consider the lowest rank theory we find with non-trivial defect group -- this happens to be the $D_{5,3}$ orbifold SCFT, a rank $r= 4$ SCFT with flavor symmetry of rank $f=3$. The group $D_{5,3}$ is generated by
    \begin{gather}
            D_{5,3}=\Bigg\langle
        \begin{pmatrix}
            \zeta_{6} & 0 & 0\\
            0 & \zeta_{6}^{-1} & 0\\
            0 & 0 & 1
        \end{pmatrix},
        \begin{pmatrix}
            0 & i & 0 \\
            i & 0 & 0 \\
            0 & 0 & 1
        \end{pmatrix}\cdot
        \begin{pmatrix}
            \zeta_{8} & 0 & 0\\
            0 & \zeta_{8} & 0\\
            0 & 0 & \zeta_{8}^{-2}
        \end{pmatrix}
    \Bigg\rangle.
\end{gather}
    We can now use the age grading of \cite{ItoReid} to understand the gauge and flavor ranks of the theory. Explicitly, we find that the number of age-1 (or `junior') classes is $7$, while the number of age-2 classes is $4$. This then gives us
    \begin{gather}
        r= b_4(X) =4, \quad f=b_2(X)-b_4(X)=3,
    \end{gather}
    where $X$ is the crepant resolution of $\bbC^3/D_{5,3}$, the $b_i(X)$ are its Betti numbers. As such, we are looking for a quiver of size $2r+f+1=12$ with an intersection pairing possessing a kernel of dimension 4. Indeed, we find the quiver in \ref{fig:D53} which possess these properties.

The $B$-matrix we obtain is
{\tiny
\begin{equation}
    \left(
\begin{array}{cccccccccccc}
 0 & 0 & 0 & 0 & 0 & 0 & 1 & -1 & 0 & 0 & 1 & -1 \\
 0 & 0 & 0 & 0 & 0 & 0 & -1 & 1 & 0 & 0 & 1 & -1 \\
 0 & 0 & 0 & -1 & 1 & 0 & 0 & 0 & -1 & 1 & 0 & 0 \\
 0 & 0 & 1 & 0 & 0 & -1 & 0 & 0 & 1 & -1 & 0 & 0 \\
 0 & 0 & -1 & 0 & 0 & 1 & 0 & 0 & 1 & -1 & 0 & 0 \\
 0 & 0 & 0 & 1 & -1 & 0 & 0 & 0 & -1 & 1 & 0 & 0 \\
 -1 & 1 & 0 & 0 & 0 & 0 & 0 & 0 & 0 & 0 & -1 & 1 \\
 1 & -1 & 0 & 0 & 0 & 0 & 0 & 0 & 0 & 0 & -1 & 1 \\
 0 & 0 & 1 & -1 & -1 & 1 & 0 & 0 & 0 & 0 & -1 & 1 \\
 0 & 0 & -1 & 1 & 1 & -1 & 0 & 0 & 0 & 0 & 1 & -1 \\
 -1 & -1 & 0 & 0 & 0 & 0 & 1 & 1 & 1 & -1 & 0 & 0 \\
 1 & 1 & 0 & 0 & 0 & 0 & -1 & -1 & -1 & 1 & 0 & 0 \\
\end{array}
\right)
\end{equation}}
From which it is easy to check the defect group for the 4d KK theory is indeed $\mathbb Z_4 \oplus \mathbb Z_4$ as expected.

    \begin{figure}
    \centering
    \includegraphics[scale=0.6]{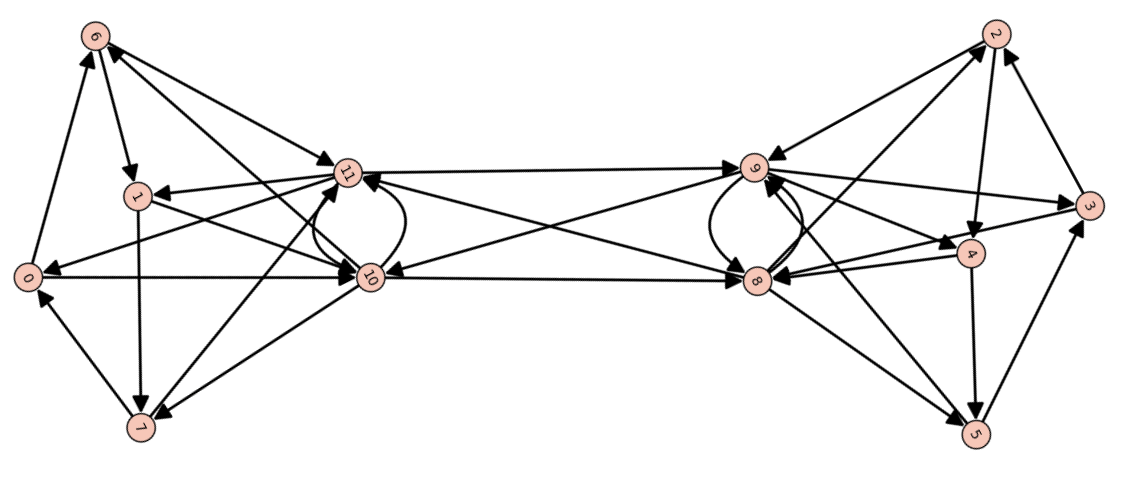}
    \caption{The orbifold quiver for the $\bbC^3/D_{5,3}$ theory.}
    \label{fig:D53}
\end{figure}

\medskip

\noindent\textbf{$T_3$ orbifold SCFT.} The $T_3$ group can be given as
    \begin{gather*}
        \Bigg\langle\begin{pmatrix}
            i & 0 & 0 \\
            0 & -i & 0 \\
            0 & 0 & 1
        \end{pmatrix},
        \begin{pmatrix}
            0 & 1 & 0 \\
            -1 & 0 & 0 \\
            0 & 0 & 1
        \end{pmatrix},
        \zeta_{18}\cdot\begin{pmatrix}
            (1+i)/2 & (-1+i)/2 & 0 \\
            (1+i)/2 & (1-i)/2 & 0 \\
            0 & 0 & \zeta_{18}^{-3}
        \end{pmatrix}\Bigg\rangle,
    \end{gather*}
    where $\zeta_{18}=e^{\pi i/9}$. From this we see that $|T_3|=72$ and that there are $21$ conjugacy classes\footnote{By the 3d McKay correspondence, this is equal to the Euler characteristic of the resolved $\bbC^3/T_3$ orbifold.}. Furthermore, the age grading gives us
    \begin{gather}
        r = b_4(X) = 9,\quad f =b_2(X)-b_4(X)= 2,
    \end{gather}
    where $X$ is the crepant resolution of $\bbC^3/T_3$, $r$ is the rank of the theory and $f$ is the rank of the flavor group. Note that these match with the constraint $2r+f+1=\chi(X)=21$. Interpreting this as quiver data, this means that we should find a quiver with $21$ nodes whose intersection pairing has a kernel of dimension $3$. Indeed, using our program to find the quiver, we obtain the quiver in \ref{fig:t3} which satisfies these conditions.

    The corresponding $B$-matrix is
    {\tiny
    \begin{equation}
        \left(
\begin{array}{ccccccccccccccccccccc}
 0 & 0 & 0 & 1 & -1 & 0 & 0 & 0 & 0 & 0 & 0 & 0 & 1 & 0 & 0 & -1 &
   0 & 0 & 0 & 0 & 0 \\
 0 & 0 & 0 & 0 & 0 & -1 & 1 & 0 & 0 & 0 & 0 & 0 & 0 & 1 & 0 & 0 &
   -1 & 0 & 0 & 0 & 0 \\
 0 & 0 & 0 & 0 & 0 & 0 & 0 & -1 & 1 & 0 & 0 & 0 & 0 & 0 & 1 & 0 & 0
   & -1 & 0 & 0 & 0 \\
 -1 & 0 & 0 & 0 & 0 & 1 & 0 & 0 & 0 & 0 & 0 & -1 & 0 & 0 & 0 & 1 &
   0 & 0 & 0 & 0 & 0 \\
 1 & 0 & 0 & 0 & 0 & 0 & 0 & 0 & -1 & 0 & 1 & 0 & -1 & 0 & 0 & 0 &
   0 & 0 & 0 & 0 & 0 \\
 0 & 1 & 0 & -1 & 0 & 0 & 0 & 0 & 0 & 0 & 0 & 1 & 0 & -1 & 0 & 0 &
   0 & 0 & 0 & 0 & 0 \\
 0 & -1 & 0 & 0 & 0 & 0 & 0 & 1 & 0 & -1 & 0 & 0 & 0 & 0 & 0 & 0 &
   1 & 0 & 0 & 0 & 0 \\
 0 & 0 & 1 & 0 & 0 & 0 & -1 & 0 & 0 & 1 & 0 & 0 & 0 & 0 & -1 & 0 &
   0 & 0 & 0 & 0 & 0 \\
 0 & 0 & -1 & 0 & 1 & 0 & 0 & 0 & 0 & 0 & -1 & 0 & 0 & 0 & 0 & 0 &
   0 & 1 & 0 & 0 & 0 \\
 0 & 0 & 0 & 0 & 0 & 0 & 1 & -1 & 0 & 0 & 0 & 0 & 0 & 0 & 1 & 0 &
   -1 & 0 & 0 & 1 & -1 \\
 0 & 0 & 0 & 0 & -1 & 0 & 0 & 0 & 1 & 0 & 0 & 0 & 1 & 0 & 0 & 0 & 0
   & -1 & 0 & 1 & -1 \\
 0 & 0 & 0 & 1 & 0 & -1 & 0 & 0 & 0 & 0 & 0 & 0 & 0 & 1 & 0 & -1 &
   0 & 0 & 0 & 1 & -1 \\
 -1 & 0 & 0 & 0 & 1 & 0 & 0 & 0 & 0 & 0 & -1 & 0 & 0 & 0 & 0 & 1 &
   0 & 0 & -1 & 0 & 1 \\
 0 & -1 & 0 & 0 & 0 & 1 & 0 & 0 & 0 & 0 & 0 & -1 & 0 & 0 & 0 & 0 &
   1 & 0 & -1 & 0 & 1 \\
 0 & 0 & -1 & 0 & 0 & 0 & 0 & 1 & 0 & -1 & 0 & 0 & 0 & 0 & 0 & 0 &
   0 & 1 & -1 & 0 & 1 \\
 1 & 0 & 0 & -1 & 0 & 0 & 0 & 0 & 0 & 0 & 0 & 1 & -1 & 0 & 0 & 0 &
   0 & 0 & 1 & -1 & 0 \\
 0 & 1 & 0 & 0 & 0 & 0 & -1 & 0 & 0 & 1 & 0 & 0 & 0 & -1 & 0 & 0 &
   0 & 0 & 1 & -1 & 0 \\
 0 & 0 & 1 & 0 & 0 & 0 & 0 & 0 & -1 & 0 & 1 & 0 & 0 & 0 & -1 & 0 &
   0 & 0 & 1 & -1 & 0 \\
 0 & 0 & 0 & 0 & 0 & 0 & 0 & 0 & 0 & 0 & 0 & 0 & 1 & 1 & 1 & -1 &
   -1 & -1 & 0 & 1 & -1 \\
 0 & 0 & 0 & 0 & 0 & 0 & 0 & 0 & 0 & -1 & -1 & -1 & 0 & 0 & 0 & 1 &
   1 & 1 & -1 & 0 & 1 \\
 0 & 0 & 0 & 0 & 0 & 0 & 0 & 0 & 0 & 1 & 1 & 1 & -1 & -1 & -1 & 0 &
   0 & 0 & 1 & -1 & 0 \\
\end{array}
\right)
    \end{equation}}

    Taking the Smith normal form we find that
    \begin{gather}
        \mathrm{coker}(B) = \bbZ^3\oplus \bbZ_9^{e}\oplus\bbZ_9^{m},
    \end{gather}
    giving us a $\bbZ_9$ electric one-form symmetry.

    \begin{figure}
        \centering
        \includegraphics[scale=0.3]{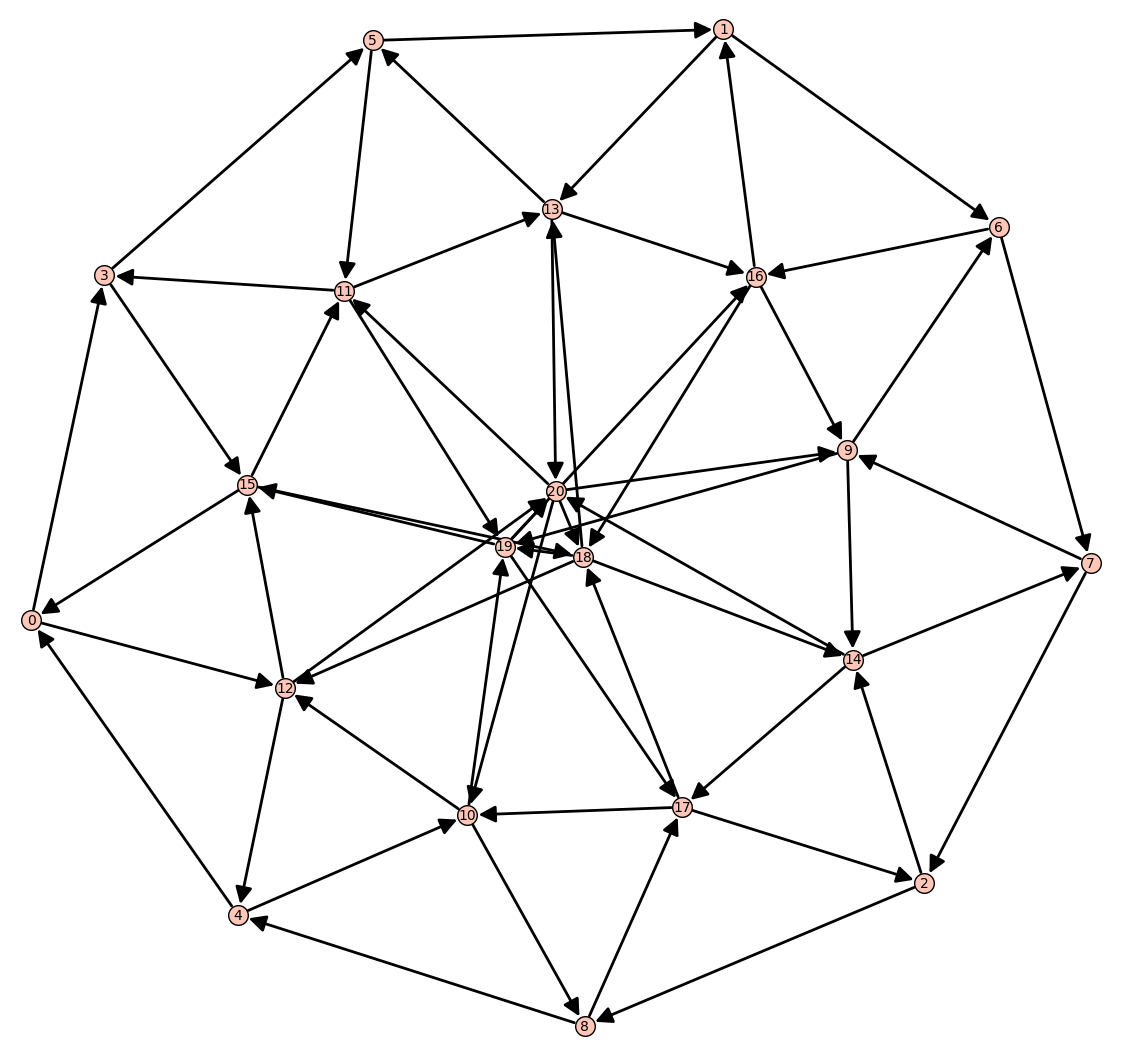}
        \caption{The quiver for the $\bbC^3/T_3$ orbifold theory.}
        \label{fig:t3}
    \end{figure}

\begin{table}
\centering
$\begin{array}{|c|c|c|c|} \hline\hline
    \Gamma & |\Gamma| & \mathbb D^{(1)}_{M2} & \text{Ab}[\Gamma] \\ \hline
    \Delta(3n^2) & 3n^2      & 0 & \bbZ_3 \times \bbZ_3 (3|n);\ \ \bbZ_3 (3 \!\!\not | n)         \\ \hline
    \Delta(6n^2) & 6n^2   & 0   &  \bbZ_2    \\ \hline
    C^{(1)}_{3l, l}\ (3|l)& 9l^2  & 0   &      \bbZ_{3} \times \bbZ_{3}        \\ \hline
    C^{(2)}_{7l, l} & 21 l^2    & 0 &  \bbZ_3 \times \bbZ_3 (3|n);\ \ \bbZ_3 (3 \!\!\not | n) \\ \hline
    D^{(1)}_{3l, l}\ (2|l)& 18l^2\    & 0 &  \bbZ_2 \times \bbZ_3       \\ \hline\hline
    H_{36}  & 108      &            0       & \bbZ_{4}          \\ \hline
    H_{72}  & 216      &            0       & \bbZ_2 \times \bbZ_2 \\ \hline
    H_{216} & 648      &            0       & \bbZ_3          \\ \hline
    H_{60}  & 60       &            0       & \mathbbm{1}  \\ \hline
    H_{168} & 168      &            0       & \mathbbm{1}   \\ \hline
    H_{360} & 1080     &            0       & \mathbbm{1}   \\ \hline
    J       & 180      &            0       & \bbZ_3   \\ \hline
    K       & 504      &            0       & \bbZ_3   \\ \hline\hline
\end{array}$
\caption{Data for orbifold theories derived from transitive subgroups of $SU(3)$. We determined the higher-form symmetry by direct computation of $\mathrm{Ab}[\pi_1(S^5 / \Gamma)]$ and from an analysis of the corresponding 5d BPS quiver. In all cases, none of these theories exhibit any one-form symmetry. For completeness, we have also included the abelianization of all these groups. Here $J$ and $K$ follows the notation of Yau and Yu \cite{yau1993gorenstein}, while we have followed the notation of \cite{Tian:2021cif} in the remaining entries. See Appendix \ref{app:ABGAMMA} for the definitions of all of these groups.}\label{tbl:transitive}
\end{table}

\medskip

\noindent\textbf{$T_5$ and $T_7$ orbifold SCFTs.} These cases can be analyzed in a similar way as above, and we again reproduce the defect groups $\mathbb Z_5^{(1)}$ for the $T_5$ orbifold SCFT and $\mathbb Z_7^{(1)}$ for the $T_7$ orbifold SCFT. We draw the corresponding quivers  in Figure \ref{fig:t5}, in a slightly different format to illustrate that these  have a box product form. We report the corresponding $B$ matrices in appendix \ref{app:Bmatrix}.

\medskip

\noindent\textbf{$O_5$ and $O_7$ orbifold SCFTs.} Also for these examples we report the relevant $B$-matrices in appendix \ref{app:Bmatrix}, which can be used to reproduce the results in table \ref{tbl:smallu2} also in these cases. We draw the corresponding quivers in Figure \ref{fig:o5} to illustrate that also these examples have quivers in a box product form.

\medskip

\noindent\textbf{General form of quivers for TOI orbifold theories.} All the other cases of orbifold 5D SCFTs with TOI orbifold groups can be analyzed similarly by extracting the corresponding quiver using the procedure of Appendix \ref{app:3dmckay}.  Based on these examples and further checks for the orbifold groups $T_m$, $O_m$ and $I_m$, we conjecture that, for suitable values of $m$, the quivers for many of the theories in this class can take the form of square tensor products too, leading to diagrams of the form
 \begin{equation}
     \hat{A}(m,0)\boxtimes \hat{E}_6\,, \qquad \hat{A}(m,0)\boxtimes \hat{E}_7\,, \qquad \hat{A}(m,0)\boxtimes \hat{E}_8,
 \end{equation}
    respectively for 5d orbifold theories of type $T_m$, $O_m$ and $I_m$. We stress this will not be the case for all values of $m$: for instance, in the case of $T_m$ theories with $3|m$, we expect to obtain quivers similar to \ref{fig:t3} consisting of an inner ring of $m$-many nodes surrounded by two rings of $3m$-many nodes connected appropriately. Moreover, these constructions will depend on choosing a suitable representative in the mutation class of the $\hat{E}_{6,7,8}$ type. Of course, for all the examples we checked, the 5d BPS quiver reproduces the result of Table \ref{tbl:smallu2}.
      \begin{figure}
        \centering
        \begin{tabular}{cc}
        \includegraphics[scale=0.4]{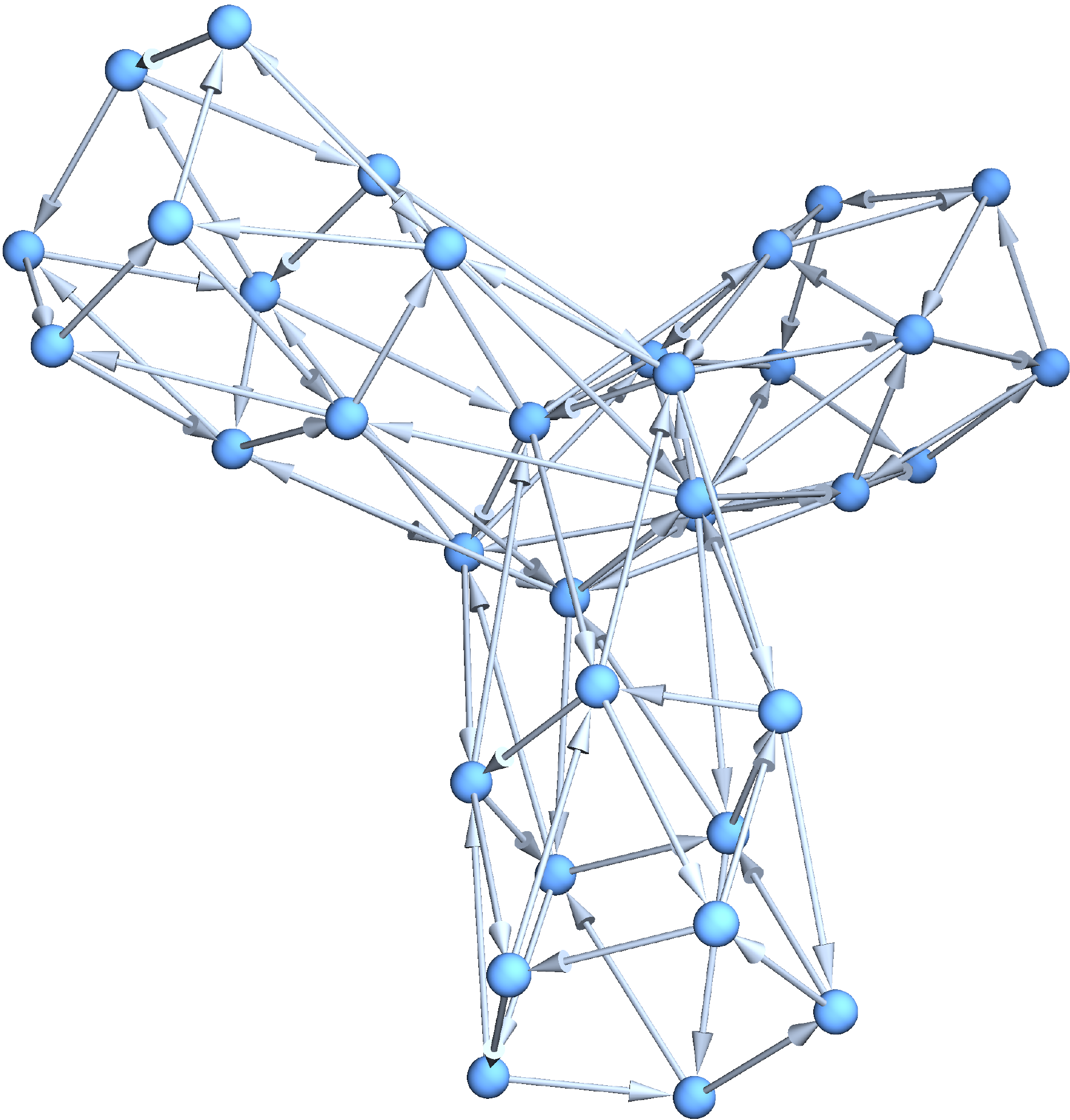}&\includegraphics[scale=0.4]{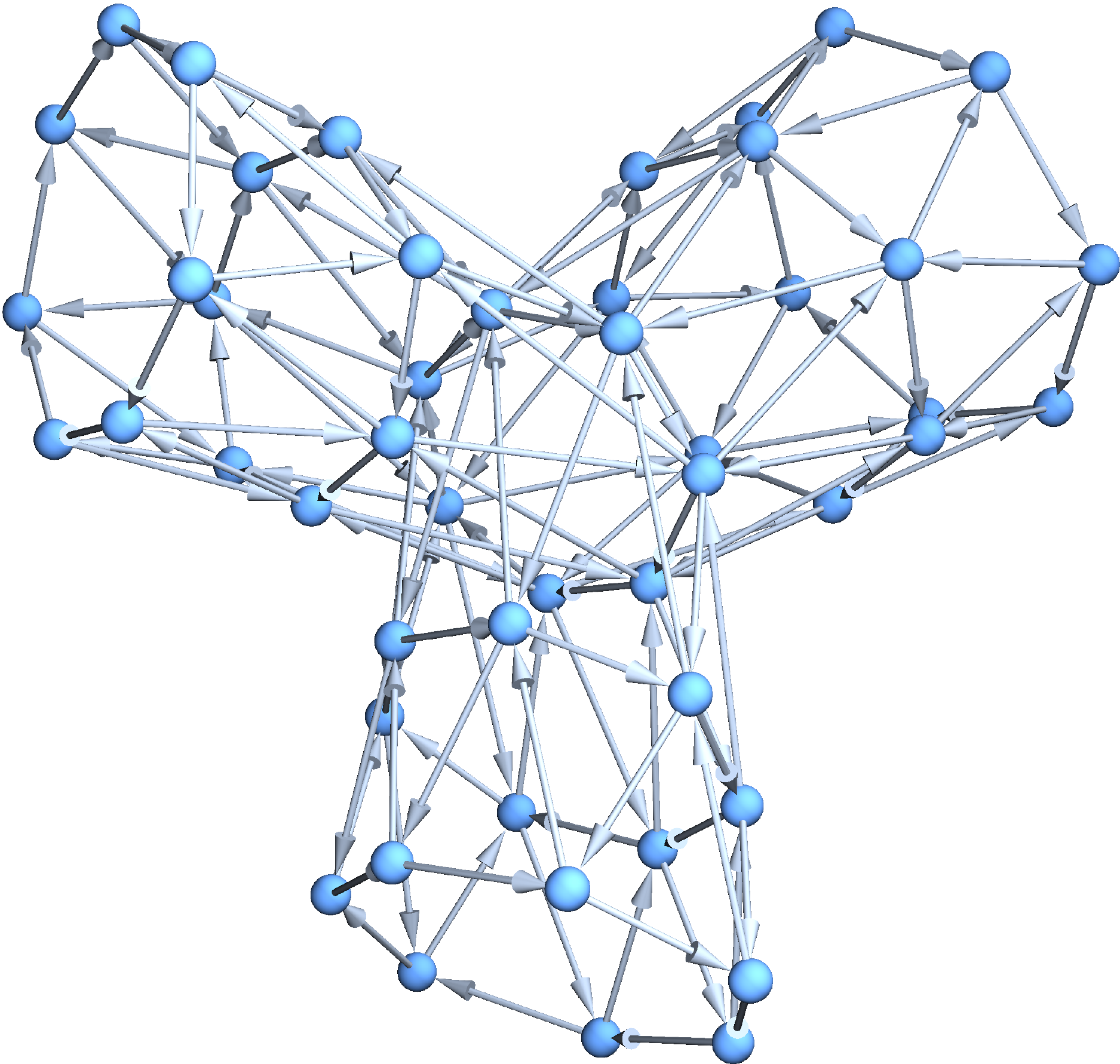}
        \end{tabular}
        \caption{\textsc{Left:} The BPS quiver corresponding to the orbifold group $T_5$; \textsc{Right:} The BPS quiver corresponding to the orbifold group $T_7$. Notice that we can recognize a box product-like structure in the quiver with an affine $\hat E_6$ structure.}\label{fig:t5}
    \end{figure}

    \begin{figure}
        \centering
        \begin{tabular}{cc}
        \includegraphics[scale=0.2]{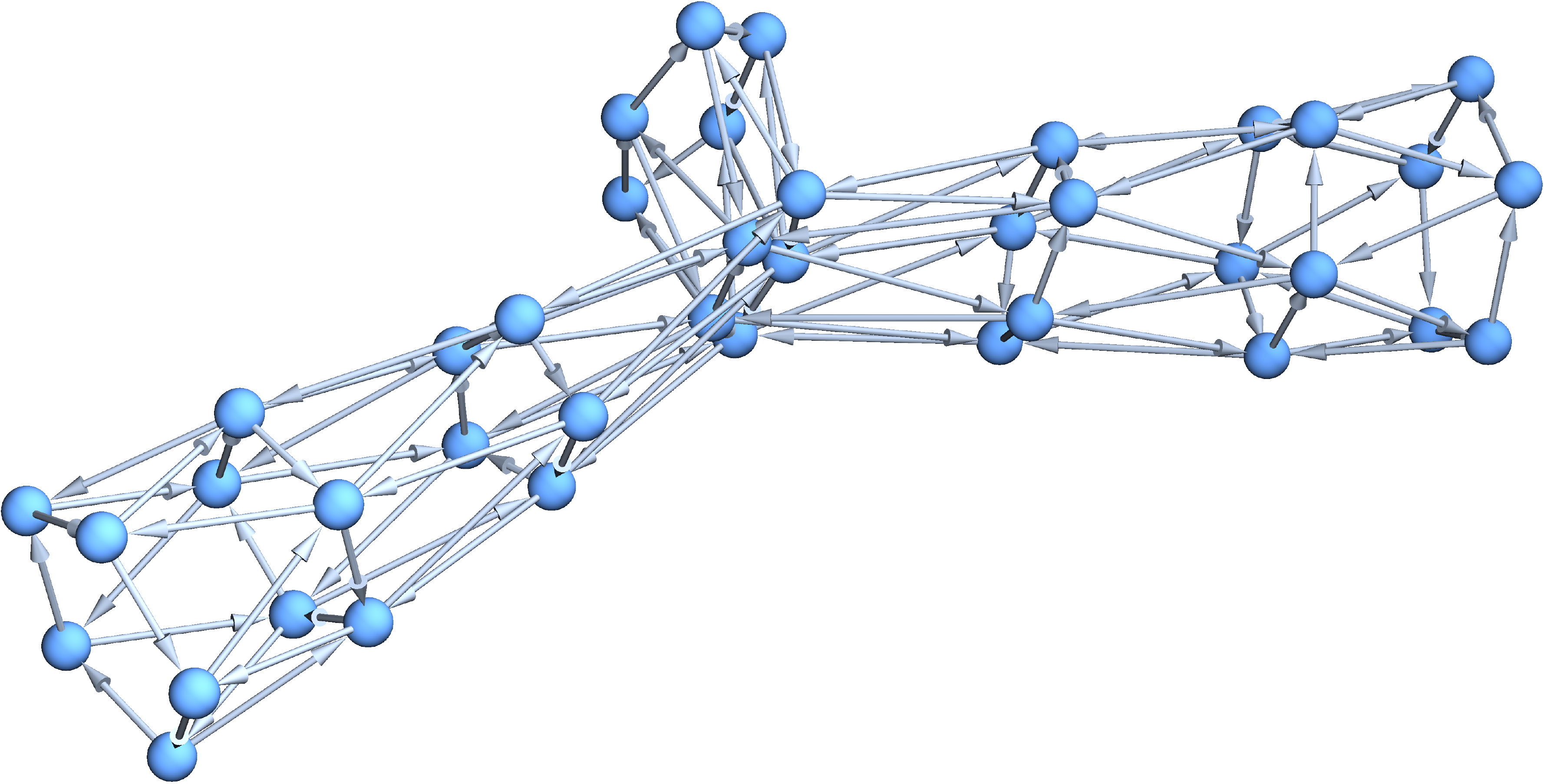}&\includegraphics[scale=0.4]{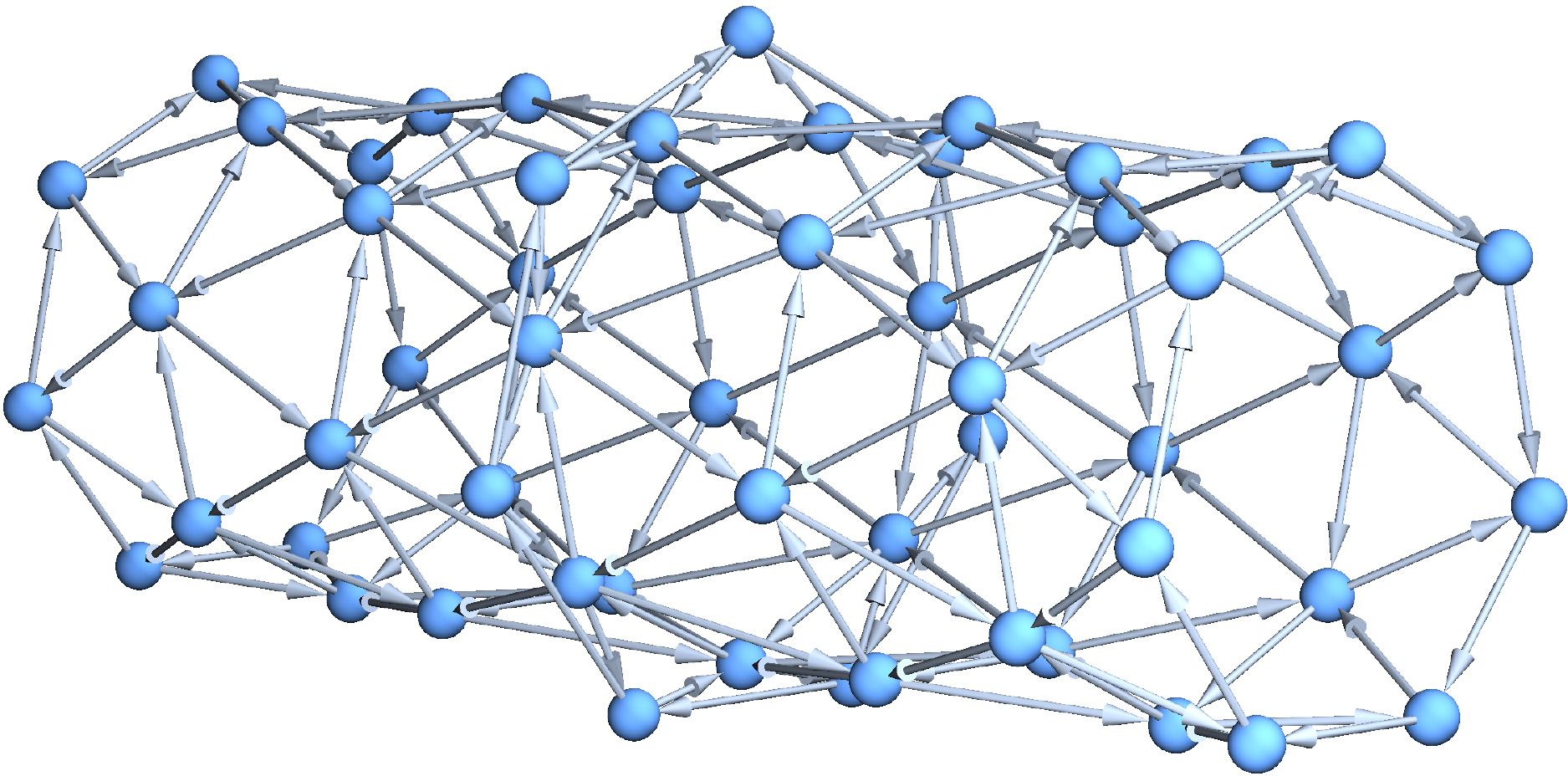}
        \end{tabular}
        \caption{\textsc{Left:} The BPS quiver corresponding to the orbifold group $O_5$; \textsc{Right:} The BPS quiver corresponding to the orbifold group $O_7$. Notice that we can recognize a box product-like structure in the quiver with an affine $\hat E_7$ structure.}\label{fig:o5}
    \end{figure}

\subsection{Larger Subgroups}

Finally, let us briefly discuss the case of ``larger subgroups,'' namely
transitive finite subgroups $\Gamma \subset SU(3)$ which are not abelian, and which are also not induced from a subgroup of $U(2)$.
In these cases, we expect that the larger size of the group correlates with a larger fixed point set in terms of a group action on $S^5$. In fact, in all examples which we have checked, we find that the resulting defect group is trivial, simply because the normal subgroup of $\Gamma$ generated by the elements which have a fixed point is simply all of $\Gamma$! We have directly checked the adjacency matrix of the corresponding BPS quiver in these cases as well, and again confirm this result, which is in accord with the statements of \cite{Tian:2021cif}. See Table \ref{tbl:transitive} for an explicit list of these examples.

\section{\texorpdfstring{$\mathrm{Ab}[\Gamma]$}{Ab[Gamma]} and 2-Group Symmetries}\label{sec:2group}

In the previous sections we saw that
the group $\mathrm{Ab}[\Gamma / H] = \mathbb D^{(1)}_{M2}$. Based on this, it is natural to ask whether
the abelianization of $\Gamma$ itself has any role to play in the 5d SCFT. Indeed, this structure
directly appears in the related context of 6d SCFTs. Recall that in the F-theory realization of 6d SCFTs, one
considers a canonical singularity of a non-compact elliptically fibered threefold $X \rightarrow B$. As shown in \cite{Heckman:2013pva},
the base $B$ is always of the form $\mathbb{C}^2 / \Gamma_{U(2)}$ for $\Gamma_{U(2)}$ a particular set of finite subgroups of $U(2)$, and
in all these cases, $\partial B = S^3 / \Gamma$. In this case, the corresponding
defect group is associated with a two-form symmetry, as specified by string-like defects
of the 6d SCFT \cite{DelZotto:2015isa} (see also \cite{GarciaEtxebarria:2019caf,Apruzzi:2020zot,Bhardwaj:2020phs,Apruzzi:2021mlh,Apruzzi:2021nmk,Bhardwaj:2021mzl}).
While we leave a more complete analysis for future work, in this section we observe that in
situations where the geometry faithfully reproduces the $0$-form symmetry of the system,
$\mathrm{Ab}[\Gamma]$ is closely correlated with the $2$-group symmetry of the 5d SCFT. Our plan in this section will be to first explain
some basic aspects of 2-group symmetries, following \cite{Benini:2018reh}, and then to turn to an analysis of a 5d SCFT where we can geometrically detect the $0$-form symmetry. For further discussion on aspects of $2$-group symmetries in 5d SCFTs, see e.g. \cite{Apruzzi:2021vcu}.

In order to investigate the potential role of  $\mathrm{Ab}[\Gamma]$ for 5d orbifold SCFTs,
we begin with the following two remarks:
\begin{enumerate}
\item Whenever $\Gamma$ acts without fixed points on $S^5$, $\mathbb D^{(1)}_{M2} \simeq \text{Ab}[\Gamma]$.
\item Whenever $\Gamma$ acts with fixed points on $S^5$, the theory $\TM$ typically has interesting 0-form symmetries, and  $\mathbb D^{(1)}_{M2}$ is a subgroup of $\text{Ab}[\Gamma]$.
\end{enumerate}
For a theory which has both 0-form symmetries and 1-form symmetries, the two can form a more interesting global categorical symmetry, that can organize into a 2-group. 2-groups are characterized by a 4-plet, consisting of a 0-form symmetry group $\mathbb F^{(0)}$, a 1-form symmetry group $\mathbb G^{(1)}$, a representation $\rho: \mathbb F^{(0)} \to \text{Aut}(\mathbb G^{(1)})$ and an element $\beta \in H^3(B\mathbb F^{(0)}, \mathbb G^{(1)})$, which characterizes the obstruction to switching on non-trivial backgrounds for $\mathbb F^{(0)}$ independently from backgrounds for $\mathbb G^{(1)}$. We can think of $\beta$ as if it is determining the 2-group structure constants, namely the extent to which the two higher symmetries mix with one another.

When $\Gamma$ acts with fixed points on $S^5$, the 5d SCFT typically has some non-trivial (non-abelian) 0-form symmetry, which can be characterized by the structure of the non-compact singularities in $\orb$. There is a ``naive'' answer dictated by lifting each simple Lie algebra factor to a simply connected Lie group, but this can in principle be quotiented to reach the true 0-form symmetry.
In such situations, $\mathbb D^{(1)}_{M2}$ is a strict subgroup of $\text{Ab}[\Gamma]$, and often the quotient can be understood in terms of the discrepancy between the naive 0-form symmetry and its quotiented counterpart. For example, in the case of the 5d $T_N$ theory we have an orbifold with structure $\mathbb C^3 / \mathbb Z_N \times \mathbb Z_N$. This geometry arises at the common intersection of three lines of singularities of the form $\mathbb C \times \mathbb C^2 / \mathbb Z_N$. In this case we have that $\mathbb D^{(1)}_{M2}$ is trivial and $\text{Ab}[\Gamma] = \mathbb Z_N \times \mathbb Z_N$ which in turn can be interpreted as the subgroup of the ``naive'' flavor symmetry $SU(N)^3$ by which we would quotient to reach the true global 0-form symmetry given by $SU(N)^3/\mathbb Z_N \times \mathbb Z_N$ (see e.g. \cite{Bhardwaj:2021ojs} for a discussion of the 
global symmetries in the closely related 4d $T_N$ theories). 
It is therefore tempting to claim that we have a nontrivial Postnikov class $\beta$ whenever the exact sequence
\begin{equation}\label{eq:thesequence}
1 \to \mathbb D^{(1)}_{M2} \to \text{Ab}[\Gamma] \to \text{Ab}[\Gamma]/\mathbb D^{(1)}_{M2} \to 1
\end{equation}
is non-split.

In order to check these general expectations, we seek a family of 5d orbifold SCFTs that have a Lagrangian interpretation and a non-trivial 1-form and 0-form symmetry. Since we also require that the geometry faithfully encodes the 0-form symmetry, an ideal class of examples in this case is provided by the abelian orbifolds $\mathbb C^3 / \mathbb Z_{2n}$ where $\mathbb Z_{2n}$ acts as $\frac{1}{2n}(1,1,2n-2)$. The latter give rise to 5d SCFTs with a gauge theory phase $SU(n)_{n}$. The 2-group structures have already been determined in \cite{Apruzzi:2021vcu}, and for $n$ even one finds that, choosing the electric global form of the theory, the 0-form symmetry of these models form a two group with the $\mathbb Z_n^{(1)}$ electric one-form symmetry. In our case since the group action is abelian we find that the sequence \eqref{eq:thesequence} reduces to
\begin{equation}
1 \to \mathbb Z_n^{(1)} \to \mathbb Z_{2n} \to \mathbb Z_2 \to 1\,.
\end{equation}
This sequence is non-split precisely when $n$ is even, which exactly reproduces the result of \cite{Apruzzi:2021vcu} obtained via other methods.

We find this remarkable, and it is natural to ask how this extends to other situations where geometry faithfully encodes the 0-form symmetry. As a further remark we stress here that if we were to choose the magnetic form of the theory, the 2-group structure disappears: this suggests that there is an interconnection between the Heisenberg algebra of non-commuting fluxes and the 2-group structure constants.

\section{Conclusions} \label{sec:CONC1}

In this chapter we have presented a general prescription for extracting the higher 1-form and 2-form symmetries
of 5d SCFTs obtained from M-theory on the orbifold $\mathbb{C}^3 / \Gamma$ with $\Gamma$ a
finite subgroup of $SU(3)$. We presented two complementary methods for extracting this data. First, building on \cite{GarciaEtxebarria:2019caf,Albertini:2020mdx,Morrison:2020ool}, we showed
how to extract it from the defining exact sequence for the defect group and the structure of $\pi_{1}(S^5 / \Gamma)$, generalizing the analysis to the case where $\Gamma$ has fixed points. Second, we showed that the same data can also be read off from the 5d BPS quiver of the corresponding SCFT, thus giving a nice consistency check to the method. We also provided some hints that the abelianization of $\Gamma$ detects the presence of a 2-group structure in such 5d SCFTs. We also remarked that the interplay with the Heisenberg algebra of non-commuting fluxes with the global form of the 5d SCFT must affect the 2-group structure in a non-trivial way. In the remainder of this section we discuss some avenues for further analysis.

Much of our analysis has focused on the computation of higher-form symmetries in these 5d SCFTs. We also saw hints of how the
2-group structure in these systems descends from the abelianization of $\Gamma$. It would be quite interesting to elucidate this structure. In particular, it would be desirable to extract the Postnikov class $\beta$ directly from the geometry of a string compactification. We think that in order to clarify this interplay it will be very fruitful to look at the symmetry TQFT \cite{Apruzzi:2021nmk} for these orbifold singularities, which arises from the reduction of the topological Chern-Simons term of M-theory on the horizon $S^5/\Gamma$.

Further compactification of these 5d SCFTs will give rise to a rich class of lower-dimensional systems. The same geometric methods
presented here can also be used to read off the corresponding higher-form symmetries of these systems. For example, compactification of our 5d SCFTs on a circle will give rise to 4d $\mathcal{N} = 2 $ SCFTs, and compactification on a $T^2$ will result in 3d $\mathcal{N} = 4$ SCFTs. Perhaps the most interesting case to study is the reduction of these 5d SCFTs on a Riemann surface $\Sigma_g$ as pioneered in \cite{Sacchi:2021wvg} to produce 3d $\mathcal N=2$ theories. In these cases the global structures we find in this chapter can give rise to more interesting effects, along the lines explored for 6d (2,0) theories in \cite{Tachikawa:2013hya}.

Although it is notoriously difficult to engineer stable non-supersymmetric backgrounds in M-theory, it is nevertheless
natural to consider more general orbifold group actions $\Gamma \subset SU(4)$.\footnote{For some recent investigations
into 5d non-supersymmetric CFTs, see e.g. references \cite{BenettiGenolini:2019zth, Bertolini:2021cew,DeCesare:2021pfb}.}
In this case, we can still read off a corresponding
non-supersymmetric quiver gauge theory, though in this case the adjacency matrix for the bosonic and fermionic degrees of freedom will be different. This serves to define two separate notions of ``defect group'' depending on the boson/fermion number of the quantity in question. It would be interesting to see whether the presence of a non-trivial defect group could be used as a way of constraining the resulting non-supersymmetric dynamics.



\chapter{6D SCFTS, CENTER-FLAVOR SYMMETRIES, AND STIEFEL-WHITNEY COMPACTIFICATIONS}

\section{Introduction}\label{sec:intro}

Since their discovery \cite{Witten:1995zh,Strominger:1995ac,Seiberg:1996qx},
6d SCFTs have been a fount of insight into the non-perturbative structure of quantum field theory in diverse dimensions.
In particular, knowledge of the six-dimensional theory and the compactification geometry can make hard-to-access non-perturbative features in lower-dimensional systems manifest. On the other hand, general arguments indicate that such theories cannot be realized via perturbations of a Gaussian fixed point, and so in this sense they are intrinsically strongly coupled \cite{Cordova:2016xhm}. This in turn complicates the construction and study of such theories.

A conjectural classification of all 6d SCFTs was proposed in \cite{Heckman:2013pva,Heckman:2015bfa} (see also \cite{DelZotto:2014hpa, Heckman:2014qba, Bhardwaj:2015xxa, Tachikawa:2015wka, Bhardwaj:2015oru, Bhardwaj:2018jgp, Heckman:2018pqx, Bhardwaj:2019hhd,Distler:2022yse}). The main idea in this classification program is to engineer such theories via F-theory backgrounds involving a non-compact elliptically fibered Calabi--Yau threefold with a canonical singularity. This has led to a vast class of new theories, and a remarkably simple unifying description of nearly all such theories on their partial branch as generalized quiver gauge theories. This perspective has been used to extract a number of calculable quantities from such systems, including, for example the anomaly polynomial \cite{Intriligator:2014eaa, Ohmori:2014pca, Ohmori:2014kda, Cordova:2018cvg}, as well as operator scaling dimensions of certain operator subsectors \cite{Heckman:2014qba, Bergman:2020bvi,Baume:2020ure, Heckman:2020otd, Baume:2022cot}. Compactification of such theories to four-dimensional systems also provides a systematic way to generate a broad class of 4d SCFTs with varying amounts of supersymmetry \cite{Gaiotto:2015usa,DelZotto:2015rca,Franco:2015jna,Coman:2015bqq,Garcia-Etxebarria:2016erx,Razamat:2016dpl,Baume:2021qho,Ohmori:2015pua,Ohmori:2015pia,Mekareeya:2017jgc,Mekareeya:2017sqh,Heckman:2016xdl,Bah:2017gph,Bourton:2017pee,Kim:2017toz,Apruzzi:2018oge,Razamat:2018zus,Kim:2018lfo,Razamat:2018gro,Kim:2018bpg,Razamat:2018gbu,Chen:2019njf,Pasquetti:2019hxf,Sela:2019nqa,Razamat:2019mdt,Razamat:2019ukg,Razamat:2020bix,Sabag:2020elc,Bourton:2020rfo,Nazzal:2021tiu,Kang:2021lic,Hwang:2021xyw,Kang:2021ccs,Bourton:2021das,Razamat:2022gpm}.\footnote{A recent overview of superconformal field theories in dimensions three to six is \cite{Argyres:2022mnu}.}

In general terms, global symmetries also play an important role in constraining correlation functions of local operators, and also figure into the analysis of higher symmetries \cite{Gaiotto:2014kfa}. This is no less true in 6d SCFTs, and also plays an important role in in the study of compactifications of such systems. As a recent example, \cite{Ohmori:2018ona} (see also \cite{Razamat:2016dpl}) demonstrated that starting from certain 6d $\mathcal{N} = (1,0)$ SCFTs, compactification on a $T^2$ in the presence of a topologically non-trivial but flat bundle associated with an 't Hooft magnetic flux can be used to generate a class of 4d $\mathcal{N} = 2$ SCFTs. In particular, this requires knowing not just the global symmetry \textit{algebra} of the 6d theory, but the actual \textit{group}.

Our aim in this chapter will be to extract the continuous zero-form group symmetries of 6d SCFTs, and to use this in the construction of 4d $\mathcal{N} = 2$ SCFTs via Stiefel--Whitney twisted compactifications. Now, although the actual method of constructing such 6d SCFTs involves the geometry of the F-theory compactification, geometry can sometimes obscure some of the symmetries \cite{Bertolini:2015bwa}. These top down considerations can often be supplemented by various bottom up considerations, including Higgsing from theories with known flavor symmetry algebras \cite{Heckman:2016ssk,Heckman:2018pqx,Hassler:2019eso,Apruzzi:2020eqi}, and thus in many cases we know the continuous global symmetry algebra.
Consequently, we can specify a corresponding ``naive'' flavor symmetry $\widetilde{G}_{\text{flavor}}$, where all simple non-Abelian factors are simply connected, and there is no finite group action on any $U(1)$ factors. One can also supplement this by the R-symmetry $SU(2)_{R}$, which is difficult to track in the F-theory construction, but which must be present in any 6d SCFT.\footnote{Recall that in a supersymmetric theory, the flavor symmetry commutes with the supercharges, whereas the R-symmetry (by definition) does not.} Since we also have the gauge symmetry on the tensor branch, there is a corresponding ``naive'' group of continuous gauge and global zero-form symmetries:
\begin{equation}
\widetilde{G}_{\text{gauge-global}} \equiv \widetilde{G}_{\text{gauge}} \times \widetilde{G}_{\text{flavor}} \times SU(2)_{R}\,,
\end{equation}
where we have kept implicit the spacetime global symmetries of the field theory. This answer is ``naive'', in the sense that the matter content and effective strings (coupling to the tensor multiplet chiral two-forms) of the effective field theory may be neutral under some subgroup of the center of $\widetilde{G}_{\text{gauge-global}}$.
Consequently, the global form may end up being quotiented by a subgroup of the center $\mathcal{C} \subset \widetilde{G}_{\text{gauge-global}}$:
\begin{equation}
G_{\text{gauge-global}} = \widetilde{G}_{\text{gauge-global}} / \mathcal{C}\,.
\end{equation}
This quotient can also act on the spacetime symmetries since the supercharges transform as spacetime spinors and R-symmetry spinors. 
The combined action on the gauge and flavor symmetry is often referred to as a ``center-gauge-flavor symmetry'', generalizing the notion of ``center-flavor symmetry" \cite{Cohen:1983sd,Aharony:2016jvv,Benini:2017dus,Cherman:2017tey,Shimizu:2017asf,Gaiotto:2017tne,Tanizaki:2017qhf,Tanizaki:2017mtm,Cordova:2018acb,Yonekura:2019vyz,Hidaka:2019jtv,Cordova:2019uob,Dierigl:2020myk,Apruzzi:2021vcu,Apruzzi:2021mlh,DelZotto:2022joo,Hubner:2022kxr,Cvetic:2022imb}.
Of course, from the perspective of the 6d SCFT, the defining data only makes reference to the global symmetries, and the same quotient, suitably projected, realizes the continuous part of the global symmetry group
\begin{equation}
G_{\text{global}} = \widetilde{G}_{\text{global}} / \mathcal{C}_{\text{global}}\,,
\end{equation}
in the obvious notation. In the conformal limit, the possible action of $\mathcal{C}_{\text{global}}$ on these internal symmetries 
can also be accompanied by a quotient on the conformal group. As already implicitly mentioned, knowing the global form of the zero-form symmetry group has important implications for the existence and structure of higher-dimensional defects in the theory, informing possible higher symmetry structures.

One of our core tasks will be to present a general algorithm for extracting $G_{\text{gauge-global}}$ and $G_{\text{global}}$ of the tensor branch effective field theory. This also amounts (upon projecting onto the global symmetry factors) to a prediction for the global continuous zero-form group of the 6d SCFT.
Extracting this data directly from the corresponding F-/M- theory background geometry \cite{Cvetic:2022imb} was recently carried out for a number of 5d supersymmetric quantum field theories obtained from circle reduction of the tensor branch of a 6d SCFT, and the ``bottom up'' approach developed here agrees with the ``top down'' results obtained in \cite{Cvetic:2022imb}. From a bottom up perspective, we simply work on
on the tensor branch where we have access to the large symmetry transformations of the system, and the correlated response from transformations on the chiral two-forms of the effective field theory. Such transformations are in turn sensitive to the global topology of background gauge/global bundle configurations \cite{Apruzzi:2020zot}.\footnote{We note that, since this analysis relies solely on the effective field theory description on the tensor branch, it can also be carried out straightforwardly for theories constructed from frozen singularities \cite{Tachikawa:2015wka, Bhardwaj:2018jgp}.} This technique has been used previously to explore some examples of non-Abelian flavor symmetry in related systems \cite{Apruzzi:2020zot, Apruzzi:2021mlh},\footnote{In gravitational theories where there are no global symmetries, the same methods give constraints on the global form of gauge symmetries \cite{Apruzzi:2020zot,Cvetic:2020kuw}, which for supergravity models in high dimensions are found to agree with patterns in string compactifications \cite{Font:2020rsk,Font:2021uyw,Cvetic:2021sjm,Cvetic:2022uuu}.} but as far as we are aware, a systematic study of all possibilities was not previously undertaken. In particular, we also show how to incorporate continuous Abelian symmetries.

Moreover, our analysis also extends to the global form of the R-symmetry, and its possible mixing with the center-flavor symmetry. This is difficult to extract from established index computations in the much-studied and related case of 4d theories (as obtained by compactification on a $T^2$ with no background bundles switched on), since in many cases, only specific R-charge sectors are counted.\footnote{We note that the possibility of mixing with the center of the R-symmetry resolves some puzzles in various claimed global forms of the flavor symmetry for certain 4d $\mathcal{N} = 2$ SCFTs which have appeared in earlier work (e.g., compare \cite{Distler:2019eky} with \cite{Bhardwaj:2021ojs}), a point we comment on in more detail later on.} In some cases, however, alternative methods have been explored for extracting the chiral ring of the corresponding Higgs branch \cite{Ferlito:2017xdq, Hanany:2018uhm}, which implicitly also determines a mod $2$ constraint on the global form of the center symmetry mixing with the R-symmetry. In these cases, we find that our analysis agrees with these constraints.

To illustrate the utility of this approach, we show in a number of examples how to extract the symmetry groups $G_{\text{gauge-global}}$ and
$G_{\text{global}}$. One large class of examples includes M5-brane probes of an ADE singularity $\mathbb{C}^2/{\Gamma_\text{ADE}}$ as well as their Higgs branch deformations. These flows are captured by group-theoretic data associated with nilpotent and semi-simple deformations of the corresponding flavor symmetry algebras. Since the corresponding tensor branch descriptions for these theories are all known, we can use our method to extract the corresponding continuous symmetry \textit{group}, including contributions from Abelian symmetry factors and mixing with the R-symmetry. Similar considerations hold for the ``orbi-instanton theories'' obtained from Higgs branch deformations of
M5-branes probing an ADE singularity $\mathbb{C}^2 / \Gamma_\text{ADE}$ wrapped by an $E_8$ nine-brane. In this case, deformations of the $E_8$ flavor symmetry factor are captured by finite group homomorphisms $\Gamma_\text{ADE} \rightarrow E_8$.

Analyzing this class of examples, we observe that many breaking patterns wind up generating a trivial quotienting subgroup $\mathcal{C}$ for the global symmetry. This occurs simply because, in many cases, there is no common center for the simply connected non-Abelian symmetry group factors. A general rule of thumb for realizing a common center-gauge-flavor symmetry is that the group-theoretic data such as a nilpotent orbit or a finite group homomorphism must have a sufficient multiplicity so that there is a non-trivial finite group action on the deformation parameter itself. This analysis also makes it clear that the vast majority of examples with non-trivial gauge-flavor symmetry mixing on the tensor branch will necessarily involve A-type symmetry algebras, simply because the corresponding Lie groups exhibit a far broader class of possible center subgroups (e.g., $SU(N)$ has center $\mathbb{Z}_N$), when compared with their non-A-type counterparts.

Once the center-flavor symmetry of a 6d SCFT is known, one can utilize it to generate a large class of lower-dimensional theories via  compactification. To illustrate, we primarily focus on the case of compactification on a $T^2$ in the presence of topologically non-trivial background bundle configurations. The corresponding 't Hooft magnetic fluxes are characterized by holonomies which commute in $G_{\text{flavor}} = \widetilde{G}_{\text{flavor}} / \mathcal{C}$, but which would not commute in $\widetilde{G}_{\text{flavor}}$ (see \cite{Witten:1997bs, Borel:1999bx}). These have been referred to as Stiefel--Whitney twisted theories in \cite{Ohmori:2018ona}. This provides a systematic way to generate a large class of 4d $\mathcal{N} = 2$ SCFTs. In particular, up to a small number of outliers, we show that after including further Higgs branch and mass deformations, this generates the full list of known rank two 4d $\mathcal{N} = 2$ SCFTs given in \cite{Martone:2021ixp}. The list of theories we generate in this way also has some overlap with other top-down constructions such as those based on D3-brane probes of $\mathcal{N} = 2$ S-folds (i.e., non-perturbative generalizations of an orientifold plane in the presence of a stack of flavor seven-branes) \cite{Apruzzi:2020pmv,Giacomelli:2020jel,Heckman:2020svr,Giacomelli:2020gee,Bourget:2020mez}. While there is indeed some overlap in 4d with suggestive evidence via string duality, we also find that there are some cases of Stiefel--Whitney twisted compactifications which resist a simple interpretation in terms of S-folds, an issue we leave for future investigations.

The rest of this chapter is organized as follows. We begin by giving a brief review of the tensor branch of a 6d SCFT, with a particular emphasis on topological terms. In Section \ref{sec:6d}, we study the global structure of the flavor symmetry group of 6d $(1,0)$ SCFTs using the tensor branch description. In particular, we extract the overall center-flavor symmetry, including Abelian factors, as well as non-trivial mixing with R-symmetry factors, illustrating with a number of examples. Section \ref{sec:e8CF} serves as an intermezzo between the 6d and 4d analysis; we extract the center-flavor symmetry for a large class of, so-called, orbi-instanton theories which we then use in the next section. In Section \ref{sec:sfolds} we turn to the resulting 4d $\mathcal{N} = 2$ SCFTs generated by Stiefel--Whitney twisted compactifications of such 6d SCFTs. This provides us with a large class of new theories, and we also comment on the similarities and differences with 4d $\mathcal{N} = 2$ S-fold constructions. In Section \ref{sec:Esfolds}, we briefly explore the DE-type generalizations of the A-type 6d and 4d SCFTs that were studied in Sections \ref{sec:e8CF} and \ref{sec:sfolds}. We present our conclusions and areas of future investigation in Section \ref{sec:conc2}. In Appendix \ref{app:JUSTHEFLUBRO}, we 
determine the continuous symmetry group for the $\mathcal{N} = (2,0)$ theories and the E-string theories. In Appendix \ref{app:ranktwo}, we show how to generate nearly all known rank two 4d $\mathcal{N}=2$ SCFTs via twisted Stiefel--Whitney compactifications, and we provide a comparison with previously obtained results in the literature in Appendix \ref{app:lit}. Appendix \ref{app:nilp} studies the nilpotent deformations in Stiefel--Whitney twisted theories inherited from the nilpotent deformations of their 6d parent theory. Finally, in Appendix \ref{app:CM}, we explore the global form of the flavor symmetry group for nilpotent deformations of conformal matter theories.

\section{Tensor Branch of 6d SCFTs}\label{sec:REVIEW}

In this section, we present a brief review of the tensor branch of a 6d SCFT, with a particular emphasis on the topological interaction terms.
Recently, much progress has been made in constructing 6d SCFTs by recasting the construction of such theories in terms of non-compact elliptically-fibered Calabi--Yau threefolds $X \rightarrow B$. In this description, one starts with a collection of curves in the base $B$, and with it a corresponding elliptic fibration. We can reach a conformal fixed point if the collection of curves can simultaneously contract to zero size. This results in a canonical singularity in the elliptic threefold (possibly partially frozen), and is the most systematic known method for realizing such theories \cite{Heckman:2013pva, Heckman:2015bfa}.

The configuration of curves prior to collapse gives a geometric realization of the so-called ``tensor branch'' of the 6d SCFT. In this regime, we have a collection of tensor multiplets, the bosonic content of each one consisting of a real scalar and an anti-chiral two-form potential. This anti-chiral two-form couples to effective strings, with tension controlled by the vacuum expectation value (vev) of the scalar.
We can potentially have 7-branes wrapped over each curve, and this results in non-Abelian gauge symmetries on the tensor branch. Collisions of 7-branes result in matter, which can include weakly coupled hypermultiplets, as well as (if we do not go to the full tensor branch) strongly coupled generalizations known as 6d conformal matter \cite{DelZotto:2014hpa,Heckman:2014qba}.

Letting $A^{ij}$ denote the intersection pairing matrix for curves in the base, a concise way to denote the tensor branch configuration is in terms of a quiver-like graph, where each node, denoted as $\overset{\fkg_i}{n_i}$, encodes the $i$th gauge algebra $\fkg_i$, whose associated tensor has self-pairing $A^{ii} = -n_i$. In what follows, we shall allow for the possibility that the gauge algebra is trivial, i.e., $\mathfrak{g}_i = \emptyset$, in which case no decoration is necessary. On the tensor branch, the condition of 6d gauge anomaly cancellation is, up to a small number of corner cases, enough to characterize the matter content of the tensor branch theory, including the spectrum of hypermultiplets.\footnote{There are a small number of cases, such as $\mathfrak{su}_6$ with $n_i = 1$ where the hypermultiplet spectrum is not uniquely fixed by the gauge algebra and self-pairing.}

Now, the F-theory model directly specifies the gauge symmetry, as associated with 7-branes wrapped on compact curves of the base $\mathcal{B}$, and this splits up into a collection of simple non-Abelian gauge symmetries:
\begin{equation}
\mathfrak{g}_{\text{gauge}} = \underset{i}{\bigoplus} \mathfrak{g}_{i}.
\end{equation}
Each factor here is a simple Lie algebra. Moreover, there are no gauged Abelian $\mathfrak{u}(1)$ factors, as follows directly from the structure of the local F-theory models \cite{Heckman:2013pva}. Turning next to the flavor symmetries of the 6d SCFT, the tensor branch description typically provides a good first approximation of the flavor symmetries of the 6d SCFT. For example, the hypermultiplets of the effective field theory often rotate under a global symmetry, and this persists at the conformal fixed point. In some cases, certain candidate flavor symmetries only become apparent once we approach the fixed point. A classic example of this phenomena is the E-string theory, namely the theory of an M5-brane probing an $E_8$ nine-brane. From the perspective of \cite{DelZotto:2014hpa,Heckman:2014qba}, the E-string, as well as other sub-configurations of matter fields can be viewed as the tensor branch of a generalized type of matter where the flavor symmetry is manifest, namely ``conformal matter''. All of this is to say that there is by now a general algorithm to read off the candidate non-Abelian flavor symmetry through a combination of the top-down F-theory geometry, and additional strong coupling enhancements (see, e.g., \cite{DelZotto:2014hpa,Heckman:2014qba,Heckman:2015bfa,Heckman:2016ssk} for some examples of such analyses). There is also a general algorithm for reading off candidate $\mathfrak{u}(1)$ symmetries which are free from mixed gauge symmetry/$\mathfrak{u}(1)$ anomalies, so-called ABJ anomalies \cite{Lee:2018ihr,Apruzzi:2020eqi}. Putting all of this together, the flavor symmetry algebra is of the general form:
\begin{equation}
\mathfrak{g}_{\text{flavor}} = \underset{a}{\bigoplus} \mathfrak{g}_{a} \oplus \underset{f}{\bigoplus} \mathfrak{u}(1)_{f},
\end{equation}
where each factor $\mathfrak{g}_{a}$ refers to a simple non-Abelian Lie algebra, and we have also included possible continuous Abelian symmetry factors. As a general point of notation, we shall distinguish the non-Abelian gauge and flavor symmetry algebras by the respective indices $i$ and $a$, while Abelian flavor symmetry algebras are indexed by $f$. Indeed, in the corresponding topological Green--Schwarz--Sagnotti--West terms, we will have couplings to both sorts of gauge bundle curvatures. Here, we have allowed for the possibility of various enhancements, as captured by working with conformal matter.
Finally, there is also the R-symmetry of the 6d SCFT, and this is also present on the tensor branch since it is unbroken. This provides an additional $\mathfrak{su}(2)_{R}$ global symmetry algebra.
Putting all of this together, the continuous global symmetry of the system is:
\begin{equation}
\mathfrak{g}_\text{gauge-global} = \mathfrak{g}_{\text{gauge}} \oplus \mathfrak{g}_{\text{flavor}} \oplus \mathfrak{su}(2)_R,
\end{equation}
where we have left implicit the spacetime symmetries. We again stress that in many cases, we can deduce the corresponding symmetry algebra from earlier work, so the main task reduces to determining the symmetry group, rather than just the algebra.

To accomplish this, we will need to know more about the topological sector of the theory. Much as in \cite{Apruzzi:2020zot}, we mainly claim that it suffices to study the topological terms of the tensor branch theory. Some of such terms are necessary for the theory to be free of gauge symmetry anomalies in the first place, while other terms inform us of global symmetry anomalies. All of these are captured by couplings between the anti-chiral two-forms and the Chern character of the non-Abelian gauge field strengths, as required to satisfy 6d anomaly cancellation via the Green--Schwarz--Sagnotti--West mechanism \cite{Green:1984bx, Sagnotti:1992qw}. Including background field strengths from global symmetries,
we get a set of topological couplings:
\begin{equation}\label{eq:green-schwarz-coupling}
    2 \pi \int_{{\cal M}_6} \Theta_i \wedge I^i \, ,
\end{equation}
where $\Theta_{i}$ refer to the anti-chiral two-forms of the $i$th tensor multiplet (denoted as $t_i$), and the $I^{i}$ are a collection of
four-forms:
\begin{align}
\begin{split}
    I^i = & - \sum_j A^{ij} \, c_2(F_j) - \sum_a B^{ia} \, c_2(F_a) + \sum_{f,f'} C^{i;f,f'} \frac{c_1(F_f) \wedge c_1(F_f')}{2} \\
          & + y^i c_2(R) - (2+A^{ii}) \tfrac14 p_1(T)\, .
\end{split}
\end{align}
Here, the $F_j$ refers to the gauge field strengths, the $F_a$ are the field strengths for the non-Abelian flavor symmetry factors, and the $F_f$ are the field strengths for Abelian symmetry factors, all of which we have expressed in terms of the corresponding Chern characters.\footnote{In our conventions, $\frac{1}{4}\text{Tr} F^2 = c_2(F)$ and $c_1(F) = \sqrt{-1} F$.} On the second line, we have also included the contribution from the R-symmetry, $c_2(R)$, as well as the first Pontryagin class of the spacetime tangent bundle.
Turning next to the coefficients appearing in $I^{i}$, the matrix $A^{ij}$ is, in our conventions, negative definite, and encodes the Dirac pairing for the effective strings, while the $B^{ia}$ are coefficients determined by the cancellation of all gauge-flavor anomalies, i.e., terms proportional to $\text{Tr}(F_i^2) \text{Tr}(F_a^2)$ in the full anomaly polynomial \cite{Ohmori:2014pca,Ohmori:2014kda,Intriligator:2014eaa,Cordova:2020tij}
\begin{equation}
    I_8 = I_\text{1-loop} + I_\text{GS} = I_\text{1-loop} - \frac12 (A^{-1})_{ij} I^i I^j \,.
\end{equation}
As an additional comment, the only $\mathfrak{u}(1)$ symmetry factors we can include are those which are free from ABJ-anomalies, which are encoded in the coefficients of $\text{Tr}(F_i^3) F_f$-terms of the anomaly polynomial.
Such terms must vanish at 1-loop for any quantum mechanically unbroken flavor $\mathfrak{u}(1)$.
For 6d SCFTs on their tensor branch, one can determine all such flavor $\mathfrak{u}(1)$s from a bottom-up approach \cite{Apruzzi:2020eqi} (see also \cite{Lee:2018ihr}). We note that, as opposed to non-Abelian flavor symmetries, these $\mathfrak{u}(1)$s are sometimes geometrically delocalized.

The main tool at our disposal for determining the global form of $G_{\text{gauge-global}}$ will be to track the global bundle structure of background field configurations using large symmetry transformations. Via the Green--Schwarz--Sagnotti--West mechanism, we know that this will also involve a non-trivial transformation from the anti-chiral two-forms $\Theta^{i}$, and the combined effect must be such that the full set of topological contributions remains invariant. We now proceed to the determination of this global structure.

\section{Topology of Global Symmetry Group for 6d SCFTs}\label{sec:6d}

In this section, we determine the global structure of the symmetry groups for 6d ${\cal N} = (1,0)$ SCFTs, based on their tensor branch characterization as a weakly-coupled gauge theory. In what follows, we assume that the symmetry algebra $\mathfrak{g}_{\text{gauge-global}}$ has already been specified. There is a corresponding ``naive'' answer for the zero-form symmetry:
\begin{equation}
\widetilde{G}_{\text{gauge-global}} = \widetilde{G}_{\text{gauge}} \times \widetilde{G}_{\text{flavor}} \times SU(2)_R \,,
\end{equation}
namely, for each non-Abelian Lie algebra, we take the corresponding simply connected Lie group, and all Abelian factors simply lift to $U(1)$. As before, we leave the spacetime symmetries implicit. The answer is naive, in the sense that this analysis does not distinguish between symmetries acting on genuine local operators, and those which are only defined as the endpoints of line operators (see \cite{Bhardwaj:2021wif, Lee:2021crt, Cvetic:2022imb, DelZotto:2022joo}). Indeed,
on general grounds, we expect that the actual zero-form symmetry group is quotiented by a subgroup of the common center for these factors. We shall refer to this as the gauge-global center symmetry, writing it as:
\begin{equation}
G_{\text{gauge-global}} = \widetilde{G}_{\text{gauge-global}} / \mathcal{C} \,.
\end{equation}
This leaves us with a residual center which is present in the actual tensor branch theory. With this in hand, we also have a candidate global symmetry for the 6d SCFT, as given by projection onto just the global symmetries of this quotient. Note that the group quotient specified by
$\mathcal{C}$ has a canonical restriction to just the global symmetries. In the obvious notation, we then have:
\begin{equation}
G_{\text{global}} = \widetilde{G}_{\text{global}} / \mathcal{C}_{\text{global}} \,.
\end{equation}

Our aim will be to extract $G_{\text{global}}$ by determining the corresponding center symmetry group $\mathcal{C}$. The analysis of this proceeds in several stages. First of all, we must require that all matter fields, including weakly coupled hypermultiplets as well as generalizations such as E-strings and conformal matter are all neutral under $\mathcal{C}$. Additionally, precisely because the group of gauge transformations in the 6d tensor branch theory also requires an accompanying transformation of the chiral two-forms of the associated tensor multiplets, we must \textit{also} require that the corresponding effective strings are neutral under $\mathcal{C}$ (see, e.g.,\cite{Apruzzi:2020zot}). In practical terms, what this amounts to is analyzing the topological sector of the tensor branch theory, and the response of the effective action under large field transformations. This leads to a non-trivial correlation between candidate 0-form symmetry bundles, which will in turn allow us to read off $G_{\text{global}}$.

In the rest of this section, we spell out the steps for extracting $G_{\text{global}}$ directly from the tensor branch. First, we begin by tracking the mixed gauge-flavor center symmetry for non-Abelian symmetry factors. We then show how to incorporate continuous Abelian symmetry factors, and then turn to possible mixing with the R-symmetry factors. The specific case of the R-symmetry group is particularly subtle, since it can evade detection via other means such as superconformal index computations. In each step, we present some illustrative examples, which we revisit to exhibit the full global symmetry structure.

\subsection{Anomalies for Center--Flavor Symmetry}\label{sec:cfanom}

We begin by considering the core example, based on mixing between the center of the gauge groups and non-Abelian symmetry factors. For now, we therefore suppress the contributions from Abelian symmetry factors as well as the R-symmetry. For a non-Abelian flavor symmetry that rotates matter charged under a gauge symmetry, a non-trivial global flavor symmetry structure generally requires a center-twisted gauge bundle that compensates the twisted flavor bundle. There is a potential obstruction to turning on such gauge and flavor bundles, which can be quantified from the tensor branch data \cite{Apruzzi:2020zot} and the topological couplings in equation \eqref{eq:green-schwarz-coupling}.

Turning on a center-twisted bundle for a simple algebra $\fkg$ (flavor or gauge), with simply-connected group $\widetilde{G}$ and center $Z(\widetilde{G})$ now leads to a fractionalization of $c_2(F)$ \cite{Kapustin:2014gua,Gaiotto:2014kfa,Gaiotto:2017yup,Cordova:2019uob}:
\begin{align}
    \tfrac14 \text{Tr}(F^2) = c_2(F) \equiv -\alpha_{\fkg} \, w(F) \cup w(F) \mod \bbZ \, .
\end{align}
Here, $w(F) \in H^2({\cal M}_6, Z(\widetilde{G}))$ is the $Z(\widetilde{G})$-valued characteristic class (the generalized Stiefel--Whitney class, also called the Brauer class in the mathematical literature) measuring the obstruction to lift a $\widetilde{G} / Z(\widetilde{G})$-bundle to a $\widetilde{G}$ bundle, and $w \cup w \equiv w^2$ is a 4-cocycle with integer periods.\footnote{To be precise, $c_2(F) \equiv \alpha_\fkg {\cal P}(w) \mod \bbZ$, where ${\cal P}$ is the Pontryagin square operation. If $w \in H^2({\cal M}, \bbZ_n)$, then for $n$ odd, ${\cal P}(w) \equiv w \cup w \in H^4({\cal M}, \bbZ_n)$; for $n$ even, ${\cal P}(w) \in H^4({\cal M}, \bbZ_{2n})$ reduces to $w \cup w$ modulo $n$.}
The fractionalization is due to the factors $\alpha_{\mathfrak{g}}$, whose fractional values depend $\fkg$ (with non-trivial center $Z(\widetilde{G})$):
\begin{equation}
\begin{aligned}
    & \mathfrak{g} = \mathfrak{su}_n \,  (\mathbb{Z}_n): \quad &&\alpha_\mathfrak{g} = \tfrac{n-1}{2n} \, , \qquad && \mathfrak{g} = \mathfrak{sp}_n \, (\mathbb{Z}_2): \quad &&\alpha_\mathfrak{g} = \tfrac{n}{4} \, , \\
    & \mathfrak{g} = \mathfrak{e}_{6} \,  (\mathbb{Z}_3): \quad &&\alpha_\mathfrak{g} = \tfrac{2}{3} \, , \qquad && \mathfrak{g} = \mathfrak{e}_{7} \, (\mathbb{Z}_2): \quad &&\alpha_\mathfrak{g} = \tfrac34 \, , \\
    & \mathfrak{g} = \mathfrak{so}_{4n+2} \, (\bbZ_4) : \quad &&\alpha_\mathfrak{g} = \tfrac{2n+1}{8} \, , \qquad && \mathfrak{g} = \mathfrak{so}_{2n+1} \, (\bbZ_2) : \quad &&\alpha_\fkg = \tfrac12 \, .
\end{aligned}
\end{equation}
In the case $\mathfrak{g} = \mathfrak{so}_{4n}$ and $\widetilde{G} = Spin(4n)$
with center $\mathbb{Z}_2^{(1)} \times \mathbb{Z}_2^{(2)}$, there are two contributions,
\begin{align}
    c_2 \equiv - \big( \tfrac{n}{4} (w^{(1)} + w^{(2)})^2 + \tfrac12 w^{(1)} \cup w^{(2)} \big) \mod \bbZ \, ,
\end{align}
originating from the center background $w^{(i)}$ of $\mathbb{Z}^{(i)}$.
In general, each $\mathbb{Z}_{\ell_s}$ factor of the full center $\prod_{i} Z(\widetilde{G}_i) \times \prod_a Z(\widetilde{G}_a) = \prod_s \bbZ_{\ell_s}$ is accompanied by a background field $w_s$. Again, the $i$ index refers to the gauge groups and the $a$ index refers to the non-Abelian flavor groups.

Because of the topological couplings in equation  \eqref{eq:green-schwarz-coupling}, a general background $\tilde{w} = (w_1 , ..., w_s, ...)$ for the center $\prod_s \bbZ_{\ell_s}$\footnote{Indexing by $s$ the individual cyclic factors distinguishes the two $\bbZ_2$ factors for a factor of $\widetilde{G}_g \cong Spin(4k_g)$.} will lead to a fractional 4-cocycle coupling to the tensor $\Theta_i$,
\begin{align}\label{eq:center_obstruction}
\begin{split}
    & \sum_j A^{ij} \, c_2(F_j) + \sum_a B^{ia} c_2(F_a) =: \sum_g {\cal A}^{ig} c_2(F_g) \\
    \equiv & \, - \sum_{\fkg_g \neq \mathfrak{so}(4n)} {\cal A}^{ig} \, \alpha_{\fkg_g} w_g^2 - \sum_{\fkg_g = \mathfrak{so}(4n_g)} {\cal A}^{ig} \left( \frac{n_g}{4} (w^{(1)}_g + w^{(2)}_g)^2 + \frac12 w^{(1)}_g \cup w^{(2)}_g \right)  \mod \bbZ \, ,
\end{split}
\end{align}
where the index $g$ runs over both gauge and flavor factors, and ${\cal A}^{ig}$ is the combined matrix of the tensor pairings $A^{ij}$ and non-Abelian flavor coefficients $B^{ia}$.
Because of this fractionalization, the action transforms anomalously under a large gauge transformation of the $i$th two-form tensor $\Theta_i$ \cite{Apruzzi:2020zot}, which poses an obstruction to turning on the corresponding twisted bundles.

However, for subgroups $Z \subset \prod_s \mathbb{Z}_{\ell_s}$, for which the $w_s$ are related to each other, it may be possible that different fractional contributions cancel, so that equation \eqref{eq:center_obstruction} is an integer class.
Concretely, for a cyclic $\bbZ_{n_r}$ subgroup with generator $(k^{(r)}_1, k^{(r)}_2, ...) \in  \prod_s \mathbb{Z}_{\ell_s}$, the corresponding center background is parametrized by $\tilde{w}^{(r)} = (k_1^{(r)} w^{(r)}, k_2^{(r)} w^{(r)}, \cdots)$, for a single independent 2-cocycle $w^{(r)}$.
If the fractionalizations vanish for a linear combination $\tilde{w} = \sum_r \tilde{w}^{(r)} = (\sum_r k_1^{(r)} w^{(r)}, \sum_r k_2^{(r)} w^{(r)}, ...)$ with generic backgrounds $w^{(r)}$ for a subgroup $Z = \prod_r \bbZ_{n_r}$, the global structure of the symmetry group is:\footnote{A short comment on notation: we reserve $\mathcal{C}$ for the full quotienting subgroup, with $Z$ the quotient on just the non-Abelian symmetry factors.}
\begin{equation}
G_{\text{gauge-global}} = \frac{\prod_i \widetilde{G}_i \times \prod_a \widetilde{G}_a}{Z}.
\end{equation}

Note that the candidate subgroups $Z$ of interest are in general severely limited by requiring that the hypermultiplet spectrum of the tensor branch theory must transform trivially under it. Any subgroup $Z$ of the full center which rotates these states by a non-trivial phase $\varphi \in U(1)$ is explicitly broken, i.e., one cannot twist the bundles by $Z$, regardless of the anomaly above.

For a simple group $G$ with $Z(G) = \bbZ_n$, a center element $x \,( \text{mod } n\bbZ) \in \bbZ_n$ acts on an irreducible representation ${\bf R}$ by the phase $\varphi^x({\bf R}) := (\varphi({\bf R}))^x$, with the phase $\varphi({\bf R})$ for the generator $1 \in \bbZ_n$ computed as follows:\footnote{In general, any irrep ${\bf R}$ of $G$ defines an element $\varphi({\bf R}) \in \text{Hom}(Z(G), U(1)) = \widehat{Z(G)} \cong Z(G)$ of the Pontryagin-dual.
The phase $\varphi^x({\bf R}) \in U(1)$ is then just the image of $x$ under $\varphi({\bf R})$.
}
\begin{itemize}
    \item for $G = SU(n)$, and ${\bf R}$ having a Young-tableaux with $m$ boxes, then $\varphi({\bf R}) = e^{2\pi i \tfrac{m}{n}}$;

    \item for $G = Sp(n)$, $\varphi({\bf R} = \text{fund}) = -1$ and $\varphi({\bf R} = \text{anti-sym}) = 1$;

    \item for $G = Spin(2n+1)$, $\varphi({\bf R} = \text{vector}) = 1$ and $\varphi({\bf R} = \text{spinor}) = -1$;

    \item for $G = Spin(4n+2)$, $\varphi({\bf R} = \text{vector}) = -1$ and $\varphi({\bf R} = \text{spinor}) = i$;

    \item for $G = E_6$, $\varphi({\bf R} = \text{fund}) = e^{\frac{2\pi i}{3}}$;

    \item for $G = E_7$, $\varphi({\bf R} = \text{fund}) = -1$.
\end{itemize}
For $G = Spin(4n)$ with $Z(G) = \mathbb{Z}_2^{(1)} \times \mathbb{Z}_2^{(2)}$, the phases associated with the generator $(1,0)$ are $\varphi^{(1)}({\bf R} = \text{vector})  = \varphi^{(1)}({\bf R} = \text{spinor}) = -1$, $\varphi^{(1)}({\bf R} = \text{co-spinor}) = 1$, and those associated with $(0,1) \in Z(G)$ are $\varphi^{(2)}({\bf R} = \text{vector})  = \varphi^{(2)}({\bf R} = \text{co-spinor}) = -1$, $\varphi^{(1)}({\bf R} = \text{spinor}) = 1$.
For a general element $(x_1, x_2) \in Z(G)$, the phase is then $\varphi^{(x_1,x_2)} ({\bf R}) = (\varphi^{(1)}({\bf R}))^{x_1} \, (\varphi^{(2)}({\bf R}))^{x_2}$.

For the center of a semi-simple group $\prod_g G_g \ni x = (x_g)$, one can analogously compute the phase from acting on a representation ${\bf R} = \bigotimes_g {\bf R}_g \equiv ({\bf R}_1, {\bf R}_2,...)$ as $\varphi^x({\bf R}) = \prod_g \varphi^{x_g}_g({\bf R})$.
Hence, for $Z \subset Z(\prod_g G_g) = \prod_g Z(G_g)$ to leave \emph{all} hypermultiplets invariant, $\varphi^x({\bf R}) = 1$ for all $x \in Z$ and all representations ${\bf R}$ that appear.
If, in addition, the obstruction in equation \eqref{eq:center_obstruction} vanishes, we propose that it is consistent to turn on the corresponding center twist (see also \cite{Apruzzi:2021mlh}).
This includes in particular the examples studied in \cite{Ohmori:2018ona}, and also agrees with expectations from explicit geometric constructions, where one can show that excitations of BPS-strings are invariant under $Z$ \cite{Apruzzi:2020zot,Bhardwaj:2020phs,Apruzzi:2021mlh}.

\subsubsection{Examples}

We now turn to examples illustrating how we extract the non-Abelian flavor symmetries. Let us also note that recently in \cite{Cvetic:2022imb}, geometric methods were developed to directly extract the global symmetry group for 5d conformal matter, i.e., the circle reduction of 6d conformal matter. Our bottom up analysis agrees with the results found there.

\paragraph{Example 1:}
Consider the SCFT with tensor branch description:
\begin{align}\label{eq:A-type_quiver_with_suN}
    [\mathfrak{su}_N^{(L)} ] \, \overset{\mathfrak{su}^{(1)}_{N}}{2} \, \overset{\mathfrak{su}^{(2)}_{N}}{2} \, \cdots \, \overset{\mathfrak{su}^{(m-1)}_{N}}{2} \, \overset{\mathfrak{su}^{(m)}_{N}}{2} \, [\mathfrak{su}_N^{(R)}] \, ,
\end{align}
which consists of $m$ gauge factors $\mathfrak{su}^{(i)}_N$ and has two $\mathfrak{su}_N^{(a)}$ ($a = L, R$) flavor factors at each end of the quiver.
The hypermultiplet spectrum consists of bifundamentals between each adjacent factor of $SU(N)_L \times \prod_i SU(N)^{(i)} \times SU(N)_R$:
\begin{align}
    {\bf R}^{(1)} = ({\bf N}, \overline{\bf N}, {\bf 1}, {\bf 1}, \ldots) \, , \quad {\bf R}^{(2)} = ({\bf 1}, {\bf N}, \overline{\bf N}, {\bf 1}, \ldots) \, , \quad \cdots \,.
\end{align}
With the tensor pairing matrix $A^{ij}$ being the negative $SU(m+1)$ Cartan matrix, the anomalies proportional to $\text{Tr}(F_i^2) \text{Tr}(F_a^2)$ are cancelled by a Green--Schwarz term with $B^{iL} = \delta^{i,1}$ and $B^{iR} = \delta^{m,R}$.
So the relevant part of the GS-coupling in equation \eqref{eq:green-schwarz-coupling} is
\begin{align}
    \Theta_i \wedge \left( -c_2(F_{i-1}) + 2 c_2(F_{i}) - c_2(F_{i+1}) \right) ,
\end{align}
where $F_{0} := F_L$ and $F_{m+1} := F_R$.

It is easy to see that the hypermultiplet spectrum is invariant under the diagonal center $\bbZ_N$ with generator
\begin{align}
    (1,1,...) \in \bbZ^{(L)}_N \times \prod_i \bbZ^{(i)}_N \times \bbZ_N^{(R)} = Z(SU(N)_L \times \prod_i SU(N)^{(i)} \times SU(N)_R) \, .
\end{align}
Since for this generator, all $-c_2(F_i) \equiv \frac{N-1}{2N} w^2$ fractionalize equally, they cancel out for each tensor multiplet $t_i$.
Therefore, the non-Abelian symmetry group is $[SU(N)_L \times \prod_i SU(N)^{(i)} \times SU(N)_R] / \bbZ_N$, and the non-Abelian flavor symmetry of the SCFT is $[SU(N)_L \times SU(N)_R]/\bbZ_N$, which agrees with known results \cite{Bah:2017gph, Cvetic:2022imb}. As an additional comment, we note that this case also has an overall $\mathfrak{u}(1)$ flavor symmetry \cite{Bah:2017gph, Apruzzi:2020eqi}, so we will revisit it when we discuss Abelian symmetry factors.

\paragraph{Example 2:}
For $N\geq 5$, there is an SCFT with tensor branch description:
\begin{align}\label{eq:single_1-curve-example}
    \underset{[\#\bigwedge^2 = 1]}{\overset{\mathfrak{su}_{N}}{1}}\, [\mathfrak{su}_{N+8}] \, ,
\end{align}
with a bifundamental hypermultiplet ${\bf R}^{(1)} = ({\bf N}, \overline{\bf N + 8})$ under $\mathfrak{su}_N \oplus \mathfrak{su}_{N+8}$, and one anti-symmetric ${\bf R}^{(2)} = (\frac{{\bf N}({\bf N -1})}{\bf 2}, {\bf 1}) = (\bigwedge^2, {\bf 1})$ that is uncharged under the $\mathfrak{su}_{N+8}$ flavor.
The Green--Schwarz four-form for the single tensor $\Theta$ of self-pairing $-1$ contains
\begin{align}
    I \supset c_2(F_N) - c_2(F_{N+8}) \,,
\end{align}
which ensures the absence of any $\text{Tr}(F_N^2) \text{Tr}(F_{N+8}^2)$ anomaly.

Some basic arithmetic reveals that there can be at most a non-trivial $\bbZ_2 \subset \bbZ_N \times \bbZ_{N+8} = Z(SU(N) \times SU(N+8))$ that acts trivially on the hypermultiplets, and that this can only occur when $N$ is even. So, for $N$ odd, there is no center-flavor symmetry. Restricting to $N$ even, the $\mathbb{Z}_2$ subgroup is generated by the element $(\tfrac{N}{2}, \tfrac{N+8}{2}) \in \bbZ_N \times \bbZ_{N+8}$.
For this candidate subgroup, the center-flavor anomaly indeed vanishes:
\begin{align}
    c_2(F_N) - c_2(F_{N+8}) \equiv \big( \underbrace{ - \tfrac{N^2}{4} \tfrac{N-1}{2N} + \tfrac{(N+8)^2}{4} \tfrac{N+7}{2N+16}}_{=7+2N} \big) w^2 \mod \bbZ \,.
\end{align}
Therefore, the faithfully acting non-Abelian symmetry group for $N$ even is $[SU(N) \times SU(N+8)]/\bbZ_2$, and the non-Abelian flavor symmetry of the SCFT is $SU(N+8)/\bbZ_2$.

\paragraph{Example 3:}
Consider next the SCFT with tensor branch description
\begin{align}\label{eq:example_3_tensor_config}
    [\mathfrak{su}_3^{(L)}] \, \overset{\mathfrak{e}_6}{3} \, 1 \, \overset{\mathfrak{su}_2}{2} \, [\mathfrak{so}_7^{(R)}] \, ,
\end{align}
which has
\begin{equation}
    -A^{ij} = \begin{pmatrix} 3 & -1 & 0 \\ -1 & 1 & -1 \\ 0 & -1 & 2 \end{pmatrix} \,,
\end{equation}
and hypermultiplets in the representations
\begin{align}
    {\bf R}^{(1)} = (\overline{\bf 3}, {\bf 27}, {\bf 1}, {\bf 1}) \, , \quad {\bf R}^{(2)} = \tfrac12 ({\bf 1}, {\bf 1}, {\bf 2}, {\bf 8}) \, ,
\end{align}
under the symmetry factors $\mathfrak{su}_3 \oplus \mathfrak{e}_6 \oplus \mathfrak{su}_2 \oplus \mathfrak{so}_7$. In the above, the ``$\frac{1}{2}$'' denotes a half-hypermultiplet, with matter in the spinor representation of $\mathfrak{spin}_7 \simeq \mathfrak{so}_7$.
Note also that this theory contains an undecorated $-1$ curve, so it provides an example where a subalgebra of the
E-string theory flavor symmetry has been gauged.

Let us now turn to the global structure of the symmetry group. The naive answer is $\widetilde{G}_{\text{gauge-global}} = SU(3) \times E_6 \times SU(2) \times Spin(7)$. Observe that the matter fields are invariant under a $\bbZ_3 \times \bbZ_2$ subgroup of the full center, where the $\bbZ_3$ is the diagonal of $Z(SU(3) \times E_6) = \bbZ_3 \times \mathbb{Z}_3$, and the $\bbZ_2$ the diagonal of $Z(SU(2) \times Spin(7)) = \bbZ_2 \times \mathbb{Z}_2$. Consider next the Green--Schwarz coupling to the tensor $\Theta_2$ of the unpaired middle node. This is an E-string not touching the flavor factors at the ends of the quiver, we find:
\begin{align}
    \Theta_2 \wedge \left( -c_2(F_{\mathfrak{e}_6}) - c_2(F_{\mathfrak{su}_2}) \right) \equiv \Theta_2 \wedge \left(\tfrac23 w_{\bbZ_3}^2 + \tfrac14 w_{\bbZ_2}^2 \right) \mod \bbZ \, ,
\end{align}
which would induce an anomaly for the large gauge transformations of $\Theta_2$.

As explained in \cite{Apruzzi:2020zot,Bhardwaj:2020phs}, the inconsistency of turning on such a twisted background, despite the absence of non-invariant hypermultiplets, can be also attributed to the excitations of the E-string, which transform in $E_8$ representations.
By decomposing the adjoint under $\mathfrak{e}_8 \supset \mathfrak{e}_6 \oplus \mathfrak{su}_3 \supset \mathfrak{e}_6 \oplus \mathfrak{su}_2$,
\begin{align}
\begin{split}
    {\bf 248} & \rightarrow ({\bf 78, 1}) \oplus ({\bf 1, 8}) \oplus  ({\bf 27, 3})  \oplus (\overline{\bf 27}, \overline{\bf 3}) \\
    & \rightarrow ({\bf 78, 1}) \oplus ({\bf 1, 3}) \oplus ({\bf 1, 2})^{\oplus 2} \oplus (({\bf 27, 2}) \oplus ({\bf 27, 1}) + \text{c.c}) \oplus ({\bf 1,1}) \, ,
\end{split}\label{eq:branching_e8_to_u1}
\end{align}
we indeed find states (the fundamentals under $E_6$ and $SU(2)$, respectively), which break the $\bbZ_3$ and $\bbZ_2$ twists, respectively.
Therefore, the non-Abelian flavor group is $SU(3) \times Spin(7)$.
However, as we will see below, the two discrete twists can be compensated if we take into the account the existence of $U(1)$ flavor factors.

\subsection{Anomalies for Center Symmetries of Abelian Factors}\label{sec:ab}

In the previous subsection we primarily focused on the non-Abelian symmetry factors. In some cases, there can also be continuous Abelian symmetry factors, which in many cases are delocalized. The procedure for extracting the global form of the center-flavor symmetry is to start with the ``naive'' gauge-global symmetry $\widetilde{G}_{\text{gauge-flavor}}$, and to then determine large symmetry transformations compatible with the presence of these $U(1)$ symmetry factors. The common center $\mathcal{C} \subset \widetilde{G}_{\text{gauge-flavor}}$ then specifies the quotient $G_{\text{gauge-flavor}} = \widetilde{G}_{\text{gauge-flavor}} / \mathcal{C} $. Note that we will also need to determine the overall normalization of $\mathfrak{u}(1)$ charges, a point we turn to shortly.

The analysis of the global form again relies on the same sort of topological terms $\Theta_i \wedge I^{i}$ encountered in our analysis of non-Abelian flavor symmetries. In the present case with Abelian symmetries, we recall that this includes:
\begin{equation}\label{eq:green-schwarz-4form_with_U1s}
    I^i = - \sum_j A^{ij} c_2(F_j) - \sum_a B^{ia} c_2(F_a) + \sum_{f,f'} C^{i;f,f'} \frac{c_1(F_f) \wedge c_1(F_f')}{2} \, .
\end{equation}
If we now activate a discrete twist $Z_f \cong \bbZ_{n_f}$ of a $U(1)_f$ bundle, then $c_1(F_f) = i F_f$ acquires a fractional part, thus affecting the large gauge transformations of the $\Theta_i$s. From this, we see that the symmetry group takes the general form
\begin{align}
    \frac{[\prod_i \widetilde{G}^{\text{gauge}}_i \times \prod_a \widetilde{G}^{\text{flavor}}_a]/Z \times \prod_f U(1)_f}{\prod_f Z_{f}} \, ,
\end{align}
where as before, the $\widetilde{G}$s refer to simply connected non-Abelian factors. In particular, notice that the quotient by $Z$, which we obtained in the previous subsection, just involves the condition of neutrality under a restricted set of center-symmetry transformations associated with the \textit{non-Abelian} symmetry factors. There can, of course, be more general symmetry transformations which involve the \textit{Abelian} factors, and this is accounted for by the $Z_f$s.

Indeed, for each $\Theta_i$, the fractionalizations of $c_2(F_i)$, $c_2(F_a)$, and $c_1(F_f)$ from the twists $Z$ and $Z_{f}$ must cancel. Note in particular that the quotienting procedure worked out for the non-Abelian symmetry factor is not contaminated by the appearance of the $U(1)$ factors. Said differently, the quotienting group $Z$ may end up only being a subgroup of the full $\mathcal{C}$ used to reach $G_{\text{gauge-global}} = \widetilde{G}_{\text{gauge-global}} / \mathcal{C}$.

To figure out the global quotient by $Z_f$, we need to know the overall normalization of the matter fields under the Abelian symmetries. This is rather subtle, because for a $U(1)$ factor, rescaling the charges is always a possibility. Importantly, such rescaling effects do not end up affecting the global form of the quotienting procedure. To demonstrate this, we now turn to an analysis of charge normalization for Abelian factors, and then illustrate how this works for hypermultiplets and E-string theories.

\subsubsection{Abelian Charge Normalization}

To determine the overall charge normalization for Abelian symmetry factors, as well as the contribution from fractional Chern classes,
it is instructive to consider bundles with structure group $U(N) = [SU(N) \times U(1)_f]/\bbZ_N$.
One can express a $U(N)$ bundle in terms of an $SU(N)/\bbZ_N$ and a $U(1)_f$ bundle, with curvatures $F$ and $F_f$ correlated via:
\begin{align}\label{eq:c1_U(N)_bundle}
    c_1(F_f) \equiv \tfrac{1}{N} w \mod \bbZ \, ,
\end{align}
where $w$ is the (generalized) Stiefel--Whitney class of the $SU(N)/\bbZ_N$ bundle.
In the language of generalized symmetries, one can think of the two 1-form center symmetries of $SU(N)$ and $U(1)_f$ being correlated through a single 2-cocycle $w$.
Namely, the background gauge field $b_e^{(2)}$ of the 1-form symmetry of $U(1)_f$ (which is itself $U(1)$-valued), which imposes $\int_{\Sigma_2} (c_1(F_f) - b_e^{(2)}) \in \bbZ$ for any 2-cycle $\Sigma_2$, is tied to the value of the $\bbZ_N$ 1-form symmetry gauge field of $SU(N)$, which in turn fixes the Stiefel--Whitney class to $w$.
We can verify explicitly that the fractional parts of the $SU(N)/\bbZ_N$ bundle, $c_2(F) \equiv - \tfrac{N-1}{2N}w^2 \mod \bbZ$, and of the $U(1)_f$ bundle, $Nc_1(F_f)^2 \equiv \tfrac{1}{N} w^2 \mod \bbZ$, cancel in $c_2(U(N)) = c_2(F) + \tfrac{N(N-1)}{2} c_1(F_f)^2$, which is indeed an integer characteristic class.

Importantly, the relation in equation \eqref{eq:c1_U(N)_bundle} holds only in a normalization of the $\mathfrak{u}(1)$ generator $\hat{q}$ where the charges span $\bbZ$, i.e., the fundamental representation of $\mathfrak{u}(N)$ has charge 1 in this normalization, and representations that are singlets under $\mathfrak{su}(N) \subset \mathfrak{u}(N)$ have charges 0 mod $N$.
In this case, the trivially acting $\bbZ_N$ center is generated by $(-1, e^{2\pi i \hat{q}/N}) \in  \bbZ_N \times U(1) = Z(SU(N) \times U(1))$.

More generally, once we fix the $U(1)_f$ charges of all representations ${\bf R}_q$ of a group $[\prod_g \widetilde{G}_g \times U(1)_f]/Z_{f}$ (with ${\bf R}$ a representation of $\prod_g \widetilde{G}_g$) to span $\bbZ$, there is no ambiguity to specify the generator of $Z_{f}$ as
\begin{align}\label{eq:normalized_center_generator_with_u1}
    (k_1,k_2,...; e^{2\pi i \hat{q} \frac{u_f}{l_f}}) \in \prod_g Z(G_g) \times U(1)_f \,,
\end{align}
where parameters must satisfy $\varphi({\bf R})^{(k_1,k_2,...)} \, \exp(2\pi i q \tfrac{u_f}{l_f}) = 1$ for any representation ${\bf R}_q$ of $[\prod_g \widetilde{G}_g \times U(1)_f]/Z_{f}$.
Then, the corresponding twist of the symmetry bundle is in terms of a 2-cocycle $w$:
\begin{align}\label{eq:1-form_twists_normalized_with_u1}
    c_1(F_f) \equiv \frac{u_f}{l_f} w \mod \bbZ \, , \quad w(F_g) = k_g w \, ,
\end{align}
with the understanding that when $\widetilde{G}_g \cong Spin(4m_g)$, we have $k_g \equiv (k_g^{(1)}, k_g^{(2)})$, and $w(F_g) \equiv (w_g^{(1)}, w_g^{(2)})$. Note that it is $c_1(F_f) c_1(F_{f'})$ that enters the four-forms $I^i$, whose fractional part,
\begin{align}
    c_1(F_f) c_1(F_{f'}) \equiv \frac{u_f u_{f'}}{l_f l_{f'}} w \cup w' + \frac{u_f}{l_f} w \cup \chi' + \frac{u_{f'}}{l_{f'}} w' \cup \chi \mod \bbZ \, ,
\end{align}
may depend on the integral parts, $\chi$ and $\chi'$, of $c_1(F_f)$ and $c_1(F_{f'})$, respectively.

Now, since we are dealing with Abelian symmetry factors, we can in principle consider rescaling the charges of the states so that we only span a rescaled subgroup of $\mathbb{Z}$, e.g., $\mathbb{Z} \rightarrow \lambda \mathbb{Z}$. Doing so has no effect on the structure of the topological Green--Schwarz couplings.\footnote{In F-theory models there is often a ``geometrically preferred'' normalization where $SU(N)$-fundamentals have charge $\frac{1}{N} \mod \bbZ$ \cite{Cvetic:2017epq}.} Indeed, on general grounds, the fractionality of $C^{i;f,f'} c_1(F_f) c_1( F_{f'})$ does not depend on the normalization.\footnote{From the formulae for $C^{i;f,f'}$ we will discuss shortly, it can be seen explicitly that the effect $c_1(F_f) \rightarrow \tfrac{1}{\lambda} c_1(F_f)$ under a charge rescaling $q_f \rightarrow \lambda q_f$ is absorbed by $C^{i;f,f'}$.}
A convenient normalization convention for $c_1(F_f)$ is to first normalize $U(1)_f$ such that the corresponding charges span $\bbZ$ (and rescale $C^{i;f,f'}$ accordingly).
We can then determine the generator in equation \eqref{eq:normalized_center_generator_with_u1} of the candidate subgroup $Z_f$ that acts trivially on all states, from which the characteristic classes in equation \eqref{eq:1-form_twists_normalized_with_u1} follow.

\paragraph{\textbf{Hypermultiplets}}

To illustrate how this works, consider the case of weakly coupled hypermultiplets charged under some $U(1)$s. Indeed, on the full tensor branch, the only source of $\text{Tr}(F_i^2) F_f F_f'$-terms in the 1-loop anomaly polynomial are the hypermultiplets in representation ${\bf R}$ under $\prod_i \widetilde{G}^{\text{gauge}}_i \times \prod_a \widetilde{G}^{\text{flavor}}_a$ and with $U(1)$-charge vector $\vec{q}$:
\begin{align}
\begin{split}
    I_\text{hyper}({\bf R}_{\vec{q}}) \supset \frac{1}{24} \text{tr}_{{\bf R}_{\vec{q}}}({\cal F}^4) \supset & \sum_{f,f',i} \frac{h_i({\bf R})}{4} \text{Tr}(F_i^2) q_f q_{f'} \, F_f F_{f'} \\
    = & - \sum_{f,f',i} h_i({\bf R}) \, q_f \, q_{f'} \, c_2(F_i) \, c_1(F_f) \, c_1(F_{f'})\, ,
\end{split}
\end{align}
where the decomposition of the curvature ${\cal F}$ of the full symmetry bundle into those of the non-Abelian (gauge) part ($F_i$) and those of the $U(1)$s ($F_f = -i c_1(F_f)$) introduces the index of the representation $h_i({\bf R})$.\footnote{For ${\bf R} = ({\bf R}^{(1)}, {\bf R}^{(2)}, \cdots)$ an irrep of a semi-simple group $\prod_g G_g$, $h_i({\bf R}) = \prod_{g\neq i} \dim({\bf R}^{(g)}) \, h_{\fkg_i}({\bf R}^{(i)})$.
In our normalization of the trace, $h_{\mathfrak{su}_N}({\bf N}) = \frac12$. For values of other representations ${\bf R}$, see \cite{Heckman:2018jxk}, where these are denoted $h_{\bf R}$.} Much as in our analysis of non-Abelian gauge and flavor anomalies, requiring the Green--Schwarz contribution $I_\text{GS} = -\frac12 (A^{-1})_{ij} I^i I^j$ to cancel the above terms of the 1-loop anomaly polynomial uniquely fixes the coefficients $C^{i;f,f'}$ in the Green--Schwarz four-forms in equation \eqref{eq:green-schwarz-4form_with_U1s}:
\begin{align}\label{eq:U(1)_anomaly_coefficients}
    C^{i;f,f'} = \sum_{{\bf R}_{\vec q}} 2 h_i({\bf R}) q_f q_{f'} \, .
\end{align}
For F-theory models (in particular, those with compact internal geometries describing 6d supergravity), where $U(1)_f$ corresponds to a rational section $\sigma_f$ (more precisely, the Shioda-map of a rational section) of the elliptic fibration, the coefficient $C^{i;f,f'}$ is the geometric intersection number of the compact curve ${\cal C}_i$ carrying the gauge algebra $\fkg_i$ with the so-called height pairing divisor $\pi(\sigma_f, \sigma_{f'})$ \cite{Park:2011ji,Morrison:2012ei}. In some cases, this structure persists even in local models \cite{Lee:2018ihr, Apruzzi:2020eqi}.

\paragraph{\textbf{E-String Contributions}}

In most cases, the $U(1)$ symmetry only acts on weakly coupled hypermultiplets. The states from an E-string sector can also be charged under a $U(1)$ factor embedded in the $E_8$ flavor symmetry factor. We can also extract the charge normalization in this case, and thus track its contribution to the global structure of the symmetry.

Concretely, consider a maximal embedding $\mathfrak{e}_8 \supset \bigoplus_{\beta \geq -d} \mathfrak{h}_\beta \oplus \bigoplus_\gamma \mathfrak{u}(1)_\gamma$ with simple algebras $\mathfrak{h}_\beta$, of which the first $d \geq 1$ factors $\mathfrak{h}_{\beta <0}$ are gauged,\footnote{At most two simple factors can be gauged, thus $d \leq 2$ \cite{Heckman:2015bfa}.} i.e., paired with tensors $\Theta_{i_\epsilon}$ ($\epsilon \in \{ -1,...,-d\}$) having $A^{\hat{\imath}, i_\epsilon} = 1$.
This leaves the commutant, $\bigoplus_{\beta \geq 0} \mathfrak{h}_\beta \oplus \bigoplus_\gamma \mathfrak{u}(1)_\gamma$, as the flavor symmetry, which receives no 1-loop anomalies from hypermultiplets (hence, in particular, no ABJ anomaly for the $\mathfrak{u}(1)$ \cite{Apruzzi:2020eqi}).\footnote{For simplicity, we will only consider rank 1 E-strings. However, the generalization to rank $Q$ is straightforward with the results from \cite{Ohmori:2014pca,Ohmori:2014kda}.}
Nevertheless, besides those of other flavor factors with labels $(a,f,f')$ in equation \eqref{eq:green-schwarz-4form_with_U1s}, there is an E-string contribution to the Green--Schwarz four-form involving the flavor backgrounds $F_{\beta \geq 0}$ and $F_\gamma$, with \cite{Ohmori:2014kda,Ohmori:2014pca,Baume:2021qho},
\begin{align}
    B^{i \beta} = \delta^{i, \hat{\imath}} \, \ell_\beta \, , \quad C^{i;\gamma,\gamma'} = -\tfrac12 \delta^{i, \hat{\imath}} \, r_{\gamma,\gamma'} \,,
\end{align}
associated to the decomposition of the trace
\begin{align}\label{eq:decompose_trace}
    \text{Tr}(F_{\mathfrak{e}_8}^2) \rightarrow \sum_{\epsilon=-d}^{-1} \text{Tr}(F_{\mathfrak{h}_\epsilon}^2) + \sum_{\beta \geq 0} \ell_\beta \text{Tr}(F_{\beta}^2) + \sum_{\gamma,\gamma'} r_{\gamma,\gamma'} F_\gamma F_{\gamma'} = \sum_{\beta \geq -d} \ell_\beta \text{Tr}(F_{\beta}^2) + \sum_{\gamma,\gamma'} r_{\gamma,\gamma'} F_\gamma F_{\gamma'} \, .
\end{align}

The coefficients $\ell_\beta$ are the Dynkin indices of $\mathfrak{h}_\beta$ associated to the embedding $\bigoplus_{\beta \geq -d} \mathfrak{h}_\beta \oplus \bigoplus_\gamma \mathfrak{u}(1)_\gamma \subset \mathfrak{e}_8$, with those of the gauged subalgebras, $\mathfrak{h}_{\beta <0}$, necessarily being 1.\footnote{For the significance of Dynkin index one embeddings in F-theory, see \cite{Esole:2020tby}.}
To compute these coefficients, we can consider the decomposition of any representation,
\begin{align}
    {\bf R} \rightarrow \bigoplus_j ({\bf R}^{(j)}_{-d}, \cdots, {\bf R}^{(j)}_{\beta},\cdots)_{q^{(j)}_1,\cdots,q^{(j)}_\gamma,\cdots} \, ,
\end{align}
interpreted as a decomposition of a vector bundle
\begin{align}
    V = \bigoplus V^{(j)} \, , \quad \text{with} \quad V^{(j)} = \bigotimes_{\beta \geq -d} U^{(j)}_\beta \otimes \bigotimes_\gamma W^{(j)}_\gamma \, ,
\end{align}
where $U_\beta^{(j)}$ is an $\mathfrak{h}_\beta$-bundle in the representation ${\bf R}_\beta^{(j)}$, and $W^{(j)}_\gamma$ a $\mathfrak{u}(1)_\gamma$-bundle in the charge $q_\gamma^{(j)}$ representation.
Using the decompositions of the Chern character, $\text{ch}(A \otimes B) = \text{ch}(A) \, \text{ch}(B)$ and $\text{ch}(A \oplus B) = \text{ch}(A) + \text{ch}(B)$, we have (here $[\cdot]_2$ extracts the degree-2 component of the total Chern character)
\begin{equation}
        \text{Tr}(F_{\mathfrak{e}_8}^2)  = \tfrac{1}{h({\bf R})} \text{tr}_{\bf R}(F_{\mathfrak{e}_8}^2) = -\tfrac{2}{h({\bf R})} [\text{ch}(V)]_2 = -\tfrac{2}{h({\bf R})} \sum_j [\text{ch}(V^{(j)})]_2 \, ,
\end{equation}
with
\begin{equation}
\begin{aligned}
       \relax[\text{ch}(V^{(j)})]_2 = &  \sum_\beta  [\text{ch}(U_\beta^{(j)})]_2 + \big( {\textstyle \prod_\beta} \text{rk}(U_\beta^{(j)}) \big) \left( \sum_{\gamma < \gamma'} [\text{ch}(W^{(j)}_\gamma)]_1 [\text{ch}(W^{(j)}_{\gamma'})]_1 + \sum_\gamma [\text{ch}(W_\gamma^{(j)})]_2 \right) \\
       = & -\sum_\beta \frac{h({\bf R}_\beta^{(j)})}{2} \text{Tr}(F_{\mathfrak{h}}^2) + \big( {\textstyle \prod_\beta} \dim({\bf R}_\beta^{(j)}) \big) \left( \sum_{\gamma < \gamma'} q^{(j)}_\gamma q^{(j)}_{\gamma'} F_\gamma F_{\gamma'} + \sum_\gamma \frac{(q^{(j)}_\gamma)^2}{2} F_\gamma^2 \right) \,.
      \end{aligned}
\end{equation}
We have thus derived the coefficients in equation \eqref{eq:decompose_trace} as
\begin{align}\label{eq:coefficients_trace_decomposition_with_u1s}
    \ell_\beta = \sum_j \frac{h({\bf R}_\beta^{(j)})}{h({\bf R})} \, , \quad r_{\gamma, \gamma'} = -\sum_j \frac{{\textstyle \prod_\beta} \dim({\bf R}_\beta^{(j)})}{h({\bf R})} q^{(j)}_\gamma q^{(j)}_{\gamma'} \, .
\end{align}
Note that the values of $\ell_\beta$ relevant to gaugings of $\mathfrak{h}_\beta \rightarrow \mathfrak{e}_8$ can be found in \cite{Baume:2021qho}.
This result does not depend on the chosen representation ${\bf R}$, as long as the decomposition is done with a fixed normalization for each $\mathfrak{u}(1)_\gamma$.

Candidate subgroups of $Z(\prod_\beta \widetilde{H}_\beta \times \prod_\gamma U(1)_\gamma)$ that can be used to twist the symmetry bundles must leave the representations resulting from decomposing the ${\bf 248}$ of $E_8$ invariant, as associated with the 
decomposition of the adjoint-valued moment map operator of the E-string theory. For this candidate subgroup, we can then verify whether the twist induces any anomaly for the large gauge transformation of the E-string tensor multiplet.

\subsubsection{Examples}

Having presented a general prescription for incorporating the contribution from continuous Abelian symmetries, we now turn to some explicit examples, focusing on the same class of examples already treated in the case of the non-Abelian flavor symmetries.
For illustrative purposes, we only consider the background field of the center-flavor symmetry involving the $U(1)$ flavor symmetry.
In all cases, it is straightforwardly verified that the fractionalizations also cancel when we turn on the previously studied center twists involving only the non-Abelian flavor factors.

\paragraph{Example 1:}
The example in equation \eqref{eq:A-type_quiver_with_suN} of a chain of $m$ $\mathfrak{su}_N$ gauge nodes provides a simple example with a $U(1)_f$ flavor symmetry:
\begin{align}
    [\mathfrak{su}_N^{(L)} ] \, \overset{\mathfrak{su}^{(1)}_{N}}{2} \, \overset{\mathfrak{su}^{(2)}_{N}}{2} \, \cdots \, \overset{\mathfrak{su}^{(m-1)}_{N}}{2} \, \overset{\mathfrak{su}^{(m)}_{N}}{2} \, [\mathfrak{su}_N^{(R)}] \, .
\end{align}
There is an overall $U(1)$ which is free from ABJ anomalies \cite{Apruzzi:2020eqi}. The $m+1$ bifundamental hypermultiplets:
\begin{align}\label{eq:hyper_spectrum_22-example}
    {\bf R}^{(1)} = ({\bf N}, \overline{\bf N}, {\bf 1}, {\bf 1}, \ldots)_1 \, , \quad {\bf R}^{(2)} = ({\bf 1}, {\bf N}, \overline{\bf N}, {\bf 1}, \ldots)_1 \, , \quad \ldots
\end{align}
have equal charge $q$, which we normalize to 1. There are 1-loop contributions to the $\text{Tr}(F_i^2) F_f^2$-terms that come from the anomaly polynomial of the hypermultiplets:
\begingroup
\allowdisplaybreaks
\begin{align}
    & I_\text{hyper}({\bf R}^{(1)}) \supset \frac14 N \, \text{Tr}_{\overline{\bf N}} (F_1^2) F_f^2 = -\frac{N}{2} c_2(F_1) c_1(F_f)^2 \, , \nonumber\\
    & \vdots \nonumber\\
    & I_\text{hyper}({\bf R}^{(i)}) \supset \frac14 (N \, \text{Tr}_{\bf N}(F_{i-1}^2) + N \, \text{Tr}_{\overline{\bf N}}(F_i^2) ) \, F_f^2 = -\frac{N}{2} (c_2(F_{i-1}) + c_2(F_{i})) \, c_1(F_f)^2 \, , \nonumber\\
    & \vdots \nonumber\\
    & I_\text{hyper}({\bf R}^{(m+1)}) \supset \frac14 N \, \text{Tr}_{\bf N}(F_m^2) \, F_f^2 = -\frac{N}{2} c_2(F_m) c_1(F_f)^2 \, , \nonumber\\
    \Longrightarrow \quad & I_\text{hypers} = \sum_{i=1}^{m+1} I_\text{hyper}({\bf R}^{(m)}) \supset -N \left( \sum_{i=1}^{m} c_2(F_i) \right) \, c_1(F_f)^2 \, .
\end{align}
\endgroup
Including the Abelian flavor backgrounds in the Green--Schwarz four-form,
\begin{align}
    I^i \supset \sum_{j=1}^m (-A^{ij})c_2(F_j) -  B^{i,L} c_2(F_L) - B^{i, R} c_2(F_R) + \tfrac12 C^{i;f,f} c_1(F_f)^2 \, ,
\end{align}
with $B^{i,L} = \delta^{i,1}$ and $B^{i,R} = \delta^{i,m}$, one can cancel the above $c_2(F_i) c_1(F_f)^2$ terms in the full anomaly polynomial $I_8 \supset I_\text{hypers} - \frac12 (A^{-1})_{ij} I^i I^j$, by fixing the coefficients $C^{i;f,f}$ to be (see equation \eqref{eq:U(1)_anomaly_coefficients})
\begin{align}
    C^{i,f,f} = 2N \, .
\end{align}
Then, the Green--Schwarz mechanism couples each tensor $\Theta_i$ to
\begin{align}\label{eq:GS-coupling_example-22}
\begin{split}
    I^i \supset -c_2(F_{i-1}) + 2 c_2(F_i) - c_2(F_{i+1})  + N c_1(F_f)^2 \, ,
\end{split}
\end{align}
again with the convention $F_{0}:= F_L$ and $F_{m+1} = F_R$ being the $\mathfrak{su}_N^{(L/R)}$ flavor backgrounds.

Before, we have seen that, with trivial $c_1(F_f)$, one finds that a $\bbZ_N$ twist is possible, leading to the non-Abelian group structure $[SU(N)^{(L)} \times \prod_i SU(N)^{(i)} \times SU(N)^{(R)}]/\bbZ_N$.
To extend the analysis to the Abelian flavor factor, we first note that the hypermultiplet charges in equation \eqref{eq:hyper_spectrum_22-example} are already properly normalized, in that the charges of all matter states span $\bbZ_N$.
The spectrum is invariant under the $\bbZ_N$ center-flavor symmetry generated by
\begin{align}
    (1,2,...,m+2; e^{\frac{2\pi i}{N} \hat{q}}) \in
    Z(SU(N)^{(L)} \times \prod_i SU(N)^{(i)} \times SU(N)^{(R)} \times U(1)) \, .
\end{align}
This means that the first Chern-class of the $U(1)_f/\bbZ_N$-bundle and the Stiefel--Whitney class of the $SU(N)/\bbZ_N$ bundles are correlated via a single 2-cocycle $w$ as
\begin{align}\label{eq:shifts_classes_example-22}
\begin{split}
    & c_1(F_f) \equiv \tfrac{1}{N} w \mod \bbZ \, , \quad w(F_i) = (i+1)w \quad (i = 0,..., m+1) \, , \\
    \Longrightarrow \quad & c_2(F_i) \equiv -(i+1)^2 \tfrac{N-1}{2N} w^2 \mod \bbZ \, , \\
    & c_1(F_f)^2 \equiv \tfrac{1}{N^2} w^2 + \tfrac{2}{N} w \cup \chi \mod \bbZ \, ,
\end{split}
\end{align}
with $\chi$ an integer 2-cocycle.
Plugging these into equation \eqref{eq:GS-coupling_example-22}, one straightforwardly verifies that the non-integer parts for the tensor couplings vanish:
\begin{align}\label{eq:u1-center-anomaly-cancellation_A-type_quiver}
    \begin{split}
        2c_2(F_i) -c_2(F_{i-1}) - c_2(F_{i+1}) + N c_1(F_f)^2 \equiv \big(\underbrace{(i^2 - 2(i+1)^2 + (i+2)^2) \tfrac{N-1}{2N} + \tfrac{1}{N}}_{=1} \big) w^2 \text{ mod } \bbZ \, .
    \end{split}
\end{align}
Hence, the structure group admits also a $\bbZ_N \cong Z_f$ quotient
\begin{equation}
    [SU(N)_L \times \prod_i SU(N)^{(i)} \times SU(N)_R \times U(1)_f]/Z_f \,.
\end{equation}
This matches the intuition from M-theory constructions \cite{Bah:2017gph}, from which one expects the flavor symmetry group of this SCFT to be $S[U(N)_L \times U(N)_R]/\bbZ_N$: the $\bbZ_N$ in this quotient is the center-flavor symmetry involving just the $SU(N) \subset U(N)$ parts, while the quotient $Z_f$ is encoded in $S[U(N) \times U(N)] \cong [SU(N) \times SU(N) \times U(1)]/\bbZ_N$.

\paragraph{Example 2:}
The theory with tensor branch description as in equation \eqref{eq:single_1-curve-example}:
\begin{align}
    \underset{[\#\bigwedge^2 = 1]}{\overset{\mathfrak{su}_{N}}{1}}\, [\mathfrak{su}_{N+8}] \, ,
\end{align}
also has a flavor $U(1)_f$ free of ABJ anomalies \cite{Apruzzi:2020eqi}, under which the hypermultiplets have the following charges:
\begin{align}
    ({\bf N}, \overline{\bf N+8}) : \quad q= N-4 \, , \quad (\textstyle{\bigwedge^2}, {\bf 1}): \quad q=-(N+8) \, .
\end{align}
By equation \eqref{eq:U(1)_anomaly_coefficients}, the Green--Schwarz mechanism couples, to the single tensor $\Theta$,
the four-form
\begin{align}
    I \supset c_2(F_N) - c_2(F_{N+8}) + \tfrac12 C^{i;f,f} c_1(F_f)^2 \, ,
\end{align}
with
\begin{align}
    \tfrac12 C^{i;f,f} = N(N-1)(N+8) \, .
\end{align}
The conditions for an element $(k_1, k_2, e^{\frac{2\pi i}{l} \hat{q}}) \in \bbZ_N \times \bbZ_{N+8} \times U(1)_f$ to act trivially on ${\bf R}$ are
\begin{align}
    \begin{aligned}
        {\bf R} &= ({\bf N}, \overline{\bf N+8}) : && \quad \tfrac{k_1}{N} - \tfrac{k_2}{N+8} + \tfrac{N-4}{l} \equiv 0 \mod \bbZ \, , \\
        {\bf R} &= (\textstyle{\bigwedge^2}, {\bf 1}): && \quad \tfrac{2k_1}{N} - \tfrac{N+8}{l} \equiv 0 \mod \bbZ \, .
    \end{aligned}
\end{align}
For odd $N$, it turns out that there is no such combined transformation leaving the hypermultiplets invariant.

For even $N$, there is always a trivially acting combination, but the general solution is cumbersome, so we will focus on an example with $N=6$.
In this case, the solution is $(k_1, k_2, l) = (3,9,14)$, so the putative quotient is a $\bbZ_{14} \simeq \mathbb{Z}_2 \times \mathbb{Z}_7$.
Notice that the charges of the hypermultiplets have a greatest common divisor of two, so, in order to be in the proper $U(1)$ normalization, we have to divide the charges by two, which means that the value of the coefficient $C^{i;f,f}$ is divided by four, $\frac12 C^{i;f,f} = 105$.
In addition, in this normalization the twist inside the $U(1)_f$ is by $e^{2\pi i/7}$.
Hence, the fractionalization of the Chern classes for this discrete twist is
\begin{align}
\begin{split}
    & c_2(F_N) \equiv -9 \times \tfrac{5}{12} w^2 \mod \bbZ \, , \quad c_2(F_{N+8}) \equiv -81 \times \tfrac{13}{28} w^2 \mod \bbZ \, , \\
    & c_1(F_f) \equiv \tfrac{1}{7} w \mod \bbZ \quad \Rightarrow \tfrac12 C^{i;f,f'} c_1(F_f)^2 \equiv 105( \tfrac{1}{49} w^2 + \tfrac{2}{7} w \cup \chi) \mod \bbZ \, ,
\end{split}
\end{align}
for an integer cocycle $\chi$.
Almost miraculously, the fractional parts cancel out in $I$, thus verifying that the full symmetry group is:
\begin{equation}
G_{\text{gauge-flavor}} = \frac{SU(6) \times SU(14) \times U(1)}{\bbZ_{14}}\, .
\end{equation}
The superconformal flavor symmetry group is then
\begin{align}
G_{\text{flavor}} = \frac{SU(14) \times U(1)}{\bbZ_{14}} \, .
\end{align}

\paragraph{Example 3:}
Finally, let us return to the example in equation \eqref{eq:example_3_tensor_config} with tensor branch configuration
\begin{align}
    [\mathfrak{su}_3^{(L)}] \, \overset{\mathfrak{e}_6}{3} \, \underset{[\mathfrak{u}(1)_f]}{1} \, \overset{\mathfrak{su}_2}{2} \, [\mathfrak{su}_3^{(R)}] \, ,
\end{align}
where we now include the Abelian flavor factor.
The $\mathfrak{u}(1)_f$ is the commutant of $\mathfrak{e}_6 \times \mathfrak{su}_2$ inside $\mathfrak{e}_8$, under which the representations resulting from the branching in equation \eqref{eq:branching_e8_to_u1} of the $E_8$-adjoint have charges
\begin{align}
    {\bf 248} \rightarrow
({\bf 78, 1})_0 \oplus ({\bf 1, 3})_0 \oplus ({\bf 1, 2})_3 \oplus ({\bf 1,2})_{-3} \oplus (({\bf 27, 2})_1 \oplus ({\bf 27, 1})_{-2} + \text{c.c}) \oplus ({\bf 1,1})_0 \, .
\end{align}
These are uncharged under the non-Abelian flavor factors at the end of the quiver. In turn, the hypermultiplets
\begin{align}
    {\bf R}^{(1)} = (\overline{\bf 3}, {\bf 27}, {\bf 1}, {\bf 1})_0 \, , \quad {\bf R}^{(2)} = \tfrac12 ({\bf 1}, {\bf 1}, {\bf 2},  {\bf 8})_0 \,,
\end{align}
are uncharged under $U(1)_f$.
From this we find that the $\bbZ_3 \times \bbZ_2 \cong \bbZ_6 \subset Z(SU(3) \times E_6 \times SU(2) \times Spin(7))$ considered previously, which leaves the hypermultiplets invariant but not the E-string states, can be compensated by a $U(1)_f$ twist, such that both sectors are invariant.
This combined $\bbZ_6$ has generator
\begin{align}\label{eq:teenchoiceawards2022}
    (2,2,1,1; e^{\frac{2\pi i \hat{q}}{6}}) \in \bbZ_3^2 \times \bbZ_2^2 \times U(1) \cong Z(SU(3)_L \times E_6 \times SU(2) \times Spin(7)_R \times U(1)_f) \, ,
\end{align}
with fractionalizations
\begin{align}
\begin{split}
    & c_2(F_{L}) \equiv -\tfrac43 w^2 \, , \quad c_2(F_{\mathfrak{e}_6}) \equiv -\tfrac83 w^2 \, , \quad c_2(F_{\mathfrak{su}_2}) \equiv -\tfrac14 w^2 \, , \quad c_2(F_{R}) \equiv -\tfrac12 w^2 \, , \\
    & c_1(F_f) \equiv \tfrac16 w \quad \Rightarrow \quad c_1(F_f)^2 \equiv \tfrac{1}{36} w^2 + \tfrac{1}{3} w \cup \chi \mod \bbZ \, .
\end{split}
\end{align}
From the above decomposition involving the $U(1)$ charges and equation \eqref{eq:coefficients_trace_decomposition_with_u1s}, we further find that
\begin{align}
    \text{Tr}(F_{\mathfrak{e}_8}^2) \rightarrow \text{Tr}(F_{\mathfrak{e}_6}^2) + \text{Tr}(F_{\mathfrak{su}_2}^2) - 12 F_f^2 \, ,
\end{align}
where the $\mathfrak{e}_6$ and $\mathfrak{su}_2$ are gauged on the left and right, respectively, of the E-string.
With the formulae from \cite{Baume:2021qho} applied to the hypermultiplets above, this gives the flavor anomaly coefficients
\begin{align}
    B^{iL} = 6\delta^{i,1}\, \quad B^{iR} = \delta^{i, 3}\, , \quad C^{i;f} = 6 \delta^{i,2} \, .
\end{align}
Together with the matrix
\begin{equation}
-A^{ij} = \begin{pmatrix} 3 & -1 & 0 \\ -1 & 1 & -1 \\ 0 & -1 & 2 \end{pmatrix}\,,
\end{equation}
we can now verify that the above $\bbZ_6$ twist does not induce any anomaly for the large gauge transformations of the tensors:
\begin{align}
    \begin{split}
        & \Theta_1 : \quad \eta^{1j} I_j^{(4)} \supset 3 c_2(F_{\mathfrak{e}_6}) - 6 c_2(F_L) \equiv (-8 + 8)w^2 \mod \bbZ \, , \\
        & \Theta_2 : \quad \eta^{2j} I_j^{(4)} \supset -c_2(F_{\mathfrak{e}_6}) - c_2(F_{\mathfrak{su}_2}) + 3c_1(F_f)^2 \equiv \big( \underbrace{ \tfrac83 + \tfrac14 + \tfrac{1}{12}}_{=3} \big) w^2 \mod \bbZ \, , \\
        & \Theta_3: \quad \eta^{3j} I_j^{(4)} \supset 2c_2(F_{\mathfrak{su}_2}) - c_2(F_R) \equiv \big(-\tfrac12 + \tfrac12 \big) w^2 \mod \bbZ \, .
    \end{split}
\end{align}
From this, we conclude that the tensor branch gauge theory has symmetry group:
\begin{equation}
G_{\text{gauge-flavor}} = \frac{SU(3) \times E_6 \times SU(2) \times Spin(7) \times U(1)}{\bbZ_6},
\end{equation}
where the group action is specified by equation \eqref{eq:teenchoiceawards2022}.
This also provides a prediction for the SCFT flavor symmetry:
\begin{equation}
G_\text{flavor} = \frac{SU(3) \times Spin(7) \times U(1)}{\bbZ_6}.
\end{equation}

\subsection{Center Twists and R-Symmetry}\label{sec:R-symmetry_twist}

In addition to the flavor symmetries, all 6d $\mathcal{N} = (1,0)$ SCFTs have an $\mathfrak{su}(2)_R$ symmetry.
This, of course, is an additional global symmetry which can in principle also mix with the center of the gauge group and flavor symmetry.
It is also worth noting that this R-symmetry is not directly manifest in the target space geometry of the corresponding F-theory models, but is realized geometrically in various M-theory constructions of 6d SCFTs.

Now, before getting to the case of center / R-symmetry mixing in 6d SCFTs, it is already instructive to note that even in the context of 4d theories, entertaining this possibility resolves some apparent puzzles, which as far as we are aware have not been previously addressed in the literature.\footnote{We thank J.~Distler for helpful correspondence.} For example, in the context of 4d $\mathcal{N} = 2$ SCFTs, the $E_6$ Minahan--Nemeschansky was argued to have a non-Abelian $E_6 / \mathbb{Z}_3$ global symmetry \cite{Bhardwaj:2021ojs}, which is also in accord with some superconformal index computations \cite{Gadde:2010te}. On the other hand, a direct analysis of BPS states would appear to detect states in the $\mathbf{27}$ of $E_6$ \cite{Distler:2019eky}. The natural resolution of this puzzle is that the center of the $E_6$ flavor symmetry mixes with the $U(1)_R$ symmetry of an $\mathcal{N} = 2$ SCFT, namely we have the global structure $[E_{6} \times U(1)_R ] / \mathbb{Z}_3$.\footnote{In this example we make no statement about the global structure involving the $SU(2)_R$ R-symmetry.} This also agrees with expectations based on the D3-brane probe of an $E_6$ 7-brane.\footnote{For the rank one theory it is possible to construct BPS states on the Coulomb branch as junctions between an $E_6$ stack of 7-branes and a D3-brane. In this scenario the states carry charge under the gauge group of the D3-brane so the symmetry group is $(E_6 \times U(1)_{\text{D3}})/\mathbb{Z}_3$. However since a $U(1)_R$ transformation in this construction is simply a rotation in the space transverse to the 7-branes it can be identified with the $U(1)$ center of mass of the D3-brane; it is possible to identify $U(1)_R \sim U(1)_{\text{D3}}$ with $U(1)_R$ being the symmetry that survives at the conformal point.}

However, this cannot be the full story, since the theory contains, for example, the supercharges which are not charged under any flavor symmetries, but do transform under a discrete R-symmetry twist.
This twist can be naturally cancelled if we include the remaining parts of the superconformal symmetry group \cite{Distler:2020tub,Manschot:2021qqe}.
We will postpone a detailed analysis of this interplay in the above 4d example, and turn our attention to 6d SCFTs for now.

For these, the supercharges are in the fundamental representation of $\mathfrak{su}(2)_R$, but otherwise uncharged under any flavor symmetry.
A natural way to cancel the effects of the $\bbZ_2 = Z(SU(2)_R)$ twist would be to activate a $\bbZ_2 = Z(Spin(1,5))$ twist in the Lorentz group which acts on spinors such as the supercharges.
Therefore, whenever we contemplate turning on an R-symmetry twist, the minimal requirement for the theory to be invariant is if it is accompanied by a $\bbZ_2$ twist of the Lorentz symmetry.

Now, we observe that our tensor branch analysis naturally incorporate both twists, since the topological Green--Schwarz couplings capture the contribution from non-trivial R-symmetry bundles, as well as the tangent bundle which is associated to Lorentz symmetry. Indeed, the Green--Schwarz four-form,
\begin{align}
    I^i \supset y^i \, c_2(R) - (2 + A^{ii}) \tfrac14 p_1(T) \,,
\end{align}
contains the second Chern-class $c_2(R)$ of the R-symmetry bundle and the first Pontryagin class $p_1(T)$ of the tangent bundle.
The coefficient $y^i \equiv h^\vee_{\fkg_i}$ is fixed to be the dual Coxeter number of the $i$th gauge algebra $\fkg_i$ by requiring the cancellation of all mixed gauge-R-symmetry anomalies; if $\fkg_i = \emptyset$ (which requires $A^{ii} = -1$ or $-2$), the coefficient is set to be $y^i = 1$.
For the R-symmetry bundle, the fractionalization is just as for any other $SU(2)/\bbZ_2$ gauge or flavor bundle, $c_2(R) \equiv -\tfrac14 w_R^2 \mod \bbZ$.
To quantify the fractionalization of the tangent bundle, we will work under the assumption that a Wick rotation to Euclidean signature does not affect the results.
Then, $p_1(T) = -\tfrac12 \text{tr}_\text{vec}({\cal R}^2)$, where the trace over the curvature ${\cal R}$ is in the vector, or anti-symmetric representation, of $Spin(6) \cong SU(4)$.
For $SU(4)$, this is the same as the 1-instanton normalized trace, so we conclude that
$\tfrac14 p_1(T) = -\tfrac12 c_2(SU(4))$.
In Euclidean signature, the corresponding $\bbZ_2$ twist (which leaves the vector representation invariant) is generated by $2 \in \bbZ_4 = Z(SU(4))$, for which the fractionalization is
\begin{align}
    \tfrac14 p_1(T) = -\tfrac12 c_2(SU(4)) \equiv \tfrac12 \times 2^2 \times \tfrac38 w_R^2 \mod \bbZ \equiv \tfrac34 w_R^2 \mod \bbZ \, .
\end{align}
Then, a twist by a center-flavor symmetry can occur if the fractionalization of the gauge, flavor, R-symmetry, and tangent bundles cancel out in $I^i$ for every $i$.

We present some examples of R-symmetry / spacetime symmetry mixing for the tensor branch of the $\mathcal{N} = (2,0)$ and E-string SCFTs in Appendix \ref{app:JUSTHEFLUBRO}. These cases are a bit special in that the tensor branch has no gauge group factors. For the sake of illustrating this general phenomenon, we now turn to some examples with center-flavor symmetry mixing, and no additional $U(1)$ factors. In this case, the gauge-global 0-form symmetry is of the general form:
\begin{equation}
G_{\text{gauge-global}} = \frac{\widetilde{G}_{\text{gauge}} \times \widetilde{G}_{\text{flavor}} \times [SU(2)_R \times Spin(1,5)]}{\mathcal{C}},
\end{equation}
where $\mathcal{C}$ is a suitably defined quotienting subgroup.
The global symmetry group that acts on spacetime scalars is then
\begin{align}
    G_\text{global} = \frac{\widetilde{G}_\text{flavor} \times SU(2)_R}{{\cal C}} \, .
\end{align}

As a first example, consider the SCFT with tensor branch description
\begin{align}\label{eq:tensor_branch_-2_example}
    \overset{\mathfrak{su}_N}{2} \, [\mathfrak{su}_{2N}] \, .
\end{align}
In this case, the $p_1(T)$-term drops out of the Green--Schwarz coupling:
\begin{align}
    \Theta \wedge \left( 2 c_2(F_{\text{gauge}}) - c_2 (F_{\text{flavor}}) + N \, c_2(R) \right) \, ,
\end{align}
which fractionalizes, for general center-twisted backgrounds, as
\begin{align}\label{eq:R-sym_example_fractionalization}
    \Theta \wedge \left( - \tfrac{N-1}{N} w_g^2 + \tfrac{2N-1}{4N} w_{f}^2 - \tfrac{N}{4} w_R^2 \right) \mod \bbZ \, .
\end{align}
The well-known flavor symmetry group $SU(2N)/\bbZ_N$ results from a combined twist of the gauge and flavor factor, with trivial R-symmetry twist:
\begin{align}
    w_{f} = -2 w_g = -2 w_N \, , \quad w_R = 0 \, ,
\end{align}
which leads to an overall integer shift in the GS-coupling.

In order to turn on a $\bbZ_2$ twist of the R-symmetry (which, as discussed above, is always accompanied by a Lorentz group twist), we must first make sure that the hypermultiplets are invariant.
Since these transform in the fundamental of $\mathfrak{su}(2)_R$, such a twist acts with a phase $(-1)$, which must be cancelled by a suitable gauge or flavor symmetry twist.
In the present example, we can turn on the $\bbZ_2 \subset \bbZ_{2N} = Z(SU(2N))$ simultaneously to achieve this.
More precisely, we claim that the theory is invariant under the central subgroup with generators
\begin{align}\label{eq:gens_quotient_R-sym_example}
    \left. \begin{array}{ll}
        \bbZ_N: & a = (1,-2,0)  \\
        \bbZ_2: & b = (0, N, 1)
    \end{array} \right\}
    \in \bbZ_N \times \bbZ_{2N} \times \bbZ_2 = Z(SU(N) \times SU(2N) \times SU(2)_R) \, .
\end{align}
At the level of background fields, these twists correspond to the following correlations,
\begin{align}
    w_g = w_N, \quad w_R =w_2, \quad w_{f} = -2 w_{N} + N w_2 \, ,
\end{align}
where $w_N$ and $w_2$ are the background fields associated to the $\bbZ_N$ and the $\bbZ_2$ generator, respectively, in equation \eqref{eq:gens_quotient_R-sym_example}.
This indeed shifts the Green--Schwarz four-form by an integer class,
\begin{align}
    - \tfrac{N-1}{N} w_g^2 + \tfrac{2N-1}{4N} w_{f}^2 - \tfrac{N}{4} w_R^2 = w_N^2 - (2N-1) w_N \cup w_{R} + \tfrac{N(N-1)}{2} w_R^2 \equiv 0 \mod \bbZ \, .
\end{align}

To write down the global symmetry group structure, note that for odd $N$, we have $\bbZ_N \times \bbZ_{2N} \times \bbZ_2 \cong \bbZ_N \times \bbZ_N \times \bbZ_2 \times \bbZ_2$, and $a$ generates the diagonal of the two $\bbZ_N$ factors, while $b$ generates the diagonal $\bbZ_2$.
For even $N = 2n$, on the other hand, we can consider the $\bbZ_2$ generator $na + b = (n, 0, 1) \in \bbZ_{N} \times \bbZ_{2N} \times \bbZ_2$, which maps trivially onto the $\bbZ_{2N}$ factor of the flavor symmetry.
Therefore, the global symmetry group compatible with the large gauge transformations of the tensor is
\begin{align}\label{eq:R-sym_example1_global_form}
\renewcommand{\arraystretch}{2.2}
    G_\text{global} = \left\{ \begin{array}{l  l}
        \displaystyle\frac{SU(2N)/\bbZ_N \times SU(2)_R}{\bbZ_2} \, , & N \ \text{odd,} \\
        \displaystyle\frac{SU(2N)}{\bbZ_N} \times \frac{SU(2)_R}{\bbZ_2} \, , & N \ \text{even.}
    \end{array} \right.
\end{align}

Let us compare this with results known from the Higgs branch chiral ring.
Elements of this ring carry representations of the global symmetry of the SCFT, so a center-flavor symmetry must leave all combinations of flavor and R-symmetry representations  that can be found in the chiral ring invariant.
For the SCFT with tensor branch description as in equation \eqref{eq:tensor_branch_-2_example}, it turns out that the chiral ring generator with non-trivial center charges has representation $(\boldsymbol\wedge^N, {\bf N + 1})$ under $SU(2N) \times SU(2)_R$ \cite{Hanany:2018vph}.
As the $N$-index anti-symmetric representation, $\boldsymbol\wedge^N$, of $SU(2N)$ picks up a phase $(-1)$ under the generator of $\bbZ_{2N} = Z(SU(2N))$, this state is clearly invariant under the $\bbZ_N$ subgroup in equation \eqref{eq:gens_quotient_R-sym_example}.
Therefore, the flavor symmetry group $SU(2N)/\bbZ_N$ is also what the Higgs branch data sees.
Moreover, $(\boldsymbol\wedge^N, {\bf N + 1})$ transforms with phase $(-1)^N$ under $N \in \bbZ_{2N}$ and with phase $(-1)^N$ under $1 \in \bbZ_2 = Z(SU(2)_R)$.
So it is also invariant under the second generator in equation \eqref{eq:gens_quotient_R-sym_example}.
Hence, the global structure in equation \eqref{eq:R-sym_example1_global_form} is also predicted from the Higgs branch chiral ring.

To illustrate the importance of the fractionalization of the tangent bundle, we consider the minimal $(D_k, D_k)$ conformal matter theory, whose tensor branch gauge theory is
\begin{align}
    \overset{\mathfrak{sp}_{k-4}}{1} \, [\mathfrak{so}_{4k}] \, ,
\end{align}
containing a half-hypermultiplet ${\bf h}$ in the bifundamental representation ${\bf R} = \tfrac12 ({\bf 2k-8, {\bf 4k}})$, with ${\bf 4k}$ the vector of $\mathfrak{so}_{4k}$. The hypermultiplet ${\bf h}$ also transforms as the fundamental of $\mathfrak{su}(2)_R$.
The Higgs branch chiral ring is generated by a moment map $\mu$ transforming in the $({\bf adj}, {\bf 3})$ of the $\mathfrak{so}_{4k} \oplus \mathfrak{su}(2)_R$, thus uncharged under the center, and a generator $\mu^+$ transforming in the $({\bf S^+}, {\bf k-1})$ representation, where ${\bf S^+}$ is one of the $\mathfrak{so}_{4k}$ spinor representations \cite{Ferlito:2017xdq,Hanany:2018uhm},\footnote{Aspects of the Higgs branch of minimal $(D_k, D_k)$ conformal matter have recently been explored from the perspective of the conformal bootstrap \cite{Baume:2021chx}.} which we pick to be of positive chirality for definiteness.\footnote{In this case, the choice of chirality of the spinor generator is irrelevant, however, it can be relevant when minimal $(D_k, D_k)$ conformal matter is used as a building block for other 6d SCFTs \cite{Distler:2022yse}.}
Therefore, the action of an element
\begin{align}
    (a_g, (a_+, a_-), a_R) \in \bbZ_2 \times (\bbZ_2^+ \times \bbZ_2^-) \times \bbZ_2 = Z(Sp(k-4)) \times Z(Spin(4k)) \times Z(SU(2)_R) \,,
\end{align}
on the representations of ${\bf h}$ and $\mu^+$ give phases
\begin{align}
    \mu^+ : \ (-1)^{a_+ + k \, a_R} \, , \quad {\bf h}: \ (-1)^{a_g + a_+ + a_- + a_R} \, ,
\end{align}
which must be trivial for any element $(a_g, (a_+, a_-), a_R)$ of the quotienting subgroup $\mathcal{C}$.
For even $k$, this requires $a_+ = 0 \ \text{mod 2}$, and $a_g + a_- + a_R = 0 \ \text{mod 2}$.
This leaves two independent generators,
\begin{align}\label{eq:DkDk_even_twists}
    (a_g, (a_+, a_-), a_R) = (1,(0,1),0) \quad \text{and} \quad (1,(0,0),1) \qquad (k \ \text{even}) \,,
\end{align}
which correspond to the diagonal $\bbZ_2$ of $Z(Sp(k-4)) \times \bbZ_2^-$ and $Z(Sp(k-4)) \times Z(SU(2)_R)$, respectively.
For odd $k$, we instead have $a_+ + a_R = 0 \ \text{mod 2}$, and $a_g + a_- = 0 \ \text{mod 2}$, which has independent solutions corresponding to the generators
\begin{align}\label{eq:DkDk_odd_twists}
    (a_g, (a_+, a_-), a_R) = (1,(0,1),0) \quad \text{and} \quad (0,(1,0),1) \qquad (k \ \text{odd}) \,,
\end{align}
of the diagonal $\bbZ_2$ factors of $Z(Sp(4-k)) \times \bbZ_2^-$ and $\bbZ_2^+ \times Z(SU(2)_R)$, respectively.
Considering the gauge invariant representations that can appear in the SCFT, the Higgs branch therefore predicts the global symmetry group
\begin{equation}\label{eqn:GHBpro}
    G_\text{global}^\text{HB} = \begin{cases}
        Spin(4k) / \mathbb{Z}_2^{-} \times (SU(2)_R/\bbZ_2) \quad &\text{if $k$ even} \,, \\[0.2em]
        [Spin(4k) / \mathbb{Z}_2^{-} \times SU(2)_R]/\mathbb{Z}_2 \quad &\text{if $k$ odd} \, .
    \end{cases}
\end{equation}

This agrees with the analysis from the Green--Schwarz coupling,
\begin{align}
    \Theta \wedge I_4 := \Theta \wedge \left( c_2(F_\mathfrak{sp}) - c_2(F_\mathfrak{so}) + h^\vee_{\mathfrak{sp}_{k-4}} c_2(R) - \tfrac14 p_1(T) \right) \, ,
\end{align}
which, with $h^\vee_{\mathfrak{sp}_{k-4}} = k-3$, fractionalizes as
\begin{align}
    I_4 \equiv - \tfrac{k}{4} w_g^2 + \tfrac{k}{4} (w_+ + w_{-})^2 + \tfrac12 w_+ \cup w_- - \tfrac{k-3}{4} w_R^2 - \tfrac34 w_R^2 \mod \bbZ \, ,
\end{align}
where $w_g$, $(w_+, w_-)$, and $w_R$ are the generic background fields for $Z(Sp(k-4))$, $Z(Spin(4k))$, and $Z(SU(2)_R)$, respectively, and we have already imposed the correlation between the R-symmetry and Lorentz group twist.
Restricting to the twists in equations \eqref{eq:DkDk_even_twists} and \eqref{eq:DkDk_odd_twists} predicted by the Higgs branch, we find
\begin{align}
    \begin{split}
        k \ \text{even:} \quad & w_g = w_- + w_R, \ w_+ = 0 \quad \Rightarrow  I_4 \equiv -\tfrac{k}{2} ( w_R^2 + w_- \cup w_R) \equiv 0 \mod \bbZ \, , \\
        k \ \text{odd:} \quad & w_g = w_-, \ w_+ = w_R \quad \Rightarrow  I_4 \equiv \tfrac{k+1}{2} w_R \cup w_- \equiv 0 \mod \bbZ \, .
    \end{split}
\end{align}

\section{Intermezzo: Orbi-Instanton Theories}\label{sec:e8CF}

Throughout Section \ref{sec:6d}, we have demonstrated that, given the quiver description of the generic point of the tensor branch of a 6d $(1,0)$ SCFT, one can determine the global structure of the flavor symmetry group. Since we also wish to generate 4d theories via Stiefel--Whitney twisted compactifications on a $T^2$, we now turn to a rich class of examples where we can systematically study possible center-flavor symmetry mixing.

The theories we now consider are Higgs branch deformations of the ``orbi-instanton theories'', as obtained from as obtained in M-theory terms from M5-branes probing an ADE singularity wrapped by an $E_8$ nine-brane \cite{DelZotto:2014hpa}. Via a process of fission and fusion, these turn out to be the progenitors for all 6d SCFTs \cite{Heckman:2018pqx} realized in a geometric phase of F-theory. As shown in \cite{Heckman:2015bfa}, a large class of Higgs branch deformations are captured by a nilpotent orbit of $\mathfrak{g}$, $\sigma$, and a homomorphism $\rho:\,\Gamma_{\mathfrak{g}} \rightarrow E_8$. We denote the resulting theories as
\begin{equation}
    \Omega_{\mathfrak{g}, N}(\rho, \sigma) \,.
\end{equation}
It is natural to ask: does the pair $(\rho, \sigma)$ capture the presence or absence of center-flavor symmetry in a straightforward manner? We assume that $N$ is sufficiently large that the Higgsing by $\rho$ and $\sigma$ are uncorrelated on the tensor branch, and in this section we will focus on the case $\mathfrak{g} = \mathfrak{su}_K$. Furthermore, we will assume that $\sigma$ is the maximal nilpotent orbit given by the trivial embedding $\mathfrak{su}_2 \rightarrow \mathfrak{g}$. As we see, the condition on $\rho$ for $\Omega_{\mathfrak{g}, N}(\rho, \sigma)$ to have a non-trivial center flavor symmetry in these cases is straightforward.

We consider the rank $N$ $(\mathfrak{e}_8, \mathfrak{su}_K)$ orbi-instanton 6d SCFT, which has the tensor branch configuration
\begin{equation}
    12\overset{\mathfrak{su}_{2}}{2}\overset{\mathfrak{su}_{3}}{2}\cdots\overset{\mathfrak{su}_{K}}{2}\underbrace{\overset{\mathfrak{su}_{K}}{2}\cdots \overset{\mathfrak{su}_{K}}{2}}_{N-1} \,.
\end{equation}
This theory can be obtained in M-theory as the worldvolume theory of a stack of $N$ M5-branes probing a $\mathbb{C}^2/\mathbb{Z}_K$ orbifold singularity and on top of an M9-plane \cite{DelZotto:2014hpa}.
The non-Abelian part of the flavor symmetry of this theory is generically
\begin{equation}
    \mathfrak{e}_8 \oplus \mathfrak{su}_K \,.
\end{equation}
A Higgsing of the $\mathfrak{e}_8$ flavor is specified by a choice of homomorphism $\rho: \, \Gamma_{\mathfrak{su}_K} \cong \mathbb{Z}_K \rightarrow E_8$. Such homomorphisms, as explained by Kac \cite{MR739850}, are captured by a weighted partition of the Dynkin labels of the affine $E_8$ Dynkin diagram into $K$:
\begin{equation}\label{eqn:Zkembed}
    (a_1, a_2, a_3, a_4, a_5, a_6, a_{4^\prime}, a_{2^\prime}, a_{3^\prime}) \,,
\end{equation}
such that
\begin{equation}\label{eqn:Zkembed2}
    a_1 + 2(a_2 + a_{2^\prime}) + 3(a_3 + a_{3^\prime}) + 4(a_4 + a_{4^\prime}) + 5a_5 + 6a_6 = K \,.
\end{equation}
We find that Higgsing the $\mathfrak{e}_8$ by a homomorphism, represented by a tuple as in equation \eqref{eqn:Zkembed} whose non-zero entries are $\{a_{i^{(')}}\}$, leads to a 6d SCFT with center-flavor symmetry
\begin{equation}
    \mathbb{Z}_\ell = \mathbb{Z}_{\gcd(\{i\})} \, .
\end{equation}
As evident from equation \eqref{eqn:Zkembed2}, this $\bbZ_\ell$ is always a subgroup of $Z(SU(K))$, consistent with the fact that the Higgsed theory has an $\mathfrak{su}_K$ flavor algebra.
For each $E_8$-homomorphism specified by equations \eqref{eqn:Zkembed} and \eqref{eqn:Zkembed2}, there exists an algorithm that determines the tensor branch configuration \cite{Mekareeya:2017jgc}. These tensor branch descriptions, for each of the putative $\mathbb{Z}_\ell$-preserving $E_8$-homomorphisms, are written in Table \ref{tbl:6dscfts};\footnote{In fact, when $t \neq 0$, the tensor branch description on the first line of Table \ref{tbl:6dscfts} corresponds to \emph{two} 6d SCFTs, depending on the choice of $\theta$-angle for the $\mathfrak{sp}_q$ gauge algebra on the $(-1)$-curve. These theories have the same central charges and flavor symmetries, but differ in the spectrum of local operators at large conformal dimension. See \cite{Mekareeya:2017jgc,Distler:2022yse} for more details; we suppress this subtlety in this chapter.} in each case one can then use the study of the large gauge transformation anomalies to verify that there is indeed a $\mathbb{Z}_\ell$ center-flavor symmetry. As the tensor branch configurations are rather involved, we explicate the analysis in one example.

\begin{sidewaystable}[]
    \centering
    \footnotesize
    \begin{threeparttable}
    \begin{tabular}{cc}
    \toprule
        $\mathbb{Z}_\ell$ & Tensor branch description of the 6d $(1,0)$ SCFT \\\midrule
        \multirow{3}{*}{$\mathbb{Z}_2$}  & $\overset{\mathfrak{sp}_{q}}{1}
        \underbrace{\overset{\mathfrak{su}_{2q + 8}}{2}
        \cdots\overset{\mathfrak{su}_{2q + 8t}}{2}}_{t}
        \underbrace{\overset{\mathfrak{su}_{2q + 8t + 6}}{2}
        \cdots\overset{\mathfrak{su}_{2q + 8t + 6u}}{2}}_{u}
        \underbrace{\overset{\mathfrak{su}_{2q + 8t + 6u + 4}}{2}
        \cdots\overset{\mathfrak{su}_{2q + 8t + 6u + 4s}}{2}}_{s}
        \underbrace{\overset{\mathfrak{su}_{2q + 8t + 6u + 4s + 2}}{2}
        \cdots\overset{\mathfrak{su}_{2q + 8t + 6u + 4s + 2p}}{2}}_p
        \underbrace{\overset{\mathfrak{su}_{2q + 8t + 6u + 4s + 2p}}{2}\cdots \overset{\mathfrak{su}_{2q + 8t + 6u + 4s + 2p}}{2}}_{N-1}$  \\

        &  $\overset{\mathfrak{su}_{2q+4}}{1}
        \underbrace{\overset{\mathfrak{su}_{2q + 12}}{2}
        \cdots\overset{\mathfrak{su}_{2q + 8t + 4}}{2}}_{t}
        \underbrace{\overset{\mathfrak{su}_{2q + 8t + 10}}{2}
        \cdots\overset{\mathfrak{su}_{2q + 8t + 6u + 4}}{2}}_{u}
        \underbrace{\overset{\mathfrak{su}_{2q + 8t + 6u + 8}}{2}
        \cdots\overset{\mathfrak{su}_{2q + 8t + 6u + 4s + 4}}{2}}_{s}
        \underbrace{\overset{\mathfrak{su}_{2q + 8t + 6u + 4s + 6}}{2}
        \cdots\overset{\mathfrak{su}_{2q + 8t + 6u + 4s + 2p + 4}}{2}}_p
        \underbrace{\overset{\mathfrak{su}_{2q + 8t + 6u + 4s + 2p + 4}}{2}\cdots \overset{\mathfrak{su}_{2q + 8t + 6u + 4s + 2p + 4}}{2}}_{N-1}$  \\\midrule

        \multirow{6}{*}{$\mathbb{Z}_3$}  & $\overset{\mathfrak{su}_{3}}{1}
        \underbrace{\overset{\mathfrak{su}_{12}}{2}
        \cdots\overset{\mathfrak{su}_{9q+3}}{2}}_{q}
        \underbrace{\overset{\mathfrak{su}_{9q+9}}{2}
        \cdots\overset{\mathfrak{su}_{9q+3+6s}}{2}}_{s}
        \underbrace{\overset{\mathfrak{su}_{9q + 6s + 6}}{2}
        \cdots\overset{\mathfrak{su}_{9q + 6s + 3p + 3}}{2}}_{p}
        \underbrace{\overset{\mathfrak{su}_{9q + 6s + 3p + 3}}{2}\cdots \overset{\mathfrak{su}_{9q + 6s + 3p + 3}}{2}}_{N-1}$ \\

         & $1 \underbrace{\overset{\mathfrak{su}_{9}}{2}
         \cdots\overset{\mathfrak{su}_{9q}}{2}}_{q}
         \underbrace{\overset{\mathfrak{su}_{9q+6}}{2}
         \cdots\overset{\mathfrak{su}_{9q+6s}}{2}}_{s}
         \underbrace{\overset{\mathfrak{su}_{9q + 6s + 3}}{2}
         \cdots\overset{\mathfrak{su}_{9q + 6s + 3p}}{2}}_{p}
         \underbrace{\overset{\mathfrak{su}_{9q + 6s + 3p}}{2}\cdots \overset{\mathfrak{su}_{9q + 6s + 3p}}{2}}_{N-1}$ \\

          & $\overset{\mathfrak{su}_{6}^\prime}{1}
         \underbrace{\overset{\mathfrak{su}_{15}}{2}
         \cdots\overset{\mathfrak{su}_{9q+6}}{2}}_{q}
         \underbrace{\overset{\mathfrak{su}_{9q + 12}}{2}
         \cdots\overset{\mathfrak{su}_{9q+6 + 6s}}{2}}_{s}
         \underbrace{\overset{\mathfrak{su}_{9q + 6s + 9}}{2}
         \cdots\overset{\mathfrak{su}_{9q + 6s + 3p + 6}}{2}}_{p}
    \underbrace{\overset{\mathfrak{su}_{9q + 6s + 6s + 3p + 6}}{2}\cdots \overset{\mathfrak{su}_{9q + 6s + 3p + 6}}{2}}_{N-1}$ \\\midrule

        \multirow{4}{*}{$\mathbb{Z}_4$}  & $\overset{\mathfrak{su}_{4}}{1} \underbrace{\overset{\mathfrak{su}_{12}}{2}
        \cdots\overset{\mathfrak{su}_{8q + 4}}{2}}_{q} \underbrace{\overset{\mathfrak{su}_{8q + 8}}{2}
        \cdots\overset{\mathfrak{su}_{8q + 4p + 4}}{2}}_{p}
        \underbrace{\overset{\mathfrak{su}_{8q + 4p + 4}}{2}\cdots \overset{\mathfrak{su}_{8q + 4p + 4}}{2}}_{N-1}$ \\

         & $1 \underbrace{\overset{\mathfrak{su}_{8}}{2}
         \cdots\overset{\mathfrak{su}_{8q}}{2}}_{q} \underbrace{\overset{\mathfrak{su}_{8q + 4}}{2}
         \cdots\overset{\mathfrak{su}_{8q + 4p}}{2}}_{p}
    \underbrace{\overset{\mathfrak{su}_{8q + 4p}}{2}\cdots \overset{\mathfrak{su}_{8q + 4p}}{2}}_{N-1}$ \\\midrule

        $\mathbb{Z}_5$  & $1\underbrace{\overset{\mathfrak{su}_{5}}{2}
        \cdots\overset{\mathfrak{su}_{5p}}{2}}_p\underbrace{\overset{\mathfrak{su}_{5p}}{2}\cdots \overset{\mathfrak{su}_{5p}}{2}}_{N-1} $ \\\midrule

        $\mathbb{Z}_6$ & $1\underbrace{\overset{\mathfrak{su}_{6}}{2}
        \cdots\overset{\mathfrak{su}_{6p}}{2}}_p\underbrace{\overset{\mathfrak{su}_{6p}}{2}\cdots \overset{\mathfrak{su}_{6p}}{2}}_{N-1} $ \\
        \bottomrule
    \end{tabular}
    \end{threeparttable}
    \caption{6d SCFTs that we consider that have discrete center-flavor symmetry. In descending order, the $E_8$-homomorphisms as in equation \eqref{eqn:Zkembed}, are $(0,p,0,s,0, u,2t+1,q,0)$, $(0,p,0,s,0, u,2t,q,0)$, $(0,0,p,0,0, s,0,0,3q+2)$, \\ $(0,0,p,0,0, s,0,0,3q+1)$, $(0,0,p,0,0, s,0,0,3q)$, $(0,0,0,p,0,0,2q+1,0,0)$, $(0,0,0,p,0,0,2q,0,0)$,\\ together with $(0,0,0,0,p,0,0,0,0)$ and $(0,0,0,0,0,p,0,0,0)$.}
    \label{tbl:6dscfts}
\end{sidewaystable}

The simplest example is the Higgsing induced by a $\bbZ_{6p} \rightarrow E_8$ homomorphism specified by $a_6 = p$, and all other labels being zero.
The resulting tensor branch gauge theory has the quiver description
\begin{equation}\label{eq:Z6-orbi-inst-example}
    [\mathfrak{su}_3 \oplus \mathfrak{su}_2] \, \, 1 \, \, \overset{\mathfrak{su}_{6}}{2} \, \, \overset{\mathfrak{su}_{12}}{2} \cdots \overset{\mathfrak{su}_{6p-6}}{2} \, \, \underset{[\mathfrak{su}_6]}{\overset{\mathfrak{su}_{6p}}{2}} \, \, \underbrace{\overset{\mathfrak{su}_{6p}}{2}\cdots \overset{\mathfrak{su}_{6p}}{2}}_{N-1} \, \, [\mathfrak{su}_{6p}] \, .
\end{equation}
Between each $\mathfrak{su}$ gauge and flavor factor on or next to $2$-nodes, there is a bifundamental hypermultiplet,
\begin{align}
\begin{split}
    & {\bf R}^{(1)} = ({\bf 6}, \overline{\bf 12})_0 \, , \quad {\bf R}^{(2)} = ({\bf 12}, \overline{\bf 18})_0 \, , \quad \cdots , \quad {\bf R}^{(p-1)} = ({\bf 6p-6}, \overline{\bf 6p})_0 \, , \\
    & {\bf R}^{(p)} = ({\bf 6p}, \overline{\bf 6})_{-p} \, ,  \quad {\bf R}^{(p+1)} = ({\bf 6p}, \overline{\bf 6p})_1 \, , \quad \cdots , \quad {\bf R}^{(p+N)} = ({\bf 6p}, \overline{\bf 6p})_1 \, .
\end{split}
\end{align}
In addition, there is a $U(1)_f$ flavor symmetry without ABJ-anomalies \cite{Apruzzi:2020eqi}, which only charges the hypermultiplets between the $\mathfrak{su}_{6p}$ factors (``the plateau'') with the charges indicated in the subscripts. The tensor pairing is the $(N+p) \times (N+p)$ matrix
\begin{align}
    A^{ij} = \begin{pmatrix}
        -1 & 1 & 0 & \cdots \\
        1 & -2 & 1 & \ddots \\
        0 & 1 & -2 & \ddots \\
        \vdots & \ddots & \ddots & \ddots
    \end{pmatrix} \, ,
\end{align}
and the anomaly coefficients of the flavor factors are
\begin{align}
\begin{split}
    & B^{i,\mathfrak{su}_3} = B^{i,\mathfrak{su}_2} = \delta^{i,1} \, , \quad B^{i,\mathfrak{su}_6} = \delta^{i,p+1} \, , \quad B^{i,\mathfrak{su}_{6p}} =  \delta^{i,N+p} \, , \\
    & C^{i;f,f} = \begin{cases}
        0 \, , & i\leq p \, , \\
        6p(p+1) \, , & i = p+1 \, , \\
        12p  \, , & i > p+1 \, .
    \end{cases}
\end{split}
\end{align}
Without taking into consideration the $U(1)_f$, one can easily verify that there is a $\bbZ_3 \times \bbZ_2 \times \bbZ_6 \cong {\bbZ_6}^{(1)} \times {\bbZ_6}^{(2)}$ center-flavor symmetry that leaves all hypermultiplets invariant. These have generators
\begin{align}
\begin{split}
    {\bbZ_6}^{(1)}: \quad & [\overset{1}{\mathfrak{su}_3} \oplus \overset{1}{\mathfrak{su}_2}] \, \, 1 \, \, \overset{\overset{0}{\mathfrak{su}_{6}}}{2} \, \, \overset{\overset{0}{\mathfrak{su}_{12}}}{2} \cdots \overset{\overset{0}{\mathfrak{su}_{6p-6}}}{2} \, \, \underset{[\underset{0}{\mathfrak{su}_6}]}{\overset{\overset{0}{\mathfrak{su}_{6p}}}{2}} \, \, \underbrace{\overset{\overset{0}{\mathfrak{su}_{6p}}}{2}\cdots \overset{\overset{0}{\mathfrak{su}_{6p}}}{2}}_{N-1} \, \, [\overset{0}{\mathfrak{su}_{6p}}] \, , \\
    {\bbZ_6}^{(2)}: \quad & [\overset{0}{\mathfrak{su}_3} \oplus \overset{0}{\mathfrak{su}_2}] \, \, 1 \, \, \overset{\overset{1}{\mathfrak{su}_{6}}}{2} \, \, \overset{\overset{2}{\mathfrak{su}_{12}}}{2} \cdots \overset{\overset{p-1}{\mathfrak{su}_{6p-6}}}{2} \, \, \underset{[\underset{1}{\mathfrak{su}_6}]}{\overset{\overset{p}{\mathfrak{su}_{6p}}}{2}} \, \, \underbrace{\overset{\overset{p}{\mathfrak{su}_{6p}}}{2}\cdots \overset{\overset{p}{\mathfrak{su}_{6p}}}{2}}_{N-1} \, \, [\overset{p}{\mathfrak{su}_{6p}}] \, ,
\end{split}
\end{align}
where we have indicated the embedding $k_g \in Z(G_g)$ by the overset $\overset{k_g}{\fkg_g}$ (or underset for the $[\mathfrak{su}_6]$ flavor factor) on each node of the quiver.
However, the presence of the E-string breaks these individual $\bbZ_6$ factors to the diagonal $\bbZ_6$, with the fractional part of Chern classes given by\footnote{By an abuse of notation, we will write $c_2(\fkg)$ for $c_2(F_\fkg)$.}
\begin{align}
    \begin{split}
        & c_2(\mathfrak{su}_3) \equiv - \tfrac13 w^2 \, , \quad c_2(\mathfrak{su}_2) \equiv -\tfrac14 w^2 \, , \quad c_2(\mathfrak{su}_{6l}) \equiv -\tfrac{l (6l-1)}{12} w^2 \quad (l=1,\ldots,p) \, .
    \end{split}
\end{align}
It is straightforward to verify that these cancel for each tensor:
\begin{align}
    \begin{aligned}
        \Theta_1 : & \qquad -c_2(\mathfrak{su}_3) - c_2(\mathfrak{su}_2) - c_2(\mathfrak{su}_6) \equiv \big( \tfrac13 + \tfrac14 + \tfrac{5}{12} \big) w^2 \equiv 0 \mod \bbZ \, , \\
        \Theta_{1+i} \, (1 \leq i < p) : & \qquad -c_2(\mathfrak{su}_{6(i-1)}) + 2c_2(\mathfrak{su}_{6i}) - c_2(\mathfrak{su}_{6(i+1)}) \\
        & \quad \equiv \tfrac{(i-1)(6i-7) -2i(6i-1) + (i+1)(6i+5)}{12} \, w^2 \equiv \tfrac{12}{12} w^2 \equiv 0 \mod \bbZ \, , \\
        \Theta_{p+1} : & \qquad -c_2(\mathfrak{su}_{6(p-1)}) + 2c_2(\mathfrak{su}_{6p}) - c_2(\mathfrak{su}_{6p}) - c_2(\mathfrak{su}_{6}) \\
        & \quad \equiv \tfrac{(p-1)(6p-7) - 2p(6p-1) + p(6p-1) -5}{12} \, w^2 \equiv \tfrac{12(p-1)}{12} \, w^2 \equiv 0 \mod \bbZ \, , \\
        \Theta_{p+i} \, (1 < i \leq N) : & \qquad -c_2(\mathfrak{su}_{6p}) + 2c_2(\mathfrak{su}_{6p}) - c_2(\mathfrak{su}_{6p}) \equiv 0 \mod \bbZ \, .
    \end{aligned}
\end{align}
Therefore, the non-Abelian structure group is
\begin{align}\label{eq:E8_Z6_example_non-Ab-flavor-group}
    \frac{[SU(3) \times SU(2)] \times \prod_{i< p} SU(6i) \times SU(6p)^N \times [SU(6)] \times [SU(6p)]}{\bbZ_6} \, ,
\end{align}
and the non-Abelian flavor symmetry of the SCFT is
\begin{equation}
    \frac{SU(3) \times SU(2) \times SU(6) \times SU(6p)}{\bbZ_6} \,.
\end{equation}
This $\bbZ_6$ center-flavor symmetry will allow us to perform a Stiefel--Whitney twisted $T^2$ compactification down to 4d, which we will turn to in the next section.

To complete the characterization of the full symmetry structure, we include possible $U(1)_f$ twists, in which case the hypermultiplets are invariant under a further center transformation of order $6p$, with generator
\begin{align}
    {\bbZ_{6p}}: \quad & [\overset{0}{\mathfrak{su}_3} \oplus \overset{0}{\mathfrak{su}_2}] \, \, 1 \, \, \overset{\overset{0}{\mathfrak{su}_{6}}}{2} \, \, \overset{\overset{0}{\mathfrak{su}_{12}}}{2} \cdots \overset{\overset{0}{\mathfrak{su}_{6p-6}}}{2} \, \, \underset{[\underset{-1}{\mathfrak{su}_6}]}{\overset{\overset{0}{\mathfrak{su}^{(0)}_{6p}}}{2}} \, \, \underbrace{\overset{\overset{1}{\mathfrak{su}^{(1)}_{6p}}}{2} \, \, \overset{\overset{2}{\mathfrak{su}^{(2)}_{6p}}}{2} \cdots \overset{\overset{N-1}{\mathfrak{su}^{(N-1)}_{6p}}}{2}}_{N-1} \, \, [\overset{N}{\mathfrak{su}^{(N)}_{6p}}] \quad \text{and} \quad e^{-\frac{2\pi i \hat{q}}{6p}} \in U(1)_f\, ,
\end{align}
where we have enumerated, for convenience, the $\mathfrak{su}_{6p}$ factors.
The resulting non-trivial Chern class fractionalizations are then
\begin{align}
    c_2(\mathfrak{su}_6) \equiv - \tfrac{5}{12}w^2 \, , \quad c_2(\mathfrak{su}_{6p}^{(k)}) \equiv -k^2 \, \tfrac{6p-1}{12p} w^2 \, , \quad c_1(F_f)^2 \equiv \tfrac{1}{36p^2} w^2 + \tfrac{1}{3p} w \cup \chi \mod \bbZ \, .
\end{align}
For the tensors of the $\mathfrak{su}_{6p}^{(k)}$ factors with $k \geq 1$, the cancellation of the fractionalizations is analogous to that appearing in equation \eqref{eq:u1-center-anomaly-cancellation_A-type_quiver} for the simple A-type quiver example. For $k=0$, the cancellation is due to
\begin{align}
\begin{split}
    &-c_2(\mathfrak{su}_{6p-6}) + 2 c_2(\mathfrak{su}_{6p}^{(0)}) - c_2(\mathfrak{su}_6) - c_2(\mathfrak{su}_{6p}^{(1)}) + 3p(p+1) c_1(F_f)^2 \\
    &\quad\equiv \,  \big( 0 + 0 + \tfrac{5}{12} + \tfrac{6p-1}{12p} + \tfrac{p+1}{12p} \big) \, w^2 \equiv 0 \mod \bbZ \, .
\end{split}
\end{align}

Lastly, we also incorporate the R-symmetry.
Again, the main constraint is to have the bifundamental hypermultiplets being invariant, which all transform in the fundamental representation of $\mathfrak{su}(2)_R$.
To cancel the phase $(-1)$ which these states acquire upon a $\bbZ_2 = Z(SU(2)_R)$ twist, we turn on a corresponding $\bbZ_2 \subset Z(SU(6k))$ in every \emph{second} $\mathfrak{su}$-factor.
These two a priori different twists are related by adding the $\bbZ_2$ subgroup of the $\bbZ_6$ center-flavor symmetry responsible for the non-Abelian flavor group structure in equation \eqref{eq:E8_Z6_example_non-Ab-flavor-group}, so do not give rise to two new and independent center-flavor symmetries when we include the R-symmetry, as expected.
For concreteness, we take the generator that compensates the $Z(SU(2)_R)$ twist to be
\begin{align}
\begin{split}
    p \ \text{even}: \quad & [\overset{0}{\mathfrak{su}_3} \oplus \overset{0}{\mathfrak{su}_2}] \, \, 1 \, \, \overset{\overset{0}{\mathfrak{su}_{6}}}{2} \, \, \overset{\overset{6}{\mathfrak{su}_{12}}}{2} \, \, 
    \overset{\overset{0}{\mathfrak{su}_{18}}}{2} \, \, 
    \overset{\overset{12}{\mathfrak{su}_{24}}}{2} \, \, \cdots \overset{\overset{0}{\mathfrak{su}_{6p-6}}}{2} \, \, \underset{[\underset{0}{\mathfrak{su}_6}]}{\overset{\overset{3p}{\mathfrak{su}_{6p}}}{2}} \, \, \underbrace{\overset{\overset{0}{\mathfrak{su}_{6p}}}{2} \, \, \overset{\overset{3p}{\mathfrak{su}_{6p}}}{2} \cdots }_{N-1} \, \, [\overset{*}{\mathfrak{su}_{6p}}] \, , \\
    p \ \text{odd}: \quad & [\overset{0}{\mathfrak{su}_3} \oplus \overset{0}{\mathfrak{su}_2}] \, \, 1 \, \, \overset{\overset{0}{\mathfrak{su}_{6}}}{2} \, \, \overset{\overset{6}{\mathfrak{su}_{12}}}{2} \, \, 
    \overset{\overset{0}{\mathfrak{su}_{18}}}{2} \, \, 
    \overset{\overset{12}{\mathfrak{su}_{24}}}{2} \, \, \cdots \overset{\overset{3p-3}{\mathfrak{su}_{6p-6}}}{2} \, \, \underset{[\underset{3}{\mathfrak{su}_6}]}{\overset{\overset{0}{\mathfrak{su}_{6p}}}{2}} \, \, \underbrace{\overset{\overset{3p}{\mathfrak{su}_{6p}}}{2} \, \, \overset{\overset{0}{\mathfrak{su}_{6p}}}{2} \cdots }_{N-1} \, \, [\overset{*}{\mathfrak{su}_{6p}}] \, ,
\end{split}
\end{align}
where the $*$ is either $3p$ if $p+N$ is even, or $0$ if $p+N$ is odd.
For this quiver, the tangent bundle enters only in the first tensor multiplet $t_1$ associated to the E-string, whose corresponding Green--Schwarz four-form contains $c_2(R)$ and $p_1(T)$:
\begin{align}
    \Theta_1 : \quad -c_2(\mathfrak{su}_3) -c_2(\mathfrak{su}_2) - c_2(\mathfrak{su}_6) + c_2(R) - \tfrac14 p_1(T) \equiv -\tfrac14 w^2 - \tfrac34 w^2 \equiv 0 \mod \bbZ \, .
\end{align}
For the other tensors, the topological coupling to the R-symmetry bundle is through the term $h^\vee c_2(R)$, where $h^\vee (\mathfrak{su}_{6k}) = 6k$.
Since these tensors all have $A^{ii}=-2$, the coupling to $p_1(T)$ is trivial.
Let us first examine those on a generic position on the ramp (i.e., a 2-node with $\mathfrak{su}_{6i<6p}$).
Here, we have
\begin{align}
    \begin{split}
        \Theta_{1+i} \, (i \ \text{even}): \quad & -c_2(\mathfrak{su}_{6(i-1)}) + 2c_2(\mathfrak{su}_{6i}) - c_2(\mathfrak{su}_{6(i+1)}) + 6i \, c_2(R) \\
        \equiv & \left( 2 \times \tfrac{9i (6i-1)}{12} +  \tfrac32 i \right) w^2 \equiv 0 \mod \bbZ \, ,
    \end{split}\\
    \begin{split}
        \Theta_{1+i} \, (i \ \text{odd}): \quad & -c_2(\mathfrak{su}_{6(i-1)}) + 2c_2(\mathfrak{su}_{6i}) - c_2(\mathfrak{su}_{6(i+1)}) + 6i \, c_2(R) \\
        \equiv & \left( - \tfrac{9(i-1)(6i-7) + 9(i+1)(6i+5)}{12} +  \tfrac32 i \right) w^2 \equiv 0 \mod \bbZ \, ,
    \end{split}
\end{align}
where the fractional part of $c_2(\mathfrak{su}_{6(i-1)})$ is automatically zero for $i=1$.
For the node that connects the ramp to the plateau (i.e., the first node with $\mathfrak{su}_{6p}$ gauge algebra), we have
\begin{align}
\begin{split}
    \Theta_{p+1} \, (p \ \text{even}) : \quad & -c_2(\mathfrak{su}_{6p-6}) + 2 c_2(\mathfrak{su}_{6p}) - c_2(\mathfrak{su}_{6p}) - c_2(\mathfrak{su}_6) + 6p \, c_2(R) \\
    \equiv & \left( 2\times \tfrac{9p(6p-1)}{12} + \tfrac{3}{2}p \right) w^2 \equiv 0 \mod \bbZ \, , 
\end{split}\\
\begin{split}
    \Theta_{p+1} \, (p \ \text{odd}) : \quad & -c_2(\mathfrak{su}_{6p-6}) + 2 c_2(\mathfrak{su}_{6p}) - c_2(\mathfrak{su}_{6p}) - c_2(\mathfrak{su}_6) + 6p \, c_2(R) \\
    \equiv & \left( - \tfrac{9(p-1)(6p-7) + 9p(6p-1) + 9 (6-1)}{12} + \tfrac{3}{2}p \right) w^2 \equiv 0 \mod \bbZ \, .
\end{split}
\end{align}
For the other nodes on the ramp, there is either a $\bbZ_2$ twist only in the corresponding gauge factor, or only in the two adjacent gauge / flavor factors:
\begin{align}
    \begin{split}
        \Theta_{p+i} \, (p+i \ \text{odd}) : \quad & -c_2(\mathfrak{su}_{6p}) + 2 c_2(\mathfrak{su}_{6p}) - c_2(\mathfrak{su}_{6p}) + 6p \, c_2(R) \equiv \\
        & \left( 2 \times \tfrac{9p(6p-1)}{12} + \tfrac32 p \right) w^2 \equiv 0 \mod \bbZ \, ,
    \end{split}\\
    \begin{split}
        \Theta_{p+i} \, (p+i \ \text{even}) : \quad & -c_2(\mathfrak{su}_{6p}) + 2 c_2(\mathfrak{su}_{6p}) - c_2(\mathfrak{su}_{6p}) + 6p \, c_2(R) \equiv \\
        & \left( - 2 \times \tfrac{9p(6p-1)}{12} + \tfrac32 p \right) w^2 \equiv 0 \mod \bbZ \, .
    \end{split}
\end{align}

Again, we omit the straightforward, but somewhat tedious crosscheck that we can activate simultaneously the $\bbZ_6$ twist in the non-Abelian flavor factors, the $\bbZ_{6p}$ twist involving the $U(1)_f$ flavor, and the $\bbZ_2$ R-symmetry twist.
From this analysis, we conclude that the 6d SCFT with tensor branch description as in equation \eqref{eq:Z6-orbi-inst-example} has global symmetry group
\begin{align}
    \frac{SU(3) \times SU(2) \times SU(6) \times SU(6p) \times U(1)_f \times SU(2)_R}{\bbZ_6 \times \bbZ_{6p} \times \bbZ_2} \,.
\end{align}
The analysis of the structure for the global symmetry for the other tensor branch descriptions in Table \ref{tbl:6dscfts} follows directly from the application of the methods described in this example.

\section{4d \texorpdfstring{\boldmath{$\mathcal{N}=2$}}{N=2} SW-folds}\label{sec:sfolds}

Having shown how to extract the global symmetry group of 6d SCFTs, we now turn to a specific application in the context of constructing 4d $\mathcal{N} = 2$ SCFTs. To reach such a theory from a 6d $\mathcal{N} = (1,0)$ SCFT, one can consider compactification on a $T^2$. Activating background gauge bundle configurations with vanishing flux provides a general template for realizing 4d $\mathcal{N} = 2$ SCFTs. In fact, one can also consider compactifications which are sensitive to the global topology of the 6d global symmetries, namely by switching on an 't Hooft magnetic flux \cite{tHooft:1979rtg} in the $T^2$ directions \cite{Ohmori:2018ona}.\footnote{One can in principle consider various generalizations, as obtained from compactifying on a more general genus $g$ Riemann surface with marked points, with non-trivial contributions from the R-symmetry bundle also switched on. In this broader setting, one would expect to get 4d $\mathcal{N} = 1$ SCFTs, along the lines of \cite{Morrison:2016nrt, Razamat:2016dpl} (see also, for example, \cite{Gaiotto:2015usa,Franco:2015jna,Coman:2015bqq,Heckman:2016xdl,Bah:2017gph,Bourton:2017pee,Kim:2017toz,Apruzzi:2018oge,Razamat:2018zus,Kim:2018lfo,Razamat:2018gro,Kim:2018bpg,Razamat:2018gbu,Chen:2019njf,Pasquetti:2019hxf,Sela:2019nqa,Razamat:2019mdt,Razamat:2019ukg,Razamat:2020bix,Sabag:2020elc,Bourton:2020rfo,Nazzal:2021tiu,Hwang:2021xyw,Bourton:2021das,Razamat:2022gpm} and references therein).}
In \cite{Ohmori:2018ona} this was referred to as a Stiefel--Whitney (SW) twisted compactification. These are a specific class of configurations involving background flat bundle configurations with non-trivial holonomies which commute in $G_{\text{global}} = \widetilde{G}_{\text{global}} / \mathcal{C}$, but which would not have commuted as holonomies of bundles with structure group $\widetilde{G}$. Treated as a bundle with structure group $\widetilde{G}$, we would have a non-zero flux valued in a subgroup of $\mathcal{C}$, i.e., the holonomies commute up to a specific element of this flux. We shall loosely speaking refer to such holonomies as being ``charged under an element of $\mathcal{C}$'' since this has a clear meaning when treating these backgrounds as $\widetilde{G}$ bundles. Owing to their similarities with S-fold constructions, we often refer to these theories as ``SW-folds'' in what follows.

We consider SW-folds obtained from the 6d theories considered in Section \ref{sec:e8CF}, namely the theories of the form $\Omega_{\mathfrak{su}_K, N}(\rho, 1)$, where $\rho$ is an $E_8$-homomorphism that leads to a $\mathbb{Z}_\ell$ center-flavor symmetry. The tensor branch descriptions of such 6d $(1,0)$ SCFTs were given in Table \ref{tbl:6dscfts}. Compactification on a torus with a $\mathbb{Z}_\ell$ Stiefel--Whitney twist then leads to the 4d SCFTs we consider herein; furthermore, many of the properties of the 4d theories can be obtained from a knowledge of the 6d $(1,0)$ parent theory. We emphasize that when we say we turn on a $\mathbb{Z}_\ell$ Stiefel--Whitney twist, we are turning on a non-commuting holonomy charged under the element $p$ of $\mathbb{Z}_\ell$ such that $\gcd(p, \ell) = 1$. All of these Stiefel--Whitney twisted theories are listed in Table \ref{tbl:genSfolds}.

As a general comment, while we could in principle extract the global symmetry \textit{group} of the resulting 4d theory, there can be additional structures which emerge from extended objects which can now wrap on the $T^2$ directions. For this reason, we primarily focus on just the global symmetry \textit{algebra} of the resulting 4d theories, leaving a more complete analysis of their global structure group to future work.

The rest of this section is organized as follows. We first explain how to extract the central charges and flavor symmetries for the resulting SW-fold theories. This is followed by an extensive list of examples, as given in Table \ref{tbl:genSfolds}. As an independent cross-check, we also directly study the Coulomb branch operator spectrum for these theories. In some cases, there are alternative ways to generate some of these theories.\footnote{A recent and detailed review of both the features of $\mathcal{N}=2$ SCFTs, and of the various different constructions, is \cite{Akhond:2021xio}.} We discuss some examples of this in the context of class $\mathcal{S}$ constructions, as well as 4d $\mathcal{N} = 2$ S-folds \cite{Apruzzi:2020pmv,  Giacomelli:2020jel, Heckman:2020svr,Giacomelli:2020gee}, and we comment on the overlap as well as differences from these other methods of generating 4d $\mathcal{N} = 2$ SCFTs.

\begin{table}[t!]
    \centering
    \renewcommand{\arraystretch}{1.4}
    \begin{threeparttable}
    \begin{tabular}{cccc}
    \toprule
         SW-fold SCFT & Orbi-instanton & $E_8$-Homomorphism & SW Twist \\\midrule
         $\mathcal{S}_2^{(N)}(p,s,u,2t+1,q)$ &  $(\mathfrak{e}_8, \mathfrak{su}_{2q + 8t + 6u + 4s + 2p + 4})$  & $(0,p,0,s,0,u,2t+1,q,0)$ & \multirow{2}{*}{$\mathbb{Z}_2$} \\
         $\mathcal{T}_2^{(N)}(p,s,u,2t,q)$ &  $(\mathfrak{e}_8, \mathfrak{su}_{2q + 8t + 6u + 4s + 2p})$  & $(0,p,0,s,0,u,2t,q,0)$ &  \\\midrule
         $\mathcal{R}_3^{(N)}(p,s,3q+2)$ &  $(\mathfrak{e}_8, \mathfrak{su}_{9q+6s+3p+6})$  & $(0,0,p,0,0,s,0,0,3q+2)$ & \multirow{3}{*}{$\mathbb{Z}_3$} \\
         $\mathcal{S}_3^{(N)}(p,s,3q+1)$ &  $(\mathfrak{e}_8, \mathfrak{su}_{9q+6s+3p+3})$  & $(0,0,p,0,0,s,0,0,3q+1)$ &  \\
         $\mathcal{T}_3^{(N)}(p,s,3q)$ &  $(\mathfrak{e}_8, \mathfrak{su}_{9q+6s+3p})$  & $(0,0,p,0,0,s,0,0,3q)$ &  \\\midrule
         $\mathcal{S}_4^{(N)}(p, 2q+1)$ &  $(\mathfrak{e}_8, \mathfrak{su}_{8q+4p+4})$  & $(0,0,0,p,0,0,2q+1,0,0)$ & \multirow{2}{*}{$\mathbb{Z}_4$} \\
         $\mathcal{T}_4^{(N)}(p, 2q)$ &  $(\mathfrak{e}_8, \mathfrak{su}_{8q+4p})$  & $(0,0,0,p,0,0,2q,0,0)$ &  \\\midrule
         $\mathcal{T}_5^{(N)}(p)$ &  $(\mathfrak{e}_8, \mathfrak{su}_{5p})$  & $(0,0,0,0,p,0,0,0,0)$ & $\mathbb{Z}_5$ \\\midrule
         $\mathcal{T}_6^{(N)}(p)$ &  $(\mathfrak{e}_8, \mathfrak{su}_{6p})$  & $(0,0,0,0,0,p,0,0,0)$ & $\mathbb{Z}_6$  \\\bottomrule
    \end{tabular}
    \end{threeparttable}
    \caption{The 4d $\mathcal{N}=2$ SW-folds that we consider in this chapter. Each SCFT is obtained by starting with the 6d rank $N$ orbi-instanton SCFT of type $(\mathfrak{e}_8, \mathfrak{su}_K)$, where $K$ is as in the second column. Higgsing the $\mathfrak{e}_8$ flavor symmetry by the homomorphism $\mathbb{Z}_K \rightarrow E_8$, given via $(a_1, a_2, a_3, a_4, a_5, a_6, a_{4^\prime}, a_{2^\prime}, a_{3^\prime})$ in the third column, yields each of the 6d SCFTs in Table \ref{tbl:6dscfts}, which have a $\mathbb{Z}_\ell$ center-flavor symmetry. Compactifying the resulting 6d SCFT on a $T^2$ with a $\mathbb{Z}_\ell$ Stiefel--Whitney twist, where $\ell$ is as in the fourth column, produces the 4d $\mathcal{N}=2$ SW-fold SCFT which we denote by the naming that appears in the first column.}
    \label{tbl:genSfolds}
\end{table}

\subsection{Central Charges and Flavor Symmetries}\label{sec:ccs}

Having specified a construction for an infinite family of 4d $\mathcal{N} = 2$, we now turn to some of their properties. As each of the SW-folds we study arises from the compactification of a 6d SCFT that is very Higgsable, we can apply the methods from \cite{Ohmori:2018ona} to determine the central charges and the flavor central charges.

To determine the central charges of the SW-folds we carry out the following procedure. First we compute the anomaly polynomial of the origin 6d SCFT, $I_8$. Next, we compute the 1-loop contribution on the full tensor branch from just the vector multiplets, tensor multiplets and hypermultiplets, and refer to this as $I_{8}^{\text{fields}}$:\footnote{We note that especially in the case of generalized quivers with conformal matter one sometimes refers to this as a ``1-loop'' contribution as well. Here, we are referring to the full tensor branch, where the conformal matter has also been decomposed into standard 6d $\mathcal{N} = (1,0)$ supermultiplets.}
\begin{equation}
I_{8}^{\text{fields}} \equiv I_{8}^{\text{1-loop,vector}} + I_{8}^{\text{1-loop,tensor}} + I_{8}^{\text{1-loop,hyper}}.
\end{equation}
Both $I_8$ and $I_{8}^{\text{fields}}$ is a formal eight-form polynomial in the characteristic classes of the symmetries of the 6d SCFT. As required, the anomaly polynomial does not contain any terms proportional to the characteristic classes of the gauge symmetries on the tensor branch, as the 6d SCFT is non-anomalous, but the quantity $I_8^\text{fields}$ does contain such gauge-anomalous terms. We write
\begin{equation}\label{eqn:fields}
    I_8 - I_8^\text{fields} = A p_1(T)^2 + B c_2(R) p_1(T) + \sum_a C_a p_1(T)\operatorname{Tr}F_a^2 + \cdots \,,
\end{equation}
where $p_1(T)$ is the first Pontryagin class of the spacetime tangent bundle, $c_2(R)$ is the $SU(2)$ R-symmetry bundle, and $\operatorname{Tr}F_a^2$ is the curvature of the flavor symmetry bundles. The sum is over the simple non-Abelian flavor symmetries of the SCFT. In terms of these quantities, $A$, $B$, and $C_a$,\footnote{As we will not include holonomies of 6d Abelian flavor factors, their anomaly coefficients $C^{i;f,f'}$ will not appear in the following, and $C_a$ will exclusively denote the coefficients in the anomaly polynomial \eqref{eqn:fields}.} we can write the central charges of the SW-fold SCFTs as
\begin{equation}\label{eqn:ack}
  \begin{aligned}
    a - a_\text{generic} &= 32\left(\frac{3}{2\ell} - \frac{3}{4}\right)A - \frac{12}{\ell}B \,, \\
    c - c_\text{generic} &= 32\left(\frac{3}{\ell} - 1\right)A - \frac{12}{\ell}B \,, \\
    \kappa_a - \kappa^a_\text{generic} &= \frac{192}{\ell} C_a I_a \,,
  \end{aligned}
\end{equation}
where $\ell$ is the order of the Stiefel--Whitney twist, and $I_a$ is the Dynkin index of the embedding of the 4d flavor symmetry as a subalgebra of the 6d flavor symmetry. Here $a_\text{generic}$, $c_\text{generic}$, and $\kappa_\text{generic}^a$ are the central charges and flavor central charges of the 4d theory at the generic point of the Coulomb branch. We can rewrite the central charges in terms of the numbers of vector and (full) hypermultiplets at the generic point of the Coulomb branch as
\begin{equation}
    a_\text{generic} = \frac{5}{24}n_V + \frac{1}{24}n_H \,, \quad c_\text{generic} = \frac{1}{6}n_V + \frac{1}{12}n_H \,.
\end{equation}
For all SW-folds the quantities $A$, $B$, $C_a$ and the generic central charges can be determined from the 6d origin and the knowledge of the $\mathbb{Z}_\ell$ center-flavor symmetry. Thus we can always determine the central charges of the SW-fold SCFT.

In this section, we are interested in specific 6d SCFTs, those that appear in Table \ref{tbl:6dscfts}, which all have tensor branch configurations of the form
\begin{equation}\label{eqn:quivquiv}
    \overset{\mathfrak{g}}{1}\overset{\mathfrak{su}_{\ell k_1}}{2}\overset{\mathfrak{su}_{\ell k_2}}{2}\cdots\overset{\mathfrak{su}_{\ell k_r}}{2} \,,
\end{equation}
where the possible choices for $\mathfrak{g}$ and the $k_i$ are specified via their 6d origin in Table \ref{tbl:6dscfts}.  For each $\ell$, we summarize the possible $\mathfrak{g}$, together with their below-mentioned numerical data, in Table \ref{tbl:onedata}. First, we will discuss the contributions to $I_8 - I_8^\text{fields}$ for such 6d SCFTs. The result differs depending on whether $\mathfrak{g}$ is trivial or not. Let us first consider the simplest case where $\mathfrak{g} \neq \varnothing$. Then
\begin{equation}\label{eqn:GS1}
    I_8 - I_8^\text{fields} = I_8^\text{GS} = -\frac{1}{2}A_{ij}I^i I^j \,,
\end{equation}
where the tensor indices $i$, $j$ run over the nodes of the tensor branch configuration in equation \eqref{eqn:quivquiv} from left to right.\footnote{We emphasize that the tensor pairing matrix $A^{ij}$, whose diagonal entries are the negative of the values attached to the nodes in equation \eqref{eqn:quivquiv}, is negative-definite.}
Recall from equation \eqref{eq:green-schwarz-coupling} that in the four-form $I^i$,
\begin{equation}
    I^i = \frac{1}{4}\bigg(-A^{ij}\operatorname{Tr} F_j^2 - B^{ia}\operatorname{Tr} F_a^2 - (2+A^{ii})p_1(T) \bigg)+ y^i c_2(R) \,.
\end{equation}
the index $i$ in the $p_1(T)$ term is not summed over.
In this case, we have $y^i = h_{\mathfrak{g}_i}^\vee$, the dual Coxeter number of the gauge algebra associated to the $i$th tensor. We see that the only contributions to $p_1(T)^2$ arise when $i = j = 1$, and thus
\begin{equation}
    A = \frac{r+1}{32} \,.
\end{equation}
The $c_2(R)p_1(T)$ term is rather more involved to determine, but it can be found to be:
\begin{equation}
  \begin{aligned}
    B &= \frac{1}{4} A_{ij} \left( 2 + A^{ii}\right)y^j  \\
    &= -\frac{1}{4}(r+1)h_{\mathfrak{g}}^\vee - \frac{1}{4}\ell \sum_{j=1}^r (r+1-j)k_j \,,
  \end{aligned}
\end{equation}
where we again need to be careful with the sum over $i$, and we emphasize that the tensor pairing matrix $A^{ij}$ and its inverse $A_{ij}$ are both symmetric. Finally, we consider the terms proportional to $p_1(T)\operatorname{Tr}F_a^2$. We find
\begin{equation}
  \begin{aligned}
    C_a &=  -\frac{1}{16} A_{ij} \left( (2 + A^{ii})B^{ja} \right) \\
    &= \frac{r+1 - k(a)}{16} \,,
  \end{aligned}
\end{equation}
where $k(a)$ is the position of the node in the tensor branch quiver diagram that ``intersects'' the flavor factor $\fkg_a$, i.e., $B^{ja} = 0$ for $j \neq k(a)$.\footnote{This would not apply to baryonic $\mathfrak{su}_2$ flavor symmetries, which we are not considering in this chapter. Such flavor factors are only relevant for very specific SW-folds, which have already been worked out in \cite{Giacomelli:2020jel}.} In all cases under consideration we have $B^{k(a)a} = 1$. Thus, we have determined $A$, $B$, and $C_a$ for tensor branch configurations of the form in equation \eqref{eqn:quivquiv} when $\mathfrak{g} \neq \varnothing$.

Let us now consider the slightly more complicated configuration where $\mathfrak{g} = \varnothing$, in which case the left-most node in equation \eqref{eqn:quivquiv} becomes an E-string.
For this configuration we have
\begin{equation}
    I_8 - I_8^\text{fields} = I_8^\text{E-string} - I_8^\text{tensor} + I_8^\text{GS} \,,
\end{equation}
where
\begin{equation}
    I_8^\text{GS} = -\frac{1}{2}\widetilde{A}_{ij} I^i I^j \,, \quad I^i = \frac{1}{4}\bigg(-\widetilde{A}^{ij}\operatorname{Tr} F_j^2 - \widetilde{B}^{ia}\operatorname{Tr} F_a^2 - (2+\widetilde{A}^{ii})p_1(T) \bigg)+ y^i c_2(R) \,.
\end{equation}
Here, the matrix of coefficients $\widetilde{A}$ and $\widetilde{B}$ can be interpreted as the contributions from a generalized quiver, where we allow conformal matter between nodes of the quiver. In this case, the indices now run over $i,j = 1, \cdots, r$. The coefficients $y^i$ remain $h_{\mathfrak{su}_{\ell k_i}}^\vee$, except for $y^1$ which is now $1 + h_{\mathfrak{su}_{\ell k_1}}^\vee$.
We can see immediately that
\begin{equation}
    A = \frac{1}{32} + \frac{r}{32} = \frac{r+1}{32} \,.
\end{equation}
Furthermore, the $c_2(R)p_1(T)$ coefficient is
\begin{equation}
  \begin{aligned}
    B &= -\frac{1}{4} + \frac{1}{4} \widetilde{A}_{ij} \left( (2 + \widetilde{A}^{ii})y^j \right) \\
    &= -\frac{1}{4} - \frac{1}{4}(r)(\ell k_1 + 1) - \frac{1}{4}\ell \sum_{j=2}^r (r+1-j) k_j \\
    &= -\frac{1}{4}(r+1) - \frac{1}{4}\ell \sum_{j=1}^r (r+1-j)k_j \,.
  \end{aligned}
\end{equation}
Finally, we need to discuss the flavor symmetry terms. The coefficient $C_a$ of $p_1(T)\operatorname{Tr}F_a^2$ is
\begin{equation}
    C_a = \frac{r+1 - k(a)}{16} \,,
\end{equation}
where $k(a)$ is the index of the quiver node that intersects the flavor symmetry indexed by $a$. Further, we have again used that, in all cases of relevance of this work, $B^{k(a)a} = 1$.

\begin{table}[ht]
    \centering
    \renewcommand{\arraystretch}{1.2}
    \begin{threeparttable}
    \begin{tabular}{cccccc}
    \toprule
         $\ell$ & $\mathfrak{g}$ & $n_V^0$ & $n_H^0$ & $d_0$ & $k_0$ \\\midrule
         \multirow{2}{*}{$2$} & $\mathfrak{sp}_{q\geq 0}$ & $\frac{q(q-1)}{2}$ & $q(q+4)$ & $q+1$ & $q$ \\
         & $\mathfrak{su}_{2q+4\geq 4}$ & $q^2 + 4q + 3$ & $\frac{3}{2}(q+2)(q+5)$ & $2q + 4$ & $q+2$ \\\midrule
         \multirow{3}{*}{$3$} & $\mathfrak{su}_6^\prime$ & $3$ & $12$ & $6$ & $2$ \\
         & $\mathfrak{su}_3$ & $0$  & $4$ & $3$ & $1$ \\
         & $\varnothing$ & $0$ & $0$ & $1$ & $0$ \\\midrule
         \multirow{2}{*}{$4$} & $\mathfrak{su}_4$ & $0$ & $3$ & $4$ & $1$\\
         & $\varnothing$ & $0$ & $0$ & $1$ & $0$ \\\midrule
         $5$ & $\varnothing$ & $0$ & $0$ & $1$ & $0$ \\\midrule
         $6$ & $\varnothing$ & $0$ & $0$ & $1$ & $0$ \\\bottomrule
    \end{tabular}
    \end{threeparttable}
    \caption{The possible decorations on the (left-most) tensor with self-pairing $1$ in equation \eqref{eqn:quivquiv}.}
    \label{tbl:onedata}
\end{table}

While it was necessary that we do the calculation slightly differently for the cases where $\mathfrak{g} \neq \varnothing$ and $\mathfrak{g} = \varnothing$, we see that the resulting coefficients appearing in $I_8 - I_8^\text{fields}$ relevant for the central charges of the compactification can be written succinctly as
\begin{equation}\label{eqn:B}
    \begin{aligned}
      A &= \frac{r+1}{32} \,, \\
      B &= -\frac{1}{4}\bigg( (r+1)d_0 + \ell \sum_{j=1}^r (r + 1 - j)k_j \bigg) \,, \\
      C_a &= \frac{r+1 - k(a)}{16} \,,
    \end{aligned}
\end{equation}
where $d_0$ is as written in Table \ref{tbl:onedata}; it is $1$ if $\mathfrak{g} = \varnothing$ and $h_{\mathfrak{g}}^\vee$ otherwise.

Next, let us determine the numbers of vector and hypermultiplets at the generic point of the 4d Coulomb branch. We recall here how the Stiefel--Whitney twist acts on the weakly coupled 6d spectrum on the tensor branch. When doing a $\mathbb{Z}_\ell$ Stiefel--Whitney twist, we need to know the following 6d $\rightarrow$ 4d transformations:
\begin{equation}\label{eqn:6d4d}
    \begin{aligned}
      \mathfrak{su}_{\ell k} \text{ vector multiplet } &\rightarrow \mathfrak{su}_{k} \text{ vector multiplet } \,, \\
      \mathfrak{su}_{\ell k_1} \oplus \mathfrak{su}_{\ell k_2} \text{ bifund. hypermultiplet } &\rightarrow \mathfrak{su}_{k_1} \oplus \mathfrak{su}_{k_2} \text{ bifund. hypermultiplet } \,, \\
      \text{ tensor multiplet } &\rightarrow \text{ vector multiplet } \,.
    \end{aligned}
\end{equation}
The subtleties arise from the possible ``decorations'' $\fkg$ on the node $\overset{\fkg}{1}$ in equation \eqref{eqn:quivquiv}, which we will often refer to as the $1$-node of the quiver.\footnote{In geometric terms that describe F-theory constructions of 6d SCFTs, such a node is usually called a $(-1)$-curve.}
How the SW-fold acts on such a tensor with the possibilities for the gauge algebra $\fkg$ has been studied in \cite{Ohmori:2018ona}. Putting all this together we see that the number of vector multiplets and hypermultiplets at the generic point of the Coulomb branch is
\begin{equation}\label{eqn:nv}
    \begin{aligned}
        n_V &= 1 + n_V^0 + \sum_{i=1}^r  k_i^2 \,, \\
        n_H &= n_H^0 + \sum_{i=1}^r k_i(2k_i - k_{i-1})  \,.
    \end{aligned}
\end{equation}
Here, $1 + n_V^0$ is the number of vector multiplets that are associated to $\fkg$ and survive the SW-fold. Similarly, $k_0$ is the dimension of the fundamental representation of this gauge algebra after SW-folding, and $k_1 + n_H^0$ is the total number of surviving hypermultiplets from the $\overset{\fkg}{1}$-node. These quantities follow directly from the action of the Stiefel--Whitney twist and they are summarized in Table \ref{tbl:onedata}. Finally, we determine the contribution to the flavor central charges at the generic point of the 4d Coulomb branch. We have
\begin{equation}
    \kappa_a^\text{generic} = 2k_{k(a)} \,,
\end{equation}
where, again, $k(a)$ is the index of the tensor that intersects the $a$th flavor factor. This follows from the existence of the bifundamental (full) hypermultiplet after the Stiefel--Whitney twist described in equation \eqref{eqn:6d4d}.\footnote{When considering the flavor algebras attached to the $\overset{\fkg}{1}$-node, the matter many not simply be a bifundamental hypermultiplet, but some other bi-representation. In these cases, the contribution from a generic hypermultiplet on the 4d Coulomb branch must be worked out individually.}
In the case of flavor symmetries that intersect the $\overset{\fkg}{1}$-node, the value of $k_0$ is written in Table \ref{tbl:onedata}; it comes from the surviving gauge algebra on that node after the Stiefel--Whitney twist.

\subsection{Examples}

Putting everything together, we can see that the 4d $\mathcal{N}=2$ SW-fold SCFT obtained via the Stiefel--Whitney twist of 6d $(1,0)$ tensor branch configuration as in equation \eqref{eqn:quivquiv} has central charges $a$, $c$, and $\kappa_a$ given as in equation \eqref{eqn:ack}. These quantities can thus be worked out for each of the theories listed in Table \ref{tbl:genSfolds}, and we now do so. The central charges $a$ and $c$ become rather lengthy expressions, especially as one decreases the order of the Stiefel--Whitney twist, $\ell$, which thus gives rise to to more parameters describing the discrete homomorphism $\bbZ_\ell \rightarrow E_8$. Therefore, we have attached a {\tt Mathematica} notebook containing these expressions to the {\tt arXiv} submission of the paper corresponding to this chapter for the ease of the reader.

\subsubsection[\texorpdfstring{$\bbZ_6$}{Z6} SW-folds: \texorpdfstring{$\mathcal{T}_6^{(N)}(p)$}{T6N(p)}]{\boldmath{$\mathbb{Z}_6$} SW-folds: \boldmath{$\mathcal{T}_6^{(N)}(p)$}}

We begin by studying the $\mathbb{Z}_6$ SW-folds: $\mathcal{T}_6^{(N)}(p)$. The 6d SCFT origin, with the flavor symmetry included, is
\begin{equation}\label{eqn:Z6TB}
[\mathfrak{su}_3]\underset{[\mathfrak{su}_2]}{1}\overbrace{\overset{\mathfrak{su}_{6}}{2}
        \cdots\underset{[\mathfrak{su}_6]}{\overset{\mathfrak{su}_{6p}}{2}}}^p\overbrace{\overset{\mathfrak{su}_{6p}}{2}\cdots \overset{\mathfrak{su}_{6p}}{2}}^{N-1} [\mathfrak{su}_{6p}] \,.
\end{equation}
As we determined in Section \ref{sec:e8CF}, the non-Abelian flavor symmetry of the 6d SCFT is generically
\begin{equation}
    G_\text{flavor} = [SU(3) \times SU(2) \times SU(6) \times SU(6p)]/\mathbb{Z}_6 \,,
\end{equation}
however in the special case where $N = 1$, the last two factors combine and one has \begin{equation}
    G_\text{flavor} = [SU(3) \times SU(2) \times SU(6(p+1))] / \mathbb{Z}_6 \,.
\end{equation}
In terms of the quiver written in equation \eqref{eqn:quivquiv}, here we have $\mathfrak{g} = \varnothing$ and
\begin{equation}\label{eqn:k6s}
    k_i = (1, 2, \cdots, p, \underbrace{\,p, \cdots, p\,}_{N-1}) \,.
\end{equation}
In this case, we shall write each of the quantities, $A$, $B$, $C_a$, $n_V$, $n_H$, and $\kappa_a^\text{generic}$ necessary to determine the central charges.
After the Stiefel--Whitney twist the only surviving flavor symmetry is either $\mathfrak{su}_p$ arising from the $\mathfrak{su}_{6p}$ in the case of generic $N$, or $\mathfrak{su}_{p+1}$ coming from the $\mathfrak{su}_{6(p+1)}$ factor when $N = 1$; as there is only one simple non-Abelian flavor algebra we shall drop the index $a$. Note, when $p = 1$ and $N > 1$ there is no surviving flavor symmetry. For the quantities determined from the 6d anomaly polynomial we find
\begin{equation}
    A = \frac{p + N}{32} \,, \quad B = -\frac{1}{4}(p^3 + 3Np^2 +  3N^2p + N) \,, \quad C = \frac{1}{16} \,.
\end{equation}
At the generic point of the 4d Coulomb branch we have
\begin{equation}
    n_V = \frac{1}{6}(6 + p - 3p^2 + 6Np^2 + 2p^3) \,, \quad n_H = \frac{1}{3}p(2 + 3Np + p^2) \,, \quad \kappa^\text{generic} = 2p \,.
\end{equation}
Plugging these values into equation \eqref{eqn:ack}, we find that the central charges of the resulting 4d $\mathcal{N}=2$ SCFTs are
\begin{align}\label{eqn:l6ccs}
      a &= \frac{1}{48}\bigg(28p^3 + 84Np^2 - 5p^2 + 72N^2p - 21p + 10 \bigg) \,, \\
      c &= \frac{1}{12}\bigg(7p^3 + 21 N p^2 - p^2 + 18 N^2 p - 5p + 2 \bigg) \,, \\
      \kappa &= 12p+2 \,.
\end{align}
We emphasize that, regardless of whether the residual flavor symmetry algebra is $\mathfrak{su}_p$ or $\mathfrak{su}_{p+1}$, the flavor central charge is identical. As it is required often throughout this section, we will explain the Dynkin indices for the special subalgebras that we consider. We have
\begin{equation}
    \mathfrak{su}_{\ell k} \rightarrow \mathfrak{su}_\ell \oplus \mathfrak{su}_k \,,
\end{equation}
such that
\begin{equation}
    \bm{\ell k} \rightarrow (\bm{\ell}, \bm{k}) \,.
\end{equation}
The index of the embedding can be worked out from this decomposition of the fundamental representation,\footnote{See \cite{Esole:2020tby} for an explanation of the embedding indices applicable to the special subalgebras.} and we find that the $\mathfrak{su}_\ell$ factor has index $k$, and the $\mathfrak{su}_k$ factor has index $\ell$.

The theory $\mathcal{T}^{(N)}_6(p=1)$ has been previously studied in \cite{Giacomelli:2020gee}. In that case, there is no remaining flavor symmetry and we can see from equation \eqref{eqn:l6ccs} that the central charges are
\begin{equation}
    a = c = \frac{1}{4}(6N + 1)(N + 1) \,.
\end{equation}
The result for this special case matches that found in \cite{Giacomelli:2020gee}.\footnote{To aid in comparison, we note that our $\mathcal{T}^{(N)}_6(p=1)$ theory is equivalent to the $\mathcal{T}^{(r+1)}_{\varnothing, 6}$ theory of \cite{Giacomelli:2020gee}.} In this particular case the central charges are equal as the theory enjoys supersymmetry enhancement, either to $\mathcal{N}=4$ supersymmetry when $N = 1$, or else to $\mathcal{N}=3$ when $N > 1$. When $p = 1$ and $N=1$ the theory is $\mathcal{N}=4$ super-Yang--Mills with gauge group $G_2$; in this case there is an $\mathfrak{su}_2$ flavor symmetry and we can see that the flavor central charge is $\kappa = 14 = \operatorname{dim}G_2$, as expected. In the generic case where $p > 1$ there is no such supersymmetry enhancement.

\subsubsection{\texorpdfstring{$\mathbb{Z}_5$}{Z5} SW-folds: \texorpdfstring{$\mathcal{T}_5^{(N)}(p)$}{T5N(p)}}

We now study the $\mathbb{Z}_5$ SW-folds: $\mathcal{T}_5^{(N)}(p)$.
The tensor branch configuration describing the 6d SCFT origins of these 4d theories are
\begin{equation}
[\mathfrak{su}_5]1\overbrace{\overset{\mathfrak{su}_{5}}{2}
        \cdots\underset{[\mathfrak{su}_5]}{\overset{\mathfrak{su}_{5p}}{2}}}^p\overbrace{\overset{\mathfrak{su}_{5p}}{2}\cdots \overset{\mathfrak{su}_{5p}}{2}}^{N-1} [\mathfrak{su}_{5p}] \,.
\end{equation}
We have here written the flavor algebras that exist for generic values of $p$ and $N$, however there is a flavor symmetry enhancement when considering a single M5-brane, $N=1$. The full global structure of the non-Abelian part of the flavor symmetry group was determined in Section \ref{sec:e8CF} and we find
\begin{equation}
    G_\text{flavor} = \begin{cases}
      (SU(5) \times SU(5(p+1)))/\mathbb{Z}_5 \qquad &\text{ when } \quad N = 1  \,, \\
      (SU(5) \times SU(5) \times SU(5p))/\mathbb{Z}_5 \qquad &\text{ when } \quad N > 1\,.
    \end{cases}
\end{equation}
To determine the central charges we must determine the $k_i$ when the tensor branch configuration is written in the form in equation \eqref{eqn:quivquiv}; observe that these $k_i$ are the same as those appearing in equation \eqref{eqn:k6s} in the $\mathcal{T}_6^{(N)}(p)$ case. Using the formula in equation \eqref{eqn:ack} leads to the following central charges:
\begin{align}
        a &= \frac{1}{240}\left(140p^3 + 420 N p^2 - 25p^2 + 360 N^2 p - 69p + 36N + 50 \right) \,, \\
        c &= \frac{1}{60}\left(35p^3 + 105N p^2 - 5p^2 + 90N^2 p- 13p + 12 N + 10 \right) \,.
\end{align}
Finally, we determine the non-Abelian flavor algebra that survives after the Stiefel--Whitney twisted compactification.
Denoting the central charges by subscripts, one finds
\begin{equation}
    \mathfrak{g}_\text{flavor}^\text{4d} = \begin{cases}
      \big(\mathfrak{su}_{p+1}\big)_{12p+2} \qquad &\text{ when } \quad N = 1 \,, \\
      \big(\mathfrak{su}_p\big)_{12p+2} \qquad &\text{ when } \quad N > 1 \,.
    \end{cases}
\end{equation}
Similarly to the $\ell = 6$ case, the theories that we have $\mathcal{T}_5^{(N)}(p=1)$ have been previously studied in \cite{Giacomelli:2020gee}, where they were referred to as the $\mathcal{T}^{(r+1)}_{\varnothing, 5}$ theories. As we can see, the $\mathbb{Z}_5$ SW-folds that are written here constitute a broad generalization of the hitherto known theories.

\subsubsection[\texorpdfstring{$\mathbb{Z}_4$}{Z4} SW-folds: \texorpdfstring{$\mathcal{T}_4^{(N)}(p, 2q)$}{T4N(p,2q)} and \texorpdfstring{$\mathcal{S}_4^{(N)}(p, 2q+1)$}{S4N(p,2q+1)}]{\boldmath{$\mathbb{Z}_4$} SW-folds: \boldmath{$\mathcal{T}_4^{(N)}(p, 2q)$} and \boldmath{$\mathcal{S}_4^{(N)}(p, 2q+1)$}}

There are two classes of $\ell = 4$ SW-folds: $\mathcal{T}_4^{(N)}(p, 2q)$ and $\mathcal{S}_4^{(N)}(p, 2q+1)$. Recall that a Higgs-branch deformation by the homomorphism $\mathbb{Z}_K \rightarrow E_8$ preserves a $\mathbb{Z}_4$ center-flavor symmetry of the 6d SCFT only if the only non-zero entries in equation \eqref{eqn:Zkembed} are $a_4$ and $a_{4^\prime}$. The distinction between the $\mathcal{T}$ and $\mathcal{S}$ theories depends on whether $a_{4^\prime}$ is even or odd, respectively. First, we consider the $\mathcal{T}_4^{(N)}(p, 2q)$ theories, which arise from 6d $(1,0)$ SCFTs with tensor branch configuration
\begin{equation}\label{eqn:T4g}
1\overbrace{\overset{\mathfrak{su}_{8}}{2}
         \cdots\underset{[\mathfrak{su}_4]}{\overset{\mathfrak{su}_{8q}}{2}}}^{q} \overbrace{\overset{\mathfrak{su}_{8q + 4}}{2}
         \cdots\underset{[\mathfrak{su}_4]}{\overset{\mathfrak{su}_{8q + 4p}}{2}}}^{p}
    \overbrace{\overset{\mathfrak{su}_{8q + 4p}}{2}\cdots \overset{\mathfrak{su}_{8q + 4p}}{2}}^{N-1} [\mathfrak{su}_{8q+4p}] \,.
\end{equation}
In this generalized quiver, we have written the flavor algebras for generic values of the parameters $p$, $q$, and $N$. From the analysis in Section \ref{sec:6d}, we can see that the non-Abelian part of the global symmetry group is
\begin{equation}
    G_\text{flavor} = \begin{cases}
    (SU(8q+8))/\mathbb{Z}_4 \qquad &\text{ when } \quad p = 0, q \geq 1, N = 1 \,, \\
    (SU(8) \times SU(8q))/\mathbb{Z}_4 \qquad &\text{ when } \quad p = 0, q \geq 1, N > 1 \,, \\
    (Spin(10) \times SU(4p+4))/\mathbb{Z}_4 \qquad &\text{ when } \quad p \geq 1, q = 0, N = 1 \,, \\
    (Spin(10) \times SU(4) \times SU(4p))/\mathbb{Z}_4 \qquad &\text{ when } \quad p \geq 1, q = 0, N > 1 \,, \\
    (SU(4) \times SU(8q + 4p + 4))/\mathbb{Z}_4 \qquad &\text{ when } \quad p \geq 1, q \geq 1, N = 1 \,, \\
    (SU(4) \times SU(4) \times SU(8q + 4p))/\mathbb{Z}_4 \qquad &\text{ when } \quad p \geq 1, q \geq 1, N > 1 \,.
    \end{cases}
\end{equation}
We can see that if we write the models in equation \eqref{eqn:T4g} in the generic form for the tensor branch configurations that we study, as in equation \eqref{eqn:quivquiv}, then the $k_i$ are given by
\begin{equation}
    k_i = (\underbrace{\,2, 4, \cdots, 2q\,}_q, \underbrace{\,2q + 1, 2q + 2, \cdots, 2q + p\,}_p,  \underbrace{\,2q + p, \cdots, 2q + p\,}_{N-1}) \,.
\end{equation}
We observe that there is a steep ramp, of length $q$, where $k_i$ increases by $2$ each step, followed by a shallower length $p$ ramp where the $k_i$ increases by $1$, and finally a plateau of length $N - 1$. Although it starts to become somewhat tedious, it is straightforward to work out the central charges using equation \eqref{eqn:ack}. We find
\begin{align}
        a &= \frac{1}{48}\bigg( 64q^3 + 192pq^2 + 192Nq^2 - 20q^2 + 168p^2 q + 336Npq - 20pq  \\&\qquad\quad + 144N^2q - 18q + 28p^3 + 84Np^2 - 5p^2 + 72N^2p - 3p + 18N + 10\bigg) \,, \cr
        c &= \frac{1}{12}\bigg( 16q^3 + 48pq^2 + 48Nq^2 - 4q^2 + 42p^2q + 84Npq - 4pq \\&\qquad\quad + 36N^2q - 2q + 7p^3 + 21Np^2 - p^2 + 18N^2p + p + 6N + 2 \bigg) \nonumber \,.
\end{align}
To determine the flavor symmetry that survives the Stiefel--Whitney twisting procedure, it is necessary to understand how, in the cases with $q=0$, the $\mathfrak{so}_{10}$ flavor algebra intersecting the E-string is acted on by the $\mathbb{Z}_4$ center-flavor symmetry. Writing
\begin{equation}\label{eqn:so10d}
    \mathfrak{so}_{10} \rightarrow \mathfrak{su}_4 \oplus \mathfrak{su}_2 \oplus \mathfrak{su}_2 \,,
\end{equation}
we can see that the $\mathbb{Z}_4$ is embedded via the generator $(1,1,0)$ inside of the combined $\mathbb{Z}_4 \times \mathbb{Z}_2 \times \mathbb{Z}_2$ center \cite{Giacomelli:2020gee}. As such, the only surviving subalgebra from the $\mathfrak{so}_{10}$ factor is an $\mathfrak{su}_2$, with embedding index $1$. The Stiefel--Whitney twisting of the remaining flavor symmetry factors can be determined as for the $\ell = 5, 6$ cases. The flavor central charges (denoted in subscript) can also be computed using the formula in equation \eqref{eqn:ack}; the result is
\begin{equation}
    \mathfrak{g}_\text{flavor}^\text{4d} = \begin{cases}
        (\mathfrak{su}_{2q + 2})_{4q + 12} &\quad p = 0, q \geq 1, N = 1 \\
        (\mathfrak{su}_{2})_{4q + 12N} \oplus (\mathfrak{su}_{2q})_{4q+12} &\quad p = 0, q \geq 1, N > 1 \\
        (\mathfrak{su}_2)_{3(p+1)} \oplus (\mathfrak{su}_{p+1})_{2p+12} &\quad p \geq 1, q = 0, N = 1 \\
        (\mathfrak{su}_2)_{3(p+N)} \oplus (\mathfrak{su}_{p})_{2p+12} &\quad p \geq 1, q = 0, N > 1 \\
        (\mathfrak{su}_{2q + p + 1})_{4q+2p+12} &\quad p \geq 1, q \geq 1, N = 1 \\
        (\mathfrak{su}_{2q + p})_{4q+2p+12} &\quad p \geq 1, q \geq 1, N > 1 \,.
    \end{cases}
\end{equation}
Again, similarly to the $\ell = 5$ and $\ell = 6$ SW-folds, the theories $\mathcal{T}_4^{(N)}(p=1, 2q=0)$ have been studied afore in \cite{Giacomelli:2020gee}, where they are called the $\mathcal{T}_{A_2, 4}^{(r+1)}$ theories.

There is another class of $\ell = 4$ SW-folds, which are obtained by starting with the 6d A-type orbi-instanton SCFT Higgsed by a $\mathbb{Z}_4$ center-flavor symmetry preserving $E_8$-homomorphism where the embedding into $a_4^\prime$, as in equation \eqref{eqn:Zkembed}, is odd. The SW-fold SCFTs obtained from the $\mathbb{Z}_4$ Stiefel--Whitney twist of these 6d SCFTs are referred to as $\mathcal{S}_\ell^{(N)}(p, 2q+1)$. The tensor branch configurations of these 6d SCFTs have the form
\begin{equation}
\overset{\mathfrak{su}_{4}}{1} \overbrace{\overset{\mathfrak{su}_{12}}{2}
        \cdots\underset{[\mathfrak{su}_4]}{\overset{\mathfrak{su}_{8q + 4}}{2}}}^{q} \overbrace{\overset{\mathfrak{su}_{8q + 8}}{2}
        \cdots\underset{[\mathfrak{su}_4]}{\overset{\mathfrak{su}_{8q + 4p + 4}}{2}}}^{p}
        \overbrace{\overset{\mathfrak{su}_{8q + 4p + 4}}{2}\cdots \overset{\mathfrak{su}_{8q + 4p + 4}}{2}}^{N-1} [\mathfrak{su}_{8q + 4p + 4}] \,,
\end{equation}
where, as usual, the flavor symmetry can enhance when the parameters $p$, $q$, and $N$ obtain their limiting values. The flavor groups, including how the $\mathbb{Z}_4$ quotient acts, can be determined from the algorithm described in Section \ref{sec:6d}. Computing the central charges is a straightforward application of equation \eqref{eqn:ack}:
\begin{align}
        a &= \frac{1}{48}\bigg( 64q^3 + 192pq^2 + 192Nq^2 + 76q^2 + 168p^2 q + 336Npq \cr&\qquad\quad+ 172pq  + 144N^2q + 192Nq   + 10q + 28p^3 + 84Np^2 \\&\qquad\quad+ 79p^2 + 72N^2p + 168Np + 35p + 72N^2 + 66N + 4\bigg) \,, \cr
        c &= \frac{1}{12}\bigg( 16q^3 + 48pq^2 + 48Nq^2 + 20q^2 + 42p^2q + 84Npq \cr&\qquad\quad + 44pq  + 36N^2q +48Nq  + 6q  + 7p^3 + 21Np^2 \\&\qquad\quad+ 20p^2 + 18N^2p + 42Np + 11p + 18N^2 + 18N + 2 \bigg) \nonumber \,.
\end{align}
Similarly, the flavor algebras that survive the Stiefel--Whitney twist can be determined, and their flavor central charges are again given by equation \eqref{eqn:ack}. We find
\begin{equation}
    \mathfrak{g}_\text{flavor}^\text{4d} = \begin{cases}
        (\mathfrak{su}_{2q+3})_{4q + 14} &\quad p = 0, q \geq 0, N = 1 \\
        (\mathfrak{su}_2)_{4q + 12N + 2} \oplus (\mathfrak{su}_{2q+1})_{4q + 14} &\quad p = 0, q \geq 0, N > 1 \\
        (\mathfrak{su}_{2q + p + 2})_{4q + 2p + 14} &\quad p \geq 1, q \geq 0, N = 1 \\
        (\mathfrak{su}_{2q + p + 1})_{4q + 2p + 14} &\quad p \geq 1, q \geq 0, N > 1 \,.
    \end{cases}
\end{equation}
The theories $\mathcal{S}_4^{(N)}(p=0, 2q+1 =1)$ were studied in \cite{Giacomelli:2020gee}, where they were called the $\mathcal{S}^{(r)}_{A_2, 4}$ theories. The central charges and flavor symmetries that we compute here agree with what was found in that particular limiting case. We have similarly labelled these generalized S-fold SCFTs by $\mathcal{S}$ and $\mathcal{T}$ to match with the notation for the special cases that have been previously studied.

In this chapter, we have mainly been concerned with the identification of the 4d non-Abelian flavor symmetry that descends from the 6d non-Abelian flavor symmetry. In fact, the 6d SCFTs under consideration also contain Abelian symmetries that arise from the ABJ-anomaly-free combinations of the $\mathfrak{u}(1)$s rotating the bifundamental hypermultiplets. Under certain circumstances, these $\mathfrak{u}(1)$s can enhance, and then we expect a further non-Abelian factor in the 4d flavor symmetry. This occurs when the 4d Coulomb branch description of the SW-fold contains a plateau of neighboring $\mathfrak{su}_2$ gauge algebras: then the $\mathfrak{u}(1)$ enhances to an $\mathfrak{su}_2$ under which the gauge bifundamentals are charged. The SW-folds with this extra, enhanced, baryonic $\mathfrak{su}_2$ flavor symmetry are
\begin{equation}\label{eqn:baryenc}
    \begin{gathered}
        \mathcal{T}_4^{(N)}(0,2) \,, \quad \mathcal{S}_4^{(N)}(1,1) \,, \quad \mathcal{T}_3^{(N)}(0,1,0) \,, \\ \mathcal{S}_3^{(N)}(1,0,1) \,, \quad \mathcal{R}_3^{(N)}(0,0,2) \,, \quad \mathcal{S}_2^{(N)}(0,0,0,1,0) \\
        \mathcal{T}_2^{(N)}(0,0,0,0,2) \,, \quad
        \mathcal{T}_2^{(N)}(0,1,0,0,0) \,, \quad
        \mathcal{T}_2^{(N)}(2,0,0,0,0) \,, \quad
        \mathcal{T}_2^{(N)}(1,0,0,0,1) \,.
    \end{gathered}
\end{equation}
In each case, we can see that they correspond to theories obtained from an orbi-instanton theory involving M5-branes probing $\mathbb{C}^2/\mathbb{Z}_{2\ell}$; after the $\mathbb{Z}_\ell$ Stiefel--Whitney twist, the orbifold is reduced to $\mathbb{C}^2/\mathbb{Z}_2$, and the additional $\mathfrak{su}_2$ global symmetry comes from the exceptional isometry of this particular orbifold. For low values of $N$ we expect that this baryonic $\mathfrak{su}_2$ can combine with other non-Abelian factors in the flavor symmetry, and cause further enhancement. In rare cases there can also be dehancement. We discuss some instances where this enhancement occurs in Section \ref{sec:classs}.

\subsubsection[\texorpdfstring{$\mathbb{Z}_3$}{Z3} SW-folds: \texorpdfstring{$\mathcal{T}_3^{(N)}(p,s,3q)$}{T3N(p,s,3q)}, \texorpdfstring{$\mathcal{S}_3^{(N)}(p,s,3q+1)$}{S3N(p,s,3q+1)}, and \texorpdfstring{$\mathcal{R}_3^{(N)}(p,s,3q+2)$}{R3N(p,s,3q+2)}]{\boldmath{$\mathbb{Z}_3$} SW-folds: \boldmath{$\mathcal{T}_3^{(N)}(p,s,3q)$}, \boldmath{$\mathcal{S}_3^{(N)}(p,s,3q+1)$}, and \boldmath{$\mathcal{R}_3^{(N)}(p,s,3q+2)$}}

There are three distinct ways that one can construct a Higgs-branch flow from the rank $N$ $(\mathfrak{e}_8, \mathfrak{su}_K)$ orbi-instanton theory such that the resulting SCFT has a $\mathbb{Z}_3$ center-flavor symmetry. From Section \ref{sec:e8CF}, we see that $K$ must be a multiple of three and the homomorphism $\mathbb{Z}_K \rightarrow E_8$ must be specified by the vector
\begin{equation}
    (a_1, a_2, a_3, a_4, a_5, a_6, a_{4^\prime}, a_{2^\prime}, a_{3^\prime}) = (0, 0, a_3, 0, 0, a_6, 0, 0, a_{3^\prime}) \,.
\end{equation}
In all cases, the resulting 6d SCFTs have tensor branch configurations of the form in equation \eqref{eqn:quivquiv}, however the algebra $\mathfrak{g}$ associated to the left-most $1$-node in equation \eqref{eqn:quivquiv}, depends on the parity modulo three of $a_{3^\prime}$. Respectively, we find that the $1$-node has no gauge algebra; $\mathfrak{su}_3$ with twelve fundamental and one antisymmetric hypermultplets; and an $\mathfrak{su}_6$ algebra with fifteen fundamental hypermultiplets and one further hypermultiplet in the triple-antisymmetric representation. As in Table \ref{tbl:genSfolds}, we label these theories by $\mathcal{T}$, $\mathcal{S}$, and $\mathcal{R}$, respectively. The tensor branch configurations for each of these configurations are shown in Table \ref{tbl:6dscfts}, and we do not repeat them here.

We begin our journey into the $\mathbb{Z}_3$ SW-folds with the $\mathcal{T}_3^{(N)}(p,s,3q)$ theory, whose 6d origins have $\fkg = \emptyset$ for the $1$-node.
The central charges can be determined straightforwardly from the tensor branch configuration by application of the formulae in equation \eqref{eqn:ack}. One finds
\begingroup
\allowdisplaybreaks
\begin{align}
        a &= \frac{1}{48} \bigg(72 N^2 p+216 N^2 q+144 N^2 s+84 N p^2+504 N p q+336 N p s \cr&\qquad\quad+324 N q^2  +576 N q s+192
   N s^2+36 N+252 p^2 q+168 p^2 s \\&\qquad\quad +28 p^3  -5 p^2+324 p q^2  +576 p q s-30 p q+192 p s^2-20 p s \cr&\qquad\quad +15 p+324
   q^2 s+108 q^3  -45 q^2+288 q s^2-60 q s-9 q+64 s^3-20 s^2+10\bigg) \,, \cr
        c &= \frac{1}{12} \bigg(18 N^2 p+54 N^2 q+36 N^2 s+21 N p^2+126 N p q+84 N p s \cr&\qquad\quad +81 N q^2  +144 N q s+48 N
   s^2+12 N+63 p^2 q+42 p^2 s+7 p^3 \\&\qquad\quad -p^2+81 p q^2  +144 p q s-6 p q+48 p s^2-4 p s+7 p+81 q^2 s \cr&\qquad\quad +27 q^3-9
   q^2+72 q s^2  -12 q s+3 q+16 s^3-4 s^2+4 s+2\bigg) \nonumber \,.
\end{align}
\endgroup
To determine the flavor algebra for the theory after Stiefel--Whitney twist, we need to understand how the $\mathbb{Z}_3$ quotient acts on the $\mathfrak{e}_6$ flavor symmetry. The decomposition is
\begin{equation}\label{eqn:e6ss}
  \begin{aligned}
    \mathfrak{e}_6 &\rightarrow \mathfrak{g}_2 \oplus \mathfrak{su}_3 \\
    \bm{27} &\rightarrow (\bm{7,3}) \oplus (\bm{1,\overline{6}}) \,,
  \end{aligned}
\end{equation}
where the $\mathbb{Z}_3$ acts on the $\mathfrak{su}_3$ factor and only the $\mathfrak{g}_2$ survives. From the decomposition of the fundamental representation in equation \eqref{eqn:e6ss}, we see that the Dynkin index of the $\mathfrak{g}_2$ subalgebra is one. Similarly, when there is an $\mathfrak{su}_3 \oplus \mathfrak{su}_2$ flavor algebra attached to the undecorated $1$-node, we note that the $\mathbb{Z}_3$ acts only on the $\mathfrak{su}_3$ factor and leaves the $\mathfrak{su}_2$ factor untouched. The remaining flavor factors are quotiented by the Stiefel--Whitney twist exactly as in the $\ell > 3$ cases that we have discussed. In the end, one discovers that the flavor symmetries, and the flavor central charges of these 4d SCFTs are:
\begin{equation}\label{eqn:T4flav}
    \begin{aligned}
        q = s = 0, p \geq 1, N \geq 1 \,&: \quad (\mathfrak{g}_2)_{4(N + p)} \oplus (\mathfrak{su}_1)_{12N+2p} \oplus (\mathfrak{su}_p)_{12+2p} \\
        q = 0, s \geq 1, p \geq 0, N \geq 1 \,&: \quad (\mathfrak{su}_2)_{4(N + p + s)} \oplus (\mathfrak{su}_1)_{12(N+p)+4s} \\ &\qquad\qquad \oplus (\mathfrak{su}_1)_{12N+4s+2p} \oplus (\mathfrak{su}_{2s+p})_{12+4s+2p} \\
        q \geq 1, s \geq 0,  p \geq 0, N \geq 1 \,&:\quad
        (\mathfrak{su}_1)_{12(N+p+s)+6q} \oplus (\mathfrak{su}_1)_{12(N+p)+6q+4s} \\ &\qquad\qquad
        \oplus (\mathfrak{su}_1)_{12N+6q+4s+2p}  \oplus (\mathfrak{su}_{3q+2s+p})_{12+6q+4s+2p}
         \,.
    \end{aligned}
\end{equation}
We have introduced a shorthand notation here as the number of combinations of $p$, $q$, $s$, and $N$ where there are flavor symmetry enhancements becomes large. In this way, if we write the flavor symmetry as $(\mathfrak{su}_{k_1})_{\kappa_1} \oplus (\mathfrak{su}_{k_2})_{\kappa_2}$ then for $\kappa_1 \neq \kappa_2$ the flavor symmetry is as written, but if $\kappa_1 = \kappa_2$ then there is an enhancement to $(\mathfrak{su}_{k_1 + k_2})_{\kappa_1}$. Of course, if there is an $\mathfrak{su}_1$ factor where the flavor central charge is such that it does not combine with another flavor symmetry factor, then that symmetry is, of course, trivial.

Next, we turn to the $\mathcal{S}_3^{(N)}(p,s,3q+1)$ SCFTs, originating from a 6d theory with $\fkg = \mathfrak{su}_3$.
From the tensor branch description of the 6d origin and the formulae in equation \eqref{eqn:ack}, one can determine the central charges. As these expressions are rather lengthy, we remind the reader that they also appear in the {\tt Mathematica} notebook attached to the {\tt arXiv} submission for the paper corresponding to this chapter. The central charges for these theories are
\begin{align}
        a &= \frac{1}{48} \bigg( 72 N^2 p+216 N^2 q+144 N^2 s+72 N^2+84 N p^2+504 N p q  +336 N p s+168 N p\nonumber\\&\qquad\quad +324 N
   q^2+576 N q s+216 N q+192 N s^2  +192 N s+72 N+252 p^2 q+168 p^2 s \nonumber \\&\qquad\quad +28 p^3+79 p^2+324 p q^2  +576 p q
   s+186 p q+192 p s^2+172 p s+41 p \\ &\qquad\quad +324 q^2 s+108 q^3 +63 q^2+288 q s^2+156 q s-3 q+64 s^3+76 s^2+16
   s+6 \bigg) \,, \nonumber\\
        c &= \frac{1}{12} \bigg( 18 N^2 p+54 N^2 q+36 N^2 s+18 N^2+21 N p^2+126 N p q  +84 N p s \cr&\qquad\quad+42 N p  +81 N q^2+144
   N q s+54 N q+48 N s^2+48 N s+21 N \\&\qquad\quad +63 p^2 q  +42 p^2 s+7 p^3  +20 p^2+81 p q^2+144 p q s+48 p q+48 p
   s^2+44 p s \cr&\qquad\quad +14 p+81 q^2 s+27 q^3  +18 q^2+72 q s^2+42 q s+6 q+16 s^3+20 s^2+9 s+3 \bigg) \nonumber \,.
\end{align}
As expected, the generic four-dimensional flavor algebra experiences enhancement at the lower limits of the parameters describing the $E_8$-homomorphism, $p$, $q$, and $s$, and also when one has only a single M5-brane, $N = 1$. The resulting flavor symmetries, together with the flavor central charges, are
\begin{equation}
    \begin{aligned}
        q,s,p \geq 0, N \geq 1 \,&:\quad
        (\mathfrak{su}_1)_{12(N+p+s)+6q+2} \oplus (\mathfrak{su}_1)_{12(N+p)+6q+4s+2} \\ &\qquad\qquad
        \oplus (\mathfrak{su}_1)_{12N+6q+4s+2p+2}  \oplus (\mathfrak{su}_{3q+2s+p+1})_{12+6q+4s+2p+2}
         \,.
    \end{aligned}
\end{equation}
Here, we use an F-theoretic convention for keeping track of trivial symmetry factors such as ``$\mathfrak{su}_1$'' since the parameters can sometimes conspire such that two of the $\mathfrak{su}$ flavor factors have the same flavor central charges. In such situations, the flavor symmetry enhances as described around equation \eqref{eqn:T4flav}.

Finally, we turn to the third class of $\ell = 3$ SW-folds, which we refer to as the $\mathcal{R}_3^{(N)}(p,s,3q+2)$ SW-folds. The central charges are again determined from the tensor branch configuration of the 6d SCFT of which these SW-folds are the $\mathbb{Z}_3$ Stiefel--Whitney twisted torus compactification. They are
\begingroup
\allowdisplaybreaks
\begin{align}\label{eqn:3Rccs}
        a &= \frac{1}{48} \bigg( 72 N^2 p+216 N^2 q+144 N^2 s+144 N^2+84 N p^2+504 N p q \cr&\qquad\quad +336 N p s  +336 N p+324 N
   q^2+576 N q s+432 N q+192 N s^2 \cr&\qquad\quad +240 N s+180 N  +252 p^2 q+168 p^2 s+28 p^3+163 p^2+324 p q^2 \\&\qquad\quad +576 p q
   s+402 p q+192 p s^2  +220 p s+139 p+324 q^2 s+108 q^3\cr&\qquad\quad+171 q^2+288 q s^2+372 q s+75 q+64 s^3+100 s^2+176 s+16 \bigg) \,, \cr
        c &= \frac{1}{12} \bigg( 18 N^2 p+54 N^2 q+36 N^2 s+36 N^2+21 N p^2+126 N p q  +84 N p s+84 N p \cr&\qquad\quad +81 N q^2+144
   N q s+108 N q+48 N s^2  +60 N s+48 N+63 p^2 q+42 p^2 s \\&\qquad\quad +7 p^3+41 p^2+81 p q^2 +144 p q s+102 p q+48 p
   s^2+56 p s+39 p+81 q^2 s \cr&\qquad\quad +27 q^3  +45 q^2+72 q s^2+96 q s+27 q+16 s^3+26 s^2+50 s+8 \bigg) \nonumber \,.
\end{align}
\endgroup
The flavor symmetries and flavor central charges can also be worked out using equation \eqref{eqn:ack}, and we find the following result:
\begin{equation}
    \begin{aligned}
        q,s,p \geq 0, N \geq 1 \,&:\quad
        (\mathfrak{su}_1)_{12(N+p+s)+6q+4} \oplus (\mathfrak{su}_1)_{12(N+p)+6q+4s+4} \\ &\qquad\qquad
        \oplus (\mathfrak{su}_1)_{12N+6q+4s+2p+4}  \oplus (\mathfrak{su}_{3q+2s+p+2})_{12+6q+4s+2p+4}
         \,.
    \end{aligned}
\end{equation}
While the $\mathcal{T}_3^{(N)}(p=1,s=0,3q=0)$ and $\mathcal{S}_3^{(N)}(p=0,s=0,3q+1=1)$ S-fold SCFTs have been studied before in \cite{Giacomelli:2020gee}, where they are referred to as the $\mathcal{T}_{D_4,3}^{(r+1)}$ and $\mathcal{S}_{D_4,3}^{(r)}$ theories, respectively, the theories $\mathcal{R}_3^{(N)}(p,s,3q+2)$ have not been studied in the context of S-folds before. The theory $\mathcal{R}_3^{(1)}(p=0,s=0,3q+2)$ has appeared previously in \cite{Ohmori:2018ona}, where the authors point out that, at the generic point of the 4d Coulomb branch, there is a half-hypermultiplet transforming in the $\bm{4}$ of the rightmost $\mathfrak{su}_2$ gauge algebra; this arises from the action of the Stiefel--Whitney twist on the triple-antisymmetric representation of the $\mathfrak{su}_6$ associated to the $1$-node. We refer the reader to Appendix \ref{app:lit}, where we list all of the limiting cases of the Stiefel--Whitney twists and S-folds that have previously appeared in the literature.

\subsubsection[\texorpdfstring{$\mathbb{Z}_2$}{Z2} SW-folds: \texorpdfstring{$\mathcal{T}_2^{(N)}(p,s,u,2t,q)$}{T2N(p,s,u,2t,q)} and \texorpdfstring{$\mathcal{S}_2^{(N)}(p,s,u,2t+1,q)$}{S2N(p,s,u,2t+1,q)}]{\boldmath{$\mathbb{Z}_2$} SW-folds: \boldmath{$\mathcal{T}_2^{(N)}(p,s,u,2t,q)$} and \boldmath{$\mathcal{S}_2^{(N)}(p,s,u,2t+1,q)$}}

The last class of SW-folds which we wish to consider are those involving a $\mathbb{Z}_2$ Stiefel--Whitney twist. These theories depend on five $E_8$-homomorphism parameters, associated to the five different nodes of the $E_8$ Dynkin diagram with even Dynkin label, and one positive integer counting the number of M5-branes. This proliferation of parameters leads to very complicated and unwieldy expressions for the central charges, which are typically degree three polynomials in these parameters. For posterity, we present these expressions here, however, we refer the reader to the attached {\tt Mathematica} notebook for a more practical format.

We begin with the $\mathcal{T}_2^{(N)}(p,s,u,2t,q)$ theories, which arise from 6d SCFTs obtained from $E_8$-homomorphisms of the rank $N$ $(\mathfrak{e}_8, \mathfrak{su}_K)$ orbi-instanton where the parameter $a_{4^\prime}$, as in equation \eqref{eqn:Zkembed} is even. The generalized quivers describing the 6d SCFTs are depicted in Table \ref{tbl:6dscfts}, and from there one can determine the central charges using equation \eqref{eqn:ack}. We find
\begingroup
\allowdisplaybreaks
\begin{align}
        a &= \frac{1}{48} \bigg( 72 N^2 p+72 N^2 q+144 N^2 s+288 N^2 t+216 N^2 u+84 N p^2+168 N p q \cr&\qquad\quad +336 N p s+672 N
   p t+504 N p u+12 N q^2+192 N q s+240 N q t+216 N q u \cr&\qquad\quad +192 N s^2+768 N s t+576 N s u+480 N t^2+864 N
   t u+324 N u^2+72 N \cr&\qquad\quad +84 p^2 q+168 p^2 s+336 p^2 t+252 p^2 u+28 p^3-5 p^2+12 p q^2+192 p q s \cr&\qquad\quad +240 p q
   t  +216 p q u-10 p q+192 p s^2+768 p s t+576 p s u-20 p s+480 p t^2 \\&\qquad\quad +864 p t u  -40 p t+324 p u^2-30 p
   u+51 p+12 q^2 s+12 q^2 t+12 q^2 u-5 q^2 \cr&\qquad\quad +96 q s^2  +240 q s t+216 q s u-20 q s+120 q t^2+240 q t u-40
   q t+108 q u^2 \cr&\qquad\quad -30 q u  +3 q  +384 s^2 t+288 s^2 u+64 s^3-20 s^2+480 s t^2+864 s t u-80 s t \cr&\qquad\quad +324 s
   u^2 -60 s u  +36 s+480 t^2 u+160 t^3-80 t^2+432 t u^2-120 t u \cr&\qquad\quad +24 t+108 u^3-45 u^2+27 u +10
   \bigg) \,, \cr
        c &= \frac{1}{12} \bigg( 18 N^2 p+18 N^2 q+36 N^2 s+72 N^2 t+54 N^2 u+21 N p^2+42 N p q  +84 N p s \cr&\qquad\quad +168 N p
   t  +126 N p u+3 N q^2+48 N q s+60 N q t+54 N q u+48 N s^2 \cr&\qquad\quad +192 N s t  +144 N s u+120 N t^2+216 N t u+81
   N u^2+24 N+21 p^2 q+42 p^2 s \cr&\qquad\quad +84 p^2 t  +63 p^2 u+7 p^3-p^2+3 p q^2+48 p q s+60 p q t+54 p q u-2 p
   q+48 p s^2 \cr&\qquad\quad +192 p s t  +144 p s u-4 p s+120 p t^2+216 p t u-8 p t+81 p u^2-6 p u+19 p  \\&\qquad\quad +3 q^2 s +3 q^2
   t  +3 q^2 u-q^2+24 q s^2+60 q s t+54 q s u-4 q s+30 q t^2+60 q t u \cr&\qquad\quad -8 q t+27 q u^2  -6 q u  + 3 q+96 s^2
   t+72 s^2 u+16 s^3-4 s^2+120 s t^2+216 s t u \cr&\qquad\quad -16 s t+81 s u^2-12 s u  +16 s+120 t^2 u+40 t^3-16
   t^2+108 t u^2 \cr&\qquad\quad -24 t u+16 t+27 u^3-9 u^2+15 u+2 \bigg) \nonumber \,.
\end{align}
\endgroup
To determine the flavor symmetries after the $\mathbb{Z}_2$ Stiefel--Whitney twist, it is necessary to know how the $\mathbb{Z}_2$ acts on the flavor factor that is attached to the $1$-node in the description of the 6d origin. First, we consider the special case where $q = 0$, then the flavor symmetry attached to the undecorated $1$-node (i.e., with $\fkg = \emptyset$) is either $\mathfrak{e}_7$, $\mathfrak{so}_{10}$, $\mathfrak{su}_3 \oplus \mathfrak{su}_2$, or $\varnothing$, depending on which combinations of parameters $t$, $u$, $s$, $p$ attain their lower limits, if any.
When we have $\mathfrak{e}_7$, we consider the special subalgebra
\begin{equation}
    \mathfrak{e}_7 \rightarrow \mathfrak{f}_4 \oplus \mathfrak{su}_2 \,,
\end{equation}
where the Dynkin index of the $\mathfrak{f}_4$ factor is one, and the $\mathbb{Z}_2$ acts as the center of the $\mathfrak{su}_2$. When the flavor algebra is $\mathfrak{so}_{10}$ we have the decomposition
\begin{equation}
    \begin{aligned}
      \mathfrak{so}_{10} &\rightarrow \mathfrak{su}_4 \oplus \mathfrak{su}_2 \oplus \mathfrak{su}_2 \\
      \bm{10} &\rightarrow (\bm{6,1,1}) \oplus (\bm{1,2,2})  \,.
    \end{aligned}
\end{equation}
The $\mathbb{Z}_2$ quotient is generated by the element $(2,0,0)$ of the combined center $\mathbb{Z}_4 \times \mathbb{Z}_2 \times \mathbb{Z}_2$ Note that this is the $\mathbb{Z}_2$ subgroup of the $\mathbb{Z}_4$ discussed around equation \eqref{eqn:so10d}. Thus we can see that the surviving flavor algebra in 4d is
\begin{equation}
    \mathfrak{su}_2^{(2)} \oplus \mathfrak{su}_2^{(1)} \oplus \mathfrak{su}_2^{(1)} \,,
\end{equation}
where we have written the Dynkin indices as superscripts, as determined from the decomposition of the vector representation. We can see from the Dynkin indices and the decomposition of the representations that the surviving flavor algebra from the $\mathfrak{so}_{10}$ after turning on this particular $\mathbb{Z}_2$ is in fact the enhanced $\mathfrak{so}_7^{(1)}$. Finally, when the 6d flavor symmetry attached to the $1$-node is $\mathfrak{su}_3 \oplus \mathfrak{su}_2$, we are considering the same decomposition as we did in the $\ell = 3$ case; the $\mathbb{Z}_2$ quotient acts on the $\mathfrak{su}_2$, and leaves the $\mathfrak{su}_3$ as a flavor symmetry of the Stiefel--Whitney twisted theory. When $q \geq 1$, the flavor symmetry attached to the $1$-node is $\mathfrak{so}_{4n}$, where $n$ is fixed in terms of the parameters $p$, $q$, $s$, $t$, $u$. The relevant decomposition appears in \cite{Ohmori:2018ona}, and we have
\begin{equation}
    \begin{aligned}
      \mathfrak{so}_{4n} &\rightarrow \mathfrak{su}_2^{(n)} \oplus \mathfrak{sp}_n^{(1)} \\
      \bm{4n} &\rightarrow (\bm{2, 2n}) \,.
    \end{aligned}
\end{equation}
The action of the $\mathbb{Z}_2$ is on the $\mathfrak{su}_2$ factor, and the flavor symmetry left after the Stiefel--Whitney twist is $\mathfrak{sp}_n$. As we can see, the embedding index of the surviving factor is one.

To proceed further, it is helpful to split up our analysis into the cases $q > 0$ and $q = 0$.

\paragraph{\boldmath{$q>0$}}

We begin by studying the flavor symmetry when $q > 0$. Due to the presence of the symplectic gauge algebra the flavor symmetries surviving after the Stiefel--Whitney twist have rather complex dependences on the $E_8$-homomorphism parameters. The most generic case occurs when $t > 0$, and we find
\begin{equation}
    \begin{aligned}
        &(\mathfrak{su}_1)_{12(N+p+s+u)+2q+8t} \oplus (\mathfrak{su}_1)_{12(N+p+s)+2q+8t+6u}
        \oplus (\mathfrak{su}_1)_{12(N+p)+2q+8t+6u+4s} \\ &\qquad \oplus
        (\mathfrak{su}_1)_{12N+2q+8t+6u+4s+2p}  \oplus (\mathfrak{su}_{q+4t+3u+2s+p})_{12+2q+8t+6u+4s+2p}
         \,.
    \end{aligned}
\end{equation}
When $t = 0$, but $u > 0$, the flavor symmetry is
\begin{equation}
    \begin{aligned}
        &(\mathfrak{sp}_1)_{6(N+p+s+u) + q} \oplus (\mathfrak{su}_1)_{12(N+p+s)+2q+6u}
        \oplus (\mathfrak{su}_1)_{12(N+p)+2q+6u+4s} \\ &\qquad \oplus
        (\mathfrak{su}_1)_{12N+2q+6u+4s+2p}  \oplus (\mathfrak{su}_{q+3u+2s+p})_{12+2q+6u+4s+2p}
         \,.
    \end{aligned}
\end{equation}
Next, we must consider the case where $t = u = 0$, but $s > 0$. The flavor symmetry becomes
\begin{equation}
    \begin{aligned}
        &(\mathfrak{sp}_2)_{6(N+p+s) + q}
        \oplus (\mathfrak{su}_1)_{12(N+p)+2q+4s}  \oplus
        (\mathfrak{su}_1)_{12N+2q+4s+2p}  \oplus (\mathfrak{su}_{q+2s+p})_{12+2q+4s+2p}
         \,.
    \end{aligned}
\end{equation}
When $t = u = s = 0$ and $p > 0$, one finds that the flavor algebra is
\begin{equation}
    \begin{aligned}
        &(\mathfrak{sp}_3)_{6(N+p) + q}
        \oplus (\mathfrak{su}_1)_{12N+2q+2p}  \oplus (\mathfrak{su}_{q+p})_{12+2q+2p}
         \,.
    \end{aligned}
\end{equation}
Finally, when all of the $E_8$-homomorphism parameters, except $q$, vanish, the flavor is
\begin{equation}\label{eqn:swiper}
    \begin{cases}
        (\mathfrak{sp}_4)_{6N + q} \oplus (\mathfrak{su}_{q})_{12+2q} \,, \quad &\text{when} \quad N > 1 \,, \\
        (\mathfrak{sp}_{q+4})_{6+q} \,, \quad &\text{when} \quad N = 1 \,.
    \end{cases}
\end{equation}

\paragraph{\boldmath{$q =0$}}
A similar analysis can be carried out when $q = 0$, and again one finds a variety of special cases. When $t > 0$ the flavor symmetry is
\begin{equation}
    \begin{aligned}
        &(\mathfrak{su}_1)_{12(N+p+s+u)+8t} \oplus (\mathfrak{su}_1)_{12(N+p+s)+8t+6u}
        \oplus (\mathfrak{su}_1)_{12(N+p)+8t+6u+4s} \\ &\qquad \oplus
        (\mathfrak{su}_1)_{12N+8t+6u+4s+2p}  \oplus (\mathfrak{su}_{4t+3u+2s+p})_{12+8t+6u+4s+2p}
         \,.
    \end{aligned}
\end{equation}
When $t = 0$, but $u > 0$, the flavor symmetry is
\begin{equation}
    \begin{aligned}
        &(\mathfrak{su}_3)_{6(N + u + s + p)} \oplus (\mathfrak{su}_1)_{12(N+p+s)+6u}
        \oplus (\mathfrak{su}_1)_{12(N+p)+6u+4s} \\ &\qquad \oplus
        (\mathfrak{su}_1)_{12N+6u+4s+2p}  \oplus (\mathfrak{su}_{3u+2s+p})_{12+6u+4s+2p}
         \,.
    \end{aligned}
\end{equation}
Next, we consider the case where $t = u = 0$, but $s > 0$. The flavor symmetry is
\begin{equation}
    \begin{aligned}
        &(\mathfrak{so}_7)_{6(N+p+s)}
        \oplus (\mathfrak{su}_1)_{12(N+p)+4s}  \oplus
        (\mathfrak{su}_1)_{12N+4s+2p}  \oplus (\mathfrak{su}_{2s+p})_{12+4s+2p}
         \,.
    \end{aligned}
\end{equation}
Finally, when $t = u = s = 0$ and $p > 0$, one finds that the flavor algebra is
\begin{equation}\label{eqn:dora}
    \begin{aligned}
        &(\mathfrak{f}_4)_{6(N+p)}
        \oplus (\mathfrak{su}_1)_{12N+2p}  \oplus (\mathfrak{su}_{p})_{12+2p}
         \,.
    \end{aligned}
\end{equation}
Note that this last expression is not valid when $p = N = 1$ due to an exceptional enhancement of the associated 6d SCFT, which we discuss below. This analysis exhausts the non-Abelian flavor symmetries of the $\mathcal{T}_2^{(N)}(p,s,u,2t,q)$ SW-fold SCFTs. In rare occasions, the flavor symmetry can be enhanced further, either because an Abelian $\mathfrak{u}(1)$ flavor symmetry of the 6d SCFT can enhance to an $\mathfrak{su}_2$ as described around equation \eqref{eqn:baryenc}, or else because the 6d SCFT has a baryonic $\mathfrak{su}_2$ flavor symmetry, in addition to the flavor symmetries that we have considered here. We discuss this latter case at the end of this subsection. Finally, it appears that in a small number of exceptional circumstances, there can also be flavor symmetry \emph{dehancement}; we explore these examples further in Section \ref{sec:classs}.

To conclude this subsection, we turn to the last class of SW-folds that we wish to consider. These are the $\mathbb{Z}_2$ SW-folds: $\mathcal{S}_2^{(N)}(p,s,u,2t+1,q)$. As usual, the central charges can be worked out from the formulae in equation \eqref{eqn:ack}, and one finds:
\begingroup
\allowdisplaybreaks
\begin{align}
        a &= \frac{1}{48} \bigg( 72 N^2 p+72 N^2 q+144 N^2 s+288 N^2 t+216 N^2 u+144 N^2+84 N p^2 \cr&\qquad\quad +168 N p q+336 N p
   s+672 N p t+504 N p u+336 N p+12 N q^2+192 N q s \cr&\qquad\quad +240 N q t+216 N q u+120 N q+192 N s^2+768 N s
   t+576 N s u+384 N s \cr&\qquad\quad +480 N t^2+864 N t u+480 N t+324 N u^2+144 N u+192 N+84 p^2 q+168 p^2 s \cr&\qquad\quad +336 p^2
   t+252 p^2 u+28 p^3+163 p^2+12 p q^2+192 p q s+240 p q t+216 p q u \\&\qquad\quad +110 p q+192 p s^2+768 p s t+576
   p s u+364 p s+480 p t^2+864 p t u+440 p t \cr&\qquad\quad +324 p u^2+114 p u+151 p + q^2 +12 q^2 s+12 q^2 t+12 q^2 u+96 q
   s^2+240 q s t \cr&\qquad\quad +216 q s u+100 q s+120 q t^2+240 q t u+80 q t+108 q u^2+98 q u+13 q+384 s^2 t \cr&\qquad\quad +288 s^2
   u+64 s^3+172 s^2+480 s t^2+864 s t u+400 s t+324 s u^2+84 s u+116 s \cr&\qquad\quad +480 t^2 u+160 t^3+160 t^2+432
   t u^2+392 t u+64 t+108 u^3 \cr&\qquad\quad +39 u^2+239 u+22 \bigg) \,, \cr
    c &= \frac{1}{12} \bigg( 18 N^2 p+18 N^2 q+36 N^2 s+72 N^2 t+54 N^2 u+36 N^2+21 N p^2+42 N p q \cr&\qquad\quad +84 N p s+168
   N p t+126 N p u+84 N p+3 N q^2+48 N q s+60 N q t+54 N q u \cr&\qquad\quad +30 N q+48 N s^2+192 N s t+144 N s u+96 N
   s+120 N t^2  +216 N t u \cr&\qquad\quad +120 N t+81 N u^2+36 N u+54 N+21 p^2 q+42 p^2 s+84 p^2 t+63 p^2 u+7 p^3 \cr&\qquad\quad +41
   p^2+3 p q^2+48 p q s+60 p q t+54 p q u+28 p q+48 p s^2+192 p s t+144 p s u \\&\qquad\quad +92 p s+120 p t^2+216 p
   t u+112 p t+81 p u^2+30 p u+45 p + \frac{1}{2}q^2 +3 q^2 s+3 q^2 t \cr&\qquad\quad +3 q^2 u+24 q s^2+60 q s t+54 q s u+26 q s+30 q
   t^2+60 q t u+22 q t+27 q u^2 \cr&\qquad\quad +28 q u+ \frac{13}{2}q+96 s^2 t+72 s^2 u+16 s^3+44 s^2+120 s t^2+216 s t u+104 s
   t \cr&\qquad\quad +81 s u^2+24 s u+38 s+120 t^2 u+40 t^3+44 t^2+108 t u^2+112 t u+30 t \cr&\qquad\quad +27 u^3+15 u^2+73 u+11 \bigg) \nonumber \,.
\end{align}
\endgroup
With generic values for the number of M5-branes and the $E_8$-homomorphism parameters, we find that the flavor symmetries of the resulting 4d $\mathcal{N}=2$ SCFTs are as follows:
\begin{equation}
    \begin{aligned}
        q > 0; t,u,s,p \geq 0; N \geq 1 \,&:\quad
        (\mathfrak{su}_1)_{12(N+p+s+u)+2q+8t+4} \oplus (\mathfrak{su}_1)_{12(N+p+s)+2q+8t+6u+4} \\ &\qquad
        \oplus (\mathfrak{su}_1)_{12(N+p)+2q+8t+6u+4s+4} \\ &\qquad \oplus
        (\mathfrak{su}_1)_{12N+2q+8t+6u+4s+2p+4}  \\&\qquad \oplus (\mathfrak{su}_{q+4t+3u+2s+p+2})_{12+2q+8t+6u+4s+2p+4}
         \,.
    \end{aligned}
\end{equation}
Again, we use a compact notation where $\mathfrak{su}_a \oplus \mathfrak{su}_b$ enhances to $\mathfrak{su}_{a+b}$ if the flavor central charges are identical.
When $q = 0$ there is an additional $\mathfrak{su}_2$ flavor symmetry in the 6d SCFT as the anti-symmetric hypermultiplet attached to the $\mathfrak{su}_4$ gauge algebra on the $1$-node is pseudo-real. The $\mathbb{Z}_2$ center-flavor symmetry does not embed inside of this $\mathfrak{su}_2$, and thus this flavor factor survives the Stiefel--Whitney twist intact. We find that the non-Abelian flavor symmetry of the SW-fold in the $q=0$ case is
\begin{equation}
    \begin{aligned}
        t,u,s,p \geq 0; N \geq 1 \,&:\quad
        (\mathfrak{su}_2)_{6(N+p+s+u+t)+3} \oplus
        (\mathfrak{su}_1)_{12(N+p+s+u)+8t+4} \\ &\qquad \oplus (\mathfrak{su}_1)_{12(N+p+s)+8t+6u+4}
        \oplus (\mathfrak{su}_1)_{12(N+p)+8t+6u+4s+4} \\ &\qquad \oplus
        (\mathfrak{su}_1)_{12N+8t+6u+4s+2p+4}  \\&\qquad \oplus (\mathfrak{su}_{4t+3u+2s+p+2})_{12+8t+6u+4s+2p+4}
         \,.
    \end{aligned}
\end{equation}
As discussed around equation \eqref{eqn:baryenc}, there can be further enhancement when $p = s = u = t = 0$, as in those cases it is expected that the baryonic $\mathfrak{u}(1)$ global symmetry enhances to an $\mathfrak{su}_2$, and this factor can further combine with the other non-Abelian factors for low values of $N$. We discuss some of these baryonic enhancements further in Section \ref{sec:classs}.

There are also two classes of $\mathbb{Z}_2$ SW-folds where the 6d SCFT origin itself has a baryonic $\mathfrak{su}_2$ flavor symmetry. These 6d theories correspond to the tensor branch configurations
\begin{equation}
    \overset{\mathfrak{sp}_1}{1}\overset{\mathfrak{su}_2}{2}\cdots\overset{\mathfrak{su}_2}{2} \,, \quad \text{ and } \quad 1\overset{\mathfrak{su}_2}{2}\cdots\overset{\mathfrak{su}_2}{2} \,.
\end{equation}
The $\mathbb{Z}_2$ center-flavor symmetry embeds trivially inside of the center of the baryonic $\mathfrak{su}_2$, and thus this additional non-Abelian factor is unbroken by the Stiefel--Whitney twist. In addition to the above flavor symmetries, these SW-folds then have the following additional flavor algebras:
\begin{equation}
    \begin{aligned}
        \mathcal{T}_2^{(N>1)}(0,0,0,0,1) &\,: \quad (\mathfrak{su}_2)_{6N^2 + N} \,, \\
        \mathcal{T}_2^{(N>1)}(1,0,0,0,0) &\,: \quad (\mathfrak{su}_2)_{6N^2 + 7N + 1} \,.
    \end{aligned}
\end{equation}
In the latter case, there is a further exceptional enhancement when $N = 1$.\footnote{For the former case, the flavor symmetry when $N=1$ is captured by equation \eqref{eqn:swiper}.} Let us discuss this special case of $\mathcal{T}_2^{(1)}(1,0,0,0,0)$, which arises from the 6d SCFT with tensor branch configuration
\begin{equation}
    1\overset{\mathfrak{su}_2}{2} \,.
\end{equation}
This 6d SCFT has an $\mathfrak{e}_7 \oplus \mathfrak{so}_7$ flavor symmetry, instead of the naively expected $\mathfrak{e}_7 \oplus \mathfrak{so}_8$ flavor algebra. Taking into account the embedding of the $\mathbb{Z}_2$ center-flavor symmetry inside of the $\mathfrak{so}_7$, we find that the flavor carried by the resulting SW-fold theory is
\begin{equation}
    (\mathfrak{f}_4)_{12} \oplus (\mathfrak{su}_{2})_{7} \oplus (\mathfrak{su}_2)_{7} \,.
\end{equation}
The SW-folds $\mathcal{T}_2^{(N)}(1,0,0,0,0)$ and $\mathcal{T}_2^{(N)}(0,0,0,0,1)$ have been discussed previously in \cite{Giacomelli:2020gee} where they were referred to as $\mathcal{T}_{E_6,2}^{(r+1)}$ and $\mathcal{S}_{E_6,2}^{(r)}$, respectively.

\subsection{Coulomb Branch Operator Spectrum}\label{sec:cb}

In \cite{Ohmori:2018ona}, the authors developed a heuristic method for determining the scaling dimensions of the Coulomb branch operators of the 4d SCFT obtained from the Stiefel--Whitney twisted compactification of a very Higgsable 6d SCFT. In this section, we test the consistency of the method described therein when applied to the SW-folds. In \cite{Ohmori:2018ona}, the authors verify their method by testing that the Coulomb branch spectrum they obtain agrees with the known spectrum from the dual twisted-class $\mathcal{S}$ theory; in our cases no such class $\mathcal{S}$ theory is known, and we rely on the weaker test that \begin{equation}
    4(2a - c) = \sum_i 2\Delta(u_i) - 1 \,,
\end{equation}
where the LHS is determined using the anomaly polynomial as in Section \ref{sec:ccs}. From the anomaly polynomial we expect that this quantity is:
\begin{equation}\label{eqn:acANOM}
    4(2a - c) = n_V - 64A - \frac{48}{\ell}B \,.
\end{equation}
We find that the method in \cite{Ohmori:2018ona} is consistent with this formula in most cases, but that it must be extended in a few special cases beyond the ones considered in \cite{Ohmori:2018ona}. These special cases occur for the E-type SW-folds that we discuss in Section \ref{sec:Esfolds}. We hope that this analysis will be useful in the determination of a closed-form top-down method for understanding the Coulomb branch spectrum from the 6d origin.

We are interested in 6d SCFTs with tensor branch configurations of the form given in equation \eqref{eqn:quivquiv}. We recall that in equation \eqref{eqn:quivquiv} we index the quiver nodes from $0$ to $r$ going from left-to-right. The algorithm presented in Appendix B of \cite{Ohmori:2018ona} for the Coulomb branch operator dimensions of the 4d $\mathcal{N}=2$ theory obtained from the $\mathbb{Z}_\ell$ Stiefel--Whitney twist of 6d SCFTs of the form in equation \eqref{eqn:quivquiv} is as follows. From each node of the form $\overset{\mathfrak{su}_{\ell k_i}}{2}$ the spectrum of operator dimensions is
\begin{equation}
    \Delta_i = \{ 6(r+1-i) \} \cup \{ 6(r+1-i) + d \,|\, d = 2, \cdots, k_i \} \,,
\end{equation}
where $i$ is the index of that quiver node. We can directly work out the contribution to $4(2a - c)$ from each of these $2$-nodes:
\begin{equation}
    4(2a - c)_i = \sum_{u \in \Delta_i} (2u - 1) = k_i^2 + 12k_i(r+1-i) - 2 \,.
\end{equation}
Summing over all contributions to $4(2a - c)$, we find
\begin{equation}\label{eqn:acCB}
    4(2a - c) = 4(2a - c)_0 + \left(1 - n_V^0 - \frac{12(r+1)d_0}{\ell}\right) + n_V  - 64A - \frac{48B}{\ell} \,,
\end{equation}
where we have used the expressions for $n_V$, $A$, and $B$ in equations \eqref{eqn:nv} and \eqref{eqn:B}. It remains for us to determine the contribution from the 1-nodes in equation \eqref{eqn:quivquiv}; there are four distinct cases which we must consider.
For each of the following $\mathfrak{g}$, the Coulomb branch operator dimensions coming from the 1-node with gauge algebra $\mathfrak{g}$ are proposed to be
\begin{equation}
    \begin{aligned}
      \mathfrak{g} = \varnothing &: \quad \left\{ \frac{6(r+1)}{\ell} \right\} \,, \\
      \mathfrak{g} = \mathfrak{su}_{\ell k_0} &: \quad \{ 6(r+1) \} \cup \{ 6(r+1) + d \,|\, d = 2, \cdots, k_0 \}  \,, \\
      \mathfrak{g} = \mathfrak{sp}_{2m+1} &: \quad \{ 6(r+1) \} \cup \{ 6(r+1) + 2d \,|\, d = 1, \cdots, m \}
      \,, \\
      \mathfrak{g} = \mathfrak{sp}_{2m} &: \quad \{ 6(r+1) \}
      \cup \{ 6(r+1) + 2d \,|\, d = 1, \cdots, m - 1 \}
      \cup \left\{ 3(r+1) + m \right\}
      \,,
    \end{aligned}
\end{equation}
where we remind the reader that the latter two options can only occur when $\ell = 2$. In each of the four cases, we can see that
\begin{equation}
    4(2a - c)_0 =  \frac{12(r+1)d_0}{\ell} + n_V^0 -1 \,,
\end{equation}
and thus we see that equation \eqref{eqn:acCB} determining $4(2a - c)$ from the putative Coulomb branch spectrum proposed in \cite{Ohmori:2018ona} matches the value of $4(2a - c)$ determined in equation \eqref{eqn:acANOM} from the 6d anomaly polynomial and tensor branch configuration.

\subsection{Alternative Constructions}

In some cases, the theories we can generate via SW-folds have alternative constructions. In this section we compare with class $\mathcal{S}$ constructions, as well as methods based 4d $\mathcal{N} = 2$ S-folds.

\subsection[Exceptional Twisted Class \texorpdfstring{$\mathcal{S}$}{S} and Flavor Symmetry]{Exceptional Twisted Class \boldmath{$\mathcal{S}$} and Flavor Symmetry}\label{sec:classs}

In contrast to constructing 4d $\mathcal{N}=2$ SCFTs via compactification of 6d $(1,0)$ SCFTs on a $T^2$, one can also explore the class $\mathcal{S}$  construction \cite{Gaiotto:2009we,Gaiotto:2009hg}. This class of theories is obtained via the twisted-compactification of the 6d $(2,0)$ SCFT of type $\mathfrak{g}$ on a punctured Riemann surface.

It has been established in \cite{Ohmori:2015pua,Ohmori:2015pia} that the 4d $\mathcal{N}=2$ SCFT that arises from compactifying minimal $(\mathfrak{e}_6, \mathfrak{e}_6)$ conformal matter on a $T^2$ with $\mathbb{Z}_3$ Stiefel--Whitney twist is dual to a class $\mathcal{S}$ theory. The latter is obtained from compactification of the 6d $(2,0)$ SCFT of type $\mathfrak{so}_8$ on a sphere with two maximal $\mathbb{Z}_3$-twisted punctures and one simple puncture. Similarly, minimal $(\mathfrak{e}_7, \mathfrak{e}_7)$ conformal matter on a $T^2$ with $\mathbb{Z}_2$ Stiefel--Whitney twist is dual to the 6d $(2,0)$ SCFT of type $\mathfrak{e}_6$ on a sphere with two maximal $\mathbb{Z}_2$-twisted punctures and one simple puncture. In rare limiting cases, some of the 6d SCFTs written in Table \ref{tbl:6dscfts} can also be obtained by starting from minimal $(\mathfrak{e}_{n}, \mathfrak{e}_{n})$ conformal matter, with $n = 6, 7$, and performing nilpotent Higgs branch deformations of the $\mathfrak{g} \oplus \mathfrak{g}$ flavor symmetry. Compactifying these theories with Stiefel--Whitney twist then gives rise to 4d $\mathcal{N}=2$ SCFTs that can also be obtained by partial closure of the maximal punctures in the aforementioned class $\mathcal{S}$ construction.\footnote{See \cite{Baume:2021qho} for an in depth analysis of the relationship between the nilpotent Higgs branch deformations and the partial closure of the punctures in the untwisted case.}

Class $\mathcal{S}$ theories of type $\mathfrak{so}_8$ with $\mathbb{Z}_3$-twisted punctures have been studied in \cite{Chacaltana:2016shw}. Similarly, class $\mathcal{S}$ of type $\mathfrak{e}_6$ with $\mathbb{Z}_2$-twisted punctures has been explored in \cite{Chacaltana:2015bna}.
We list the SW-fold SCFTs from Table \ref{tbl:genSfolds} that can be realized in class $\mathcal{S}$, as described, in Table \ref{tbl:cs}. In all cases, bar one, the flavor symmetry determined from the dual class $\mathcal{S}$ construction matches the flavor symmetry that was determined from the Stiefel--Whitney twisted description in Section \ref{sec:ccs}. There is one special case, the SW-fold theory $\mathcal{T}_2^{(1)}(0,1,0,0,0)$, for which the analysis in Section \ref{sec:ccs} predicts that the non-Abelian flavor symmetry should be
\begin{equation}
    (\mathfrak{so}_7)_{12} \oplus (\mathfrak{so}_7)_{16} \,,
\end{equation}
but the dual class $\mathcal{S}$ theory has non-Abelian flavor algebra
\begin{equation}
    (\mathfrak{so}_7)_{12} \oplus (\mathfrak{g}_2)_{16} \,.
\end{equation}
This kind of dehancement occurs in the context of 6d SCFTs when one has an $\mathfrak{su}_2$ gauge algebra associated to a tensor with self-pairing 2 \cite{Morrison:2016djb}. In this case, the Coulomb branch description of the 4d SW-fold SCFT has a single $\mathfrak{su}_2$ gauge algebra, and a parallel argument to that in 6d may explain why the flavor symmetry is smaller than expected. We would similarly suspect that the SW-fold SCFTs $\mathcal{T}_4^{(1)}(0,1)$ and $\mathcal{T}_3^{(1)}(0,1,0)$, whose 4d Coulomb branch descriptions also involve a single $\mathfrak{su}_2$ gauge algebra coming from a 2-node decorated algebra in 6d, to evince similar dehancement. It would be interesting to understand the physical mechanism behind this rare but curious effect.

\begin{table}[ht]
    \centering
    \begin{threeparttable}
    \begin{tabular}{ccccc}
    \toprule
        SW-fold & SW Twist
        & Class $\mathcal{S}$ Type & Punctures & Flavor
        \\\midrule
        $\mathcal{T}_2^{(1)}(2,0,0,0,0)$ & \multirow{9}{*}{$\mathbb{Z}_2$} & \multirow{9}{*}{$\mathfrak{e}_6$}
        & $[0, A_2]_I$ & $(\mathfrak{f}_4)_{18} \oplus (\mathfrak{su}_3)_{16}$ 
        \\
        $\mathcal{T}_2^{(2)}(1,0,0,0,0)$ & &
        & $[0, A_2 + \widetilde{A}_1]_I$ & $(\mathfrak{f}_4)_{18} \oplus (\mathfrak{su}_2)_{39}$  
        \\
        $\mathcal{T}_2^{(1)}(1,0,0,0,1)$ & &
        & $[A_2, A_1]_I$ &  $(\mathfrak{sp}_3)_{13} \oplus (\mathfrak{su}_3)_{16}$ 
        \\
        $\mathcal{T}_2^{(2)}(0,0,0,0,1)$ & &
        & $[A_2 + \widetilde{A}_1, A_1]_I$ & $(\mathfrak{sp}_4)_{13} \oplus (\mathfrak{su}_2)_{26}$ 
        \\
        $\mathcal{T}_2^{(1)}(0,1,0,0,0)$ &  &
        & $[A_2, \widetilde{A}_1]_I$ & $(\mathfrak{so}_7)_{12} \oplus (\mathfrak{g}_2)_{16}$ 
        \\
        $\mathcal{T}_2^{(1)}(1,0,0,0,0)$ & &
        & $[A_2 + \widetilde{A}_1, \widetilde{A}_1]_M$ & $(\mathfrak{f}_4)_{12} \oplus 2(\mathfrak{su}_2)_7$ 
        \\
        $\mathcal{S}_2^{(1)}(0,0,0,0,0)$ & &
        & $[A_2, A_1 + \widetilde{A}_1]_M$ & $(\mathfrak{su}_6)_{16} \oplus (\mathfrak{su}_2)_9$ 
        \\
        $\mathcal{T}_2^{(1)}(0,0,0,0,1)$ & &
        & $[A_2 + \widetilde{A}_1, A_1 + \widetilde{A}_1]_M$ & $(\mathfrak{sp}_5)_7$ 
        \\\midrule
        $\mathcal{T}_3^{(1)}(1,0,0)$  & \multirow{2}{*}{$\mathbb{Z}_3$} & \multirow{2}{*}{$\mathfrak{so}_8$} & $[0, A_1]_I$ & $(\mathfrak{g}_2)_{8} \oplus (\mathfrak{su}_2)_{14}$ 
        \\
        $\mathcal{S}_3^{(1)}(0,0,1)$  & & & $[A_1, A_1]_I$ & $(\mathfrak{su}_4)_{14}$ 
        \\\bottomrule
    \end{tabular}
    \end{threeparttable}
    \caption{Twisted punctures are usually denoted with an \underline{underline}, however, since all of the punctures that we write in this table are twisted, we have chosen to drop this notational feature. The subscripts $I$ (interacting) and $M$ (mixed) denote whether the class $\mathcal{S}$ theory is an interacting SCFT, or whether it is coupled to free hypermultiplets, respectively. In the latter case, the SW-fold SCFT matches the interacting part of the class $\mathcal{S}$ theory. In the flavor column we write the non-Abelian flavor algebra as determined from the class $\mathcal{S}$ perspective.}
    \label{tbl:cs}
\end{table}

\subsubsection{Relation to 4d \texorpdfstring{$\mathcal{N} = 2$}{N=2} S-fold Theories}\label{sec:seqsw}

In the previous sections we studied the properties of 4d $\mathcal{N} = 2$ SW-folds, and we also observed that in some cases, the resulting theories can be realized via 4d $\mathcal{N} = 2$ S-fold theories. In this section we discuss some suggestive hints that such a top-down correspondence may be at work, but leave a more complete treatment for future work.

To frame the discussion to follow, recall that an S-fold in Type IIB / F-theory backgrounds
is a non-perturbative generalization of an orientifold plane in which a quotient on the target space is combined with a group action from the
$SL(2,\mathbb{Z})$ duality group of Type IIB string theory. Now, for such a quotient to exist we must work at specific values of the axio-dilaton compatible with this group action, e.g. $\tau = i$ and $\tau = \exp(2 \pi i / 6)$. In the presence of a probe D3-brane, this can be used to realize $\mathcal{N} = 3$ SCFTs, as noted in \cite{Garcia-Etxebarria:2015wns} (see also \cite{Aharony:2016kai}).
One can also introduce 7-branes provided they are compatible with a specific value of $\tau$, and this leads to 4d $\mathcal{N} = 2$ S-folds. D3-brane probes of such systems then realize 4d $\mathcal{N} = 2$ SCFTs \cite{Apruzzi:2020pmv,Giacomelli:2020jel,Heckman:2020svr,Giacomelli:2020gee,Bourget:2020mez}. As a general comment, the global symmetry in these systems also depends on the presence (or absence) of a torsional flux, and this effect can be detected via open string junctions which extend from the D3-brane to the 7-brane flavor stack \cite{Heckman:2020svr}.

As we now explain, there are reasons to suspect that the 4d SW-fold theories considered in this chapter, and 4d S-fold theories are potentially related by a chain of dualities. To see why, it is helpful to first consider some of the different top-down realizations of the rank $N$ $E_8$ Minahan--Nemeschansky theory \cite{Minahan:1996fg, Minahan:1996cj}. One way is to first start with the rank $N$ E-string theory 6d SCFT. Compactification on a $T^2$ then yields the 4d $\mathcal{N} = 2$ SCFT. Observe that in M-theory, this is engineered from the $T^2$ compactification of $N$ M5-branes probing an $E_8$ nine-brane in M-theory. On the other hand, we can also directly relate this to Type IIB/F-theory backgrounds with $N$ D3-branes probing an $E_8$ seven-brane. Intuitively, there is a generalized notion of T-duality at play which allows us trade the $E_8$ nine-brane of M-theory for the $E_8$ seven-brane of F-theory.\footnote{Indeed, this figures prominently in the standard Fourier--Mukai transformation of heterotic vector bundles on an elliptically fibered Calabi--Yau threefold and their characterization in the associated spectral cover construction for gauge theory on the base K\"ahler surface (see, e.g., \cite{Donagi:2000fw}).}

There is a natural extension of this generalized T-duality which makes any proposed correspondence quite suggestive. On the M-theory side, our 6d SCFT orbi-instanton theories were realized by small instantons probing an ADE singularity wrapped by an $E_8$ nine-brane. Likewise, we note that D3-branes probing an ADE singularity wrapped by an $E_8$ seven-brane will give rise to 4d $\mathcal{N} = 2$ SCFTs. In both cases, the worldvolume theory of the probe brane is specified as an instanton solution in the directions filled by the ambient brane. As such, we can generate a wide class of examples by specifying the boundary data of a flat connection at the boundary $S^3 / \Gamma_\text{ADE}$, which are in turn captured by discrete group homomorphisms $\text{Hom}(\Gamma_\text{ADE} \rightarrow E_8)$ \cite{DelZotto:2014hpa,Heckman:2015bfa}. So, from this perspective, we see that the $T^2$ compactification of the 6d orbi-instanton theories provides us with a direct way to match the two sets of theories.

So far, our discussion has made no reference to switching on an SW-fold on the orbi-instanton side of this correspondence. Now, on the SW-fold side we consider a pair of holonomies which commute in $\widetilde{G}$ up to a flux valued in the quotienting subgroup. These profiles make direct reference to the $T^2$ direction on which we have compactified the orbi-instanton theory. To make sense of such deformations in the D3-brane probe theories, we would need to have a notion of generalized T-duality which extends to such configurations as well. The fact that there are known examples where SW-folds and S-folds produce the same conformal fixed point is of course suggestive \cite{Giacomelli:2020jel}, but without
a suitable generalization of T-duality, it is unclear whether it should be expected to persist for all SW-folds and S-folds, or just some subset. Exploring this issue further would be of great interest and would likely lead to a better understanding of both sorts of constructions.

\section{SW-folds of Type DE}\label{sec:Esfolds}

In Section \ref{sec:e8CF}, we have enumerated the 6d SCFTs that have non-trivial center-flavor symmetry and arise from a Higgsing, by homomorphisms $\mathbb{Z}_K \rightarrow E_8$, of the rank $N$ orbi-instanton theory of type $(\mathfrak{e}_8, \mathfrak{su}_K)$. In Section \ref{sec:sfolds}, we considered the compactification of the 6d SCFTs found in Section \ref{sec:e8CF} on a $T^2$ together with a Stiefel--Whitney twist in the center-flavor symmetry. We refer to the resulting 4d $\mathcal{N}=2$ SCFTs as the A-type SW-folds, due to the $\mathfrak{su}_K$ factor in the orbi-instanton origin. In this section we consider the rank $N$ orbi-instanton theories of type $(\mathfrak{e}_8, \mathfrak{g})$, where $\mathfrak{g}$ is any ADE Lie algebra. We consider homomorphisms $\Gamma \rightarrow E_8$, where $\Gamma$ is the finite ADE group of the same type as $\mathfrak{g}$, and such that the Higgsed 6d SCFT has a non-trivial center-flavor symmetry. For generic values of $N$, this center-flavor symmetry can be, at most
\begin{equation}
    \begin{aligned}
      \mathbb{Z}_4 \qquad &\text{ for } \qquad \mathfrak{g} = \mathfrak{so}_{4k+2} \,, \\
      \mathbb{Z}_2 \times \mathbb{Z}_2 \qquad &\text{ for } \qquad  \mathfrak{g} = \mathfrak{so}_{4k} \,, \\
      \mathbb{Z}_3 \qquad &\text{ for } \qquad  \mathfrak{g} = \mathfrak{e}_6 \,, \\
      \mathbb{Z}_2 \qquad &\text{ for } \qquad \mathfrak{g} = \mathfrak{e}_7 \,.
    \end{aligned}
\end{equation}
We can now consider the $T^2$ compactifications of these 6d SCFTs with a non-trivial Stiefel--Whitney class inside of the center-flavor symmetry turned on. This opens up a vast new vista of D-type and E-type SW-folds. We will not consider all such families of SW-folds here, but we highlight a few choice examples; the remaining cases can be determined straightforwardly from the methods utilized throughout this chapter.

\subsection[\texorpdfstring{$E_6$}{E6}-type SW-folds]{\boldmath{$E_6$}-type SW-folds}\label{sec:e6sfolds}

We begin with the $(\mathfrak{e}_8, \mathfrak{e}_6)$ orbi-instanton, of rank $N$. The tensor branch configuration has the form
\begin{equation}
    1 \, 2 \overset{\mathfrak{su}_2}{2} \overset{\mathfrak{g}_2}{3} 1 \overset{\mathfrak{f}_4}{5} 1 \overset{\mathfrak{su}_3}{3} 1 \overset{\mathfrak{e}_6}{6}
      1 \overset{\mathfrak{su}_3}{3} 1
      \overbrace{\overset{\mathfrak{e}_6}{6} 1 \overset{\mathfrak{su}_3}{3} 1 \cdots \overset{\mathfrak{e}_6}{6} 1 \overset{\mathfrak{su}_3}{3} 1}^{N-1} \,,
\end{equation}
which has an $\mathfrak{e}_8 \oplus \mathfrak{e}_6$ flavor symmetry, and no center-flavor symmetry. We will consider the 6d SCFTs obtained by the finite group homomorphism:
\begin{equation}
    \Gamma_{\mathfrak{e}_6} \rightarrow E_8 \,,
\end{equation}
with $\Gamma_{E_6}$ the binary tetrahderal finite subgroup of $SU(2)$. The Higgs branch flows induced by such homomorphisms have been studied in \cite{Frey:2018vpw}. There are fifty-two such SCFTs, however we are only interested in those that have a non-trivial center-flavor symmetry. There are only seven such Higgsings which give rise to a center-flavor symmetry, which is always a $\mathbb{Z}_3$.\footnote{Much as in Section \ref{sec:e8CF}, we emphasize that if $N = 1$ then many more of the $E_8$-homomorphisms lead to theories with center-flavor symmetry, and it is not restricted to be $\mathbb{Z}_3$.} These correspond to the seven tensor branch geometries
\begingroup
\allowdisplaybreaks
\begin{align}
        \overset{\mathfrak{su}_3}{3} 1 \overset{\mathfrak{su}_3}{3} 1 \overset{\mathfrak{e}_6}{6}
      1 \overset{\mathfrak{su}_3}{3} 1
      \underbrace{\overset{\mathfrak{e}_6}{6} 1 \overset{\mathfrak{su}_3}{3} 1 \cdots \overset{\mathfrak{e}_6}{6} 1 \overset{\mathfrak{su}_3}{3} 1}_{N-1} &\,, \label{eqn:e61} \\
      \overset{\mathfrak{su}_3}{2} \overset{\mathfrak{su}_3}{2} 1 \overset{\mathfrak{e}_6}{6}
      1 \overset{\mathfrak{su}_3}{3} 1
      \underbrace{\overset{\mathfrak{e}_6}{6} 1 \overset{\mathfrak{su}_3}{3} 1 \cdots \overset{\mathfrak{e}_6}{6} 1 \overset{\mathfrak{su}_3}{3} 1}_{N-1} &\,, \\
      1\overset{\mathfrak{su}_3}{3} 1 \underset{1}{\overset{\mathfrak{e}_6}{6}}
      1 \overset{\mathfrak{su}_3}{3} 1
      \underbrace{\overset{\mathfrak{e}_6}{6} 1 \overset{\mathfrak{su}_3}{3} 1 \cdots \overset{\mathfrak{e}_6}{6} 1 \overset{\mathfrak{su}_3}{3} 1}_{N-1} &\,, \\
      \overset{\mathfrak{su}_3}{2} 1 \underset{1}{\overset{\mathfrak{e}_6}{6}}
      1 \overset{\mathfrak{su}_3}{3} 1
      \underbrace{\overset{\mathfrak{e}_6}{6} 1 \overset{\mathfrak{su}_3}{3} 1 \cdots \overset{\mathfrak{e}_6}{6} 1 \overset{\mathfrak{su}_3}{3} 1}_{N-1} &\,, \\
      1 \overset{1}{\underset{1}{\overset{\mathfrak{e}_6}{6}}}
      1 \overset{\mathfrak{su}_3}{3} 1
      \underbrace{\overset{\mathfrak{e}_6}{6} 1 \overset{\mathfrak{su}_3}{3} 1 \cdots \overset{\mathfrak{e}_6}{6} 1 \overset{\mathfrak{su}_3}{3} 1}_{N-1} &\,, \\
      \overset{\mathfrak{e}_6}{3}
      1 \overset{\mathfrak{su}_3}{3} 1
      \underbrace{\overset{\mathfrak{e}_6}{6} 1 \overset{\mathfrak{su}_3}{3} 1 \cdots \overset{\mathfrak{e}_6}{6} 1 \overset{\mathfrak{su}_3}{3} 1}_{N-1} &\,, \\
      \overset{\mathfrak{su}_6}{2}
      \overset{\mathfrak{su}_3}{2} 1
      \underbrace{\overset{\mathfrak{e}_6}{6} 1 \overset{\mathfrak{su}_3}{3} 1 \cdots \overset{\mathfrak{e}_6}{6} 1 \overset{\mathfrak{su}_3}{3} 1}_{N-1} &\,. \label{eqn:e67}
\end{align}
\endgroup
To work out the central charges of the 4d $\mathcal{N}=2$ SW-folds obtained from the Stiefel--Whitney twisted compactification of these 6d SCFTs we will again use the formulae of \cite{Ohmori:2018ona}, which we have summarized in equations \eqref{eqn:fields} and \eqref{eqn:ack}.

We consider the compactification of the tensor branch configuration in equation \eqref{eqn:e61} in detail. The 6d SCFT has an $(SU(3)^2 \times E_6)/\mathbb{Z}_3$ flavor symmetry group, and after the Stiefel--Whitney twist there remains only a $G_2$ subgroup of the $E_6$. For the 6d SCFT from which this SW-fold originates, the relevant terms in the anomaly polynomial are
\begin{equation}
    I_8 \supset \frac{1}{24}\left(\left(\frac{7N}{8} + \frac{35}{12} \right)p_1(T)^2 - \left(72N^2 + 209N + 102\right)c_2(R)p_1(T)\right) + \frac{3}{16}p_1(T)\operatorname{Tr}F^2 \,,
\end{equation}
where we have only written the mixed-gravitational-flavor anomaly for the $\mathfrak{e}_6$ flavor algebra. Next, we find that the contribution from the weakly-coupled multiplets is
\begin{equation}
    I_8^\text{fields} \supset \left(-\frac{17N}{192} - \frac{1}{288} \right)p_1(T)^2 + \left(-\frac{41N}{24} - \frac{1}{4} \right)c_2(R)p_1(T) \,.
\end{equation}
Putting this altogether we find that
\begin{equation}
    a - a_\text{generic} = 12N^2 + 27N + 15 \,, \quad c - c_\text{generic} = 12N^2 + 28N + 16 \,, \quad \kappa - \kappa_\text{generic} = 12 \,,
\end{equation}
where we have used that the Dynkin index of the $\mathfrak{g}_2$ inside of the $\mathfrak{e}_6$ is one, as explained around equation \eqref{eqn:e6ss}.
It remains for us to determine what the contributions to the central charges are from the 4d theory at the generic point of the Coulomb branch. The $\mathbb{Z}_3$ Stiefel--Whitney twist breaks the $\mathfrak{su}_3$ gauge algebras completely, and it breaks each $\mathfrak{e}_6$ down to a $\mathfrak{g}_2$. As such, at the generic point of the Coulomb branch, we have $(4(N-1)+8)$ vector multiplets from the 6d tensors, and $N\operatorname{dim}(\mathfrak{g}_2)$ vector multiplets from the surviving $\mathfrak{g}_2$ gauge symmetries. We end up with
\begin{equation}
    a_\text{generic} = \frac{5}{24}(18N + 4) \,, \qquad c_\text{generic} = \frac{1}{6}(18N + 4) \,.
\end{equation}
Furthermore, since there are no hypermultiplets charged under the residual $\mathfrak{g}_2$ flavor symmetry we find that
\begin{equation}
    \kappa_\text{generic} = 0 \,.
\end{equation}
The central charges of the SW-fold are thus:
\begin{equation}\label{eqn:acE6}
    a = 12N^2 + \frac{123N}{4} + \frac{95}{6} \,, \qquad c = 12N^2 + 31N + \frac{50}{3} \,,
\end{equation}
and the flavor symmetry and flavor central charge is
\begin{equation}
    (\mathfrak{g}_2)_{12} \,.
\end{equation}
The central charges from each of the 6d SCFTs with tensor branch descriptions given in equations \eqref{eqn:e61} to \eqref{eqn:e67} can be determined, and we do not belabor the computation here. The central charges, flavor symmetries, and flavor central charges for the seven families of $E_6$-type $\mathbb{Z}_3$ SW-folds are given in Table \ref{tbl:e6Sfolds}.

\begin{table}[ht]
    \centering
    \small
    \begin{threeparttable}
    \begin{tabular}{cccc}
    \toprule
         6d Origin & $a$ & $c$ & Flavor \\\midrule
         $\overset{\mathfrak{su}_3}{3} 1 \overset{\mathfrak{su}_3}{3} 1 \overset{\mathfrak{e}_6}{6}
      1 \overset{\mathfrak{su}_3}{3} 1
      \underbrace{\overset{\mathfrak{e}_6}{6} 1 \overset{\mathfrak{su}_3}{3} 1 \cdots \overset{\mathfrak{e}_6}{6} 1 \overset{\mathfrak{su}_3}{3} 1}_{N-1}$ & $12N^2 + \frac{123N}{4} + \frac{95}{6}$ & $12N^2 + 31N + \frac{50}{3}$ & $(\mathfrak{g}_2)_{12}$ \\

      $\overset{\mathfrak{su}_3}{2} \overset{\mathfrak{su}_3}{2} 1 \overset{\mathfrak{e}_6}{6}
      1 \overset{\mathfrak{su}_3}{3} 1
      \underbrace{\overset{\mathfrak{e}_6}{6} 1 \overset{\mathfrak{su}_3}{3} 1 \cdots \overset{\mathfrak{e}_6}{6} 1 \overset{\mathfrak{su}_3}{3} 1}_{N-1}$ & $12N^2 + \frac{111N}{4} + \frac{97}{8}$ & $12N^2 + 28N + 13$ & $(\mathfrak{g}_2)_{12}$ \\

      $1\overset{\mathfrak{su}_3}{3} 1 \underset{\displaystyle 1}{\overset{\mathfrak{e}_6}{6}}
      1 \overset{\mathfrak{su}_3}{3} 1
      \underbrace{\overset{\mathfrak{e}_6}{6} 1 \overset{\mathfrak{su}_3}{3} 1 \cdots \overset{\mathfrak{e}_6}{6} 1 \overset{\mathfrak{su}_3}{3} 1}_{N-1}$ & $12N^2 + \frac{91N}{4} + \frac{47}{6}$ & $12N^2 + 23N + \frac{26}{3}$ & $(\mathfrak{g}_2)_{12} \oplus (\mathfrak{g}_2)_{4N+8}$ \\

      $\overset{\mathfrak{su}_3}{2} 1 \underset{\displaystyle 1}{\overset{\mathfrak{e}_6}{6}}
      1 \overset{\mathfrak{su}_3}{3} 1
      \underbrace{\overset{\mathfrak{e}_6}{6} 1 \overset{\mathfrak{su}_3}{3} 1 \cdots \overset{\mathfrak{e}_6}{6} 1 \overset{\mathfrak{su}_3}{3} 1}_{N-1}$ & $12N^2 + \frac{87N}{4} + \frac{145}{24}$ & $12N^2 + 22N + \frac{41}{6}$ & $(\mathfrak{g}_2)_{12} \oplus (\mathfrak{su}_2)_{12N+20}$  \\

      $1 \overset{\displaystyle 1}{\underset{\displaystyle 1}{\overset{\mathfrak{e}_6}{6}}}
      1 \overset{\mathfrak{su}_3}{3} 1
      \underbrace{\overset{\mathfrak{e}_6}{6} 1 \overset{\mathfrak{su}_3}{3} 1 \cdots \overset{\mathfrak{e}_6}{6} 1 \overset{\mathfrak{su}_3}{3} 1}_{N-1}$ & $12N^2 + \frac{75N}{4} + \frac{23}{8}$ & $12N^2 + 19N + \frac{7}{2}$ & $(\mathfrak{g}_2)_{12}$  \\

      $\overset{\mathfrak{e}_6}{3}
      1 \overset{\mathfrak{su}_3}{3} 1
      \underbrace{\overset{\mathfrak{e}_6}{6} 1 \overset{\mathfrak{su}_3}{3} 1 \cdots \overset{\mathfrak{e}_6}{6} 1 \overset{\mathfrak{su}_3}{3} 1}_{N-1}$ & $12N^2 + \frac{63N}{4} + \frac{7}{12}$ & $12N^2 + 16N + \frac{7}{6}$ & $(\mathfrak{g}_2)_{12}$  \\

      $\overset{\mathfrak{su}_6}{2}
      \overset{\mathfrak{su}_3}{2} 1
      \underbrace{\overset{\mathfrak{e}_6}{6} 1 \overset{\mathfrak{su}_3}{3} 1 \cdots \overset{\mathfrak{e}_6}{6} 1 \overset{\mathfrak{su}_3}{3} 1}_{N-1}$ & $12N^2 + \frac{27N}{4} - \frac{31}{12}$ & $12N^2 + 7N - \frac{5}{3}$ & $(\mathfrak{g}_2)_{12} \oplus (\mathfrak{su}_3)_{12N + 16}$ \\\bottomrule
    \end{tabular}
    \end{threeparttable}
    \caption{In this table, we write the central charges, non-Abelian flavor algebras, and flavor central charges of the $E_6$-type SW-folds.}
    \label{tbl:e6Sfolds}
\end{table}

\subsection{Coulomb Branch Scaling Dimensions}\label{sec:cbe6}

In Section \ref{sec:cb}, we determined the conformal dimensions of the spectrum of Coulomb branch operators of the 4d $\mathcal{N}=2$ SCFTs arising from the Stiefel--Whitney twisted torus compactifications of the (Higgsed) rank $N$ $(\mathfrak{e}_8, \mathfrak{su}_K)$ orbi-instanton theories. In such cases, the Coulomb branch spectrum was determined by following the heuristic proposal in Appendix B of \cite{Ohmori:2018ona}; therein the scaling dimensions were determined in terms of each curve/algebra combination, $\overset{\mathfrak{g}}{m}$, in the tensor branch description, together with the knowledge of the residual gauge algebra after the $\mathbb{Z}_\ell$ Stiefel--Whitney twist. The contributions were proposed on a case-by-case basis for certain combinations of $(\mathfrak{g}, \ell)$, however, theories involving $(\mathfrak{e}_6, 3)$ and $(\mathfrak{e}_7, 2)$ were not explored in \cite{Ohmori:2018ona}.

When studying the E-type SW-folds, as we are doing here, it is necessary to extend the proposal of \cite{Ohmori:2018ona} to include the $(\mathfrak{e}_6, 3)$ and $(\mathfrak{e}_7, 2)$ cases. We make the following, again heuristic, proposal for the Coulomb branch scaling dimensions of the operators that arise in the $(\mathfrak{e}_6, 3)$ case\footnote{The quantity $J$ is a number associated to each curve in the tensor branch configuration which, roughly, counts where that curves lies in the order of blow-downs required to reach the origin of the tensor branch. This was referred to as $n$ in Appendix B of \cite{Ohmori:2018ona}, and we refer the reader there for the definition.}
\begin{equation}\label{eqn:6curveCOR}
    6J \quad 6J \times 1 + 2 \quad 6J \times 2 + 6 \,.
\end{equation}
Here, the multiplicative factors of $1$ and $2$ that we have introduced are the comarks of the residual $\mathfrak{g}_2$ gauge algebra; furthermore, the additive factors of $2$ and $6$ are the degrees of the Casimir invariants of $\mathfrak{g}_2$.\footnote{We note that, because we are only checking the matching of $4(2a - c)$, which is given by equation \eqref{eqn:2ac}, then $6J$, $6J \times 2 + 2$, and $6J \times 1 + 6$ also work equally well.} Similarly, when the 6d tensor branch contains a curve/algebra combination of the form $\overset{\mathfrak{e}_7}{8}$, then, after a $\mathbb{Z}_2$ SW-twist one obtains a residual $\mathfrak{f}_4$ gauge algebra on the Coulomb branch. We propose that the contribution from this curve/algebra combination to the 4d Coulomb branch consists of five operators with scaling dimensions:
\begin{equation}\label{eqn:f4comark}
    6J \quad 6J \times 1 + 2 \quad 6J \times 2 + 6 \quad 6J \times 3 + 8 \quad 6J \times 2 + 12
\end{equation}
Here the multiplicative factors $1$, $2$, $3$, and $2$ are the comarks, and $2$, $6$, $8$, and $12$ are the degrees of the Casimir invariants, of the surviving $\mathfrak{f}_4$ gauge algebra.\footnote{Again, we emphasize that the level of analysis here is insensitive to which comark is paired with which Casimir degree.} 

We first consider the 6d $(1,0)$ SCFT with tensor branch configuration as given in equation \eqref{eqn:e61}. We have determined that the central charges of the $\mathbb{Z}_3$ SW-twisted torus compactification satisfy
\begin{equation}\label{eqn:2acanom}
    4(2a - c) = 48N^2 + 122N + 60 \,.
\end{equation}
This quantity can also be recovered from the scaling dimensions of the Coulomb branch operators:
\begin{equation}\label{eqn:2ac}
    4(2a - c) = \sum_{i = 1}^r (2D(u_i) - 1) \,,
\end{equation}
where $r$ is the rank of the Coulomb branch and $u_i$ are the Coulomb branch operators. Combining the analysis in Appendix B of \cite{Ohmori:2018ona} with our proposal in equation \eqref{eqn:6curveCOR}, we conjecture that the Coulomb branch operators dimensions are
\begin{equation}\label{eqn:e6CB}
    \begin{aligned}
        &\left.6\right. \\
        &\left.12\right. \\
        &\left.8\right. \\
        &\left.\begin{aligned}
        &6J \quad 6J \times 1 + 2 \quad 6J \times 2 + 6 \\
        &6J + 6 \\
        &12J + 12 \\
        &6J + 8 \\
        \end{aligned}\,\, \right\} \quad J = 1, \cdots, N \\
        &\left.6(N+1)\right. \,.
    \end{aligned}
\end{equation}
In this way, we find that $4(2a-c)$ as worked out from the anomaly, as written in Table \ref{tbl:e6Sfolds}, matches with $4(2a - c)$ as worked out from the Coulomb branch spectrum using equation \eqref{eqn:2ac}. In fact, this matching occurs for all of the SW-twisted theories appearing in Table \ref{tbl:e6Sfolds}. Unfortunately, we do not know of any dual class $\mathcal{S}$ description of a 4d $\mathcal{N}=2$ SCFT obtained from a $\mathbb{Z}_3$ Stiefel--Whitney twist of the 6d theory containing such an $\mathfrak{e}_6$ algebra, and thus we do not have any independent verification of the proposal given in equation \eqref{eqn:6curveCOR}.

To further explore the association between the tensor branch configuration and the dimensions of the Coulomb branch operators of the Stiefel--Whitney twisted theory, we now study one example of an $E_7$-type SW-fold. The tensor branch configuration
\begin{equation}
    1 \overset{\mathfrak{su}_2}{2} \overset{\mathfrak{so}_7}{3} \overset{\mathfrak{su}_2}{2} 1 \underset{\displaystyle 1}{\overset{\mathfrak{e}_7}{8}} 1 \overset{\mathfrak{su}_2}{2} \overset{\mathfrak{so}_7}{3} \overset{\mathfrak{su}_2}{2} 1
      \underbrace{\overset{\mathfrak{e}_7}{8} 1 \overset{\mathfrak{su}_2}{2} \overset{\mathfrak{so}_7}{3} \overset{\mathfrak{su}_2}{2} 1  \cdots \overset{\mathfrak{e}_7}{8} 1 \overset{\mathfrak{su}_2}{2} \overset{\mathfrak{so}_7}{3} \overset{\mathfrak{su}_2}{2} 1}_{N-1} \,,
\end{equation}
has a $\mathbb{Z}_2$ center-flavor symmetry, and arises via Higgsing the $\mathfrak{e}_8$ flavor symmetry of the rank $N$ $(\mathfrak{e}_8, \mathfrak{e}_7)$ orbi-instanton by a homomorphism $\Gamma_{\mathfrak{e}_7} \rightarrow E_8$. The flavor group of this 6d SCFT is $(E_7 \times E_7 \times SU(2))/\mathbb{Z}_2$. Using the anomaly polynomial of the 6d SCFT associated to this tensor branch configuration, and the Coulomb branch theory in 4d after $\mathbb{Z}_2$ Stiefel--Whitney twist, one finds from equation \eqref{eqn:ack} that
\begin{equation}\label{eqn:e7sfoldanom}
    4(2a - c) = 144N^2 + 316N + 120 \,.
\end{equation}
We propose that the scaling dimensions of the Coulomb branch operators are
\begin{equation}
    \begin{aligned}
        &\left.9N + 9\right. \\
        &\left.12N + 12\right. \\
        &\left.6N + 6 \quad 6N + 8 \quad 3N + 5\right. \\
        &\left.6N + 12\right. \\
        &\left.3N + 9\right. \\
        &\left.\begin{aligned}
        &6J \quad 6J \times 1 + 2 \quad 6J \times 2 + 6 \quad 6J \times 3 + 8 \quad 6J \times 2 + 12 \\
        &12J + 6 \\
        &18J + 6 \\
        &12J \quad 12J + 2 \quad 6J + 2 \\
        &18J \\
        &12J \\
        \end{aligned}\,\, \right\} \quad J = 1, \cdots, N \\
        &\left.3(N+1)\right. \,,
    \end{aligned}
\end{equation}
where we have written the contributions from different curves on different lines. Here, we have used our proposal in equation \eqref{eqn:f4comark} for the curves with residual $\mathfrak{f}_4$ gauge algebras. We can see that
\begin{equation}
    \sum_u 2D(u) - 1 = 144N^2 + 316N + 120 \,,
\end{equation}
where the sum is taken over all of the Coulomb branch operators. As we can see, this matches the anomaly polynomial result in equation \eqref{eqn:e7sfoldanom}.

A uniform expression for the Coulomb branch scaling dimensions associated to a pair $(\mathfrak{g}, \ell)$, combining the proposals in Appendix B of \cite{Ohmori:2018ona} and equations \eqref{eqn:6curveCOR} and \eqref{eqn:f4comark}, has been observed in \cite{YTpriv}.\footnote{We thank Y.~Tachikawa for sharing this observation, and for encouraging us to include it here.} As discussed in \cite{Ohmori:2018ona}, when considering a $\mathbb{Z}_\ell$ Stiefel--Whitney twist that breaks the gauge algebra, $\mathfrak{g}$, to a residual gauge algebra, $\mathfrak{h}$, then the coefficients that appear in the Coulomb branch scaling dimensions may be expected to be some $\widetilde{c}_i$ satisfying
\begin{equation}
    1 + \sum_{i=1}^{\operatorname{rank}(\mathfrak{h})} \widetilde{c}_i = \frac{h_\mathfrak{g}^\vee}{\ell} \,.
\end{equation}
For a pair $(\mathfrak{g}, \ell)$, such $\widetilde{c}_i$ have been studied from the perspective of the supersymmetric index of 4d $\mathcal{N}=1$ pure Yang--Mills in \cite{Witten:2000nv}, where the mathematical results on almost commuting holonomies for compact Lie groups\cite{Borel:1999bx} were utilized, which we now review briefly.

Consider $(\mathfrak{g}, \ell)$, where $\mathbb{Z}_\ell$ is a subgroup of the center of the simply-connected Lie group $\widetilde{G}$ associated to $\mathfrak{g}$. The subgroup $\mathbb{Z}_\ell$ can be identified with a particular graph automorphism of the extended Dynkin diagram of $\mathfrak{g}$, $\Gamma$. One can construct a second extended Dynkin diagram, $\Gamma^\prime$, via the action of $\mathbb{Z}_\ell$ on $\Gamma$; each collection of nodes of $\Gamma$ lying within the same $\mathbb{Z}_\ell$ orbit maps to the same node of $\Gamma^\prime$. Furthermore, each node of $\Gamma^\prime$ has a ``generalized comark'', obtained by summing the comarks of all the nodes of $\Gamma$ which map to that particular node of $\Gamma^\prime$. These $\Gamma^\prime$ together with the generalized comarks are shown explicitly in the appendix of \cite{Borel:1999bx}. We refer to these generalized comarks as $c_i$, and then $\widetilde{c}_i = c_i/\ell$. The uniform expression for the dimensions of the Coulomb branch operators arising from the residual gauge algebra is then
\begin{equation}\label{eqn:YTconj}
    6J \times \frac{c_1}{\ell} + d_1 \quad \cdots \quad 6J \times \frac{c_r}{\ell} + d_r \,,
\end{equation}
where $d_i$ are the Casimirs of the residual gauge algebra.\footnote{Recall that there, in addition, exists a Coulomb branch operator arising from the torus reduction of the 6d tensor multiplet associated to each curve.} We note that, while the $c_i/\ell$ sometimes are identical with the comarks of the residual gauge algebra, this is not always the case. Equation \eqref{eqn:YTconj} appears to produce the correct answer in all known cases of Stiefel--Whitney twisted torus compactifications of very Higgsable 6d $(1,0)$ SCFTs; we consider it an interesting open question to understand such a formula from a top-down perspective.

\section{Conclusion}\label{sec:conc2}

The global symmetries of a quantum field theory constitute some of its most basic data. In this chapter we have presented a general prescription for reading off the continuous zero-form symmetry group for 6d SCFTs based on the topological structure of the effective field theory on the tensor branch. Using this, we have determined the continuous part of the zero-form symmetry group on the tensor branch, including the center-flavor symmetry, the contribution from Abelian symmetry factors, as well as possible mixing with R-symmetry factors. Using this data, we have also determined the continuous zero-form symmetry group for a large class of orbi-instanton theories as obtained from small instantons probing an $E_8$ nine-brane. Making use of this global structure, we have also shown that such theories provide a fruitful starting point for generating a large class of 4d $\mathcal{N} = 2$ SCFTs via Stiefel--Whitney twisted compactifications on a $T^2$. In the remainder of this section we discuss some avenues of future investigation.

In this work we have primarily focused on the structure of the global zero-form symmetries, but one can in principle also study higher symmetries that act on extended objects. For example, the one-form symmetries of some 6d SCFTs were recently studied in \cite{Bhardwaj:2020phs, Hubner:2022kxr}, and the corresponding 0-form, 1-form and 2-group symmetries of the 5d theories obtained from a reduction on an $S^1$ were recently calculated using the geometry of the associated non-compact elliptically fibered Calabi--Yau threefold \cite{Cvetic:2022imb}. Some aspects of these issues have also been explored in \cite{Cvetic:2021sxm,Apruzzi:2021mlh,Apruzzi:2022dlm}. It would be interesting to use our bottom up approach based on the effective field theory on the tensor branch to provide an independent cross-check on these results.

One of the operating assumptions in much of our work is that the effective field theory on the tensor branch provides an accurate characterization of the resulting flavor symmetries of an SCFT. In some cases, the SCFT may have enhanced flavor symmetry, and in others, there can even be a dehancement. For example, $\mathfrak{su}_2$ gauge theory on a $-2$ curve with eight half hypermultiplets in the fundamental representation has an $\mathfrak{so}_8$ flavor symmetry algebra on the tensor branch, but only a $\mathfrak{so}_7$ flavor symmetry at the fixed point (see, e.g., \cite{Heckman:2015bfa,Ohmori:2015pia,Morrison:2016djb,Hanany:2018vph,Baume:2021qho}). Similarly, when one of the half-hypermultiplets is eaten up by a neighboring undecorated self-pairing $2$ tensor, the naive $\mathfrak{so}_7$ flavor symmetry is dehanced to a $\mathfrak{g}_2$. Geometrically, this curiosity is related to the complicated nature of the $I_0^*$ singular fiber, which engineers $\mathfrak{so}_8$, $\mathfrak{so}_7$, and $\mathfrak{g}_2$ algebras; some of the geometric properties and subtleties in these cases have been studied in \cite{Bertolini:2015bwa,Esole:2017qeh,Esole:2018mqb}. In this chapter, we have seen evidence that 6d SCFTs of the form
\begin{equation}\label{eqn:app1}
    \cdots \overset{\mathfrak{su}_{2\ell}}{2} \,,
\end{equation}
compactified on a $T^2$ with a $\mathbb{Z}_\ell$ Stiefel--Whitney twist also feature these type of dehancements. These observations have mainly come from dual class $\mathcal{S}$ descriptions, as in Section \ref{sec:classs}, where there are alternative methods to calculate the exact superconformal flavor symmetry; in configurations of the form in equation \eqref{eqn:app1} without a known class $\mathcal{S}$ dual, the flavor symmetry is at present not convincingly known. It would be worthwhile to understand both the field theoretic and the geometric origin of these rare and exceptional dehancements directly from a 6d or F-theory perspective.

We have explicitly shown that the global form of the R-symmetry can potentially mix with the flavor symmetries of a 6d SCFT.
Now, in the context of compactification to lower-dimensional spaces, a partial topological twist is often used to correctly
capture the resulting supersymmetries which are retained. It would be quite interesting to track this data in the resulting compactifications of theories. Indeed, in the broader context of generating 4d SCFTs from compactification 6d SCFTs, it is natural to consider Stiefel--Whitney twists on a genus $g$ curve with marked points. Here, we can in principle consider more than just a single pair of holonomies which commute up to a center-valued flux in $\widetilde{G}$. Since we now have a large class of 6d SCFTs which can generate such theories, it is natural to consider this more general situation.

\chapter{THE BRANES BEHIND GENERALIZED SYMMETRY OPERATORS}

\section{Introduction}

One of the important recent advances in the study of quantum field theory
(QFT) has been the appreciation that generalized symmetries can often be
better understood in terms of corresponding topological operators
\cite{Gaiotto:2014kfa}. For a $d$-dimensional QFT, an $m$-form symmetry
naturally acts on $m$-dimensional defects. There is a corresponding
dimension $d-m-1$ generalized symmetry operator which is topological, i.e., it is unchanged by local perturbations to
its shape \cite{Gaiotto:2014kfa, Gaiotto:2010be,Kapustin:2013qsa,Kapustin:2013uxa,Aharony:2013hda}.
This includes higher-form symmetries, their entwinement via higher-groups,
as well as more general categorical structures.\footnote{For recent work in
this direction, see e.g.,
\cite{Gaiotto:2014kfa,Gaiotto:2010be,Kapustin:2013qsa,Kapustin:2013uxa,Aharony:2013hda,
DelZotto:2015isa,Sharpe:2015mja, Heckman:2017uxe, Tachikawa:2017gyf,
Cordova:2018cvg,Benini:2018reh,Hsin:2018vcg,Wan:2018bns,
Thorngren:2019iar,GarciaEtxebarria:2019caf,Eckhard:2019jgg,Wan:2019soo,Bergman:2020ifi,Morrison:2020ool,
Albertini:2020mdx,Hsin:2020nts,Bah:2020uev,DelZotto:2020esg,Hason:2020yqf,Bhardwaj:2020phs,
Apruzzi:2020zot,Cordova:2020tij,Thorngren:2020aph,DelZotto:2020sop,BenettiGenolini:2020doj,
Yu:2020twi,Bhardwaj:2020ymp,DeWolfe:2020uzb,Gukov:2020btk,Iqbal:2020lrt,Hidaka:2020izy,
Brennan:2020ehu,Komargodski:2020mxz,Closset:2020afy,Thorngren:2020yht,Closset:2020scj,
Bhardwaj:2021pfz,Nguyen:2021naa,Heidenreich:2021xpr,Apruzzi:2021phx,Apruzzi:2021vcu,
Hosseini:2021ged,Cvetic:2021sxm,Buican:2021xhs,Bhardwaj:2021zrt,Iqbal:2021rkn,Braun:2021sex,
Cvetic:2021maf,Closset:2021lhd,Thorngren:2021yso,Sharpe:2021srf,Bhardwaj:2021wif,Hidaka:2021mml,
Lee:2021obi,Lee:2021crt,Hidaka:2021kkf,Koide:2021zxj,Apruzzi:2021mlh,Kaidi:2021xfk,Choi:2021kmx,
Bah:2021brs,Gukov:2021swm,Closset:2021lwy,Yu:2021zmu,Apruzzi:2021nmk,Beratto:2021xmn,Bhardwaj:2021mzl,
Debray:2021vob, Wang:2021vki,
Cvetic:2022uuu,DelZotto:2022fnw,Cvetic:2022imb,DelZotto:2022joo,
DelZotto:2022ras,Bhardwaj:2022yxj,Hayashi:2022fkw,
Kaidi:2022uux,Roumpedakis:2022aik,Choi:2022jqy,
Choi:2022zal,Arias-Tamargo:2022nlf,Cordova:2022ieu, Bhardwaj:2022dyt,
Benedetti:2022zbb, Bhardwaj:2022scy,Antinucci:2022eat,Carta:2022spy,
Apruzzi:2022dlm, Heckman:2022suy, Baume:2022cot, Choi:2022rfe,
Bhardwaj:2022lsg, Lin:2022xod, Bartsch:2022mpm, Apruzzi:2022rei,
GarciaEtxebarria:2022vzq, Cherman:2022eml}. For a recent overview of generalized symmetries,
see reference \cite{Cordova:2022ruw}.} One of the general aims in this
direction is to use such topological structures to gain access to
non-perturbative information on various QFTs. This is especially important in
the context of strongly coupled systems where one typically does not have
access to a practically useful Lagrangian description of the system.

In this vein, one of the lessons from recent work in stringy realizations of QFT is
that there are large families of QFTs which do not have a (known) Lagrangian
description. This includes, for example, all 6D superconformal field theories
(SCFTs), as well as many compactifications of these theories.\footnote{See
\cite{Witten:1995zh,Strominger:1995ac,Seiberg:1996qx} for early examples, and
for recent work on the construction and study of such theories, see
\cite{Heckman:2015bfa,Tachikawa:2015wka,Heckman:2013pva,DelZotto:2014hpa, Heckman:2014qba, Intriligator:2014eaa,
Ohmori:2014pca, Ohmori:2014kda, DelZotto:2014fia,
Bhardwaj:2015xxa, DelZotto:2015isa, Bhardwaj:2015oru,
Cordova:2018cvg, Bhardwaj:2018jgp, Heckman:2018pqx, Bhardwaj:2019hhd,
Bergman:2020bvi,Baume:2020ure, Heckman:2020otd,Baume:2021qho, Distler:2022yse,
Heckman:2022suy} as well as \cite{Heckman:2018jxk, Argyres:2022mnu} for recent
reviews.} More broadly, one can consider the QFT limit of \textit{any} string
background $X$, as obtained by decoupling gravity. In this context, it is
natural to expect that the extra-dimensional geometry directly encodes these
generalized symmetries.

This expectation is, to a large extent, borne out by the explicit construction
of the defect operators of these systems. In the stringy setting, we can
generate defects, i.e., non-dynamical objects with formally infinite tension
by wrapping branes on non-compact cycles of $X$. The resulting higher
symmetries act on these objects, but can be partially screened by dynamical
degrees of freedom wrapped on compact cycles of the geometry. This generalized
screening argument \`{a} la 't Hooft was used in \cite{DelZotto:2015isa}
to define the ``defect group'' of a 6D SCFT. As noted in
\cite{GarciaEtxebarria:2019caf, Albertini:2020mdx, Morrison:2020ool},
specifying a polarization of the defect group amounts to determining the
electric / magnetic higher-form symmetries of the system. This perspective has
by now been generalized in a number of directions, and has reached the stage
where there are explicit algorithms for reading off generalized symmetries for
a large number of geometries \cite{DelZotto:2015isa, Albertini:2020mdx, DelZotto:2020esg, Gukov:2020btk, DelZotto:2020sop, Apruzzi:2020zot, Bhardwaj:2020phs, Bhardwaj:2021pfz, Apruzzi:2021vcu, Agrawal:2015dbf, Cvetic:2021maf, Apruzzi:2021mlh,
Apruzzi:2021nmk, Tian:2021cif, Bhardwaj:2021mzl, DelZotto:2022fnw, Hubner:2022kxr, Cvetic:2022imb, DelZotto:2022joo, Heckman:2022suy}.

One of the puzzling features of these analyses is that the topological
operators of reference \cite{Gaiotto:2014kfa} are in some sense only
implicitly referenced in such stringy constructions. The absence of an explicit brane
realization of these symmetry topological operators makes it challenging to
access some features of generalized symmetries in these systems. For example, it is
well-known in various weakly coupled examples that generalized symmetry operators
can support a topological field theory, and that in the context of theories
with non-invertible symmetries, these can also produce a non-trivial fusion algebra.

In this note we present a general prescription for how to construct
topological operators in the context of geometric engineering. We mainly focus
on the tractable case of 2-form symmetries for 6D SCFTs and their
compactification, as engineered via F-theory backgrounds. In these cases, the
generalized symmetry operators arise from D3-branes wrapped on boundary torsional cycles.
We find that when the $SL(2,\mathbb{Z})$ bundle of the F-theory model is non-trivial, these models
generically have a non-invertible symmetry simply because the fusion algebra for the
generalized symmetry operators contains multiple summands. This is quite analogous
to what has been observed in the context of various field theoretic constructions
\cite{Thorngren:2019iar,Komargodski:2020mxz, Gaiotto:2020iye, Nguyen:2021naa, Heidenreich:2021xpr, Thorngren:2021yso, Agrawal:2015dbf, Robbins:2021ibx, Robbins:2021xce, Sharpe:2021srf, Koide:2021zxj, Huang:2021zvu,
Inamura:2021szw, Cherman:2021nox, Kaidi:2021xfk, Choi:2021kmx, Wang:2021vki, Bhardwaj:2022yxj,
Hayashi:2022fkw, Sharpe:2022ene, Choi:2022zal, Kaidi:2022uux, Choi:2022jqy, Cordova:2022ieu,
Bashmakov:2022jtl, Inamura:2022lun, Damia:2022bcd, Choi:2022rfe, Lin:2022dhv, Bartsch:2022mpm,
Lin:2022xod, Cherman:2022eml, Burbano:2021loy, Damia:2022rxw} as well as some recent holographic models \cite{Apruzzi:2022rei, GarciaEtxebarria:2022vzq}.

We emphasize, however, that the construction we present can be applied to
essentially any QFT which can be engineered via a string / M- / F-theory
compactification. We expect that experts may already be aware of various aspects of this construction, but as far as we are aware, the closest analog of our construction only appeared a few weeks ago in the context of some specific holographic constructions \cite{Apruzzi:2022rei,GarciaEtxebarria:2022vzq}.

\section{Branes and Generalized Symmetry Operators}\label{sec:branesandgensym}

Our interest will be in understanding the brane realization of generalized
symmetry operators. To frame the discussion to follow, let us first briefly
recall how defects are engineered in such systems. We begin by considering a QFT
engineered via a string / M-theory background of the form $\mathbb{R}%
^{d-1,1}\times X$ where $X$ is taken to be a non-compact
$D$-dimensional geometry ($d+D=10$ for a string background and $d+D=11$ for an
M-theory background). We get a QFT by introducing branes and / or localized
singularities at a common point of $X$. These singularities need not be
isolated, and can in principle extend all the way out to the boundary
$\partial X$. Gravity is decoupled because $X$ is non-compact.
This provides a general template for engineering a wide range of (typically
supersymmetric) QFTs.

We obtain supersymmetric defects by wrapping BPS branes on non-compact cycles
of $X$ which extend from the localized singularity out to the boundary. As
explained in \cite{DelZotto:2015isa, GarciaEtxebarria:2019caf, Morrison:2020ool,
Albertini:2020mdx} a screening argument \`{a} la 't Hooft
then tells us that there is a corresponding set of unscreened defects:
\begin{equation}
\mathbb{D=}\underset{m}{\mathbb{%
{\displaystyle\bigoplus}
}}\mathbb{D}^{(m)}\text{ \ \ with \ \ }\mathbb{D}^{(m)}=\underset{p-k=m-1}{%
{\displaystyle\bigoplus}
}\frac{H_{k}(X,\partial X)}{H_{k}(X)},
\end{equation}
where in the above, the superscript $m$ references an $m$-dimensional defect
acted on by an $m$-form symmetry, we sum over supersymmetric $p$-branes and $k$ is a cycle
dimension. Specifying a polarization of $\mathbb{D}$ picks an electric /
magnetic basis of operators and also dictates the higher-form symmetries via
the Pontryagin dual. In the $d$-dimensional QFT, a $p$-brane wrapped on a
$k$-cycle will fill out $p+1-k$ spacetime dimensions. We indicate this by
saying that the brane fills the space:\footnote{We reserve untilded quantities for the
generalized symmetry operators.}
\begin{equation}
\widetilde{S}_{p+1}=\widetilde{M}_{p+1-k}\times\widetilde{\Sigma}_{k}= \widetilde{M}_{m}\times \widetilde{\Sigma}_{k},
\end{equation}
where $\widetilde{\Sigma}_{k} = \mathrm{Cone}(\widetilde{\gamma}_{k-1})$ is the cone generated by extending the
boundary cycle $\widetilde{\gamma}_{k-1} \in H_{k-1}(\partial X)$ from infinity to the tip of the cone.
A final comment is that the torsional factors of $\mathbb{D}^{(m)}$ will define discrete
higher-form symmetries. Non-torsional generators instead label continuous symmetries.

In the $d$-dimensional QFT, the appearance of an $m=(p+1-k)$-dimensional defect
implicitly means there are also corresponding operators $\mathcal{O}_{m}$
with support on an $m$-dimensional subspace. The $m$-form symmetry acts on the defects and operators by
passing these operators through a topological operator $\mathcal{U}(M_n)$ with support on an
$n$-dimensional subspace. To link with the defect, we have:%
\begin{equation}
m+n=d-1.
\end{equation}

In seeking out an extra-dimensional origin for these operators, we
first observe that the defect embeds in spacetime,
and extends along the radial direction which starts at the tip of the
singularity and goes all the way to the boundary $\partial X$, wrapping a boundary cycle of
$H_{k-1}(\partial X)$, namely it is specified by an element of $H_{k}(X, \partial X)$.
Our general proposal is that the topological operator
which links with this object is given by a magnetic dual brane which links with the original
brane in both $X$ as well as the spacetime $\mathbb{R}^{d-1,1}$. In
particular, we can wrap a $q$-brane on a cycle of the form:%
\begin{equation}
S_{q+1}= M_{q+1-(D-1-k)}\times \gamma_{D-1-k}%
= M_{n}\times \gamma_{D-1-k},
\end{equation}
where $\gamma_{D-1-k}$ is a cycle in $H_{D-1-k}(\partial X)$ and $M_{q+1-(D-1-k)}$ is a subspace
of the $d$-dimensional spacetime. Observe that in
$X$, the cycle does not fill the \textquotedblleft radial
direction\textquotedblright. Rather, it always \textquotedblleft sits at
infinity\textquotedblright. Now, for this to be a topological operator which
properly links with the defect, we also require:%
\begin{equation}
(p+1-k)+(q+1-(D-1-k))=d-1,
\end{equation}
or equivalently:%
\begin{equation}
p+q=D+d-2.
\end{equation}
But observe that this is just the requirement that in the full string / M-theory background,
our sought after $q$-brane is simply the magnetic dual $p$-brane!
See figure \ref{fig:TopOpLinking} for a depiction.

\begin{figure}[t!]
\begin{center}
\includegraphics[scale = 1.0, trim = {0cm 0.0cm 0cm 0.0cm}]{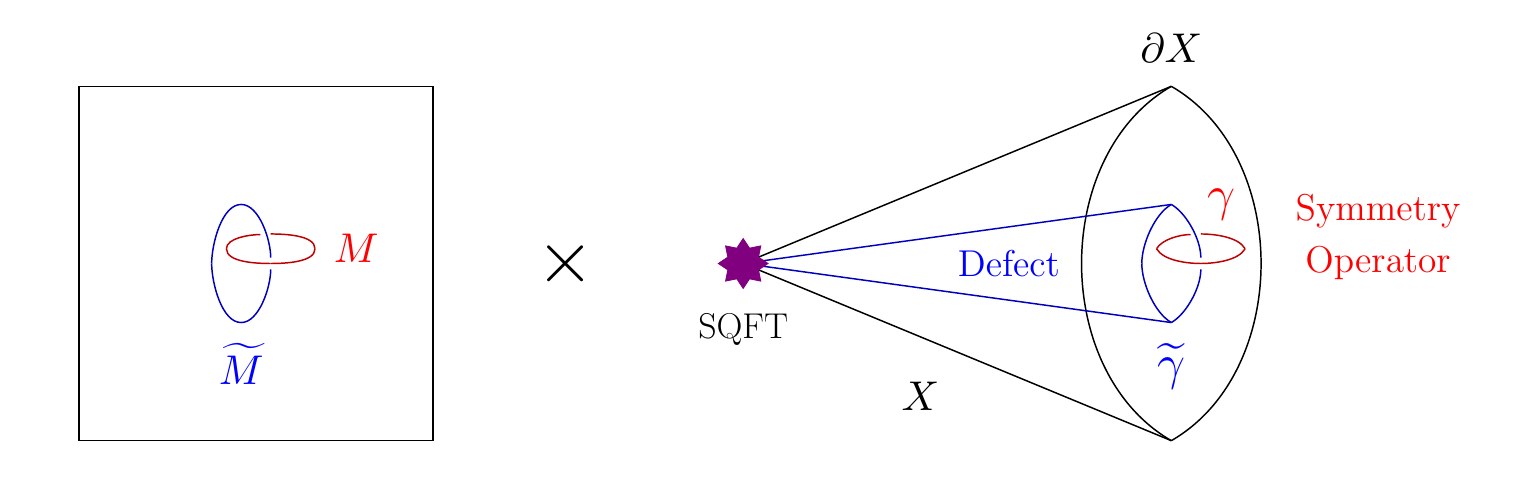}
\caption{Depiction of a generalized symmetry operator realized via a wrapped brane. We consider a
supersymmetric QFT (SQFT) localized at the tip of a conical geometry $X$. At the boundary $\partial X$, we have
bounding cycles $\widetilde{\gamma}$ and $\gamma$ which link in $\partial X$. Wrapping a $p$-brane on the cone over $\widetilde{\gamma}$ yields a
defect for an $m$-form symmetry, and wrapping the magnetic dual $q$-brane on a linking cycle $\gamma$
realizes a generalized symmetry operator. The spacetime loci $\widetilde{M}$ of the $p$-brane and $M$ of the $q$-brane also link with each other. The wrapped $q$-brane is automatically topological since it sits at infinity in the radial geometry.}
\label{fig:TopOpLinking}
\end{center}
\end{figure}

We now argue that wrapping a $q$-brane on the cycle at infinity
$S_{q+1} = M_{q+1 - (D-1-k)} \times \gamma_{D-1-k}$
can be viewed as inserting a topological operator for the $m$-form symmetry.
Along these lines, recall that for a (supersymmetric) $p$-brane, there is a corresponding
$(p+1)$-form potential $\widetilde{C}_{p+1}$ which couples to this object, and thus a
$(p+2)$-form field strength $\widetilde{F}_{p+2}$. In the full
higher-dimensional geometry, we also can speak of the dual field strength
$\ast \widetilde{F}_{p+2}= F_{q+2}$, as sourced by a magnetic dual $(q+1)$-form potential
$C_{q+1}$. The presence of a $p$-brane signals that there is a modified Bianchi identity:
\begin{equation}
d F_{q+2} = \delta_{\widetilde{S}_{p+1}},
\end{equation}
namely a delta function supported contribution to the flux. Of course, this is just indicating that we have
a suitable symmetry current which couples to the magnetic dual potential $C_{q+1}$. Following the general
reasoning in \cite{Gaiotto:2014kfa}, we conclude that we can represent our topological symmetry operator
by exponentiating the integral of $F_{q+2}$ over the $(q+2)$-chain $Z_{q+2} = M_{q+1 - (D-1-k)} \times \mathrm{Cone}(\gamma_{D-1-k})$:
\begin{equation}
\mathcal{U}(Z) = \exp \left( 2 \pi i \int_{Z} F_{q + 2} \right).
\end{equation}
In particular, if we wish to know the action of $\mathcal{U}(Z)$ on any correlation function,
we can simply insert it into the path integral of the $(D+d)$-dimensional supergravity theory.

On the other hand, since we are integrating $F_{q + 2}$ over a chain, we can reduce
this to the integral of $C_{q+1}$ over the
bounding geometry $S_{q+1} = M_{q+1 - (D-1-k)} \times \gamma_{D-1-k}$. Since this is an insertion in the path integral, we can view this is as just specifying a ``brane at infinity'', namely a $(q+1)$-brane wrapped on the prescribed cycle. Indeed, integrating $C_{q+1}$ over this cycle is just one of the worldvolume couplings of the $q$-brane action. Note also that by wrapping the $q$-brane on a linking cycle at infinity, it is automatically topological from the perspective of the $d$-dimensional QFT. Indeed, any localized fluctuation of such a brane is completely decoupled from the dynamics of the field theory localized at the singularity. The only remaining structure which could in principle persist
is purely topological in nature.\footnote{Another way to arrive at the same conclusion is 
to consider localized fluctuations from the singularity. Any correlation function involving 
operators of the theory will be---up to topological couplings---completely decoupled from the ``brane at infinity''. 
Thus, the only possible remnant of the brane at infinity on the localized dynamics could be topological in nature.}

Summarizing, our proposal is that for a wrapped $p$-brane which produces a defect, the corresponding
generalized symmetry operators which act on these defects are realized
by magnetic dual $q$-branes wrapped on linking cycles of the geometry.
This is compatible with the holographic discussion considered a few weeks ago in \cite{Apruzzi:2022rei, GarciaEtxebarria:2022vzq},
which considers the case of QFTs engineered via D3-brane probes of appropriate singularities. In that setting, the near horizon geometry is of the form $AdS_5 \times Y$ where $Y$ can be viewed as the asymptotic geometry $\partial X = Y$ probed by the D3-brane. Indeed, as noted in \cite{GarciaEtxebarria:2022vzq}, defects arise from branes which fill the radial direction of the $AdS_5$, while the symmetry operators arise from branes wrapped on a cycle of $Y$ and sitting at a point of the conformal boundary $\partial AdS_5$. It is important to emphasize that precisely because we are dealing with a conformal boundary the construction presented in the holographic setting is indeed compatible with the perspective developed here.

Observe that we can also read off the corresponding topological field theory
(TFT) localized on this symmetry operator. Starting from
$S_{q+1} = M_{q+1-(D-1-k)}\times \gamma_{D-1-k}$, we consider the topological
couplings on the worldvolume theory of our $q$-brane. Roughly speaking, we can
integrate this theory along $\gamma_{D-1-k}$ and arrive at a TFT\ on
$M_{n} = M_{q+1-(D-1-k)}$. To see this procedure through from start to finish, then,
we need to know the topological couplings on the original brane, as well as a
technique to dimensionally reduce along $\gamma_{D-1-k}$.

As a further abstraction, now that we have a method for realizing generalized symmetry operators, we can in principle
just consider branes wrapped on torsional cycles ``at infinity''. In particular, there is a
priori no need for there to exist explicit defects of the appropriate codimension which link
with these branes.\footnote{This, for example, happens in various 3D Chern-Simons-like theories with charge conjugation,
i.e., there is a (non-invertible) 0-form symmetry which acts on no local operators,
but line operators do transform non-trivially
in passing through the wall (see e.g., \cite{Seiberg:2020bhn, Choi:2022zal}).}

\section{Example: 2-Form Symmetries of 6D SCFTs}

To illustrate the above considerations, we now show how this works in practice for
(discrete) 2-form symmetries of 6D\ SCFTs. All known 6D SCFTs can be engineered
via F-theory on an elliptically fibered Calabi-Yau threefold with base $\mathcal{B}$ such that the threefold
has a canonical singularity \cite{Heckman:2013pva, DelZotto:2014hpa, Heckman:2015bfa}.
In the SCFT\ limit, all of the bases take the form $\mathcal{B}=\mathbb{C}^{2}%
/\Gamma$ for $\Gamma$ an appropriate finite subgroup of $U(2)$ (see
\cite{Heckman:2013pva} for the classification of all such $\Gamma$).
The defect group for the 2-form symmetry is $\mathrm{Ab}[\Gamma]$,
the abelianization of $\Gamma$ (see reference \cite{DelZotto:2015isa}).
Some basic features of these orbifold singularities are summarized in table \ref{tab:defectgrps}.

\begin{table}
\begin{center}
\renewcommand{\arraystretch}{1.25}
\begin{tabular}{|| c | c |  c  ||}
\hline
   $\Gamma$ & $\mathbb{D}^{(2)}$ & $L_\Gamma$  \\ [0.5ex]
 \hline\hline $\mathbb{Z}_{N}$ & $\mathbb{Z}_{N}$ & $1/N$ \\
\hline $D_{2N}$ & $\mathbb{Z}_2\times \mathbb{Z}_2$ & $\frac{1}{2}\begin{pmatrix}
  N & N-1 \\
  N-1 & N
\end{pmatrix} $\\
\hline $ D_{2N+1}$ & $\mathbb{Z}_4$ & $\frac{2N-1}{4}$ \\
\hline $2T$ & $\mathbb{Z}_3$ & $1/3$\\
\hline $2O$ & $\mathbb{Z}_2$ & $1/2$\\
\hline $2I$ & 1 & 1\\
\hline $\mathbb{Z}_p(q)$ & $\mathbb{Z}_p$ & $-q/p$\\
\hline $D_{p+q,q}$ & $\begin{array}{c}
   \mathbb{Z}_{2p}\times \mathbb{Z}_2~ \textnormal{($q$ even)} \\
      \mathbb{Z}_{2p} ~ \textnormal{($q$ odd)}
\end{array} $
& (\textnormal{See main text})\\
\hline
\end{tabular}
\end{center}
 \caption{In the left column we list out all of the families of finite subgroups of $U(2)$ associated to 6D SCFTs. Here $D_{k}$ means the dicyclic groups of order $2k$, and $2T$, $2O$, and $2I$ denote the binary tetrahedral, octahedral, and icosahedral groups respectively. $\mathbb{Z}_p(q)$ denotes a $\mathbb{Z}_p$ subgroup of $U(2)$ generated by an action $(z_1,z_2)\mapsto (\zeta_p z_1,\zeta^q_p z_2)$ ($p$ and $q$ coprime). Finally, $D_{p+q,q}$ is a $U(2)$ subgroup which generalizes the dicyclic group (see \cite{Heckman:2013pva, DelZotto:2015isa} and references therein for more details). }\label{tab:defectgrps}
\end{table}

It is helpful to decompose the base geometry as a fibration $S^{3}%
/\Gamma\rightarrow\mathcal{B}\rightarrow\mathbb{R}_{\geq0}$ in which the SCFT
sits at the point $r=0$ in $\mathbb{R}_{\geq0}$ where the $S^{3}/\Gamma$
collapses to zero size. We can introduce a defect by wrapping a D3-brane on
the radial direction of $\mathbb{R}_{\geq0}$ as well as a torsional 1-cycle of
$S^{3}/\Gamma$. In this case, we expect the topological operator which acts on
such defects to be given by a D3-brane which wraps the boundary torsional
1-cycle as well as a three-dimensional subspace $M_{3}$\footnote{We take $M_3$ to be connected throughout.} of the 6D spacetime.

The procedure for how to work out the TFT generated by our wrapped D3-brane
follows similar steps to those developed in
\cite{GarciaEtxebarria:2022vzq}. Starting from the
D3-brane worldvolume theory, we have the topological
couplings \cite{Douglas:1995bn, Minasian:1997mm}:%
\begin{equation}
\mathcal{S}_{\text{top}}^{D3}=2\pi i \int_{S} \exp(\mathcal{F}%
_{2})\sqrt{\frac{\widehat{A}(T S)}{\widehat{A}(N S)}%
}\left(  C_{0}+C_{2}+C_{4}\right)  ,
\end{equation}
where here, $\mathcal{F}_{2}=F_{2}-B_{2}$, with $F_{2}$ the $U(1)$ gauge field
strength of the D3-brane, and $B_{2}$ the (pullback of) the NS-NS\ 2-form
potential. Additionally, the $C_{m}$ are the pullbacks of RR potentials onto
the worldvolume of the D3-brane. Due to $SL(2,\mathbb{Z})$ duality covariance, we will label the 2-form curvatures as $F_3:= dB_2$ and $F^D_3:=dC_2$. Expanding out, the relevant couplings for us
are, expressed in differential cohomology (see e.g.,
\cite{Freed:2006ya, Apruzzi:2021nmk}),:\footnote{There is a subtlety here due to the fact that the 5-form
field strength is self-dual. For additional discussion on this point, see
e.g., \cite{Belov:2006jd, Belov:2006xj, Heckman:2017uxe}.}%
\begin{equation}\label{eq:topaction}
\mathcal{S}_{\text{top}}^{D3}=2\pi i \int_{S} \breve{F}_{5}%
+\breve{F}^D_{3}\star\mathcal{\breve{F}}_{2}+\breve{F}_{1}\star\left(  \frac
{1}{2}\mathcal{\breve{F}}_{2}\star\mathcal{\breve{F}}_{2}+\frac{1}{24}%
\breve{e}\right),
\end{equation}
with $\breve{e}$ the Euler class of $S$. Comparing with \cite{GarciaEtxebarria:2022vzq},
it will turn out to be important to also track the term involving $\breve{F}_{1} \star \mathcal{\breve{F}}_{2}\star\mathcal{\breve{F}}_{2}$. On the other hand,
the contribution from the Euler term will play little role in the present analysis.\footnote{It can
play a role in situations where we demand specific Spin / Pin structures for $M_3$.}

One might ask how a term involving $F_1$ arises from purely field theoretic considerations.
Indeed, because there are no continuous marginal parameters in 6D SCFTs \cite{Louis:2015mka, Cordova:2016xhm},
one might be tempted to conclude that no such dependence could be present. Observe, however, that the
6D SCFT admits 4D defects (i.e., codimension two defects) given by D3-brane probes of the local model. The
worldvolume of this D3-brane contains a continuous parameter $\tau$ which is precisely what is also entering
in our generalized symmetry operators.

To proceed further, we need to dimensionally reduce the WZ terms of the D3-brane wrapped
on a torsional cycle $\gamma$ of the extra-dimensional geometry. There is a subtlety here in cases where the $SL(2,\mathbb{Z})$ bundle of an F-theory model
is non-trivial because these duality transformations act on the axio-dilaton as well as the
doublet of 2-form potentials of the IIB background.\footnote{Indeed, even in configurations
where the axio-dilaton is constant, there can still be a non-trivial action on the
2-form potentials of the IIB background.} Consequently, the first case we consider involves
the 6D SCFTs with $\mathcal{N} = (2,0)$ supersymmetry. In these cases the elliptic fibration
is completely trivial, which simplifies the analysis of the D3-brane topological terms.
We next treat the case of single curve non-Higgsable cluster
theories \cite{Morrison:2012np}. In this case, the presence of a non-trivial duality bundle leads us to a
discrete Chern-Simons gauge theory on the generalized symmetry operator, which potentially coupled to background fields. We expect similar considerations to hold
in any background where the axio-dilaton is constant.
In all these cases, we find that 3D defects charged under a suitable 3-form symmetry detect a non-invertible symmetry,
namely the fusion algebra for the symmetry operators contains multiple summands.

The most general situation in which the axio-dilaton is position dependent is, by the same reasoning, expected to also lead to non-invertible symmetries. We anticipate that more general possibilities can arise once we consider topological operators which are also fused with those associated with the 0-form and 1-form symmetries of these 6D SCFTs. These generically arise once we take into account the contributions from flavor 7-branes (see e.g., \cite{Hubner:2022kxr, Cvetic:2022imb, Heckman:2022suy}).

Though we leave the details for future work, it is also clear that we can apply the
same methodology when we compactify a 6D SCFT on a background manifold $Q$ of dimension $l$.
Indeed, all that is required is that we also wrap the topological operator on the relevant cycle (possibly
torsional)\ of $Q$, and again perform the appropriate dimensional reduction.

\subsection{6D \texorpdfstring{$\mathcal{N}=(2,0)$}{N=(2,0)} Theories}\label{ssec:6d2comma0}

As a first class of examples, consider the 6D $\mathcal{N} = (2,0)$ SCFTs as engineered by
type IIB on an ADE singularity $\mathbb{C}^2 / \Gamma$ with $\Gamma$ a finite subgroup of $SU(2)$.
We begin by considering the case $\Gamma$ a cyclic group and then turn to the case of $\Gamma$ non-abelian.

\paragraph{$\Gamma$ Cyclic}
Consider first the case where $\Gamma$ is a cyclic group. The topological field theory of the operator constructed from the D3-brane is then derived by reduction in differential cohomology on the quotient $S^3/\Gamma$. Let us denote the cohomology generators of $S^3/\Gamma$ by $1,u_2,\textnormal{vol}$ in degree 0,2,3 and their lift to differential cohomology by $\breve{1},\breve{u}_2,\textnormal{\u{v}ol}$. We expand as
\begin{equation}\label{eq:expansion}
\begin{aligned}
    \breve{F}_5&=\breve{a}_2 *\textnormal{\u{v}ol}+\breve{a}_3 * \breve{u}_2+\breve{a}_5 * \breve{1}   \\
    \breve{F}_3&= \breve{b}_0 * \textnormal{\u{v}ol}+\breve{b}_1 * \breve{u}_2+\breve{b}_3 * \breve{1} \\
    \breve{F}_2&= \breve{c}_0 * \breve{u}_2+\breve{c}_2 * \breve{1} \\
    \breve{F}_1&= \breve{d}_1 * \breve{1}
    \end{aligned}
\end{equation}
and similarly for $\breve{F}_2^{D},\breve{F}_3^{D}$, where the ``D'' superscript refers to the field strength obtained under an S-duality transformation. The coefficients multiplying $\breve{u}$ are background fields for the discrete symmetries
\begin{equation}\label{eq:sym1}
\begin{aligned}
\breve{a}_3 \leftrightarrow \mathbb{Z}_N^{(2)}\,, \qquad \breve{b}_1 \leftrightarrow \mathbb{Z}_N^{(0)} \,, \qquad \breve{c}_0 \leftrightarrow \mathbb{Z}_N^{(-1)} \,,
\end{aligned}
\end{equation}
while those multiplying $ \breve{1}$ are field strengths for the continuous abelian symmetries
\begin{equation}\label{eq:sym2}
\begin{aligned}
\breve{b}_3 \leftrightarrow U(1)^{(1)} \,, \qquad \breve{c}_2 \leftrightarrow U(1)^{(0)} \,, \qquad \breve{d}_1 \leftrightarrow U(1)^{(-1)}
\end{aligned}
\end{equation}
In the above, the superscripts $(s)$ refer to the corresponding $s$-form symmetry. The expansion along $\textnormal{\u{v}ol}$ is a standard reduction and as $S^3/\Gamma$ has formally infinite volume $\breve{a}_2,\breve{b}_0$ are non-dynamical, measuring fluxes which are absent in the purely geometric background (and therefore vanish). The self-duality of $\breve{F}_5$ implies the vanishing of $\breve{a}_5$. Regarding the axio-dilaton, the curvature of  $\breve{d}_1$ is identified with\footnote{The righthand side is not exact since $\tau$ is not single-valued.} $R(\breve{d}_1)=d(\mathrm{Re}(\tau))\in H_1(M_3,\mathbb{Z})$\footnote{The map $R$ on the differential cohomology group $\breve{H}^p(M_3)$ is part of the short exact sequence $0\rightarrow H^{p-1}(M_3,U(1))\rightarrow \breve{H}^p(M_3)\xrightarrow{R} \Omega^p_\mathbb{Z}(M_3)\rightarrow 0$ where $\Omega^p_\mathbb{Z}(M_3)$ denotes $p$-forms on $M_3$ with integer periods, i.e. where standard $U(1)$ fluxes live. For more details on the basics of differential cohomology geared towards physicists see \cite{Freed:2006ya, Freed:2006yc}, as well as section 2 of the recent paper \cite{Apruzzi:2021nmk}.} and when this class is trivial, then the data contained in the differential cohomology class $\breve{d}_1$ is simply $\tau$. With this in mind, we will also employ a slight redefinition of the $\breve{F}_3$ fields to get rid of the cumbersome $\tau$ factors in the D3 topological action which is $F^{new}_3\equiv \frac{1}{\tau} F^{old}_3$ and $(F^D)^{new}_3\equiv \tau (F^D_3)^{old}$. Consistency with Dirac quantization follows from lifting these fields to M-theory on a torus fibration in the standard duality dictionary, i.e., we interpret the type IIB $SL(2,\mathbb{Z})$ covariant 3-form flux as an M-theory 4-form flux reduced on the elliptic fiber.

We emphasize now that in this case there is no flux non-commutativity contrary to the setup in \cite{GarciaEtxebarria:2019caf}. To see why, consider the Hamiltonian formulation by writing $M_3 = N_2 \times \mathbb{R}_t$. We get the pair of electric and magnetic flux operator valued in the $\text{Tor}\,H^2(N_2 \times \gamma; \mathbb{Z})$ as $\Phi_e(b_0 \star u_2)$ and $\Phi_m(c_0 \star u_2)$. Now on $N_2$, the Poincar\'{e} duals $PD[b_0]$ and $PD[c_0]$ do not intersect for degree reasons, so $\Phi_e(b_0 \star u_2)$ commutes with $\Phi_m(c_0 \star u_2)$ and there are no terms involving co-boundaries giving rise to non-commutativity upon quantization as in \cite{GarciaEtxebarria:2022vzq} that describes a discrete gauge theory in the sense of \cite{Banks:2010zn}.

We insert the expansion \eqref{eq:expansion} into our expression for the topological action to find:
\begin{equation}\label{eq:SD3top}
\begin{aligned}
\mathcal{S}_{\textnormal{top}}^{\textnormal{D3}}=\frac{2\pi i}{N}\int_{M_3} \bigg( a_3 + c_0 \cup b^D_3 + c_0^D\cup b_3+c_2 \cup b^D_1 + c_2^D \cup b_1   \\
-c_0\cup b_3-c^D_0\cup b^D_3-c_2\cup b_1-c^D_2\cup b^D_1\bigg),
 \end{aligned}
\end{equation}
where we have added terms derived from similar expansions for $\breve{F}_2^{D}$ to restore invariance under S-duality. Notice that terms coming from expanding $\breve{F}_3^{D}$ are already present in the $\frac{1}{2}\breve{F}_1\star\mathcal{\breve{F}}_2\star \mathcal{\breve{F}}_2$ term in \eqref{eq:topaction}. The above action simplifies after taking linear combinations given by $b_1'\equiv b^D_1-b_1$ (and similarly for $b_3$, $c_0$, $c_2$ and their duals) after which the topological action is just
\begin{equation}\label{eq:SD3top2}
\begin{aligned}
\mathcal{S}_{\textnormal{top}}^{\textnormal{D3}}=\frac{2\pi i}{N}\int_{M_3} \left( a_3 + c'_0 \cup b'_3 +c'_2 \cup b'_1\right).
 \end{aligned}
\end{equation}

Recall that $\breve{F}_2$ and $\breve{F}_2^D$ are worldvolume field strengths on the D3-brane at infinity and therefore $c_0,c_2$ and their dual partners are path-integrated over. The topological operator therefore takes the form:
\begin{equation}
\begin{aligned}
\mathcal{U}(M_3)= \frac{1}{\mathcal{K}}\int Dc'_0 Dc'_2  \exp\left(  \mathcal{S}_{\textnormal{top}}^{\textnormal{D3}}\right)
 \end{aligned}
\end{equation}
where $\mathcal{K}$ is a normalization constant we determine shortly. In our definition of $\mathcal{U}(M_3)$,
we have left implicit the dependence on the torsional 1-cycle of the boundary geometry (to avoid cluttering
notation). At this point, unless otherwise stated, we assume that this torsional 1-cycle is a generator of
$H_1(S^3 / \Gamma, \mathbb{Z})$. The topological operator $\mathcal{U}(M_3)$ is a product of the operator
\begin{equation}
\begin{aligned}
\mathcal{U}_0=\exp\left( \frac{2 \pi i}{N} \int_{M_3} a_3 \right)\,,
\end{aligned}
\end{equation}
which is the standard flux operator for surface defects of the SCFT, and
\begin{equation}
\begin{aligned}
\mathcal{U}_1&=\frac{1}{|H_1(M_3, \mathbb{Z}_N)|}\int Dc'_2 \exp\left( \frac{2 \pi i}{N} \int_{M_3} c'_2\cup b'_1 \right)\,, \\
\mathcal{U}_3&=\frac{1}{N}\int Dc'_0 \exp\left( \frac{2 \pi i}{N} \int_{M_3} c'_0\cup b'_3 \right)\,.
\end{aligned}
\end{equation}
So altogether we have
\begin{equation}
    \mathcal{U}(M_3) = \mathcal{U}_0 \mathcal{U}_1 \mathcal{U}_3 \,.
\end{equation}
which sets the normalization constant $\mathcal{K}$. When the $C_2$ and $B_2$ backgrounds are turned off we have, $\mathcal{U}(M_3) = \mathcal{U}_0$.

Let us now study the fusion algebra. Note that all operators except $\mathcal{U}_0$ are condensation operators since they specify a 3-gauging of a $U(1)^{(3)}$ or $\mathbb{Z}^{(4)}_N$ symmetry along the $M_3$ worldvolume. Moreover, we show that these operators satisfy the fusion algebra of projections
\begin{equation}
    \mathcal{U}_i \mathcal{U}_i = \mathcal{U}_i\,, \; \; \; \; \textnormal{($i=1,3$)},
\end{equation}
and so formally speaking are non-invertible. That being said, they are invertible when
restricted to their image where they equate to the identity operator. This follows
for instance for $\mathcal{U}_1$ by the manipulations
\begin{equation}
\begin{aligned}
    \mathcal{U}_1&=\frac{1}{|H_1(M_3, \mathbb{Z}_N)|}\int Dc'_2 \exp\left( \frac{2 \pi i}{N} \int_{M_3} c'_2\cup b'_1 \right) \\
    &=\frac{1}{|H_1(M_3, \mathbb{Z}_N)|}\sum_{\ell \in H_1(M_3, \mathbb{Z}_N)} \exp\left( \frac{2 \pi i}{N} \int_\ell b'_1 \right)\\
     &=\frac{1}{|H_1(M_3, \mathbb{Z}_N)|}\prod_{\ell'} \left(\sum_{k=0}^{N-1} \exp\left( \frac{2 \pi i k}{N} \int_{\ell'} b'_1 \right)\right)
    \end{aligned}
\end{equation}
together with the integrality of the periods of $b'_1$. Here $\{\ell'\}$ are a generating set for the lattice $H_1(M_3, \mathbb{Z}_N)$. So whenever such periods are non-vanishing we have a vanishing sum of roots of unity.
From this we also see that $\mathcal{U}_i= \mathcal{U}_i^\dagger$ for $i\neq 0$ follows from relabeling $k\rightarrow -k$. The normalization is now explicitly $\mathcal{K}=N|H_1(M_3, \mathbb{Z}_N)|$. On the other hand the operator $\mathcal{U}_0$
displays a cyclic fusion ring
\begin{equation}
\begin{aligned}
\mathcal{U}_0 \mathcal{U}_0^\dagger =1\,, \qquad \mathcal{U}_0^n =\exp\left(\frac{2 \pi i n}{N} \int_{M_3} a_3 \right).
\end{aligned}
\end{equation}

Concerning the operators charged under $\mathcal{U}(M_3)$, these include the surface operators of the defect group $\mathbb{D}$ constructed from D3-branes wrapped on relative 2-cycles of the F-theory base $\mathcal{B}$.
The operators $\mathcal{U}_1, \mathcal{U}_3$ do not act on elements of $\mathbb{D}$ since they carry no charge under the symmetries of \eqref{eq:sym1} and \eqref{eq:sym2} other than $\mathbb{Z}_N^{(2)}$. Therefore the restriction of $\mathcal{U}(M_3)$ on $\mathbb{D}$ is given by $\mathcal{U}_0$, namely the standard flux operator. However, as mentioned at the end of Section \ref{sec:branesandgensym}, $\mathcal{U}(M_3)$ can act on operators with spacetime dimension other than $2$ as well. The $\mathcal{U}_1$ piece acts on local operators of the 6D SCFT that originate from $D1$ and $F1$ strings wrapping a torsional 1-cycle in the boundary $S^3/\Gamma$ times the radial direction of $\mathbb{C}^2/\Gamma$, while the $\mathcal{U}_3$ piece acts on line operators that wrap a point in $S^3/\Gamma$ times the radial direction. The actions of $\mathcal{U}_{1}$ and $\mathcal{U}_{3}$ on these operators is almost trivial: it multiplies by zero on any operators with non-zero charge under the symmetry groups $\mathbb{Z}^{(0)}_N$ and $U(1)^{(1)}$ respectively.

\paragraph{$\Gamma$ Non-Abelian}
Consider next the case where $\Gamma$ is non-abelian. As far as the defect group is concerned, the relevant data is captured by the abelianization $\mathrm{Ab}[\Gamma]$. Returning to the entries of Table \ref{tab:defectgrps}, we see that in nearly all cases, we again have a single cyclic group factor so the analysis proceeds much as we already presented. On the other hand, for some $D$-type subgroups,
$\mathrm{Ab}[\Gamma]$ has two cyclic group factors. For this reason, we now focus on this case.

Proceeding more generally, when we insert the above expansion (\ref{eq:expansion}) into (\ref{eq:topaction}),
the overall coefficient we obtain in the exponential is given by the canonical link pairing in first homology:
\begin{equation}
    L_\Gamma: \; H_1(S^3/\Gamma)\times H_1(S^3/\Gamma)\rightarrow \mathbb{Q}/\mathbb{Z}.
\end{equation}
This is because given $\breve{t}^i_2$ such that $I(\breve{t}^i_2)=t^i_2\in H^2(S^3/\Gamma,\mathbb{Z})\simeq \mathbb{Z}_{n_i}\times \mathbb{Z}_{n_j}$ we have that
\begin{equation}
\int_{S^3/\Gamma}\breve{t}^i_2 * \breve{t}^j_2 =L^{ij}_\Gamma.
\end{equation}
Table \ref{tab:defectgrps} gives the explicit linking pairing for all $\Gamma$ a finite subgroup of $SU(2)$.\footnote{In the more general case where $\Gamma$ is a $D_{p+q,q}$ subgroup of $U(2)$ and $\mathrm{Ab}[\Gamma]$ has two cyclic group factors, determining the linking pairing is somewhat dependent on the divisibility properties of $p$, $q$ and $p+q$. The linking pairing was worked out on a case by case basis in some examples in reference \cite{GarciaEtxebarria:2019caf}. There, one can see that $L^{ij}_\Gamma$ can be recast as an intersection pairing of certain non-compact 2-cycles in a blow-up of $\mathbb{C}^2/\Gamma$. It is tempting to speculate that one can use a quiver-based method to directly read off this data, much as in \cite{DelZotto:2022fnw}.}

Let us turn next to the TFT obtained from wrapping a D3-brane on a torsional cycle of $S^3 / \Gamma$. When $\breve{H}^{2}(S^3 / \Gamma)$ has more than one generator, the previously considered $\breve{a}_3$, $\breve{b}_1$, and $\breve{c}_0$ each pick up an index. In determining the spectrum of topological operators, it is enough to consider D3-branes wrapping $\gamma = \nu^{1} \gamma_{1} + \nu^{2} \gamma_{2}$, with $\gamma_{i}$ a primitive generators of $H_{1}(S^3 / \Gamma)$. The action is now (reverting back to the original duality basis for clarity):
\begin{equation}
    \begin{aligned}
\mathcal{S}_{\textnormal{top}}^{\textnormal{D3}}=2\pi \sqrt{-1} \nu^{i} (L_\Gamma)_{ij} \underset{M_3}{\int} \bigg( a^j_3 + c^j_0 \cup b^D_3 + (c_0^D)^j\cup b_3+c_2 \cup (b_1^D)^j  + c_2^D \cup b^j_1  \\
-c^j_0 \cup b_3 - (c_0^D)^j\cup b^D_3-c_2 \cup b^j_1 - c_2^D \cup (b_1^D)^j
\bigg).
 \end{aligned}
\end{equation}
Just as in the case of $\Gamma = \mathbb{Z}_N$, we observe that the fusion rules for these topological operators produce an invertible symmetry when the background $C_2$ and $B_2$ fields are switched off.

\subsection{6D NHC Theories} \label{sec:6d10}

We now turn to rank one 6D $\mathcal{N}  = (1,0)$ theories in which the axio-dilaton is constant
but the duality bundle of the F-theory model is still non-trivial. In particular, we consider the
case of the single curve non-Higgsable clusters (NHCs) of reference \cite{Morrison:2012np}
in which the base of the F-theory
model supports a curve of self-intersection $-n$ for $n=3,4,6,8,12$. These models
can be written as $(\mathbb{C}^2 \times T^2) / \mathbb{Z}_n$, where the action on the $\mathbb{C}^2$
base is by a common $n^{th}$ primitive root of unity \cite{Witten:1996qb, Heckman:2013pva}.
On the tensor branch, these theories are characterized by a 6D gauge theory coupled to a tensor multiplet with charge
prescribed by the self-intersection number. With notation as in \cite{Heckman:2013pva}, we have:
\begin{equation}
    \overset{\mathfrak{su}(3)}{3}, \;\;  \overset{\mathfrak{so}(8)}{4}, \;\;  \overset{\mathfrak{e}_6}{6}, \;\;  \overset{\mathfrak{e}_7}{8}, \;\;  \overset{\mathfrak{e}_8}{12}
\end{equation}
where $\overset{\mathfrak{g}}{n}$ refers to a $(-n)$-curve with a ADE singularity of type $\mathfrak{g}$ wrapping it.
These can all be presented as F-theory backgrounds $(\mathbb{C}^2\times T^2)/\mathbb{Z}_n$ (see \cite{Witten:1996qb, Heckman:2013pva}) where the quotient is defined by the group action:
\begin{equation}
    (z_1,z_2,z_3)\rightarrow (\zeta_n z_1,\zeta_n z_2, \zeta^{-2}_n w )
\end{equation}
where $w$ is the torus-fiber coordinate. In the nomenclature of table \ref{tab:defectgrps}, $\Gamma=\mathbb{Z}_n{(1)}$ (i.e., $p = n$ and $q = 1$), and thus the link pairing is $L_{\Gamma}=1/n$.

To build $\mathcal{U}(M_3)$, we again wrap a D3 on $M_3\times \gamma$ where $\gamma$ is a generating 1-cycle with boundary homology class $\gamma \in H_1( S^3/\Gamma)$, but now there is a non-trivial $SL(2, \mathbb{Z})$ monodromy for $n=3,\; 4, \; 6, \; 8$ and $12$. This clearly modifies the expansion of $\breve{F}_3$, $\breve{F}_2$, and their duals in such a way that one must generally consider the vectors $(F_{2},F^D_{2}), (F_{3},F^D_{3})$ modulo some relations as well-defined objects rather than the individual components. More precisely, we need to expand these fields in cohomology with the twisted coefficient module
\begin{equation}\label{eq:twistcoef}
    (\mathbb{Z} \oplus \mathbb{Z})_\rho
\end{equation}
where $\rho$ is the $SL(2, \mathbb{Z})$ monodromy of order $k$ when going around $\gamma$ as given in table \ref{tab:TopData}.

\begin{table}
\begin{center}
\renewcommand{\arraystretch}{1.25}
\begin{tabular}{|| c | c | c | c | c | c ||}
 \hline
$n$ & \textnormal{Kodaira Type} & Monodromy $\rho$ & $k= \mathrm{ord}(\rho)$ & Tor\,$H_1(T_3)$ &  $L_\gamma^t$ \\ [0.5ex]
 \hline\hline
 3 & $IV$ &  $\left(\begin{array}{cc}  0 & 1 \\ -1 & -1 \end{array}\right)$ & 3 &  $\mathbb{Z}_3$ & 1/3  \\
 \hline
 4 & $I_0^*$ & $\left(\begin{array}{cc}  -1 & 0 \\ 0 & -1 \end{array}\right)$ & 2  &  $\mathbb{Z}_2 \oplus \mathbb{Z}_2$ & $\left(\begin{array}{cc}  0 & 1/2 \\ 1/2 & 0 \end{array}\right)$ \\
 \hline
 6 & $IV^*$ &  $\left(\begin{array}{cc}  -1 & -1 \\ 1 & 0 \end{array}\right)$ & 3  &  $\mathbb{Z}_3$ & 2/3 \\
 \hline
 8 & $III^*$ & $\left(\begin{array}{cc}  0 & -1 \\ 1 & 0 \end{array}\right)$ & 4 &  $\mathbb{Z}_2$ & 1/2  \\
 \hline
 12 & $II^*$ & $\left(\begin{array}{cc}  0 & -1 \\ 1 & 1 \end{array}\right)$ & 6 &  $0$ & $0$  \\
 \hline
\end{tabular}
\end{center}
\caption{Topological data for asymptotic geometries of 6D $\mathcal{N}=(1,0)$ SCFTs. The F-theory geometry consists of an elliptic fibration over a Lens space base containing a torsional 1-cycle $\gamma$. The torus fibration restricted to $\gamma$ gives a three-manifold $T_3$, a torus bundle more precisely, whose torsion and linking forms $L^t_\gamma$ we list, see \cite{Cvetic:2021sxm}. D3 branes wrapped on $\gamma$ map under M-/F-theory duality to M5-branes wrapped on $T_3$. }
\label{tab:TopData}
\end{table}

We begin by computing the twisted cohomology of the boundary of the base space $S^3/\Gamma = \partial(\mathbb{C}^2/\Gamma)$. Via an identical computation\footnote{Let $A = \mathbb{Z}^2$ be a $\mathbb{Z}_n$ module, then $H^*(S^{2r-1}/\mathbb{Z}_n, A)$ is computed by taking the cohomology of the cochain complex:
\begin{equation}
\mathbb{Z}^2 \xrightarrow[]{\,1-\rho\,} \mathbb{Z}^2 \xrightarrow[]{\,1+\rho+\cdots+\rho^{n-1}\,}  \mathbb{Z}^2 \xrightarrow[]{\,1-\rho\,} \cdots \xrightarrow[]{\,1+\rho+\cdots+\rho^{n-1}\,} \mathbb{Z}^2 \xrightarrow[]{\,1-\rho\,} \mathbb{Z}^2\,.
\end{equation}
} to that given in section 3.2 of \cite{Aharony:2016kai}, we get
\begin{align}
    H^*(S^3/\mathbb{Z}_n; (\mathbb{Z} \oplus \mathbb{Z})_\rho) = \{0, G_k, 0, G_k\}
\end{align}
where
\begin{equation}\label{eq:discgrps}
    G_k = \left\{
   \begin{array}{lll}
    \mathbb{Z}_2 \oplus \mathbb{Z}_2 &\quad k=2  &\quad (n = 4) \\
    \mathbb{Z}_3 &\quad k=3  &\quad  (n = 3,6) \\
    \mathbb{Z}_2 &\quad k=4 &\quad  (n = 8) \\
        1 &\quad k=6 &\quad  (n = 12)
    \end{array}
    \right.
\end{equation}

Now, $\breve{F}_5$ associated with D3 branes should be reduced on untwisted differential cocycles $\breve{u}_i \in \breve{H}^i(S^3/\Gamma; \mathbb{Z})$:
\begin{equation}
        \breve{F}_5 = \breve{a}_2 \star \textnormal{\u{v}ol} + \breve{a}_3 \star \breve{u}_2 + \breve{a}_5 \star \breve{1}, \\
\end{equation}
whereas $\breve{F}_3, \breve{F}_3^{(D)}$ associated with D1 and F1 strings and the worldvolume  $\breve{F}_2, \breve{F}_2^{(D)}$ should be reduced on twisted differential cocycles $\breve{t}_i \in \breve{H}^i(S^3/\Gamma; (\mathbb{Z} \oplus \mathbb{Z})_\rho)$. The reduction goes as follows: when $k\neq 2$ (here $\rho$ in the superscript stands for self-dual operators that are compatible with the $\rho$ twisting):
\begin{align}
\label{eq:Monodromy}
    \begin{split}
     \breve{F}_3^{\rho} = \left(\begin{array}{c} \breve{F}_3 \\ \breve{F}_3^{(D)} \end{array}\right)/\mathrm{Im}(\rho - 1) &= \breve{b}_2^{\rho} \star \breve{t}_1 + \breve{b}_0^{\rho} \star \breve{t}_3  \\
    \breve{F}_2^{\rho} = \left(\begin{array}{c} \breve{F}_2 \\ \breve{F}_2^{(D)} \end{array}\right)/\mathrm{Im}(\rho - 1) &= \breve{c}_1^{\rho} \star \breve{t}_1\,.
    \end{split}
\end{align}
Similar expansions hold for $ \breve{b}_0^{\rho}$ and $\breve{b}_2^{\rho}$. The notation of the lefthand side denotes the reduction of $(\breve{F}_2, \breve{F}_2^{(D)})$ modulo $SL(2, \mathbb{Z})$ monodromy. The fields $c_{n}$ and $b_{m}$ are discrete $G_k$ valued $n$-cocycles and $m$-cocycles where $G_k=\mathbb{Z}_2,\mathbb{Z}_3$ as in \eqref{eq:discgrps} respectively.

When $k=2$ we have the decomposition
\begin{equation}
H^i(S^3/\Gamma; (\mathbb{Z} \oplus \mathbb{Z})_\rho)
=H^i(S^3/\Gamma; (\mathbb{Z})_\rho)^{(e)}\oplus H^i(S^3/\Gamma; (\mathbb{Z})_\rho)^{(m)}\end{equation}
with $\rho=-1$ for each coefficient ring. Consequently the expansion is then
\begin{equation}
    \breve{F}_2 =\breve{c}_1 \star \breve{t}_{(e),1} \,, \qquad
    \breve{F}_2^{(D)}= \breve{c}_1^D \star  \breve{t}_{(m),1}
\end{equation}
where $I(\breve{t}_{(e,m),1})=t_{(e,m),1}$ generates $H^1(S^3/\Gamma; (\mathbb{Z})_\rho)^{(e,m)}$. Similar expansions hold for $ \breve{b}_0$ and $\breve{b}_2$. In this case all fields are discrete $\mathbb{Z}_2$ co-cycles with the degree as indicated by their index.

When reducing the $\breve{F}_5$ term, the non-zero term comes from
\begin{equation}
     \int_{S^3/\Gamma} \breve{u}_2 \star \breve{u}_2 \equiv L_\Gamma(\breve{u}_2)
\end{equation}
on $S^3/\Gamma$. This gives the contribution to the action of
\begin{equation}
    \exp \left( 2\pi i L_\Gamma(\breve{u}_2) \int_{M_3} a_3  \right).
\end{equation}

For $\breve{F}_3, \breve{F}_3^{(D)}, \breve{F}_2, \breve{F}_2^{(D)}$, on the other hand, the non-zero terms can be evaluated from the pairing of the twisted cohomology classes on $\gamma \in H_1(S^3/\Gamma; \mathbb{Z})$:
\begin{equation}\begin{aligned}\label{eq:Pairing}
 k=3,4\,:\qquad &\int_{\gamma} \breve{t}_1 \star \breve{t}_1 \equiv L_\gamma^{t}(\breve{t}_1)\\ k=2\,:\qquad &\int_{\gamma} \breve{t}_{(e),1} \star \breve{t}_{(m),1} \equiv L_\gamma^{t}(\breve{t}_{(e),1},\breve{t}_{(m),1})
\end{aligned}\end{equation}
which we  both denote by $L_\gamma^t$ whenever the context is clear. The self-pairing of $\breve{t}_{(e,m),1}$ vanishes as we shortly argue.

The pairing between twisted classes in differential cohomology generalizing torsional linking are computed using the methods in reference \cite{Bott1982DifferentialFI}. M-/F-theory duality gives a natural relation of such pairings to linking forms in ordinary singular homology. We now explain this relation as we perform our computations from the latter perspective.

To frame the discussion we introduce the torus bundle three-manifold $T_3$ as the restriction of the S-duality torus bundle to the 1-cycle $\gamma$ wrapped by the D3 brane. As all three-manifolds its homology groups are fully determined by $H_1(T_3;\mathbb{Z})$ which is computed by application of the Mayer-Vietoris sequence to
\begin{equation}
    H_1(T_3)=\mathbb{Z}\oplus \textnormal{coker}\,(\rho-1)
\end{equation}
where $\rho$ is the $SL(2,\mathbb{Z})$ monodromy matrix acting on 1-cycles upon traversing $\gamma$. The torsional subgroups are listed in table \ref{tab:TopData}. The linking form on $T_3$ follows from similar considerations \cite{Cvetic:2021sxm}. Let us denote the torsional generators of $\textnormal{Tor}\,H^2(T_3;\mathbb{Z})$ by $t_2$ which by Poincar\'e duality and the universal coefficient theorem is dual to the generators $\ell_1$ of $\textnormal{Tor}\,H_1(T_3;\mathbb{Z})$. When the monodromy matrix is of type $I_0^*$ $(k=2)$ both $t_2$ and $\ell_1$ are further indexed by $(e,m)$ distinguishing the factors in $\textnormal{Tor}\,H_1(T_3;\mathbb{Z})\cong \mathbb{Z}_2 \oplus \mathbb{Z}_2$ for that case.

Now note that M-/F-theory duality maps a D3 brane wrapped on $\gamma\times M_3$ to an M5-brane wrapped on $T_3\times M_3$ where $M_3$ is the space-time submanifold supporting the topological operator. The Wess-Zumino-Witten term of the M5-brane contains the term \cite{Aharony:1996wp,  Bandos:1997ui}
\begin{equation}\label{eq:WZWM5}\begin{aligned}   \mathcal{S}_{\text{top}}^{M5} \supset  2\pi i \int_{M_3\times T_3}  \breve{F}_7 + \frac{1}{2} \breve{F}_3 \star \breve{F}_4
\end{aligned}\end{equation}
where $\breve{F}_3$ is the anti-self-dual 3-form field strength (of the anti-chiral 2-form field on the M-theory worldvolume), and $\breve{F}_7$ is the pullback of the magnetic dual 7-form field strength.
We can therefore equivalently compute the topological field theory on $M_3$ starting from the action \eqref{eq:WZWM5}. This approach however expresses the coefficient of the topological theory via geometric data of $T_3$ and avoids $SL(2,\mathbb{Z})$ twisted cohomology classes. We therefore conjecture that the pairing \eqref{eq:Pairing} is geometrized to a link pairing on the torus bundle
\begin{equation}\label{eq:Conjecture}\begin{aligned}
   k=3,4\,:\qquad &L_\gamma^{t}(\breve{t}_1)= \int_\gamma \breve{t}_1\star\breve{t}_1=\int_{T_3} \breve{t}_2\star\breve{t}_2\\
      k=2\,:\qquad &L_\gamma^{t}(\breve{t}_{(e),1},\breve{t}_{(m),1})= \int_\gamma \breve{t}_{(e),1}\star\breve{t}_{(m),1}=\int_{T_3} \breve{t}_{(e),2}\star\breve{t}_{(m),2}=\frac{1}{2}
   \end{aligned}
\end{equation}
where the righthand side is computed by the linking pairing given in table \ref{tab:TopData}. We will have more to say on the M-theory perspective in section \ref{sec:Mtheory}. Evidence for the identity \eqref{eq:Conjecture} is already given in \cite{GarciaEtxebarria:2022vzq} which considers a setup with $-1\in SL(2;\mathbb{Z})$ monodromy along $\gamma$ and where the case $k=2$ in \eqref{eq:Conjecture} was found to hold.

Before writing down the full topological action of our D3-brane, we must first comment on the expected non-commutativity of flux operators in this scenario. Due to the presence of a non-trivial duality bundle, there is a mixing between the electric and magnetic dynamical two-form curvatures on the D3 worldvolume gauge theory, so considering the on-shell relation $F^D_2=*F_2$ (see for instance \cite{Polchinski:2014mva}) we must quantize these fields as a self-dual Maxwell theory.\footnote{Note that our worldvolume theory is Euclidean.} This is especially clear in the M5-brane picture where the flux quantization is already that of anti-self-dual fields and the torsion homology of the $T^2$-bundle precisely descends to the torsion in the twisted homology that the D3-brane wraps. Depending on the value of $k$, it is occasionally possible to have a canonical splitting of electric and magnetic fluxes on the D3.

 Now expanding on the treatment of \cite{GarciaEtxebarria:2022vzq} to examine the non-commutativity of fluxes on the D3 worldvolume in our cases, we first assume that $M_3=N_2\times \mathbb{R}_t$ to employ a Hamiltonian formalism. The Hilbert space associated to the spatial manifold $N_2 \times \gamma$ of the D3-brane worldvolume will then be a representation a Heisenberg algebra, the details of which depend on the value of $k$. The Heisenberg algebra is generated by non-commuting electric and magnetic flux operators $\Phi_e,\Phi_m$ respectively detecting fluxes through torsional cycles. The cases are:\footnote{We thank I. Garcia Etxebarria for a question which prompted this clarification. See also \cite{GarciaHosseini} for a related discussion.}
\begin{itemize}
    \item For $k=2$, we have a pair of non-commuting electric and magnetic fluxes associated with $\mathrm{Tor}\, H^2(N_2 \times \gamma)$:
    \begin{equation}
        \Phi_e( \breve{c}_1 \star \breve{t}_{(e),1}) \Phi_m(\breve{c}_1^{(D)} \star \breve{t}_{(m),1}) = \exp\left( 2\pi iL_\gamma^{t}(\breve{t}_{(e),1},\breve{t}_{(m),1}) \right) \Phi_m(\breve{c}_1^{(D)} \star \breve{t}_{(m),1})  \Phi_e( \breve{c}_1 \star \breve{t}_{(e),1})
    \end{equation}
    we thus get a $\mathbb{Z}_2$ gauge theory as in \cite{Freed:2006ya}, with the action given by
    \begin{equation}
    \mathcal{S}_{\mathbb{Z}_2} =  \pi i \int_{M_3}c_1 \cup \delta c_1^{(D)}
    \end{equation}
    where $(c_1, c_1^{(D)})$ are a pair of discrete gauge fields which together are valued in $G_2=\mathbb{Z}_2 \times \mathbb{Z}_2$. In other words, $c_1$ and $c_1^{(D)}$ are each separately $\mathbb{Z}_2$ valued discrete gauge fields normalized such that $\int c_1=\ell \; \mathrm{mod}\; 2$. 

    \item For $k=3, 4,6$, we have a pair of non-commuting self-dual fluxes associated
    with $\mathrm{Tor}\,H^2(N_2 \times \gamma)$.
    Now, extra care has to be taken since the electric and magnetic field has to be the same \cite{GarciaEtxebarria:2019caf}. For $x, y \in H^{1}(\gamma; (\mathbb{Z} \oplus \mathbb{Z})_\rho)$
    \begin{equation}
        \Phi_a(\breve{c}_1^{\rho} \star \breve{x}) \Phi_b(\breve{c}_1^{\rho} \star \breve{y}) = \exp\left( 2\pi i L_\gamma^{t}(\breve{x}, \breve{y}) \right)   \Phi_b(\breve{c}_1^{\rho} \star \breve{y}) \Phi_a(\breve{c}_1^{\rho} \star \breve{x})
    \end{equation}
    here $L^t_\gamma(x, y)$ is the bilinear form of the twisted linking pairing. 
    Thus, we need to include a discrete CS gauge theory of the form
    \begin{equation}
        \mathcal{S} = 2 \pi i L^t_{\gamma}(\breve{t}_1) \int_{M_3} c_1^\rho \cup \delta c_1^\rho,
    \end{equation}
    where $c^\rho_1$ is a discrete gauge field valued in $G_k$ as in (\ref{eq:discgrps}) normalized such that $\int c^\rho_1=\ell \; \mathrm{mod}\; \mathrm{2\;  or\;  3}$. 
\end{itemize}

Furthermore, the topological actions generating the link pairing above produce an additional term in the effective action of the D3 brane reduced on the twisted 1-cycle, the middle term(s) in both lines of \eqref{eq:SD3toptwist}, because the effective action must be a functional of the gauge invariant combination $\delta c^\rho_1-b^\rho_2$ where $b^\rho_2$ is defined in the first line of equation \eqref{eq:Monodromy}.\footnote{This follows from the fact that the standard D3 topological action must be a functional of the gauge invariant combination $F-B_2$ and its S-dual completion $F^D-C_2$.}

To summarize then, we get the action of the topological operator (where again we keep the dependence on the torsional 1-cycle implicit, include a normalization factor $\mathcal{K}^{-1}$, and leave the cup products implicit) $\mathcal{U}(M_3) =\mathcal{K}^{-1} \exp(\mathcal{S}_{\textnormal{top}}^{\textnormal{D3}} )$ as:
\begin{equation} \label{eq:SD3toptwist}
\begin{aligned}
    k=2 &: \qquad  \mathcal{S}_{\textnormal{top}}^{\textnormal{D3}} =
    2\pi i\int_{M_3} \bigg( -\tfrac{1}{4} a_3 - \tfrac{1}{2}b^{(D)}_2 c_1 - \tfrac{1}{2} b_2 c_1^{(D)} + \tfrac{1}{2} c_1 \delta c_1^{(D)} \bigg)
    \\
    k= 3, 4, 6 &:   \qquad \mathcal{S}_{\textnormal{top}}^{\textnormal{D3}} = 2\pi i\int_{M_3} \bigg( L_\Gamma a_3 - L_\gamma^t  b_2^{\rho} c_1^{\rho} + L_{\gamma}^t c_1^{\rho} \delta c_1^{\rho} \bigg)
\end{aligned}
\end{equation}
where for $k = 3, 4, 6$, the path integral is written as $\mathcal{K}^{-1}\int Dc_1^{\rho}  \exp\left(  \mathcal{S}_{\textnormal{top}}^{\textnormal{D3}}\right)$ with the implicit understanding that a delta--function relating $c_1, c_1^{(D)}$ has been inserted to gauge fix the monodromy relations of line \eqref{eq:Monodromy}. For the values of both $L_\Gamma=L_\Gamma(\breve{u}_2) = \frac{1}{n}$ and $L_{\gamma}^t=L_\gamma^t(\breve{t}_1)$ see table \ref{tab:TopData}. Also, in this subsection the normalization factor will be $\mathcal{K}=(|H_2(M_3,G_k)|)^{1/2}$ for reasons that will be clear in what follows.


Notice that due to the coupling to the various $b_2$ fields in line \eqref{eq:SD3toptwist}, we see again, just as in subsection \ref{ssec:6d2comma0}, that the symmetry operator $\mathcal{U}(M_3)$ acts on more than just dimension-2 operators in the defect group $\mathbb{D}$. The $b_2$'s are background fields for a discrete $G_k^{(1)}$-symmetry and are sourced by defects constructed from $(p,q)$-5-branes wrapping homology classes\footnote{Said differently, a 2-cycle with $(p,q)$-charge of the brane is being measured modulo $\textnormal{Im}\,(\rho-1)$.} in $H_2(S^3/\Gamma, (\mathbb{Z}\oplus \mathbb{Z})_\rho)=\mathrm{Hom}(G_k,U(1))$ times the radial direction of $\mathbb{C}^2/\Gamma$.

 Notice that due to the presence of terms like $c\delta c$ in \eqref{eq:SD3toptwist}, we see that even if we ignore terms involving $b_2$ fields (and hence the effect of $\mathcal{U}(M_3)$ on the line operators mentioned in the previous paragraph) our 2-form symmetry operators are tensored with discrete topological gauge theories, namely some level-N Dijkgraaf-Witten theories with gauge group $G_k$, $\mathcal{T}^{(N,G_k)}$. The levels of these gauge theories are classified by by $H^4(G_k,\mathbb{Z})$ \cite{Dijkgraaf:1989pz}, and the possible levels relevant to single node NHCs are $H^4(\mathbb{Z}_N,\mathbb{Z})=\mathbb{Z}_N$ and $H^4(\mathbb{Z}^2_2,\mathbb{Z})=\mathbb{Z}^3_2$. Our action (\ref{eq:SD3toptwist}) thus predicts the levels of these discrete gauge theories living on the defect group symmetry operators and notice that the operator fusion simply adds together the cyclically defined Chern-Simons levels, so this confirms that that $\mathcal{U}(M_3)$ appears to be an invertible operator when linking with operators with trivial charge under the $G^{(1)}_k$ symmetry.

\textbf{Fusion Rules}
We now turn to the fusion rules for our symmetry operators, namely we
compute $\mathcal{U}(M_3)\times \mathcal{U}^\dagger(M_3)$. We find that
in the presence of the $G^{(1)}_k$ background field $b_2$ that the fusion rule typically
contains multiple summands, i.e., the hallmark of a non-invertible symmetry.\footnote{The non-invertible fusion is in fact not very surprising considering that the terms in the actions of (\ref{eq:SD3toptwist}) not involving $a_3$ are
a discrete analog of the 3D actions one would write for the standard fractional Quantum Hall effect (FQHE). This is similar to what was found in the analysis of ABJ anomalies in references \cite{Choi:2022jqy,Cordova:2022ieu}.} This
is detected by 3D defects sourcing such backgrounds and linked by surface operators which in turn are produced in the fusion.
\begin{itemize}
\item For the simplest case of $k = 6$, the symmetry operator associated with a generator of $H_1(S^3 / \mathbb{Z}_{12}, \mathbb{Z})$ takes the form
\begin{equation}
    \mathcal{U}(M_3) = \exp \left( \frac{2\pi i}{12}  \int_{M_3} a_3 \right)
\end{equation}
which simply reproduces the $\mathbb{Z}_{12}$ defect group of a $(-12)$-curve NHC.
\item For $k = 2$ (i.e. the $-4$ NHC theory), we can decompose $\mathcal{U}(M_3) = \mathcal{U}_0(M_3) \mathcal{U}_1(M_3)$. $\mathcal{U}_0(M_3) = \left( \frac{2\pi i}{4}  \int_{M_3} \breve{a}_3 \right)$ generates a $\mathbb{Z}_4$ defect group, whereas $\mathcal{U}_1(M_3) $ is non-invertible when $b_2$ or $b^D_2$ is turned on. Physically, this means
3D defects which are charged under the 1-form symmetry (with background field $b_2$) will detect
this non-invertible fusion rule.

Because our topological action exactly matches that of equation (3.14) of \cite{GarciaEtxebarria:2022vzq} up to an overall irrelevant minus sign, we can borrow the result to state
\begin{equation}\label{eq:kequal2condensation}
 \mathcal{U}_1(M_3)\times  \mathcal{U}_1(M_3)=\frac{1}{|H_2(M_3,\mathbb{Z}_2)|^2}\sum_{\sigma, \sigma'\in H_2(M_3,\mathbb{Z}_2)} \exp{\bigg(\pi i \int_\sigma b_2\bigg)}\cdot \exp{\bigg(\pi i \int_{\sigma'} b^{(D)}_2\bigg)}
\end{equation}
where $\sigma$ and $\sigma'$ are generators of $H_2(M_3, \mathbb{Z}_2)$. Notice that these exponents are symmetry operators for a discrete $\mathbb{Z}_2\times \mathbb{Z}_2$ 1-form symmetry in the 6D SCFT. In the language of \cite{Roumpedakis:2022aik} this sum over symmetry operators restricted to lie in $M_3$ means that this is a 3-gauging of the $\mathbb{Z}_2\times \mathbb{Z}_2$ 3-form symmetry along $M_3$. This is commonly known as a condensation operator.


\item For $k = 3,4$, we can similarly decompose $\mathcal{U}(M_3) = \mathcal{U}_0(M_3) \mathcal{U}_1(M_3)$ where the invertible piece $\mathcal{U}_0(M_3) = \left( \frac{2\pi i}{n}  \int_{M_3} \breve{a}_3 \right)$ produces a $\mathbb{Z}_6$ and $\mathbb{Z}_8$ algebra respectively.\footnote{Recall that $n=6$ for $k=3$ and $n=8$ for $k=4$.} $\mathcal{U}_1(M_3)$ is again non-invertible which we can see from the fact that the fusion product $\mathcal{U}_1(M_3)\times \mathcal{U}^\dagger_1(M_3)\neq \mathbf{1}$ (for $k=2$, $\mathcal{U}_1(M_3)=\mathcal{U}^\dagger_1(M_3)$).
Calculating the total fusion (leaving the correct normalization until the final result),
\begin{align}
    & \mathcal{U}(M_3) \times \mathcal{U}^\dagger(M_3) = \int Dc_1^\rho D c_1^{\prime \rho}  \exp\bigg( 2\pi i L^t_\gamma \int_{M_3} \big( c_1^{\rho} \cup\delta c_1^{\rho} - c_1^{\prime\rho} \cup \delta c_1^{\prime\rho}+b^\rho_2\cup(c^{\prime\rho}_1-c_1^{\rho}) \big) \bigg)
\end{align}
and after substituting $\hat{c}^\rho_1:=c^{\prime\rho}_1-c^{\rho}_1$ and integrating a term by parts we find
\begin{align}\label{eq:uudaggerk34}
   &\mathcal{U}(M_3)\times \mathcal{U}^\dagger(M_3) =\int Dc_1^{\prime \rho} D \hat{c}_1^{ \rho}  \exp\bigg( 2\pi i L^t_\gamma \int_{M_3} \big(\hat{c}_1^{\rho} \cup \delta \hat{c}_1^{\rho}+b^\rho_2\cup\hat{c}^\rho_1 \big) \bigg).
\end{align}
This is slightly different than the $k=2$ case where there was no analog of the middle term above. Comparing to equation (B.15) of \cite{Choi:2022jqy}, we see indeed that the left-hand side is still a condensation operator and the coefficient in front of the middle term is interpreted as discrete theta angle given by the identity element in $H^3(G_k,U(1))\simeq G_k$ given the coefficient of the middle term.\footnote{More generally, the relevant discrete theta angles of this $G_k$ gauge theory is given by $\mathrm{Hom}\big(\mathrm{Tor}\;\Omega^{Spin}_3,U(1) \big)$ when $M_3$ is a spin manifold. We will leave this more refined consideration of the structure of $M_3$ to future work.}\footnote{Notice that (B.15) of \cite{Choi:2022jqy} is written in terms of $U(1)$ valued forms, where the purpose of their first term is to constrain the gauge field to be discretely valued.} Explicitly we have (restoring the correct normalization),
\begin{align}
     &\mathcal{U}(M_3)\times \mathcal{U}^\dagger(M_3) =\frac{1}{|H_2(M_3,G_k)|}\sum_{\sigma \in H_2(M_3,G_k)} \epsilon(M_3,\sigma) \exp{\bigg(2\pi i L^t_\gamma \int_\sigma b^\rho_2\bigg)}
\end{align}
where $\epsilon(M_3,\sigma)$ is a discrete torsion term and following \cite{Choi:2022jqy}, we see that the right-hand side is equivalent to a level-1 Dijkgraaf-Witten theory with gauge group $G_k$ coupled to a 1-form electric background field $b^\rho_2$. In other words,
\begin{align}
     &\mathcal{U}(M_3)\times \mathcal{U}^\dagger(M_3) =\mathcal{T}^{(1,G_k)}_{DW}(M_3,b^\rho_2)
\end{align}
in the obvious notation. Note that this is again a 3-gauging of a 3-form symmetry along $M_3$.

\end{itemize}



\subsection{More General 6D SCFTs}

We can extend our discussion in a few different directions. One can also consider 
more general F-theory backgrounds with constant axio-dilaton \cite{Heckman:2013pva, DelZotto:2014hpa, Heckman:2014qba, Heckman:2015bfa, DelZotto:2017pti}).\footnote{This includes, for example, rank $N$ conformal matter of type $(G,G)$ \cite{DelZotto:2014hpa, Heckman:2014qba}.} 
Even though the axio-dilaton is constant, the duality bundle can still be non-trivial. 
The base of the model is again a generalized ADE-type singularity, and 
has boundary torsional 1-cycles on which we can wrap D3-branes.

More broadly speaking, whenever we have a non-trivial defect group we anticipate that a similar structure persists.
When we have a position dependent axio-dilaton profile at the boundary of the base geometry, it appears simplest to extract the relevant topological terms for the generalized symmetry operators by starting with the topological terms of an M5-brane and dimensionally reducing along the (torsional) 3-cycle obtained by fibering the F-theory torus over the torsional 1-cycle of the base.

It is also natural to treat the effects of the 0-form and 1-form symmetries by explicitly tracking the profile of flavor 7-branes in the system. One way to proceed is to pass to the M-theory limit by compactifying on a further circle. So long as the generalized symmetry operator does not wrap this circle, we can then analyze these effects in purely geometric terms using \cite{Cvetic:2022imb}. Alternatively, we can use the known structure of topological terms on the tensor branch of these 6D theories to extract the same data from a ``bottom up'' perspective \cite{Apruzzi:2020zot, Apruzzi:2021mlh, Hubner:2022kxr, Heckman:2022suy}.

In both situations, however, the appearance of a non-trivial $SL(2,\mathbb{Z})$ bundle in the F-theory background
is a strong indication that the resulting generalized symmetry operators will have a fusion algebra which is not captured
by a group law. Said differently, we expect that generically, these 6D SCFTs will have non-invertible symmetries which act on 3D defects sourcing a background for $G^{(1)}_k$.

We leave a more systematic analysis of these cases for future work.

\section{M-theory Examples} \label{sec:Mtheory}

Although we have focussed on IIB / F-theory backgrounds, the same
considerations clearly hold more broadly. For example, 5D\ SCFTs
engineered via M-theory on Calabi-Yau canonical singularities can also
support various defects \cite{Albertini:2020mdx, Morrison:2020ool,
Tian:2021cif, DelZotto:2022fnw}. To set notation, let $X$ denote a non-compact
Calabi-Yau threefold which generates a 5D\ SCFT. We can get defects by
wrapping M2-branes and M5-branes on non-compact cycles which extend to the
boundary $\partial X$. The corresponding topological operators are obtained by
wrapping magnetic dual branes on the appropriate cycles:%
\begin{align}
\underset{\text{Line Defect}}{\underbrace{\text{M2 on }\widetilde{M}_{1}%
\times\mathbb{R}_{\geq0}\times \widetilde{\gamma}_{1}}}  & \leftrightarrow
\underset{\text{Gen. Symm. Membrane}}{\underbrace{\text{M5 on }{M}%
_{3}\times\gamma_{3}}}\\
\underset{\text{Surface Defect}}{\underbrace{\text{M5 on }\widetilde{M}_{2}%
\times\mathbb{R}_{\geq0}\times \widetilde{\gamma}_{3}}}  & \leftrightarrow
\underset{\text{Gen. Symm. Defect}}{\underbrace{\text{M2 on }{M}%
_{2}\times\gamma_{1}}}\\
\underset{\text{Wall Defect}}{\underbrace{\text{M5 on }\widetilde{M}_{4}\times
\mathbb{R}_{\geq0}\times \widetilde{\gamma}_{1}}}  & \leftrightarrow
\underset{\text{Gen. Symm. Point}}{\underbrace{\text{M2 on pt}\times\gamma_{3}%
}}.
\end{align}
Reduction of the topological terms on the worldvolume of these branes then produces the
corresponding TFT concentrated on our symmetry operator, see \eqref{eq:WZWM5}.

As a final comment on this example, we note that 5D\ SCFTs sometimes also
enjoy flavor symmetries as realized by various discrete symmetries as well as
\textquotedblleft flavor 6-branes\textquotedblright\ (namely
ADE\ singularities). One can in principle consider wrapping such ``6-branes'' on torsional cycles
of the boundary geometry. This can be viewed as introducing a singular profile for the M-theory
metric in the asymptotic geometry. Wrapping such a 6-brane on a torsional 3-cycle would result
in a generalized symmetry operator for a 0-form symmetry (as it is codimension 1 in the 5D spacetime).
Clearly, this case is a bit more subtle to treat, but it is so intriguing that we leave it as an
avenue to pursue in future work.


\section{Further Generalizations}

So far, we have mainly explained how to lift various ``bottom up'' field theory structures to explicit string constructions. This is
already helpful because it provides us with a machine for extracting the corresponding worldvolume TFT on these generalized symmetry operators, as well as the resulting fusion rules.

But the stringy perspective provides us with even more. For one thing, it makes clear the ultimate fate of these ``topological'' operators once we recouple to gravity. Indeed, once we couple to gravity, $X$ no longer has a boundary, and so all of our wrapped branes will again become dynamical. Moreover, we can also see that in many cases, these generalized symmetries automatically trivialize in compact geometries.

Reinterpreting generalized symmetry operators in terms of wrapped branes also suggests a further ``categorical'' generalization of the standard generalized symmetries paradigm. Indeed, it has been appreciated for some time that at least in type II backgrounds on a Calabi-Yau threefold, the spectrum of topological branes is captured, in the case of the topological B-model by the (bounded) derived category of coherent sheaves and in the mirror A-model by the
triangulated Fukaya category.\footnote{See e.g., \cite{Kontsevich:1994dn, Douglas:2000gi, Aspinwall:2001pu} and \cite{Aspinwall:2004jr} for a review.} The important point here is that for these more general objects, simply working in terms of ``branes wrapped on cycles'' is often inadequate. This in turn suggests that instead of assigning a generalized symmetry operator to a sub-manifold of the $d$-dimensional spacetime, it is more appropriate to work in terms of a complex of objects (in the appropriate derived category). Note also that because these derived categories are monoidal, there is also a notion of fusion in this setting.

\chapter{TOP DOWN APPROACH TO TOPOLOGICAL DUALITY DEFECTS}

\section{Introduction}

Dualities sit at the heart of some of the deepest insights into the non-perturbative dynamics of
quantum fields and strings. In the case of quantum field theories (QFTs) engineered via string theory,
these dualities can often be recast in terms of specific geometric transformations of the extra-dimensional
geometry. A particularly notable example of this sort is the famous $SL(2,\mathbb{Z})$ duality symmetry of
type IIB string theory which descends to a duality action on the QFTs realized on the worldvolume of
probe D3-branes.

Recently it has been appreciated that symmetries themselves can be generalized in a number of different
ways. In particular, in \cite{Gaiotto:2014kfa} it was argued that symmetries can be understood in
terms of corresponding topological operators (see also \cite{Gaiotto:2010be,Kapustin:2013qsa,Kapustin:2013uxa,Aharony:2013hda}).\footnote{For a partial list 
of recent work in this direction see e.g.,
\cite{Gaiotto:2014kfa,Gaiotto:2010be,Kapustin:2013qsa,Kapustin:2013uxa,Aharony:2013hda,
DelZotto:2015isa,Sharpe:2015mja, Heckman:2017uxe, Tachikawa:2017gyf,
Cordova:2018cvg,Benini:2018reh,Hsin:2018vcg,Wan:2018bns,
Thorngren:2019iar,GarciaEtxebarria:2019caf,Eckhard:2019jgg,Wan:2019soo,Bergman:2020ifi,Morrison:2020ool,
Albertini:2020mdx,Hsin:2020nts,Bah:2020uev,DelZotto:2020esg,Hason:2020yqf,Bhardwaj:2020phs,
Apruzzi:2020zot,Cordova:2020tij,Thorngren:2020aph,DelZotto:2020sop,BenettiGenolini:2020doj,
Yu:2020twi,Bhardwaj:2020ymp,DeWolfe:2020uzb,Gukov:2020btk,Iqbal:2020lrt,Hidaka:2020izy,
Brennan:2020ehu,Komargodski:2020mxz,Closset:2020afy,Thorngren:2020yht,Closset:2020scj,
Bhardwaj:2021pfz,Nguyen:2021naa,Heidenreich:2021xpr,Apruzzi:2021phx,Apruzzi:2021vcu,
Hosseini:2021ged,Cvetic:2021sxm,Buican:2021xhs,Bhardwaj:2021zrt,Iqbal:2021rkn,Braun:2021sex,
Cvetic:2021maf,Closset:2021lhd,Thorngren:2021yso,Sharpe:2021srf,Bhardwaj:2021wif,Hidaka:2021mml,
Lee:2021obi,Lee:2021crt,Hidaka:2021kkf,Koide:2021zxj,Apruzzi:2021mlh,Kaidi:2021xfk,Choi:2021kmx,
Bah:2021brs,Gukov:2021swm,Closset:2021lwy,Yu:2021zmu,Apruzzi:2021nmk,Beratto:2021xmn,Bhardwaj:2021mzl,
Debray:2021vob, Wang:2021vki,
Cvetic:2022uuu,DelZotto:2022fnw,Cvetic:2022imb,DelZotto:2022joo,DelZotto:2022ras,Bhardwaj:2022yxj,Hayashi:2022fkw,
Kaidi:2022uux,Roumpedakis:2022aik,Choi:2022jqy,
Choi:2022zal,Arias-Tamargo:2022nlf,Cordova:2022ieu, Bhardwaj:2022dyt,
Benedetti:2022zbb, Bhardwaj:2022scy,Antinucci:2022eat,Carta:2022spy,
Apruzzi:2022dlm, Heckman:2022suy, Baume:2022cot, Choi:2022rfe,
Bhardwaj:2022lsg, Lin:2022xod, Bartsch:2022mpm, Apruzzi:2022rei,
GarciaEtxebarria:2022vzq, Cherman:2022eml, Heckman:2022muc, Lu:2022ver, Niro:2022ctq, Kaidi:2022cpf,
Mekareeya:2022spm, vanBeest:2022fss, Antinucci:2022vyk, Giaccari:2022xgs, Bashmakov:2022uek,Cordova:2022fhg,
GarciaEtxebarria:2022jky, Choi:2022fgx, Robbins:2022wlr, Bhardwaj:2022kot, Bhardwaj:2022maz, Bartsch:2022ytj, Gaiotto:2020iye,Agrawal:2015dbf, Robbins:2021ibx, Robbins:2021xce,Huang:2021zvu,
Inamura:2021szw, Cherman:2021nox,Sharpe:2022ene,Bashmakov:2022jtl, Inamura:2022lun, Damia:2022bcd, Lin:2022dhv,Burbano:2021loy, Damia:2022rxw} and \cite{Cordova:2022ruw} for a recent review.}

The fact that generalized symmetry operators are topological also suggests that the ``worldvolume'' itself may support a 
non-trivial topological field theory. One striking consequence of this fact is that the product of 
two generalized symmetry operators need not produce a third symmetry operator. Instead, there can be a non-trivial 
fusion category (i.e., multiple summands in the product), and this is closely tied to the appearance of ``non-invertible'' symmetry operators (see e.g., \cite{Thorngren:2019iar,Komargodski:2020mxz, Gaiotto:2020iye, Nguyen:2021naa, Heidenreich:2021xpr, Thorngren:2021yso, Agrawal:2015dbf, Robbins:2021ibx, Robbins:2021xce, Sharpe:2021srf, Koide:2021zxj, Huang:2021zvu,
Inamura:2021szw, Cherman:2021nox, Kaidi:2021xfk, Choi:2021kmx, Wang:2021vki, Bhardwaj:2022yxj,
Hayashi:2022fkw, Sharpe:2022ene, Choi:2022zal, Kaidi:2022uux, Choi:2022jqy, Cordova:2022ieu,
Bashmakov:2022jtl, Inamura:2022lun, Damia:2022bcd, Choi:2022rfe, Lin:2022dhv, Bartsch:2022mpm,
Lin:2022xod, Cherman:2022eml, Burbano:2021loy, Damia:2022rxw, Apruzzi:2022rei, GarciaEtxebarria:2022vzq, Heckman:2022muc, 
Niro:2022ctq, Kaidi:2022cpf, Mekareeya:2022spm, Antinucci:2022vyk, Giaccari:2022xgs, Bashmakov:2022uek,Cordova:2022fhg,
GarciaEtxebarria:2022jky, Choi:2022fgx, Bhardwaj:2022kot, Bhardwaj:2022maz, Bartsch:2022ytj}).

Now, one of the notable places where non-invertible symmetries have been observed is in the context of certain ``duality / triality defects''.\footnote{A word on terminology: in this chapter we use the stringy notion of a non-abelian $SL(2,\mathbb{Z})$ duality. In particular, we will be interested in operations which generate subgroups of $SL(2,\mathbb{Z})$ with order different than two. In what follows we shall sometimes 
refer to all of these as dualities even if the order is different than two. That being said, in our field theory examples we will explain when we are dealing with a specific duality / triality defect.}
At generic points of parameter space, a duality interchanges one description of a field theory with another. However, at special points in the parameter space (such as the critical point of the 2D Ising model), this duality operation simply sends one back to the same theory. In that case, one has a 0-form symmetry, and therefore one expects a codimension one generalized symmetry operator \cite{Kaidi:2021xfk, Choi:2021kmx, Choi:2022zal, Kaidi:2022uux, Kaidi:2022cpf}. 
An intriguing feature of these symmetry operators is that they can sometimes have a non-trivial fusion rule, indicating the presence of a non-invertible symmetry. In many cases, one can argue for the existence of such a non-invertible symmetry, even without knowing the full structure of the TFT localized on a duality defect.

Given the fact that many field theoretic dualities have elegant geometric characterizations, it is natural to ask whether these topological duality defects can be directly realized in terms of objects in string theory. In particular, one might hope that performing this analysis could provide additional insight into the associated worldvolume TFTs, and provide a systematic method for extracting the corresponding fusion rules for these generalized symmetry operators. A related point is that in many QFTs of interest, a weakly coupled Lagrangian description may be unavailable and so one must seek out alternative (often geometric) characterizations of these systems.

Along these lines, it was recently shown in \cite{Heckman:2022muc} (see also \cite{Apruzzi:2022rei, GarciaEtxebarria:2022vzq}) that for QFTs engineered via localized singularities / branes, generalized symmetry operators are obtained from ``branes at infinity''. The resulting defects are topological in the sense that they do not contribute to the stress energy tensor of the localized QFT. Starting from the topological terms of a brane ``at infinity'', one can then extract the resulting TFT on its worldvolume, and consequently, extract the resulting fusion rules for the associated generalized symmetry operators.

Our aim in this note will be to use this perspective to propose a general prescription for duality defects where the duality of the QFT is inherited from the $SL(2,\mathbb{Z})$ duality of type IIB strings. In particular, we focus on the case of 4D QFTs realized from D3-branes probing a localized singularity of a non-compact Calabi-Yau threefold $X$ which we assume has a conical topology, namely it can be written as a cone over $\partial X$: $\mathrm{Cone}(\partial X) = X$. Such QFTs have a marginal parameter $\tau$ descending from the axio-dilaton of type IIB string theory, and 2-form potentials of a bulk 5D TFT descending from the $SL(2,\mathbb{Z})$ doublet of 2-form potentials (RR and NS-NS) which governs the 1-form electric and magnetic symmetries of the 4D probe theory.

In this setting, the ``branes at infinity'' which implement a duality transformation are simply given by specific bound states of $(p,q)$ 7-branes. 
In a general IIB / F-theory background, a bound state of $(p,q)$ 7-branes acts on the axio-dilaton and $SL(2,\mathbb{Z})$ doublet of 2-form 
potentials $\mathcal{B}^j = (C_2 , B_2)$ as:
\begin{equation}
\tau \mapsto \frac{a \tau + b}{c \tau + d} \,\,\, \text{and} 
\left[
\begin{array}
[c]{c}%
C_{2}\\
B_{2}%
\end{array}
\right]  \mapsto\left[
\begin{array}
[c]{cc}%
a & b\\
c & d
\end{array}
\right]  \left[
\begin{array}
[c]{c}%
C_{2}\\
B_{2}%
\end{array}
\right] 
\end{equation}
in the obvious notation. At the level of topology this monodromy can be localized to a branch cut whose endpoints are physical, namely the locus of a bound state of 7-branes. 

Of particular significance are the specific monodromy transformations which leave fixed particular values of $\tau$. Geometrically, these are specified by constant axio-dilaton profiles for 7-branes, which are in turn given by specific Kodaira fibers which specify how the elliptic fiber of F-theory degenerates on the locus of the 7-brane. The full list is $II, III, IV, I_{0}^{\ast}, IV^{\ast}, III^{\ast}, II^{\ast}$, which respectively support the gauge algebras $\mathfrak{su}_1, \mathfrak{su}_{2}, \mathfrak{su_3}, \mathfrak{so}_8, \allowbreak \mathfrak{e}_6, \mathfrak{e}_7, \mathfrak{e}_8$. Putting all of this together, it is natural to expect that 
the duality defects of the QFT simply lift to appropriate 7-branes 
wrapped on all of $\partial X$.

Our main claim is that wrapping 7-branes on a ``cycle at infinity'' leads to topological duality / triality defects in the 
4D worldvolume theory of the probe D3-brane. One way to see this is to consider the dimensional reduction on the boundary five-manifold $\partial X$. This results in the 5D symmetry TFT of the 4D field theory (see \cite{Apruzzi:2021nmk} as well as \cite{Aharony:1998qu, Heckman:2017uxe}). In this 5D theory, 7-branes wrapped on $\partial X$ specify codimension two defects which fill out a three-manifold in the 4D spacetime. In this 5D TFT limit where all metric data has been decoupled, the reduction of the 7-brane ``at infinity'' can be pushed into the interior, and can equivalently be viewed as specifying a codimension two defect in the bulk. In particular, as codimension two objects, they come with a branch cut structure, and this in turn impacts the structure of anomalies both in the 5D bulk as well as the 3D TFT localized on the topological defect.

In particular, we find that the choice of where to terminate the other end of the branch cut emanating from the 7-branes has a non-trivial impact on the resulting structure of the TFT. For each choice of branch cut, we get a corresponding anomaly inflow to the 7-brane defect. Doing so, we show that one choice of a branch cut descends to the Kramers- Wannier-like duality defects of \cite{Kaidi:2021xfk, Kaidi:2022cpf, Antinucci:2022vyk}, while another choice produces the half-space gauging construction of \cite{Choi:2021kmx, Choi:2022zal}.
One can also entertain ``hybrid'' configurations of branch cuts, and these also produce duality / triality defects. In an Appendix we also show how these considerations are compatible with dimensional reduction of topological terms present in the 8D worldvolume of the 7-branes. We emphasize that while these analyses also make use of the 5D symmetry TFT, our analysis singles out the role of codimension two objects (and their associated branch cuts) which descend from wrapped 7-branes. Indeed, this top down perspective allows us to unify different construction techniques.

In the field theory literature, the main examples of duality / triality defects have centered on $\mathcal{N} = 4$ SYM theory and closely related examples. In the present context where this QFT arises from D3-branes probing $\mathbb{C}^3$, we see that the main ingredients for duality / triality defects readily generalize to $\mathcal{N} = 1$ SCFTs as obtained from D3-branes probing $X$ a Calabi-Yau cone with a singularity. In that setting, the IIB duality group corresponds to a duality action which is present at a specific (tuned) subspace of the conformal manifold of the SCFT. In particular, dimensional reduction of 7-branes on $\partial X$ leads to precisely the same topological defects, and thus provides us with a generalization to QFTs with less supersymmetry. On the other hand, the full 5D symmetry TFT will in this case be more involved simply because the topology $\partial X$ can in general support more kinds of objects. For example, other 0-form symmetries are present in such systems, and crossing the associated local defects charged under these discrete 0-form symmetries  through a duality / triality wall leads to non-trivial transformation rules.

The rest of this chapter is organized as follows. In section \ref{sec:SETUP} we present our general setup involving probe
D3-branes in a Calabi-Yau threefold. In particular, we show how boundary conditions ``at infinity'' specify the global form of the theory, and how duality / triality defects arise from 7-branes wrapped on the boundary geometry. In section \ref{sec:DefectTFT} we consider the 5D symmetry TFT obtained from dimensional reduction on the boundary $\partial X$. The 7-branes descend to codimension two objects with branch cuts, and the choice of how to terminate these branch cuts leads to different implementations of duality / triality defects. After this, in section \ref{sec:N4} we show that our top down considerations are compatible with the bottom up analyses in the field theory literature. Section \ref{sec:N=1} shows how these considerations generalize to systems with minimal supersymmetry. We present our conclusions and some directions for future work in section \ref{sec:CONC3}. In Appendix \ref{app:other} we show how the various defects considered in the main body are implemented in other top down constructions. In Appendix \ref{app:minimalTFT7branes} we give a proposal for the relevant topological terms of a non-perturbative 7-brane which reduce to a suitable 3D TFT (after reduction on $\partial X$). Finally, in Appendix \ref{app:orbo} we give some further details on the special case of D3-branes probing $\mathbb{C}^{3} / \mathbb{Z}_3$.

\section{General Setup}\label{sec:SETUP}

We now present the general setup for implementing duality / triality interfaces and defects in the context of brane probes of singularities. The construction we present produces supersymmetric 4D quantum field theories $\mathfrak{T}^{(N)}_X$ realized as the world-volume theory of a stack of $N$ D3-branes probing a non-compact Calabi-Yau threefold $X$. 


The Calabi-Yau threefolds $X$ we are considering are of conical topology
\be 
X=\mathrm{Cone } \left( \partial X \right)
\ee 
with link $\partial X$, the asymptotic boundary of $X$. The topology of $\partial X$ therefore determines the topology of $X$ fully. The apex of the cone supports a real codimension six singularity. We introduce the radial coordinate $r\in \mathbb{R}_{\geq 0}$ so that the singularity sits at $r=0$ and the asymptotic boundary sits at $r=\infty$.

For example, $X=\mathbb{C}^3$ determines $\mathfrak{T}^{(N)}_X$ to be 4D $\mathcal{N}=4$ supersymmetric Yang-Mills theory. In cases of reduced holonomy, for example $X=\mathbb{C}^3/\Gamma$ with $\Gamma\subset SU(3)$, we preserve $\mathcal{N}=1$ supersymmetry. In all cases, $\mathfrak{T}^{(N)}_X$ is some quiver gauge theory, with quiver nodes specified by a basis of ``fractional branes'' which can be visualized at large volume (i.e., away from the orbifold point of moduli space) as a collection of D3-, D5- and D7-branes and their anti-brane counterparts wrapped on cycles in a resolution of $X$ \cite{Klebanov:1998hh, Uranga:1998vf,Aharony:1997ju}. Nodes are connected by oriented arrows which should be viewed as open strings stretching between the fractional branes. The gauge theory characterization is especially helpful at weak coupling, and serves to define the QFT in the first place. The quiver gauge theory comes with a collection of marginal couplings and we can consider tuning these parameters to ``strong coupling''. At such points in the conformal manifold, the gauge theory description is less useful, but we can still speak of the SCFT defined by the probe D3-branes.

In the quiver gauge theory, the IIB axio-dilaton descends to a particular choice of marginal couplings. Moreover, the celebrated $SL(2,\mathbb{Z})$ duality of IIB strings\footnote{The precise form of the duality group and its actions on fermions leads to some additional subtleties. For example, taking into account fermions, there is the metaplectic cover of $SL(2,\mathbb{Z})$ \cite{Pantev:2009de}, and taking into account reflections on the F-theory torus (associated with worldsheet orientation reversal and $(-1)^{F_L}$ parity, this enhances to the $\mathsf{Pin}^{+}$ cover of $GL(2,\mathbb{Z})$ \cite{Tachikawa:2018njr} (see also \cite{Debray:2021vob, Dierigl:2022reg}). These subtleties can appear if one carefully tracks the boson / fermion number of extended operators but in what follows we neglect this issue.} descends to a duality transformation at a specific point in the conformal manifold of the 4D SCFT \cite{Lawrence:1998ja, Kachru:1998ys} (for a recent discussion see, e.g., \cite{Garcia-Etxebarria:2016bpb}).

The other bulk supergravity fields of type IIB also play an important role in specifying the global structures of the field theory. Boundary conditions $P$ for such bulk fields at $\partial X$ determine an absolute theory $\mathfrak{T}^{(N)}_{X,P}$ from the relative theory $\mathfrak{T}^{(N)}_{X}$. In particular, such boundary conditions also determine the spectrum of extended objects ending at or contained within $\partial X$ which specify the defects and generalized symmetry operators of $\mathfrak{T}^{(N)}_{X,P}$ \cite{Gaiotto:2014kfa,GarciaEtxebarria:2019caf, Bhardwaj:2021mzl}.

For example, there is an $SL(2,\mathbb{Z})$ doublet $(C_2, B_2) = \mathcal{B}^{j}$ (RR and NS) of 2-form potentials which couple to D1- and F1-strings of the IIB theory respectively. Wrapping bound states of these objects compatible with $P$ along the radial direction in $X$ leads to heavy line defects of the 4D quiver gauge theory $\mathfrak{T}^{(N)}_{X,P}$. The spectrum of line defects then fixes the global form of the quiver gauge group.

\begin{figure}[t]
    \centering
    \scalebox{0.8}{
    \begin{tikzpicture}
	\begin{pgfonlayer}{nodelayer}
		\node [style=none] (0) at (-2, 2) {};
		\node [style=none] (1) at (-2, -2) {};
		\node [style=none] (2) at (-3, 2) {$r=\infty$};
		\node [style=none] (3) at (-3, -2) {$r=0$};
		\node [style=none] (10) at (0, 2.5) {$\partial X$};
		\node [style=none] (11) at (-3, -1.75) {};
		\node [style=none] (12) at (-3, 1.75) {};
		\node [style=none] (13) at (-3.75, 0) {$\mathbb{R}_{\geq0}$};
		\node [style=none] (15) at (2, 2) {};
		\node [style=none] (16) at (2, -2) {};
		\node [style=none] (17) at (-1.75, -2.5) {};
		\node [style=none] (18) at (1.75, -2.5) {};
		\node [style=none] (19) at (0, -3) {$\mathbb{R}_\perp$};
		\node [style=none] (20) at (3, -2) {$\ket{\mathfrak{T}^{(N)}_X}$};
		\node [style=none] (21) at (3, 2) {$\ket{P,D}$};
        \node [style=none] (22) at (0, 0) {\large $\frac{N}{2\pi} \int_{M_4\times \mathbb{R}_{\geq 0}}B_2\wedge dC_2$};
	\end{pgfonlayer}
	\begin{pgfonlayer}{edgelayer}
		\draw [style=ArrowLineRight] (11.center) to (12.center);
		\draw [style=ThickLine] (0.center) to (15.center);
		\draw [style=ThickLine] (1.center) to (16.center);
		\draw [style=ArrowLineRight] (17.center) to (18.center);
	\end{pgfonlayer}
\end{tikzpicture}}
    \caption{Sketch of the symmetry TFT \eqref{eq:5d TFT terms1}. We depict the half-plane $\mathbb{R}_{\geq 0}\times\mathbb{R}_\perp$ with coordinates $(r,x_\perp)$ where $\mathbb{R}_\perp$ is some direction parallel to the D3-brane worldvolume. The boundary conditions for the symmetry TFT are denoted $\ket{\mathfrak{T}^{(N)}_X},\ket{P,D}$ respectively. }
    \label{fig:SetUp0}
\end{figure}

The possible boundary conditions $P$ are determined by the symmetry TFT which follows by reduction of the 10D Chern-Simons term in IIB supergravity,\footnote{Strictly speaking, one also needs to utilize the self-dual condition for $F_5=dC_4$ in order to get the correct coefficient in \eqref{eq:5d TFT terms1}. 
For further discussion on this point, see e.g., \cite{Belov:2006jd, Belov:2006xj}.} much as in references \cite{Aharony:1998qu, Apruzzi:2021nmk} (see also \cite{Heckman:2017uxe}):
\begin{equation}
    S_{\mathrm{(CS)}}=-\frac{1}{4\kappa^2}\int_{M_4\times X}C_4\wedge dB_2\wedge dC_2.
\end{equation}
The stack of D3-branes source $N$ units of 5-form flux threading $\partial X$ and therefore the symmetry TFTs of all quiver gauge theories under consideration contain the universal term
\begin{equation}\label{eq:5d TFT terms1}
    S_{(\mathrm{SymTFT}),0}= -\frac{N}{4\pi}\int_{M_4\times \mathbb{R}_{\geq 0}} \epsilon_{ij} \mathcal{B}^{i} \cup d \mathcal{B}^{j} \,,
\end{equation}
where we have integrated over the link $\partial X$. Here we have introduce a manifestly $SL(2,\mathbb{Z})$ invariant presentation of the action using the 
two-index tensor $\epsilon_{ij}$ to raise and lower doublet indices. In our conventions, $\epsilon_{21}=-\epsilon_{12}=1$. In terms of the individual components of this $SL(2,\mathbb{Z})$ doublet, 
the equations of motion for the action \eqref{eq:5d TFT terms1} are
\begin{equation}\label{eq:ZN}
    N dB_2 = N dC_2 = 0,
\end{equation}
which constrains $B_2$ and $C_2$ to be $\mathbb{Z}_N$-valued 1-form symmetry background fields.\footnote{Note that these steps are identical to the derivation of a bulk topological term in $AdS_5$ \cite{Witten:1998wy}, while here the term lives along $M_4\times \mathbb{R}_{\geq 0}$.} In general, we will denote $\mathbb{Z}_N$-valued fields using the same notation as their $U(1)$ counterparts but are related by a rescaling. For example, in conventions where the NS-NS flux $\frac{1}{2\pi}H_3$ is integrally quantized we have\footnote{Another natural choice would be to take $\int_{Q_3}H_3\in \mathbb
{Z}$ for all 3-manfolds $Q_3$ in which case we would drop the factor of $2\pi$ on the RHS of \eqref{eq:rescaling}.}
\begin{equation}\label{eq:rescaling}
    B^{U(1)}_2=\frac{2\pi}{N}B^{\mathbb{Z}_N}_2
\end{equation}
where the holonomies $\int_{\Sigma_2}B^{\mathbb{Z}_N}_2=k \; \mathrm{mod}\; N$ for some Riemann surface $\Sigma$ and integer $k$. Notice that since \eqref{eq:rescaling} is only valid when the holonomies of the $U(1)$ field are $N^{\mathrm{th}}$ roots of unity, for a $\mathbb{Z}_N$-valued field one is free to take either the LHS or RHS as normalizations. We will drop the superscripts in the future making clear which convention we are using for discrete fields when it arises. For additional details on the structure of the defect group in this theory (via related top down constructions) see Appendix \ref{app:other}. 

The relative theory $\mathfrak{T}^{(N)}_X$ sets enriched Neumann boundary conditions at $r=0$ while at $r=\infty$ we have mixed Neumann-Dirichlet boundary conditions for the fields of the symmetry TFT.  These are respectively denoted as
\begin{equation}
      \ket{\mathfrak{T}^{(N)}_X}\,, \quad \ket{P,D}\,, \qquad \quad \braket{ P,D\,|\,\mathfrak{T}^{(N)}_X}=Z_{\mathfrak{T}^{(N)}_{X,P}}(D)
\end{equation}
and contract to give the partition function of the absolute theory $\mathfrak{T}^{(N)}_{X,P}$ with background fields determined by $P$ set to the values $D$. Here $D$ is a form profile and in particular does not carry $SL(2,\mathbb{Z})$ indices (see figure \ref{fig:SetUp0}).

Consider for example $X=\mathbb{C}^3$ in which case \eqref{eq:5d TFT terms1} describes the full symmetry TFT. First note that \eqref{eq:ZN} makes it clear that we are discussing a theory with gauge algebra $\mathfrak{su}(N)$ rather than $\mathfrak{u}(N)$. This $U(1)$ factor is lifted via a Stueckelberg mechanism.\footnote{Intuitively, when we pick an origin for $\mathbb{C}^3$ we put the whole system in a ``box'' with a conformal boundary. This removes the center of mass degree of freedom for the system.} A standard set of boundary conditions include a purely electric or purely magnetic polarization via the boundary conditions:
\begin{align}\label{eq:electricbc}
 \textnormal{$B_2|_{\partial X}$ Dirichlet, $C_2|_{\partial X}$ Neumann} \quad  &\longleftrightarrow \quad \textnormal{Global electric 1-form symmetry}\\ \label{eq:magneticbc} 
 \textnormal{$B_2|_{\partial X}$ Neumann, $C_2|_{\partial X}$ Dirichlet} \quad  &\longleftrightarrow \quad \textnormal{Global magnetic 1-form symmetry}.
\end{align}
Concretely, we are considering $\mathcal{N} = 4$ SYM theory with gauge algebra $\mathfrak{su}(N)$. The electric polarization produces gauge group $SU(N)$ while the magnetic polarization produces gauge group $PSU(N) = SU(N) / \mathbb{Z}_N$. Given electric/magnetic boundary conditions we can stretch F1/D1 strings between the D3-branes and the asymptotic boundary to construct line defects in the 4D worldvolume theory (see figure \ref{fig:F1D1}).

\begin{figure}[t]
    \centering
    \scalebox{0.8}{
\begin{tikzpicture}
	\begin{pgfonlayer}{nodelayer}
		\node [style=none] (0) at (-3, 2) {};
		\node [style=none] (1) at (8.75, 2) {};
		\node [style=none] (2) at (-3, -2) {};
		\node [style=none] (3) at (8.75, -2) {};
		\node [style=none] (4) at (-4, 2) {$r=\infty$};
		\node [style=none] (5) at (-4, -2) {$r=0$};
		\node [style=none] (6) at (-1.25, 0) {};
		\node [style=none] (7) at (-0.75, 0) {};
		\node [style=none] (8) at (-1.25, -2) {};
		\node [style=none] (9) at (-0.75, -2) {};
		\node [style=none] (10) at (-1.25, 2) {};
		\node [style=none] (11) at (-0.75, 2) {};
		\node [style=none] (12) at (6.5, 0) {};
		\node [style=none] (13) at (7, 0) {};
		\node [style=none] (14) at (7, -2) {};
		\node [style=none] (15) at (6.5, -2) {};
		\node [style=none] (16) at (6.5, 2) {};
		\node [style=none] (17) at (7, 2) {};
		\node [style=none] (18) at (-1, 2.5) {$\partial X$};
		\node [style=none] (20) at (-4, -1.625) {};
		\node [style=none] (21) at (-4, 1.625) {};
		\node [style=none] (22) at (-4.75, 0) {$\mathbb{R}_{\geq0}$};
		\node [style=none] (23) at (-2, 0) {F1};
		\node [style=none] (24) at (5.75, 0) {D1};
		\node [style=none] (25) at (1, 2) {};
		\node [style=none] (26) at (1, -2) {};
		\node [style=none] (27) at (4.75, 2) {};
		\node [style=none] (28) at (4.75, -2) {};
		\node [style=none] (29) at (6.75, 2.5) {$\partial X$};
		\node [style=none] (31) at (-2.75, -2.5) {};
		\node [style=none] (32) at (0.75, -2.5) {};
		\node [style=none] (33) at (-1, -3) {$\mathbb{R}_\perp$};
		\node [style=none] (34) at (5, -2.5) {};
		\node [style=none] (35) at (8.5, -2.5) {};
		\node [style=none] (36) at (6.75, -3) {$\mathbb{R}_\perp$};
		\node [style=none] (37) at (1.75, -2) {$\ket{\mathcal{T}^{(N)}_{X}}$};
		\node [style=none] (38) at (1.8675, 2) {$\ket{P_1,D_1}$};
		\node [style=none] (39) at (9.5, -2) {$\ket{\mathcal{T}^{(N)}_{X}}$};
		\node [style=none] (40) at (9.55, 2) {$\ket{P_2,D_2}$};
	\end{pgfonlayer}
	\begin{pgfonlayer}{edgelayer}
		\draw [style=ArrowLineRed] (8.center) to (6.center);
		\draw [style=ArrowLineRed] (11.center) to (7.center);
		\draw [style=RedLine] (10.center) to (8.center);
		\draw [style=RedLine] (11.center) to (9.center);
		\draw [style=ArrowLineBlue] (17.center) to (13.center);
		\draw [style=ArrowLineBlue] (15.center) to (12.center);
		\draw [style=BlueLine] (16.center) to (15.center);
		\draw [style=BlueLine] (17.center) to (14.center);
		\draw [style=ArrowLineRight] (20.center) to (21.center);
		\draw [style=ThickLine] (0.center) to (25.center);
		\draw [style=ThickLine] (2.center) to (26.center);
		\draw [style=ThickLine] (28.center) to (3.center);
		\draw [style=ThickLine] (27.center) to (1.center);
		\draw [style=ArrowLineRight] (31.center) to (32.center);
		\draw [style=ArrowLineRight] (34.center) to (35.center);
	\end{pgfonlayer}
\end{tikzpicture}
    }
    \caption{Boundary conditions and defects for $\mathfrak{T}_X^{(N)}$. We sketch the half-plane $\mathbb{R}_{\geq 0}\times \mathbb{R}_\perp$ parametrized by $(r,x_\perp)$. The polarization $P_1,P_2$ determine that Dirichlet boundary conditions are set for $B_2,C_2$ respectively $B_2|_{\partial X}=D_1$ and $C_2|_{\partial X}=D_2$. Line defects are realized by F1/D1-strings and correspond to Wilson and 't Hooft lines respectively. Our conventions are such that the left, radially outgoing strings are of charge $[0,-1]$ and $[-1,0]$ and the right, incoming strings are of charge $[0,1]$ and $[1,0]$ respectively.}
    \label{fig:F1D1}
\end{figure}

One can also consider more general mixed boundary conditions. In general, $SL(2,\mathbb{Z}_N)$ duality transformations
\begin{equation}\label{eq:Mono}
~\begin{bmatrix} C_2 \\B_2 \end{bmatrix}\rightarrow \begin{bmatrix} a&b\\c&d \end{bmatrix}\begin{bmatrix} C_2 \\B_2 \end{bmatrix}, \qquad \mathbb{S}=\begin{bmatrix} 0&1\\-1&0 \end{bmatrix}\,, \quad \mathbb{T}=\begin{bmatrix} 1&1\\0&1 \end{bmatrix}
\end{equation}
where $ad-bc=1$, map between boundary conditions and group these into orbits. In the case $\mathcal{N}=4$ SYM with gauge algebra $\mathfrak{su}(N)$ and when $N$ has no square divisors,\footnote{See \cite{Bergman:2022otk} for details when dropping this assumption.} one can generate all possible mixed boundary conditions for $B_2$ and $C_2$, and thus all forms of the gauge group.

More generally, given a quiver gauge theory, the individual gauge group factors are all correlated due to bifundamental states which are charged under different centers of the gauge group. Indeed, note also that we can always move the D3-brane away from the singularity (i.e., to finite $r > 0$), and the center of this gauge group in the infrared will need to be compatible with the boundary conditions specified at $r = \infty$.

Let us also record our $SL(2,\mathbb{Z})$ conventions for strings and 5-branes here. The charge vector $Q$ of a $(p,q)$-string or $(p,q)$-5-brane has components, with $\epsilon_{12}=\epsilon_{21}=-1$, following conventions laid out in \cite{Weigand:2018rez},
\begin{equation}
    Q_i=\epsilon_{ij}Q^j=[q,p]\,, \qquad Q^i=\begin{bmatrix}
        p \\ -q
    \end{bmatrix}\,,
\end{equation}
in particular S-duality maps Wilson lines $W$ and 't Hooft lines $H$ on the D3-brane worldvolume as $(W,H)\rightarrow (H,-W)$.

We further lay out our conventions for symmetry TFTs following \cite{Kaidi:2022uux,Kaidi:2022cpf}. The enriched Neumann boundary condition at $r=0$ is expanded as
\begin{equation}
    \ket{\mathfrak{T}_X^{(N)}}=\sum_{\mathbf{a}\in P} Z_{\mathfrak{T}_{X,P}^{(N)}}(a)\ket{\mathbf{a}}
\end{equation}
where $Z_{\mathfrak{T}_{P,X}^{(N)}}(a)$ is the partition function of the absolute theory derived from $\mathfrak{T}_X^{(N)}$ by choice of polarization $P$ and $a$ is a background field profile for the corresponding higher symmetry. In this chapter we are mainly concerned with 1-form symmetries of gauge theories and here $P$ fixes the global form of the gauge group and $a$ is a 1-form symmetry background field. Further we denote a background field configuration by a vector $\mathbf{a}$ which is oriented in the corresponding defect group and has a form profile of $a$. Topological Dirichlet and Neumann boundary conditions at $r=\infty$ in 4D are respectively
\begin{equation}
\begin{aligned}    \ket{P,D}_{\mathrm{Dirichlet}}&=\sum_{\mathbf{a}\in P}\delta(D-a)\ket{\mathbf{a}} \\  \ket{P,E}_{\mathrm{Neumann}}&=\sum_{\mathbf{a}\in P}\exp\left(\frac{2\pi i}{N} \int E\cup a\right)\ket{\mathbf{a}}
    \end{aligned}
\end{equation}
where we have normalized fields to take values in $\mathbb{Z}_N$. We will mainly work with Dirichlet boundary conditions throughout and omit the index `Dirichlet' when it causes no confusion. Whenever two polarizations $P,P'$ are related by a discrete Fourier transform or equivalently by gauging we have the pairing
\begin{equation}
    \braket{\mathbf{a}|\mathbf{b}}=\exp\left(\frac{2\pi i}{N}\int a \cup b \right)\qquad \forall\, \mathbf{a}\in P,\mathbf{b}\in P'\,.
\end{equation}
More generally, boundary conditions in 4D can be stacked with counterterms, so we define 
\begin{equation}
\begin{aligned}    \ket{P_{G_r},D}_{\mathrm{Dirichlet}}&=\sum_{\mathbf{a}\in P_{G_k}}\delta(D-a)\exp\left(\frac{2\pi ir}{N}\int \frac{\mathcal{P}(D)}{2}\right)\ket{\mathbf{a}} 
    \end{aligned}
\end{equation}
where $\mathcal{P}$ is the Pontryagin square. Here we have labelled a polarization $P$ by the global form of the gauge group $G$ it realizes, and the subscript $r$ counts the number of stacked counterterms. For example $SU(2)_r$ denotes $SU(2)$ theory stacked with $r$ couterterms.

\subsection{Proposal for Topological Duality Interfaces/Operators}
\label{sec:Proposal}

{\renewcommand{\arraystretch}{1.35}
\begin{table}[t]
    \centering
    \begin{center}
\begin{tabular}{||c | c | c | c | c | c ||}
 \hline
 Fiber Type $\mathfrak{F}$ & Lines & Monodromy $ \rho $ & Refined Linking $p/2k$ & $(k,m)$ & $\tau$ \\ [0.5ex] 
 \hline\hline
 $II,\,\mathfrak{su}(1)$ & $-$ & {\footnotesize $\left(\begin{array}{cc}
     0 &  1 \\
     -1  & 1
 \end{array}\right)$ } & $-$ & $-$ & $e^{i\pi/3}$\\ 
 \hline
 $III,\,\mathfrak{su}(2)$ & $\mathbb{Z}_2$ & {\footnotesize $\left(\begin{array}{cc}
     0 &  1 \\
     -1  & 0
 \end{array}\right)$ } &  $\frac{3}{4}$ & $(2,3)$ & $e^{i\pi/2}$ \\
 \hline
 $IV,\,\mathfrak{su}(3)$ & $\mathbb{Z}_3$ & {\footnotesize $\left(\begin{array}{cc}
     0 &  1 \\
     -1  & -1
 \end{array}\right)$ } &    $\frac{4}{6}$ & $(3,4)$ &$e^{i\pi/3}$ \\
 \hline
 $I_{0}^{\ast},\,\mathfrak{so}(8)$ & $\mathbb{Z}_2\oplus \mathbb{Z}_2$ &  {\footnotesize $\left(\begin{array}{cc}
     -1 &  0 \\
     0  & -1
 \end{array}\right)$ } & {\footnotesize $\left(\begin{array}{cc}
     2/4 &  3/4 \\
     3/4  & 2/4
 \end{array}\right)$ } & $-$  & $\tau$ \\
 \hline
 $IV^{\ast},\,\mathfrak{e}_6$ & $\mathbb{Z}_3$ & {\footnotesize $\left(\begin{array}{cc}
     -1 &  -1 \\
     1  & 0
 \end{array}\right)$ } &  $\frac{2}{6}$ & $(3,2)$ &$e^{i\pi/3}$ \\
 \hline
 $III^{\ast},\,\mathfrak{e}_7$ & $\mathbb{Z}_2$ & {\footnotesize $\left(\begin{array}{cc}
     0 &  -1 \\
     1  & 0
 \end{array}\right)$ } &   $\frac{1}{4}$ & $(2,1)$ &$e^{i\pi/2}$ \\
 \hline
 $II^{\ast},\,\mathfrak{e}_8$ & $-$ &  {\footnotesize $\left(\begin{array}{cc}
     1 &  -1 \\
     1  & 0
 \end{array}\right)$ } & $-$ & $-$ & $e^{i\pi/3}$ \\
  \hline
\end{tabular}
\end{center}
    \caption{Elliptic data of 7-brane profiles with constant axio-dilaton $\tau$. Their group of lines is isomorphic to $\textnormal{coker}(\rho-1)$ which is isomorphic to $\mathbb{Z}_k$ except for fiber type $I_0^*,II,II^*$. The label $m$ of these lines is determined from the refined self-linking numbers $m/2k$ which gives the spin of non-trivial lines. The refined self-linking numbers compute via the Gordon-Litherland approach laid out in \cite{Apruzzi:2021nmk} employing the divisors (Kodaira thimbles) computed in \cite{Hubner:2022kxr} or alternatively via the quadratic refinement laid out in \cite{Gukov:2020btk} and the linking number computations in \cite{Cvetic:2021sxm}. Note in particular that in all cases $\textnormal{gcd}(k,m)=1$ and $mk\in 2\mathbb{Z}$.}
    \label{tab:Fibs}
\end{table}}

Consider the spacetime $M_4=M_3\times \mathbb{R}_\perp$ with $\mathbb{R}_\perp$ parametrized by the coordinate $x_\perp$. We now argue that 7-branes wrapped on $M_3\times\partial X$ at some point $\bar x_{\perp}\in \mathbb{R}_\perp$ realize topological duality/triality interfaces and operators. A subset of our constructions work only for 7-branes with constant axio-dilaton profile and we list these in table \ref{tab:Fibs} together with their topological data. 

\begin{figure}[t]
    \centering
    \scalebox{0.8}{\begin{tikzpicture}
	\begin{pgfonlayer}{nodelayer}
		\node [style=none] (0) at (-6, 1.5) {};
		\node [style=none] (1) at (-1, 1.5) {};
		\node [style=none] (2) at (-6, -1.5) {};
		\node [style=none] (3) at (-1, -1.5) {};
		\node [style=none] (4) at (0, 0) {$=$};
		\node [style=none] (5) at (1, 1.5) {};
		\node [style=none] (6) at (6, 1.5) {};
		\node [style=none] (7) at (1, -1.5) {};
		\node [style=none] (8) at (6, -1.5) {};
		\node [style=Star] (9) at (-3.5, 1.5) {};
		\node [style=Star] (10) at (3.5, 0) {};
		\node [style=none] (11) at (1, 0) {};
		\node [style=none] (12) at (2.25, -0.5) {Branch cut};
		\node [style=none] (13) at (3.5, 0.5) {7-branes};
		\node [style=none] (14) at (-3.5, 1) {7-branes};
		\node [style=none] (15) at (-5.25, 2) {$\ket{P_1,D_1}$};
		\node [style=none] (16) at (-1.75, 2) {$\ket{P_2,D_2}$};
		\node [style=none] (17) at (1.75, 2) {$\ket{P_2,D_2}$};
		\node [style=none] (18) at (5.25, 2) {$\ket{P_2,D_2}$};
		\node [style=none] (19) at (-7, -1.5) {$r=0$};
		\node [style=none] (20) at (-7, 1.5) {$r=\infty$};
		\node [style=none] (21) at (-7, 1) {};
		\node [style=none] (22) at (-7, -1) {};
		\node [style=none] (23) at (-5, -2) {};
		\node [style=none] (24) at (-2, -2) {};
		\node [style=none] (25) at (-3.5, -2.5) {$x_\perp$};
		\node [style=none] (26) at (2, -2) {};
		\node [style=none] (27) at (5, -2) {};
		\node [style=none] (28) at (3.5, -2.5) {$x_\perp$};
        \node [style=none] (29) at (1.5, 0.45) {$\mathbb{H}_{\leftarrow}$};
	\end{pgfonlayer}
	\begin{pgfonlayer}{edgelayer}
		\draw [style=ThickLine] (0.center) to (1.center);
		\draw [style=ThickLine] (3.center) to (2.center);
		\draw [style=ThickLine] (5.center) to (6.center);
		\draw [style=ThickLine] (8.center) to (7.center);
		\draw [style=DottedLine] (11.center) to (10);
		\draw [style=ArrowLineRight] (22.center) to (21.center);
		\draw [style=ArrowLineRight] (23.center) to (24.center);
		\draw [style=ArrowLineRight] (26.center) to (27.center);
	\end{pgfonlayer}
\end{tikzpicture}
}
    \caption{Case\,(1), 7-branes wrapped on $M_3\times \partial X$, we sketch the plane $\mathbb{R}_{\geq0}\times \mathbb{R}_\perp$. The topological boundary conditions $\ket{P_1,D_1}$ are the monodromy transform of the boundary conditions $\ket{P_2,D_2}$ and result from stacking the branch cut with the asymptotic boundary. The branch cut is supported on $\mathbb{H}_{\leftarrow}\times \partial X$ and runs parallel to the D3-branes. Conventions are such that the monodromy matrix $\rho$ acts crossing the branch cut top to bottom. }
    \label{fig:7BraneInfinity2}
\end{figure}

\begin{figure}[t]
    \centering
    \scalebox{0.8}{
    \begin{tikzpicture}
	\begin{pgfonlayer}{nodelayer}
		\node [style=none] (0) at (-6, 1.5) {};
		\node [style=none] (1) at (-1, 1.5) {};
		\node [style=none] (2) at (-6, -1.5) {};
		\node [style=none] (3) at (-1, -1.5) {};
		\node [style=none] (4) at (0, 0) {$=$};
		\node [style=none] (5) at (1, 1.5) {};
		\node [style=none] (6) at (6, 1.5) {};
		\node [style=none] (7) at (1, -1.5) {};
		\node [style=none] (8) at (6, -1.5) {};
		\node [style=Star] (9) at (-3.5, 1.5) {};
		\node [style=Star] (10) at (3.5, 0) {};
		\node [style=none] (11) at (3.5, -1.5) {};
		\node [style=none] (12) at (4.125, -0.75) {$\mathbb{H}_{\downarrow}$};
		\node [style=none] (13) at (3.5, 0.5) {7-brane};
		\node [style=none] (14) at (-3.5, 2) {7-brane};
		\node [style=none] (15) at (-5.5, 2) {$\ket{P_1,D_1}$};
		\node [style=none] (16) at (-1.5, 2) {$\ket{P_1,D_1}$};
		\node [style=none] (18) at (3.5, 2) {$\ket{P_1,D_1}$};
		\node [style=none] (19) at (-7, -1.5) {$r=0$};
		\node [style=none] (20) at (-7, 1.5) {$r=\infty$};
		\node [style=none] (21) at (-7, 1) {};
		\node [style=none] (22) at (-7, -1) {};
		\node [style=none] (23) at (-5, -2) {};
		\node [style=none] (24) at (-2, -2) {};
		\node [style=none] (25) at (-3.5, -2.5) {$x_\perp$};
		\node [style=none] (26) at (2, -2) {};
		\node [style=none] (27) at (5, -2) {};
		\node [style=none] (28) at (3.5, -2.5) {$x_\perp$};
		\node [style=Circle] (30) at (-3.5, -1.5) {};
		\node [style=Circle] (31) at (3.5, -1.5) {};
		\node [style=none] (32) at (-4.125, -1) {$M_3'$};
		\node [style=none] (33) at (2.875, -1) {$M_3'$};
        \node [style=none] (34) at (-2.875, 0) {$\mathbb{H}_{\downarrow}$};
	\end{pgfonlayer}
	\begin{pgfonlayer}{edgelayer}
		\draw [style=ThickLine] (0.center) to (1.center);
		\draw [style=ThickLine] (3.center) to (2.center);
		\draw [style=ThickLine] (5.center) to (6.center);
		\draw [style=ThickLine] (8.center) to (7.center);
		\draw [style=DottedLine] (11.center) to (10);
		\draw [style=ArrowLineRight] (22.center) to (21.center);
		\draw [style=ArrowLineRight] (23.center) to (24.center);
		\draw [style=ArrowLineRight] (26.center) to (27.center);
		\draw [style=DottedLine] (9) to (30);
	\end{pgfonlayer}
\end{tikzpicture}
    }
    \caption{Case\,(2), 7-branes wrapped on $M_3\times \partial X$, we sketch the plane $\mathbb{R}_{\geq0}\times \mathbb{R}_\perp$. There is a single set of boundary conditions $\ket{P_1,D_1}$. The branch cut is supported on $\mathbb{H}_{\downarrow}\times \partial X$ and runs perpendicular to the D3-branes. Conventions are such that the monodromy matrix $\rho$ acts crossing the branch cut left to right.}
    \label{fig:Case2}
\end{figure}

\begin{figure}
    \centering
    \scalebox{0.8}{
    \begin{tikzpicture}
	\begin{pgfonlayer}{nodelayer}
		\node [style=none] (0) at (-6, 1.5) {};
		\node [style=none] (1) at (-1, 1.5) {};
		\node [style=none] (2) at (-6, -1.5) {};
		\node [style=none] (3) at (-1, -1.5) {};
		\node [style=none] (4) at (0, 0) {$=$};
		\node [style=none] (5) at (1, 1.5) {};
		\node [style=none] (6) at (6, 1.5) {};
		\node [style=none] (7) at (1, -1.5) {};
		\node [style=none] (8) at (6, -1.5) {};
		\node [style=Star] (9) at (-3.5, 1.5) {};
		\node [style=Star] (10) at (3.5, 0) {};
		\node [style=none] (11) at (3.5, 1.5) {};
		\node [style=none] (12) at (4, 0.75) {$\mathbb{H}_{\uparrow}$};
		\node [style=none] (13) at (3.5, -0.5) {7-branes};
		\node [style=none] (14) at (-3.5, 2) {$M_3''$};
		\node [style=none] (15) at (-5.25, 2) {$\ket{P_1,D_1}$};
		\node [style=none] (16) at (-1.75, 2) {$\ket{P_2,D_2}$};
		\node [style=none] (18) at (-7, -1.5) {$r=0$};
		\node [style=none] (19) at (-7, 1.5) {$r=\infty$};
		\node [style=none] (20) at (-7, 1) {};
		\node [style=none] (21) at (-7, -1) {};
		\node [style=none] (22) at (-5, -2) {};
		\node [style=none] (23) at (-2, -2) {};
		\node [style=none] (24) at (-3.5, -2.5) {$x_\perp$};
		\node [style=none] (25) at (2, -2) {};
		\node [style=none] (26) at (5, -2) {};
		\node [style=none] (27) at (3.5, -2.5) {$x_\perp$};
		\node [style=Circle] (29) at (3.5, 1.5) {};
		\node [style=none] (30) at (-3.5, 1) {7-branes};
		\node [style=none] (31) at (3.5, 2) {$M_3''$};
		\node [style=none] (34) at (1.75, 2) {$\ket{P_1,D_1}$};
		\node [style=none] (35) at (5.25, 2) {$\ket{P_2,D_2}$};
	\end{pgfonlayer}
	\begin{pgfonlayer}{edgelayer}
		\draw [style=ThickLine] (0.center) to (1.center);
		\draw [style=ThickLine] (3.center) to (2.center);
		\draw [style=ThickLine] (5.center) to (6.center);
		\draw [style=ThickLine] (8.center) to (7.center);
		\draw [style=DottedLine] (11.center) to (10);
		\draw [style=ArrowLineRight] (21.center) to (20.center);
		\draw [style=ArrowLineRight] (22.center) to (23.center);
		\draw [style=ArrowLineRight] (25.center) to (26.center);
	\end{pgfonlayer}
\end{tikzpicture}
    }
    \caption{Case\,(3), 7-branes wrapped on $M_3\times \partial X$, we sketch the plane $\mathbb{R}_{\geq0}\times \mathbb{R}_\perp$. The 7-brane insertion gives rise to two boundary conditions $\ket{P_1,D_1},\ket{P_2,D_2}$. The branch cut is supported on $\mathbb{H}_{\uparrow} \times \partial X$ and runs perpendicular to the D3-branes. Conventions are such that the monodromy matrix $\rho$ acts crossing the branch cut right to left.}
    \label{fig:Case3}
\end{figure}

\begin{figure}
    \centering
    \scalebox{0.8}{\begin{tikzpicture}
	\begin{pgfonlayer}{nodelayer}
		\node [style=none] (0) at (-6, 1.5) {};
		\node [style=none] (1) at (-1, 1.5) {};
		\node [style=none] (2) at (-6, -1.5) {};
		\node [style=none] (3) at (-1, -1.5) {};
		\node [style=none] (4) at (0, 0) {$=$};
		\node [style=none] (5) at (1, 1.5) {};
		\node [style=none] (6) at (6, 1.5) {};
		\node [style=none] (7) at (1, -1.5) {};
		\node [style=none] (8) at (6, -1.5) {};
		\node [style=Star] (9) at (-3.5, 1.5) {};
		\node [style=Star] (10) at (3.5, 0) {};
		\node [style=none] (11) at (3.5, 1.5) {};
		\node [style=none] (13) at (4.5, 0.5) {7-branes};
		\node [style=none] (14) at (-3.5, 2) {7-branes};
		\node [style=none] (15) at (-5.25, 2) {$\ket{P_1,D_1}$};
		\node [style=none] (16) at (-1.75, 2) {$\ket{P_2,D_2}$};
		\node [style=none] (17) at (-7, -1.5) {$r=0$};
		\node [style=none] (18) at (-7, 1.5) {$r=\infty$};
		\node [style=none] (19) at (-7, 1) {};
		\node [style=none] (20) at (-7, -1) {};
		\node [style=none] (21) at (-5, -2.5) {};
		\node [style=none] (22) at (-2, -2.5) {};
		\node [style=none] (23) at (-3.5, -3) {$x_\perp$};
		\node [style=none] (24) at (2, -2.5) {};
		\node [style=none] (25) at (5, -2.5) {};
		\node [style=none] (26) at (3.5, -3) {$x_\perp$};
		\node [style=none] (28) at (-4, 1) {$M_3'$};
		\node [style=none] (29) at (3.5, 2) {$M_3''$};
		\node [style=none] (30) at (0, 2.5) {};
		\node [style=none] (31) at (1.75, 2) {$\ket{P_3,D_3}$};
		\node [style=none] (32) at (5.25, 2) {$\ket{P_4,D_4}$};
		\node [style=Circle] (33) at (3.5, 1.5) {};
		\node [style=Circle] (34) at (-3.5, -1.5) {};
		\node [style=none] (35) at (1, 0) {};
		\node [style=none] (36) at (3.5, -1.5) {};
		\node [style=none] (37) at (6, 0) {};
		\node [style=none] (38) at (3, -1) {$\rho_{\downarrow}$};
		\node [style=none] (39) at (5.5, -0.5) {$\rho_{\rightarrow}$};
		\node [style=none] (40) at (3, 1) {$\rho_{\uparrow}$};
		\node [style=none] (41) at (3.5, -2) {$M_3'$};
		\node [style=Circle] (42) at (3.5, -1.5) {};
		\node [style=none] (43) at (1.5, -0.5) {$\rho_{\leftarrow}$};
	\end{pgfonlayer}
	\begin{pgfonlayer}{edgelayer}
		\draw [style=ThickLine] (0.center) to (1.center);
		\draw [style=ThickLine] (3.center) to (2.center);
		\draw [style=ThickLine] (5.center) to (6.center);
		\draw [style=ThickLine] (8.center) to (7.center);
		\draw [style=DottedLine] (11.center) to (10);
		\draw [style=ArrowLineRight] (20.center) to (19.center);
		\draw [style=ArrowLineRight] (21.center) to (22.center);
		\draw [style=ArrowLineRight] (24.center) to (25.center);
		\draw [style=DottedLine] (9) to (34);
		\draw [style=DottedLine] (35.center) to (10);
		\draw [style=DottedLine] (10) to (37.center);
		\draw [style=DottedLine] (10) to (36.center);
	\end{pgfonlayer}
\end{tikzpicture}}
    \caption{Case\,(4) hybrid case of cases 1,2,3, the 7-branes wrapped on $M_3\times \partial X$, we sketch the plane $\mathbb{R}_{\geq0}\times \mathbb{R}_\perp$. The overall monodromy is $\rho=\rho_{\leftarrow}\rho_{\uparrow}\rho_{\rightarrow}\rho_{\downarrow}$. Each monodromy factor $\rho_{\bullet}$ has its own branch cut separately. The branch cuts labelled $\rho_{\downarrow},\rho_{\uparrow}$ intersect the D3-brane worldvolume and asymptotic boundary in $M_3',M_3''$, respectively.}
    \label{fig:Case4}
\end{figure}

\begin{figure}
    \centering
    \scalebox{0.8}{\begin{tikzpicture}
	\begin{pgfonlayer}{nodelayer}
		\node [style=none] (0) at (-2.5, 1.5) {};
		\node [style=none] (1) at (2.5, 1.5) {};
		\node [style=none] (2) at (-2.5, -1.5) {};
		\node [style=none] (3) at (2.5, -1.5) {};
		\node [style=none] (6) at (-1.75, 2) {$\ket{P_1,D_1}$};
		\node [style=none] (7) at (1.75, 2) {$\ket{P_2,D_2}$};
		\node [style=none] (8) at (-3.5, -1.5) {$r=0$};
		\node [style=none] (9) at (-3.5, 1.5) {$r=\infty$};
		\node [style=none] (10) at (-3.5, 1) {};
		\node [style=none] (11) at (-3.5, -1) {};
		\node [style=none] (12) at (-1.5, -2.5) {};
		\node [style=none] (13) at (1.5, -2.5) {};
		\node [style=none] (14) at (0, -3) {$x_\perp$};
		\node [style=none] (15) at (-0.5, 1) {$M_3''$};
		\node [style=none] (17) at (-0.5, -1) {$M_3'$};
		\node [style=Circle] (18) at (0, 1.5) {};
		\node [style=Circle] (19) at (0, -1.5) {};
	\end{pgfonlayer}
	\begin{pgfonlayer}{edgelayer}
		\draw [style=ThickLine] (0.center) to (1.center);
		\draw [style=ThickLine] (3.center) to (2.center);
		\draw [style=ArrowLineRight] (11.center) to (10.center);
		\draw [style=ArrowLineRight] (12.center) to (13.center);
		\draw [style=DottedLine] (18) to (19);
	\end{pgfonlayer}
\end{tikzpicture}}
    \caption{S-duality is realized by cutting the symmetry TFT along the dotted line and gluing the pieces according to the desired S-duality transformation. No 7-branes are inserted. }
    \label{fig:Sduality}
\end{figure}

First consider a 7-brane wrapped at $r=\infty$ following the prescription in \cite{Heckman:2022muc}. This gives rise to a topological interface / symmetry operator in the 4D theory as the 7-brane is formally at infinite distance wrapped on a cycle of infinite volume. This decouples the non-topological interactions between the D3 and 7-brane and the non-topological degrees of freedom on the 7-brane worldvolume respectively \cite{Heckman:2022muc}. In this topological limit, for the 5D bulk theory, we can consider adding codimension two defects, i.e., the remnants of these 7-branes at infinity. Since everything is now treated as topological, we are free to insert these 7-branes anywhere in the interior, and as such we have different choices for where to extend the branch cut of this defect. We consider four distinct choices for the $SL(2,\mathbb{Z})$ monodromy branch cut:
\begin{enumerate}
    \item The branch cut is supported on $\mathbb{H}_{\leftarrow}\times \partial X$ with $\partial \mathbb{H}_{\leftarrow} =M_3$ and is oriented along $x_\perp$ parallel to the D3-branes with $x_\perp < \bar x_\perp$ along the asymptotic boundary. See the left subfigure in figure \ref{fig:7BraneInfinity2}. The branch cut ends at infinity. 
    \item The branch cut is supported on $\mathbb{H}_{\downarrow}\times \partial X$ with $\partial \mathbb{H}_{\downarrow}=M_3-M_3'$ (treated as a 3-chain) and is oriented radially inwards along $x_\perp=\bar x_\perp$ perpendicular to the D3-branes. See the left subfigure in figure \ref{fig:Case2}. The branch cut ends on $M_3'$, which is a subset of the D3-brane worldvolume, and supports an operator $\mathcal{D}(M_3')$ possibly coupled to background fields. While cuts can normally only begin / end on 7-branes, this can be made precise via a method of images procedure. The 7-brane has constant axio-dilaton profile.
    \item The branch cut is supported on $\mathbb{H}_{\uparrow}\times \partial X$ with $\partial \mathbb{H}_{\uparrow}= M_3- M_3''$ (treated as a 3-chain) and is oriented radially outwards along $x_\perp=\bar x_\perp$ perpendicular to the D3-branes (see figure \ref{fig:Case3}). The branch cut ends on $M_3''$ which is contained in the asymptotic boundary. The 7-brane has constant axio-dilaton profile.
    \item Whenever the 7-brane monodromy matrix is not prime over $\mathbb{Z}_N$, i.e., it can be factored into more than one non-trivial factor in $SL(2,\mathbb{Z}_N)$, we can consider separate branch cuts for each factor. We can then
    consider hybrid configurations of cases 1,2,3 with each branch cut realizing one of the previous setups (see figure \ref{fig:Case4}). The 7-brane has constant axio-dilaton profile.
\end{enumerate}
The cases differ in the structure of the boundary conditions and cases 2,3,4 include additional operators supported on $M_3',M_3''$ absorbing the branch cut.\footnote{Branch cuts are not physical, but their endpoints are. The branch cut arises from a choice of gauge for the background connection $A_1$ of the $SL(2,\mathbb{Z})$ duality bundle as we explain momentarily. This localizes an anomaly flow which we can either absorb at finite distance (case 2,3) or divert to infinite distance (case 1).} In this picture, S-duality is realized by a vertically running branch cut without 7-brane insertions. (see figure \ref{fig:Sduality}). We discuss each case in turn. Of course we can also consider multiple 7-brane insertions.

First consider the setup for case 1. The branch cut is stacked with part of the asymptotic boundary, let us therefore move the topological 7-brane into the bulk. See the right subfigure in figure \ref{fig:7BraneInfinity2}. In the symmetry TFT formalism these two configurations are topologically equivalent. There is now a single 5D boundary condition $\ket{P_2,D_2}$ along the asymptotic boundary at $r=\infty$. Here $P_2$ denotes the choice of polarization, i.e., which combination of $B_2,C_2$ have Dirichlet boundary conditions imposed and $D_2$ denotes the boundary value for the condition, i.e., $(p_iB_2+q_iC_2)|_{r=\infty}=D_2$ for some collection of integer pairs $(p_i,q_i)$ specified by $P_2$.

Let us denote the monodromy matrix of the 7-brane by $\rho$. Crossing the branch cut, a string labelled by charges $[q,p]$ is transformed to a string with charges $[q,p]\rho$. Given a single topological boundary condition $\ket{P_2,D_2}$ which permits strings of charge $[q,p]$ to end on the boundary, it now follows that in the half-spaces $\bar x_\perp< x_\perp$ and $\bar x_\perp> x_\perp$ of the D3-brane worldvolume we have line defects whose charges are multiples of $[q,p]$ and $[q,p]\rho$, respectively (see figure \ref{fig:halfspacegauging2}).  Stacking the branch cut as shown in figure \ref{fig:7BraneInfinity2} (right to left) now clearly alters the boundary conditions by a monodromy transformation, more precisely its mod $N$ reduction $\rho\in SL(2,\mathbb{Z}_N)$. This change in boundary condition is a topological manipulation at the boundary, such as half-space gauging or the stacking of a counterterm
along $x_\perp<\bar x_\perp$ in $M_4$ and we propose:
\begin{equation}
    \textnormal{Topological Duality Interface $\mathcal{I}(M_3,\mathfrak{F})$}~~ \longleftrightarrow  
    \begin{array}{c}
    \textnormal{7-brane of Type $\mathfrak{F}$ on} \\ \textnormal{ $M_3\times \partial X$ with cut $\mathbb{H}_{\leftarrow}\times \partial X$} 
    \end{array}
\end{equation}
The interface $\mathcal{I}(M_3,\mathfrak{F})$ promotes to a topological defect operator $\mathcal{U}(M_3,\mathfrak{F})$ when the theories separated by the defect are dual. Due to the branch cut not intersecting the D3-brane stack we can employ arbitrary 7-branes in this construction.

\begin{figure}[t]
    \centering
    \scalebox{0.8}{\begin{tikzpicture}
	\begin{pgfonlayer}{nodelayer}
		\node [style=none] (0) at (-2.5, 1.5) {};
		\node [style=none] (1) at (2.5, 1.5) {};
		\node [style=none] (2) at (-2.5, -1.5) {};
		\node [style=none] (3) at (2.5, -1.5) {};
		\node [style=Star] (4) at (0, 0) {};
		\node [style=none] (5) at (-2.5, 0) {};
		\node [style=none] (7) at (0, 0.5) {7-branes};
		\node [style=none] (8) at (0, 2) {$\ket{P,D}$};
		\node [style=none] (10) at (-1.5, -2) {};
		\node [style=none] (11) at (1.5, -2) {};
		\node [style=none] (12) at (0, -2.5) {$x_\perp$};
		\node [style=none] (13) at (1, 1.5) {};
		\node [style=none] (14) at (1, -1.5) {};
		\node [style=none] (15) at (-1, 1.5) {};
		\node [style=none] (16) at (-1, -1.5) {};
		\node [style=none] (17) at (-1, 0) {};
		\node [style=none] (18) at (2, 0) {$[q,p]$};
		\node [style=none] (19) at (-1.75, 0.75) {$[q,p]$};
		\node [style=none] (20) at (-1.75, -0.75) {$[q,p]\rho$};
		\node [style=none] (21) at (-3.5, -1.5) {$r=0$};
		\node [style=none] (22) at (-3.5, 1.5) {$r=\infty$};
		\node [style=none] (23) at (-3.5, 1) {};
		\node [style=none] (24) at (-3.5, -1) {};
	\end{pgfonlayer}
	\begin{pgfonlayer}{edgelayer}
		\draw [style=ThickLine] (0.center) to (1.center);
		\draw [style=ThickLine] (3.center) to (2.center);
		\draw [style=DottedLine] (5.center) to (4);
		\draw [style=ArrowLineRight] (10.center) to (11.center);
		\draw [style=RedLine] (13.center) to (14.center);
		\draw [style=RedLine] (15.center) to (17.center);
		\draw [style=BlueLine] (17.center) to (16.center);
		\draw [style=ArrowLineRight] (24.center) to (23.center);
	\end{pgfonlayer}
\end{tikzpicture}
}
    \caption{Half-space gauging via 7-brane insertion. Given a boundary condition $\ket{P,D}$ such that strings of type $[q,p]$ can terminate on $\partial X$ we find the admissible line defects on the D3-branes (located at $r=0$) to differ between the left and right hand side by a monodromy transformation. The D3-brane stack therefore experiences an effective change of polarization which amounts to a half-space gauging.}
    \label{fig:halfspacegauging2}
\end{figure} 

Let us now consider the setup for case 2. As above, we can move the topological 7-brane into the bulk giving a topologically equivalent configuration, see figure \ref{fig:Case2} left to right. There is a single 5D boundary condition $\ket{P_1,D_1}$. Defects running between the D3 stack and the topological boundary do not cross the branch cut and the effective polarization on the brane is identical along $\bar x_\perp < x_\perp$ and $\bar x_\perp > x_\perp$. The branch cut intersecting the D3 worldvolume gives rise to a 0-form operator realizing an S-duality transformation on local\footnote{Standard S-duality acts on both local and non-local operators and is realized as in figure \ref{fig:Sduality} by a completely vertical branch cut. Such a branch cut cannot be deformed into the 5D bulk and connects two distinct boundary conditions $\ket{P_i,D_i}$ with $i=1,2$.} operators in 4D. Therefore, upon collapsing the 5D slab to 4D we obtain for all choices of 7-branes with constant axio-dilaton profile a topological defect operator in a given theory. We propose: 
\begin{equation}
    \textnormal{Topological Defect Operator $\mathcal{V}(M_3,\mathfrak{F})$}~~ \longleftrightarrow  
    \begin{array}{c}
    \textnormal{7-brane of Type $\mathfrak{F}$ on} \\ \textnormal{ $M_3\times \partial X$ with cut $\mathbb{H}_{\downarrow}\times \partial X$} 
    \end{array}
\end{equation}


Let us now consider the setup for case 3. Again we can move the topological 7-brane into the bulk giving a topologically equivalent configuration, see figure \ref{fig:Case3} left to right. There are now two boundary conditions $\ket{P_1,D_1},\ket{P_2,D_2}$ giving rise to distinct polarizations in $\bar{x}_\perp < x_\perp$ and $\bar{x}_\perp > x_\perp$. This change in polarization is realized by a codimension one operator in the asymptotic boundary. This operator acts on symmetry operators according to the monodromy matrix of the 7-brane. It does not act on local operators and we will not have much to say about this operator.

Next we discuss common properties of $\mathcal{I},\mathcal{V}$, independent of the choice of branch cuts, depending exclusively on the type $\mathfrak{F}$ of the 7-brane inserted. Hanany-Witten brane creation predicts the creation of a topological symmetry operator when line defects are dragged through $\mathcal{I}$, see figure \ref{fig:HW}. Genuine defects constructed from asymptotic $[q,p]$ strings transform upon crossing $M_3$ to non-genuine defects with identical $[q,p]$ charge but with a topological operator of charge $[q,p](\rho-1)$ attached. The latter results from the corresponding string which attaches to the 7-branes stack and is otherwise embedded in the asymptotic boundary making it topological \cite{Heckman:2022muc}. Identical considerations hold for the operator $\mathcal{V}$.

\begin{figure}[t]
    \centering
    \scalebox{0.8}{\begin{tikzpicture}
	\begin{pgfonlayer}{nodelayer}
		\node [style=none] (0) at (-6.5, 1.5) {};
		\node [style=none] (1) at (-1.5, 1.5) {};
		\node [style=none] (2) at (-6.5, -1.5) {};
		\node [style=none] (3) at (-1.5, -1.5) {};
		\node [style=Star] (4) at (-4, 0) {};
		\node [style=none] (5) at (-6.5, 0) {};
		\node [style=none] (6) at (-4, 0.5) {7-branes};
		\node [style=none] (7) at (-4, 2) {$\ket{P,D}$};
		\node [style=none] (9) at (-5.5, -2) {};
		\node [style=none] (10) at (-2.5, -2) {};
		\node [style=none] (11) at (-4, -2.5) {$x_\perp$};
		\node [style=none] (14) at (-5.25, 1.5) {};
		\node [style=none] (15) at (-5.25, -1.5) {};
		\node [style=none] (16) at (-5.25, 0) {};
		\node [style=none] (18) at (-6, 0.75) {$[q,p]$};
		\node [style=none] (19) at (-6, -0.75) {$[q,p]\rho$};
		\node [style=none] (20) at (-7.5, -1.5) {$r=0$};
		\node [style=none] (21) at (-7.5, 1.5) {$r=\infty$};
		\node [style=none] (22) at (-7.5, 1) {};
		\node [style=none] (23) at (-7.5, -1) {};
		\node [style=none] (24) at (1.5, 1.5) {};
		\node [style=none] (25) at (6.5, 1.5) {};
		\node [style=none] (26) at (1.5, -1.5) {};
		\node [style=none] (27) at (6.5, -1.5) {};
		\node [style=Star] (28) at (3, 0) {};
		\node [style=none] (29) at (1.5, 0) {};
		\node [style=none] (30) at (3, 0.5) {7-branes};
		\node [style=none] (31) at (4, 2) {$\ket{P,D}$};
		\node [style=none] (33) at (2.5, -2) {};
		\node [style=none] (34) at (5.5, -2) {};
		\node [style=none] (35) at (4, -2.5) {$x_\perp$};
		\node [style=none] (38) at (5.5, 1.5) {};
		\node [style=none] (39) at (5.5, -1.5) {};
		\node [style=none] (40) at (5.5, 0) {};
		\node [style=none] (44) at (6.25, 0.75) {$[q,p]$};
		\node [style=none] (45) at (4.25, -0.5) {$[q,p]\rho$};
		\node [style=none] (46) at (6.75, -0.5) {$[q,p](\rho-1)$};
		\node [style=none] (47) at (0, 0) {$=$};
	\end{pgfonlayer}
	\begin{pgfonlayer}{edgelayer}
		\draw [style=ThickLine] (0.center) to (1.center);
		\draw [style=ThickLine] (3.center) to (2.center);
		\draw [style=DottedLine] (5.center) to (4);
		\draw [style=ArrowLineRight] (9.center) to (10.center);
		\draw [style=RedLine] (14.center) to (16.center);
		\draw [style=BlueLine] (16.center) to (15.center);
		\draw [style=ArrowLineRight] (23.center) to (22.center);
		\draw [style=ThickLine] (24.center) to (25.center);
		\draw [style=ThickLine] (27.center) to (26.center);
		\draw [style=DottedLine] (29.center) to (28);
		\draw [style=ArrowLineRight] (33.center) to (34.center);
		\draw [style=RedLine] (38.center) to (40.center);
		\draw [style=BlueLine] (40.center) to (39.center);
		\draw [style=ThickLine] (40.center) to (28);
	\end{pgfonlayer}
\end{tikzpicture}
}
    \caption{Boundary condition $\ket{P,D}$ admitting $[q,p]$ strings to terminate at infinity. Passing a 4D line defect labelled by $[q,p]\rho$ through the symmetry defect $\mathcal{U}(M_3,\mathfrak{F})$ from right to left creates a topological surface operator of charge $[q,p](\rho-1)$ via Hanany-Witten brane creation. }
    \label{fig:HW}
\end{figure}

Let us now quantify the qualitative discussion above. The insertion of 7-branes turns on a non-trivial flat background for the $SL(2,\mathbb{Z})$ duality bundle. The topological nature of the symmetry TFT allows us to localize this background onto a choice of branch cut and in terms of the $SL(2,\mathbb{Z})$ doublet $(\mathcal{B}^i) = (C_2,B_2)$ then
\begin{equation}\label{eq:symtftcoupling}\begin{aligned}
    \mathcal{S}_{\mathrm{(SymTFT),0}}&=-\frac{N}{4\pi}\int_{M_5} \epsilon_{ij}\mathcal{B}^i \wedge  d \mathcal{B}^j \\[1em] 
     ~\xrightarrow[~\text{Insertion}~]{7-\text{brane}}~\qquad \mathcal{S}_{\mathrm{(SymTFT),1}}&=\mathcal{S}_{\mathrm{(SymTFT),0}}+\mathcal{S}_{(\mathrm{cut})} + \mathcal{S}_{\textnormal{(Defects)}}\,.
    \end{aligned}
\end{equation}
Here the term $\mathcal{S}_{\textnormal{(Defects)}}$ denotes the 3D TFT supported on $M_3\subset M_4$ resulting from wrapping 7-branes on $M_3\times \partial X$ in case 1 and in cases 2 and 3 it includes possibly an additional TFT supported on $M_3',M_3''$. We postpone the analysis of $\mathcal{S}_{\textnormal{(Defects)}}$ together with the discussion of boundary conditions for the bulk fields of the 5D symmetry TFT along such defects. The term $\mathcal{S}_{(\mathrm{cut})}$ denotes a counterterm localized to the branch cut $\mathbb{H}$ with $M_3 \subset \partial \mathbb{H}$.

\begin{figure}[t]
    \centering
    \scalebox{0.8}{
    \begin{tikzpicture}
	\begin{pgfonlayer}{nodelayer}
		\node [style=none] (0) at (-2, 0) {};
		\node [style=none] (1) at (2, 0) {};
		\node [style=none] (4) at (-3.25, 0) {Branch cut};
		\node [style=none] (5) at (0, 2) {};
		\node [style=none] (6) at (0, 0) {};
		\node [style=none] (7) at (-1, -2) {};
		\node [style=none] (8) at (1, -2) {};
		\node [style=none] (13) at (0.75, 2) {$S_{[q,p]}$};
		\node [style=none] (19) at (-1.75, -2) {$S_{[q,p]}$};
		\node [style=none] (20) at (2.25, -2) {$S_{[q,p](\rho-1)}$};
        \node [style=none] (21) at (0, -2.5) {};
	\end{pgfonlayer}
	\begin{pgfonlayer}{edgelayer}
		\draw [style=DottedLine] (0.center) to (1.center);
		\draw [style=RedLine] (5.center) to (6.center);
		\draw [style=RedLine] (6.center) to (7.center);
		\draw [style=ThickLine] (6.center) to (8.center);
	\end{pgfonlayer}
\end{tikzpicture}
    }
    \caption{5D surface symmetry operator stretching across the branch cut. $S_{[q,p]}$ transforms into $S_{[q,p]\rho}$ which can be decomposed into two surface symmetry operators as shown.}
    \label{fig:Surface}
\end{figure}

We now derive the 4D action $\mathcal{S}_{(\mathrm{cut})}$. Let us begin by determining the action on the branch cut attaching to a single D7-brane. The monodromy matrix is denoted $\mathbb{T}$ in \eqref{eq:Mono}. Clearly, a $B_2$ profile remains unchanged when crossing the branch cut and we now claim
\begin{equation}\label{eq:Cut3}
    \mathcal{S}_{(\mathrm{cut})}^{\mathrm{D7}}=\frac{2\pi i}{N}\int_\mathbb{H} \frac{\mathcal{P}(B_2)}{2} 
\end{equation}
where $\mathbb{H}$ denotes the branch cut. For a $(p,q)$ 7-brane we replace $B_2$ by $B_2^{[q,p]}=pB_2+qC_2$. Stacks of $k$ D7-branes then have action $k \mathcal{S}_{(\mathrm{cut})}^{\mathrm{D7}}$ and similarly for stacks of $(p,q)$ 7-branes.

We argue for \eqref{eq:Cut3} by considering its action on topological boundary conditions and via anomaly considerations. The former amounts to checking that the branch cut action indeed realizes the 0-form symmetry with monodromy matrix $\rho= \mathbb{T}$ on symmetry operators and defect operators in 5D. The latter shows that the branch cut carries off an anomaly sourced by the 7-brane insertion. In \cite{Antinucci:2022vyk} the action \eqref{eq:Cut3} is derived via condensation.

As a check, consider Dirichlet boundary conditions $\ket{P_{G_k},D}$ realizing $B_2|_{r=\infty}=D$ with global form $G$ and $k$ counterterms proportional to $\mathcal{P}(D)$ stacked. Then colliding the branch cut with the topological boundary amounts to acting as
\begin{equation}
    \ket{P_{G_k},D} \rightarrow \exp(\mathcal{S}_{(\mathrm{cut})})\ket{P_{G_{k}},D}=\ket{P_{G_{k+1}},D}
\end{equation}
which here stacks the boundary condition with the counterterm $(2\pi i/N)\mathcal{P}(D)/2$ without changing the polarization. This is consistent with the fact that defects for $P$ are Wilson lines constructed from fundamental strings which are not acted on upon crossing the branch cut. Therefore the polarization must remain unchanged and stacking the branch cut with the boundary can at most add counterterms. More generally this follows from T-duality \cite{Aharony:2013hda} and other cases are argued for identically, we give explicit computations in section \ref{sec:N4}.

\begin{figure}
    \centering
    \scalebox{0.8}{
    \begin{tikzpicture}
	\begin{pgfonlayer}{nodelayer}
		\node [style=none] (0) at (-2.5, 1.5) {};
		\node [style=none] (1) at (2.5, 1.5) {};
		\node [style=none] (2) at (-2.5, -1.5) {};
		\node [style=none] (3) at (2.5, -1.5) {};
		\node [style=none] (4) at (0, 2) {$\ket{P,D}$};
		\node [style=none] (6) at (-3.5, -1.5) {$r=0$};
		\node [style=none] (7) at (-3.5, 1.5) {$r=\infty$};
		\node [style=none] (8) at (-3.5, 1) {};
		\node [style=none] (9) at (-3.5, -1) {};
		\node [style=none] (10) at (-1.5, -2.5) {};
		\node [style=none] (11) at (1.5, -2.5) {};
		\node [style=none] (12) at (0, -3) {$x_\perp$};
		\node [style=Star] (13) at (1, 0) {};
		\node [style=none] (14) at (0, 0.5) {};
		\node [style=none] (16) at (0, -0.5) {};
		\node [style=none] (17) at (-2, -0.5) {};
		\node [style=none] (18) at (-2, 0) {};
		\node [style=none] (19) at (-2, 0.5) {};
		\node [style=none] (20) at (-2.5, -0.5) {$A^l$};
		\node [style=none] (21) at (-2.5, 0) {$B^r$};
		\node [style=none] (22) at (-2.5, 0.5) {$C^s$};
		\node [style=none] (23) at (1.75, 0.5) {7-brane};
	\end{pgfonlayer}
	\begin{pgfonlayer}{edgelayer}
		\draw [style=ThickLine] (0.center) to (1.center);
		\draw [style=ThickLine] (3.center) to (2.center);
		\draw [style=ArrowLineRight] (9.center) to (8.center);
		\draw [style=ArrowLineRight] (10.center) to (11.center);
		\draw [style=DottedLine] (19.center) to (14.center);
		\draw [style=DottedLine] (17.center) to (16.center);
		\draw [style=DottedLine] (14.center) to (13);
		\draw [style=DottedLine] (16.center) to (13);
		\draw [style=DottedLine] (18.center) to (13);
	\end{pgfonlayer}
\end{tikzpicture}
    }
    \caption{The branch cut of attaching to any 7-brane insertion can be decomposed into branch cuts individually associated with $(p,q)$ 7-branes. }
    \label{fig:BCsplit}
\end{figure}

With this we can give the branch cut term for any 7-brane insertion (see figure \ref{fig:BCsplit}). Any 7-brane can be represented as a supersymmetric bound state of certain $(p,q)$ 7-branes denoted $A,B,C$, their $(p,q)$ charges are $(1,0),(3,1),(1,1)$ respectively. All 7-branes are then of the form $A^lB^rC^s$ and their branch cut therefore supports the operator
\begin{equation}\label{eq:BranchCutOperator}
\mathcal{O}_{(l,r,s)}=\mathcal{O}_A^l\mathcal{O}_B^r\mathcal{O}_C^s
\end{equation}
which is purely determined by the monodromy of the 7-brane and where
\begin{equation}\begin{aligned}
    \mathcal{O}_A&=\exp\left( \frac{2\pi i}{N}\int_{\mathbb{H}_A} \frac{\mathcal{P}(B_2)}{2}\right)\\
    \mathcal{O}_B&=\exp\left( \frac{2\pi i}{N}\int_{\mathbb{H}_B} \frac{\mathcal{P}(3B_2+C_2)}{2}\right)\\
    \mathcal{O}_C&=\exp\left( \frac{2\pi i}{N}\int_{\mathbb{H}_C} \frac{\mathcal{P}(B_2+C_2)}{2}\right)\,.
\end{aligned}\end{equation}
Here $B_2,C_2$ are normalized to $\mathbb{Z}_N$-valued forms and $\mathbb{H}_A,\mathbb{H}_B,\mathbb{H}_C$ denote the branch cuts localizing the monodromy of the respective brane stack. Note that it is not possible in general to present $\mathcal{O}_{(l,r,s)}$ as a single exponential, the fields $B_2,C_2$ are conjugate and do not commute. However, it is possible to present it as a TFT \cite{Antinucci:2022vyk}.

Let us discuss possible anomalies. Note first that we have the boundary condition $B_2|_{\mathrm{D7}}=0$ at the D7 locus. This follows from noting that $B_2|_{\mathrm{D7}}\neq 0$ implies that along infinitesimal loops linking the D7-brane we have $C_2\rightarrow C_2'\approx C_2+B_2|_{\mathrm{D7}}$ leading to a discontinuous, ill-defined profile for $C_2$ contradicting the D7-brane solution. In the M-theory dual picture this is equivalent to the torus cycle corresponding to $B_2$ shrinking at the D7-brane locus. Backgrounds with only $C_2$ turned on are invariant under monodromy and therefore we have no constraint on the $C_2$ profile along the D7-brane.

When considering a stack of $k$ D7-branes the boundary condition is $kB_2|_{\mathrm{D7}}=0$ and together with $B_2$ taking values in $\mathbb{Z}_N$ we have $\textnormal{gcd}(k,N)B_2|_{\mathrm{D7}}=0$. Now $B_2|_{\mathrm{D7}}$ no longer vanishes generically but takes values which are multiples of $N/\textnormal{gcd}(k,N)$. The $C_2$ profile remains unconstrained. 

Next, consider the background transformation $\mathcal{B}\rightarrow \mathcal{B}+d\lambda$ where $\lambda$ is twisted by the monodromy of the 7-brane, i.e. $d\lambda$ is subject to the same boundary conditions along the D7-brane as $\mathcal{B}_2$. The Pontryagin square term $\mathcal{P}(B_2)$ gives rise to a boundary term iff $\textnormal{gcd}(k,N)\neq 1$. This anomaly must be absorbed by the 3D TFT associated with the wrapped D7-branes. 

The 3D TFT has non-trivial lines precisely when $\textnormal{gcd}(k,N)\neq 1$. For this consider the dual M-theory geometry of a D7-brane whose normal geometry is given by an elliptic fibration $Z$ over $\mathbb{C}$ with one Kodaira $I_k$ singularity. The boundary $\partial Z$ of this normal geometry is elliptically fibered over a circle and has first homology groups $H_1(\partial Z)=\mathbb{Z}\oplus\mathbb{Z}\oplus \mathbb{Z}_k$. One free factor corresponds to the base circle and can be neglected in our discussion. This leaves us with $\mathbb{Z}\oplus \mathbb{Z}_k$. Of these, only the torsional generator of $\mathbb{Z}_k$ collapses at the $I_k$ singularity, sweeping out a non-compact two-cycle in the process. Wrapping an M2 brane on this cycle constructs a Wilson line. In 5D we work modulo $N$ and overall this gives a defect group of lines isomorphic to $\mathbb{Z}_{\textnormal{gcd}(k,N)}$ for the 3D TFT. In section \ref{sec:DefectTFT} we show that these lines organize into a 3D TFT with an anomaly sourcing the one carried by the branch cut. 

More generally the Pontryagin square terms on individual elementary branch cuts are anomalous under background transformations and we have a local anomaly flowing along each cut. Summing over branch cuts, the net anomaly is then necessarily absorbed by the 7-brane which can be viewed as an edge mode to the theories localized on the branch cuts. Alternatively, the overall anomaly is non-vanishing whenever the 7-brane sources an anomaly.

Let us therefore discuss when 7-brane insertions source anomalies in greater generality by studying the branch cut terms. For this, let us first consider in the 5D symmetry TFT the surface symmetry operators acting on the surface defects constructed from $(p,q)$-strings. They are:\footnote{Our definition differs from the definition $S_{(p,q)}'(\Sigma_2)\equiv S_{(p,0)}(\Sigma_2)S_{(0,q)}(\Sigma_2)$ as given in \cite{Kaidi:2022cpf} by a phase with argument proportional to the self-linking number of the surface $\Sigma_2$. This relative phase follows from the Baker-Campbell-Hausdorff formula and noting that $B_2,C_2$ are conjugate variables in the 5D TFT.}
\begin{equation}
    S_{[q,p]}(\Sigma_2)=\exp\left(2\pi i\oint_{\Sigma_2} (pB_2+qC_2)\right)\,, 
\end{equation}
with integers $p,q$ modulo $N$. Similarly to strings they are transformed when stretching across the $SL(2,\mathbb{Z})$ branch cut. Consider a surface $\Sigma_2$ separated by the branch cut $\mathbb{H}$ into two components $\Sigma_2 =\Sigma_2^+\cup \Sigma_2^-$, then
\begin{equation}
    S_{[q,p]}(\Sigma_2^{+})S_{[q,p]\rho}(\Sigma_2^{-})=S_{[q,p]}(\Sigma_2)S_{[q,p](\rho-1)}(\Sigma_2^{-})
\end{equation}
whenever the self-linking number of $\Sigma_2$ in the 5D bulk vanishes, see figure \ref{fig:Surface}. This is interpreted as the intersection of $S_{[q,p]}(\Sigma_2)$ with the branch cut sourcing $S_{[q,p](\rho-1)}(\Sigma_2^{-})$. Correspondingly, in terms of the $\mathbb{Z}_N$-valued 1-form symmetry background fields this relation can be expressed as
\begin{equation}\label{eq:dualityznaction}
    \delta \left( (\rho-1) \begin{bmatrix}
        C_2 \\ B_2
    \end{bmatrix}\right) = A_1\cup \begin{bmatrix}
        C_2 \\ B_2
    \end{bmatrix}
\end{equation}
where $A_1$ is the background field of the $SL(2,\mathbb{Z})$ bundle proportional to the Poincar\'e dual of the branch cut. A priori, $A_1$ takes values in $\mathbb{Z}_n$ where $n$ is the order of $\rho\in SL(2,\mathbb{Z})$, or $U(1)$ when $n$ is infinite, but in \eqref{eq:dualityznaction} this $A_1$ takes values in in $\mathbb{Z}_{\mathrm{gcd}(n,N)}$. In other words, $A_1$ takes values in $\mathbb{Z}_{\mathrm{gcd}(4,N)}$ or $\mathbb{Z}_{\mathrm{gcd}(3,N)}$ for duality\footnote{One might ask whether one should call this order four operation a ``quadrality'' operator. On local operators it is indeed order two, which accounts for the terminology. This subtlety between $2$ versus $4$ will show up later in our analysis of mixed anomalies.} and triality defects respectively\footnote{There are also hexality defects which are furnished by 7-branes of Type $II$ or $II^*$ whose background field in the 4D relative theory is $\mathbb{Z}_{\mathrm{gcd}(6,N)}$-valued. We come back to these theories at the end of the section to show that their mixed anomalies with the 1-form symmetry is always trivial.}, and similar to \eqref{eq:rescaling} we have a relation between normalizations of discrete-valued fields as (assuming $n$ is finite)
\begin{equation}\label{eq:rescaling2}
    A^{\mathbb{Z}_n}_1=\frac{1}{\mathrm{gcd}(n,N)}A^{\mathbb{Z}_{\mathrm{gcd}(k,N)}}_1
\end{equation}
where $A^{\mathbb{Z}_\ell}_1$ has holonomies that take values $p \; \mathrm{mod}\; \ell$.

Two related points to take into account are that one must first choose a polarization to have an invertible anomaly TFT, and in general $\rho-1$ does not have an inverse in $\mathbb{Z}_N$ coefficients.\footnote{In the cases when $(\rho-1)$ does have an inverse we have that $\mathrm{gcd}(2,N)=0$ or $\mathrm{gcd}(3,N)=0$ making the anomaly trivial in the first place.} We are motivated then to consider polarization choices that are duality/triality invariant, which is consistent with the fact that otherwise, the topological defect implementing the duality or triality defect is an interface between separate theories rather than a symmetry operator which may have a mixed 't Hooft anomaly. Recall that the background 1-form field for a given polarization spans a 1-dimensional subspace of $\mathrm{Span}(C_2,B_2)^\mathrm{T}$ and we say that a polarization is duality or triality invariant if there are non-trivial solutions to the following equation, posed over $\mathbb{Z}_N$,
\begin{equation}\label{eq:Eigenvalue}
    (\rho-1)\mathcal{B}_2=0\,
\end{equation}
which only has non-trivial solutions in $\mathbb{Z}_K$ subgroups of $\mathbb{Z}_N$ where $K=\mathrm{gcd}(k,N)$ and $k$ is the coefficient appearing in Table \ref{tab:Fibs}.\footnote{There are no solutions if $K$ and $N$ are coprime.} We denote such eigenvector solutions as $\mathcal{B}_2^\rho$ which is the product of an $SL(2,\mathbb{Z})$ vector and the form profile $B_2^\rho$. 

Consider for example the case of dualities that $N$ is even. We denote the lift to the full $\mathbb{Z}^{(1)}_N$-background field by $B^\rho_2$ as well where the two are related as
\begin{equation}\label{eq:rescale3}
    (B^\rho_2)^{\mathbb{Z}_{N}}=\frac{1}{2}(B^\rho_2)^{\mathbb{Z}_{2}}
\end{equation}
when the LHS only takes values in the $\mathbb{Z}_2$ subgroup of $\mathbb{Z}_N$ generated by $N/2 \; \mathrm{mod} \; N$. It will be clear that the anomaly will be invariant under this choice of lift.

We can decompose the vector space as $\mathrm{ker}(\rho-1)\oplus \mathrm{ker}(\rho-1)^\perp$ where the latter is generated by vectors $v$ such that $(\rho-1)v=v$. This allows us to define $(\mathcal{B}^\rho_2)^\perp$ which due to the $SL(2,\mathbb{Z})$ invariant pairing $\epsilon_{ij}$ in \eqref{eq:symtftcoupling}, allows us to rewrite that coupling schematically as $\int \mathcal{B}_2^\rho \cup \delta (\mathcal{B}^\rho_2)^\perp$. Substituting \eqref{eq:dualityznaction} then allows us to derive the mixed 't Hooft anomalies between 0-form duality/triality symmetries and $\mathbb{Z}^{(1)}_N$ 1-form symmetries:\footnote{In $\mathcal{S}_{\mathrm{duality}}$, observe that we take a gcd with respect to $2$ rather then $N$ because the operation is order two on local operators.}
\begin{equation}\label{eq:mixed anomaly term in the bulk}
    \begin{aligned}
       \mathcal{S}_{\mathrm{duality}}&= \frac{2\pi}{\mathrm{gcd}(2,N)}\int A_1\cup \frac{\mathcal{P}(B^\rho_2)}{2}  \\
    \mathcal{S}_{\mathrm{triality}} &= \frac{2\pi}{\mathrm{gcd}(3,N)}\int A_1\cup \frac{\mathcal{P}(B^\rho_2)}{2}
    \end{aligned}
\end{equation}
where we implemented the refinement $B_2^\rho \cup B_2^\rho \rightarrow \mathcal{P}(B_2^\rho)/2$ with $\mathcal{P}$ the Pontryagin square operation following \cite{Gaiotto:2014kfa,Kapustin:2014gua,Benini:2018reh}. The normalization of discrete-valued fields is motivated to match with \cite{Kaidi:2021xfk} which, in our presentation, means that we are using the LHS of \eqref{eq:rescaling2} for the normalization of $A_1$ and the RHS of \eqref{eq:rescaling} for the normalization of the $\mathbb{Z}_2$ or $\mathbb{Z}_3$-valued field $B^\rho_2$.\footnote{For the duality case, only the image of $A_1$ in the quotient $\mathbb{Z}_4/\mathbb{Z}_2\simeq \mathbb{Z}_2$ couples to $B^\rho_2$ which explains the factor $\mathrm{gcd}(2,N)$ rather than $\mathrm{gcd}(4,N)$.} 

We have that $A_1$ is Poincar\'e dual to the branch cut $\mathbb{H}$ and therefore
\begin{equation}\label{eq:CT}
    \begin{aligned}
    \textnormal{Duality Interface:}&\quad   S_{\mathrm{cut}}= \frac{2\pi}{\mathrm{gcd}(2,N)}\int_\mathbb{H} \frac{\mathcal{P}(B^\rho_2)}{2}  \\
    \textnormal{Triality Interface:}&\quad S_{\mathrm{cut}}= \frac{2\pi}{\mathrm{gcd}(3,N)}\int_\mathbb{H} \frac{\mathcal{P}(B^\rho_2)}{2}\,.
    \end{aligned}
\end{equation}
We conclude that upon inserting a 7-brane giving rise to the anomaly \eqref{eq:mixed anomaly term in the bulk}, we can localize this anomaly to a single branch cut with action \eqref{eq:CT}. Conversely, in cases with no net anomaly the anomaly localized to individual branch cuts cancels and therefore the branch cut term does not admit a presentation as a simple exponential.

\begin{figure}
    \centering
    \scalebox{0.8}{\begin{tikzpicture}
	\begin{pgfonlayer}{nodelayer}
		\node [style=none] (0) at (-2.5, 1.5) {};
		\node [style=none] (1) at (2.5, 1.5) {};
		\node [style=none] (2) at (-2.5, -1.5) {};
		\node [style=none] (3) at (2.5, -1.5) {};
		\node [style=none] (7) at (0, 2) {$\ket{P,D}$};
		\node [style=none] (13) at (-3.5, -1.5) {$r=0$};
		\node [style=none] (14) at (-3.5, 1.5) {$r=\infty$};
		\node [style=none] (15) at (-3.5, 1) {};
		\node [style=none] (16) at (-3.5, -1) {};
		\node [style=none] (17) at (-2.5, 0) {};
		\node [style=none] (18) at (2.5, 0) {};
		\node [style=none] (19) at (3.25, 0) {$\mathcal{O}^{[q,p]}$};
		\node [style=none] (20) at (0, -0.5) {$M_4'$};
		\node [style=none] (21) at (0, 0.25) {};
		\node [style=none] (22) at (0, 1.25) {};
		\node [style=none] (23) at (0, -2) {$\ket{\mathfrak{T}^{(N)}_X}$};
	\end{pgfonlayer}
	\begin{pgfonlayer}{edgelayer}
		\draw [style=ThickLine] (0.center) to (1.center);
		\draw [style=ThickLine] (3.center) to (2.center);
		\draw [style=ArrowLineRight] (16.center) to (15.center);
		\draw [style=DottedLine] (17.center) to (18.center);
		\draw [style=ArrowLineRight] (21.center) to (22.center);
	\end{pgfonlayer}
\end{tikzpicture}
}
    \caption{The operator $\mathcal{O}^{[q,p]}$ is defined on $M_4'$ in the 5D bulk which is homotopic to either boundary. The operator acts on the topological boundary condition $\ket{P,D}$ by colliding $M_4'$ with the corresponding boundary component. }
    \label{fig:BulkOps}
\end{figure}

Let us now discuss another class of codimension one bulk operators of the 5D symmetry TFT. We use this to generalize the discussion of counterterm actions considered above. Along these lines, fix $B_2^{[q,p]}$ as the combination of doublet fields which pulls back to a $(p,q)$ 7-brane (i.e., is compatible with this choice of monodromy). 
Then, introduce the operator:
\begin{equation}
 \mathcal{O}^{[q,p]}=\exp\left( \frac{2\pi i}{N}\int_{M_4'}\frac{\mathcal{P}(B_2^{[q,p]})}{2}\right)\,.
\end{equation}
where $M_4'$ a copy of $M_4$ deformed into the bulk of the 5D TFT. It runs horizontally, parallel to the D3-brane worldvolume. More generally we could consider four-manifolds with boundary in conjunction with various edge modes.  Here $B_2^{[q,p]}=pB_2+qC_2$ is the form supported on the branch cut of a $(p,q)$ 7-brane. When contracting the 5D slab to 4D the branch cut is layered with the topological boundary condition and so the unitary operator $\mathcal{O}^{[q,p]}$ realizes an isomorphism on the vector space of boundary conditions generated by $\{\ket{P_i,D_i}\}$ (see figure \ref{fig:BulkOps}). Let us write
\begin{equation}\label{eq:Opq}
    \bra{P_i,D_i}\mathcal{O}^{[q,p]}= \bra{P_i^{[q,p]},D_i}\,.
\end{equation}
Note that whenever the $SL(2,\mathbb{Z})$ vector $[q,p]$ is contained in the polarization $P$ then the corresponding topological boundary condition is an eigenvector and $\mathcal{O}^{[q,p]}$ acts by stacking the boundary condition with a counterterm whose profile is determined by the Dirichlet boundary condition. However, more generally $\mathcal{O}^{[q,p]}$ is not diagonal, in particular there is no operator $\mathcal{O}^{[q,p]}$ which acts via counterterm stacking on all topological boundary conditions. 

We close this section by emphasizing that nothing about this discussion requires a large $N$ limit, the existence of a holographic dual, or having $\mathcal{N} = 4$ supersymmetry.

\section{Defect TFT}
\label{sec:DefectTFT}

We now discuss the 3D TFT supported on the defect constructed by wrapping some 7-brane on $ \partial X$ and its coupling to the bulk 5D symmetry TFT. The main idea will be to treat the wrapped 7-branes as codimension two defects in the 5D theory. See Appendix \ref{app:minimalTFT7branes} for a discussion of how dimensional reduction of the 7-brane on $\partial X$ can result in such topological terms.

\subsection{Minimal Abelian TFT from Defects}\label{ssec:linkingminimaltheory}

Consider the setup of the previous section consisting of a stack of 7-branes of type $\mathfrak{F}$ with monodromy matrix $\rho$ compactified on $\partial X$ to 3D in the presence of $N$ D3-branes located at the apex of $X$, the cone over $\partial X$. This produces a 3D TFT $\mathcal{T}$ supported on $M_3$ coupled to the worldvolume of the D3-brane stack. 

Let us begin by considering case 1 with a horizontal branch cut (along $\mathbb{H}_{\leftarrow}$) for which there is a single topological boundary condition along the asymptotic boundary. The 7-brane is located at $\bar x_\perp$ and separates the D3-worldvolume into two half-spaces $\bar x_\perp>  x_\perp$ and $\bar x_\perp<  x_\perp$.

Previously we studied, via the Hanany-Witten transition, the consequence of dragging lines in $P_i$ through $M_3$ for $i=1,2$. Now consider two lines of equal but opposite asymptotic charge in each of the half-spaces and collide these. This produces a line operator and is interpreted as an open string running between the 7-branes and the D3-branes. If the lines have charges $\pm[q,p]$ then the latter is a string of charge $[q,p](1-\rho)$ (see figure \ref{fig:Screening}). These are precisely the lines not inherent to the 3D TFT $\mathcal{T}$, they can be moved off $M_3$. The inherent lines of the 3D TFT are associated with open strings running between the 7-branes and D3-branes modulo this screening effect.

\begin{figure}[t]
    \centering
    \scalebox{0.8}{
    \begin{tikzpicture}
	\begin{pgfonlayer}{nodelayer}
		\node [style=none] (0) at (-6.5, 1.5) {};
		\node [style=none] (1) at (-1.5, 1.5) {};
		\node [style=none] (2) at (-6.5, -1.5) {};
		\node [style=none] (3) at (-1.5, -1.5) {};
		\node [style=Star] (4) at (-4, 0) {};
		\node [style=none] (5) at (-6.5, 0) {};
		\node [style=none] (6) at (-4, 0.5) {7-branes};
		\node [style=none] (7) at (-4, 2) {$\ket{P,D}$};
		\node [style=none] (8) at (-5.5, -2) {};
		\node [style=none] (9) at (-2.5, -2) {};
		\node [style=none] (10) at (-4, -2.5) {$x_\perp$};
		\node [style=none] (11) at (-5.5, 1.5) {};
		\node [style=none] (12) at (-5.5, -1.5) {};
		\node [style=none] (13) at (-5.5, 0) {};
		\node [style=none] (14) at (-6.25, 0.75) {$[q,p]$};
		\node [style=none] (15) at (-6.25, -0.75) {$[q,p]\rho$};
		\node [style=none] (16) at (-8, -1.5) {$r=0$};
		\node [style=none] (17) at (-8, 1.5) {$r=\infty$};
		\node [style=none] (18) at (-8, 1) {};
		\node [style=none] (19) at (-8, -1) {};
		\node [style=none] (21) at (-2.5, 1.5) {};
		\node [style=none] (22) at (-2.5, -1.5) {};
		\node [style=none] (23) at (-1.75, 0.75) {$[q,p]$};
		\node [style=none] (24) at (-2.5, 0) {};
		\node [style=none] (25) at (1.5, 1.5) {};
		\node [style=none] (26) at (6.5, 1.5) {};
		\node [style=none] (27) at (1.5, -1.5) {};
		\node [style=none] (28) at (6.5, -1.5) {};
		\node [style=Star] (29) at (4, 0) {};
		\node [style=none] (30) at (1.5, 0) {};
		\node [style=none] (32) at (4, 2) {$\ket{P,D}$};
		\node [style=none] (33) at (2.5, -2) {};
		\node [style=none] (34) at (5.5, -2) {};
		\node [style=none] (35) at (4, -2.5) {$x_\perp$};
		\node [style=none] (36) at (4.5, 1.5) {};
		\node [style=none] (37) at (3.5, -1.5) {};
		\node [style=none] (38) at (3.5, 0) {};
		\node [style=none] (39) at (2.75, 0.75) {$[q,p]$};
		\node [style=none] (40) at (2.75, -0.75) {$[q,p]\rho$};
		\node [style=none] (46) at (4.5, -1.5) {};
		\node [style=none] (47) at (5.5, 0) {$[q,p]$};
		\node [style=none] (48) at (4.5, 0) {};
		\node [style=none] (49) at (0, 0) {$=$};
		\node [style=none] (50) at (4, 1) {};
	\end{pgfonlayer}
	\begin{pgfonlayer}{edgelayer}
		\draw [style=ThickLine] (0.center) to (1.center);
		\draw [style=ThickLine] (3.center) to (2.center);
		\draw [style=DottedLine] (5.center) to (4);
		\draw [style=ArrowLineRight] (8.center) to (9.center);
		\draw [style=RedLine] (11.center) to (13.center);
		\draw [style=BlueLine] (13.center) to (12.center);
		\draw [style=ArrowLineRight] (19.center) to (18.center);
		\draw [style=RedLine] (21.center) to (22.center);
		\draw [style=ArrowLineBlue] (12.center) to (13.center);
		\draw [style=ArrowLineRed] (21.center) to (24.center);
		\draw [style=ThickLine] (25.center) to (26.center);
		\draw [style=ThickLine] (28.center) to (27.center);
		\draw [style=DottedLine] (30.center) to (29);
		\draw [style=ArrowLineRight] (33.center) to (34.center);
		\draw [style=BlueLine] (38.center) to (37.center);
		\draw [style=ArrowLineBlue] (37.center) to (38.center);
		\draw [style=RedLine] (46.center) to (48.center);
		\draw [style=RedLine, in=0, out=90] (48.center) to (50.center);
		\draw [style=RedLine, in=90, out=180] (50.center) to (38.center);
		\draw [style=ArrowLineRed, in=90, out=0] (50.center) to (48.center);
	\end{pgfonlayer}
\end{tikzpicture}
    }
    \caption{Collision of line defect of charge $[q,p]$ with line defect $-[q,p]\rho$. Conversely, line defects of charge $[q,p](1-\rho)$ are lines not inherent to the 3D TFT $\mathcal{T}$.}
    \label{fig:Screening}
\end{figure}

With this the line defects of $\mathcal{T}$ are simply the line defects of the 7-brane stack modulo $N$. Let us study the lines of $\mathcal{T}$ in absence of the D3-brane flux, i.e., we do not impose screening modulo $N$.

In order to study the spin (in the sense of \cite{Hsin:2018vcg, Gukov:2020btk}) of such lines, consider the purely geometric M-/F-theory dual setup of an isolated stack of such 7-branes consisting of a local K3 surface $Z\rightarrow \mathbb{C}$. The boundary $\partial Z\rightarrow S^1$ has first homology $H_1(\partial Z)\cong \mathbb{Z}\oplus \textnormal{coker}(\rho-1)$ and line defects charged under the center symmetry of the system are constructed by wrapping M2-branes on cones over torsional one-cycles in $H_1(\partial Z)$. Let us assume that this homology group  is isomorphic to $\mathbb{Z}_k$ with generator $\gamma$. This torsional one-cycle is contained in the elliptic fiber, we write $\gamma=r \sigma_{a}+ s \sigma_{b}$ (in the obvious notation) for one of its representatives. This is dual to a string of charge $Q=[s,r]$. We have $k\gamma=0$ and $k$ copies of the corresponding string are screened in the F-theory setup. 

The spin of the lines associated with $\gamma$ is now determined by the refined self-linking number of $\gamma$ in $\partial Z$ as given by
\begin{equation}
    \ell(\gamma,\gamma)=\frac{1}{2k}\;\!\gamma \cdot \Sigma_\gamma\quad \mathrm{mod~1}
    \end{equation}
where $\partial \Sigma_\gamma=k\gamma$, determines the spin $h[L_\gamma]=\ell(L_\gamma,L_\gamma)=m/2k$ of the line $L_\gamma$ (see table \ref{tab:Fibs}). For a stack of $k$ $(p,q)$ 7-branes we have $m=k-1$.

Let us now take the screening effects due to the D3-brane flux into account. In this case the lines of 3D TFT $\mathcal{T}$ trivialize modulo $N$ and $k$, and therefore give rise to a 1-form symmetry $\mathbb{Z}_K$ with charged lines $L_\gamma$ and $K=\textnormal{gcd}(k,N)$. Whenever $mK\in 2\mathbb{Z}$ and $\textnormal{gcd}(m,K)=1$ it follows from the general discussion in \cite{Hsin:2018vcg} that the lines $L_\gamma$ form a consistent minimal abelian TFT denoted $\mathcal{A}^{K,m}$ \footnote{$\mathcal{A}^{K,m}$ is defined as a minimal 3D TFT with $\mathbb{Z}_K^{(1)}$ symmetry, whose symmetry lines have spins given by $h[a^s] \equiv \frac{ms^2}{2K} \mod 1$, where $a$ is a generating symmetry line such that $a^K = 1$.}. For branes of constant axio-dilaton and stacks of $(p,q)$ 7-branes we have $\textnormal{gcd}(k,m)=1$ and therefore $\textnormal{gcd}(K,m)=1$ follows. When these two conditions are met we have
\begin{equation}\label{eq:TFT}
    \mathcal{T}[B]=\mathcal{A}^{K,m}[B] \otimes \mathcal{T}'
\end{equation}
where $\mathcal{T}'$ is a decoupled TFT with lines neutral under the $\mathbb{Z}_K$ 1-form symmetry and $B$ is a 2-form background field for the 1-form symmetry which follows from the coupling of the open strings running between the D3-branes and 7-branes to $(B_2,C_2)$ and is $B=Q_i \mathcal{B}^i|_{\textnormal{7-brane}}$ which is $SL(2,\mathbb{Z})$ invariant where $Q$ is the charge vector of the strings.

\begin{figure}[t]
    \centering
    \scalebox{0.8}{\begin{tikzpicture}
	\begin{pgfonlayer}{nodelayer}
		\node [style=none] (0) at (-2.5, 1.5) {};
		\node [style=none] (1) at (2.5, 1.5) {};
		\node [style=none] (2) at (-2.5, -1.5) {};
		\node [style=none] (3) at (2.5, -1.5) {};
		\node [style=Star] (4) at (0, 0) {};
		\node [style=none] (5) at (-2.5, 0) {};
		\node [style=none] (6) at (0, 2) {$\ket{P,D}$};
		\node [style=none] (7) at (-1.5, -2) {};
		\node [style=none] (8) at (1.5, -2) {};
		\node [style=none] (9) at (0, -2.5) {$x_\perp$};
		\node [style=none] (10) at (0, 0.5) {7-branes};
		\node [style=none] (17) at (0, -1.5) {};
		\node [style=none] (18) at (-3.5, -1.5) {$r=0$};
		\node [style=none] (19) at (-3.5, 1.5) {$r=\infty$};
		\node [style=none] (20) at (-3.5, 1) {};
		\node [style=none] (21) at (-3.5, -1) {};
		\node [style=none] (22) at (0.5, -0.75) {$L_\gamma$};
	\end{pgfonlayer}
	\begin{pgfonlayer}{edgelayer}
		\draw [style=ThickLine] (0.center) to (1.center);
		\draw [style=ThickLine] (3.center) to (2.center);
		\draw [style=DottedLine] (5.center) to (4);
		\draw [style=ArrowLineRight] (7.center) to (8.center);
		\draw [style=RedLine] (4) to (17.center);
		\draw [style=ArrowLineRight] (21.center) to (20.center);
	\end{pgfonlayer}
\end{tikzpicture}
}
    \caption{Line operators of $\mathcal{T}$ coupled to the 4D theory are open strings running between the 7-branes and the D3-brane stack. The lines $L_\gamma^n$ organize into the minimal abelian TFT $\mathcal{A}^{K,m}$ where $K,m$ follow from the elliptic data of the 7-brane.}
    \label{fig:MinimalLines}
\end{figure}

The charge vector $Q_i$ associated with $\gamma$ is defined modulo vectors in the image of $\rho-1$. Consider the charge vector $Q'_i=Q_i+(q (\rho-1))_i$ in the same coset. Then, $B$ changes as
\begin{equation}
    B=Q_i\mathcal{B}^i~\rightarrow ~ Q_i'\mathcal{B}^i=Q_i\mathcal{B}^i+(q(\rho-1))_i\mathcal{B}^i=Q_i\mathcal{B}^i+q_i((\rho-1)\mathcal{B})^i
\end{equation}
and for this coupling to be well-defined we require the profile of $\mathcal{B}$ to lie in the kernel of $\rho-1$ modulo $K$. 

For example, consider $N=2$ with insertion of a 7-brane of type $\mathfrak{F}=III^{\ast}$. We compute $k=2$ and therefore $K=2$. The eigenvector is $\mathcal{B}_2^\rho=B_2^\rho[1,1]^t$ and the background in \eqref{eq:TFT} is $B=B_2^\rho$. We have $\mathcal{T}[B_2^\rho]=\mathcal{A}^{2,1}[B_2^\rho]\otimes \mathcal{T}'$.

When $K=1$ no eigenvector exists, $B_2^\rho$ vanishes and $\mathcal{T}$ has no lines coupling to the bulk. In this case $\mathcal{T}$ does not absorb an anomaly. 

We note that this discussion of the TFT $\mathcal{T}$ is independent of branch cut choice and therefore extends to cases 2,3,4. 

Next, let us discuss case 2 with vertically running branch cut terminating on the 0-form operator $\mathcal{D}(M_3
',B_2^\rho)$ contained in the D3-worldvolume. This is another defect possible supporting a 3D TFT. We claim that for all parameter values $\mathcal{D}(M_3
',B_2^\rho)$ does not support a TFT interacting with the bulk. As discussed previously, the operator $\mathcal{D}(M_3
',B_2^\rho)$ realizes a gluing condition on the enriched Neumann boundary condition set by the D3-brane stack and, is not realized by a 7-brane. Therefore, no strings end on $M_3'$ and it does not support a defect group of its own, in contrast to the 7-brane insertion. With this we conjecture that the TFT living at the intersection is trivial or at least does not interact with the bulk.

\section{Example: \texorpdfstring{$\mathcal{N} = 4$}{N=4} SCFTs}
\label{sec:N4}
In the previous section we presented a general discussion of how 7-branes implement duality defects in systems obtained from D3-branes probing an isolated Calabi-Yau singularity. In particular, we saw that the structure of the branch cuts leads to distinct implementations for various sorts of duality defects (as well as triality defects). In this section we show that this matches to the available results in the literature for $\mathcal{N} = 4$ SYM theory, in particular the case where the gauge algebra is $\mathfrak{su}(2)$. Our construction readily generalizes to higher rank Lie algebras, and (deferring the classification of possible global realizations of the gauge group) this provides a uniform perspective for duality defects in other $\mathcal{N} = 4$ SCFTs realized by probe D3-branes. Combining this with the discussion in Appendix \ref{app:orbo}, we anticipate that the same considerations will also apply to the full set of $\mathcal{N} = 4$ SCFTs.

Before proceeding, let us briefly spell out our notational conventions, which essentially follow those of references 
\cite{Aharony:2013hda,Kaidi:2022uux}. All possible global forms are given by $SU(2)_i$ and $SO(3)_{\pm, i}$, $i = 0, 1$. Here $SU(2)$ is the electric polarization where only the Wilson lines with $(z_e, z_m) = (1, 0)$ are present; $SO(3)_+$ is the global form in which only the 't Hooft lines with $(z_e, z_m) = (0, 1)$ are present, and for $SO(3)_-$ only the dyonic lines with $(z_e, z_m) = (1, 1)$ are present. On top of that, we use $i = 0, 1$ to specify the absence or presence of a counterterm $\delta S = -\int \frac{\mathcal{P}(B)}{2}$, where $B$ is the background gauge field for the $\mathbb{Z}_2^{(1)}$ 1-form symmetry.

\subsection{Duality Defects in \texorpdfstring{$\mathfrak{su}(2)$}{} SYM\texorpdfstring{$_{\mathcal{N} = 4}$}{} Theory}

Duality defects for 4D $\mathcal{N}=4$ $\mathfrak{su}(2)$ supersymmetric Yang-Mills theory can be constructed field theoretically via Kramers-Wannier-like constructions \cite{Kaidi:2021xfk} and half-space gauging \cite{Choi:2022zal}. We present the string theory construction for both emphasizing differences and similarities between the two approaches.

\subsection*{Kramers-Wannier-like construction}
Let us first discuss the Kramers-Wannier-like construction of duality defects and its string theory realization. We start with a lighting review of the field-theoretic construction and refer the reader to  \cite{Kaidi:2021xfk} for more details.  Consider the $SO(3)_-$ theory, with 1-form symmetry background field $B$ and mixed anomaly
\begin{equation}
    \pi \int_{M_5} A^{(1)}\cup \frac{\mathcal{P}(B)}{2}
\end{equation}
where $A^{(1)}$ is the background field for a $\mathbb{Z}_2$ 0-form symmetry, here S-duality on local operators at $\tau=i$ . Denote by $\mathcal{D}(M_3,B)$ the codimension one topological operator realizing this $\mathbb{Z}_2$ symmetry operator in the presence of a 1-form background $B$. The mixed anomaly implies that
\begin{equation}
    \mathcal{D}(M_3',B)\exp\left( i\pi \int_{\mathbb{\mathbb{H}}'}\frac{\mathcal{P}(B)}{2} \right)
\end{equation}
is invariant under background transformations of $B$. Here, $\mathbb{H}'\subset M_4$ is a half-space of the spacetime $M_4$ with $\partial \mathbb{H}'=M_3'$. Similarly, the minimal abelian TFT $\mathcal{A}^{2,1}$ is an edge mode, and the combination:
\begin{equation}
    \mathcal{A}^{2,1}(M_3,B)\exp\left( i\pi \int_{\mathbb{H}}\frac{\mathcal{P}(B)}{2} \right)
\end{equation}
is also invariant under background transformations of $B$, here $\partial \mathbb{H}=M_3$. We can therefore consider
\begin{equation}\label{eq:Fun}
    \mathcal{A}^{2,1}(M_3,B)\exp\left( i\pi \int_{\mathbb{H}''}\frac{\mathcal{P}(B)}{2} \right) \mathcal{D}(M_3',B)
\end{equation}
with $\partial \mathbb{H}''=M_3-M_3'$ and an invertible codimension one defect in $SO(3)_-$ theory is constructed contracting $\mathbb{H}''$ and setting $M_3=M_3'$. Gauging $B$ to the global form $SU(2)$, this defect becomes non-invertible.

Now we turn to the string theory realization of the duality defects. We introduce a 7-brane of type $III^*$ wrapped on $M_3\times S^5$ with the branch cut intersecting with D3-branes at $M_3'$, separating the 4D spacetime into two parts. From table \ref{tab:Fibs}, $\tau=i$ is the fixed value of $III^*$ monodromy so we are able to have $\tau=i$ on the full worldvolume of the D3-branes and realize the operator $\mathcal{D}(M_3',B_2^\rho)$ on $M_3'$. The 3D TFT on $M_3$ is determined by the type of 7-brane, which can be read from table \ref{tab:Fibs} for type $III^*$ is $\mathcal{A}^{2,1}[B_2^\rho] \otimes \mathcal{T}'$. The left picture in figure \ref{fig:KOZ} illustrates this construction in terms of the 5D symmetry TFT slab.

\begin{figure}
    \centering
    \scalebox{0.8}{
    \begin{tikzpicture}
	\begin{pgfonlayer}{nodelayer}
		\node [style=none] (0) at (-2.5, 1.5) {};
		\node [style=none] (1) at (2.5, 1.5) {};
		\node [style=none] (2) at (-2.5, -1.5) {};
		\node [style=none] (3) at (2.5, -1.5) {};
		\node [style=Star] (4) at (0, 0) {};
		\node [style=none] (5) at (0, -1.5) {};
		\node [style=none] (7) at (0.75, -0.25) {$III^*$};
		\node [style=none] (8) at (0, 2) {$\ket{P_{SU(2)_0},D}$};
		\node [style=none] (9) at (-1.5, -2.5) {};
		\node [style=none] (10) at (1.5, -2.5) {};
		\node [style=none] (11) at (0, -3) {$x_\perp$};
		\node [style=Circle] (12) at (0, -1.5) {};
		\node [style=none] (13) at (0, -2) {$\mathcal{D}(M_3',B_2^\rho)$};
		\node [style=none] (14) at (-3.5, -1.5) {$r=0$};
		\node [style=none] (15) at (-3.5, 1.5) {$r=\infty$};
		\node [style=none] (16) at (-3.5, 1) {};
		\node [style=none] (17) at (-3.5, -1) {};
        \node [style=none] (29) at (0, 0.5) {$\mathcal{A}^{2,1}(B_2^\rho)\otimes \mathcal{T}'$};
        \node [style=none] (18) at (3.5, 0) {};
		\node [style=none] (19) at (5.5, 0) {};
		\node [style=none] (20) at (6.5, 0) {};
		\node [style=none] (21) at (11.5, 0) {};
		\node [style=none] (22) at (9, 0) {};
		\node [style=none] (23) at (4.5, -0.5) {5D $\rightarrow$ 4D};
		\node [style=none] (24) at (9, -0.5) {$\mathcal{N}(B_2^\rho)\otimes \mathcal{T}'$};
		\node [style=CircleRed] (25) at (9, 0) {};
		\node [style=none] (26) at (7.75, 0.5) {$Z_{SU(2)_0}(D)$};
		\node [style=none] (28) at (10.25, 0.5) {$Z_{SU(2)_0}(D)$};
	\end{pgfonlayer}
	\begin{pgfonlayer}{edgelayer}
		\draw [style=ThickLine] (0.center) to (1.center);
		\draw [style=ThickLine] (3.center) to (2.center);
		\draw [style=DottedLine] (5.center) to (4);
		\draw [style=ArrowLineRight] (9.center) to (10.center);
		\draw [style=ArrowLineRight] (17.center) to (16.center);
        \draw [style=ThickLine] (20.center) to (21.center);
		\draw [style=ArrowLineRight] (18.center) to (19.center);
	\end{pgfonlayer}
\end{tikzpicture}
    }
    \caption{Insertion of a 7-brane of type $III^*$ into the 5D symmetry TFT of 4D $\mathcal{N}=4$ SYM with vertically running branch cut and global structure $SU(2)_0$ constructs the symmetry operator $\mathcal{N}(M_3,B_2^\rho)=\mathcal{A}^{2,1}(M_3,B_2^\rho)\otimes \mathcal{D}(M_3,B_2^\rho)$ introduced in \cite{Kaidi:2021xfk}. Here tensor product denotes stacking as a consequence of contracting the branch cut. The background $B_2^\rho$ is the background field for the global form $SO(3)_-$. In particular it is not the background field for $SU(2)$ and therefore $\mathcal{N}$ couples dynamically to the theories on both half-spaces.}
    \label{fig:KOZ}
\end{figure}

Contracting the 5D slab to 4D then stacks $\mathcal{A}^{2,1}(M_{3}, B_{2}^{\rho})$ and $\mathcal{D}(M_{3}^{\prime}, B_{2}^{\rho})$, which gives rise to the 3D duality defect $\mathcal{N}(M_{3},B_2^\rho)$ within the 4D spacetime. Note that the whole construction so far is independent of the choice of the global structure of the theory. In other words, one can choose any boundary condition at $r=\infty$ for the 5D symmetry TFT and then contract the slab. In the case of $SU(2)$ and $SO(3)_+$, the $B_2^\rho$ is a dynamical field in the 4D bulk so $\mathcal{N}(B_2^\rho)$ is a non-invertible defect since it non-trivially couples to the 4D theory. In the case of $SO(3)_-$, $B_2^\rho$ is a background and non-dynamical, therefore $\mathcal{N}(B_2^\rho)$ is invertible. This perfectly matches the result from the field theory perspective and is identified as a non-intrinsic defect in \cite{Kaidi:2022cpf}.

\subsection*{Half-space gauging construction}\label{subsec:half-space gauging su(2) duality}
Let us now give the string theoretic setup for the half-space gauging construction in \cite{Choi:2022zal}. For example, we insert a 7-brane of type $\mathfrak{F}=III^*$ into the bulk and with a horizontally running branch cut which funnels the anomaly sourced by this insertion to infinity (see figure \ref{fig:ShaoEtAl}).

\begin{figure}
    \centering
    \scalebox{0.8}{\begin{tikzpicture}
	\begin{pgfonlayer}{nodelayer}
		\node [style=none] (5) at (-6, 1.5) {};
		\node [style=none] (6) at (-1, 1.5) {};
		\node [style=none] (7) at (-6, -1.5) {};
		\node [style=none] (8) at (-1, -1.5) {};
		\node [style=Star] (10) at (-2, 0) {};
		\node [style=none] (11) at (-6, 0) {};
		\node [style=none] (13) at (-1.5, 0.5) {$III^*$};
		\node [style=none] (18) at (-3.5, 2) {$\bra{P_{SU(2)_{+,0}},D}$};
		\node [style=none] (20) at (-7, 1.5) {$r=\infty$};
		\node [style=none] (21) at (-7, 1) {};
		\node [style=none] (22) at (-7, -1) {};
		\node [style=none] (26) at (-5, -2) {};
		\node [style=none] (27) at (-2, -2) {};
		\node [style=none] (28) at (-3.5, -2.5) {$x_\perp$};
		\node [style=none] (29) at (-4, -0.5) {$\mathbb{H}$};
		\node [style=none] (31) at (0, 0) {};
		\node [style=none] (32) at (2, 0) {};
		\node [style=none] (33) at (3, 0) {};
		\node [style=none] (34) at (8, 0) {};
		\node [style=none] (35) at (5.5, 0) {};
		\node [style=none] (36) at (1, -0.5) {5D $\rightarrow$ 4D};
		\node [style=none] (37) at (5.5, -0.5) {$\mathcal{A}^{2,1}\otimes \mathcal{T}'$};
		\node [style=CircleRed] (38) at (5.5, 0) {};
		\node [style=none] (39) at (4.25, 0.5) {$Z_{P_{SO(3)_{+,0}}}(D)$};
		\node [style=none] (40) at (6.75, 0.5) {$Z_{SU(2)_{0}}(D)$};
		\node [style=none] (41) at (-4, 0.5) {$e^{i\pi \int P(B_2^\rho)/2}$};
	\end{pgfonlayer}
	\begin{pgfonlayer}{edgelayer}
		\draw [style=ThickLine] (5.center) to (6.center);
		\draw [style=ThickLine] (8.center) to (7.center);
		\draw [style=DottedLine] (11.center) to (10);
		\draw [style=ArrowLineRight] (22.center) to (21.center);
		\draw [style=ArrowLineRight] (26.center) to (27.center);
		\draw [style=ThickLine] (33.center) to (34.center);
		\draw [style=ArrowLineRight] (31.center) to (32.center);
	\end{pgfonlayer}
\end{tikzpicture}
    }
    \caption{Insertion of a 7-brane of type $III^*$ into the 5d symmetry TFT of 4D $\mathcal{N}=4$ SYM with horizontal running branch cut and topological boundary condition $SU(2)_0$ gives rise to half-space gauging as in \cite{Choi:2022zal}.}
    \label{fig:ShaoEtAl}
\end{figure}

\noindent We claim that this realizes half-space gauging. This follows since $B_2^\rho$ is oriented along the polarization of the global form $SO(3)_-$, we therefore have:
\begin{equation}\label{eq:LongComp}\begin{aligned}
&~~~\, \bra{P_{SU(2)_0},D}\exp\left( i\pi \int \mathcal{P}(B_2^\rho)/2 \right)\\  &=  \sum_d \braket{P_{SU(2)_0},D\,|\,P_{SO(3)_{-,0}},d}\bra{P_{SO(3)_{-,0}},d}\exp\left( i\pi \int \mathcal{P}(d)/2 \right)\\
 &=  \sum_d \braket{P_{SU(2)_1},D\,|\,P_{SO(3)_{-,0}},d}\bra{P_{SO(3)_{-,0}},d}\exp\left( i\pi \int \mathcal{P}(D)/2+\mathcal{P}(d)/2 \right)\\
  &=  \sum_d \bra{P_{SO(3)_{-,0}},d}\exp\left( i\pi \int \mathcal{P}(D)/2 +D\cup d+\mathcal{P}(d)/2 \right)\\
&=\sum_d \bra{P_{SO(3)_{-,1}},d}\exp\left( i\pi \int D\cup d\right)\exp\left( i\pi \int \mathcal{P}(D)/2 \right)
  \\
   &= \bra{P_{SO(3)_{+,1}},D}\exp\left( i\pi \int \mathcal{P}(D)/2 \right)\\
    &=\bra{P_{SO(3)_{+,0}},D} \\
\end{aligned}
\end{equation}
as anticipated by noting that the S-duality branch cut interchanges electric and magnetic lines. Here the sums run over $\mathbb{Z}_2$-valued 2-forms $d$ (see figure \ref{fig:ShaoEtAl}). 

\begin{figure}
    \centering
    \scalebox{0.8}{\begin{tikzpicture}
	\begin{pgfonlayer}{nodelayer}
		\node [style=none] (0) at (-3, 4) {$SU(2)_1$};
		\node [style=none] (1) at (0, 4) {$SO(3)_{+,1}$};
		\node [style=none] (2) at (-3, 1.5) {$SU(2)_0$};
		\node [style=none] (3) at (0, 1.5) {$SO(3)_{+,0}$};
		\node [style=none] (4) at (3, 1.5) {$SO(3)_{-,0}$};
		\node [style=none] (5) at (3, 4) {$SO(3)_{-,1}$};
		\node [style=none] (12) at (-3, 3.5) {};
		\node [style=none] (13) at (0, 3.5) {};
		\node [style=none] (14) at (3, 3.5) {};
		\node [style=none] (15) at (-3, 2) {};
		\node [style=none] (16) at (0, 2) {};
		\node [style=none] (17) at (3, 2) {};
		\node [style=none] (18) at (-3, 0) {$SU(2)_1$};
		\node [style=none] (19) at (0, 0) {$SO(3)_{+,1}$};
		\node [style=none] (20) at (-3, -2.5) {$SU(2)_0$};
		\node [style=none] (21) at (0, -2.5) {$SO(3)_{+,0}$};
		\node [style=none] (22) at (3, -2.5) {$SO(3)_{-,0}$};
		\node [style=none] (23) at (3, 0) {$SO(3)_{-,1}$};
		\node [style=none] (24) at (-3, -0.5) {};
		\node [style=none] (25) at (0, -0.5) {};
		\node [style=none] (26) at (3, -0.5) {};
		\node [style=none] (27) at (-3, -2) {};
		\node [style=none] (28) at (0, -2) {};
		\node [style=none] (29) at (3, -2) {};
		\node [style=none] (30) at (-3, -4) {$SU(2)_1$};
		\node [style=none] (31) at (0, -4) {$SO(3)_{+,1}$};
		\node [style=none] (32) at (-3, -6.5) {$SU(2)_0$};
		\node [style=none] (33) at (0, -6.5) {$SO(3)_{+,0}$};
		\node [style=none] (34) at (3, -6.5) {$SO(3)_{-,0}$};
		\node [style=none] (35) at (3, -4) {$SO(3)_{-,1}$};
		\node [style=none] (38) at (3, -4.5) {};
		\node [style=none] (41) at (3, -6) {};
		\node [style=none] (42) at (-5.5, 2.75) {$\mathcal{O}^{[1,0]}:$};
		\node [style=none] (43) at (-5.5, -1.25) {$\mathcal{O}^{[0,1]}:$};
		\node [style=none] (44) at (-5.5, -5.25) {$\mathcal{O}^{[1,1]}:$};
		\node [style=none] (45) at (0, 2.75) {};
		\node [style=none] (48) at (-3, -1.25) {};
		\node [style=none] (49) at (1.5, -1.5) {};
		\node [style=none] (50) at (3, -5.25) {};
		\node [style=none] (51) at (1.5, -1) {};
		\node [style=none] (58) at (-3, -4.5) {};
		\node [style=none] (59) at (0, -4.5) {};
		\node [style=none] (60) at (-3, -6) {};
		\node [style=none] (61) at (0, -6) {};
		\node [style=none] (62) at (-1.5, -5.5) {};
		\node [style=none] (63) at (-1.5, -5) {};
		\node [style=none] (66) at (4.5, 4.5) {};
		\node [style=none] (67) at (4.5, 1) {};
		\node [style=none] (68) at (-6.5, 4.5) {};
		\node [style=none] (69) at (-6.5, 1) {};
		\node [style=none] (70) at (4.5, 0.5) {};
		\node [style=none] (71) at (4.5, -3) {};
		\node [style=none] (72) at (-6.5, 0.5) {};
		\node [style=none] (73) at (-6.5, -3) {};
		\node [style=none] (74) at (4.5, -3.5) {};
		\node [style=none] (75) at (4.5, -7) {};
		\node [style=none] (76) at (-6.5, -3.5) {};
		\node [style=none] (77) at (-6.5, -7) {};
	\end{pgfonlayer}
	\begin{pgfonlayer}{edgelayer}
		\draw [style=ArrowLineRight] (45.center) to (13.center);
		\draw [style=ArrowLineRight] (45.center) to (16.center);
		\draw [style=ArrowLineRight] (48.center) to (24.center);
		\draw [style=ArrowLineRight] (48.center) to (27.center);
		\draw [style=ArrowLineRight] (50.center) to (38.center);
		\draw [style=ArrowLineRight] (50.center) to (41.center);
		\draw [style=ArrowLineRight] (45.center) to (12.center);
		\draw [style=ArrowLineRight] (45.center) to (17.center);
		\draw [style=ArrowLineRight] (45.center) to (14.center);
		\draw [style=ArrowLineRight] (45.center) to (15.center);
		\draw [style=ArrowLineRight, in=60, out=-180, looseness=0.75] (49.center) to (28.center);
		\draw [style=ArrowLineRight, in=120, out=0, looseness=0.75] (49.center) to (29.center);
		\draw [style=ArrowLineRight, in=-120, out=0, looseness=0.75] (51.center) to (26.center);
		\draw [style=ArrowLineRight, in=-60, out=180, looseness=0.75] (51.center) to (25.center);
		\draw [style=ArrowLineRight, in=60, out=-180, looseness=0.75] (62.center) to (60.center);
		\draw [style=ArrowLineRight, in=120, out=0, looseness=0.75] (62.center) to (61.center);
		\draw [style=ArrowLineRight, in=-120, out=0, looseness=0.75] (63.center) to (59.center);
		\draw [style=ArrowLineRight, in=-60, out=180, looseness=0.75] (63.center) to (58.center);
		\draw (68.center) to (66.center);
		\draw (66.center) to (67.center);
		\draw (68.center) to (69.center);
		\draw (69.center) to (67.center);
		\draw (72.center) to (70.center);
		\draw (70.center) to (71.center);
		\draw (72.center) to (73.center);
		\draw (73.center) to (71.center);
		\draw (76.center) to (74.center);
		\draw (74.center) to (75.center);
		\draw (76.center) to (77.center);
		\draw (77.center) to (75.center);
	\end{pgfonlayer}
\end{tikzpicture}    }
    \caption{Operators $\mathcal{O}^{[q,p]}$ for 4D $\mathcal{N}=4$ $\mathfrak{su}(2)$ gauge theory.}
    \label{fig:Opq}
\end{figure}

Note that the operation of half-space gauging is not realized universally by one type of 7-brane. For example, in the above $\bra{P_{SO(3)_{-,r}}}$ is mapped to $\bra{P_{SO(3)_{-,r+1}}}$, with index $r$ mod 2, by branch cut stacking and the global structure is preserved. 

For this reason, it is instructive to study the other possible operators of type $\mathcal{O}^{[q,p]}$ as introduced in \eqref{eq:Opq}. The full generating set is: 
\begin{equation}\label{eq:definition of su(2) branch cut operators}
    \begin{aligned}
        \mathcal{O}^{[1,0]}&=\exp\left( i\pi \int \mathcal{P}(C_2)/2 \right)\\
        \mathcal{O}^{[0,1]}&=\exp\left( i\pi \int \mathcal{P}(B_2)/2 \right)\\
        \mathcal{O}^{[1,1]}&=\exp\left( i\pi \int \mathcal{P}(B_2+C_2)/2 \right)\\
    \end{aligned}
\end{equation}
and for the case of 7-branes of type $\mathfrak{F}=III^*$ we have $\mathcal{O}^{[1,1]}$ realized on the branch cut. Repeating computations similar to \eqref{eq:LongComp}, we find the results displayed in figure \ref{fig:Opq}. The global forms related by gauging are \cite{Kaidi:2022uux}:
\begin{equation}\label{eq:GlobalFormPair}\begin{aligned}
    SU(2)_0~&\leftrightarrow~ SO(3)_{+,0}\\
    SU(2)_1~&\leftrightarrow~ SO(3)_{-,0}\\
    SO(3)_{+,1}~&\leftrightarrow~ SO(3)_{-,1}
    \end{aligned}
\end{equation}

We can now give the completely general procedure. Pick a pair of global forms in \eqref{eq:GlobalFormPair} to be realized on two half-spaces. Then, determine the operator $\mathcal{O}^{[q,p]}$ connecting these
\begin{equation}\label{eq:GlobalFormPair2}\begin{aligned}
    \mathcal{O}^{[1,1]}\,:&\qquad SU(2)_0~\!\!\!\!\!\!&&\leftrightarrow~ SO(3)_{+,0}\\
    \mathcal{O}^{[1,0]}\,:&\qquad SU(2)_1~\!\!\!\!\!\!&&\leftrightarrow~ SO(3)_{-,0}\\
    \mathcal{O}^{[0,1]}\,:&\qquad SO(3)_{+,1}~\!\!\!\!\!\!&&\leftrightarrow~ SO(3)_{-,1}
    \end{aligned}
    \end{equation}
Next determine the 7-brane which supports $\mathcal{O}^{[q,p]}$ on their branch cut. These are for example $\mathfrak{F}=III^*,  I_1^{[0,1]},I_1^{[1,0]}$ respectively where the latter two are D7- and $[0,1]$-7-branes. Note that in all cases $B_2^\rho$ is neither the background field for the left nor for the right global form on each half-space. Consequently, the minimal abelian TFT supported on the 7-brane interacts with the degrees of freedom in both half-spaces. This leads to the defect realizing a non-invertible symmetry.

\subsection{Triality Defects in \texorpdfstring{$\mathfrak{su}(2)$}{} SYM\texorpdfstring{$_{\mathcal{N} = 4}$}{} Theory}
Let us move to triality defects in the $\mathfrak{su}(2)$ theory. We will see that in this case, only the half-space gauging construction works, which aligns with the fact that triality defects for $\mathfrak{su}(2)$ are intrinsic. 

\subsection*{Kramers-Wannier-like construction}

We first consider the Kramers-Wannier-like construction and show it does not work. In this case the branch cut of the 7-brane is vertical and intersects with the worldvolume of the D3-branes. 
Therefore, we are restricted in our construction to 7-branes whose monodromy has fixed points at $\tau=\frac{i\pi}{3}$ or $\tau=\frac{2\pi i}{3}$, which are values necessary for triality defects \cite{Choi:2022zal, Kaidi:2022uux}. 

Let us take the type $IV^*$ 7-brane as an example. Consider a similar setup as in Figure \ref{fig:KOZ}, but with the type $IV^*$ instead of $III^*$ 7-brane. Naively, one may expect 3D TFTs $\mathcal{A}^{3,2}(B_2^\rho)$ and $\mathcal{D}(M_3',B_2^\rho)$ living on the two ends of the branch cut, respectively. However, in the case of $\mathfrak{su}(2)$ theory, where $B_2$ and $C_2$ are both $\mathbb{Z}_2$ fields in the bulk, there is in fact no non-trivial $B_2^\rho$ preserved by the type $IV^*$ 7-brane monodromy and hence the 7-brane does not source an anomaly. So the setup with inserted 7-branes reduces to studying the branch cut. This is nicely aligned with our discussion on the mixed anomalies shown in \eqref{eq:mixed anomaly term in the bulk}. For triality defects of $\mathfrak{su}(2)$ theory, the mixed anomaly between the triality symmetry and the 1-form symmetry is trivial.

\subsection*{Half-space gauging construction}

\begin{figure}
    \centering
    \scalebox{0.8}{\begin{tikzpicture}
	\begin{pgfonlayer}{nodelayer}
		\node [style=none] (0) at (-2.5, 1.5) {};
		\node [style=none] (1) at (2.5, 1.5) {};
		\node [style=none] (2) at (-2.5, -1.5) {};
		\node [style=none] (3) at (2.5, -1.5) {};
		\node [style=Star] (4) at (0.75, 0.25) {};
		\node [style=none] (6) at (0, 2) {$\bra{P_{SO(3)_{-,0}},D}$};
		\node [style=none] (8) at (-3.5, -1.5) {$r=0$};
		\node [style=none] (9) at (-3.5, 1.5) {$r=\infty$};
		\node [style=none] (10) at (-3.5, 1) {};
		\node [style=none] (11) at (-3.5, -1) {};
		\node [style=Star] (15) at (-0.75, -0.25) {};
		\node [style=none] (16) at (-2.5, 0.25) {};
		\node [style=none] (17) at (-2.5, -0.25) {};
		\node [style=none] (18) at (1.375, 0.25) {$I_1^{[1,0]}$};
		\node [style=none] (19) at (-0.125, -0.25) {$I_1^{[0,1]}$};
		\node [style=none] (20) at (-2, 0.75) {$\mathcal{O}^{[1,0]}$};
		\node [style=none] (21) at (-2, -0.75) {$\mathcal{O}^{[0,1]}$};
		\node [style=none] (22) at (3.5, 0) {};
		\node [style=none] (23) at (5.5, 0) {};
		\node [style=none] (24) at (6.5, 1.5) {};
		\node [style=none] (25) at (14, 1.5) {};
		\node [style=none] (26) at (10.25, 1.5) {};
		\node [style=none] (27) at (4.5, 0.5) {5D $\rightarrow$ 4D};
		\node [style=none] (28) at (9, 1) {$\mathcal{N}_L\otimes\mathcal{T}'_L$};
		\node [style=CircleRed] (29) at (9, 1.5) {};
		\node [style=none] (30) at (7.25, 2) {$Z_{SU(2)_0}(D)$};
		\node [style=none] (31) at (13.25, 2) {$Z_{SO(3)_{-,0}}(D)$};
		\node [style=CircleRed] (32) at (11.5, 1.5) {};
		\node [style=none] (33) at (10.25, 2) {$Z_{SU(2)_1}(D)$};
		\node [style=none] (34) at (11.5, 1) {$\mathcal{N}_R\otimes\mathcal{T}'_R$};
		\node [style=none] (35) at (10.25, 0) {$=$};
		\node [style=none] (36) at (6.5, -1.5) {};
		\node [style=none] (37) at (14, -1.5) {};
		\node [style=none] (38) at (10.25, -1.5) {};
		\node [style=none] (39) at (10.25, -2) {$\mathcal{N}_3\otimes\mathcal{T}'$};
		\node [style=CircleRed] (40) at (10.25, -1.5) {};
		\node [style=none] (41) at (7.75, -1) {$Z_{SU(2)_0}(D)$};
		\node [style=none] (42) at (12.75, -1) {$Z_{SO(3)_{-,0}}(D)$};
        \node [style=none] (43) at (-1.5, -2) {};
		\node [style=none] (44) at (1.5, -2) {};
		\node [style=none] (45) at (0, -2.5) {$x_\perp$};
	\end{pgfonlayer}
	\begin{pgfonlayer}{edgelayer}
		\draw [style=ThickLine] (0.center) to (1.center);
		\draw [style=ThickLine] (3.center) to (2.center);
		\draw [style=ArrowLineRight] (11.center) to (10.center);
		\draw [style=DottedLine] (16.center) to (4);
		\draw [style=DottedLine] (17.center) to (15);
		\draw [style=ThickLine] (24.center) to (25.center);
		\draw [style=ArrowLineRight] (22.center) to (23.center);
		\draw [style=ThickLine] (36.center) to (37.center);
        \draw [style=ArrowLineRight] (43.center) to (44.center);
	\end{pgfonlayer}
\end{tikzpicture}}
    \caption{Left: sketch of the 5D symmetry TFT for 4D $\mathcal{N}=4$ $\mathfrak{su}(2)$ theory. With two 7-brane insertions of type $\mathfrak{F}=I_1^{[1,0]},I_1^{[0,1]}$ and branch cut operators $\mathcal{O}^{[1,0]},\mathcal{O}^{[0,1]}$ respectively. Right: sketch of the 5D slab contracted to 4D, top and bottom are equivalent. The top, bottom figure shows the 4D theory when the 7-brane insertions are displaced, aligned along $x_\perp$ respectively. The latter results in the triality defect $\mathcal{N}_3$.   }
    \label{fig:Triality}
\end{figure}

All ingredients we need to build triality defects are already introduced in Section \ref{subsec:half-space gauging su(2) duality}. Let us take $SO(3)_{-,0}$ theory as an example. From (\ref{eq:GlobalFormPair2}) we know that $SO(3)_{-,0}$ turns into $SU(2)_1$ under half-space gauging, which can be realized by acting with the operator $\mathcal{O}^{[1,0]}$ living on the horizontal branch cut off a $(0,1)$-7-brane. We can then insert a D7-brane into the bulk, so that the operator $\mathcal{O}^{[0,1]}$ on the horizontal branch cut is introduced. This leads to adding a counterterm for the $SU(2)_1$ theory, so that it becomes the $SU(2)_0$ theory. According to \cite{Aharony:2013hda}, the $SU(2)_0$ theory is indeed dual to $SO(3)_{-,0}$ via the modular transformation $\mathbb{T}\cdot \mathbb{S}$. Therefore by contracting the 5D TFT slab, we indeed get a triality defect $\mathcal{N}_3$.

Furthermore, in this case we are able to determine the lines of the TFT $\mathcal{N}_3$. First, note that the insertion of a single $(p,q)$ 7-brane gives a 3D defect with no lines of its own. See the discussion in subsection \ref{sec:Proposal}. However, when inserting multiple 7-branes, as shown in figure \ref{fig:Triality}, the combined system can contain line defects which are constructed by $(p,q)$ string junctions terminating at the 7-brane insertion and the D3-brane locus. This generalizes the setup displayed in figure \ref{fig:MinimalLines}. These lines constitute the lines of the 3D TFT obtain from the fusion of the TFT supported at the individual 7-brane insertions. In the case of the triality defect we may have Y-shaped string junctions ending on $\mathcal{N}_{L},\mathcal{N}_{R}$ and at $r=0$ and these descend to the lines of the triality defect $\mathcal{N}_3$, the fusion of $\mathcal{N}_{L}$ with $\mathcal{N}_{R}$. 

The line defects of $\mathcal{N}_3$ are therefore determined by the total monodromy $\rho=\rho_L\rho_R$. In the case where $\mathcal{N}_{L},\mathcal{N}_{R}$ are respectively engineered by $I_1^{[0,1]},I_1^{[1,0]}$ type fibers the overall monodromy has trivial cokernel $\textnormal{coker}(\rho-1)$ and therefore there are no additional lines coupling to the fields $B_2,C_2$.


Triality defects for other global structures of $\mathfrak{su}(2)$ can be realized following the same steps. Figure \ref{fig:Trialityforallsu2} illustrates the generic construction, for which we specify ingredients for all cases in the following table.
{\renewcommand{\arraystretch}{1.35}
\begin{table}[H]
\begin{center}
\begin{tabular}{|c|c|c|}
\hline
   $P_{G_m}$  & $\mathfrak{F}_R, \mathfrak{F}_L$ & $Z_{P_L}(D), Z_{P'}(D)$, $Z_{P_R}(D)$  \\
   \hline 
   $SU(2)_0$  & $III^*, I_1^{[1,0]}$ & $Z_{SO(3)_{+,1}}(D), Z_{SO(3)_{+,0}}(D)$, $Z_{SU(2)_0}(D)$  \\
   \hline 
   $SU(2)_1$  & $I_1^{[1,0]},III^*$ & $Z_{SO(3)_{-,1}}(D), Z_{SO(3)_{-,0}}(D)$, $Z_{SU(2)_1}(D)$  \\
   \hline
   $SO(3)_{+,0}$  & $III^*,I_1^{[0,1]}$ & $Z_{SU(2)_1}(D), Z_{SU(2)_0}(D)$, $Z_{SO(3)_{+,0}}(D)$ \\
   \hline
   $SO(3)_{+,1}$  & $I_1^{[0,1]},III^*$ & $Z_{SO(3)_{-,0}}(D), Z_{SO(3)_{-,1}}(D)$, $Z_{SO(3)_{+,1}}(D)$\\
   \hline 
   $SO(3)_{-,0}$  & $I_1^{[1,0]},I_1^{[0,1]}$ & $Z_{SU(2)_{0}}(D), Z_{SU(2)_{1}}(D)$, $Z_{SO(3)_{-,0}}(D)$ \\
   \hline
   $SO(3)_{-,1}$  & $I_1^{[0,1]},I_1^{[1,0]}$ & $Z_{SO(3)_{+,0}}(D), Z_{SO(3)_{+,1}}(D)$, $Z_{SO(3)_{-,1}}(D)$ \\
   \hline
\end{tabular}
\end{center}
\caption{Triality defect constructions for all global structures of $\mathfrak{su}(2)$, with all ingredients illustrated in Figure \ref{fig:Trialityforallsu2}.}
    \label{tab:all triality defects for su(2)}
\end{table}}

\begin{figure}
    \centering
    \scalebox{0.8}{\begin{tikzpicture}
	\begin{pgfonlayer}{nodelayer}
		\node [style=none] (0) at (-2.5, 1.5) {};
		\node [style=none] (1) at (2.5, 1.5) {};
		\node [style=none] (2) at (-2.5, -1.5) {};
		\node [style=none] (3) at (2.5, -1.5) {};
		\node [style=Star] (4) at (0.75, 0.25) {};
		\node [style=none] (6) at (0, 2) {$\bra{P,D}$};
		\node [style=none] (8) at (-3.5, -1.5) {$r=0$};
		\node [style=none] (9) at (-3.5, 1.5) {$r=\infty$};
		\node [style=none] (10) at (-3.5, 1) {};
		\node [style=none] (11) at (-3.5, -1) {};
		\node [style=Star] (15) at (-0.75, -0.25) {};
		\node [style=none] (16) at (-2.5, 0.25) {};
		\node [style=none] (17) at (-2.5, -0.25) {};
		\node [style=none] (18) at (1.375, 0.25) {$\mathfrak{F}_R$};
		\node [style=none] (19) at (-0.125, -0.25) {$\mathfrak{F}_L$};
		\node [style=none] (20) at (-2, 0.75) {$\mathcal{O}^{\mathfrak{F}_R}$};
		\node [style=none] (21) at (-2, -0.75) {$\mathcal{O}^{\mathfrak{F}_L}$};
		\node [style=none] (22) at (3.5, 0) {};
		\node [style=none] (23) at (5.5, 0) {};
		\node [style=none] (24) at (6.5, 1.5) {};
		\node [style=none] (25) at (14, 1.5) {};
		\node [style=none] (26) at (10.25, 1.5) {};
		\node [style=none] (27) at (4.5, 0.5) {5D $\rightarrow$ 4D};
		\node [style=none] (28) at (9, 1) {$\mathcal{N}_L\otimes\mathcal{T}'_L$};
		\node [style=CircleRed] (29) at (9, 1.5) {};
		\node [style=none] (30) at (7.25, 2) {$Z_{P_L}(D)$};
		\node [style=none] (31) at (13.25, 2) {$Z_{P_R}(D)$};
		\node [style=CircleRed] (32) at (11.5, 1.5) {};
		\node [style=none] (33) at (10.25, 2) {$Z_{P'}(D)$};
		\node [style=none] (34) at (11.5, 1) {$\mathcal{N}_R\otimes\mathcal{T}'_R$};
		\node [style=none] (35) at (10.25, 0) {$=$};
		\node [style=none] (36) at (6.5, -1.5) {};
		\node [style=none] (37) at (14, -1.5) {};
		\node [style=none] (38) at (10.25, -1.5) {};
		\node [style=none] (39) at (10.25, -2) {$\mathcal{N}_3\otimes\mathcal{T}'$};
		\node [style=CircleRed] (40) at (10.25, -1.5) {};
		\node [style=none] (41) at (7.75, -1) {$Z_{P_L}(D)$};
		\node [style=none] (42) at (12.75, -1) {$Z_{P_R}(D)$};
        \node [style=none] (43) at (-1.5, -2) {};
		\node [style=none] (44) at (1.5, -2) {};
		\node [style=none] (45) at (0, -2.5) {$x_\perp$};
	\end{pgfonlayer}
	\begin{pgfonlayer}{edgelayer}
		\draw [style=ThickLine] (0.center) to (1.center);
		\draw [style=ThickLine] (3.center) to (2.center);
		\draw [style=ArrowLineRight] (11.center) to (10.center);
		\draw [style=DottedLine] (16.center) to (4);
		\draw [style=DottedLine] (17.center) to (15);
		\draw [style=ThickLine] (24.center) to (25.center);
		\draw [style=ArrowLineRight] (22.center) to (23.center);
		\draw [style=ThickLine] (36.center) to (37.center);
        \draw [style=ArrowLineRight] (43.center) to (44.center);
	\end{pgfonlayer}
\end{tikzpicture}}
    \caption{Left: Triality defects construction for 4D $\mathcal{N}=4$ $\mathfrak{su}(2)$ theory from 
    the 5D symmetry TFT. With two 7-brane insertions of type $\mathfrak{F}_L,\mathfrak{F}_R$ and branch cut operators $\mathcal{O}^{\mathfrak{F}_L},\mathcal{O}^{\mathfrak{F}_R}$ respectively. }
    \label{fig:Trialityforallsu2}
\end{figure}


\subsection{Defects in the \texorpdfstring{$\mathfrak{su}(N)$}{} Case}
With these examples in hand, we now generalize our discussion to 
construct duality and triality defects for $\mathfrak{su}(N)$ theories with $\mathcal{N} = 4$ supersymmetry. 
In this case, a complete treatment would first require the classification of possible global forms of the theory. 
Rather than proceed in this way, we mainly focus on how things work in the purely electric case, where the gauge group is $SU(N)$, as well as the purely magnetic case, where the gauge group is $SU(N) / \mathbb{Z}_N$.

First, consider the Kramers-Wannier-like construction. For duality/triality defects of $\mathfrak{su}(N)$ theories, if there is a non-trivial $B_2^\rho$ living on the vertical branch cut, one can always choose a topological boundary condition for the 5D TFT, i.e., a global structure of the 4D gauge theory, so that $B_2^\rho$ is the non-dynamical background gauge field. One then ends up with an invertible duality/triality defect $\mathcal{N}(M_3, B_2^\rho)$ in this duality frame, so it is a non-intrinsic duality/triality defect \cite{Kaidi:2022cpf}. On the other hand, if there is no non-trivial $B_2^\rho$ that can be defined on the branch cut, the mixed anomaly is trivial. In that case, the Kramers-Wannier-like construction does not work. This leads to an intrinsic duality/triality defect \cite{Kaidi:2022cpf} which can only be realized by half-space gauging.

From now on, we focus on providing a universal realization for the half-space gauging construction with $N$ generic.

\subsection*{Duality defects}
As we already discussed, 7-branes with the horizontal branch cut in the 5D TFT give rise to topological manipulations possibly changing the global structure, and / or adding counter-terms for the 4D theory. In order to build duality defects, we need 7-branes with topological manipulations which can be compensated exactly by that of the modular $\mathbb{S}$ or $\mathbb{S}^{-1}$ transformation. Recall the $\mathbb{S}$ transformation is defined by 
\begin{equation}
    \mathbb{S}: [q,p]\rightarrow [q,p]\begin{pmatrix} 0&1\\-1&0 \end{pmatrix}, \quad \tau \rightarrow -\frac{1}{\tau}
\end{equation}
where $p$ and $q$ are electric and magnetic charges respectively. In our string theory construction, $p$ and $q$ are charges for F1- and D1-strings respectively. Therefore, the candidate 7-branes for duality defects are type $III$ and $III^*$, whose monodromy matrices $\rho$ have same actions on the $[q,p]$ charges as $\mathbb{S}$ and $\mathbb{S}^{-1}$. Therefore, by inserting $III$ or $III^*$ 7-branes and contracting the 5D TFT slab, the 4D theory in the half-space acted by the branch cut is dual  to the original theory in the other half-space via a modular $\mathbb{S}$ or $\mathbb{S}^{-1}$ transformation. This realizes duality defects at $\tau=i$. See the left picture of Figure \ref{fig:III*resolution}.

\begin{figure}
    \centering
    \scalebox{0.8}{\begin{tikzpicture}
	\begin{pgfonlayer}{nodelayer}
		\node [style=none] (0) at (-6.5, 1.5) {};
		\node [style=none] (1) at (-1.5, 1.5) {};
		\node [style=none] (2) at (-6.5, -1.5) {};
		\node [style=none] (3) at (-1.5, -1.5) {};
		\node [style=none] (4) at (-4, 2) {$\ket{P,D}$};
		\node [style=none] (5) at (-7.5, -1.5) {$r=0$};
		\node [style=none] (6) at (-7.5, 1.5) {$r=\infty$};
		\node [style=none] (7) at (-7.5, 1) {};
		\node [style=none] (8) at (-7.5, -1) {};
		\node [style=Star] (12) at (-4, 0) {};
		\node [style=none] (16) at (-6, 0) {};
		\node [style=none] (19) at (-6.5, 0) {};
		\node [style=none] (21) at (-3.25, 0.25) {$III^*$};
		\node [style=none] (22) at (1.5, 1.5) {};
		\node [style=none] (23) at (6.5, 1.5) {};
		\node [style=none] (24) at (1.5, -1.5) {};
		\node [style=none] (25) at (6.5, -1.5) {};
		\node [style=none] (26) at (4, 2) {$\ket{P,D}$};
		\node [style=Star] (34) at (5, 0) {};
		\node [style=none] (35) at (4, 0.5) {};
		\node [style=none] (36) at (4, -0.5) {};
		\node [style=none] (37) at (2.25, -0.5) {};
		\node [style=none] (38) at (2.25, 0) {};
		\node [style=none] (39) at (2.25, 0.5) {};
		\node [style=none] (40) at (1.5, -0.5) {$\mathcal{O}_A^6$};
		\node [style=none] (41) at (1.5, 0) {$\mathcal{O}_B$};
		\node [style=none] (42) at (1.5, 0.5) {$\mathcal{O}_C^2$};
		\node [style=none] (44) at (5.5, 0.5) {$III^*$};
		\node [style=none] (45) at (0, 0) {$=$};
		\node [style=none] (46) at (-4, -2) {};
		\node [style=none] (47) at (4, -2) {};
		\node [style=none] (48) at (-2, -3.5) {};
		\node [style=none] (49) at (2.5, -3.5) {};
		\node [style=none] (50) at (-4, -4.25) {};
		\node [style=none] (51) at (4, -4.25) {};
		\node [style=none] (52) at (0.25, -4.25) {};
		\node [style=CircleRed] (54) at (0, -4.25) {};
		\node [style=none] (55) at (-2, -4.75) {$Z_{P'}(D)=\mathbb{S}^{-1}Z_{P}(D)$};
		\node [style=none] (58) at (2, -4.75) {$Z_{P}(D)$};
		\node [style=none] (59) at (-5.25, -4.25) {$4D:$};
	\end{pgfonlayer}
	\begin{pgfonlayer}{edgelayer}
		\draw [style=ThickLine] (0.center) to (1.center);
		\draw [style=ThickLine] (3.center) to (2.center);
		\draw [style=ArrowLineRight] (8.center) to (7.center);
		\draw [style=DottedLine] (16.center) to (12);
		\draw [style=ThickLine] (22.center) to (23.center);
		\draw [style=ThickLine] (25.center) to (24.center);
		\draw [style=DottedLine] (39.center) to (35.center);
		\draw [style=DottedLine] (37.center) to (36.center);
		\draw [style=DottedLine] (35.center) to (34);
		\draw [style=DottedLine] (36.center) to (34);
		\draw [style=DottedLine] (38.center) to (34);
		\draw [style=ArrowLineRight] (46.center) to (48.center);
		\draw [style=ArrowLineRight] (47.center) to (49.center);
		\draw [style=ThickLine] (50.center) to (51.center);
	\end{pgfonlayer}
\end{tikzpicture}}
    \caption{Type $III^*$ 7-brane with monodromy inverse to S-duality. Left: the monodromy is localized onto a single branch cut. Right: the monodromy is localized onto three distinct branch cuts associated with stacks of $(p,q)$ 7-branes. Bottom: contraction of the 5D slab to 4D. }
    \label{fig:III*resolution}
\end{figure}

Let us now discuss topological operators living on the branch cut. From now on, we will focus on the case of the type $III^\ast$ 7-brane, but the following discussion also works for the type $III$ 7-brane similarly. For generic values of $N$, i.e., $B_2$ and $C_2$ both $\mathbb{Z}_N$-valued, it is not always possible to define a $B_2^\rho$ which is invariant under the $III^\ast$ monodromy matrix. So it seems unclear how to build topological operators living on the branch cut. However, it is always possible to factorize the monodromy matrix into elements in $SL(2,\mathbb{Z}_N)$, such that the branch cut is correspondingly separated for each element where a non-trivial $B_2^\rho$ can be defined. In fact, the type $III^\ast$ 7-brane can be constructed by three simple types of $(p,q)$ 7-branes: \footnote{We remark that the convention in this chapter follows that of \cite{Weigand:2018rez}, which is different from all other chapters of this dissertation.}
\begin{equation}\label{eq:definition ABC brane}
    A:(1,0), ~B:(3,1), ~C:(1,1),
\end{equation}
as 
\begin{equation}
    III^*: A^6BC^2.
\end{equation}
The monodromy matrix for a $(p,q)$ 7-brane is given by 
\begin{equation}
    \rho_{(p,q)}=\begin{pmatrix}1+pq&p^2\\-q^2&1-pq\end{pmatrix},
\end{equation}
under which $pB_2+qC_2$ is obviously always invariant. Therefore, based on our discussion in Section \ref{sec:SETUP}, on the branch cut with monodromy matrix $\rho_{(p,q)}$, we can define a 4D topological operator: 
\begin{equation}
    \mathcal{O}^{(p,q)}=\exp \left( {\frac{2\pi i}{N}\int \frac{\mathcal{P}(pB_2+qC_2)}{2}}\right)\,.
\end{equation}

Now we can separate the branch cut of the $III^*$ 7-brane into multiple ones corresponding to the monodromy matrices for $C^2,B$ and $A^6$, respectively. See the right picture of figure \ref{fig:III*resolution}. Stacking topological operators living on all branch cuts together, we end up with the topological operator for the type $III^\ast$ 7-brane:
\begin{equation}\label{eq:branchcut operator for e7}
    \mathcal{O}^{III^*}=\exp \left( \frac{2\pi i\times 6}{N}\int \frac{\mathcal{P}(B_2)}{2}  \right) \exp \left( \frac{2\pi i}{N}\int \frac{\mathcal{P}(3B_2+C_2)}{2}  \right)\exp \left( \frac{2\pi i\times 2}{N}\int \frac{\mathcal{P}(B_2+C_2)}{2}  \right)
\end{equation}
This is well-defined for $\mathbb{Z}_N$-valued $B_2$ and $C_2$ with $N$ generic. 

For the $\mathfrak{su}(2)$ theory, the first and the third factors in (\ref{eq:branchcut operator for e7}) are both trivial. The second term becomes 
$\exp \left( i\pi \int \mathcal{P}(B_2+C_2)/2  \right)$ because $B_2$ is $\mathbb{Z}_2$-valued. Now $\mathcal{O}^{III^*}$ exactly reduces to the operator $\mathcal{O}^{[1,1]}$ we discussed around (\ref{eq:definition of su(2) branch cut operators}), realizing non-invertible duality defects for $SU(2)$ and $SO(3)_+$. For the $\mathfrak{su}(3)$ theory, the operator reduces to 
\begin{equation}
    \mathcal{O}^{III^*}_{N=3}=\exp \left( \frac{2\pi i}{3}\int \frac{\mathcal{P}(C_2)}{2}  \right)\exp \left( \frac{4\pi i}{3}\int \frac{\mathcal{P}(B_2+C_2)}{2}  \right). 
\end{equation}
Acting with this operator on, for example, the boundary conditions of the 5D TFT which result in the $SU(3)_0$ theory, one can do a  similar computation as in (\ref{eq:LongComp}) and reach the $\overline{PSU(3)}_{0,0}$ theory\footnote{Here we denote global structures with the notation in \cite{Kaidi:2022cpf}, with the two sub-indices for different background gauge fields and counterterms respectively. The overline means an opposite sign for the background field compared to that without an overline. Explicitly, $\overline{PSU(3)}_{0,0}$ means background field purely magnetic $-C_2$ without counterterm.}, which is dual to $SU(3)_0$ theory via the modular $\mathbb{S}$ transformation \cite{Kaidi:2022cpf}. Hence, this realizes non-invertible duality defects. Note the fact that duality defects for $\mathfrak{su}(2)$ and $\mathfrak{su}(3)$ are respectively non-intrinsic and intrinsic, but our construction provides a simple unified realization for them. In fact, our construction works for all values of $N$ regardless of its divisors.

\subsection*{Triality defects}
Triality defects from 7-branes can be built following similar steps as in duality defects. In order to find an order 3 topological manipulation, we need to consider 7-branes with monodromy matrices acting on $[q,p]$-string charges which can be compensated by modular transformations  $\mathbb{S}^{m}\cdot \mathbb{T}^{n}$ or $\mathbb{T}^{m}\cdot \mathbb{S}^{n}$, with $m=\pm1, n=\pm1$. The candidate 7-branes with this desired property include the Kodaiar types $II, II^\ast, IV$ and $IV^\ast$.

As in the case of duality defects, we will focus on one certain type of 7-brane and all other candidates work in a similar way. Let us take the $IV^\ast$ 7-brane as an example. In order to  define the topological operator living on its branch cut consistently for all $N$, we separate its branch cut into multiple ones. Each of these branch cuts provides a monodromy as an element in $SL(2,\mathbb{Z})$. In IIB string/F-theory, the $IV^\ast$ 7-brane admits a standard decomposition  
\begin{equation}
    \mathrm{Type} \, IV^\ast : A^5BC^2,
\end{equation}
where $A,B$ and $C$ are elementary $(p,q)$ 7-branes defined in (\ref{eq:definition ABC brane}). Therefore, we can now separate the branch cut of the $IV^*$ 7-brane into those corresponding to the monodromy matrices for $C^2, B$ and $A^5$ 7-branes, respectively. Stacking topological operators living on all branch cuts together, we end up with the operator for the type $IV^\ast$ 7-brane as: 
\begin{equation}\label{eq:branchcut operator for e6}
    \mathcal{O}^{IV^*}=\exp \left( \frac{2\pi i \times 5}{N}\int \frac{\mathcal{P}(B_2)}{2}  \right) \exp \left( \frac{2\pi i }{N}\int \frac{\mathcal{P}(3B_2+C_2)}{2}  \right)\exp \left( \frac{2\pi i \times 2}{N}\int \frac{\mathcal{P}(B_2+C_2)}{2}  \right).
\end{equation}

For the $\mathfrak{su}(2)$ theory, the third factor in (\ref{eq:branchcut operator for e6}) degenerates, so the operator reduces to 
\begin{equation}
    \mathcal{O}^{IV^\ast}_{N=2}=\exp \left( \pi i\int \frac{\mathcal{P}(B_2)}{2}  \right) \exp \left( \pi i \int \frac{\mathcal{P}(B_2+C_2)}{2}  \right),
\end{equation}
which reads $\mathcal{O}^{[0,1]}\cdot \mathcal{O}^{[1,1]}$ under the definition (\ref{eq:definition of su(2) branch cut operators}) for $\mathfrak{su}(2)$. One can easily see from Figure \ref{fig:Opq} that this corresponds to triality defects. For example, starting from $SO(3)_{-,0}$, this topological operator corresponds to first adding a counterterm and then gauging. The resulting theory is $SO(3)_{+,1}$, which is dual to $SO(3)_{-,0}$ via modular $\mathbb{S}\cdot \mathbb{T}$ transformation, thus realizing triality defects. For the $\mathfrak{su}(3)$ theory, the operator for $IV^\ast$ 7-brane reduces to
\begin{equation}
    \mathcal{O}^{IV^\ast}_{N=3}=\exp \left( \frac{4\pi i}{3}\int \mathcal{P}(B_2)/2  \right) \exp \left( \frac{2\pi i}{3}\int \mathcal{P}(C_2)/2  \right)\exp \left( \frac{4\pi i}{3}\int \mathcal{P}(B_2+C_2)/2  \right).
\end{equation}
Acting with this operator on, for example, the $SU(3)_0$ theory, one can do a similar computation as in (\ref{eq:LongComp}) and reach the $\overline{PSU(3)}_{1,0}$ theory, which is dual to $SU(3)_0$ theory via a modular $\mathbb{S}\cdot \mathbb{T}^{-1}$ transformation \cite{Kaidi:2022cpf}, thus realizing non-invertible duality defects. As in the case of duality defects, triality defects for $\mathfrak{su}(2)$ and $\mathfrak{su}(3)$ are of different types, namely intrinsic for $\mathfrak{su}(2)$ and non-intrinsic for $\mathfrak{su}(3)$. However, our construction does not depend on this difference and provides a simple unified realization for them. In fact, our construction works for all values of $N$ regardless of its divisors.

\section{Example: \texorpdfstring{$\mathcal{N} = 1$}{N=1} SCFTs}\label{sec:N=1}

In this section we show that the considerations of the previous sections readily extend to a broad class of $\mathcal{N} = 1$ SCFTs. We focus on the case of D3-brane probes of an isolated Calabi-Yau threefold singularity. It is well-known that this can be characterized in terms of a quiver gauge theory. In this characterization, the IIB axio-dilaton descends to a specific marginal coupling of the gauge theory. In particular, the IIB $SL(2,\mathbb{Z})$ duality group descends to a specific non-abelian duality group action at this special point in the moduli space. In particular, at the special points in moduli space given by $\tau_{IIB} = i$ and $\tau_{IIB} = \exp(2 \pi i / 6)$ we expect there to be duality / triality defects which act on the resulting $\mathcal{N} = 1$ SCFTs. Of course, at these points of strong coupling the sense in which the weakly coupled Lagrangian description in terms of a quiver gauge theory is actually available is less clear, but the definition of the field theory is still clear. An important feature of this more general class of SCFTs is that we typically have more 0-form symmetries, and we can also consider passing the associated defects charged under these symmetries through the  duality / triality defects.

To begin, let us recall some basic properties of these D3-brane probes of Calabi-Yau threefold singularities.\footnote{There is a vast literature on the subject of D-brane probes of singularities, see e.g., \cite{Douglas:1996sw, Franco:2005rj, Yamazaki:2008bt} for reviews.} In general terms, we get quiver gauge theories with gauge algebra $\mathfrak{su}(N_1)\times \mathfrak{su}(N_2)\times \cdots \times \mathfrak{su}(N_M)$, with $M$ the number of quiver gauge factors. Each gauge group factor has a complexified gauge coupling $\tau_i$ which specifies a marginal coupling of the SCFT. In an SCFT realized by a probe D3-brane, there exists a linear combination of these $\tau_{i}$ which corresponds to the IIB axio-dilaton:
\begin{equation}
\tau_{IIB} = \underset{i}{\sum} n_i\tau_{i},
\end{equation}
where the $n_i$ are positive integers which depend on the details of the geometry. There is a special subspace in the conformal manifold at which the $SL(2,\mathbb{Z})$ duality group action of type IIB strings descends to the SCFT. For example, in the special case where $n_i = 1$ for all $i$, this is given by taking $\tau_{i} = \tau$.

In any case, the IIB $SL(2,\mathbb{Z})$ duality descends to a field-theoretic duality, much as in the $\mathcal{N}=4$ case, and just as there, this also changes the global structure of the theory. The 1-form symmetry depends on the global structure of the theory. For example, with gauge group $SU(N)^M$, the $\mathcal{N}=1$ theory has a purely electric $\mathbb{Z}_N$ 1-form symmetry, which is the diagonal center of all $SU(N)$ groups. The charged Wilson lines can be built, as in the $\mathcal{N}=4$ case, by F1-strings stretched between D3-branes and the asymptotic boundary $\partial X$ of Calabi-Yau threefold $X$.

The construction of duality / triality defects in $\mathcal{N}=4$ theories readily generalizes to $\mathcal{N}=1$ theories. Indeed, our analysis carries over essentially unchanged, with 7-branes wrapped on the boundary $\partial X$ of the Calabi-Yau cone $X$ probed by the stack of D3-branes. In this sense, we automatically implement duality / triality defects in these $\mathcal{N} = 1$ SCFTs just from the top down origin as D3-brane probe theories. 

However, this is not the end of the story. Compared to the $\mathcal{N} = 4$ case where $X = \mathbb{C}^3$, in the $\mathcal{N} = 1$ case, the boundary topology $\partial X$ will in general be more intricate. In particular, dimensional reduction of the 10D supergravity on $\partial X$ will result in a Symmetry TFT with additional fields and interactions terms. These additional fields indicate the presence of more symmetries for the 4D field theory, and these can also be stacked with the duality defect. Here we explain some of the main elements of this construction, focusing on how 0-form symmetry topological operators behave under crossing a duality defect.

\subsection{Symmetry TFTs via IIB String Theory}

We will now present a top down approach to extracting symmetry TFTs for $\mathcal{N}=1$ SCFTs via IIB on a Calabi-Yau threefold $X$. We will reduce the topological term of IIB action on the asymptotic boundary $\partial X$ of the internal manifold $X$, which is a Sasaki-Einstein 5-manifold. This procedure involves treating the various IIB supergravity fluxes as elements in differential cohomology (see e.g., \cite{Freed:2006yc, Freed:2006ya}). Reduction of the linking 5-manifold leads to a symmetry TFT, as explained in  \cite{Apruzzi:2021nmk} (see also \cite{Aharony:1998qu, Heckman:2017uxe}).
Compared with the $\mathcal{N}=4$ case, this reduction leads to additional fields and interaction terms.

More explicitly, consider D3-branes probing a non-compact Calabi-Yau threefold $X$ with an isolated singularity at the tip of the cone. For ease of exposition, we assume that the asymptotic boundary $\partial X$ has the cohomology classes:
\begin{equation}\label{eq:cohomology class of E-S}
    H^*(\partial X)=\{ \mathbb{Z},0,\mathbb{Z}^{b_2}\oplus \text{Tor}H^2(\partial X),\mathbb{Z}^{b_2},\text{Tor}H^4(\partial X),\mathbb{Z} \},
\end{equation}
which covers many cases of interest. In the above,
$b_2$ is the second Betti number. Although not exhaustive, the cases covered by line (\ref{eq:cohomology class of E-S}) includes an infinite family of Calabi-Yau singularities, including $\mathbb{C}^3/\Gamma$ with isolated singularities, as well as complex cones over del Pezzo surfaces. The relevant term in the IIB supergravity action descends from the IIB topological term:\footnote{Here we neglect a subtlety with the 5-form field strength being self-dual. This amounts to a choice of quadratic refinement in an auxiliary 11D spacetime. See \cite{Belov:2006jd, Belov:2006xj, Monnier:2012xd, Heckman:2017uxe}.}
\begin{equation}\label{eq:IIB TFT differential cohomology}
     -\int_{M_4\times X}C_4\wedge dB_2 \wedge dC_2 \rightarrow -\int_{M_4\times  X} \breve{F}_5\star \breve{H}_3\star \breve{G}_3.
\end{equation}
Based on (\ref{eq:cohomology class of E-S}), we can expand $\breve{F}_5, \breve{H}_3$ and $\breve{G}_3$ as 
\begin{equation}
\begin{split}
    \breve{F}_5=&\breve{f}_5\star \breve{1}+\sum_{\alpha=1}^{b_2}\breve{f}_3^{(\alpha)}\star \breve{u}_{2(\alpha)}+\sum_{\alpha=1}^{b_2}\breve{f}_{2(\alpha)}\star \breve{u}_3^{(\alpha)}+N\mathrm{\breve{v}ol}
    +\sum_{i}\breve{E}_3^{(i)}\star \breve{t}_{2(i)}+\sum_{i}\breve{E}_{1(i)}\star \breve{t}_4^{(i)},\\
    \breve{H}_3=&\breve{h}_3\star \breve{1}+\sum_{\alpha=1}^{b_2}\breve{h}_1^{(\alpha)}\star \breve{u}_{2(\alpha)}+\sum_{\alpha=1}^{b_2}\breve{h}_{0(\alpha)}\star \breve{u}_3^{(\alpha)}
    +\sum_{i}\breve{B}_1^{(i)}\star \breve{t}_{2(i)},\\
    \breve{G}_3=&\breve{g}_3\star \breve{1}+\sum_{\alpha=1}^{b_2}\breve{g}_1^{(\alpha)}\star \breve{u}_{2(\alpha)}+\sum_{\alpha=1}^{b_2}\breve{g}_{0(\alpha)}\star \breve{u}_3^{(\alpha)}
    +\sum_{i}\breve{C}_1^{(i)}\star \breve{t}_{2(i)}.
\end{split}
\end{equation}
In the above equations, $\breve{1}, \mathrm{\breve{v}ol}$ and $\breve{u}_{2(\alpha)},\breve{u}_{3}^{(\alpha)}$ correspond to the free part $\mathbb{Z}$ and $\mathbb{Z}^{b_2}$ of cohomology classes, respectively, whereas $\breve{t}_{2(i)}$ and $\breve{t}_4^{(i)}$ correspond to the torsional part $\text{Tor}H^2(\partial X)=\text{Tor}H^4(\partial X)$ respectively and $i$ runs over its generators. The index placement of $\breve{u}_{2(\alpha)}$ compared to $  \breve{u}_3^{(\alpha)}$ (and $\breve{t}_{2(i)}$ compared to $\breve{t}_{4}^{(i)}$) indicate that their star product have non-trivial integral over $\partial X$. 

In particular, since $\partial X$ has infinite volume, any fields arising as coefficients of free cocycles that are dual to free cycles of positive degree should be interpreted as infinitely massive dynamical fields, and thus should be set to zero:
\begin{equation}
  \breve{f}_3^{(\alpha)} = \breve{f}_{2(\alpha)} = \breve{h}_1^{(\alpha)} = \breve{h}_{0(\alpha)} =
    \breve{g}_1^{(\alpha)} = \breve{g}_{0(\alpha)} = 0.
\end{equation}
The IIB topological term (\ref{eq:IIB TFT differential cohomology}) can then be expanded as
\begin{equation}\label{eq:DC expansion of IIB TFT}
\begin{split}
    &-\int_{M_4\times X} \breve{F}_5\star \breve{H}_3\star \breve{G}_3 \\
    &=-\int_{\partial X} \mathrm{\breve{v}ol}\star \breve{1} \star \breve{1} \int_{M_4\times \mathbb{R}_{\ge 0}}N\breve{h}_3\star \breve{g}_3\\
    &-\sum_{i,j,k}\int_{\partial X}\breve{t}_{2(i)}\star \breve{t}_{2(j)}\star \breve{t}_{2(k)}\int_{M_4\times \mathbb{R}_{\ge 0}}\breve{E}_3^{(i)}\star \breve{B}_1^{(j)}\star \breve{C}_1^{(k)}\\
    &-\sum_{i,j}\int_{\partial X}\breve{t}_{2(i)}\star \breve{t}_4^{(j)}\int_{M_4\times \mathbb{R}_{\ge 0}} \breve{E}_{1(j)}\star \left( \breve{B}_1^{(i)}\star \breve{g}_3+\breve{h}_3\star \breve{C}_1^{(i)} \right).
\end{split}
\end{equation}

Carrying out the reduction on $\partial X$ now involves integrating over this space in differential cohomology. 
For non-torsional classes, we find:
\begin{equation}\label{eq:integral over vol and dual forms}
\int_{\partial X}\mathrm{\breve{v}ol}\star \breve{1}\star \breve{1}=1,
\end{equation}
since $\mathrm{\breve{v}ol}$ is the volume form of $\partial X$. Integrals of torsional generators over $\partial X$ are determined by linking numbers which can be derived from intersection numbers between divisors of $X$:
\begin{equation}\label{eq:integral from linking pairing}
\begin{split}
    &\mathcal{C}_{ijk}\equiv \int_{\partial X}\breve{t}_{2(i)}\star \breve{t}_{2(j)}\star \breve{t}_{2(k)},\\
    &\mathcal{C}_i^j\equiv \int_{\partial X}\breve{t}_{2(i)}\star \breve{t}_4^{(j)}.
\end{split}
\end{equation}

The IIB topological term then reduces to the 5D Symmetry TFT:
\begin{equation}\label{eq:TFT for generic N=1}
    \begin{split}
       \mathcal{S}_{5D}&=-\underset{M_4\times \mathbb{R}_{\ge 0}}{\int} \bigg\{ N\breve{h}_3\star \breve{g}_3 - \sum_{i,j,k}\mathcal{C}_{ijk}\breve{E}_3^{(i)}\star \breve{B}_1^{(j)}\star \breve{C}_1^{(k)}-\sum_{i,j}\mathcal{C}_{i}^j\breve{E}_{1(j)}\star \left( \breve{B}_1^{(i)}\star \breve{g}_3+\breve{h}_3\star \breve{C}_1^{(i)} \right) \bigg\}.
    \end{split}
\end{equation}
Here fields denoted by capital letters are from torsional classes and are background fields for discrete symmetries:
\begin{equation}
    \breve{E}_3^{(i)} \leftrightarrow G^{(2)},\quad  \breve{B}_1^{(i)} \leftrightarrow G^{(0)},\quad  \breve{C}_1^{(i)} \leftrightarrow \tilde{G}^{(0)}
\end{equation}
where $G^{(2)}  \cong G^{(0)} \cong \text{Tor} H^2(\partial X) \cong \text{Tor} H^4(\partial X)$. On the other hand, fields denoted by lowercase letters correspond to field strengths of background fields for $U(1)$ symmetries. However, due to the presence of the coefficient $N$ in the first term, they are effectively $\mathbb{Z}_N$ symmetries
\begin{equation}
    \breve{g}_3 \leftrightarrow \mathbb{Z}_{N(m)}^{(1)},\quad  \breve{h}_3 \leftrightarrow \mathbb{Z}_{N(e)}^{(1)},
\end{equation}

The correspondence between fields, global symmetries and charged operators from wrapped branes are presented in table \ref{tab:charged operators of N=1 scfts}.

\begin{table}[t!]
    \centering
    \begin{tabular}{|c|c|c|}
\hline Fields&Global symmetries&Charged operators\\
\hline $\breve{h}_3$&$\mathbb{Z}_N^{(1)}$&F1-strings along $\mathbb{R}_{\geq 0}$, D3-branes wrapping $\mathbb{R}_{\geq 0}\times \sigma_2$\\
\hline $\breve{g}_3$&$\mathbb{Z}_N^{(1)}$&D1-strings along $\mathbb{R}_{\geq 0}$, D3-branes wrapping $\mathbb{R}_{\geq 0}\times \sigma_2$\\
\hline $\breve{E}_3^{(i)}$&$[\text{Tor}H^4(\partial X)]^{(2)}$&D3-branes wrapping $\mathbb{R}_{\geq 0}\times \gamma_1^{(i)}$\\
\hline $\breve{E}_{1(i)}$&$[\text{Tor}H^2(\partial X)]^{(0)}$&D3-branes wrapping $\mathbb{R}_{\geq 0}\times \gamma_{3(i)}$\\
\hline $\breve{B}_{1}^{(i)}$&$[\text{Tor}H^4(\partial X)]^{(0)}$&F1-string wrapping $\mathbb{R}_{\geq 0}\times \gamma_1^{(i)}$\\
\hline $\breve{C}_{1}^{(i)}$&$[\text{Tor}H^4(\partial X)]^{(0)}$&D1-string wrapping $\mathbb{R}_{\geq 0}\times \gamma_1^{(i)}$\\
\hline
\end{tabular}
    \caption{Fields in the 5D TFT and their corresponding global symmetries in 4D SCFTs. The charged operators composed of various types of branes are also presented. $\sigma$ and $\gamma$ denote non-torsional and torsional cycles, respectively. We use an upper index for the 1-cycles and a lower index for the 3-cycles.}
    \label{tab:charged operators of N=1 scfts}
\end{table}

The first term in (\ref{eq:TFT for generic N=1}) is the same as (\ref{eq:5d TFT terms1}), and so in this sense, all of our analysis of the $\mathcal{N} = 4$ SYM case carries over directly to this more general setting. In the $\mathcal{N}=1$ quiver SCFTs with gauge algebra $\mathfrak{su}(N)^K$, $\breve{h}_3$ and $\breve{g}_3$ are field strengths of gauge fields for the diagonal $\mathbb{Z}_N$ electric and magnetic 1-form symmetries, respectively.  The second term encodes the mixed anomaly between a 2-form symmetry and two 0-form symmetries. The last two terms encode mixed anomalies between 1-form symmetries and two 0-form symmetries.

These additional contributions beyond $\breve{h}_3$ and $\breve{g}_3$ are the main distinction from the $\mathcal{N} = 4$ case. We now discuss in further detail their stringy origins as well some of their properties.

\subsection{Discrete 0-form symmetries}
Let us now investigate the discrete 0-form symmetries more carefully. Denote the three classes of 0-form symmetries as $\mathbf{E}_{(i)}, \mathbf{B}^{(i)}$ and $\mathbf{C}^{(i)}$, with respective background fields $\breve{E}_{1(i)},\breve{B}^{1(i)}$ and $\breve{C}^{1(i)}$. As we list in table \ref{tab:charged operators of N=1 scfts}, $\mathbf{B}$ and $\mathbf{C}$ act on charged local operators in the 4D SCFT constructed respectively by F1- and D1-strings wrapping the cone over a torsional one-cycle $\gamma_1^{(i)}$. According to \cite{Heckman:2022muc}, we can build their symmetry operators with the magnetic dual branes, i.e., NS5- and D5-branes wrapping $M_3\times \gamma_{3{(i)}}$ where $M_3\subset M_4$ are 3D symmetry defects inside the 4D spacetime. The symmetry operators are then given by:
\begin{equation}\label{eq:symmetry operators for BC}
\begin{split}
     \mathcal{U}_{\mathbf{B}^{(i)}}=\exp \left(  i \int_{M_3\times \gamma_{3(i)}} B_6+\cdots \right),\\
    \mathcal{U}_{\mathbf{C}^{(i)}}=\exp \left(  i \int_{M_3\times \gamma_{3(i)}} C_6 +\cdots \right),
\end{split}
\end{equation}
where the $\cdots$ indicate additional terms in the respective Wess-Zumino actions. We defer a treatment of these other terms to future work. These two 0-form symmetry operators are related to each other under the IIB $SL(2,\mathbb{Z})$ duality. For example, the leading order terms $B_6$ and $C_6$ transform as:
\begin{equation}\label{eq:sl2z for b6c6}
\left[
    \begin{array}
[c]{c}%
B_{6}\\
C_{6}%
\end{array}
\right]  \mapsto\left[
\begin{array}
[c]{cc}%
a & b\\
c & d
\end{array}
\right]  \left[
\begin{array}
[c]{c}%
B_{6}\\
C_{6}%
\end{array}
\right] ,
\end{equation}
which gives rise to
\begin{equation}\label{eq:s-duality between 0-form symmetries}
    \mathcal{U}_{\mathbf{B}^{(i)}}\rightarrow \mathcal{U}_{\mathbf{B}^{(i)}}^a\mathcal{U}_{\mathbf{C}^{(i)}}^b, ~\mathcal{U}_{\mathbf{C}^{(i)}}\rightarrow \mathcal{U}_{\mathbf{B}^{(i)}}^c\mathcal{U}_{\mathbf{C}^{(i)}}^d.
\end{equation}
The symmetry operator of $\mathbf{E}_{(i)}$ is built by D3-branes wrapping torsional 1-cycles:
\begin{equation}\label{eq:symmetry operator for E}
    \mathcal{U}_{\mathbf{E}_{(i)}}=\exp \left( i \int_{M_3\times \gamma_1^{(i)}}C_4 +\cdots \right),
\end{equation}
which is obviously self-dual under S-duality due to the self-dual property of D3-branes.

We comment that similar transformation rules were worked out in \cite{Gukov:1998kn} in the special case $X = \mathbb{C}^3 / \mathbb{Z}_3$ with boundary topology $S^5 / \mathbb{Z}_3$. See Appendix \ref{app:orbo} for further discussion on this point, and the relation between the present chapter and this analysis. As noted in \cite{Gukov:1998kn}, the symmetry generators for $\mathbf{B}^{(i)}$ and $\mathbf{C}^{(i)}$ do not quite commute in the presence of a D3-brane wrapped on a torsional 1-cycle. Essentially the same arguments used in this special case carry over to this more general case as well, and we refer the interested reader to \cite{Gukov:1998kn} for further details.

Let us now turn to the interplay of these 0-form symmetries with our duality defects. We have already seen that the $\mathbf{B}^{(i)}$ and $\mathbf{C}^{(i)}$ symmetry generators transform non-trivially under IIB dualities. As such, we expect that when passing the corresponding operators charged under these symmetries through a duality wall that the transformation will be non-trivial.

Recall that charged operators for $\mathbf{B}^{(i)}$ and $\mathbf{C}^{(i)}$ are respectively non-compact D3-branes, F1- and D1-strings which stretch along the radial direction of the Calabi-Yau cone. To answer what happens when we pull such a brane through the duality wall, it is enough to track the transformation of the respective branes. The action of duality defects on these charged local operators in the 4D SCFT can be read from the Hanany-Witten transition. For example, insert a 7-brane as the duality defect in 4D theory, and consider a string wrapping Cone$(\gamma_1)$ as a local operator with charged  respectively under $\mathbf{B}^{(i)}$ and $\mathbf{C}^{(i)}$. The Hanany-Witten transition between the 7-brane and the string creates a new string attaching them, which is a 1D symmetry operator in the 4D SCFT. See figure \ref{fig:HW1} for an illustration in the case of half-space gauging construction.

\begin{figure}[t!]
    \centering
    \scalebox{0.8}{\begin{tikzpicture}
	\begin{pgfonlayer}{nodelayer}
		\node [style=none] (0) at (-6.5, 1.5) {};
		\node [style=none] (1) at (-1.5, 1.5) {};
		\node [style=none] (2) at (-6.5, -1.5) {};
		\node [style=none] (3) at (-1.5, -1.5) {};
		\node [style=Star] (4) at (-4, 0) {};
		\node [style=none] (5) at (-6.5, 0) {};
		\node [style=none] (6) at (-4, 0.5) {7-branes};
		\node [style=none] (7) at (-4, 2) {$\ket{P,D}$};
		\node [style=none] (9) at (-5.5, -2) {};
		\node [style=none] (10) at (-2.5, -2) {};
		\node [style=none] (11) at (-4, -2.5) {$x_\perp$};
		\node [style=none] (14) at (-5.25, 1.5) {};
		\node [style=none] (15) at (-5.25, -1.5) {};
		\node [style=none] (16) at (-5.25, 0) {};
		\node [style=none] (18) at (-6, 0.75) {$[q,p]$};
		\node [style=none] (19) at (-6, -0.75) {$[q,p]\rho$};
		\node [style=none] (20) at (-7.5, -1.5) {$r=0$};
		\node [style=none] (21) at (-7.5, 1.5) {$r=\infty$};
		\node [style=none] (22) at (-7.5, 1) {};
		\node [style=none] (23) at (-7.5, -1) {};
		\node [style=none] (24) at (1.5, 1.5) {};
		\node [style=none] (25) at (6.5, 1.5) {};
		\node [style=none] (26) at (1.5, -1.5) {};
		\node [style=none] (27) at (6.5, -1.5) {};
		\node [style=Star] (28) at (3, 0) {};
		\node [style=none] (29) at (1.5, 0) {};
		\node [style=none] (30) at (3, 0.5) {7-branes};
		\node [style=none] (31) at (4, 2) {$\ket{P,D}$};
		\node [style=none] (33) at (2.5, -2) {};
		\node [style=none] (34) at (5.5, -2) {};
		\node [style=none] (35) at (4, -2.5) {$x_\perp$};
		\node [style=none] (38) at (5.5, 1.5) {};
		\node [style=none] (39) at (5.5, -1.5) {};
		\node [style=none] (40) at (5.5, 0) {};
		\node [style=none] (44) at (6.25, 0.75) {$[q,p]$};
		\node [style=none] (45) at (4.25, -0.5) {$[q,p]\rho$};
		\node [style=none] (46) at (6.75, -0.5) {$[q,p](\rho-1)$};
		\node [style=none] (47) at (0, 0) {$=$};
	\end{pgfonlayer}
	\begin{pgfonlayer}{edgelayer}
		\draw [style=ThickLine] (0.center) to (1.center);
		\draw [style=ThickLine] (3.center) to (2.center);
		\draw [style=DottedLine] (5.center) to (4);
		\draw [style=ArrowLineRight] (9.center) to (10.center);
		\draw [style=RedLine] (14.center) to (16.center);
		\draw [style=BlueLine] (16.center) to (15.center);
		\draw [style=ArrowLineRight] (23.center) to (22.center);
		\draw [style=ThickLine] (24.center) to (25.center);
		\draw [style=ThickLine] (27.center) to (26.center);
		\draw [style=DottedLine] (29.center) to (28);
		\draw [style=ArrowLineRight] (33.center) to (34.center);
		\draw [style=RedLine] (38.center) to (40.center);
		\draw [style=BlueLine] (40.center) to (39.center);
		\draw [style=ThickLine] (40.center) to (28);
	\end{pgfonlayer}
\end{tikzpicture}
}
    \caption{A $[q,p]$ string wrapping the cone of a torsional 1-cycle at $r=\infty$ corresponds to a local operator of the 4D SCFT at $r=0$. Passing the  duality defect $\mathcal{U}(M_3,\mathfrak{F})$ through the local operator from right to left creates a symmetry line operator of charge $[q,p](\rho-1)$ via a Hanany-Witten transition.}
    \label{fig:HW1}
\end{figure}

This is very similar to our discussion around Figure \ref{fig:HW} how charged line operators transformed through duality defects, which from the string theory point of view obviously have the same origin as the Hanany-Witten transition. As it is for local operators, this is also analogous to the case in 2D CFTs, e.g. the critical Ising model, where the non-invertible symmetry line maps the local spin operator to the disorder operator which lives at the edge of another symmetry line \cite{Verlinde:1988sn}.

One can in principle also consider passing the topological operators for these 0-form symmetries through a duality / triality defect as well. This case is more subtle since it involves an order of limits issue concerning how fast we send the 7-branes to infinity, and how the 5-branes (by)pass the corresponding branch cuts. It would be interesting to study this issue, but it is beyond the scope of the present chapter.

\section{Conclusions}
\label{sec:CONC3}
In this chapter we have presented a top down construction of duality defects for 4D QFTs engineered via D3-brane probes of isolated Calabi-Yau singularities. In this description, the duality group of type IIB strings descends to a duality of the localized QFT at a special point of the conformal manifold, and topological duality defects are implemented by suitable 7-branes wrapped ``at infinity'' which implement field theoretic duality topological symmetry operators. The branch cuts of the 7-branes directly descend to branch cuts in the 5D symmetry TFT which governs the anomalies of the 4D theory. This provides a uniform perspective on various ``bottom up'' approaches to realizing duality / triality defects such as Kramers-Wannier-like defects and half-space gauging constructions. This uniform perspective applies both to $\mathcal{N} = 4$ SYM theory as well as a number of $\mathcal{N} = 1$ quiver gauge theories tuned to a point of strong coupling. In the remainder of this section we discuss a few avenues which would be natural to consider further.

One of the main items in our analysis is the important role of the branch cuts present in 7-branes, and how these dictate the structure of anomalies in the 5D symmetry TFT, as well as the resulting 3D TFTs localized on the duality defects. In Appendix \ref{app:minimalTFT7branes} we take some preliminary steps in reading this data off directly from dimensional reduction of a 7-brane. It would be interesting to perform further checks on this proposal, perhaps by considering dimensional reduction on other ``internal'' manifolds.

In our extension to $\mathcal{N} = 1$ quiver gauge theories, we focused on the special case where the singularity of the Calabi-Yau cone is isolated. This was more to obtain some technical simplifications rather than there being any fundamental obstacle to performing the same computations. It would be interesting to study the structure of the resulting 5D symmetry TFT, as well as the interplay between 7-branes (and their branch cuts) with non-isolated singularities.

It would also be quite interesting to consider other special points in the conformal manifold of 4D SCFTs realized via D3-brane probes of singularities. If these special points admit a non-trivial automorphism, one could hope to similarly lift this automorphism to a geometric object in a string construction, and thus realize a broader class of symmetry operators.

Our primary focus has been on QFTs realized via D3-branes at singularities, but we have also indicated in Appendix \ref{app:other} how these considerations can be generalized to other top down constructions. In particular, it would be interesting to study the structure of duality defects in theories which cannot be obtained from D3-brane probes of singularities.




\clearpage \phantomsection \addcontentsline{toc}{chapter}{CONCLUSIONS}

\textbf{\huge CONCLUSIONS}

\textbf{\Large Supergravity and String Universality}

We analyzed the consistency of string vacua and string universality in high spacetime dimensions. On the one hand, we sharpened our understanding of discrete aspects of the string landscape in 7, 8, and 9 spacetime dimensions, especially on their generalized global symmetries and frozen singularities. On the other hand, we identified an explicit consistency condition that applies to any 8D quantum gravity theories (i.e., we do not require the existence of a string construction), that a specific mixed anomaly of its center 1-form symmetry has to vanish. Combining these two directions allows us to rule out a large family of 8D $\mathcal{N} = 1$ supergravity theories that do not come from string theory, thus giving strong evidence to string 8D universality. 

In \cite{Cvetic:2020kuw}, we consider 8D $\mathcal{N} = 1$ supergravity theories. Based on no global symmetry conjecture, a global symmetry must be broken or gauged. We can identify some (unbroken) center 1-form symmetries. We identify a mixed 't Hooft anomaly of such center 1-form symmetries. However, to ensure they can be consistently gauged, we must require a mixed anomaly to vanish. In this way, we find a non-trivial swampland constraint on the consistency of 8D supergravity theories. Such a constraint rules out many supergravity theories without any known string constructions. Therefore, we obtain strong evidence for string universality in 8 dimensions.

Afterward, we classify the string landscape in eight and nine dimensions. In \cite{Cvetic:2021sjm} and \cite{Cvetic:2022uuu}, we explicitly determined all 8D supergravities that could be constructed from string theory. Specifically, in \cite{Cvetic:2021sjm}, we analyzed the landscape of 8D CHL strings. In these cases, non-simply-laced gauge algebras are involved, so the determination of the global structure of the gauge group requires extra care, both algebraically and physically. After taking into account such complications, we gave a full list of all maximally-enhanced gauge groups in 8D CHL strings. Later in \cite{Cvetic:2022uuu}, we managed to generalize and unify the 8D string landscape. The first step in obtaining this list is to re-interpret the known cases in non-perturbative IIB superstring theory with (p,q) 7-branes and string junctions. The next step is to understand the important role played by the $O7^+$ brane (which is the only frozen singularity in F-theory) in reducing the rank of the gauge group and possible patterns that string junctions can end on (p,q) 7-branes. In the end, we are able to determine the global structure of the gauge group in general. By introducing a pair of $O7^+$ planes, we obtained the first non-perturbative description of the lowest rank gauge group of 8D string theory, and our classification implicitly incorporates two disconnected branches of it (see \cite{Montero:2022vva} for elaboration and extensive study of these theories). Moreover, for all the global structures that we obtained, we are able to verify that they satisfy the mixed anomaly constraint that we mentioned above.

 In \cite{Cvetic:2021sxm}, we extend the analysis of generalized global symmetries down to 7D, focusing on local models. Via F-/M-theory duality, we determined the generalized symmetries of 7D M-theory vacua and 8D F-theory vacua via geometry, which also incorporated a local string junction treatment on the ``F-theory" side. In addition, for these string-constructed vacua, we also re-derived the general mixed 1-form anomaly in 7D by dimensionally reducing the Chern-Simons term of M-theory.

In upcoming work, we will also give a local treat for the 7D frozen singularities in M-theory. These cases are M-theory on an ALE space with fractional $C_3$ holonomy on its boundary lens space. We focus on better understanding the freezing mechanism from a microscopic perspective. Using a known duality \cite{Tachikawa:2015wka} between 7D M-theory frozen singularity and 7D F-theory geometric vacua, we are able to explicitly derive the freezing rule at the level of wrapped string states. This requires us to first re-phrase the 7D F-theory vacua via $(p, q)$ 7-branes and string junctions on non-geometric quotients $S^1 \times \bbP^2$ and then imposing consistency conditions on string junctions living in such a space. The resulting freezing rule makes explicit reference to simple roots in a succinct fashion. A natural follow-up is to understand all 7D M-theory vacua with frozen singularities in this fashion.

In the long term, one would wish to push this classification program of supergravity theories down to lower dimensions, which further opens up the possibility of having less supersymmetries. In addition, from the supergravity perspective, a long-term goal is to identify more consistency conditions at high energy and to understand their interconnections. 

\newpage

\textbf{\Large Geometric Engineering and SCFTs}

Strongly-coupled QFTs are QFTs that do not admit a conventional Lagrangian description, and they have important applications in describing low-dimensional condensed matter systems. All of the known ways of engineering such strongly-coupled SCFTs have some underlying string-theoretic origin: they come from either direct compactification of string theories or the worldvolume theories of branes in string theory on a particular geometric background. One could also construct them from further compactifications of higher-dimensional strongly-coupled SCFTs, which are, in turn, built from string theory. Hence, such strongly-coupled QFTs intrinsically come from string theory.

A general theme of \cite{DelZotto:2022fnw}, \cite{Heckman:2022suy}, \cite{Heckman:2022muc} and \cite{Heckman:2022xgu} is to study the generalized symmetries of some QFTs engineered by string theory, and to characterize them in terms of the geometric and topological properties of the internal geometry. 

In \cite{DelZotto:2022fnw}, we studied M-theory compactified on a $\mathbb{C}^3/\Gamma$ orbifold geometry where the $\Gamma$ is a finite subgroup of $SU(3)$. The generalized global symmetries in such 5D QFTs depend on both $\Gamma$ as a group and its action on the $\mathbb{C}^3$ internal space. Out of these theories, we made an illuminating conjecture on how the 2-group global symmetries in such M-theory compactifications could arise from geometry.

In \cite{Heckman:2022suy}, we perform exhaustive classification of center-flavor symmetries associated with the center-flavor symmetry of 6D SCFTs, based on their known geometric classification from F-theory. We then constructed new families of 4D SCFTs by further compactifying those 6D SCFTs on a torus with flat connections associated with the center-flavor symmetry as mentioned above. As a byproduct, we also identified the 6D origins of some 4D theories. The properties of the new and old 4D SCFTs can be computed from 6D, which is consistent with the results of intrinsic 4D methods.

Motivated by recent papers that work in the setup of holographic duality, we found in \cite{Heckman:2022muc} that for \textit{all} string-constructed QFTs via geometrical engineering, the \textit{boundary} of the internal geometry controls their higher symmetries in a surprisingly general manner. The \textit{symmetry operators} associated with these symmetries can be directly obtained from wrapping branes on cycles on the boundary of the internal geometry. As an example, we examined a family of geometrically engineered 6D SCFTs and explicitly identified the symmetry operators of non-invertible 2-form symmetries in these theories. We also pointed out several other applications of this proposal.

In a follow-up work \cite{Heckman:2022xgu}, we constructed various duality defects in 4D SCFTs by wrapping 7-branes on the boundary of the internal geometry. This construction is motivated by the crucial fact that 7-brane itself carries a non-trivial $SL(2, \mathbb{Z})$ monodromy in IIB. According to different places that the endpoint of the 7-brane attaches to in the bulk 5D topological field theory (TFT), we unified two field-theoretic constructions. Vertically arranging the branch cut gives the one based on gauging a field with a mixed 't Hooft anomaly \cite{Kaidi:2021xfk}, while horizontally arranging the branch cut gives another construction based on half-space gauging in even dimensions \cite{Choi:2021kmx}. To illustrate that our construction is not limited to 4D $N=4$ SYM theories, we also constructed such non-invertible topological defects in 4D $N=1$ theories, and gave a prescription of how their non-invertibility is encoded in the symmetry TFT of such theories in general.

A broader goal of this investigation is to understand how the global categorical symmetries of a string-constructed QFT can be obtained from the information of the internal geometry, possibly equipped with the non-geometrical fluxes. In particular, in type II string theories, it has been known for more than a decade that the behavior of D-brane wrapped on cycles has sophisticated categorical descriptions, where homological cycles are already not the correct objects to talk about. A similar description in M-/F-theory remains an open question. With such a description of categorical symmetric via internal space at hand, a longer-term objective is to take different dual string-theoretic construction of the same theory and compare their global categorical symmetries. By demanding that the global categorical symmetries should match across different descriptions, we have a great opportunity to sharpen our (yet incomplete) understanding of string/M-/F-theory.

\end{mainf}


\begin{append}
\chapter{\MakeUppercase{Chapter 1 Appendix}}

\section{Bounds on Coefficients}
\label{app:m}

In this appendix we briefly explain how anomaly inflow arguments can restrict the integer coefficients $m_i$ in \eqref{eq:anomcoup}. From \cite{Montero:2020icj} one knows that consistent theories in 8d have a gauge group with rank $\in \{18,10,2\}$. This means, that for a fixed rank the non-Abelian part of the gauge group captured by $G$ has to be supplemented by 
\begin{align}
n_{U(1)} = \text{rank} - \text{rank} (G) \,,
\end{align}
Abelian $U(1)$ factors. With the gauge algebra specified, one then has to restrict the prefactors $m_i$ such that the bound on the left-moving central charge \eqref{eq:anominfl} is satisfied.

\subsection{Rank 18}

For the case of a gauge group of rank 18, we note that the anomaly contribution is bounded from below by the rank of the corresponding group factor
\begin{align}
\frac{m_i \, \text{dim}(G_i)}{m_i + h_i} \geq \text{rank}(G_i) \,,
\end{align}
with the inequality satisfied for simply-laced groups at $m_i = 1$. This implies
\begin{align}
c_L \geq 18 \geq \sum \frac{m_i \, \text{dim}(G_i)}{m_i + h_i} + n_{U(1)} \geq \text{rank} = 18 \,,
\label{eq:ineqbound}
\end{align}
from which we see that for rank-18 cases the only gauge factors allowed are simply-laced $G_i$ with
\begin{align}
m_i = 1 \,,
\end{align}
i.e., the prefactors are fixed uniquely.

For reference, the values of $h$ and $\text{dim}(G)$ for allowed simple Lie groups $G$ in 8d are:
\begin{align}\label{tab:coxeter}
	\renewcommand{\arraystretch}{1.3}
	\begin{array}{|c||c|c|c|c|c|c|}
		\hline
		G & SU(n\geq 2) & Spin(n \geq 8) & Sp(n) & E_6 & E_7 & E_8 \\ \hline \hline 
		\text{dim} & n^2 - 1 & \tfrac{n(n-1)}{2} & n(2n+1) & 78 & 133 & 248\\ \hline
		h & n & n-2 & n+1 & 12 & 18 & 30 \\ \hline
	\end{array}
\end{align}

\subsection{Lower Ranks}

For lower ranks the possibilities for $(G_i, m_i)$ combinations increase since now the right-hand side of \eqref{eq:ineqbound} is smaller than 18. 
In particular, for smaller gauge groups or the case $\text{rank} = 2$ the combinatorics become more involved, which requires a case by case analysis beyond the scope of this work, that will be treated in more detail elsewhere.
Here, we make some simple observations about rank-10 cases that saturate the bound, which are also known to arise from the CHL string in 8d.

For the first example we consider the CHL string with non-Abelian gauge group $E_8 \times U(1) \times U(1)$.
This model has an $E_8$ current algebra at level $2$, and therefore
\begin{align}
\frac{m \, \text{dim}(G)}{m + h} + n_{U(1)}  = \tfrac{35}{2} \,,
\end{align}
which basically saturates the bound of 18, as there are no possible group/level contributing with $1/2$.
Higher levels of the current algebra with $m \geq 3$ are forbidden in this case.
Note that for the typical product subgroups with non-trivial quotient structure, such as $[E_7 \times SU(2)]/\bbZ_2$ or $[E_6 \times SU(3)]/\bbZ_3$, the anomaly \eqref{eq:anomaly} of the gauged center is indeed trivial.


In the second example we consider the CHL string with its maximal symplectic group, which is $Sp(10)$. The current algebra is at level 1, and one finds
\begin{align}
\frac{m \, \text{dim}(G)}{m + h} = \tfrac{35}{2} \,,
\end{align}
again saturating the allowed upper bound and prohibiting $m \geq 2$.
8d theories with level 1 groups in \eqref{eq:allowed_groups2} can potentially arise as subgroups, and will be investigated in detail in future work.

\section{Mordell--Weil Torsion and the Gauge Group in F-theory}
\label{secapdx:geometry}

8d ${\cal N}=1$ string compactifications with total gauge rank 18 can be described by F-theory compactified on elliptic K3 surfaces.
We refer to reviews \cite{Taylor:2011wt, Weigand:2018rez, Cvetic:2018bni} for a broader introduction, and focus in the following on two aspects key to discussion of global gauge group structure and center anomalies.
First, the non-Abelian gauge algebras $\mathfrak{g}_i$ (associated to simply-connected groups $G_i$) are captured by reducible Kodaira-fibers of ADE-type $\mathfrak{g}_i$ \cite{Vafa:1996xn,Morrison:1996na,Morrison:1996pp}.
Second, the global structure of the gauge group, $[\prod_i G_i]/Z$, is determined by the torsional part $Z$ of the Mordell--Weil group of sections \cite{Aspinwall:1998xj} (see especially \cite{Cvetic:2018bni} for a pedagogical introduction for this).

In general, the notation $G/Z$ requires a specification of the subgroup $Z \subset Z(G) = \prod_i Z(G_i)$ of the full center $Z(G)$.
In F-theory, this is determined by the intersection pattern between the generating sections of the Mordell--Weil group and the components of the $\mathfrak{g}_i$-Kodaira fiber that form the affine Dynkin diagram of $\mathfrak{g}_i$ \cite{Mayrhofer:2014opa,Cvetic:2017epq}.

For definiteness, we restrict ourselves to compactifications with $\mathfrak{g}_i = \mathfrak{su}_{n_i}$, $i=1,...,s$, realized by K3 surfaces with only $I_{n_i}$ fibers.
Each reducible $I_{n_i}$ fiber consists of $n_i$ irreducible components that form a loop, reflecting the structure of the affine $\mathfrak{su}_{n_i}$ Dynkin diagram.
Starting with the affine node (determined by intersection with the zero-section) we label the components by $0,...,n_i-1$ as we go around the loop of the $i$-th fiber.
Then, an $\ell$-torsional section $\tau$ is \emph{uniquely} characterized by the tuple $(k_1,...,k_s)$ which labels the $k_i$-th component in the $i$-th fiber met by $\tau$ \cite{MirandaPersson}.
Moreover, one has $k_i \ell \equiv 0 \mod n_i$.

As explained in \cite{Mayrhofer:2014opa,Cvetic:2017epq}, $\tau = (k_1,...,k_s)$ corresponds precisely to the order $\ell$ element $(k_1,...,k_s) \in \prod_i \bbZ_{n_i} = Z(\prod_i SU(n_i))$.
This element acts trivially on all matter states of the F-theory compactification, hence giving rise to the gauge group $G/\langle \tau \rangle \cong G/\bbZ_\ell$.

The allowed combinations of $G$ and $\bbZ_\ell$ is heavily constrained geometrically by the following fact pertaining to intersection patterns between torsional sections and fiber components.
For a K3 $X$ with only $I_{n_i}$ fibers, the non-affine components of each fiber span a sublattice $R \subset H_2(X,\bbZ)$ with $\text{rank}(R) \leq 18$.
Then, one can associate to $R$ a so-called discriminant-form group \cite{MirandaPersson}
\begin{align}
	G_R \cong \bbZ_{n_1} \times ... \times \bbZ_{n_s} \, .
\end{align}
This group inherits from the lattice structure of $H_2(X,\bbZ)$ a quadratic form
\begin{align}\label{eq:quadratic_form}
	q: G_R \rightarrow \mathbb{Q}/\bbZ \, , \quad (x_1, ..., x_s) \mapsto \sum_{i=1}^s \frac{1-n_i}{2n_i} x_i^2 \mod \bbZ \, .
\end{align}
Notice that $G_R = \prod_{i=1}^s Z(SU(n_i)) = Z(G)$.
Then, by regarding a torsional section $\tau = (k_1,...,k_s)$ as an element of $G_R$, it can be shown \cite{MirandaPersson} that
\begin{align}\label{eq:isotropy_torsion_section}
	q(k_1,...,k_s) = \sum_{i=1}^s  \frac{1-n_i}{2n_i} k_i^2 \equiv 0 \mod \bbZ \,.
\end{align}
This precisely reproduces the constraint for the absence of the anomaly \eqref{eq:anomaly} involving the $\bbZ_\ell$ center 1-form symmetry with $m_i = 1$ for all $G_i$.

\section{A Heterotic Case Study}
\label{app:heterotic_example}

In this appendix, we study the global gauge group structure of a rank 20 heterotic model, with $\mathfrak{g} = \mathfrak{su}(2)^2 \oplus \mathfrak{su}(4)^2 \oplus \mathfrak{so}(20)$.

We choose a presentation of the Narain lattice $\Lambda_N$ and its vector space $V_N = \Lambda_N \otimes \mathbb{R}$ as
\begin{align}
    V_N \ni {\bf v}^{(\ell)} =  (l^{(\ell)}_1, l^{(\ell)}_2, n^{(\ell)}_1, n^{(\ell)}_2; s^{(\ell)}_1, \dots, s^{(\ell)}_{16} ) \, ,
\end{align}
with pairing
\begin{equation}
    \langle \mathbf{v}^{(1)}, \mathbf{v}^{(2)} \rangle = l^{(1)}_1 n^{(2)}_1 + l^{(2)}_1 n^{(1)}_1 + l^{(1)}_2 n^{(2)}_2 + l^{(2)}_2 n^{(1)}_2 + \sum_{j = 1}^{16} s_j^{(1)}  s_j^{(2)} \, .
\end{equation}
Then, vectors in $\Lambda_N = \Lambda_N^* \cong U \oplus U \oplus \Gamma_{16}$ are characterized by
\begin{align}
    l^{(\ell)}_i, n^{(\ell)}_i \in \bbZ \, , \quad (s_1,...,s_{16})\in \tfrac12 \bbZ \ \ \text{with} \ \ \sum_{j=1}^{16} s_j \in 2\bbZ \, , \ \ s_j - s_k \in \bbZ \ \ \forall j, k \, .
\end{align}

The explicit embedding of the $\mathfrak{g}$ root lattice $\Lambda_\text{r}^{\mathfrak{g}}$ into $\Lambda_N$ is given as:
\begingroup\makeatletter\def\f@size{8}\check@mathfonts
\def\maketag@@@#1{\hbox{\m@th\large\normalfont#1}}%
\begin{align}
\left[\begin{array}{cccc|cccccccccccccccc}
1 & 4 & -1&-3 & 0 &-2 &-2 &-1 &-1 &-1 &-1 &-4 & 0 & 0 & 0 & 0 & 0 & 0 & 0 & 0 \\\hline
1 & 2 &-1 &-3 &-1 &-1 &-1 &-1 &-1 &-1 &-1 &-3 & 0 & 0 & 0 & 0 & 0 & 0 & 0 & 0 \\ \hline
0 & 2 & 0 &-2 & 0 &-2 &-1 &-1 &-1 &-1 &-1 &-1 & 0 & 0 & 0 & 0 & 0 & 0 & 0 & 0 \\
0 & 0 & 0 & 0 & 0 & 1 &-1 & 0 & 0 & 0 & 0 & 0 & 0 & 0 & 0 & 0 & 0 & 0 & 0 & 0 \\
0 &-2 & 0 & 1 & 0 & 0 & 1 & 1 & 1 & 1 & 1 & 1 & 0 & 0 & 0 & 0 & 0 & 0 & 0 & 0 \\\hline
0 & 0 & 0 & 0 & 0 & 0 & 0 & 0 & 0 & 1 &-1 & 0 & 0 & 0 & 0 & 0 & 0 & 0 & 0 & 0 \\
0 & 0 & 0 & 0 & 0 & 0 & 0 & 0 & 1 &-1 & 0 & 0 & 0 & 0 & 0 & 0 & 0 & 0 & 0 & 0 \\
0 &-2 & 0 & 1 & 1 & 1 & 1 & 0 & 0 & 1 & 1 & 1 & 0 & 0 & 0 & 0 & 0 & 0 & 0 & 0 \\\hline
1 & 0 & 1 & 0 & 0 & 0 & 0 & 0 & 0 & 0 & 0 & 0 & 0 & 0 & 0 & 0 & 0 & 0 & 0 & 0 \\
0 & 4 &-1 &-3 &-1 &-2 &-2 &-2 &-1 &-1 &-1 &-3 &-1 & 0 & 0 & 0 & 0 & 0 & 0 & 0 \\
0 & 0 & 0 & 0 & 0 & 0 & 0 & 0 & 0 & 0 & 0 & 0 & 1 &-1 & 0 & 0 & 0 & 0 & 0 & 0 \\
0 & 0 & 0 & 0 & 0 & 0 & 0 & 0 & 0 & 0 & 0 & 0 & 0 & 1 &-1 & 0 & 0 & 0 & 0 & 0 \\
0 & 0 & 0 & 0 & 0 & 0 & 0 & 0 & 0 & 0 & 0 & 0 & 0 & 0 & 1 &-1 & 0 & 0 & 0 & 0 \\
0 & 0 & 0 & 0 & 0 & 0 & 0 & 0 & 0 & 0 & 0 & 0 & 0 & 0 & 0 & 1 &-1 & 0 & 0 & 0 \\
0 & 0 & 0 & 0 & 0 & 0 & 0 & 0 & 0 & 0 & 0 & 0 & 0 & 0 & 0 & 0 & 1 &-1 & 0 & 0 \\
0 & 0 & 0 & 0 & 0 & 0 & 0 & 0 & 0 & 0 & 0 & 0 & 0 & 0 & 0 & 0 & 0 & 1 &-1 & 0 \\
0 & 0 & 0 & 0 & 0 & 0 & 0 & 0 & 0 & 0 & 0 & 0 & 0 & 0 & 0 & 0 & 0 & 0 & 1 &-1 \\
0 & 0 & 0 & 0 & 0 & 0 & 0 & 0 & 0 & 0 & 0 & 0 & 0 & 0 & 0 & 0 & 0 & 0 & 1 & 1 
\end{array}\right],
\end{align}\endgroup
whose rows we label by ${\boldsymbol\mu}_1, ..., {\boldsymbol\mu}_{18}$. Here ${\boldsymbol\mu}_1$ and ${\boldsymbol\mu}_2$ are the roots of two $\mathfrak{su}(2)$'s, $({\boldsymbol\mu}_3, {\boldsymbol\mu}_4, {\boldsymbol\mu}_5$ and $({\boldsymbol\mu}_{6}, {\boldsymbol\mu}_7, {\boldsymbol\mu}_{8})$ are the roots of two $\mathfrak{su}(4)$'s. 
$({\boldsymbol\mu}_9, ..., {\boldsymbol\mu}_{18})$ are the roots of $\mathfrak{so}(20)$, with ${\boldsymbol\mu}_{17}, {\boldsymbol\mu}_{18}$ the two branched nodes.
Since these are all ADE-systems, we have ${\boldsymbol\mu}_i = {\boldsymbol\mu}_i^\vee$.

The coweight lattice $\Lambda_\text{cw}^\mathfrak{g}$ is spanned by the coweights
\begin{equation}
    \overline{\bf w}_i = (C^{-1})_{ij} {\boldsymbol\mu}_j \, , \quad \text{with} \quad C_{ij} = \langle {\boldsymbol\mu}_i , {\boldsymbol\mu}_j \rangle \, .
\end{equation}
Note that $C$ is simply the block-diagonal sum of the Cartan matrices of each simple factor in $\mathfrak{g}$.
Now we examine the $F$ plane --- the orthogonal subspace to $E := \Lambda_\text{r}^\mathfrak{g} \otimes \mathbb{R}$, which is two-dimensional in this case. 
Its generators can be chosen to be:
\begin{align}
    \begin{split}
    \boldsymbol\xi_1 &= (-2, 0, 2, 1;0, 0, 0, 0, 0, 0, 0, 2, 0, 0, 0, 0, 0, 0, 0, 0) \, , \\
    \boldsymbol\xi_2 &= (2, 14, -2, -11; -3, -7, -7,  -6, -4, -4, -4, -11, 0, 0, 0, 0, 0, 0, 0, 0) \, ,\\
    \boldsymbol\xi_1^2 &= \boldsymbol\xi_2^2 = -4, \ \ \boldsymbol\xi_1 \cdot \boldsymbol\xi_2 = 0 \, .
    \end{split}
\end{align}

With this basis, a general element $\overline{\bf v}$ of $\Lambda_N^* = \Lambda_N$ can be written as a linear combination of coweights and the $U(1)$ generators:
\begin{equation}
    \overline{\bf v} = (l_1, l_2, n_1, n_1; s_1, \dots, s_{16}) = \sum_{j = 1}^{18} k_j \overline{\bf w}_{j} + m_1 \boldsymbol\xi_1 + m_2 \boldsymbol\xi_2,\ \ \ k_j \in \bbZ \, .
\end{equation}
Modulo the (co-)roots $\Lambda_\text{cr}^\mathfrak{g} = \Lambda_\text{r}^\mathfrak{g}$, we find two independent basis vectors of $\Lambda_N \cap E = \Lambda_\text{cc}^G$:
\begin{align}
    \begin{split}
    \hat{\bf c}_1  &= (1, 5, -1, -5; -\tfrac{3}{2}, -\tfrac{5}{2}, -\tfrac{7}{2}, -\tfrac{5}{2}, -\tfrac{3}{2}, -\tfrac{3}{2}, -\tfrac{3}{2}, -\tfrac{9}{2}, \tfrac{1}{2},  \tfrac{1}{2}, \tfrac{1}{2}, \tfrac{1}{2}, \tfrac{1}{2}, \tfrac{1}{2}, \tfrac{1}{2}, -\tfrac{1}{2}) \\
    &= \overline{\bf w}_2 +  \overline{\bf w}_4 + \overline{\bf w}_{17} \, , \\
    \hat{\bf c}_2  &= (1, 1, 0, -1;  \tfrac{1}{2}, -\tfrac{1}{2}, -\tfrac{1}{2}, -\tfrac{1}{2}, \tfrac{1}{2}, -\tfrac{1}{2}, -\tfrac{1}{2}, -\tfrac{3}{2}, -\tfrac{1}{2},  -\tfrac{1}{2}, -\tfrac{1}{2}, -\tfrac{1}{2}, -\tfrac{1}{2}, -\tfrac{1}{2}, -\tfrac{1}{2}, \tfrac{1}{2}) \\
    & = \overline{\bf w}_1 + \overline{\bf w}_7 + \overline{\bf w}_{9} - \overline{\bf w}_{17} \, ,
    \end{split}
\end{align}
each of which defines an order two element, i.e., generates a $\bbZ_2 \subset Z(SU(2)^2 \times SU(4)^2 \times Spin(20)) = \bbZ_2 \times \bbZ_2 \times \bbZ_4 \times \bbZ_4 \times (\bbZ_2 \times \bbZ_2)$, via the the embeddings
\begin{align}
    z(\hat{\bf c}_1) = (0, 1, 2, 0, (1,0)) \, , \quad z(\hat{\bf c}_2) = (1, 0, 0, 2, (0,1)) \, .
\end{align}

The two generators of $\Lambda_N$ that are not within $\Lambda_N \cap E$ are:
\begin{align}
    \begin{split}
    \hat{\bf c}_3 & = (0, 1, 0, -2; 0, -1, -1, -1, 0, -1, -1, -1, 0, 0, 0, 0, 0, 0, 0, 0)= \tfrac{1}{4}\boldsymbol\xi_1 + \overline{\bf w}_2 + \overline{\bf w}_{3}  + \overline{\bf w}_{7} \, , \\
    \hat{\bf c}_4 & = (1, 4, -1,-4; 0, -2, -3, -2, -1, -1, -1, -4, 0, 0, 0, 0, 0, 0, 0, 0)= \tfrac{1}{4}\boldsymbol\xi_2 + \overline{\bf w}_1 + \overline{\bf w}_4 + \overline{\bf w}_{8} \, .
    \end{split}
\end{align}
Their projection under $\pi_E$ onto $\Lambda_\text{cw}^\mathfrak{g}$ define the following equivalence classes in $Z(\widetilde{G}):$
\begin{equation}
    z(\hat{\bf c}_3) = (0, 1, 1, 2, (0, 0)) \, \quad z(\hat{\bf c}_4) = (1, 0, 2, 1, (0, 0)) \, .
\end{equation}
In summary, we find that the full gauge group is
\begin{equation}
    \frac{[(SU(2)^2 \times SU(4)^2 \times Spin(20))/(\bbZ_2 \times \bbZ_2)] \times U(1)^2}{\bbZ_4 \times \bbZ_4} \, .
\end{equation}

\chapter{\MakeUppercase{Chapter 2 Appendix}}

\section{Defect Group Structures of M-theory on Elliptic Fibrations}
\label{app:defect_group_elliptic}

In this appendix, we compute the defect group structure for the 7d theories listed in Table \ref{tab:fibers_and_kappa}.
The strategy is to first determine a representative of the $N_i$-torsional 1-cycles ${\cal C}_i$ in $H_1(\partial Y)_\text{tors} \cong \text{coker}(\kappa)$ as a linear combination of the ${\cal A} = (1,0)$ and ${\cal B} = (0,1)$ cycles, which intersect on $T^2$ as ${\cal A}^2 = {\cal B}^2 = 0$ and ${\cal A} \cdot {\cal B} = - {\cal B} \cdot {\cal A} = 1$.
This can be done by a Smith decomposition on $\kappa$.
Then, there is a 2-chain $\Sigma_i \subset \partial Y$ with $\partial \Sigma_i = N_i {\cal C}_i$.
The linking pairing $L: \text{Tor}(H_1(\partial Y)) \times \text{Tor}(H_1(\partial Y)) \rightarrow \mathbb{Q}/\bbZ$ is then computed as
\begin{align}
  L({\cal C}_i, {\cal C}_j) = \frac{1}{N_i} \Sigma_i \cdot {\cal C}_j \mod \bbZ \, .
\end{align}

\paragraph{$I_N$ fiber:}
The standard representation of the monodromy around an $I_N$ fiber is given by $K = \left( \begin{smallmatrix} 1 & -N \\ 0 & 1 \end{smallmatrix} \right)$.
It is easy to see that
\begin{align}\label{eq:gamma_I_N_fiber}
	\kappa = \begin{pmatrix} 1 & 0 \\ 0 & 1 \end{pmatrix} - \begin{pmatrix} 1 &-N \\ 0 & 1 \end{pmatrix} = \begin{pmatrix} 0 & N \\ 0 & 0 \end{pmatrix}
\end{align}
has $\text{im}(\kappa) \cong \bbZ$, generated by the 1-cycle $\kappa {\cal B} = \left( \begin{smallmatrix} N \\ 0 \end{smallmatrix} \right) = N \times {\cal A}$.
Therefore,
\begin{align}
	\text{coker}(\kappa) = \frac{H^1(T^2) }{\text{im}(\kappa)} \cong \frac{\langle \left(\begin{smallmatrix} 1 \\ 0 \end{smallmatrix}\right) \rangle_\bbZ \times \langle \left(\begin{smallmatrix} 0 \\ 1 \end{smallmatrix}\right) \rangle_\bbZ }{\langle \left(\begin{smallmatrix} N \\ 0 \end{smallmatrix}\right) \rangle_\bbZ} \cong \bbZ_N \times \bbZ \, ,
\end{align}
where $\langle v \rangle_\bbZ$ denotes the $\bbZ$-span of a vector $v$.
Here, the $\bbZ_N$ factor is generated by the ${\cal A}$-cycle, and the free $\bbZ$ factor by the ${\cal B}$-cycle.
Another way to read \eqref{eq:gamma_I_N_fiber} is that the 2-cycle $\Sigma \subset \partial Y$, obtained by tracing the ${\cal B}$-cycle over any point $p \in S^1_\text{base}$ once around the circle $S^1_\text{base}$ back to $p$, has boundary $\partial \Sigma = K{\cal B} - {\cal B} = -\kappa {\cal B} = -N \, {\cal A}$.
This allows us to compute the linking pairing:
\begin{align}\label{eq:defect_pairing_I_N}
	L({\cal A}, {\cal A}) = -\tfrac{1}{N} \Sigma \cdot {\cal A} \mod \bbZ = \tfrac{1}{N} \mod \bbZ \, ,
\end{align}
where we have used the specific realization of $\Sigma$ as the ${\cal B}$-cycle fibered over $S^1$, which intersects the ${\cal A}$-cycle only in one fiber, where ${\cal B} \cdot_{T^2} {\cal A} = -1$.
This result agrees with the physical expectation that the $I_N$ fiber realizes an $\mathfrak{su}(N)$ gauge symmetry as a 7d M-theory compactification, whose electric 1-form and magnetic 4-form symmetry are both $\mathbb{Z}_N$, with the defect group pairing \eqref{eq:defect_pairing_I_N}.

\paragraph{$I^*_{N-4}$ fiber:} The monodromy around an $I_{N-4}^*$, $N\geq 4$, type fiber, in the above $(\cal A, B)$ basis, is $K = \left( \begin{smallmatrix} -1 & N-4 \\ 0 & -1 \end{smallmatrix} \right)$.
To compute $\text{coker}(\kappa)$, it is easiest to distinguish between even and odd $N$.

For even $N \equiv 2n$, $n \geq 2$, we have
\begin{align}\label{eq:smith_decomp_D2n_monodromy}
	\kappa =\begin{pmatrix} 1& 0 \\ 0 & 1 \end{pmatrix} - \begin{pmatrix} -1 & 2n - 4 \\ 0 & -1 \end{pmatrix} = \begin{pmatrix} 2 & 4-2n\\ 0 & 2 \end{pmatrix} = \begin{pmatrix} 1& 2-n \\ 0 & 1 \end{pmatrix} \begin{pmatrix} 2 & 0 \\ 0 & 2 \end{pmatrix} \begin{pmatrix} 1 & 0 \\ 0 & 1 \end{pmatrix} \, ,
\end{align}
where the last equality can be interpreted as a Smith decomposition, showing in particular that
\begin{align}
	\text{coker}(\kappa) \cong \frac{\langle \left(\begin{smallmatrix} 1 \\ 0 \end{smallmatrix}\right) \rangle_\bbZ \times \langle \left(\begin{smallmatrix} 0 \\ 1 \end{smallmatrix}\right) \rangle_\bbZ}{\langle \left( \begin{smallmatrix} 2 \\ 0 \end{smallmatrix} \right) \rangle_\bbZ \times \langle \left( \begin{smallmatrix} 0 \\ 2 \end{smallmatrix} \right) \rangle_\bbZ } \cong \bbZ_2 \times \bbZ_2 \, .
\end{align}
Now consider the 2-chain $\Sigma_1 \subset \partial Y$ obtained from fibering the 1-cycle $-{\cal A} \equiv \left( \begin{smallmatrix} -1 \\ 0 \end{smallmatrix} \right)$ once over the boundary circle back to a reference point $p \in S^1_\text{base}$.
Then we have $\partial \Sigma_1 = -\kappa (-{\cal A}) = 2 {\cal A}$, showing that ${\cal A}$ is a 2-torsional 1-cycle in $\partial Y$.
Likewise, we can also see that ${\cal B}$ is a 2-torsional cycle in $\partial Y$, as $-\kappa \left( \begin{smallmatrix} 2-n \\ -1 \end{smallmatrix} \right) = \left( \begin{smallmatrix} 0 \\ 2 \end{smallmatrix} \right)$, i.e., $2{\cal B}$ is the boundary of the 2-chain $\Sigma_2$ swept out by moving the 1-cycle $\left( \begin{smallmatrix} 2-n \\ -1 \end{smallmatrix} \right)$ around $S^1_\text{base}$ once.
To compute the linking pairing, first note that 
\begin{align}
	\begin{aligned}
		& \Sigma_1 \cdot {\cal A} = -{\cal A} \cdot_{T^2} {\cal A} = 0 \, , \quad && \Sigma_1 \cdot {\cal B} = -{\cal A} \cdot_{T^2} {\cal B} = -1 \, , \\
		& \Sigma_2 \cdot {\cal A} = ((2-n) {\cal A} - {\cal B}) \cdot_{T^2} {\cal A} = -1\, , \quad && {\cal B} \cdot \Sigma_2 = {\cal B} \cdot_{T^2} ((2-n) {\cal A} - {\cal B} = 2-n \, .
	\end{aligned}
\end{align}
Thus, the linking pairing, in the basis $({\cal A}, {\cal B})$, is (with $\tfrac12 = -\tfrac12 \mod \bbZ$)
\begin{align}\label{eq:linking_pairing_Deven}
	\text{even $N$} : \quad L_{I^*_{N-4}} = \begin{pmatrix} 0 & \tfrac12 \\ \tfrac12 & \tfrac{N}{4} \end{pmatrix} \mod \bbZ \, .
\end{align}

For odd $N = 2n+1$, $n\geq 2$, we have
\begin{align}\label{eq:smith_decomp_Dodd_monodromy}
	\gamma =\begin{pmatrix} 1& 0 \\ 0 & 1 \end{pmatrix} - \begin{pmatrix} -1 & 2n - 3 \\ 0 & -1 \end{pmatrix} = \begin{pmatrix} 2 & 3-2n\\ 0 & 2 \end{pmatrix} = \begin{pmatrix} 1& n-1 \\ -2 & 3-2n \end{pmatrix} \begin{pmatrix} 1 & 0 \\ 0 & 4 \end{pmatrix} \begin{pmatrix} 0 & -1 \\ 1 & 2 \end{pmatrix} \, ,
\end{align}
implying $\text{coker}(\kappa) \cong \bbZ_4$.
A generating 4-torsional 1-cycle is given by $\left( \begin{smallmatrix} 1-n \\ 1 \end{smallmatrix} \right) = \tfrac14 \times (- \kappa \left( \begin{smallmatrix} 1 \\ -2 \end{smallmatrix} \right))$, i.e., four copies of it is the boundary of the 2-chain $\Sigma$ obtained from moving $\left( \begin{smallmatrix} 1 \\ -2 \end{smallmatrix} \right)$ around $S^1_\text{base}$ once.
From $\Sigma \cdot \left( \begin{smallmatrix} 1-n \\ 1 \end{smallmatrix} \right) = \left( \begin{smallmatrix} 1 \\ -2 \end{smallmatrix} \right) \cdot_{T^2} \left( \begin{smallmatrix} 1-n \\ 1 \end{smallmatrix} \right) = 3 -2n = 4 -N$, we find the linking pairing
\begin{align}\label{eq:linking_pairing_Dodd}
	\text{odd $N$}: \quad L_{I^*_{N-4}} = \tfrac{-N}{4} \mod \bbZ \, .
\end{align}

Again, this reproduces the physical expectations, namely that $I^*_{N-4}$ realizes an $\mathfrak{so}(2N)$ gauge symmetry.
For even $N = 2n$, the electric and magnetic symmetries are $\bbZ_2 \times \bbZ_2$, whose defect group pairing is \eqref{eq:linking_pairing_Deven}.\footnote{
Note that this deviates from \cite{Albertini:2020mdx} for odd $n$, or $N \not\equiv 0 \mod 4$, where the pairing matrix is given by $\left( \begin{smallmatrix} 1/2 & 0\\ 0 & 1/2\end{smallmatrix} \right)$.
One can easily check that we arrive in this form by a simply basis change $({\cal A}, {\cal B}) \rightarrow ({\cal A + B}, {\cal B})$.}
Meanwhile, for odd $N=2n+1$, $\mathfrak{so}(2N)$ has $\bbZ_4$ center symmetry, which pairs with its magnetically dual $\bbZ_4$ symmetry as \eqref{eq:linking_pairing_Dodd}.

\paragraph{$IV$ fiber:} The monodromy is given by $K=\left( \begin{smallmatrix} -1 & -1\\ 1& 0\end{smallmatrix} \right)$, so
\begin{align}
	\kappa = \begin{pmatrix} 2 & 1 \\ -1 & 1 \end{pmatrix} = \begin{pmatrix} 2 & -1 \\ -1 & 1 \end{pmatrix} \begin{pmatrix} 1 & 0 \\ 0 & 3 \end{pmatrix} \begin{pmatrix} 1 & 2 \\ 0 & 1 \end{pmatrix} \, .
\end{align}
Therefore $\text{coker}(\kappa) \cong \bbZ_3$, whose generator can be represented by the ${\cal B}$-cycle, since $3{\cal B} = \left( \begin{smallmatrix} 0 \\ 3 \end{smallmatrix} \right) = -\kappa \left( \begin{smallmatrix} -1 \\ 2 \end{smallmatrix} \right)$ is the boundary of $\Sigma$ obtained from fibering $\left( \begin{smallmatrix} -1 \\ 2 \end{smallmatrix} \right)$ over $S^1_\text{base}$.
From $\Sigma \cdot {\cal B} = \left( \begin{smallmatrix} -1 \\ 2 \end{smallmatrix} \right) \cdot_{T^2} \left( \begin{smallmatrix} 0 \\ 1 \end{smallmatrix} \right) = -1$, the linking pairing is $L({\cal B}, {\cal B}) = -\tfrac13 \mod \bbZ$.
This defines the linking pairing for the $\bbZ_3$ 1-form electric / 4-form magnetic symmetry of an $\mathfrak{su}(3)$ gauge symmetry, which is different than an $\mathfrak{su}(3)$ realized via an $I_3$ fiber.

\paragraph{$IV^*$ fiber:} In this case, the monodromy is $K=\left( \begin{smallmatrix} -1 & 1\\ -1& 0\end{smallmatrix} \right)$, so
\begin{align}
  \kappa = \begin{pmatrix} 2 & -1 \\ 1 & 1 \end{pmatrix} = \begin{pmatrix} 2 & -1 \\ 1 & 0 \end{pmatrix} \begin{pmatrix} 1 & 0 \\ 0 & 3 \end{pmatrix} \begin{pmatrix} 1 & 1 \\ 0 & 1 \end{pmatrix} \, .
\end{align}
Hence coker$(\kappa) \cong \bbZ_3$, which can be generated by the ${\cal A}$-cycle, since $3{\cal A} = \left( \begin{smallmatrix} 3 \\ 0 \end{smallmatrix} \right) = -\kappa \left( \begin{smallmatrix} -1 \\ 1 \end{smallmatrix} \right) = \partial \Sigma$, which also defines $\Sigma$ to be the 2-chain obtained by fibering $\left( \begin{smallmatrix} -1 \\ 1 \end{smallmatrix} \right)$ over $S^1_{\text{base}}$.
Then $\Sigma \cdot {\cal A} = \left( \begin{smallmatrix} -1 \\ 1 \end{smallmatrix} \right) \cdot_{T^2} \left( \begin{smallmatrix} 1 \\ 0 \end{smallmatrix} \right) = -1$, so the linking pairing is $L({\cal A}, {\cal A}) = -\tfrac13 = \frac23 \mod \bbZ$
Physically, this agrees with the expectation from an 7d $\mathfrak{e}_6$ gauge symmetry from an $IV^*$ fiber, which has 1-form electric / 4-form magnetic $\bbZ_3$ symmetry, and defect group pairing $-\tfrac13 = \tfrac23 \mod \bbZ$ \cite{Albertini:2020mdx}.

\paragraph{$III$ fiber:} The monodromy of this fiber is $K=\left( \begin{smallmatrix} 0 & -1\\ 1& 0\end{smallmatrix} \right)$, so
\begin{align}
	\kappa = \begin{pmatrix} 1 & 1 \\ -1 & 1 \end{pmatrix} = \begin{pmatrix} 1 & 0 \\ -1 & 1 \end{pmatrix} \begin{pmatrix} 1 & 0 \\ 0 & 2 \end{pmatrix} \begin{pmatrix} 1 & 1 \\ 0 & 1 \end{pmatrix} \, .
\end{align}
Therefore $\text{coker}(\kappa) \cong \bbZ_2$, which is generated by the ${\cal B}$-cycle, since $2{\cal B} = \left( \begin{smallmatrix} 0 \\ 2 \end{smallmatrix} \right) = -\kappa \left( \begin{smallmatrix} 1 \\ -1 \end{smallmatrix} \right)$ is the boundary of $\Sigma$ obtained from fibering $\left( \begin{smallmatrix} 1 \\ -1 \end{smallmatrix} \right)$ over $S^1_\text{base}$.
From $\Sigma \cdot {\cal B} = \left( \begin{smallmatrix} 1 \\ -1 \end{smallmatrix} \right) \cdot_{T^2} \left( \begin{smallmatrix} 0 \\ 1 \end{smallmatrix} \right) = 1$, the linking pairing is $L({\cal B}, {\cal B}) = \tfrac12 \mod \bbZ$.
Physically, this agrees with the expectation from a 7d theory with an $\mathfrak{su}(2)$ gauge symmetry from a type $III$ fiber.

\paragraph{$III^*$ fiber:} The monodromy is $K=\left( \begin{smallmatrix} 0 & 1\\ -1& 0\end{smallmatrix} \right)$, so
\begin{align}
  \kappa = \begin{pmatrix} 1 & -1 \\ 1 & 1 \end{pmatrix} = \begin{pmatrix} 1 & -1 \\ 1 & 0 \end{pmatrix} \begin{pmatrix} 1 & 0 \\ 0 & 2 \end{pmatrix} \begin{pmatrix} 1 & 1 \\ 0 & 1 \end{pmatrix} \, .
\end{align}
The coker$(\kappa) \cong \bbZ_2$ can be represented by the ${\cal A}$-cycle, which is the boundary of $\Sigma$ obtained by fibering $\left( \begin{smallmatrix} -1 \\ 1 \end{smallmatrix} \right)$ around $S^1_\text{base}$: $-\kappa \left( \begin{smallmatrix} -1 \\ 1 \end{smallmatrix} \right) = \left( \begin{smallmatrix} 2 \\ 0 \end{smallmatrix} \right)$.
The defect group pairing is determined from $\Sigma \cdot {\cal A} = \left( \begin{smallmatrix} -1 \\ 1 \end{smallmatrix} \right) \cdot_{T^2} \left( \begin{smallmatrix} 1 \\0 \end{smallmatrix} \right) = -1$ to be $L({\cal A}, {\cal A}) = -\frac12 = \frac12 \mod \bbZ$, which agrees with that of the 1-form electric / 4-form magnetic symmetry of a 7d $\mathfrak{e}_7$ theory \cite{Albertini:2020mdx}.

\paragraph{$II$ and $II^*$ fibers:} Lastly, these fiber have monodromy $K_{II} = \left( \begin{smallmatrix} 1 & -1\\ 1& 0\end{smallmatrix} \right)$ and $K_{II^*} = \left( \begin{smallmatrix} 1 & 1\\ -1& 0\end{smallmatrix} \right)$.
Since, in these cases,
\begin{align}
	\kappa_{II} = \begin{pmatrix} 0 & 1 \\ -1 & 1 \end{pmatrix} \, , \quad \kappa_{II^*} = \begin{pmatrix} 0 & -1 \\ 1 & 1 \end{pmatrix}
\end{align}
is unimodular, the cokernels are trivial, conforming with expectations of neither a trivial nor an $\mathfrak{e}_8$ gauge symmetry have any higher-form electric and magnetic symmetries.

\section{Instanton Fractionalization in 5d \texorpdfstring{\boldmath{$SU(N)_k$}}{SU(N)[k]} Theories}\label{app:smith_decomp_suN_5d}

In this appendix, compute the 't Hooft anomaly between the instanton $U(1)_I$ 0-form symmetry and the $\bbZ_{\text{gcd}(N,k)}$ 1-form symmetry of a 5d $SU(N)_k$ gauge symmetry.
The Calabi--Yau threefold geometry is the local neighborhood of the surface configuration \cite{Bhardwaj:2019ngx}
\begin{align}\label{eq:surface_config_suN}
\begin{split}
	\bbF_{N-k-2} \quad _{e_1} \overset{\line(2,0){10}}{\;}_{h_2} & \quad \bbF_{N-k-4} \quad _{e_2}\overset{\line(2,0){5}}{\;} \ldots \overset{\line(2,0){5}}{\;}_{h_m} \quad \bbF_{N-k-2m} \quad _{e_m} \overset{\line(2,0){10}}{\;}_{e_{m+1}} \quad \bbF_{-N+k+2(m+1)} \quad _{h_{m+1}} \overset{\line(2,0){5}}{\;} \ldots \\
	\ldots \overset{\line(2,0){5}}{\;}_{e_{N-2}} & \quad \bbF_{N+k-4} \quad _{h_{N-2}} \overset{\line(2,0){10}}{\;}_{e_{N-1}} \quad \bbF_{N+k-2} \, ,
\end{split}
\end{align}
where the degrees $n_a = N-k-2a$ for $a \leq m$, and $n_a = -(N-k-2a)$ for $a>m$, are all positive.
Each of these (complex) surfaces have a basis of 2-cycles, $(e_a, f_a)$, with $h_a := e_a + n_a \, f_a$, which inside $\bbF_{n_a}$ intersect as $e_a^2 = -n_a$, $f_a^2 = 0$, $e_a \cdot f_a = 1$.
The set of all $(e_a, f_a)$ generate $H_2(Y_6)$.
However, they are not all independent, since the intersections $_{C_a} \overset{\line(2,0){10}}{\;}_{C_{a+1}}$ between $\sigma_a$ and $\sigma_{a+1}$, as indicated in \eqref{eq:surface_config_suN}, impose the gluing condition $C_1 = C_2$ in $Y_6$.\footnote{Note that in \cite{Morrison:2020ool}, the degrees of $\bbF_{n_a}$ are uniformly denoted by $n_a = N-k-2a$ and hence can go negative, with the understanding that in $\bbF_{-n} \cong \bbF_{n}$, the role of $e$ and $h = e+n\,f$ are exchanged \cite{Bhardwaj:2019ngx}.
}

This allows to pick a basis $\{\gamma_i\}$ for $H_2(Y)$, which we set to be
\begin{align}\label{eq:2-cycle_basis_suN}
	\gamma_a = f_a \; (1 \leq a \leq N-1) \, , \quad \gamma_N = e_1 \, .
\end{align}
The other curves $e_a$, $a\geq 2$, are then determined by the gluing conditions $h_a = e_{a-1}$ for $a\leq m$, $h_a = e_{a+1}$ for $a>m$:
\begin{align}\label{eq:gluing_conditions}
	\begin{split}
		e_a & = e_1 - \sum_{b=2}^{a} n_b \, f_b \quad \text{for} \quad 2 \leq a \leq m < N-1\, , \\
		e_{m+1} & = e_m = e_1 - \sum_{b=2}^{m} n_b \, f_b \, , \\
		e_a & = e_1 - \sum_{b=2}^m n_b\,f_b + \sum_{b=m+1}^{a-1} n_b\,f_b \quad \text{for} \quad m+2 \leq a \leq N-1 \, .
	\end{split}
\end{align}
The intersection matrix is
\begin{align}\label{eq:intersection_matrix_suN_5d}
	M_{ai} = \langle \sigma_a, \gamma_i \rangle = 
	\begin{pmatrix}
		-2 & 1 & 0 & \ldots & N-k-4 \\
		1 & -2 & \ddots &  & 2+k - N\\
		0 & \ddots & \ddots & 1 & 0\\
		0 & & 1 & -2 & 0
	\end{pmatrix}
	\equiv 
	\left( \begin{array}{@{}l | r@{}}
		C & \vec{v}
	\end{array} \right) \, ,
\end{align}
where $C_{ab} = \langle \sigma_a, f_b \rangle$ is the (negative) Cartan matrix of $SU(N)$ (see \eqref{eq:suN_cartan_matrix_7d}), and $v_a = \langle \sigma_a, \gamma_N \rangle = \langle \sigma_a , e_1 \rangle$.\footnote{One can infer the intersection pairing $\langle \cdot, \cdot \rangle$ in the Calabi--Yau threefold $Y_6$ from the intersection of 2-cycles $C_1 \cdot C_2$ in $\sigma \cong \bbF_{n}$.
Let $\gamma \subset \sigma$, then $\langle\sigma , \gamma\rangle = -(2e + (n+2)f) \cdot \gamma$.
If there is another 4-cycle $\sigma'$ such that $\sigma$ and $\sigma'$ intersect along $C \subset \sigma$, then $\langle \sigma', \gamma \rangle = C \cdot \gamma$.
}
Note that this intersection matrix does not depend on the different presentation of the surface configuration \eqref{eq:surface_config_suN} compared to \cite{Morrison:2020ool}.
Hence, the subsequent computations of this appendix proceed analogously.

Since $C$ has the known Smith decomposition \eqref{eq:smith_decomp_suN_cartan}, we have
\begin{align}
	M = S \left( \begin{array}{@{}l|r @{}} D & \vec{w} \end{array}  \right) \left(
	\begin{array}{c|c}
		T & 0 \\ \hline
		0 & 1
	\end{array}
	\right) \, , \quad \vec{w} \equiv S^{-1}\vec{v} = (-2, k-N, \ldots, k-N)^t \, ,
\end{align}
where the $(N-1) \times (N-1)$ matrices $(S,D,T)$ are given as in \eqref{eq:smith_decomp_suN_cartan}.
With the first $N-2$ diagonal entries of $D$ being 1, one further can eliminate the first $N-2$ components of $\vec{w}$ with elementary column operations that add multiples of the first $N-2$ columns of $D$ to $\vec{w}$:
\begin{align}
  (D \, | \, \vec{w}) \left( \begin{array}{ccc|cc}
     &  & & 0 & -w_1 \\
     & \mathbbm{1}_{N-2} & & \vdots & \vdots \\
    &  & & 0 & -w_{N-2} \\ \hline
    & 0 & & \multicolumn{2}{c}{\multirow{2}{*}{$\mathbbm{1}_2$}} \\
    & 0 & & 
  \end{array} \right) =
  \begin{pmatrix}
    1 & 0 &  & & 0 \\
    0 & \ddots & & & \vdots \\
    & & 1 & 0 & \vdots \\
    & & & N & k-N
  \end{pmatrix} \equiv ( D' \, | \, \vec{w}') \, .
\end{align}
For the last two entries, $(N, k-N)$, we can use an extended Euclidean algorithm, such that there is an invertible integer $2 \times 2$ matrix $Q^{-1}$ with
\begin{align}\label{eq:Q_matrix_def}
\begin{split}
  & \underbrace{\det(Q)}_{=\pm 1} \left( Q_{22} N - Q_{21} (k-N)), (k-N) Q_{11}-Q_{12} N \right) \equiv (N, k-N) \, Q^{-1} = \det(Q) (\text{gcd}(N,k), 0) \\
  & \Longrightarrow \quad  Q_{11} = \frac{N}{\text{gcd}(N,k)} =: \ell \, , \quad Q_{12} = \frac{k-N}{\text{gcd}(N,k)} =: \tilde\ell \, , \quad Q_{22} = \frac{\det{Q} + \tilde\ell \, Q_{21}}{\ell} \in \bbZ \, .
\end{split}
\end{align}
This can be summarized as
\begin{align}
\begin{split}
	& \left( \begin{array}{@{}l|r @{}} D' & \vec{w}' \end{array}  \right) 
	\left( \begin{array}{ccc|cc}
		 &  & & 0 & 0 \\
		 & \mathbbm{1}_{N-2} & & \vdots & \vdots \\
		&  & & 0 & 0 \\ \hline
		& 0 & & \multicolumn{2}{c}{\multirow{2}{*}{$Q^{-1}$}} \\
		& 0 & & 
	\end{array} \right)
	=
	\begin{pmatrix}
		1 & 0 &  & & 0 \\
		0 & \ddots & & & \vdots \\
		& & 1 & 0 & \vdots \\
		& & & \text{gcd}(N,k) & 0
	\end{pmatrix} 
	\equiv \left( D'' \,  |  \, 0 \right) \, ,
\end{split}
\end{align}
which shows explicitly that $H_4(Y_6, \partial Y_6)/\text{im}(\jmath_4) \cong \bbZ \oplus \bbZ_{\text{gcd}(N,k)}$ \cite{Morrison:2020ool}.
This means that the Smith decomposition of \eqref{eq:intersection_matrix_suN_5d} takes the form
\begin{align}
\begin{split}
	M & = S \, \left( D'' \, | \, 0 \right) 
  \left( \begin{array}{ccc|cc}
     &  & & 0 & 0 \\
     & \mathbbm{1}_{N-2} & & \vdots & \vdots \\
    &  & & 0 & 0 \\ \hline
    & 0 & & \multicolumn{2}{c}{\multirow{2}{*}{$Q^{-1}$}} \\
    & 0 & & 
  \end{array} \right)^{-1}
  \left( \begin{array}{ccc|cc}
     &  & & 0 & -w_1 \\
     & \mathbbm{1}_{N-2} & & \vdots & \vdots \\
    &  & & 0 & -w_{N-2} \\ \hline
    & 0 & & \multicolumn{2}{c}{\multirow{2}{*}{$\mathbbm{1}_2$}} \\
    & 0 & & 
  \end{array} \right)^{-1} \left(\begin{array}{c|c}
		T & 0 \\ \hline
		0 & 1
	\end{array}
	\right) \\
	& = S \, \left( D'' \, | \, 0 \right) \left( \begin{array}{c|c}
		\mathbbm{1}_{N-2} & * \\ \hline
		0 & Q
	\end{array} \right)
	\left( 
		\begin{array}{c|c}
			\mathbbm{1}_{N-2} & *' \\ \hline
			0 & \mathbbm{1}_{2}
		\end{array}
	\right) = S \, \left( D'' \, | \, 0 \right) 
		\underbrace{\left( \begin{array}{c|c}
		\mathbbm{1}_{N-2} & \tilde{*} \\ \hline
		0 & Q
	\end{array} \right)}_{\equiv \tilde{T}} \, ,
\end{split}
\end{align}
where we have used the schematic form \eqref{eq:smith_decomp_suN_cartan} of the $(N-1) \times (N-1)$ matrix $T = \left( \begin{smallmatrix} \mathbbm{1}_{N-2} & \ast' \\ 0 & 1\end{smallmatrix} \right)$.
We are omitting the details on the upper right part of the matrices, because these ultimately specify only the generators of $H_4(Y_6, \partial Y_6)$ that lie in $\text{im}(\jmath_4)$ in terms of the dual basis of $\gamma_i$; these are projected out in the quotient that determines the global and 1-form symmetries.
The decomposition also tells us that the $\text{gcd}(N,k)$-torsional boundary flux may be represented as
\begin{align}
\frac{1}{\text{gcd}(N,k)} \sum_a (S^{-1})_{N-1,a} \, \jmath_4(\sigma_a) = \frac{1}{\text{gcd}(N,k)} \sum_a a \, \jmath_4(\sigma_a) \, .
\end{align}
Moreover, we have the generator
\begin{align}\label{eq:def_top_U1_suN_5d_example}
\begin{split}
	& \epsilon_I = \sum_{i} \tilde{T}_{N,i} \eta_i = Q_{21} \, \eta_{N-1} + Q_{22} \, \eta_{N} \in \text{Hom}(H_2(Y_6), \bbZ) \, : \\
	& \sum_{i=1}^N \mu_i \gamma_i \equiv \sum_{i=1}^{N-2} \mu_i \gamma_i + \mu_{N-1} \, f_{N-1} + \mu_N \, e_1 \mapsto Q_{21} \, \mu_{N-1} + Q_{22} \, \mu_N \, .
\end{split}
\end{align}

Furthermore, the intersections $\gamma_{ab} = \sigma_a \cdot \sigma_b$ with $a \neq b$, as indicated by \eqref{eq:surface_config_suN}, can be expressed in the basis $\gamma_i \in \{f_1, f_2,..., f_{N-1}, e_1\}$ via \eqref{eq:gluing_conditions}:
$\gamma_{ab}$ is either trivial (if $|a-b|>1$), or one of the gluing curves in \eqref{eq:surface_config_suN}, whose homology class takes the form $e_l = e_1 + \sum_{c} (\pm n_c) f_c$.
Omitting the precise expression, the important point becomes that the sum \emph{never} contains $f_{N-1}$ because of \eqref{eq:gluing_conditions}.
For $\gamma_{aa} = \sigma_a \cdot \sigma_a = -(2e_a + n_a f_a) = -2e_1 + \sum_b y_b f_b$, the sum also does not contain $f_{N-1}$ for $a<N-1$.
For $a=N-1$, we have
\begin{align}
	\gamma_{N-1, N-1} = -(2e_{N-1} + (n_{N-1}+2) f_{N-1}) = -2e_1 + \sum_{b=2}^{N-2} y_b' f_b - (N+k) f_{N-1}
\end{align}
for some $y_b'$.
Therefore, we have
\begin{align}
\begin{split}
	K_{Iab} & = \epsilon_I(\sigma_a \cdot \sigma_b) = \begin{cases}
		-2Q_{22} - \delta_{a,N-1} (N+k) Q_{21} \, , & \text{if } \, a = b \, ,\\
		Q_{22} \, , & \text{if } \, b=a+1 \, ,\\
		Q_{22} \, , & \text{if } \, b=a-1 \, ,\\
		0 \, & \text{otherwise} \, .
	\end{cases} \\
	& = Q_{22} \, (-C_{ab}) -  \delta_{a,N-1}\delta_{b,N-1} (N+k) \, Q_{21} \, .
\end{split}
\end{align}
Therefore, in the presence of the $\bbZ_{\text{gcd}(N,k)}$ 1-form symmetry background, the cross-term's coefficients in \eqref{eq:fractional_shift_CS-terms_5d} are
\begin{align}
\begin{split}
	\sum_{a} K_{Iab} \frac{\lambda_a}{\text{gcd}(N,k)} & = \frac{1}{\text{gcd}(N,k)} \sum_a a  \left( Q_{22} \, (-C_{ab})  - \delta_{a,N-1}\delta_{b,N-1} (N+k) Q_{21} \right)\\
	& = 
	\begin{cases}
		0 \, , & \text{if } \, b < N-1 \, , \\
		\frac{Q_{22} \, N - (N-1)(N+k) Q_{21}}{\text{gcd}(N,k)} \, , & \text{if } \, b = N-1 \, ,
	\end{cases}
\end{split}
\end{align}
where we have used \eqref{eq:sum_cartan_a} for the particular weighted sum over Cartan matrix entries of $SU(N)$.
Since both $N$ and $(N+k)$ divide $\text{gcd}(N,k)$, these coefficients are indeed integer.

Finally, we can compute the fractional shift of the instanton density,
\begin{align}
\begin{split}
	& \frac12 \sum_{a,b} \int_{M_5}  K_{Iab} \, A_I \wedge \left( F_a \wedge F_b  + \frac{\lambda_a \lambda_b}{\text{gcd}(N,k)^2} B \wedge B \right) \\
	= & \int_{M_5} \left( \sum_{a,b}  \frac{Q_{22}}{2} (-C_{ab}) A_I \wedge F_a \wedge F_b - \frac{(N+k)Q_{21}}{2} A_I \wedge F_{N-1} \wedge F_{N-1} \right) \\
	+ & \int_{M_5} \frac{1}{2 \, \text{gcd}(N,k)^2} \left( Q_{22} \, N(N-1) - Q_{21} (N+k) (N-1)^2 \right) B \wedge B \, .
\end{split} 
\end{align}
For the last term, we define $\ell = N/\text{gcd}(N,k)$; we can view the $\bbZ_{\text{gcd}(N,k)}\subset \bbZ_{N}$ subgroup of the center of $SU(N)$ being generated by $\ell \mod N$.
Then
\begin{align}
\begin{split}
	& \frac{1}{2\text{gcd}(N,k)^2} \left( Q_{22} \, N(N-1) - Q_{21} (N+k) (N-1)^2 \right) \\
	= &\frac{N-1}{2N} \ell^2 \left( Q_{22}  - Q_{21} (N+k) \frac{N-1}{N} \right) \, .
\end{split}
\end{align}

\chapter{\MakeUppercase{Chapter 3 Appendix}}

\section{Global Gauge Group of Maximally Enhanced 8d CHL Vacua}
\label{app:big_table}

In this appendix, we present the non-Abelian gauge group $G = \widetilde{G} / {\cal Z}$ of maximally enhanced 8d CHL vacua, i.e., with rank$(G) = 10$.
There are 61 of them \cite{Font:2021uyw,Hamada:2021bbz}, listed in the same order as \cite{Hamada:2021bbz}.
We determined these from their ``parent'' heterotic models, as described in Section \ref{sec:3}.
The global structure of these theories can be obtained with various methods, including that of \cite{Font:2020rsk}.
In practice, we use a generalization of string junctions techniques \cite{Guralnik:2001jh}, which will be elaborated in our upcoming work \cite{Cvetic:2021sjm}.
There, we will also compute the full global gauge group, including the $U(1)$s.

From the embeddings ${\cal Z} \hookrightarrow Z(\widetilde{G})$, one can explicitly verify that all non-trivial gauge groups are consistent with the vanishing of the mixed 1-form center anomaly \cite{Cvetic:2020kuw}.
We have also checked that the two cases, \#24 and \#52, whose character lattice contains only real representations, satisfy the constraint $\dim(G) + \text{rank}(G) = 0\mod 8$ \cite{Montero:2020icj}:
\begin{equation}
\begin{split}
    \#24:& \quad \text{dim}(Spin(12)) + \text{dim}(Sp (4)) +\text{rank}(Spin(12)) + \text{rank}(Sp (4)) = 112 = 0 \text{ mod } 8 \,, \\
    \#52:& \quad \text{dim}(Spin(16)) + 2 \, \text{dim}(SU (2)) +\text{rank}(Spin(16)) + 2 \, \text{rank}(SU (2)) = 136 = 0 \text{ mod } 8 \,.
\end{split}    
\end{equation}

\renewcommand{\arraystretch}{1.3}
\begin{longtable}{|c|c|c|c|c|}
\caption{All 61 maximally enhanced CHL vacua, together with the simply-connected cover $\widetilde{G} = \prod_i \widetilde{G}_i$ of their non-Abelian gauge group $G = G/{\cal Z}$.
The embedding ${\cal Z} \hookrightarrow Z(\widetilde{G})$ is specified by expressing the generator(s) of ${\cal Z}$ via a tuple $(k_i) \in \prod_i Z(\widetilde{G}_i)$.
If $\widetilde{G}_i = Spin(4n)$, then $k_i = (k_i^{(1)}, k_i^{(2)}) \in Z(Spin(4n)) \cong \bbZ_2 \times \bbZ_2$.
All ADE-factors have Kac-Moody level 2, while the $Sp(n)$ factors have level 1.
Note that, while $Sp(1) \cong SU(2)$ as Lie groups, we will use $Sp(1)$ if the gauge factor is at level 1, and $SU(2)$ if it is at level 2.
\label{tab:big_table}
} \\
\hline
\#  & $\widetilde{G}$    & $\mathcal{Z}$   &  $\mathcal{Z} \hookrightarrow Z(\widetilde{G})$\\ \hline 
1   & $E_8 \times Sp(2)$        &  0   & -  \\  \hline
2   & $E_8 \times Sp(1) \times SU(2)$  &  0   & -   \\  \hline
3   & $E_7 \times Sp(3)$        &  0   & - \\  \hline
4   & $E_7 \times Sp(2) \times SU(2)$  &  $\bbZ_2$  &    $(1, 1, 0)$     \\  \hline
5   & $E_7 \times Sp(1) \times SU(3)$  &  0   & -  \\  \hline
6   & $E_7 \times SU(3) \times SU(2)$  &  $\bbZ_2$  &   $(1, 0, 1)$     \\  \hline
7   & $E_6 \times Sp(4)$        &  0   & -  \\   \hline
8   & $E_6 \times Sp(3) \times SU(2)$  &  0   & -  \\ \hline
9   & $E_6 \times Sp(1) \times SU(4)$  &  0   & -   \\ \hline
10  & $E_6 \times Sp(1) \times SU(3) \times SU(2)$    &  $0$  &  -   \\ \hline
11  & $E_6 \times SU(5)$    &   $0$  &  -  \\ \hline
12  & $Sp(10)$    &   $0$  & -  \\ \hline
13  & $Sp(9) \times SU(2)$    &   $0$  & -  \\ \hline
14  & $Sp(8) \times SU(3)$    &   $\bbZ_2$  &   $(1, 0)$    \\ \hline
15  & $Sp(8) \times SU(2)^2$     &   $\bbZ_2$  &   $(1, 0, 0)$    \\ \hline
16  & $Sp(7) \times SU(3) \times SU(2)$ &  $0$       & -   \\ \hline
17  & $Sp(6) \times SU(5)$    &   $0$  & -  \\ \hline
18  & $Sp(6) \times SU(4) \times SU(2)$    &   $\bbZ_2$  &   $(1, 0, 1)$    \\ \hline
19  & $Sp(6) \times SU(3)^2$     &  $0$      &  -      \\ \hline
20  & $Sp(6) \times SU(3) \times SU(2)^2$  & $\bbZ_2$    &  $(1, 0, 1, 0)$        \\ \hline
21  & $Sp(5) \times Spin(10)$     &  $0$      &  -     \\ \hline
22  & $Sp(5) \times SU(6)$      &  $0$      &  -     \\ \hline
23  & $Sp(5) \times SU(5) \times SU(2)$  &  $0$     &  -     \\ \hline
24  & $Sp(4) \times Spin(12)$     & $\bbZ_2$    &   $(1, (1,1))$        \\ \hline
25  & $Sp(4) \times Spin(10) \times SU(2)$ & $\bbZ_2$    &  $(1, 2, 0)$        \\ \hline
26  & $Sp(4) \times SU(5) \times SU(2)^2$  & $\bbZ_2$    &  $(1, 0, 1, 1)$        \\ \hline
27  & $Sp(4) \times SU(4) \times SU(3) \times SU(2)$   & $\bbZ_2$    &    $(1, 2, 0, 0)$          \\ \hline
28  & $Sp(4) \times SU(3)^2 \times SU(2)^2$   & $\bbZ_2$    &  $(1, 0, 0, 1, 1)$          \\ \hline
29  & $Sp(3) \times SU(7) \times SU(2)$    &  $0$      & -       \\ \hline
30  & $Sp(3) \times SU(6) \times SU(3)$    &  $0$      & -       \\ \hline
31  & $Sp(3) \times SU(5) \times SU(3) \times SU(2)$    &  $0$      & -       \\ \hline
32  & $Sp(3) \times SU(4) \times SU(3)^2$    &  $0$      & -       \\ \hline
33  & $Sp(2) \times Spin(14) \times SU(2)$    &  $\bbZ_2$      &   $(1, 2, 1)$         \\ \hline
34  & $Sp(2) \times Spin(12) \times SU(3)$    &  $\bbZ_2$      &   $(1, (1, 0), 0) $         \\ \hline
35  & $Sp(2) \times Spin(10) \times SU(3) \times SU(2)$    &  $\bbZ_2$      &    $(1, 2, 0, 1)$         \\ \hline
36  & $Sp(2) \times SU(9)$    &  $0$      & -       \\ \hline
37  & $Sp(2) \times SU(7) \times SU(3)$    &  $0$      & -       \\ \hline
38  & $Sp(2) \times SU(6) \times SU(4)$    &  $\bbZ_2$   &   $(1, 3, 0)$         \\ \hline
39  & $Sp(2) \times SU(6) \times SU(2)^3$   &  $\bbZ_2 \times \bbZ_2$ &  \tabincell{l}{$(1, 3, 0, 0, 0)$,\\ $(1, 0, 1, 1, 1)$}          \\ \hline
40  & $Sp(2) \times SU(5)^2$   &  $0$      & -       \\ \hline
41  & $Sp(2) \times SU(5) \times SU(4) \times SU(2)$ & $\bbZ_2$   &    $(1, 0, 2, 1)$          \\ \hline
42  & $Sp(2) \times SU(4)^2 \times SU(2)^2$   &   $\bbZ_2 \times \bbZ_2$ &  \tabincell{l}{$(1, 2, 0, 1, 0)$,\\ $(1, 0, 2, 0, 1)$}        \\ \hline
43  & $Sp(1) \times Spin(18)$    &  $0$      & -       \\ \hline
44  & $Sp(1) \times Spin(10) \times SU(5)$    &  $0$   & -       \\ \hline
45  & $Sp(1) \times SU(10)$       &  $0$      & -       \\ \hline
46  & $Sp(1) \times SU(9) \times SU(2)$       &  $0$      & -       \\ \hline
47  & $Sp(1) \times SU(8) \times SU(2)^2$    & $\bbZ_2$   &  $(0, 4, 1, 1)$          \\ \hline
48  & $Sp(1) \times SU(7) \times SU(3) \times SU(2)$   &  $0$      & -       \\ \hline
49  & $Sp(1) \times SU(6) \times SU(5)$    &  $0$      & -       \\ \hline
50  & $Sp(1) \times SU(6) \times SU(4) \times SU(2)$  &  $\bbZ_2$   &    $(0, 3, 2, 1)$        \\ \hline
51  & $Sp(1) \times SU(5) \times SU(3)^2 \times SU(2)$   &  $0$      & -       \\ \hline
52  & $Spin(16) \times SU(2)^2$   &  $\bbZ_2 \times \bbZ_2$ &   \tabincell{l}{$((1, 1), 1, 1)$,\\ $((0, 1), 0, 0 )$}        \\ \hline
53  & $Spin(12) \times SU(4) \times SU(2)$  &  $\bbZ_2 \times \bbZ_2$ &   \tabincell{l}{$((1, 1), 2, 0)$, \\ $((1, 0), 0, 1)$ }          \\ \hline
54  & $Spin(10)^2$    &     $\bbZ_2$   &  $(2, 2)$          \\ \hline
55  & $SU(10) \times SU(2)$  &     $\bbZ_2$   &  $(5, 1)$         \\ \hline
56  & $SU(8) \times SU(3) \times SU(2)$  &     $\bbZ_2$   &  $(4, 0, 0)$         \\ \hline
57  & $SU(7) \times SU(3)^2$       &  $0$      & -       \\ \hline
58  & $SU(6)^2$  &     $\bbZ_2$   &    $(3, 3)$        \\ \hline
59  & $SU(6) \times SU(5) \times SU(2)$  &     $\bbZ_2$   &   $(3, 0, 1)$          \\ \hline
60  & $SU(6) \times SU(4) \times SU(2)^2$  &  $\bbZ_2 \times \bbZ_2$ &  \tabincell{l}{$(0, 2, 1, 1)$,\\  $(3, 0, 1, 0)$}         \\ \hline
61  & $SU(4)^2 \times SU(3)^2$  &  $\bbZ_2$   &  $(2, 2, 0, 0)$          \\ \hline
\end{longtable}

\chapter{\MakeUppercase{Chapter 4 Appendix}}

\section[Deriving the evenness condition on O7\texorpdfstring{\boldmath{$^+$}}{+} via 8d CHL strings]{Deriving the evenness condition on O7\boldmath{$^+$} via 8d CHL strings} \label{apdx:evenness}

In this appendix, we present a derivation of the evenness condition for string junctions the O7$^+$-plane in 8d rank $(2,10)$ models.
This proof utilizes the known equivalence between the heterotic Narain lattice \eqref{eq:Narainlattice} and the junction lattice (modulo null junctions) on 24 ordinary 7-branes, and the construction $\Lambda_\text{Mikhailov} \hookrightarrow \Lambda_\text{Narain}$ of the Mikhailov lattice, describing states of rank $(2,10)$ vacua, as a sublattice \cite{Mikhailov:1998si}, also known as ``freezing'' \cite{Font:2021uyw,Fraiman:2021hma}.

Assuming that the freezing mechanism in the IIB / 7-brane picture is a local operation, ${\bf D}_8 = \bA^8 \bB \bC \rightarrow {\bf O7}^+$, we show that the evenness condition discussed in Section \ref{subsec:junctions_on_O7} is the necessary and sufficient condition for the string junction lattice on the O7$^+$ and the unaffected 7-branes to agree with the Mikhailov lattice.

To this end, first recall that there is a particular $\mathfrak{so}_{16}$ root lattice $(- \text{D}_8) \subset \Lambda_\text{Narain}$ along which one defines an orthogonal projection $P$ \cite{Mikhailov:1998si} (see also \cite{Cvetic:2021sjm}).
Since $P(\text{D}_8) = 0$, this $\mathfrak{so}_{16}$ is interpreted as projected out from, or ``frozen'' inside the heterotic model.
Then, $\Lambda_\text{Mikhailov} \subset \Lambda_\text{Narain}$ is the image of $P$ \emph{inside} $\Lambda_\text{Narain}$, i.e.,
\begin{align}
    [{\bf j}] \in \Lambda_\text{Mikhailov} \Leftrightarrow \exists [\hat{\bf j}] \in \Lambda_\text{Narain} : [{\bf j}] = P([\hat{\bf j}]) \in \Lambda_\text{Narain} \, .
\end{align}
This is a non-trivial condition on the choice of $[\hat{\bf j}]$, since not all elements of $\Lambda_\text{Narain}$ map to integer lattice points under $P$.

To make contact with the junction description, it is important to remember that any elemnt of the Narain lattice corresponds to an equivalence class of physical junctions modulo null junctions.
Therefore, we first identify $\hat{\bf j} \in J_\text{phys}^\text{el}$ as a physical string junction in a rank $(2,18)$ 7-brane configuration with a ${\bf D}_8$ stack.
Now, as explained in Section \ref{subsec:global_junciton_lattice_null_junctions}, the junction $\hat{\bf j}$, prior to freezing, enjoys a decomposition into an integer linear combination,
\begin{align}\label{eq:appB1}
    \hat{\bf j} = \sum_{i=1}^8 a^i {\bf w}_i + a^p \boldsymbol\omega_p^{\mathfrak{so}_{16}} + a^q \boldsymbol\omega_q^{\mathfrak{so}_{16}} + \hat{\bf j}'' \, ,
\end{align}
where $\hat{\bf j''}$ has no prongs on ${\bf D}_8$.
This decomposition is unique only up to the addition of physical null junctions $\boldsymbol\delta^N_{(p,q)}$; however, because such junctions carry no physical charge, their prongs on the ${\bf D}_8$ stack must induce no $\mathfrak{so}_{16}$ center charges, which, according to \eqref{eq:center_charge_table_ADE}, requires even multiples of $\boldsymbol\omega_p^{\mathfrak{so}_{16}}$ and $\boldsymbol\omega_q^{\mathfrak{so}_{16}}$.
Hence, any representative $\hat{\bf j}$ of the equivalence class $[\hat{\bf j}]$ modulo null junctions takes the form
\begin{align}
    \hat{\bf j} + \boldsymbol\delta^N = \sum_{i=1}^8 a^i {\bf w}_i + (a^p + 2n^p) \boldsymbol\omega_p^{\mathfrak{so}_{16}} + (a^q + 2n^q) \boldsymbol\omega_q^{\mathfrak{so}_{16}} + \hat{\bf j}' \, ,
\end{align}
for some junction $\hat{\bf j}'$ that has no prongs on the ${\bf D}_8$.

As this stack will be replaced with the O7$^+$, the D$_8$ root lattice defining the projection in the momentum lattice description is identified with the root junction lattice of this stack.
The representative for $P([\hat{\bf j}])$ is then
\begin{align}
    P(\hat{\bf j}) := (a^p + 2n^p) \boldsymbol\omega_p^{\mathfrak{so}_{16}} + (a^q + 2n^q) \boldsymbol\omega_q^{\mathfrak{so}_{16}} + \hat{\bf j}' \, , \quad a^p, n^p, a^q, n^q \in \bbZ \, .
\end{align}
Therefore, the condition $P([\hat{\bf j}]) \in \Lambda_\text{Narain}$ translates into $P(\hat{\bf j}) \in J_\text{phys}^\text{el}$, i.e., its prongs on the ${\bf D}_8$ stack must satisfy the physicality conditions.
Since, by construction, $\hat{\bf j}'$ has no prongs on the ${\bf D}_8$ stack, this means that $a^p, a^q \in 2\bbZ$.
As we identify the extended weights of $\mathfrak{so}_{16}$ with those of the O7$^+$ after freezing (see Section \ref{subsec:junctions_on_O7}), we conclude that the junction $P(\hat{\bf j})$ representing an element of $\Lambda_\text{Mikhailov}$ must have even $\colpq{p}{q}$-charge.

Finally, it is straightforward to verify the condition for the magnetically dual 5-brane junctions from the lattice description of the dual Mikhailov lattice, which in terms of the above projection map is given by $\Lambda_\text{Mikhailov}^* = P(\Lambda_\text{Narain})$ \cite{Mikhailov:1998si,Cvetic:2021sxm}.
Since the projection simply removes the terms proportional to the $\mathfrak{so}_{16}$ weights in \eqref{eq:appB1} from any physical junction $\hat{\bf j}$ in the rank $(2,18)$ configuration, it is obvious that one ends up with any integer-valued $a^p,a^q$.

\section{Embedding O7\texorpdfstring{\boldmath{$^+$}}{+} into \texorpdfstring{\boldmath{$\widehat{\text{E}}_8$}}{affine E8}}
\label{app:O7_into_E8}

In this appendix, we discuss the junctions resulting from embedding an O7$^+$ into an $\widehat{\textbf{E}}_8$ stack.
First describe the embedding $\mathfrak{so}_{16} \hookrightarrow \widehat{\mathfrak{e}}_8$ in terms of 7-branes.
To this end, we use the equivalence $\widehat{\textbf{E}}_8 \cong \textbf{E}_9$ of 7-brane stacks \cite{DeWolfe:1998yf,DeWolfe:1998pr}, where $\textbf{E}_9$ is conjugate to ${\bf A}^8 {\bf B} {\bf C}^2$, see \eqref{eq:ADE_brane_stacks}.
In this presentation, it is straightforward to identify the $\mathfrak{so}_{16}$ subalgebra as the $\textbf{D}_8 = {\bf A}^8 {\bf B} {\bf C}$ part.
By the ``freezing'' procedure, the $\textbf{E}_9$ stack becomes an ${\bf O7}^+ {\bf C}$ stack.
Now, due to the evenness condition discussed in Section \ref{sec:local_analysis}, only a subset of string junctions that were allowed to end on $\textbf{E}_9$ prior to freezing are allowed in the presence of the O7$^+$.

One such junction that we will focus on in the following is the $\boldsymbol\epsilon$-junction given in \eqref{eqn:1dP9_a}.
This junction has a unit $\colpq{3}{1}$-prong on the ${\bf X}_{[3,1]}$-brane that affinizes the stack --- however, this is in the realization $\widehat{\textbf{E}}_8$!
To connect the two descriptions, we repeatedly use brane moves \eqref{eq:brane_move_left-to-right} and \eqref{eq:brane_move_right-to-left}, to obtain
\begin{align}
\begin{split}
    \widehat{\textbf{E}}_8 = {\bf A}^7 {\bf B} {\bf C}^2 {\bf X}_{[3,1]} \, & \rightarrow  {\bf A}^7 {\bf B} {\bf C} {\bf B} {\bf C} \\
    & \rightarrow {\bf B} {\bf X}_{[0,1]}^7 {\bf C}{\bf B}{\bf C} \rightarrow {\bf B} {\bf C} {\bf A}^7 {\bf B} {\bf C}\\
    & \rightarrow {\bf C} {\bf X}_{[3,1]} {\bf A}^7 {\bf B} {\bf C} \rightarrow {\bf C} {\bf A}^7 {\bf X}_{[-4,1]} {\bf B} {\bf C} \rightarrow {\bf C} {\bf A}^7  {\bf B} {\bf X}_{[-1,-2]} {\bf C}\\
    & \rightarrow {\bf C} {\bf A}^7 {\bf B}{\bf C} {\bf X}_{[0,-1]} \rightarrow {\bf C} {\bf A}^7 {\bf B} {\bf X}_{[0,1]} {\bf X}_{[1,2]} \rightarrow {\bf C} {\bf A}^7 {\bf X}_{[0,1]} {\bf A} {\bf X}_{[1,2]}\\
    & \rightarrow {\bf X}_{[0,1]}^7 {\bf C} {\bf X}_{[0,1]} {\bf A} {\bf X}_{[1,2]} \rightarrow {\bf X}_{[0,1]}^8 {\bf X}_{[1,2]} {\bf A} {\bf X}_{[1,2]}\\
    & \rightarrow \underbrace{ {\bf X}^8_{[0,1]} {\bf X}_{[1,4]} {\bf X}^{(1)}_{[1,2]} }_{= \textbf{D}_8'} {\bf X}^{(2)}_{[1,2]} = \textbf{E}_9'\, .
\end{split}
\end{align}
In each step, it is easy to track the changes of the prongs of the $\boldsymbol\epsilon$-junction that starts out with a unit $\bx_{[3,1]}$-prong, simply by requiring that the prongs on the two moving branes change in such a way that the net $\colpq{p}{q}$-charge remains invariant.
For example, after the first step, we have $\bx_{[3,1]} \rightarrow \bb_2 + 2\bc_2$.
After the whole process, we end up with
\begin{align}
    \bx_{[3,1]} \rightarrow -\bx_{[0,1]}^{(8)} - 2 \bx_{[1,4]} + 4 \bx_{[1,2]}^{(1)} + \bx_{[1,2]}^{(2)} \, .
\end{align}
Since the first three summands end on the ${\bf D}_8'$ stack, one can decompose their sum using the extended weights of $\mathfrak{so}_{16}$ and the weight junctions; the important thing to track here is that the net $\colpq{p}{q}$-charge of this part is $\colpq{2}{-1}$.
However, after introducing the O7$^+$, i.e., replace ${\bf D}_8' \rightarrow {{\bf O7}^+}'$, which removes the $\mathfrak{so}_{16}$ weights, there is an odd $q$-charge emanating via this junction from the orientifold, which is not allowed for a string junction.
Indeed, it is easy to check that any physical string junction leaving the $\widehat{\textbf{E}}_8 \rightarrow {{\bf O7}^+}' \bX_{[1,2]}$ stack must have even $q$-charge.
Therefore, only even multiples of $\boldsymbol\epsilon$ are physical string junctions after freezing.
On the other hand, this prong, and therefore also $\boldsymbol\epsilon$ would be acceptable as a 5-brane junction.

\section{All 8d supergravity vacua via \texorpdfstring{\boldmath{$[p,q]$}}{[p,q]}-7-branes}
\label{app:results}

In this appendix, we give the full catalog of maximally-enhanced 7-brane configurations realizing 8d string vacua of maximal non-Abelian rank for all three classes of models, i.e. total rank $(2,18)$, $(2,10)$, and $(2,2)$. We further determine global structure of their non-Abelian subgroup given by $\mathcal{Z}$ and the explicit realization of the fracrtional null junction.

Before we provide the classification we further describe a procedure that allows to incorporate the non-maximally-enhanced cases, with additional $\mathfrak{u}(1)$ factors.

\subsection{Non-maximally-enhanced cases}

In principle, the process of obtaining the global gauge group topology for the non-maximally enhanced cases is equivalent to what was described in the main text: First one obtains the associated brane configuration, with which one has access to the discrete quotients $\mathcal{Z}$ via the fractional null junctions as in \ref{eq:juncZ} as well as $\mathcal{Z}'$ as in \ref{eq:abelian_quotient_from_junctions}. Even though one cannot avoid repeating the computations of $\mathcal{Z}$ and $\mathcal{Z}'$, one fortunately can take a shortcut of finding the corresponding brane configurations (which is technically the most challenging step) by starting from the maximally-enhanced setups and suitably splitting the brane stacks.

Here we stress that, given a single non-Abelian brane stack, all of the natural brane splittings corresponds to Higgs transition with W-boson vacuum expectation values that decrease the rank by 1. Adjoint Higgsing that preserves the rank (such as $\mathfrak{e}_8 \rightarrow \mathfrak{so}_{16}$), on the other hand, are not guaranteed to admit a realization in a specific brane configuration. Even in brane configurations where such adjoint Higgsings are possible, it would necessarily involve not only the constituent branes in the stack but also some additional branes ($\mathbf{E}_8 \cong \bA^7 \bB \bC^2$ and $\mathbf{D}_8 \cong \bA^8 \bB \bC$ in the example). For this reason, we focus on the W-boson Higgsing, which is guaranteed to have a straightforward brane realizations.

\begin{itemize}

    \item{Splitting $\mathfrak{su}_{k}$:} $\quad$ $\mathfrak{su}_{k} \rightarrow \mathfrak{su}_{k^{\prime}} \oplus \mathfrak{su}_{k-k^{\prime}}$: $\bA^{k} \rightarrow \bA^{k' } + \bA^{k - k'}$
    
    \item{Splitting $\mathfrak{sp}_l$:} $\quad$ $\mathfrak{sp}_{l} \rightarrow \mathfrak{su}_{l^{\prime}} \oplus \mathfrak{sp}_{l-l^{\prime}}$: $\bA^{l} {\bf O7}^+ \rightarrow \bA^{l' } + \bA^{l - l'} {\bf O7}^+$
    
    \item{Splitting $\mathfrak{so}_{2m}$:} 
    \begin{itemize}
        \item{$\mathfrak{so}_{2m} \rightarrow \mathfrak{su}_{m}$}: $\bA^m \bB \bC \rightarrow \bA^m + \bB \bC$
        \item{$\mathfrak{so}_{2m} \rightarrow 2\mathfrak{su}_2 \oplus \mathfrak{su}_{m-2}$}: $\bA^m \bB \bC \simeq \bA^{m-2} \bN^2 \bC^2 \rightarrow \bA^{m-2} + \bN^2 + \bC^2 $
        \item{$\mathfrak{so}_{2m} \rightarrow \mathfrak{su}_4 \oplus \mathfrak{su}_{m-3}$}: $\bA^m \bB \bC  \simeq \bA^{m-2} \bN^2 \bC^2 \simeq \bA^{m-3} \bC^4 \bX_{[3, 2]} \rightarrow \bA^{m-3} + \bC^4 + \bX_{[3, 2]}$ 
        \item{$\mathfrak{so}_{2m} \rightarrow \mathfrak{so}_{2m'} \oplus \mathfrak{su}_{m - m'}$ ($4 \leq m' \leq m - 1$)}: $\bA^m \bB \bC \rightarrow \bA^{m'}  \bB \bC + \bA^{m - m'}$
    \end{itemize}

    \item{Splitting $\mathfrak{e}_n$:} 
    \begin{itemize}
        \item{$\mathfrak{e}_n \rightarrow \mathfrak{su}_{n}$}: $\bA^{n-1} \bB \bC^2 \simeq \bA^n \bX_{[3, -1]} \bN \rightarrow \bA^n + \bX_{[3, -1]} + \bN$ (see (2.12) of \cite{DeWolfe:1998eu})
        \item{$\mathfrak{e}_n \rightarrow \mathfrak{so}_{2n-2}$}:  $\bA^{n-1} \bB \bC^2 \rightarrow \bA^{n-1} \bB \bC + \bC$
        \item{$\mathfrak{e}_n \rightarrow \mathfrak{su}_2 \oplus \mathfrak{su}_{n-1}$}: $\bA^{n-1} \bB \bC^2 \rightarrow \bA^{n-1} + \bB + \bC^2$
        \item{$\mathfrak{e}_n \rightarrow \mathfrak{su}_2 \oplus \mathfrak{su}_3 \oplus \mathfrak{su}_{n-3} (\simeq \mathfrak{e}_3 \oplus \mathfrak{su}_{n-3})$}: $\bA^{n-1} \bB \bC^2 \rightarrow \bA^{n-3} + \bA^2 \bB \bC^2 \simeq \bA^{n-3} + \bC \bA^2 \bC^2 \simeq \bA^{n-3} + \bN^2 \bC^3 \rightarrow \bA^{n-3} + \bN^2 + \bC^3$
        \item{$\mathfrak{e}_n \rightarrow \mathfrak{su}_5 \oplus \mathfrak{su}_{n-4} (\simeq \mathfrak{e}_4 \oplus \mathfrak{su}_{n-4})$}: $\bA^{n-1} \bB \bC^2 \rightarrow \bA^{n-4} + \bA^3 \bB \bC^2 \simeq \bA^{n-4} + \bX_{[1, 2]} \bC^5 \rightarrow \bA^{n-4} + \bX^{[1, 2]} + \bC^5$
        \item{$\mathfrak{e}_n \rightarrow \mathfrak{so}_{10} \oplus \mathfrak{su}_{n-5} (\simeq \mathfrak{e}_5 \oplus \mathfrak{su}_{n-5})$}: $\bA^{n-1} \bB \bC^2 \rightarrow  \bA^{n-5} + \bA^4 \bB \bC^2 \simeq \bA^{n-5} + \bC^5 \bA \bX_{[1, 2]}$ (see (2.11) of \cite{DeWolfe:1998eu}).
        \item{$\mathfrak{e}_n \rightarrow \mathfrak{e}_{n'} \oplus \mathfrak{su}_{n-n'}\ \  (6 \leq n' \leq n-1)$}: $\bA^{n-1} \bB \bC^2 \rightarrow \bA^{n - n'} + \bA^{n'-1} \bB \bC^2$
    \end{itemize}
\end{itemize}

These splittings matches with the ``substitution rules" as given in Table 2.2 of \cite{shimada2000}.

\newpage 

\begin{landscape}

\subsection{No \texorpdfstring{O7$^+$}{O7+}}

We give all possible brane configurations with rank $(2,18)$ realizing maximally-enhanced non-Abelian gauge algebras. Our list reproduces the mathematical classification of \cite{shimada2000} of the ADE-singularities of elliptically-fibered K3 surfaces. This is an expected result, since the junctions describe the same physics in a type IIB perspective. For each brane configuration, we give not only the non-Abelian fundamental group $\pi_1(G_{\text{nA}}) = \mathcal{Z}$ but also its particular embedding into the center $\pi_1(G_{\text{nA}}) \hookrightarrow Z(\widetilde{G})$ using string junctions, where $\widetilde{G}$ is the simply-connected cover of the no-Abelian gauge algebra $G_{\text{nA}} = \widetilde{G}_{\text{nA}}/\mathcal{Z}$. 

\captionsetup{width=23cm}
\begin{longtable}{|c|c|c||c|c|c|} 
\caption{All 8d maximally-enhanced rank $(2,18)$ brane configurations. For each entry, we present the brane configuration, the non-Abelian gauge algebra $\mathfrak{g}_\text{nA}$ and $\mathcal{Z} = \pi_1(G_{\text{nA}})$, generators of its associated fractional null junctions (FNJ), and the embedding $\pi_1(G_{\text{nA}}) \hookrightarrow Z(\widetilde{G}_{\text{nA}})$ of the fundamental group into the center of the simply-connected cover $\widetilde{G}_{\text{nA}}$ of the non-Abelian symmetry. For the last entry, the values are ordered in alignment with the ``brane configuration" column rather than the $\mathfrak{g}$ column.}
\label{tab:rank20_braneConfigs} \\
\hline
No. &  $\mathfrak{g}$ & $\pi_1(G_{\text{nA}})$                                              &  Brane Config.  & FNJ & \tabincell{l}{$\pi_1(G_{\text{nA}}) \hookrightarrow Z(\widetilde{G}_\text{nA})$ \\ \tiny{(Order via Brane Config.)}}\\ \hline\hline
1   & $6\mathfrak{su}_4$                            &      $\bbZ_4 \times \bbZ_4$       &     $\bA^4 \bB^4 \bN^4 \bX_{[1, 2]}^4 \bC^4 \bX_{[2, 1]}^4$ &   \tabincell{l}{$\boldsymbol\delta^N_{(1,0)}/4$, \\ $\boldsymbol\delta^N_{(0,1)}/4$}    &    \tabincell{l}{$(0, 3, 1, 2, 1, 1),$\\ $(3, 3, 0, 3, 3, 2)$}    \\\hline \hline
2   & $2\mathfrak{su}_2 \oplus 4\mathfrak{su}_5$                     &      $\bbZ_5$                     &   $\bA^5 \bN^5 \bX_{[1,4]}^5 \bX_{[1,2]}^2 \bC^5 \bX_{[3, 1]}^2$  &  $\boldsymbol\delta^N_{(3,1)}/5$     & $(4, 3, 1, 0, 2, 0)$ \\ \hline
3   & $2\mathfrak{su}_3 \oplus 2\mathfrak{su}_4 \oplus 2\mathfrak{su}_5$              &      -                            & $\bA^3 \bN^5 \bX_{[1,4]}^5 \bX_{[1,3]}^3 \bX_{[1,2]}^4 \bC^4$ & - & -\\
\hline
\end{longtable}



\subsection{One \texorpdfstring{O7$^+$}{O7+}}

We proceed in this part to give the full list to brane configurations with a single O7$^+$ realizing maximally-enhanced 8d vacua of rank $(2,10)$. This list precisely matches our previous results in 8d CHL strings in Appendix B of \cite{Cvetic:2021sjm}. For each such brane configuration, in addition to giving all the information as provided in the previous table, we also refer to its particular ``uplift'' to rank $(2,18)$, namely the rank $(2,18)$ configuration that one gets by unfreezing the $\bA^n \mathbf{O7}^+$ stack into a $\bA^{n+8} \bB \bC$ stack.

\begin{longtable}{|c|c|c|c||c|c|c|} 
\caption{All 8d maximally-enhanced rank $(2,10)$ brane configurations, in similar convention as the above rank $(2,18)$ catalog.}
\label{tab:rank12_braneConfigs}\\
\hline
\#  &  $\#_{rk\ 20}$ & $\widehat{\mathfrak{g}}$ ($\mathfrak{g}$) & $\pi_1(G_{\text{nA}})$ &  Brane Config.  & FNJ &  $\pi_1(G_{\text{nA}}) \hookrightarrow Z(\widetilde{G}_\text{nA})$   \\ \hline 
1   &320 & ${\mathfrak e}_{8} \oplus \mathfrak{sp}_{2}/(\mathfrak{so}_{20})$    &  0   &  $(\bA^2 {\bf O7}^+)\bC (\bA^7 \bB \bC^2) \bX_{[4, 1]}$  & -  & - \\ \hline
2   &319 & $\mathfrak{e}_8 \oplus \mathfrak{sp}_{1}/(\mathfrak{so}_{18}) \oplus \mathfrak{su}_2$  &  0  & $(\bA {\bf O7}^+) \bC \bX_{[3,1]}^2 (\bX_{[3,1]}^7 \bX_{[13, 4]} \bX_{[7,2]}^2)$ & - & -  \\ \hline
3   &291 & ${\mathfrak e}_{7} \oplus \mathfrak{sp}_{3}/(\mathfrak{so}_{22})$    &  0  & $(\bA^3 {\bf O7}^+) \bC (\bA^6 \bN \bX_{[2,1]}^2) \bX_{[5, 1]}$ & - & -  \\ \hline
4   &290 & ${\mathfrak e}_{7} \oplus \mathfrak{sp}_{2}/(\mathfrak{so}_{20}) \oplus \mathfrak{su}_2$    &  $\bbZ_2$  &  $(\bA^2 {\bf O7}^+) \bB (\bA^6\bB\bC^2)\bX_{[3, 1]}^2$ &  $\boldsymbol\delta^N_{(1,1)}/2$  &  $(1, 1, 0)$ \\ \hline
5   &289 & ${\mathfrak e}_{7} \oplus \mathfrak{sp}_{1}/(\mathfrak{so}_{18}) \oplus \mathfrak{su}_3$    & 0  & $(\bA {\bf O7}^+) \bC \bX_{[3, 1]}^3(\bX_{[3, 1]}^6 \bX_{[13, 4]} \bX_{[7, 2]}^2)$ & - & - \\ \hline
6   &288 & $(\mathfrak{so}_{16} \oplus ) {\mathfrak e}_{7} \oplus \mathfrak{su}_3 \oplus \mathfrak{su}_2$    &  $\bbZ_2$  & $(\bA^6\bB\bC^2) {\bf O7}^+ \bN^2\bC^3$ &  $\boldsymbol\delta^N_{(1,1)}/2$  & $(1, 1, 0)$  \\ \hline
7   &255 & ${\mathfrak e}_{6} \oplus \mathfrak{sp}_{4}/(\mathfrak{so}_{24})$    &  $0$  & $(\bA^4 {\bf O7}^+) (\bA^5 \bC \bX_{[3,1]}^2) \bX_{[6, 1]}$ & - & - \\   \hline
8   &254 & ${\mathfrak e}_{6} \oplus \mathfrak{sp}_{3}/(\mathfrak{so}_{22}) \oplus \mathfrak{su}_2$ & $0$   & $(\bA^5 \bB \bC^2) (\bA^3 {\bf O7}^+) \bX_{[2,-1]} \bN^2  $ & - & - \\ \hline
9   &253 & ${\mathfrak e}_{6} \oplus \mathfrak{sp}_{1}/(\mathfrak{so}_{18}) \oplus \mathfrak{su}_4$  & $0$ & $(\bA {\bf O7}^+) \bC \bX_{[3, 1]}^4 (\bX_{[3, 1]}^5 \bX_{[13, 4]} \bX_{[7, 2]}^2)$ & - & - \\ \hline
10  &252 & ${\mathfrak e}_{6} \oplus \mathfrak{sp}_{1}/(\mathfrak{so}_{18}) \oplus \mathfrak{su}_3 \oplus \mathfrak{su}_2$    &  $0$  & $(\bA^5 \bB \bC^2) (\bA {\bf O7}^+) \bB^2 \bN^3$ & - & - \\ \hline
11  &251 & $(\mathfrak{so}_{16} \oplus ){\mathfrak e}_{6} \oplus \mathfrak{su}_5$    &   $0$  & $(\bA^5 \bB \bC^2) {\bf O7}^+ \bN^5 \bX_{[1,3]}$ & - & - \\ \hline
12  &218 & $\mathfrak{sp}_{10}/(\mathfrak{so}_{36})$    &   $0$  &  $ (\bA^{10} {\bf O7}^+) \bC \bX_{[4, 1]} \bX_{[8,1]} \bX_{[11,1]}$ & - & -  \\ \hline
13  &217 & $\mathfrak{sp}_{9}/(\mathfrak{so}_{34}) \oplus \mathfrak{su}_2$    &   $0$  & $(\bA^9 {\bf O7}^+) \bX_{[5,-1]} \bX_{[2, -1]} \bC^2 \bX_{[3,1]}$ & - & - \\ \hline
14  &216 & $\mathfrak{sp}_{8}/(\mathfrak{so}_{32}) \oplus \mathfrak{su}_3$    &   $\bbZ_2$    &  $ (\bA^8 {\bf O7}^+) \bX_{[-4, 1]} \bN \bX_{[4, 1]} \bX_{[6, 1]}^3 $ & $\boldsymbol\delta^N_{(0,1)}/2$  &  $(1, 0)$  \\ \hline
15  &215 & $\mathfrak{sp}_{8}/(\mathfrak{so}_{32}) \oplus 2\mathfrak{su}_2$     &   $\bbZ_2$     &  $ (\bA^8 {\bf O7}^+) \bX_{[-4, 1]} \bX_{[2, 1]} \bX_{[4, 1]}^2\bX_{[6, 1]}^2 $ & $\boldsymbol\delta^N_{(0,1)}/2$  &  $(1, 0, 0)$ \\ \hline
16  &214 & $\mathfrak{sp}_{7}/(\mathfrak{so}_{30}) \oplus \mathfrak{su}_3 \oplus \mathfrak{su}_2$ &  $0$       &  $(\bA^7 {\bf O7}^+) \bC \bX_{[5,1]}^2 \bX_{[6,1]}^3 \bX_{[9, 1]}$ & - & - \\ \hline
17  &213 & $\mathfrak{sp}_{6}/(\mathfrak{so}_{28}) \oplus \mathfrak{su}_5$    &   $0$  & $(\bA^6 {\bf O7}^+) \bC \bX_{[9,2]} \bX_{[5, 1]}^5 \bX_{[8, 1]}$ & - & - \\ \hline
18  &212 & $\mathfrak{sp}_{6}/(\mathfrak{so}_{28}) \oplus \mathfrak{su}_4 \oplus \mathfrak{su}_2$    &   $\bbZ_2$   &   $ (\bA^6 {\bf O7}^+) \bX_{[-4, 1]} \bX_{[2, 1]} \bX_{[4, 1]}^4\bX_{[9, 2]}^2 $ & $\boldsymbol\delta^N_{(0,1)}/2$  &   $(1, 0, 1)$  \\ \hline
19  &210 & $\mathfrak{sp}_{6}/(\mathfrak{so}_{28}) \oplus 2\mathfrak{su}_3$     &  $0$      & $(\bA^6 {\bf O7}^+) \bX_{[2,-1]} \bN^3 \bC^3 \bX_{[3,1]}$ & - & - \\ \hline
20  &211 & $\mathfrak{sp}_{6}/(\mathfrak{so}_{28}) \oplus \mathfrak{su}_3 \oplus 2\mathfrak{su}_2$  & $\bbZ_2$    &   $ (\bA^6 {\bf O7}^+) \bX_{[-4, 1]} \bC^2 \bX_{[2, 1]}^3\bX_{[4, 1]}^2 $   &  $\boldsymbol\delta^N_{(0,1)}/2$  &  $(1, 1, 0, 0)$ \\ \hline
21  &209 & $\mathfrak{sp}_{5}/(\mathfrak{so}_{26}) \oplus \mathfrak{so}_{10}$     &  $0$      &  $(\bA^5\bB\bC)(\bN^{13}\bA\bX_{[1,-2]})\bX_{[2,15]} \bX_{[1,6]}$  & - & - \\ \hline
22  &208 & $\mathfrak{sp}_{5}/(\mathfrak{so}_{26}) \oplus \mathfrak{su}_6$      &  $0$      & $(\bA^5 {\bf O7}^+) \bB \bX_{[2,3]} \bC^6 \bX_{[3,1]}$ & - & - \\ \hline
23  &207 & $\mathfrak{sp}_{5}/(\mathfrak{so}_{26}) \oplus \mathfrak{su}_5 \oplus \mathfrak{su}_2$  &  $0$     &   $(\bA^5 {\bf O7}^+) \bC (\bX_{[11,2]} \bX_{[6,1]}^5) \bX_{[8, 1]}^2$ & - & - \\ \hline
24  &206 & $\mathfrak{sp}_{4}/(\mathfrak{so}_{24}) \oplus \mathfrak{so}_{12}$     & $\bbZ_2$    &   $ (\bA^4 {\bf O7}^+) (\bX_{[2, -1]}^6 \bX_{[1, -1]} \bX_{[3, -1]}) \bN \bX_{[4, 1]} $   & $\boldsymbol\delta^N_{(0,1)}/2$ & $(1, (1,1))$  \\ \hline
25  &205 & $\mathfrak{sp}_{4}/(\mathfrak{so}_{24}) \oplus \mathfrak{so}_{10} \oplus \mathfrak{su}_2$ & $\bbZ_2$   &  $ (\bA^4 {\bf O7}^+) (\bX_{[2, -1]}^5 \bX_{[1, -1]} \bX_{[3, -1]}) \bN^2 \bX_{[2, 1]} $  &  $\boldsymbol\delta^N_{(0,1)}/2$   &  $(1, 2, 0)$  \\ \hline
26  &204 & $\mathfrak{sp}_{4}/(\mathfrak{so}_{24}) \oplus \mathfrak{su}_5 \oplus 2\mathfrak{su}_2$  & $\bbZ_2$  &   $ (\bA^4 {\bf O7}^+) \bX_{[-4, 1]} \bC^2 \bX_{[2, 1]}^5\bX_{[5, 2]}^2 $ &  $\boldsymbol\delta^N_{(0,1)}/2$  &  $(1, 1, 0, 1)$  \\ \hline
27  &203 & $\mathfrak{sp}_{4}/(\mathfrak{so}_{24}) \oplus \mathfrak{su}_4 \oplus \mathfrak{su}_3 \oplus \mathfrak{su}_2$   & $\bbZ_2$   &    $ (\bA^4 {\bf O7}^+) \bN^2 \bC^4 \bX_{[2, 1]}^3\bX_{[4, 1]} $  & $\boldsymbol\delta^N_{(0,1)}/2$  &   $(1, 0, 2, 0)$ \\ \hline
28  &202 & $\mathfrak{sp}_{4}/(\mathfrak{so}_{24}) \oplus 2\mathfrak{su}_3 \oplus 2\mathfrak{su}_2$   & $\bbZ_2$    &   $ (\bA^4 {\bf O7}^+) \bN^3 \bC^2 \bX_{[5, 1]}^2 \bX_{[6, 1]}^3 $   &  $\boldsymbol\delta^N_{(0,1)}/2$  &  $((1, 0, 1, 1, 0)$  \\ \hline
29  &201 & $\mathfrak{sp}_{3}/(\mathfrak{so}_{22}) \oplus \mathfrak{su}_7 \oplus \mathfrak{su}_2$    &  $0$ & $(\bA^3 {\bf O7}^+) \bC (\bX_{[11,2]} \bX_{[6,1]}^7) \bX_{[13, 2]}^2$ & - & - \\ \hline
30  &200 & $\mathfrak{sp}_{3}/(\mathfrak{so}_{22}) \oplus \mathfrak{su}_6 \oplus \mathfrak{su}_3$    &  $0$ & $(\bA^3 {\bf O7}^+) \bC \bX_{[9,2]} \bX_{[5, 1]}^6 \bX_{[6, 1]}^3$ & - & - \\ \hline
31  &199 & $\mathfrak{sp}_{3}/(\mathfrak{so}_{22}) \oplus \mathfrak{su}_5 \oplus \mathfrak{su}_3 \oplus \mathfrak{su}_2$    &  $0$ & $(\bA^3 {\bf O7}^+)\bX_{[2,3]}\bC^5\bX_{[5, 1]}^2\bX_{[6, 1]}^3$ & -  & - \\ \hline
32  &198 & $\mathfrak{sp}_{3}/(\mathfrak{so}_{22}) \oplus \mathfrak{su}_4 \oplus 2\mathfrak{su}_3$    &  $0$ & $(\bA^3 {\bf O7}^+) (\bN^4 \bC^3 \bX_{[2,1]}^3) \bX_{[5, 1]}$ & - & - \\ \hline
33  &197 & $\mathfrak{sp}_{2}/(\mathfrak{so}_{20}) \oplus \mathfrak{so}_{14} \oplus \mathfrak{su}_2$    &  $\bbZ_2$   &  $ (\bA^2 {\bf O7}^+) \bX_{[3, -1]}^2 \bN (\bX_{[2, 1]}^7 \bX_{[1, 1]} \bX_{[5, 3]})$   & $\boldsymbol\delta^N_{(0,1)}/2$   & $(1, 1, 2)$  \\ \hline
34  &196 & $\mathfrak{sp}_{2}/(\mathfrak{so}_{20}) \oplus \mathfrak{so}_{12} \oplus \mathfrak{su}_3$    &  $\bbZ_2$   &   $ (\bA^2 {\bf O7}^+) (\bX_{[2, -1]}^6 \bX_{[1, -1]} \bX_{[3, -1]}) \bB^3 \bC $   &  $\boldsymbol\delta^N_{(1,1)}/2$  &  $(1, (1, 0), 0) $ \\ \hline
35  &195 & $\mathfrak{sp}_{2}/(\mathfrak{so}_{20}) \oplus \mathfrak{so}_{10} \oplus \mathfrak{su}_3 \oplus \mathfrak{su}_2$    &  $\bbZ_2$   &   $ (\bA^2 {\bf O7}^+) (\bX_{[2, -1]}^5 \bX_{[1, -1]} \bX_{[3, -1]}) \bN^3 \bC^2 $  &  $\boldsymbol\delta^N_{(0,1)}/2$  &  $(1, 2, 0, 1)$ \\ \hline
36  &194 & $\mathfrak{sp}_{2}/(\mathfrak{so}_{20}) \oplus \mathfrak{su}_9$    &  $0$ & $(\bA^2 {\bf O7}^+) \bX_{[4,-1]} \bX_{[1,-2]} \bN^9 \bX_{[1, 3]} $ & - & - \\ \hline
37  &193 & $\mathfrak{sp}_{2}/(\mathfrak{so}_{20}) \oplus \mathfrak{su}_7 \oplus \mathfrak{su}_3$    &  $0$ & $(\bA^2 {\bf O7}^+) (\bX_{[3,4]} \bC^7 \bX_{[2,1]}^3) \bX_{[5, 1]}$ & - & - \\ \hline
38  &192 & $\mathfrak{sp}_{2}/(\mathfrak{so}_{20}) \oplus \mathfrak{su}_6 \oplus \mathfrak{su}_4$    &  $\bbZ_2$   &   $ (\bA^2 {\bf O7}^+) \bB \bX_{[3, 5]} \bC^4 \bX_{[2, 1]}^6 $   & $\boldsymbol\delta^N_{(1,1)}/2$  &   $(1, 0, 3)$  \\ \hline
39  &191 & $\mathfrak{sp}_{2}/(\mathfrak{so}_{20}) \oplus \mathfrak{su}_6 \oplus 3\mathfrak{su}_2$   &  $\bbZ_2 \times \bbZ_2$  &  $ (\bA^2 {\bf O7}^+) \bN^6 \bX_{[1, 5]}^2 \bX_{[1, 3]}^2 \bC^2$  & \tabincell{l}{ $\boldsymbol\delta^N_{(1,1)}/2$, \\ $\boldsymbol\delta^N_{(0,1)}/2$} &  \tabincell{l}{$(1, 3, 0, 0, 0)$,\\ $(1, 0, 1, 1, 1)$}  \\ \hline
40  &189 & $\mathfrak{sp}_{2}/(\mathfrak{so}_{20}) \oplus 2\mathfrak{su}_5$   &  $0$ & $(\bA^2 {\bf O7}^+)\bX_{[2,3]}\bC^5\bX_{[5, 1]}^5\bX_{[16, 3]}$ & - & - \\ \hline
41  &190 & $\mathfrak{sp}_{2}/(\mathfrak{so}_{20}) \oplus \mathfrak{su}_5 \oplus \mathfrak{su}_4 \oplus \mathfrak{su}_2$ & $\bbZ_2$  &    $ (\bA^2 {\bf O7}^+) \bN^4 \bX_{[1,2]}^2 \bC^5 \bX_{[3, 1]} $  &  $\boldsymbol\delta^N_{(1,1)}/2$   &  $(1, 0, 2, 1)$   \\ \hline
42  &188 & $\mathfrak{sp}_{2}/(\mathfrak{so}_{20}) \oplus 2\mathfrak{su}_4 \oplus 2\mathfrak{su}_2$   &   $\bbZ_2 \times \bbZ_2$  &  $ (\bA^2 {\bf O7}^+) \bN^4 \bX_{[1, 3]}^4 \bX_{[2, 5]}^2 \bC^2$  & \tabincell{l}{$\boldsymbol\delta^N_{(1,1)}/2$, \\ $\boldsymbol\delta^N_{(0,1)}/2$}   & \tabincell{l}{$(1, 2, 0, 1, 0)$,\\ $(1, 0, 2, 0, 1)$} \\ \hline
43  &179 & $\mathfrak{sp}_{1}/(\mathfrak{so}_{18}) \oplus \mathfrak{so}_{18}$    &  $0$ & $(\bA {\bf O7}^+) \bB (\bC^9 \bA \bX_{[1, 2]}) \bX_{[3, 2]}$ & - & - \\ \hline
44  &187 & $\mathfrak{sp}_{1}/(\mathfrak{so}_{18}) \oplus \mathfrak{so}_{10} \oplus \mathfrak{su}_5$    &  $0$ &  $(\bA {\bf O7}^+)(\bX_{[2, -1]}^5 \bB \bX_{[3, -1]}) \bN^5 \bX_{[1,2]}$ & - & - \\ \hline
45  &186 & $\mathfrak{sp}_{1}/(\mathfrak{so}_{18}) \oplus \mathfrak{su}_{10}$       &  $0$ &  $(\bA {\bf O7}^+) \bC  (\bX_{[3, 1]}^{10} \bX_{[19, 6]} \bX_{[10, 3]})$ & - & - \\ \hline
46  &185 & $\mathfrak{sp}_{1}/(\mathfrak{so}_{18}) \oplus \mathfrak{su}_9 \oplus \mathfrak{su}_2$       &  $0$ &  $(\bA {\bf O7}^+) \bC \bX_{[3, 1]}^9 \bX_{[13, 4]} \bX_{[7, 2]}^2$ & - & - \\ \hline
47  &184 & $\mathfrak{sp}_{1}/(\mathfrak{so}_{18}) \oplus \mathfrak{su}_8 \oplus 2\mathfrak{su}_2$    & $\bbZ_2$  &    $(\bA {\bf O7}^+) \bN^8 \bX_{[1, 4]} \bX_{[2, 5]}^2 \bC^2$  &  $\boldsymbol\delta^N_{(1,0)}/2$  &   $(0, 4, 1, 1)$ \\ \hline
48  &183 & $\mathfrak{sp}_{1}/(\mathfrak{so}_{18}) \oplus \mathfrak{su}_7 \oplus \mathfrak{su}_3 \oplus \mathfrak{su}_2$   &  $0$ & $(\bA {\bf O7}^+) \bC \bX_{[3, 1]}^7 \bX_{[10, 3]}^2 \bX_{[7,2]}^3$ & - & - \\ \hline
49  &182 & $\mathfrak{sp}_{1}/(\mathfrak{so}_{18}) \oplus \mathfrak{su}_6 \oplus \mathfrak{su}_5$    &  $0$ & $(\bA {\bf O7}^+) \bC (\bX_{[3, 1]}^6 \bX_{[17, 5]} \bX_{[7, 2]}^5)$ & - & - \\ \hline
50  &181 & $\mathfrak{sp}_{1}/(\mathfrak{so}_{18}) \oplus \mathfrak{su}_6 \oplus \mathfrak{su}_4 \oplus \mathfrak{su}_2$  &  $\bbZ_2$ &   $(\bA {\bf O7}^+) \bN^6 \bX_{[1, 5]}^4 \bX_{[2, 9]}^2 \bX_{[1, 2]}$  &  $\boldsymbol\delta^N_{(1,0)}/2$  &  $(0, 3, 2, 1)$  \\ \hline
51  &180 & $\mathfrak{sp}_{1}/(\mathfrak{so}_{18}) \oplus \mathfrak{su}_5 \oplus 2\mathfrak{su}_3 \oplus \mathfrak{su}_2$   &  $0$ &  $(\bA {\bf O7}^+) \bX_{[1,2]}^5 \bX_{[5,9]}^3 \bX_{[3, 5]}^2 \bC^3$ & - & - \\ \hline
52  &169 & $(\mathfrak{so}_{16} \oplus ) \mathfrak{so}_{16} \oplus 2\mathfrak{su}_2$   &  $\bbZ_2 \times \bbZ_2$  &  ${\bf O7}^+ (\bX_{[2, -1]}^8 \bX_{[1,-1]} \bX_{[3, -1]}) \bB^2 \bC^2$   &  \tabincell{l}{$\boldsymbol\delta^N_{(0,1)}/2$,\\ $\boldsymbol\delta^N_{(1,1)}/2$} &  \tabincell{l}{$((1, 1), 1, 1)$,\\ $((0, 1), 0, 0 )$}  \\ \hline
53  &178 & $(\mathfrak{so}_{16} \oplus ) \mathfrak{so}_{12} \oplus \mathfrak{su}_4 \oplus \mathfrak{su}_2$  &  $\bbZ_2 \times \bbZ_2$  &    ${\bf O7}^+ (\bX_{[2, -1]}^6 \bX_{[1,-1]} \bX_{[3, -1]}) \bB^4 \bN^2$ &  \tabincell{l}{$\boldsymbol\delta^N_{(0,1)}/2$, \\ $\boldsymbol\delta^N_{(1,1)}/2$} & \tabincell{l}{$((1, 1), 2, 0)$, \\ $((1, 0), 0, 1)$ }   \\ \hline
54  &177 & $(\mathfrak{so}_{16} \oplus ) 2\mathfrak{so}_{10}$    &     $\bbZ_2$  &  ${\bf O7}^+ (\bX_{[2, -1]}^5 \bX_{[1,-1]} \bX_{[3, -1]}) (\bN^5 \bC \bB)$  &  $\boldsymbol\delta^N_{(0,1)}/2$ &  $(2, 2)$ \\ \hline
55  &176 & $(\mathfrak{so}_{16} \oplus ) \mathfrak{su}_{10} \oplus \mathfrak{su}_2$  &     $\bbZ_2$  &  ${\bf O7}^+ \bX_{[1,2]} \bC^{10} \bX_{[3, 2]} \bX^2_{[3, 1]} $    &  $\boldsymbol\delta^N_{(1,0)}/2$  &  $(5, 1)$  \\ \hline
56  &175 & $(\mathfrak{so}_{16} \oplus ) \mathfrak{su}_8 \oplus \mathfrak{su}_3 \oplus \mathfrak{su}_2$  &     $\bbZ_2$   &   ${\bf O7}^+ \bB^2 \bX_{[1, -3]} \bN^8 \bC^2$   & $\boldsymbol\delta^N_{(1,1)}/2$   & $(0, 4, 0)$  \\ \hline
57  &174 & $(\mathfrak{so}_{16} \oplus ) \mathfrak{su}_7 \oplus 2\mathfrak{su}_3$       &  $0$ &  ${\bf O7}^+ \bB^3 \bN^3 \bC^7 \bX_{[3,2]}$ & - & - \\ \hline
58  &171 & $(\mathfrak{so}_{16} \oplus ) 2\mathfrak{su}_6$  &     $\bbZ_2$   &    ${\bf O7}^+ \bB^6 \bX_{[1, -2]} \bC^6 \bX_{[3, 2]}$   & $\boldsymbol\delta^N_{(1,0)}/2$    &  $(3, 3)$  \\ \hline
59  &173 & $(\mathfrak{so}_{16} \oplus ) \mathfrak{su}_6 \oplus \mathfrak{su}_5 \oplus \mathfrak{su}_2$  &     $\bbZ_2$   &    ${\bf O7}^+ \bB^6 \bX_{[2, -3]} \bN^5 \bC^2$   & $\boldsymbol\delta^N_{(0,1)}/2$   &  $(3, 0, 1)$  \\ \hline
60  &172 & $(\mathfrak{so}_{16} \oplus ) \mathfrak{su}_6 \oplus \mathfrak{su}_4 \oplus 2\mathfrak{su}_2$  &  $\bbZ_2 \times \bbZ_2$ &  ${\bf O7}^+ \bB^4 \bX_{[1, -2]}^2 \bN^6 \bC^2$   &  \tabincell{l}{$\boldsymbol\delta^N_{(0,1)}/2$,\\ $\boldsymbol\delta^N_{(1,1)}/2$}  &  \tabincell{l}{$(2, 1, 0, 1)$,\\  $(0, 1, 3, 0)$}   \\ \hline
61  &170 & $(\mathfrak{so}_{16} \oplus ) 2\mathfrak{su}_4 \oplus 2\mathfrak{su}_3$  &  $\bbZ_2$  &  ${\bf O7}^+ \bB^3 \bN^4 \bX_{[1, 2]}^4 \bC^3$ & $\boldsymbol\delta^N_{(0,1)}/2$   &   $(0, 2, 2, 0)$ \\ \hline
\end{longtable}  

\subsection{Two \texorpdfstring{O7$^+$}{O7+}'s}
\label{subapdx:rank4_catalog}

Finally, we give all six vacua in the rank $(2,2)$ branch via brane configurations, this time no longer restricting to maximally-enhanced cases. This is the first time that the global structure of such string vacua without a heterotic or CHL description has been computed.

\renewcommand{\arraystretch}{1.4}

\begin{table}[H]
    \centering
    \begin{tabular}{|c|c|c||c|c|c|c|} \hline
        $\#$  &  $\#_{\text{rk} 12}$   &  $\#_{\text{rk} 20}$   & $(\mathfrak{g}, Z)_{(2,18)}$ & Brane Config. & FNJ & $\pi_1(G_{\text{nA}}) \hookrightarrow Z(\widetilde{G}_\text{nA})$ \\ \hline \hline
        1  &  -  &   -   &   $(2\mathfrak{so}_{16}, \bbZ_2)$  & ${\bf O7}^+ \bB {\bf O7}^+ \bX_{[3, -1]} \bB \bC$ &- &- \\ \hline
        2  &  -  &     -   &   $(2\mathfrak{so}_{16}, -)$  & ${\bf O7}^+ \bB\bC {\bf O7}^+_{[[1,1],[0,1]]} \bA\bX_{[1, 2]}$ &- &-  \\ \hline
        3  &  -  &   -   &   $(2\mathfrak{so}_{16} \oplus \mathfrak{su}_2, \mathbb{Z}_2)$  & ${\bf O7}^+ \bB {\bf O7}^+ \bB\bC^2$ &- &- \\ \hline
        4  &  52 &   169   &   $(2\mathfrak{so}_{16} \oplus 2\mathfrak{su}_2, \mathbb{Z}_2 \times \mathbb{Z}_2)$  & ${\bf O7}^+ {\bf O7}^+_{[[2, -1],[1,0]]} \bB^2 \bC^2$ & $\boldsymbol\delta^N_{(1,1)}/2$ & (1,1) \\ \hline
        5  &  -  &     -   &   $(\mathfrak{so}_{18} \oplus \mathfrak{so}_{16}, -)$  & $\bA {\bf O7}^+ \bB (\bC\ {\bf O7}^+_{[[1, 1],[0,1]]}) \bX_{[3, 2]}$ &- &- \\ \hline
        6  &  43 &   179   &   $(2\mathfrak{so}_{18}, -)$  & $(\bA\ {\bf O7}^+)\bB (\bC\  {\bf O7}^+_{[[1, 1],[0,1]]}) \bX_{[3 ,2]}$ &- &- \\ \hline
    \end{tabular}
    \caption{All 6 vacua of 8d rank $(2,2)$ with two O7$^+$-planes, not limited to maximally-enhanced vacua. We give their rank $(2,18)$ and rank $(2,10)$ uplift for the two maximally-enhanced vacua. Here the two ${\bf O7}^+$ could be mutually non-locally, and in this case the subscript $[[p, q],[r, s]]$ (with $ps - qr = 1$) stands for the $SL(2, \bbZ)$ transformation that ones need to transform a standard ${\bf O7}^+$ into a ${\bf O7}^+_{[[p, q],[r, s]]}$.}
    \label{tab:rank4}
\end{table}

\renewcommand{\arraystretch}{1}

\end{landscape}

\chapter{\MakeUppercase{Chapter 5 Appendix}}

\section{Properties and Computation of \texorpdfstring{$\text{Ab}[\Gamma]$}{Ab[Gamma]} and \texorpdfstring{$\text{Ab}[\Gamma/H]$}{Ab[Gamma/H]}}\label{app:ABGAMMA}

In this Appendix, we present some computations used in the main body concerning the abelianization of $\Gamma \subset SU(3)$,
as well as its quotient $\Gamma / H$ by $H$ the normal subgroup generated by those elements of $\Gamma$ which have a fixed point locus on
$S^5 = \partial \mathbb{C}^3$.

The Appendix is organized as follows. First, in Appendix \ref{subapdx:ab_properties} we give the definition of the abelianization $\text{Ab}[\Gamma]$ of a finite group $\Gamma$ together with its properties. In Appendix \ref{subapdx:ab_algorithm} we give the algorithm of computing the abelianization via \texttt{Sage} \cite{sagemath}. The code requires as input the presentation of the finite group, which we give for all finite subgroups of $SU(3)$ in Appendix \ref{subapdx:presentations}. Finally in Appendix \ref{subapdx:ab_quotient_results}, we explain the algorithm for computing $\text{Ab}[\Gamma/H]$.

\subsection{General Aspects of Abelianization} \label{subapdx:ab_properties}
Recall that for a finite group $\Gamma$, the {\it abelianization} $\mathrm{Ab}[\Gamma]$ is defined as
\begin{gather}
    \mathrm{Ab}[\Gamma]=\frac{\Gamma}{[\Gamma,\Gamma]},
\end{gather}
where $[\Gamma,\Gamma]$ is the commutator subgroup, or derived subgroup, of $\Gamma$. A group that satisfies $\mathrm{Ab}[G]=\mathbbm{1}$ is called {\it perfect} while an abelian group clearly satisfies $\mathrm{Ab}[\Gamma]=\Gamma$.

A useful fact about $\mathrm{Ab}[\Gamma]$ is that any homomorphism $\varphi$ from $\Gamma$ to an abelian group $A$ factors through $\mathrm{Ab}[\Gamma]$, which is often referred to as the ``universal property of abelianizations". By this we mean that there exists a unique homomorphism $\psi:\mathrm{Ab}[\Gamma]\rightarrow A$ such that following diagram commutes
\begin{gather*}
    \xymatrix{\Gamma\ar[r]^{\varphi} \ar[d]_{\mathrm{quot.}} & A \\
    \mathrm{Ab}[\Gamma]\ar[ru]^{\psi}}
\end{gather*}
Using this property, we can set up the following diagram
\begin{gather*}
	\xymatrix{
		\Gamma \ar[r]^{q_1} \ar[d]^{p} & \Gamma/H \ar[r]^{q_2} & \mathrm{Ab}[\Gamma/H] \\
		 	\mathrm{Ab}[\Gamma]\ar[urr]_{u}
	}
\end{gather*}
where $N$ is a normal subgroup of $\Gamma$ and $p,q_1, q_2$ are all the natural quotient maps. From this it follows that $u$ must be surjective and, by the first isomorphism theorem, we get
\begin{gather}
    \mathrm{Ab}[\Gamma/H]\cong \frac{\mathrm{Ab}[\Gamma]}{\ker u}.
\end{gather}
Since quotients of abelian groups are isomorphic to subgroups of that same group, we get that $\mathrm{Ab}[\Gamma/H]$ is isomorphic to a subgroup of $\mathrm{Ab}[\Gamma]$.

\subsection{Algorithm} \label{subapdx:ab_algorithm}

If we can find a presentation of these groups, then \texttt{Sage} computes for us the abelianization (via \texttt{G.abelian\_invariants()}), where $\texttt{G}$ is a \texttt{FreeGroup} quotiented by equivalence relations. We can check that this works for $SU(2)$ subgroups where both the presentations and $\text{Ab}[\Gamma]$ are known.

For non-abelian subgroups of $SU(3)$, the presentation is not known explicitly. We take the following approach when finding them.
\begin{itemize}
    \item We start from the sets of matrix generators as given in \cite{Tian:2021cif}, and use them to construct a free group. We then find as many relations among them (as can be checked explicitly) as possible that we mod out by.

    \item If we do not yet have a complete set of defining relations, then the group we get $\hat{\Gamma}$ (either finite or infinite) would have $\Gamma$ as a non-trivial subgroup. In this case we go back to the previous step. If, on the other hand, we find a finite group with the correct order, then we have obtained a correct presentation of this $\Gamma \subset SU(3)$.
\end{itemize}
In the end, we use a \texttt{Sage} function to compute the abelianization of the group.

An alternative method of computing $\text{Ab}[\Gamma]$ is to input the groups as a matrix group and then use the {\tt as\_permutation\_group()} and {\tt as\_finitely\_presented\_group()} functions to convert the matrix groups into a form for which Sage can compute the abelianization readily. The advantage of this method is that we no longer need to find a presentation of the group. However, the running time of the \texttt{as\_permutation\_group()} function grows significantly as $|\Gamma|$ increases, so we have primarily used the previously listed method.

\subsection{Presentations of \texorpdfstring{$SU(3)$}{SU(3)} Subgroups} \label{subapdx:presentations}

In this Appendix we give the explicit presentations of various finite subgroups of $SU(3)$ used in the main text.

\subsubsection{Discrete Subgroups of \texorpdfstring{$U(2)$}{U(2)}}

This case is organized using rather different notation in \cite{Tian:2021cif} when compared to \cite{Carrasco2014}. We will start by doing most of the cases with the \cite{Tian:2021cif} notation, while shifting to \cite{Carrasco2014} notation when looking at ``sporadic subgroups".

\begin{itemize}

\item{$G_m$} $\quad a^{2m} = b^2 = c^2 = aba^{-1}b^{-1} = (ac)^2 = (bc)^2a^m = 1$, where $a = M_1, b = M_2, c = M_3$ as in (3.19) of \cite{Tian:2021cif}.

\item{$G_{p, q}$} $\quad a^{2pq} = b^{2q} = c^2 = aba^{-1}b^{-1} = (ac)^2 = (bc)^{4q} = (bc)^2a^p b^{-2} = 1$, where $a = M_1, b = M_2, c = M_3$ as in (3.26) of \cite{Tian:2021cif}.

\item{$G'_{m}$}
    \begin{itemize}
    \item{$m$ even} $\quad a^4 = b^{4m} = (ab)^2 = a^2 b^{2m}$, where $a = M_1, b = M_2$ as in (3.34) of \cite{Tian:2021cif}
    \item{$m$ odd} $\quad a^4 = b^{4m} = c^4 = (ab)^2 = a^2 b a^{-2} b^{-1} = a^2 b^{2m} = bcb^{-1}c^{-1} = (ac)^4 = (bc)^m$, with $a = M_1, b = M_2, c = M_3$ as in (3.37) of \cite{Tian:2021cif}
    \end{itemize}

\item{``Sporadic cases"}
    \begin{itemize}
        \item{$E^{(1)}$:} $a^3 b^{-3} = b^3 c^{-2} = abc^{-1} = d^3 a^4 = ada^{-1}d^{-1} = bdb^{-1}d^{-1} = cdc^{-1} d^{-1} = 1 $
        \item{$E^{(2)}$:} $a^4 = b^2 = aba^{-1}b^{-1} = c^3 =  cac^{-1}a^{-1} = cbc^{-1}b^{-1} = 1$
        \item{$E^{(3)}$:} $ a^4 b^{-3} = b^3 c^{-2} = abc^{-1} = d^2 a^4 = ada^{-1}d^{-1} = bdb^{-1}d^{-1} = cdc^{-1} d^{-1} = 1$
        \item{$E^{(4)}$:} $c^3 = bc^{-1}a^{-1}c = (da)^2 = c^{-1}bca^{-1}b = d^{-1}ba^{-2}d^{-1} = c^{-1}a^{-1}d^{-1}b^{-1}c^{-1}d^{-1} = 1$
        \item{$E^{(5)}$:} $ a^3 b^{-3} = b^3 c^{-2} = abc^{-1} = d^2 c^3 = ada^{-1}d^{-1} = bdb^{-1}d^{-1} = cdc^{-1} d^{-1} = 1 $
        \item{$E^{(6)}$:} $ a^3 b^{-3} = b^3 c^{-2} = abc^{-1} = d^2 a^3 = ada^{-1}d^{-1} = bdb^{-1}d^{-1} = cdc^{-1} d^{-1} = 1$
        \item{$E^{(7)}$:} $a^4 b^{-3} = b^3 c^{-2} = abc^{-1} = d^3 a^4 = ada^{-1}d^{-1} = bdb^{-1}d^{-1} = cdc^{-1} d^{-1} = 1$
        \item{$E^{(8)}$:} $ a^4 b^{-3} = b^3 c^{-2} = abc^{-1} = d^4 a^4 = ada^{-1}d^{-1} = bdb^{-1}d^{-1} = cdc^{-1} d^{-1} = 1$
        \item{$E^{(9)}$:} $ a^5 b^{-3} = b^3 c^{-2} = abc^{-1} = d^2 c^5 = = ada^{-1}d^{-1} = bdb^{-1}d^{-1} = cdc^{-1} d^{-1} = 1 $
        \item{$E^{(10)}$:} $ a^5 b^{-3} = b^3 c^{-2} = abc^{-1} = d^3 c^5 = = ada^{-1}d^{-1} = bdb^{-1}d^{-1} = cdc^{-1} d^{-1} = 1 $
        \item{$E^{(11)}$:} $ a^5 b^{-3} = b^3 c^{-2} = abc^{-1} = d^5 c^5 = = ada^{-1}d^{-1} = bdb^{-1}d^{-1} = cdc^{-1} d^{-1} = 1 $
    \end{itemize}

\item{$D_{n, q}$ cases}

    There are all small subgroups of $U(2)$ whose presentation and abelianization have been presented in (4.26) - (4.27) of \cite{DelZotto:2015isa}.)

\item{$T_m, O_m, I_m$ cases}

    We notice that $O_m = O \times \bbZ_m$ for all allowed $m$ (namely $(m, 6 = 1)$), $I_m = I \times \bbZ_m$ for all allowed $m$ (namely $(m, 30) = 1$), and $T_m = T \otimes \bbZ_m$ for $m \equiv 1, 5 \mod\ 6$, where $T, O, I \subset SU(2)$ are the tetrahedral, octahedral and isocahedral groups, respectively. So for these cases, the abelianization follows from those of $SU(2)$ finite subgroups, and for the latter we refer to \cite{DelZotto:2015isa}.

    For $T_m$ where $m = 6k+3$, the above direct product expression no longer holds. Instead, the presentation is given by $ bab^{-1}a = a^4 = ab^{-2}a = acbc^{-1} = bcabc^{-1} = c^{3m} = 1 $.
\end{itemize}

\subsubsection{Transitive subgroups of \texorpdfstring{$SU(3)$}{SU(3)}}

\begin{itemize}
    \item{$\Delta$ series}
    \begin{itemize}
    \item{$\Delta(3n^2)$} $\quad a^n = b^n = aba^{-1}b^{-1} = afb^{-1}f^{-1} = (af)^3$, where $a = L_n, f = E$ as in (3.56) of \cite{Tian:2021cif}, and $b = \text{diag}\{1, \omega_n, \omega_n^{-1}\}$.
    \item{$\Delta(6n^2)$} generated with $a, b, f$ as in $\Delta(3n^2)$ with the above relations, and $h = I$ as in (3.62) of \cite{Tian:2021cif} such that $h^2 = (hf)^2 = hahb = 1.$
    \end{itemize}
    \item{$C^{(k)}_{n, l}$}
    \begin{itemize}
        \item $(r, k, l) = (3, 1, l),\ 3 | l$  $\quad$ defined by $a^{3l} = b^l = f^3 = afa^{-1}f^{-1} = (af)^3 = faf^{-1}a^{-1}b^{-1} = 1$, where $a = B_{9, 1}, b = G_{7, 1}, f = E$ as in (3.83) of \cite{Tian:2021cif}.
        \item $(r, k, l) = (7, 2, l)$ $\quad$ defined by $a^{7l} = b^l = f^3 = afa^{-1}f^{-1} = (af)^3 = afa^{-1}f^{-1}a b^{-1} = 1$, where $a = B_{7, 2}, b = G_{7l, 7}, f = E$ as in (3.90) of \cite{Tian:2021cif}
    \end{itemize}
    \item{$D^{(1)}_{3l, l}, \ 2 | l$} $\quad$ generated with $a = B_{3l, l}, b = G_{7, 7}, f = E, h = I$ as in (3.62) and (3.83) such that $a^{3l} = b^l = f^3 = h^2 = aba^{-1}b^{-1} = (af)^3 = faf^{-1}a^{-1}b^{-1} = h a h b a^2 = (fh)^2 = afba^2f^{-1} = (ba^2f)^3 = 1$
    \item{Exceptional subgroups}
    \begin{itemize}
    \item{$H_{36}$:} $\quad a^3 = b^3 = f^3 = z^4 = aba^{-1}b^{-1} = afb^{-1}f^{-1}b^{-1}a = azafz^{-1}f^{-1}a^{-1}b = azfz^{-1} = 1$,  where $a = M_1, f = M_2, z = M_3, b = \text{diag}\{\omega_3, \omega_3^2, 1\}$
    \item{$H_{72}$:} $\quad a^3 = b^3 = f^3 = z^4 = w^4 = aba^{-1}b^{-1} = afb^{-1}f^{-1}b^{-1}a = azfz^{-1} = z^2aw^2 = z^2 f^2 z w^3 z f b^{-1} a w^{-1} = 1$, where $a, b, f, z$ as defined in $H_{36}$, and $w = M_4$ as in (3.125) of \cite{Tian:2021cif}.
    \item{$H_{216}$:} $\quad$ defined with $a, b, f, w, z$ as in $H_{72}$ satisfying all the above relations, and $y = M_4$ in (3.137) of \cite{Tian:2021cif} with the extra relation that $y^9 = yay^{-1}a^{-1} = (yz)^3 = wyz^{-1}y^{-1} = 1$
    \item{$H_{60}$:} $\quad a^5 = b^2 = c^2 = (ab)^3 = (ac)^2 = (abc)^5 = (bc)^2 = 1$, where $a = H_1, b = H_2, c = H_3$ according to section 3.4.1 of \cite{Carrasco2014}.
    \item{$H_{360}$} Defined with $a, b, c$ as in $H_{60}$ with the above relations, and $d = M_4$ in (3.145) of \cite{Tian:2021cif} with extra relations $d^2 = (ad)^3 = (cd)^2 = (adbdc)^3 = (aba^2cad)^3 = 1$
    \item{$H_{168}$} $a^7 = b^7 = aba^{-1}b^{-1} = E^3 = aEb^{-1}E^{-1} = Q^2 = (Qa)^4 = (Qb)^7 = (QE)^2 = (Qab)^3 = 1$, where $a = M_1, E = M_2, Q = M_3$ as in (3.132) of \cite{Tian:2021cif}, and $b = \text{diag}\{\omega_7^4, \omega_7, \omega_7^2\}$ is a suitable permutation of the diagonal entries in $a$.
    \item{$J$} This is a case with order 180 that was filled in by Yau and Yu \cite{yau1993gorenstein}, which is isomorphic to $H_{60} \times \bbZ_3$, so $\text{Ab}[J] = \text{Ab}[H_{60}] \times \bbZ_3$.
    \item{$K$} Similar to $J$, this is a case with order 504 such that $K = H_{168} \times \bbZ_3$, so $\text{Ab}[K] = \text{Ab}[H_{168}] \times \bbZ_3$.
    \end{itemize}

\end{itemize}

\subsection{Computation of \texorpdfstring{$\text{Ab}[\Gamma/H]$}{Ab[Gamma/H]}}\label{subapdx:ab_quotient_results}

Given a explicit quotient singularity $\bbC^3/\Gamma$ specified by a discrete group $\Gamma$ and its explicit action on $\bbC^3$, we now give a \texttt{Sage} algorithm to compute the normal subgroup $H$ generated by all elements inside $\Gamma$ whose action has fixed points, and eventually $\text{Ab}[\Gamma/H]$.

An action of $\Gamma$ on $\bbC^3$ is determined by a representation $\rho: g \rightarrow GL(3, \bbC)$ that assign to an element $g \in \Gamma$ a 3-by-3 matrix $\rho(g)$ with complex entries. $\rho(g)$ will has a fixed element $\mathbf{v} \in \bbC^3$ if and only if
\begin{equation}
    \exists \mathbf{v}  \in \bbC^3\backslash\{0\} \ \text{s.t.}\ \rho(g) \mathbf{v} = \mathbf{v} \ \ \Leftrightarrow\ \  |\rho(g) - I| = 0,
\end{equation}
where $I$ is the 3-by-3 identity matrix. So our task is to ask \texttt{Sage} to compute $|\rho(g) - I|$ for all $g \in \Gamma$, and then determine the normal subgroup $H \trianglelefteq \Gamma$ that these elements generate.

\medskip

\noindent \textbf{Computing $H$.} Technically, one can simplify this by noticing that $\forall h \in \Gamma, |\rho(hgh^{-1} - I)| = |\rho(h)||\rho(g) - I||\rho^{-1}(h)| = |\rho(g) - I|$, so  $|\rho(g) - I|$ only depends on the conjugacy class that $g$ sits in. So we need these steps to compute $H$:
\begin{itemize}
    \item Determine the list of conjugacy classes of $\Gamma$
    \item Find an element $g$ for each such conjugacy class and compute its $|\rho(g) - I|$, and thus determining $|\rho(g) - I|$ for this entire conjugacy class
    \item Take the union of all conjugacy classes that has $|\rho(g) - I| = 0$, and compute the subgroup $H$ which they generate.
\end{itemize}

\medskip

\noindent \textbf{Computing $\text{Ab}|\Gamma/H|$.} Having both $\Gamma$ and $H$ explicitly, we can then use \texttt{Sage} to compute $\Gamma/H$ as well as $\mathrm{Ab}[\Gamma/H]$.

\section{Group Theory of \texorpdfstring{$D_{n,q}, T_m, O_m$ and $I_m$}{Dnq, Tm, Om and Im}}\label{app:conv}
In this Appendix we present the explicit generators for the finite subgroups of $SU(3)$ induced by finite subgroups of $U(2)$ specified
by $D_{n,q}, T_m, O_m$ and $I_m$. These are essentially just twists by an additional cyclic subgroup of the familiar $D$ and $E$-series finite subgroups of $SU(2)$. Our discussion follows that in reference \cite{yau1993gorenstein}.

\subsection{\texorpdfstring{$D_{n,q}$}{Dnq}}
This group is built from a small representation of an extension of the binary dihedral group $\mathbf{BD_n}$. To ensure the smallness of this representation, we require $1<q<n$ and $(n,q)=1$. Furthermore, we must split this case into two subclasses.

Taking $m=n-q$ to be odd, we can generate $D_{n,q}$ by
\begin{gather}
    D_{n,q}=\Bigg\langle
        \begin{pmatrix}
            \zeta_{2q} & 0 & 0\\
            0 & \zeta_{2q}^{-1} & 0\\
            0 & 0 & 1
        \end{pmatrix},
        \begin{pmatrix}
            0 & i & 0 \\
            i & 0 & 0 \\
            0 & 0 & 1
        \end{pmatrix},
        \begin{pmatrix}
            \zeta_{2m} & 0 & 0\\
            0 & \zeta_{2m} & 0\\
            0 & 0 & \zeta_{2m}^{-2}
        \end{pmatrix}
    \Bigg\rangle.
\end{gather}
where $\zeta_k=e^{2\pi i/k}$. Now taking $m=n-q$ even, we generate the group as
\begin{gather}
    D_{n,q}=\Bigg\langle
        \begin{pmatrix}
            \zeta_{2q} & 0 & 0\\
            0 & \zeta_{2q}^{-1} & 0\\
            0 & 0 & 1
        \end{pmatrix},
        \begin{pmatrix}
            0 & i & 0 \\
            i & 0 & 0 \\
            0 & 0 & 1
        \end{pmatrix}\cdot
        \begin{pmatrix}
            \zeta_{4m} & 0 & 0\\
            0 & \zeta_{4m} & 0\\
            0 & 0 & \zeta_{4m}^{-2}
        \end{pmatrix}
    \Bigg\rangle.
\end{gather}
Note that in the even case, there are only two generators.
\subsection{\texorpdfstring{$T_m$}{Tm}}
This group is built from a small representation of an extension of the binary tetrahedral group $\mathbf{BT}$. The smallness condition in this case is that $m$ must be odd. Again, we must split this into two further cases.

Taking $m=1$ or $5$ (mod $6)$, we generate the group as
\begin{gather}
    \Bigg\langle
    \begin{pmatrix}
        i & 0 & 0 \\
        0 & -i& 0 \\
        0 & 0 & 1
    \end{pmatrix},
    \begin{pmatrix}
        0 & i & 0 \\
        i & 0 & 0 \\
        0 & 0 & 1
    \end{pmatrix},
    \begin{pmatrix}
        (1+i)/2 & (-1+i)/2 & 0 \\
        (1+i)/2 & (1-i)/2 & 0 \\
        0 & 0 & 1
    \end{pmatrix},
    \begin{pmatrix}
        \zeta_{2m} & 0 & 0 \\
        0 & \zeta_{2m} & 0 \\
        0 & 0 & \zeta_{2m}^{-2}
    \end{pmatrix}\Bigg\rangle.
\end{gather}
Taking $m=3$ (mod $6)$, we generate the group by
\begin{gather}
    \Bigg\langle
    \begin{pmatrix}
        i & 0 & 0 \\
        0 & -i& 0 \\
        0 & 0 & 1
    \end{pmatrix},
    \begin{pmatrix}
        0 & i & 0 \\
        i & 0 & 0 \\
        0 & 0 & 1
    \end{pmatrix},
    \zeta_{6m}\cdot\begin{pmatrix}
        (1+i)/2 & (-1+i)/2 & 0 \\
        (1+i)/2 & (1-i)/2 & 0 \\
        0 & 0 & \zeta_{6m}^{-3}
    \end{pmatrix}\Bigg\rangle.
\end{gather}
\subsection{\texorpdfstring{$O_m$}{Om}}
This group is built from a small representation of an extension of the binary octahedral group $\mathbf{BO}$. To ensure smallness, we impose $(m,6)=1$. This is enough to give one set of generators for any valid $m$. The group is therefore always generated by
\begin{gather}
    \Bigg\langle
    \begin{pmatrix}
        \zeta_8 & 0 & 0\\
        0 & \zeta_8^{-1} & 0 \\
        0 & 0 & 1
    \end{pmatrix},
    \begin{pmatrix}
        0 & i & 0 \\
        i & 0 & 0 \\
        0 & 0 & 1
    \end{pmatrix},
    \begin{pmatrix}
        (1+i)/2 & (-1+i)/2 & 0 \\
        (1+i)/2 & (1-i)/2 & 0 \\
        0 & 0 & 1
    \end{pmatrix},
    \begin{pmatrix}
        \zeta_{2m} & 0 & 0 \\
        0 & \zeta_{2m} & 0 \\
        0 & 0 & \zeta_{2m}^{-2}
    \end{pmatrix}\Bigg\rangle.
\end{gather}
\subsection{\texorpdfstring{$I_m$}{Im}}
This group is built from a small representation of an extension of the binary icosahedral group $\mathbf{BI}$. The smallness condition is given by $(m,30)=1$. The group can always be generated as
\begin{gather}
    \Bigg\langle\begin{pmatrix}
        \zeta_{2m} & 0 & 0 \\
        0 & \zeta_{2m} & 0 \\
        0 & 0 & \zeta_{2m}^{-2}
    \end{pmatrix},
    \begin{pmatrix}
        0 & -1 & 0 \\
        1 & 0 & 0 \\
        0 & 0 & 1
    \end{pmatrix},
    \begin{pmatrix}
        \zeta_5^3 & 0 & 0 \\
        0 & \zeta_5^2 & 0 \\
        0 & 0 & 1
    \end{pmatrix},
    \frac{1}{\sqrt{5}}
    \begin{pmatrix}
        \zeta_5^4-\zeta_5 & \zeta_5^2-\zeta_5^3 & 0 \\
        \zeta_5^2-\zeta_5^3 & \zeta_5-\zeta_5^4 & 0\\
        0 & 0 & \sqrt{5}
    \end{pmatrix}\Bigg\rangle.
\end{gather}



\section{3d McKay Correspondence}\label{app:3dmckay}

In this Appendix we review the construction of the BPS quiver obtained from D-branes probing an orbifold singularity. In the main body of the text, this is used to compute the BPS quiver of the 5d SCFT, and is equivalent to extracting the worldvolume theory of a probe D0-brane of the orbifold singularity. With this in mind, it suffices to consider a T-dual description as obtained from 4d $\mathcal{N}=4$ Super-Yang Mills theory with gauge group $U(n)$. We then can apply the general orbifold prescription of \cite{Douglas:1996sw} as described in references \cite{Hanany:1998sd,Lawrence:1998ja}.

To begin, recall that the matter content of 4d $\mathcal{N}=4$ SYM is given by adjoint valued fields in the singlet, fundamental and two-index anti-symmetric representation of the R-symmetry $SU(4)$, respectively describing the vector bosons, fermions, and scalars of the theory:
\begin{center}
	\begin{tabular}{|c|c|c|}
		\hline
		&$U(n)$ gauge &  SU(4) R-symmetry \\
		\hline
		$A_{IJ}$ & Adj. & {\bf 1} \\ \hline
		$\psi^\alpha_{IJ}$ & Adj. & {\bf 4}  \\ \hline
		$\Phi^\alpha_{IJ}$ & Adj. & {\bf 6}  \\ \hline
	\end{tabular}
\end{center}

To construct the orbifolded theory, we can consider $\Gamma$, a finite subgroup of $SU(3)$, which is in turned embedded in the $SU(4)_R$ symmetry so that the fundamental decomposes as $\mathbf{4} \rightarrow \mathbf{3} \oplus \mathbf{1}$.
Under the orbifold action, the indices of the vector boson break into various representation of $\Gamma$, $\gamma_i$, such that
\begin{align}
	U(N) \to \prod_i U(N_i),
\end{align}
where $N_i=n \text{ dim}(\gamma_i)$.

Indeed, we can write adjoint fields of $U(N)$ as $ Hom(\mathbb{C}^N,\mathbb{C}^N)$.
When we take the orbifold quotient, we keep only the invariant homomorphisms
under the action of irreducible representations (irreps) of $\Gamma$
\begin{align}
	(\Hom(\mathbb{C}^N,\mathbb{C}^N))^\Gamma=\bigoplus_{i}\Hom(\mathbb{C}^{N_i},\mathbb{C}^{N_i}).
\end{align}
Since we are dealing with a brane probe theory which preserves 4d $\mathcal{N} = 1$ supersymmetry, it suffices to consider the fermions, which will be paired with scalar degrees of freedom.

The $\psi^\alpha_{IJ}$ transforms as ${\bf 4}_R \otimes \Hom(\mathbb{C}^N,\mathbb{C}^N)$. When we quotient we get:
\begin{align}
	({\bf 4}_R \otimes \Hom(\mathbb{C}^N,\mathbb{C}^N))^\Gamma=\bigoplus_{i,j} a^{\bf 4}_{i j }\Hom(\mathbb{C}^{N_i},\mathbb{C}^{N_j}).
\end{align}
The fermions are now bifundamentals charged under the various $U(N_i)$, and the matrix $a^{\bf 4}_{i j }$ gives the adjacency matrix for the quiver describing the theory. To be precise,  $a^{\bf 4}_{i j }$ gives the number of arrows from node $i$ to node $j$ in the quiver.
To compute $a^{\bf 4}_{i j }$, we use the following decomposition:
\begin{align}
	{\bf 4}_R \otimes \gamma_i = \oplus_{j} a^{\bf 4}_{i j } \gamma_j.
\end{align}
We now trace this relation to have a relation between characters of irreducible representations:
\begin{align}
	\chi({\bf 4}_R)^\alpha \chi(\gamma_i)^\alpha = \sum_j a^{\bf 4}_{i j } \chi(\gamma_j)^\alpha,
\end{align}
where $\alpha$ indicates the conjugacy class. Using the orthogonality of the characters we can express $a^{\bf 4}_{i j }$ as
\begin{align}\label{aijalpha}
	a^{\bf 4}_{i j }=\frac{1}{\vert \Gamma \vert} \sum_\alpha r_\alpha \chi({\bf 4}_R)^\alpha  \chi(\gamma_i)^\alpha \overline{\chi(\gamma_j)^\alpha},
\end{align}
where $r_\alpha$ counts the dimension of the $\alpha$ conjugacy class and the bar means complex conjugate.

What we need to specify now is $\chi({\bf 4}_R)^\alpha$. Using the decomposition ${\bf 4_R} \rightarrow \mathbf{3} \oplus \mathbf{1}$, for the fundamental of $SU(4)_R$ into $SU(3)$, the character becomes
\begin{align}
	\chi({\bf 4}_R)^\alpha=\chi({\bf 1})^\alpha+\chi({\bf 3})^\alpha=1+\chi({\bf 3})^\alpha.
\end{align}
This tells us that $\Gamma$ acts on fermions with a three dimensional representation, which needs not to be irreducible. We have the following possible decompositions for three dimensional representations:
\begin{align} \label{irrepdecomp}
	{\bf 1'}\oplus{\bf 1''}\oplus{\bf 1'''} &\to \chi({\bf 1'})^\alpha +\chi({\bf 1''})^\alpha+ \chi({\bf 1'''})^\alpha \nonumber  \\
	{\bf 1'}\oplus{\bf 2} &\to \chi({\bf 1'})^\alpha + \chi({\bf 2})^\alpha \nonumber \\
	{\bf 3} &\to \chi({\bf 3})^\alpha,
\end{align}
where ${\bf 1'}$ is a one-dimensional, possibly non-trivial, irreducible representation.

In order to choose a consistent decomposition of a three-dimensional representation in term of irreducible representations,
we require that the product of the determinant of all irreducible representations be unity.
This means that the one-dimensional representation must be chosen such that
\begin{align}
	\prod_\alpha \chi(\bf 1')^\alpha\chi(\bf 1'')^\alpha\chi(\bf 1''')^\alpha=1
\end{align}
since the character table of one dimensional irreps corresponds with the representations itself. For the ${\bf 1'}\oplus{\bf 2}$ we can also work out the determinant of the two-dimensional irreducible representations using the Adams Operations.

\subsection{Computing the Defect Group}

As briefly discussed in section \ref{sec:PRESCRIPTION},
from the matrix $a_{ij}^{\bf 4}$ we can compute the Dirac pairing
\begin{align}
    B_{ij}=a_{ij}^{\bf 4}-a_{ji}^{\bf 4},
\end{align}
and using the results of \cite{Caorsi:2017bnp} (see also \cite{Albertini:2020mdx,Hosseini:2021ged}), we can extract the defect group of the theory.

Since $B$ is a $n \times n$ matrix with entries in $\mathbb{Z}$, one can decompose it into Smith Normal Form (SNF). This amounts to finding
invertible matrices $S$ and $T$ over $\mathbb{Z}$ such that $B=SB_{SNF}T$, this is a change of base of $B$. In the new frame, $B_{SNF}=\text{diag}\{a_1,a_2,a_3,\hdots,a_m,0,\hdots,0\}$, such that $a_i$ are integers and $a_i$ divides $a_{i+1}$ for each $i < m$.

The matrices $B$ and $B_{SNF}$ have the same cokernel given by
\begin{equation}
    \text{coker} (B)=\text{coker} (B_{SNF})=\mathbb{Z}^l\oplus \mathbb{Z}^m / (a_1\mathbb{Z}\oplus  a_2\mathbb{Z}\oplus\hdots\oplus a_m\mathbb{Z})\, ,
\end{equation}
with $l$ the number of zero diagonal elements of $B_{SNF}$, corresponding to vectors which lie in $\mathrm{ker}(Q)$. We also have that the $a_i$ comes in equal pair, corresponding to electric and magnetic defect charges.

As discussed in \cite{Albertini:2020mdx}, the cokernel of $B$ gives the defect group of the theory. In particular, we have that
\begin{equation}
   \text{Tor } \bbD^{(1)} = \text{Tor}(\text{coker}(B))=(\bbZ_{a_1}\oplus \bbZ_{a_1}) \oplus \hdots \oplus (\bbZ_{a_{m/2}} \oplus \bbZ_{a_{m/2}}),
\end{equation}
where the each pair in parenthesis represent a couple of non local defect charges. This information is not enough to fully determine the Heisenberg algebra of non-commuting fluxes. The latter can be reconstructed exploiting the prescription on the Weyl pairing discussed in \cite{Caorsi:2017bnp}, to which we refer our readers.

\section{Selected B-matrices}\label{app:Bmatrix}

\subsection{\texorpdfstring{$T_5$}{T5} orbifold SCFT}

\scalebox{0.4}{$
    \left(
\begin{array}{ccccccccccccccccccccccccccccccccccc}
 0 & 1 & 0 & 0 & -1 & 0 & 0 & 0 & 0 & 0 & 0 & 0 & 0 & 0 & 0 & 0 & 0
   & 0 & 0 & 0 & 0 & 0 & 0 & 0 & 0 & 0 & 0 & 1 & -1 & 0 & 0 & 0 & 0
   & 0 & 0 \\
 -1 & 0 & 1 & 0 & 0 & 0 & 0 & 0 & 0 & 0 & 0 & 0 & 0 & 0 & 0 & 0 & 0
   & 0 & 0 & 0 & 0 & 0 & 0 & 0 & 0 & 0 & 0 & 0 & 1 & -1 & 0 & 0 & 0
   & 0 & 0 \\
 0 & -1 & 0 & 1 & 0 & 0 & 0 & 0 & 0 & 0 & 0 & 0 & 0 & 0 & 0 & -1 &
   0 & 0 & 0 & 0 & 0 & 0 & 0 & 0 & 0 & 0 & 0 & 0 & 0 & 1 & 0 & 0 &
   0 & 0 & 0 \\
 0 & 0 & -1 & 0 & 1 & 0 & 0 & 0 & 0 & 0 & 0 & 0 & 0 & 0 & 0 & 1 & 0
   & 0 & 0 & 0 & 0 & 0 & 0 & 0 & 0 & 0 & -1 & 0 & 0 & 0 & 0 & 0 & 0
   & 0 & 0 \\
 1 & 0 & 0 & -1 & 0 & 0 & 0 & 0 & 0 & 0 & 0 & 0 & 0 & 0 & 0 & 0 & 0
   & 0 & 0 & 0 & 0 & 0 & 0 & 0 & 0 & 0 & 1 & -1 & 0 & 0 & 0 & 0 & 0
   & 0 & 0 \\
 0 & 0 & 0 & 0 & 0 & 0 & 0 & 1 & 0 & 0 & -1 & 0 & 0 & 0 & 0 & 0 & 0
   & 0 & 0 & 1 & -1 & 0 & 0 & 0 & 0 & 0 & 0 & 0 & 0 & 0 & 0 & 0 & 0
   & 0 & 0 \\
 0 & 0 & 0 & 0 & 0 & 0 & 0 & 0 & 0 & 0 & 0 & 1 & 0 & 0 & -1 & 0 & 0
   & 0 & 0 & 0 & 0 & 0 & 0 & 1 & -1 & 0 & 0 & 0 & 0 & 0 & 0 & 0 & 0
   & 0 & 0 \\
 0 & 0 & 0 & 0 & 0 & -1 & 0 & 0 & 1 & 0 & 0 & 0 & 0 & 0 & 0 & 0 & 0
   & 0 & 0 & 0 & 1 & -1 & 0 & 0 & 0 & 0 & 0 & 0 & 0 & 0 & 0 & 0 & 0
   & 0 & 0 \\
 0 & 0 & 0 & 0 & 0 & 0 & 0 & -1 & 0 & 1 & 0 & 0 & 0 & 0 & 0 & 0 &
   -1 & 0 & 0 & 0 & 0 & 1 & 0 & 0 & 0 & 0 & 0 & 0 & 0 & 0 & 0 & 0 &
   0 & 0 & 0 \\
 0 & 0 & 0 & 0 & 0 & 0 & 0 & 0 & -1 & 0 & 1 & 0 & 0 & 0 & 0 & 0 & 1
   & 0 & -1 & 0 & 0 & 0 & 0 & 0 & 0 & 0 & 0 & 0 & 0 & 0 & 0 & 0 & 0
   & 0 & 0 \\
 0 & 0 & 0 & 0 & 0 & 1 & 0 & 0 & 0 & -1 & 0 & 0 & 0 & 0 & 0 & 0 & 0
   & 0 & 1 & -1 & 0 & 0 & 0 & 0 & 0 & 0 & 0 & 0 & 0 & 0 & 0 & 0 & 0
   & 0 & 0 \\
 0 & 0 & 0 & 0 & 0 & 0 & -1 & 0 & 0 & 0 & 0 & 0 & 1 & 0 & 0 & 0 & 0
   & 0 & 0 & 0 & 0 & 0 & 0 & 0 & 1 & -1 & 0 & 0 & 0 & 0 & 0 & 0 & 0
   & 0 & 0 \\
 0 & 0 & 0 & 0 & 0 & 0 & 0 & 0 & 0 & 0 & 0 & -1 & 0 & 1 & 0 & 0 & 0
   & -1 & 0 & 0 & 0 & 0 & 0 & 0 & 0 & 1 & 0 & 0 & 0 & 0 & 0 & 0 & 0
   & 0 & 0 \\
 0 & 0 & 0 & 0 & 0 & 0 & 0 & 0 & 0 & 0 & 0 & 0 & -1 & 0 & 1 & 0 & 0
   & 1 & 0 & 0 & 0 & 0 & -1 & 0 & 0 & 0 & 0 & 0 & 0 & 0 & 0 & 0 & 0
   & 0 & 0 \\
 0 & 0 & 0 & 0 & 0 & 0 & 1 & 0 & 0 & 0 & 0 & 0 & 0 & -1 & 0 & 0 & 0
   & 0 & 0 & 0 & 0 & 0 & 1 & -1 & 0 & 0 & 0 & 0 & 0 & 0 & 0 & 0 & 0
   & 0 & 0 \\
 0 & 0 & 1 & -1 & 0 & 0 & 0 & 0 & 0 & 0 & 0 & 0 & 0 & 0 & 0 & 0 & 0
   & 0 & 0 & 0 & 0 & 0 & 0 & 0 & 0 & 0 & 1 & 0 & 0 & -1 & 0 & 0 & 1
   & -1 & 0 \\
 0 & 0 & 0 & 0 & 0 & 0 & 0 & 0 & 1 & -1 & 0 & 0 & 0 & 0 & 0 & 0 & 0
   & 0 & 1 & 0 & 0 & -1 & 0 & 0 & 0 & 0 & 0 & 0 & 0 & 0 & 0 & 0 & 1
   & -1 & 0 \\
 0 & 0 & 0 & 0 & 0 & 0 & 0 & 0 & 0 & 0 & 0 & 0 & 1 & -1 & 0 & 0 & 0
   & 0 & 0 & 0 & 0 & 0 & 1 & 0 & 0 & -1 & 0 & 0 & 0 & 0 & 0 & 0 & 1
   & -1 & 0 \\
 0 & 0 & 0 & 0 & 0 & 0 & 0 & 0 & 0 & 1 & -1 & 0 & 0 & 0 & 0 & 0 &
   -1 & 0 & 0 & 1 & 0 & 0 & 0 & 0 & 0 & 0 & 0 & 0 & 0 & 0 & 0 & 0 &
   0 & 1 & -1 \\
 0 & 0 & 0 & 0 & 0 & -1 & 0 & 0 & 0 & 0 & 1 & 0 & 0 & 0 & 0 & 0 & 0
   & 0 & -1 & 0 & 1 & 0 & 0 & 0 & 0 & 0 & 0 & 0 & 0 & 0 & -1 & 0 &
   0 & 0 & 1 \\
 0 & 0 & 0 & 0 & 0 & 1 & 0 & -1 & 0 & 0 & 0 & 0 & 0 & 0 & 0 & 0 & 0
   & 0 & 0 & -1 & 0 & 1 & 0 & 0 & 0 & 0 & 0 & 0 & 0 & 0 & 1 & -1 &
   0 & 0 & 0 \\
 0 & 0 & 0 & 0 & 0 & 0 & 0 & 1 & -1 & 0 & 0 & 0 & 0 & 0 & 0 & 0 & 1
   & 0 & 0 & 0 & -1 & 0 & 0 & 0 & 0 & 0 & 0 & 0 & 0 & 0 & 0 & 1 &
   -1 & 0 & 0 \\
 0 & 0 & 0 & 0 & 0 & 0 & 0 & 0 & 0 & 0 & 0 & 0 & 0 & 1 & -1 & 0 & 0
   & -1 & 0 & 0 & 0 & 0 & 0 & 1 & 0 & 0 & 0 & 0 & 0 & 0 & 0 & 0 & 0
   & 1 & -1 \\
 0 & 0 & 0 & 0 & 0 & 0 & -1 & 0 & 0 & 0 & 0 & 0 & 0 & 0 & 1 & 0 & 0
   & 0 & 0 & 0 & 0 & 0 & -1 & 0 & 1 & 0 & 0 & 0 & 0 & 0 & -1 & 0 &
   0 & 0 & 1 \\
 0 & 0 & 0 & 0 & 0 & 0 & 1 & 0 & 0 & 0 & 0 & -1 & 0 & 0 & 0 & 0 & 0
   & 0 & 0 & 0 & 0 & 0 & 0 & -1 & 0 & 1 & 0 & 0 & 0 & 0 & 1 & -1 &
   0 & 0 & 0 \\
 0 & 0 & 0 & 0 & 0 & 0 & 0 & 0 & 0 & 0 & 0 & 1 & -1 & 0 & 0 & 0 & 0
   & 1 & 0 & 0 & 0 & 0 & 0 & 0 & -1 & 0 & 0 & 0 & 0 & 0 & 0 & 1 &
   -1 & 0 & 0 \\
 0 & 0 & 0 & 1 & -1 & 0 & 0 & 0 & 0 & 0 & 0 & 0 & 0 & 0 & 0 & -1 &
   0 & 0 & 0 & 0 & 0 & 0 & 0 & 0 & 0 & 0 & 0 & 1 & 0 & 0 & 0 & 0 &
   0 & 1 & -1 \\
 -1 & 0 & 0 & 0 & 1 & 0 & 0 & 0 & 0 & 0 & 0 & 0 & 0 & 0 & 0 & 0 & 0
   & 0 & 0 & 0 & 0 & 0 & 0 & 0 & 0 & 0 & -1 & 0 & 1 & 0 & -1 & 0 &
   0 & 0 & 1 \\
 1 & -1 & 0 & 0 & 0 & 0 & 0 & 0 & 0 & 0 & 0 & 0 & 0 & 0 & 0 & 0 & 0
   & 0 & 0 & 0 & 0 & 0 & 0 & 0 & 0 & 0 & 0 & -1 & 0 & 1 & 1 & -1 &
   0 & 0 & 0 \\
 0 & 1 & -1 & 0 & 0 & 0 & 0 & 0 & 0 & 0 & 0 & 0 & 0 & 0 & 0 & 1 & 0
   & 0 & 0 & 0 & 0 & 0 & 0 & 0 & 0 & 0 & 0 & 0 & -1 & 0 & 0 & 1 &
   -1 & 0 & 0 \\
 0 & 0 & 0 & 0 & 0 & 0 & 0 & 0 & 0 & 0 & 0 & 0 & 0 & 0 & 0 & 0 & 0
   & 0 & 0 & 1 & -1 & 0 & 0 & 1 & -1 & 0 & 0 & 1 & -1 & 0 & 0 & 1 &
   0 & 0 & -1 \\
 0 & 0 & 0 & 0 & 0 & 0 & 0 & 0 & 0 & 0 & 0 & 0 & 0 & 0 & 0 & 0 & 0
   & 0 & 0 & 0 & 1 & -1 & 0 & 0 & 1 & -1 & 0 & 0 & 1 & -1 & -1 & 0
   & 1 & 0 & 0 \\
 0 & 0 & 0 & 0 & 0 & 0 & 0 & 0 & 0 & 0 & 0 & 0 & 0 & 0 & 0 & -1 &
   -1 & -1 & 0 & 0 & 0 & 1 & 0 & 0 & 0 & 1 & 0 & 0 & 0 & 1 & 0 & -1
   & 0 & 1 & 0 \\
 0 & 0 & 0 & 0 & 0 & 0 & 0 & 0 & 0 & 0 & 0 & 0 & 0 & 0 & 0 & 1 & 1
   & 1 & -1 & 0 & 0 & 0 & -1 & 0 & 0 & 0 & -1 & 0 & 0 & 0 & 0 & 0 &
   -1 & 0 & 1 \\
 0 & 0 & 0 & 0 & 0 & 0 & 0 & 0 & 0 & 0 & 0 & 0 & 0 & 0 & 0 & 0 & 0
   & 0 & 1 & -1 & 0 & 0 & 1 & -1 & 0 & 0 & 1 & -1 & 0 & 0 & 1 & 0 &
   0 & -1 & 0 \\
\end{array}
\right)
$}

\subsection{\texorpdfstring{$T_7$}{T7} orbifold SCFT}
\scalebox{0.4}{$\left(
\begin{array}{ccccccccccccccccccccccccccccccccccccccccccccccccc}
 0 & 1 & 0 & 0 & 0 & 0 & -1 & 0 & 0 & 0 & 0 & 0 & 0 & 0 & 0 & 0 & 0
   & 0 & 0 & 0 & 0 & 0 & 0 & 0 & 0 & 0 & 0 & 0 & 0 & 0 & 0 & 0 & 0
   & 0 & 0 & 0 & 0 & 0 & 1 & -1 & 0 & 0 & 0 & 0 & 0 & 0 & 0 & 0 & 0
   \\
 -1 & 0 & 1 & 0 & 0 & 0 & 0 & 0 & 0 & 0 & 0 & 0 & 0 & 0 & 0 & 0 & 0
   & 0 & 0 & 0 & 0 & 0 & 0 & 0 & 0 & 0 & 0 & 0 & 0 & 0 & 0 & 0 & 0
   & 0 & 0 & 0 & 0 & 0 & 0 & 1 & -1 & 0 & 0 & 0 & 0 & 0 & 0 & 0 & 0
   \\
 0 & -1 & 0 & 1 & 0 & 0 & 0 & 0 & 0 & 0 & 0 & 0 & 0 & 0 & 0 & 0 & 0
   & 0 & 0 & 0 & 0 & 0 & 0 & 0 & 0 & 0 & 0 & 0 & 0 & 0 & 0 & 0 & 0
   & 0 & 0 & 0 & 0 & 0 & 0 & 0 & 1 & -1 & 0 & 0 & 0 & 0 & 0 & 0 & 0
   \\
 0 & 0 & -1 & 0 & 1 & 0 & 0 & 0 & 0 & 0 & 0 & 0 & 0 & 0 & 0 & 0 & 0
   & 0 & 0 & 0 & 0 & -1 & 0 & 0 & 0 & 0 & 0 & 0 & 0 & 0 & 0 & 0 & 0
   & 0 & 0 & 0 & 0 & 0 & 0 & 0 & 0 & 1 & 0 & 0 & 0 & 0 & 0 & 0 & 0
   \\
 0 & 0 & 0 & -1 & 0 & 1 & 0 & 0 & 0 & 0 & 0 & 0 & 0 & 0 & 0 & 0 & 0
   & 0 & 0 & 0 & 0 & 1 & 0 & 0 & 0 & 0 & 0 & 0 & 0 & 0 & 0 & 0 & 0
   & 0 & 0 & 0 & -1 & 0 & 0 & 0 & 0 & 0 & 0 & 0 & 0 & 0 & 0 & 0 & 0
   \\
 0 & 0 & 0 & 0 & -1 & 0 & 1 & 0 & 0 & 0 & 0 & 0 & 0 & 0 & 0 & 0 & 0
   & 0 & 0 & 0 & 0 & 0 & 0 & 0 & 0 & 0 & 0 & 0 & 0 & 0 & 0 & 0 & 0
   & 0 & 0 & 0 & 1 & -1 & 0 & 0 & 0 & 0 & 0 & 0 & 0 & 0 & 0 & 0 & 0
   \\
 1 & 0 & 0 & 0 & 0 & -1 & 0 & 0 & 0 & 0 & 0 & 0 & 0 & 0 & 0 & 0 & 0
   & 0 & 0 & 0 & 0 & 0 & 0 & 0 & 0 & 0 & 0 & 0 & 0 & 0 & 0 & 0 & 0
   & 0 & 0 & 0 & 0 & 1 & -1 & 0 & 0 & 0 & 0 & 0 & 0 & 0 & 0 & 0 & 0
   \\
 0 & 0 & 0 & 0 & 0 & 0 & 0 & 0 & 0 & 1 & 0 & 0 & 0 & 0 & -1 & 0 & 0
   & 0 & 0 & 0 & 0 & 0 & 0 & 0 & 0 & 0 & 0 & 0 & 0 & 0 & 0 & 0 & 1
   & -1 & 0 & 0 & 0 & 0 & 0 & 0 & 0 & 0 & 0 & 0 & 0 & 0 & 0 & 0 & 0
   \\
 0 & 0 & 0 & 0 & 0 & 0 & 0 & 0 & 0 & 0 & 0 & 0 & 0 & 0 & 0 & 1 & 0
   & 0 & 0 & 0 & -1 & 0 & 0 & 0 & 0 & 0 & 1 & -1 & 0 & 0 & 0 & 0 &
   0 & 0 & 0 & 0 & 0 & 0 & 0 & 0 & 0 & 0 & 0 & 0 & 0 & 0 & 0 & 0 &
   0 \\
 0 & 0 & 0 & 0 & 0 & 0 & 0 & -1 & 0 & 0 & 1 & 0 & 0 & 0 & 0 & 0 & 0
   & 0 & 0 & 0 & 0 & 0 & 0 & 0 & 0 & 0 & 0 & 0 & 0 & 0 & 0 & 0 & 0
   & 1 & -1 & 0 & 0 & 0 & 0 & 0 & 0 & 0 & 0 & 0 & 0 & 0 & 0 & 0 & 0
   \\
 0 & 0 & 0 & 0 & 0 & 0 & 0 & 0 & 0 & -1 & 0 & 1 & 0 & 0 & 0 & 0 & 0
   & 0 & 0 & 0 & 0 & 0 & 0 & 0 & 0 & 0 & 0 & 0 & 0 & 0 & 0 & 0 & 0
   & 0 & 1 & -1 & 0 & 0 & 0 & 0 & 0 & 0 & 0 & 0 & 0 & 0 & 0 & 0 & 0
   \\
 0 & 0 & 0 & 0 & 0 & 0 & 0 & 0 & 0 & 0 & -1 & 0 & 1 & 0 & 0 & 0 & 0
   & 0 & 0 & 0 & 0 & 0 & 0 & -1 & 0 & 0 & 0 & 0 & 0 & 0 & 0 & 0 & 0
   & 0 & 0 & 1 & 0 & 0 & 0 & 0 & 0 & 0 & 0 & 0 & 0 & 0 & 0 & 0 & 0
   \\
 0 & 0 & 0 & 0 & 0 & 0 & 0 & 0 & 0 & 0 & 0 & -1 & 0 & 1 & 0 & 0 & 0
   & 0 & 0 & 0 & 0 & 0 & 0 & 1 & 0 & 0 & 0 & 0 & 0 & 0 & -1 & 0 & 0
   & 0 & 0 & 0 & 0 & 0 & 0 & 0 & 0 & 0 & 0 & 0 & 0 & 0 & 0 & 0 & 0
   \\
 0 & 0 & 0 & 0 & 0 & 0 & 0 & 0 & 0 & 0 & 0 & 0 & -1 & 0 & 1 & 0 & 0
   & 0 & 0 & 0 & 0 & 0 & 0 & 0 & 0 & 0 & 0 & 0 & 0 & 0 & 1 & -1 & 0
   & 0 & 0 & 0 & 0 & 0 & 0 & 0 & 0 & 0 & 0 & 0 & 0 & 0 & 0 & 0 & 0
   \\
 0 & 0 & 0 & 0 & 0 & 0 & 0 & 1 & 0 & 0 & 0 & 0 & 0 & -1 & 0 & 0 & 0
   & 0 & 0 & 0 & 0 & 0 & 0 & 0 & 0 & 0 & 0 & 0 & 0 & 0 & 0 & 1 & -1
   & 0 & 0 & 0 & 0 & 0 & 0 & 0 & 0 & 0 & 0 & 0 & 0 & 0 & 0 & 0 & 0
   \\
 0 & 0 & 0 & 0 & 0 & 0 & 0 & 0 & -1 & 0 & 0 & 0 & 0 & 0 & 0 & 0 & 1
   & 0 & 0 & 0 & 0 & 0 & 0 & 0 & 0 & 0 & 0 & 1 & -1 & 0 & 0 & 0 & 0
   & 0 & 0 & 0 & 0 & 0 & 0 & 0 & 0 & 0 & 0 & 0 & 0 & 0 & 0 & 0 & 0
   \\
 0 & 0 & 0 & 0 & 0 & 0 & 0 & 0 & 0 & 0 & 0 & 0 & 0 & 0 & 0 & -1 & 0
   & 1 & 0 & 0 & 0 & 0 & 0 & 0 & 0 & 0 & 0 & 0 & 1 & -1 & 0 & 0 & 0
   & 0 & 0 & 0 & 0 & 0 & 0 & 0 & 0 & 0 & 0 & 0 & 0 & 0 & 0 & 0 & 0
   \\
 0 & 0 & 0 & 0 & 0 & 0 & 0 & 0 & 0 & 0 & 0 & 0 & 0 & 0 & 0 & 0 & -1
   & 0 & 1 & 0 & 0 & 0 & -1 & 0 & 0 & 0 & 0 & 0 & 0 & 1 & 0 & 0 & 0
   & 0 & 0 & 0 & 0 & 0 & 0 & 0 & 0 & 0 & 0 & 0 & 0 & 0 & 0 & 0 & 0
   \\
 0 & 0 & 0 & 0 & 0 & 0 & 0 & 0 & 0 & 0 & 0 & 0 & 0 & 0 & 0 & 0 & 0
   & -1 & 0 & 1 & 0 & 0 & 1 & 0 & -1 & 0 & 0 & 0 & 0 & 0 & 0 & 0 &
   0 & 0 & 0 & 0 & 0 & 0 & 0 & 0 & 0 & 0 & 0 & 0 & 0 & 0 & 0 & 0 &
   0 \\
 0 & 0 & 0 & 0 & 0 & 0 & 0 & 0 & 0 & 0 & 0 & 0 & 0 & 0 & 0 & 0 & 0
   & 0 & -1 & 0 & 1 & 0 & 0 & 0 & 1 & -1 & 0 & 0 & 0 & 0 & 0 & 0 &
   0 & 0 & 0 & 0 & 0 & 0 & 0 & 0 & 0 & 0 & 0 & 0 & 0 & 0 & 0 & 0 &
   0 \\
 0 & 0 & 0 & 0 & 0 & 0 & 0 & 0 & 1 & 0 & 0 & 0 & 0 & 0 & 0 & 0 & 0
   & 0 & 0 & -1 & 0 & 0 & 0 & 0 & 0 & 1 & -1 & 0 & 0 & 0 & 0 & 0 &
   0 & 0 & 0 & 0 & 0 & 0 & 0 & 0 & 0 & 0 & 0 & 0 & 0 & 0 & 0 & 0 &
   0 \\
 0 & 0 & 0 & 1 & -1 & 0 & 0 & 0 & 0 & 0 & 0 & 0 & 0 & 0 & 0 & 0 & 0
   & 0 & 0 & 0 & 0 & 0 & 0 & 0 & 0 & 0 & 0 & 0 & 0 & 0 & 0 & 0 & 0
   & 0 & 0 & 0 & 1 & 0 & 0 & 0 & 0 & -1 & 0 & 0 & 0 & 1 & -1 & 0 &
   0 \\
 0 & 0 & 0 & 0 & 0 & 0 & 0 & 0 & 0 & 0 & 0 & 0 & 0 & 0 & 0 & 0 & 0
   & 1 & -1 & 0 & 0 & 0 & 0 & 0 & 1 & 0 & 0 & 0 & 0 & -1 & 0 & 0 &
   0 & 0 & 0 & 0 & 0 & 0 & 0 & 0 & 0 & 0 & 0 & 0 & 0 & 1 & -1 & 0 &
   0 \\
 0 & 0 & 0 & 0 & 0 & 0 & 0 & 0 & 0 & 0 & 0 & 1 & -1 & 0 & 0 & 0 & 0
   & 0 & 0 & 0 & 0 & 0 & 0 & 0 & 0 & 0 & 0 & 0 & 0 & 0 & 1 & 0 & 0
   & 0 & 0 & -1 & 0 & 0 & 0 & 0 & 0 & 0 & 0 & 0 & 0 & 1 & -1 & 0 &
   0 \\
 0 & 0 & 0 & 0 & 0 & 0 & 0 & 0 & 0 & 0 & 0 & 0 & 0 & 0 & 0 & 0 & 0
   & 0 & 1 & -1 & 0 & 0 & -1 & 0 & 0 & 1 & 0 & 0 & 0 & 0 & 0 & 0 &
   0 & 0 & 0 & 0 & 0 & 0 & 0 & 0 & 0 & 0 & 0 & 0 & 0 & 0 & 1 & -1 &
   0 \\
 0 & 0 & 0 & 0 & 0 & 0 & 0 & 0 & 0 & 0 & 0 & 0 & 0 & 0 & 0 & 0 & 0
   & 0 & 0 & 1 & -1 & 0 & 0 & 0 & -1 & 0 & 1 & 0 & 0 & 0 & 0 & 0 &
   0 & 0 & 0 & 0 & 0 & 0 & 0 & 0 & 0 & 0 & 0 & 0 & 0 & 0 & 0 & 1 &
   -1 \\
 0 & 0 & 0 & 0 & 0 & 0 & 0 & 0 & -1 & 0 & 0 & 0 & 0 & 0 & 0 & 0 & 0
   & 0 & 0 & 0 & 1 & 0 & 0 & 0 & 0 & -1 & 0 & 1 & 0 & 0 & 0 & 0 & 0
   & 0 & 0 & 0 & 0 & 0 & 0 & 0 & 0 & 0 & -1 & 0 & 0 & 0 & 0 & 0 & 1
   \\
 0 & 0 & 0 & 0 & 0 & 0 & 0 & 0 & 1 & 0 & 0 & 0 & 0 & 0 & 0 & -1 & 0
   & 0 & 0 & 0 & 0 & 0 & 0 & 0 & 0 & 0 & -1 & 0 & 1 & 0 & 0 & 0 & 0
   & 0 & 0 & 0 & 0 & 0 & 0 & 0 & 0 & 0 & 1 & -1 & 0 & 0 & 0 & 0 & 0
   \\
 0 & 0 & 0 & 0 & 0 & 0 & 0 & 0 & 0 & 0 & 0 & 0 & 0 & 0 & 0 & 1 & -1
   & 0 & 0 & 0 & 0 & 0 & 0 & 0 & 0 & 0 & 0 & -1 & 0 & 1 & 0 & 0 & 0
   & 0 & 0 & 0 & 0 & 0 & 0 & 0 & 0 & 0 & 0 & 1 & -1 & 0 & 0 & 0 & 0
   \\
 0 & 0 & 0 & 0 & 0 & 0 & 0 & 0 & 0 & 0 & 0 & 0 & 0 & 0 & 0 & 0 & 1
   & -1 & 0 & 0 & 0 & 0 & 1 & 0 & 0 & 0 & 0 & 0 & -1 & 0 & 0 & 0 &
   0 & 0 & 0 & 0 & 0 & 0 & 0 & 0 & 0 & 0 & 0 & 0 & 1 & -1 & 0 & 0 &
   0 \\
 0 & 0 & 0 & 0 & 0 & 0 & 0 & 0 & 0 & 0 & 0 & 0 & 1 & -1 & 0 & 0 & 0
   & 0 & 0 & 0 & 0 & 0 & 0 & -1 & 0 & 0 & 0 & 0 & 0 & 0 & 0 & 1 & 0
   & 0 & 0 & 0 & 0 & 0 & 0 & 0 & 0 & 0 & 0 & 0 & 0 & 0 & 1 & -1 & 0
   \\
 0 & 0 & 0 & 0 & 0 & 0 & 0 & 0 & 0 & 0 & 0 & 0 & 0 & 1 & -1 & 0 & 0
   & 0 & 0 & 0 & 0 & 0 & 0 & 0 & 0 & 0 & 0 & 0 & 0 & 0 & -1 & 0 & 1
   & 0 & 0 & 0 & 0 & 0 & 0 & 0 & 0 & 0 & 0 & 0 & 0 & 0 & 0 & 1 & -1
   \\
 0 & 0 & 0 & 0 & 0 & 0 & 0 & -1 & 0 & 0 & 0 & 0 & 0 & 0 & 1 & 0 & 0
   & 0 & 0 & 0 & 0 & 0 & 0 & 0 & 0 & 0 & 0 & 0 & 0 & 0 & 0 & -1 & 0
   & 1 & 0 & 0 & 0 & 0 & 0 & 0 & 0 & 0 & -1 & 0 & 0 & 0 & 0 & 0 & 1
   \\
 0 & 0 & 0 & 0 & 0 & 0 & 0 & 1 & 0 & -1 & 0 & 0 & 0 & 0 & 0 & 0 & 0
   & 0 & 0 & 0 & 0 & 0 & 0 & 0 & 0 & 0 & 0 & 0 & 0 & 0 & 0 & 0 & -1
   & 0 & 1 & 0 & 0 & 0 & 0 & 0 & 0 & 0 & 1 & -1 & 0 & 0 & 0 & 0 & 0
   \\
 0 & 0 & 0 & 0 & 0 & 0 & 0 & 0 & 0 & 1 & -1 & 0 & 0 & 0 & 0 & 0 & 0
   & 0 & 0 & 0 & 0 & 0 & 0 & 0 & 0 & 0 & 0 & 0 & 0 & 0 & 0 & 0 & 0
   & -1 & 0 & 1 & 0 & 0 & 0 & 0 & 0 & 0 & 0 & 1 & -1 & 0 & 0 & 0 &
   0 \\
 0 & 0 & 0 & 0 & 0 & 0 & 0 & 0 & 0 & 0 & 1 & -1 & 0 & 0 & 0 & 0 & 0
   & 0 & 0 & 0 & 0 & 0 & 0 & 1 & 0 & 0 & 0 & 0 & 0 & 0 & 0 & 0 & 0
   & 0 & -1 & 0 & 0 & 0 & 0 & 0 & 0 & 0 & 0 & 0 & 1 & -1 & 0 & 0 &
   0 \\
 0 & 0 & 0 & 0 & 1 & -1 & 0 & 0 & 0 & 0 & 0 & 0 & 0 & 0 & 0 & 0 & 0
   & 0 & 0 & 0 & 0 & -1 & 0 & 0 & 0 & 0 & 0 & 0 & 0 & 0 & 0 & 0 & 0
   & 0 & 0 & 0 & 0 & 1 & 0 & 0 & 0 & 0 & 0 & 0 & 0 & 0 & 1 & -1 & 0
   \\
 0 & 0 & 0 & 0 & 0 & 1 & -1 & 0 & 0 & 0 & 0 & 0 & 0 & 0 & 0 & 0 & 0
   & 0 & 0 & 0 & 0 & 0 & 0 & 0 & 0 & 0 & 0 & 0 & 0 & 0 & 0 & 0 & 0
   & 0 & 0 & 0 & -1 & 0 & 1 & 0 & 0 & 0 & 0 & 0 & 0 & 0 & 0 & 1 &
   -1 \\
 -1 & 0 & 0 & 0 & 0 & 0 & 1 & 0 & 0 & 0 & 0 & 0 & 0 & 0 & 0 & 0 & 0
   & 0 & 0 & 0 & 0 & 0 & 0 & 0 & 0 & 0 & 0 & 0 & 0 & 0 & 0 & 0 & 0
   & 0 & 0 & 0 & 0 & -1 & 0 & 1 & 0 & 0 & -1 & 0 & 0 & 0 & 0 & 0 &
   1 \\
 1 & -1 & 0 & 0 & 0 & 0 & 0 & 0 & 0 & 0 & 0 & 0 & 0 & 0 & 0 & 0 & 0
   & 0 & 0 & 0 & 0 & 0 & 0 & 0 & 0 & 0 & 0 & 0 & 0 & 0 & 0 & 0 & 0
   & 0 & 0 & 0 & 0 & 0 & -1 & 0 & 1 & 0 & 1 & -1 & 0 & 0 & 0 & 0 &
   0 \\
 0 & 1 & -1 & 0 & 0 & 0 & 0 & 0 & 0 & 0 & 0 & 0 & 0 & 0 & 0 & 0 & 0
   & 0 & 0 & 0 & 0 & 0 & 0 & 0 & 0 & 0 & 0 & 0 & 0 & 0 & 0 & 0 & 0
   & 0 & 0 & 0 & 0 & 0 & 0 & -1 & 0 & 1 & 0 & 1 & -1 & 0 & 0 & 0 &
   0 \\
 0 & 0 & 1 & -1 & 0 & 0 & 0 & 0 & 0 & 0 & 0 & 0 & 0 & 0 & 0 & 0 & 0
   & 0 & 0 & 0 & 0 & 1 & 0 & 0 & 0 & 0 & 0 & 0 & 0 & 0 & 0 & 0 & 0
   & 0 & 0 & 0 & 0 & 0 & 0 & 0 & -1 & 0 & 0 & 0 & 1 & -1 & 0 & 0 &
   0 \\
 0 & 0 & 0 & 0 & 0 & 0 & 0 & 0 & 0 & 0 & 0 & 0 & 0 & 0 & 0 & 0 & 0
   & 0 & 0 & 0 & 0 & 0 & 0 & 0 & 0 & 0 & 1 & -1 & 0 & 0 & 0 & 0 & 1
   & -1 & 0 & 0 & 0 & 0 & 1 & -1 & 0 & 0 & 0 & 1 & 0 & 0 & 0 & 0 &
   -1 \\
 0 & 0 & 0 & 0 & 0 & 0 & 0 & 0 & 0 & 0 & 0 & 0 & 0 & 0 & 0 & 0 & 0
   & 0 & 0 & 0 & 0 & 0 & 0 & 0 & 0 & 0 & 0 & 1 & -1 & 0 & 0 & 0 & 0
   & 1 & -1 & 0 & 0 & 0 & 0 & 1 & -1 & 0 & -1 & 0 & 1 & 0 & 0 & 0 &
   0 \\
 0 & 0 & 0 & 0 & 0 & 0 & 0 & 0 & 0 & 0 & 0 & 0 & 0 & 0 & 0 & 0 & 0
   & 0 & 0 & 0 & 0 & 0 & 0 & 0 & 0 & 0 & 0 & 0 & 1 & -1 & 0 & 0 & 0
   & 0 & 1 & -1 & 0 & 0 & 0 & 0 & 1 & -1 & 0 & -1 & 0 & 1 & 0 & 0 &
   0 \\
 0 & 0 & 0 & 0 & 0 & 0 & 0 & 0 & 0 & 0 & 0 & 0 & 0 & 0 & 0 & 0 & 0
   & 0 & 0 & 0 & 0 & -1 & -1 & -1 & 0 & 0 & 0 & 0 & 0 & 1 & 0 & 0 &
   0 & 0 & 0 & 1 & 0 & 0 & 0 & 0 & 0 & 1 & 0 & 0 & -1 & 0 & 1 & 0 &
   0 \\
 0 & 0 & 0 & 0 & 0 & 0 & 0 & 0 & 0 & 0 & 0 & 0 & 0 & 0 & 0 & 0 & 0
   & 0 & 0 & 0 & 0 & 1 & 1 & 1 & -1 & 0 & 0 & 0 & 0 & 0 & -1 & 0 &
   0 & 0 & 0 & 0 & -1 & 0 & 0 & 0 & 0 & 0 & 0 & 0 & 0 & -1 & 0 & 1
   & 0 \\
 0 & 0 & 0 & 0 & 0 & 0 & 0 & 0 & 0 & 0 & 0 & 0 & 0 & 0 & 0 & 0 & 0
   & 0 & 0 & 0 & 0 & 0 & 0 & 0 & 1 & -1 & 0 & 0 & 0 & 0 & 1 & -1 &
   0 & 0 & 0 & 0 & 1 & -1 & 0 & 0 & 0 & 0 & 0 & 0 & 0 & 0 & -1 & 0
   & 1 \\
 0 & 0 & 0 & 0 & 0 & 0 & 0 & 0 & 0 & 0 & 0 & 0 & 0 & 0 & 0 & 0 & 0
   & 0 & 0 & 0 & 0 & 0 & 0 & 0 & 0 & 1 & -1 & 0 & 0 & 0 & 0 & 1 &
   -1 & 0 & 0 & 0 & 0 & 1 & -1 & 0 & 0 & 0 & 1 & 0 & 0 & 0 & 0 & -1
   & 0 \\
\end{array}
\right)$}

\subsection{\texorpdfstring{$O_5$}{O5} orbifold SCFT}
\scalebox{0.35}{$\left(
\begin{array}{cccccccccccccccccccccccccccccccccccccccc}
 0 & 0 & 0 & 0 & 0 & 0 & 1 & 0 & 0 & -1 & 0 & 0 & 0 & 0 & 0 & 0 & 0
   & 0 & 0 & 0 & 1 & 0 & -1 & 0 & 0 & 0 & 0 & 0 & 0 & 0 & 0 & 0 & 0
   & 0 & 0 & 0 & 0 & 0 & 0 & 0 \\
 0 & 0 & 1 & 0 & 0 & -1 & 0 & 0 & 0 & 0 & 0 & 0 & 0 & 0 & 0 & 0 & 0
   & 0 & 0 & 1 & 0 & -1 & 0 & 0 & 0 & 0 & 0 & 0 & 0 & 0 & 0 & 0 & 0
   & 0 & 0 & 0 & 0 & 0 & 0 & 0 \\
 0 & -1 & 0 & 1 & 0 & 0 & 0 & 0 & 0 & 0 & 0 & 0 & 0 & 0 & 0 & 0 & 0
   & 0 & 0 & 0 & 0 & 1 & 0 & -1 & 0 & 0 & 0 & 0 & 0 & 0 & 0 & 0 & 0
   & 0 & 0 & 0 & 0 & 0 & 0 & 0 \\
 0 & 0 & -1 & 0 & 1 & 0 & 0 & 0 & 0 & 0 & 0 & 0 & 0 & 0 & 0 & -1 &
   0 & 0 & 0 & 0 & 0 & 0 & 0 & 1 & 0 & 0 & 0 & 0 & 0 & 0 & 0 & 0 &
   0 & 0 & 0 & 0 & 0 & 0 & 0 & 0 \\
 0 & 0 & 0 & -1 & 0 & 1 & 0 & 0 & 0 & 0 & 0 & 0 & 0 & 0 & 0 & 1 & 0
   & -1 & 0 & 0 & 0 & 0 & 0 & 0 & 0 & 0 & 0 & 0 & 0 & 0 & 0 & 0 & 0
   & 0 & 0 & 0 & 0 & 0 & 0 & 0 \\
 0 & 1 & 0 & 0 & -1 & 0 & 0 & 0 & 0 & 0 & 0 & 0 & 0 & 0 & 0 & 0 & 0
   & 1 & 0 & -1 & 0 & 0 & 0 & 0 & 0 & 0 & 0 & 0 & 0 & 0 & 0 & 0 & 0
   & 0 & 0 & 0 & 0 & 0 & 0 & 0 \\
 -1 & 0 & 0 & 0 & 0 & 0 & 0 & 1 & 0 & 0 & 0 & 0 & 0 & 0 & 0 & 0 & 0
   & 0 & 0 & 0 & 0 & 0 & 1 & 0 & -1 & 0 & 0 & 0 & 0 & 0 & 0 & 0 & 0
   & 0 & 0 & 0 & 0 & 0 & 0 & 0 \\
 0 & 0 & 0 & 0 & 0 & 0 & -1 & 0 & 1 & 0 & 0 & 0 & 0 & 0 & 0 & 0 &
   -1 & 0 & 0 & 0 & 0 & 0 & 0 & 0 & 1 & 0 & 0 & 0 & 0 & 0 & 0 & 0 &
   0 & 0 & 0 & 0 & 0 & 0 & 0 & 0 \\
 0 & 0 & 0 & 0 & 0 & 0 & 0 & -1 & 0 & 1 & 0 & 0 & 0 & 0 & 0 & 0 & 1
   & 0 & -1 & 0 & 0 & 0 & 0 & 0 & 0 & 0 & 0 & 0 & 0 & 0 & 0 & 0 & 0
   & 0 & 0 & 0 & 0 & 0 & 0 & 0 \\
 1 & 0 & 0 & 0 & 0 & 0 & 0 & 0 & -1 & 0 & 0 & 0 & 0 & 0 & 0 & 0 & 0
   & 0 & 1 & 0 & -1 & 0 & 0 & 0 & 0 & 0 & 0 & 0 & 0 & 0 & 0 & 0 & 0
   & 0 & 0 & 0 & 0 & 0 & 0 & 0 \\
 0 & 0 & 0 & 0 & 0 & 0 & 0 & 0 & 0 & 0 & 0 & 1 & 0 & 0 & -1 & 0 & 0
   & 0 & 0 & 0 & 0 & 0 & 0 & 0 & 0 & 0 & 0 & 0 & 0 & 0 & 0 & 0 & 0
   & 0 & 0 & 0 & 0 & 1 & -1 & 0 \\
 0 & 0 & 0 & 0 & 0 & 0 & 0 & 0 & 0 & 0 & -1 & 0 & 1 & 0 & 0 & 0 & 0
   & 0 & 0 & 0 & 0 & 0 & 0 & 0 & 0 & 0 & 0 & 0 & 0 & 0 & 0 & 0 & 0
   & 0 & 0 & 0 & 0 & 0 & 1 & -1 \\
 0 & 0 & 0 & 0 & 0 & 0 & 0 & 0 & 0 & 0 & 0 & -1 & 0 & 1 & 0 & 0 & 0
   & 0 & 0 & 0 & 0 & 0 & 0 & 0 & 0 & 0 & 0 & 0 & 0 & 0 & 0 & 0 & 0
   & 0 & 0 & -1 & 0 & 0 & 0 & 1 \\
 0 & 0 & 0 & 0 & 0 & 0 & 0 & 0 & 0 & 0 & 0 & 0 & -1 & 0 & 1 & 0 & 0
   & 0 & 0 & 0 & 0 & 0 & 0 & 0 & 0 & 0 & 0 & 0 & 0 & 0 & 0 & 0 & 0
   & 0 & 0 & 1 & -1 & 0 & 0 & 0 \\
 0 & 0 & 0 & 0 & 0 & 0 & 0 & 0 & 0 & 0 & 1 & 0 & 0 & -1 & 0 & 0 & 0
   & 0 & 0 & 0 & 0 & 0 & 0 & 0 & 0 & 0 & 0 & 0 & 0 & 0 & 0 & 0 & 0
   & 0 & 0 & 0 & 1 & -1 & 0 & 0 \\
 0 & 0 & 0 & 1 & -1 & 0 & 0 & 0 & 0 & 0 & 0 & 0 & 0 & 0 & 0 & 0 & 0
   & 1 & 0 & 0 & 0 & 0 & 0 & -1 & 0 & 0 & 0 & 0 & 0 & 0 & 0 & 0 & 1
   & -1 & 0 & 0 & 0 & 0 & 0 & 0 \\
 0 & 0 & 0 & 0 & 0 & 0 & 0 & 1 & -1 & 0 & 0 & 0 & 0 & 0 & 0 & 0 & 0
   & 0 & 1 & 0 & 0 & 0 & 0 & 0 & -1 & 0 & 0 & 0 & 1 & -1 & 0 & 0 &
   0 & 0 & 0 & 0 & 0 & 0 & 0 & 0 \\
 0 & 0 & 0 & 0 & 1 & -1 & 0 & 0 & 0 & 0 & 0 & 0 & 0 & 0 & 0 & -1 &
   0 & 0 & 0 & 1 & 0 & 0 & 0 & 0 & 0 & 0 & 0 & 0 & 0 & 0 & 0 & 0 &
   0 & 1 & -1 & 0 & 0 & 0 & 0 & 0 \\
 0 & 0 & 0 & 0 & 0 & 0 & 0 & 0 & 1 & -1 & 0 & 0 & 0 & 0 & 0 & 0 &
   -1 & 0 & 0 & 0 & 1 & 0 & 0 & 0 & 0 & 0 & 0 & 0 & 0 & 1 & -1 & 0
   & 0 & 0 & 0 & 0 & 0 & 0 & 0 & 0 \\
 0 & -1 & 0 & 0 & 0 & 1 & 0 & 0 & 0 & 0 & 0 & 0 & 0 & 0 & 0 & 0 & 0
   & -1 & 0 & 0 & 0 & 1 & 0 & 0 & 0 & 0 & -1 & 0 & 0 & 0 & 0 & 0 &
   0 & 0 & 1 & 0 & 0 & 0 & 0 & 0 \\
 -1 & 0 & 0 & 0 & 0 & 0 & 0 & 0 & 0 & 1 & 0 & 0 & 0 & 0 & 0 & 0 & 0
   & 0 & -1 & 0 & 0 & 0 & 1 & 0 & 0 & -1 & 0 & 0 & 0 & 0 & 1 & 0 &
   0 & 0 & 0 & 0 & 0 & 0 & 0 & 0 \\
 0 & 1 & -1 & 0 & 0 & 0 & 0 & 0 & 0 & 0 & 0 & 0 & 0 & 0 & 0 & 0 & 0
   & 0 & 0 & -1 & 0 & 0 & 0 & 1 & 0 & 0 & 1 & 0 & 0 & 0 & 0 & -1 &
   0 & 0 & 0 & 0 & 0 & 0 & 0 & 0 \\
 1 & 0 & 0 & 0 & 0 & 0 & -1 & 0 & 0 & 0 & 0 & 0 & 0 & 0 & 0 & 0 & 0
   & 0 & 0 & 0 & -1 & 0 & 0 & 0 & 1 & 1 & 0 & -1 & 0 & 0 & 0 & 0 &
   0 & 0 & 0 & 0 & 0 & 0 & 0 & 0 \\
 0 & 0 & 1 & -1 & 0 & 0 & 0 & 0 & 0 & 0 & 0 & 0 & 0 & 0 & 0 & 1 & 0
   & 0 & 0 & 0 & 0 & -1 & 0 & 0 & 0 & 0 & 0 & 0 & 0 & 0 & 0 & 1 &
   -1 & 0 & 0 & 0 & 0 & 0 & 0 & 0 \\
 0 & 0 & 0 & 0 & 0 & 0 & 1 & -1 & 0 & 0 & 0 & 0 & 0 & 0 & 0 & 0 & 1
   & 0 & 0 & 0 & 0 & 0 & -1 & 0 & 0 & 0 & 0 & 1 & -1 & 0 & 0 & 0 &
   0 & 0 & 0 & 0 & 0 & 0 & 0 & 0 \\
 0 & 0 & 0 & 0 & 0 & 0 & 0 & 0 & 0 & 0 & 0 & 0 & 0 & 0 & 0 & 0 & 0
   & 0 & 0 & 0 & 1 & 0 & -1 & 0 & 0 & 0 & 0 & 1 & 0 & 0 & -1 & 0 &
   0 & 0 & 0 & 0 & 0 & 1 & -1 & 0 \\
 0 & 0 & 0 & 0 & 0 & 0 & 0 & 0 & 0 & 0 & 0 & 0 & 0 & 0 & 0 & 0 & 0
   & 0 & 0 & 1 & 0 & -1 & 0 & 0 & 0 & 0 & 0 & 0 & 0 & 0 & 0 & 1 & 0
   & 0 & -1 & 0 & 0 & 1 & -1 & 0 \\
 0 & 0 & 0 & 0 & 0 & 0 & 0 & 0 & 0 & 0 & 0 & 0 & 0 & 0 & 0 & 0 & 0
   & 0 & 0 & 0 & 0 & 0 & 1 & 0 & -1 & -1 & 0 & 0 & 1 & 0 & 0 & 0 &
   0 & 0 & 0 & 0 & 0 & 0 & 1 & -1 \\
 0 & 0 & 0 & 0 & 0 & 0 & 0 & 0 & 0 & 0 & 0 & 0 & 0 & 0 & 0 & 0 & -1
   & 0 & 0 & 0 & 0 & 0 & 0 & 0 & 1 & 0 & 0 & -1 & 0 & 1 & 0 & 0 & 0
   & 0 & 0 & -1 & 0 & 0 & 0 & 1 \\
 0 & 0 & 0 & 0 & 0 & 0 & 0 & 0 & 0 & 0 & 0 & 0 & 0 & 0 & 0 & 0 & 1
   & 0 & -1 & 0 & 0 & 0 & 0 & 0 & 0 & 0 & 0 & 0 & -1 & 0 & 1 & 0 &
   0 & 0 & 0 & 1 & -1 & 0 & 0 & 0 \\
 0 & 0 & 0 & 0 & 0 & 0 & 0 & 0 & 0 & 0 & 0 & 0 & 0 & 0 & 0 & 0 & 0
   & 0 & 1 & 0 & -1 & 0 & 0 & 0 & 0 & 1 & 0 & 0 & 0 & -1 & 0 & 0 &
   0 & 0 & 0 & 0 & 1 & -1 & 0 & 0 \\
 0 & 0 & 0 & 0 & 0 & 0 & 0 & 0 & 0 & 0 & 0 & 0 & 0 & 0 & 0 & 0 & 0
   & 0 & 0 & 0 & 0 & 1 & 0 & -1 & 0 & 0 & -1 & 0 & 0 & 0 & 0 & 0 &
   1 & 0 & 0 & 0 & 0 & 0 & 1 & -1 \\
 0 & 0 & 0 & 0 & 0 & 0 & 0 & 0 & 0 & 0 & 0 & 0 & 0 & 0 & 0 & -1 & 0
   & 0 & 0 & 0 & 0 & 0 & 0 & 1 & 0 & 0 & 0 & 0 & 0 & 0 & 0 & -1 & 0
   & 1 & 0 & -1 & 0 & 0 & 0 & 1 \\
 0 & 0 & 0 & 0 & 0 & 0 & 0 & 0 & 0 & 0 & 0 & 0 & 0 & 0 & 0 & 1 & 0
   & -1 & 0 & 0 & 0 & 0 & 0 & 0 & 0 & 0 & 0 & 0 & 0 & 0 & 0 & 0 &
   -1 & 0 & 1 & 1 & -1 & 0 & 0 & 0 \\
 0 & 0 & 0 & 0 & 0 & 0 & 0 & 0 & 0 & 0 & 0 & 0 & 0 & 0 & 0 & 0 & 0
   & 1 & 0 & -1 & 0 & 0 & 0 & 0 & 0 & 0 & 1 & 0 & 0 & 0 & 0 & 0 & 0
   & -1 & 0 & 0 & 1 & -1 & 0 & 0 \\
 0 & 0 & 0 & 0 & 0 & 0 & 0 & 0 & 0 & 0 & 0 & 0 & 1 & -1 & 0 & 0 & 0
   & 0 & 0 & 0 & 0 & 0 & 0 & 0 & 0 & 0 & 0 & 0 & 1 & -1 & 0 & 0 & 1
   & -1 & 0 & 0 & 1 & 0 & 0 & -1 \\
 0 & 0 & 0 & 0 & 0 & 0 & 0 & 0 & 0 & 0 & 0 & 0 & 0 & 1 & -1 & 0 & 0
   & 0 & 0 & 0 & 0 & 0 & 0 & 0 & 0 & 0 & 0 & 0 & 0 & 1 & -1 & 0 & 0
   & 1 & -1 & -1 & 0 & 1 & 0 & 0 \\
 0 & 0 & 0 & 0 & 0 & 0 & 0 & 0 & 0 & 0 & -1 & 0 & 0 & 0 & 1 & 0 & 0
   & 0 & 0 & 0 & 0 & 0 & 0 & 0 & 0 & -1 & -1 & 0 & 0 & 0 & 1 & 0 &
   0 & 0 & 1 & 0 & -1 & 0 & 1 & 0 \\
 0 & 0 & 0 & 0 & 0 & 0 & 0 & 0 & 0 & 0 & 1 & -1 & 0 & 0 & 0 & 0 & 0
   & 0 & 0 & 0 & 0 & 0 & 0 & 0 & 0 & 1 & 1 & -1 & 0 & 0 & 0 & -1 &
   0 & 0 & 0 & 0 & 0 & -1 & 0 & 1 \\
 0 & 0 & 0 & 0 & 0 & 0 & 0 & 0 & 0 & 0 & 0 & 1 & -1 & 0 & 0 & 0 & 0
   & 0 & 0 & 0 & 0 & 0 & 0 & 0 & 0 & 0 & 0 & 1 & -1 & 0 & 0 & 1 &
   -1 & 0 & 0 & 1 & 0 & 0 & -1 & 0 \\
\end{array}
\right)$}

\subsection{\texorpdfstring{$O_7$}{O7} orbifold SCFT}

\scalebox{0.35}{$\left(
\begin{array}{ccccccccccccccccccccccccccccccccccccccccccccccccccccc
   ccc}
 0 & 0 & 0 & 0 & 0 & 0 & 0 & 0 & 1 & 0 & 0 & 0 & 0 & -1 & 0 & 0 & 0
   & 0 & 0 & 0 & 0 & 0 & 0 & 0 & 0 & 0 & 0 & 0 & 1 & 0 & -1 & 0 & 0
   & 0 & 0 & 0 & 0 & 0 & 0 & 0 & 0 & 0 & 0 & 0 & 0 & 0 & 0 & 0 & 0
   & 0 & 0 & 0 & 0 & 0 & 0 & 0 \\
 0 & 0 & 1 & 0 & 0 & 0 & 0 & -1 & 0 & 0 & 0 & 0 & 0 & 0 & 0 & 0 & 0
   & 0 & 0 & 0 & 0 & 0 & 0 & 0 & 0 & 0 & 0 & 1 & 0 & -1 & 0 & 0 & 0
   & 0 & 0 & 0 & 0 & 0 & 0 & 0 & 0 & 0 & 0 & 0 & 0 & 0 & 0 & 0 & 0
   & 0 & 0 & 0 & 0 & 0 & 0 & 0 \\
 0 & -1 & 0 & 1 & 0 & 0 & 0 & 0 & 0 & 0 & 0 & 0 & 0 & 0 & 0 & 0 & 0
   & 0 & 0 & 0 & 0 & 0 & 0 & 0 & 0 & 0 & 0 & 0 & 0 & 1 & 0 & -1 & 0
   & 0 & 0 & 0 & 0 & 0 & 0 & 0 & 0 & 0 & 0 & 0 & 0 & 0 & 0 & 0 & 0
   & 0 & 0 & 0 & 0 & 0 & 0 & 0 \\
 0 & 0 & -1 & 0 & 1 & 0 & 0 & 0 & 0 & 0 & 0 & 0 & 0 & 0 & 0 & 0 & 0
   & 0 & 0 & 0 & 0 & 0 & 0 & 0 & 0 & 0 & 0 & 0 & 0 & 0 & 0 & 1 & 0
   & -1 & 0 & 0 & 0 & 0 & 0 & 0 & 0 & 0 & 0 & 0 & 0 & 0 & 0 & 0 & 0
   & 0 & 0 & 0 & 0 & 0 & 0 & 0 \\
 0 & 0 & 0 & -1 & 0 & 1 & 0 & 0 & 0 & 0 & 0 & 0 & 0 & 0 & 0 & 0 & 0
   & 0 & 0 & 0 & 0 & -1 & 0 & 0 & 0 & 0 & 0 & 0 & 0 & 0 & 0 & 0 & 0
   & 1 & 0 & 0 & 0 & 0 & 0 & 0 & 0 & 0 & 0 & 0 & 0 & 0 & 0 & 0 & 0
   & 0 & 0 & 0 & 0 & 0 & 0 & 0 \\
 0 & 0 & 0 & 0 & -1 & 0 & 1 & 0 & 0 & 0 & 0 & 0 & 0 & 0 & 0 & 0 & 0
   & 0 & 0 & 0 & 0 & 1 & 0 & -1 & 0 & 0 & 0 & 0 & 0 & 0 & 0 & 0 & 0
   & 0 & 0 & 0 & 0 & 0 & 0 & 0 & 0 & 0 & 0 & 0 & 0 & 0 & 0 & 0 & 0
   & 0 & 0 & 0 & 0 & 0 & 0 & 0 \\
 0 & 0 & 0 & 0 & 0 & -1 & 0 & 1 & 0 & 0 & 0 & 0 & 0 & 0 & 0 & 0 & 0
   & 0 & 0 & 0 & 0 & 0 & 0 & 1 & 0 & -1 & 0 & 0 & 0 & 0 & 0 & 0 & 0
   & 0 & 0 & 0 & 0 & 0 & 0 & 0 & 0 & 0 & 0 & 0 & 0 & 0 & 0 & 0 & 0
   & 0 & 0 & 0 & 0 & 0 & 0 & 0 \\
 0 & 1 & 0 & 0 & 0 & 0 & -1 & 0 & 0 & 0 & 0 & 0 & 0 & 0 & 0 & 0 & 0
   & 0 & 0 & 0 & 0 & 0 & 0 & 0 & 0 & 1 & 0 & -1 & 0 & 0 & 0 & 0 & 0
   & 0 & 0 & 0 & 0 & 0 & 0 & 0 & 0 & 0 & 0 & 0 & 0 & 0 & 0 & 0 & 0
   & 0 & 0 & 0 & 0 & 0 & 0 & 0 \\
 -1 & 0 & 0 & 0 & 0 & 0 & 0 & 0 & 0 & 1 & 0 & 0 & 0 & 0 & 0 & 0 & 0
   & 0 & 0 & 0 & 0 & 0 & 0 & 0 & 0 & 0 & 0 & 0 & 0 & 0 & 1 & 0 & -1
   & 0 & 0 & 0 & 0 & 0 & 0 & 0 & 0 & 0 & 0 & 0 & 0 & 0 & 0 & 0 & 0
   & 0 & 0 & 0 & 0 & 0 & 0 & 0 \\
 0 & 0 & 0 & 0 & 0 & 0 & 0 & 0 & -1 & 0 & 1 & 0 & 0 & 0 & 0 & 0 & 0
   & 0 & 0 & 0 & 0 & 0 & 0 & 0 & 0 & 0 & 0 & 0 & 0 & 0 & 0 & 0 & 1
   & 0 & -1 & 0 & 0 & 0 & 0 & 0 & 0 & 0 & 0 & 0 & 0 & 0 & 0 & 0 & 0
   & 0 & 0 & 0 & 0 & 0 & 0 & 0 \\
 0 & 0 & 0 & 0 & 0 & 0 & 0 & 0 & 0 & -1 & 0 & 1 & 0 & 0 & 0 & 0 & 0
   & 0 & 0 & 0 & 0 & 0 & -1 & 0 & 0 & 0 & 0 & 0 & 0 & 0 & 0 & 0 & 0
   & 0 & 1 & 0 & 0 & 0 & 0 & 0 & 0 & 0 & 0 & 0 & 0 & 0 & 0 & 0 & 0
   & 0 & 0 & 0 & 0 & 0 & 0 & 0 \\
 0 & 0 & 0 & 0 & 0 & 0 & 0 & 0 & 0 & 0 & -1 & 0 & 1 & 0 & 0 & 0 & 0
   & 0 & 0 & 0 & 0 & 0 & 1 & 0 & -1 & 0 & 0 & 0 & 0 & 0 & 0 & 0 & 0
   & 0 & 0 & 0 & 0 & 0 & 0 & 0 & 0 & 0 & 0 & 0 & 0 & 0 & 0 & 0 & 0
   & 0 & 0 & 0 & 0 & 0 & 0 & 0 \\
 0 & 0 & 0 & 0 & 0 & 0 & 0 & 0 & 0 & 0 & 0 & -1 & 0 & 1 & 0 & 0 & 0
   & 0 & 0 & 0 & 0 & 0 & 0 & 0 & 1 & 0 & -1 & 0 & 0 & 0 & 0 & 0 & 0
   & 0 & 0 & 0 & 0 & 0 & 0 & 0 & 0 & 0 & 0 & 0 & 0 & 0 & 0 & 0 & 0
   & 0 & 0 & 0 & 0 & 0 & 0 & 0 \\
 1 & 0 & 0 & 0 & 0 & 0 & 0 & 0 & 0 & 0 & 0 & 0 & -1 & 0 & 0 & 0 & 0
   & 0 & 0 & 0 & 0 & 0 & 0 & 0 & 0 & 0 & 1 & 0 & -1 & 0 & 0 & 0 & 0
   & 0 & 0 & 0 & 0 & 0 & 0 & 0 & 0 & 0 & 0 & 0 & 0 & 0 & 0 & 0 & 0
   & 0 & 0 & 0 & 0 & 0 & 0 & 0 \\
 0 & 0 & 0 & 0 & 0 & 0 & 0 & 0 & 0 & 0 & 0 & 0 & 0 & 0 & 0 & 1 & 0
   & 0 & 0 & 0 & -1 & 0 & 0 & 0 & 0 & 0 & 0 & 0 & 0 & 0 & 0 & 0 & 0
   & 0 & 0 & 0 & 0 & 0 & 0 & 0 & 0 & 0 & 0 & 0 & 0 & 0 & 0 & 0 & 0
   & 0 & 0 & 0 & 1 & -1 & 0 & 0 \\
 0 & 0 & 0 & 0 & 0 & 0 & 0 & 0 & 0 & 0 & 0 & 0 & 0 & 0 & -1 & 0 & 1
   & 0 & 0 & 0 & 0 & 0 & 0 & 0 & 0 & 0 & 0 & 0 & 0 & 0 & 0 & 0 & 0
   & 0 & 0 & 0 & 0 & 0 & 0 & 0 & 0 & 0 & 0 & 0 & 0 & 0 & 0 & 0 & 0
   & 0 & 0 & 0 & 0 & 1 & -1 & 0 \\
 0 & 0 & 0 & 0 & 0 & 0 & 0 & 0 & 0 & 0 & 0 & 0 & 0 & 0 & 0 & -1 & 0
   & 1 & 0 & 0 & 0 & 0 & 0 & 0 & 0 & 0 & 0 & 0 & 0 & 0 & 0 & 0 & 0
   & 0 & 0 & 0 & 0 & 0 & 0 & 0 & 0 & 0 & 0 & 0 & 0 & 0 & 0 & 0 & 0
   & 0 & 0 & 0 & 0 & 0 & 1 & -1 \\
 0 & 0 & 0 & 0 & 0 & 0 & 0 & 0 & 0 & 0 & 0 & 0 & 0 & 0 & 0 & 0 & -1
   & 0 & 1 & 0 & 0 & 0 & 0 & 0 & 0 & 0 & 0 & 0 & 0 & 0 & 0 & 0 & 0
   & 0 & 0 & 0 & 0 & 0 & 0 & 0 & 0 & 0 & 0 & 0 & 0 & 0 & 0 & 0 & 0
   & -1 & 0 & 0 & 0 & 0 & 0 & 1 \\
 0 & 0 & 0 & 0 & 0 & 0 & 0 & 0 & 0 & 0 & 0 & 0 & 0 & 0 & 0 & 0 & 0
   & -1 & 0 & 1 & 0 & 0 & 0 & 0 & 0 & 0 & 0 & 0 & 0 & 0 & 0 & 0 & 0
   & 0 & 0 & 0 & 0 & 0 & 0 & 0 & 0 & 0 & 0 & 0 & 0 & 0 & 0 & 0 & 0
   & 1 & -1 & 0 & 0 & 0 & 0 & 0 \\
 0 & 0 & 0 & 0 & 0 & 0 & 0 & 0 & 0 & 0 & 0 & 0 & 0 & 0 & 0 & 0 & 0
   & 0 & -1 & 0 & 1 & 0 & 0 & 0 & 0 & 0 & 0 & 0 & 0 & 0 & 0 & 0 & 0
   & 0 & 0 & 0 & 0 & 0 & 0 & 0 & 0 & 0 & 0 & 0 & 0 & 0 & 0 & 0 & 0
   & 0 & 1 & -1 & 0 & 0 & 0 & 0 \\
 0 & 0 & 0 & 0 & 0 & 0 & 0 & 0 & 0 & 0 & 0 & 0 & 0 & 0 & 1 & 0 & 0
   & 0 & 0 & -1 & 0 & 0 & 0 & 0 & 0 & 0 & 0 & 0 & 0 & 0 & 0 & 0 & 0
   & 0 & 0 & 0 & 0 & 0 & 0 & 0 & 0 & 0 & 0 & 0 & 0 & 0 & 0 & 0 & 0
   & 0 & 0 & 1 & -1 & 0 & 0 & 0 \\
 0 & 0 & 0 & 0 & 1 & -1 & 0 & 0 & 0 & 0 & 0 & 0 & 0 & 0 & 0 & 0 & 0
   & 0 & 0 & 0 & 0 & 0 & 0 & 1 & 0 & 0 & 0 & 0 & 0 & 0 & 0 & 0 & 0
   & -1 & 0 & 0 & 0 & 0 & 0 & 0 & 0 & 0 & 0 & 0 & 0 & 1 & -1 & 0 &
   0 & 0 & 0 & 0 & 0 & 0 & 0 & 0 \\
 0 & 0 & 0 & 0 & 0 & 0 & 0 & 0 & 0 & 0 & 1 & -1 & 0 & 0 & 0 & 0 & 0
   & 0 & 0 & 0 & 0 & 0 & 0 & 0 & 1 & 0 & 0 & 0 & 0 & 0 & 0 & 0 & 0
   & 0 & -1 & 0 & 0 & 0 & 0 & 1 & -1 & 0 & 0 & 0 & 0 & 0 & 0 & 0 &
   0 & 0 & 0 & 0 & 0 & 0 & 0 & 0 \\
 0 & 0 & 0 & 0 & 0 & 1 & -1 & 0 & 0 & 0 & 0 & 0 & 0 & 0 & 0 & 0 & 0
   & 0 & 0 & 0 & 0 & -1 & 0 & 0 & 0 & 1 & 0 & 0 & 0 & 0 & 0 & 0 & 0
   & 0 & 0 & 0 & 0 & 0 & 0 & 0 & 0 & 0 & 0 & 0 & 0 & 0 & 1 & -1 & 0
   & 0 & 0 & 0 & 0 & 0 & 0 & 0 \\
 0 & 0 & 0 & 0 & 0 & 0 & 0 & 0 & 0 & 0 & 0 & 1 & -1 & 0 & 0 & 0 & 0
   & 0 & 0 & 0 & 0 & 0 & -1 & 0 & 0 & 0 & 1 & 0 & 0 & 0 & 0 & 0 & 0
   & 0 & 0 & 0 & 0 & 0 & 0 & 0 & 1 & -1 & 0 & 0 & 0 & 0 & 0 & 0 & 0
   & 0 & 0 & 0 & 0 & 0 & 0 & 0 \\
 0 & 0 & 0 & 0 & 0 & 0 & 1 & -1 & 0 & 0 & 0 & 0 & 0 & 0 & 0 & 0 & 0
   & 0 & 0 & 0 & 0 & 0 & 0 & -1 & 0 & 0 & 0 & 1 & 0 & 0 & 0 & 0 & 0
   & 0 & 0 & 0 & 0 & 0 & 0 & 0 & 0 & 0 & 0 & 0 & 0 & 0 & 0 & 1 & -1
   & 0 & 0 & 0 & 0 & 0 & 0 & 0 \\
 0 & 0 & 0 & 0 & 0 & 0 & 0 & 0 & 0 & 0 & 0 & 0 & 1 & -1 & 0 & 0 & 0
   & 0 & 0 & 0 & 0 & 0 & 0 & 0 & -1 & 0 & 0 & 0 & 1 & 0 & 0 & 0 & 0
   & 0 & 0 & 0 & 0 & 0 & 0 & 0 & 0 & 1 & -1 & 0 & 0 & 0 & 0 & 0 & 0
   & 0 & 0 & 0 & 0 & 0 & 0 & 0 \\
 0 & -1 & 0 & 0 & 0 & 0 & 0 & 1 & 0 & 0 & 0 & 0 & 0 & 0 & 0 & 0 & 0
   & 0 & 0 & 0 & 0 & 0 & 0 & 0 & 0 & -1 & 0 & 0 & 0 & 1 & 0 & 0 & 0
   & 0 & 0 & 0 & -1 & 0 & 0 & 0 & 0 & 0 & 0 & 0 & 0 & 0 & 0 & 0 & 1
   & 0 & 0 & 0 & 0 & 0 & 0 & 0 \\
 -1 & 0 & 0 & 0 & 0 & 0 & 0 & 0 & 0 & 0 & 0 & 0 & 0 & 1 & 0 & 0 & 0
   & 0 & 0 & 0 & 0 & 0 & 0 & 0 & 0 & 0 & -1 & 0 & 0 & 0 & 1 & 0 & 0
   & 0 & 0 & -1 & 0 & 0 & 0 & 0 & 0 & 0 & 1 & 0 & 0 & 0 & 0 & 0 & 0
   & 0 & 0 & 0 & 0 & 0 & 0 & 0 \\
 0 & 1 & -1 & 0 & 0 & 0 & 0 & 0 & 0 & 0 & 0 & 0 & 0 & 0 & 0 & 0 & 0
   & 0 & 0 & 0 & 0 & 0 & 0 & 0 & 0 & 0 & 0 & -1 & 0 & 0 & 0 & 1 & 0
   & 0 & 0 & 0 & 1 & 0 & 0 & 0 & 0 & 0 & 0 & -1 & 0 & 0 & 0 & 0 & 0
   & 0 & 0 & 0 & 0 & 0 & 0 & 0 \\
 1 & 0 & 0 & 0 & 0 & 0 & 0 & 0 & -1 & 0 & 0 & 0 & 0 & 0 & 0 & 0 & 0
   & 0 & 0 & 0 & 0 & 0 & 0 & 0 & 0 & 0 & 0 & 0 & -1 & 0 & 0 & 0 & 1
   & 0 & 0 & 1 & 0 & -1 & 0 & 0 & 0 & 0 & 0 & 0 & 0 & 0 & 0 & 0 & 0
   & 0 & 0 & 0 & 0 & 0 & 0 & 0 \\
 0 & 0 & 1 & -1 & 0 & 0 & 0 & 0 & 0 & 0 & 0 & 0 & 0 & 0 & 0 & 0 & 0
   & 0 & 0 & 0 & 0 & 0 & 0 & 0 & 0 & 0 & 0 & 0 & 0 & -1 & 0 & 0 & 0
   & 1 & 0 & 0 & 0 & 0 & 0 & 0 & 0 & 0 & 0 & 1 & -1 & 0 & 0 & 0 & 0
   & 0 & 0 & 0 & 0 & 0 & 0 & 0 \\
 0 & 0 & 0 & 0 & 0 & 0 & 0 & 0 & 1 & -1 & 0 & 0 & 0 & 0 & 0 & 0 & 0
   & 0 & 0 & 0 & 0 & 0 & 0 & 0 & 0 & 0 & 0 & 0 & 0 & 0 & -1 & 0 & 0
   & 0 & 1 & 0 & 0 & 1 & -1 & 0 & 0 & 0 & 0 & 0 & 0 & 0 & 0 & 0 & 0
   & 0 & 0 & 0 & 0 & 0 & 0 & 0 \\
 0 & 0 & 0 & 1 & -1 & 0 & 0 & 0 & 0 & 0 & 0 & 0 & 0 & 0 & 0 & 0 & 0
   & 0 & 0 & 0 & 0 & 1 & 0 & 0 & 0 & 0 & 0 & 0 & 0 & 0 & 0 & -1 & 0
   & 0 & 0 & 0 & 0 & 0 & 0 & 0 & 0 & 0 & 0 & 0 & 1 & -1 & 0 & 0 & 0
   & 0 & 0 & 0 & 0 & 0 & 0 & 0 \\
 0 & 0 & 0 & 0 & 0 & 0 & 0 & 0 & 0 & 1 & -1 & 0 & 0 & 0 & 0 & 0 & 0
   & 0 & 0 & 0 & 0 & 0 & 1 & 0 & 0 & 0 & 0 & 0 & 0 & 0 & 0 & 0 & -1
   & 0 & 0 & 0 & 0 & 0 & 1 & -1 & 0 & 0 & 0 & 0 & 0 & 0 & 0 & 0 & 0
   & 0 & 0 & 0 & 0 & 0 & 0 & 0 \\
 0 & 0 & 0 & 0 & 0 & 0 & 0 & 0 & 0 & 0 & 0 & 0 & 0 & 0 & 0 & 0 & 0
   & 0 & 0 & 0 & 0 & 0 & 0 & 0 & 0 & 0 & 0 & 0 & 1 & 0 & -1 & 0 & 0
   & 0 & 0 & 0 & 0 & 1 & 0 & 0 & 0 & 0 & -1 & 0 & 0 & 0 & 0 & 0 & 0
   & 0 & 0 & 0 & 1 & -1 & 0 & 0 \\
 0 & 0 & 0 & 0 & 0 & 0 & 0 & 0 & 0 & 0 & 0 & 0 & 0 & 0 & 0 & 0 & 0
   & 0 & 0 & 0 & 0 & 0 & 0 & 0 & 0 & 0 & 0 & 1 & 0 & -1 & 0 & 0 & 0
   & 0 & 0 & 0 & 0 & 0 & 0 & 0 & 0 & 0 & 0 & 1 & 0 & 0 & 0 & 0 & -1
   & 0 & 0 & 0 & 1 & -1 & 0 & 0 \\
 0 & 0 & 0 & 0 & 0 & 0 & 0 & 0 & 0 & 0 & 0 & 0 & 0 & 0 & 0 & 0 & 0
   & 0 & 0 & 0 & 0 & 0 & 0 & 0 & 0 & 0 & 0 & 0 & 0 & 0 & 1 & 0 & -1
   & 0 & 0 & -1 & 0 & 0 & 1 & 0 & 0 & 0 & 0 & 0 & 0 & 0 & 0 & 0 & 0
   & 0 & 0 & 0 & 0 & 1 & -1 & 0 \\
 0 & 0 & 0 & 0 & 0 & 0 & 0 & 0 & 0 & 0 & 0 & 0 & 0 & 0 & 0 & 0 & 0
   & 0 & 0 & 0 & 0 & 0 & 0 & 0 & 0 & 0 & 0 & 0 & 0 & 0 & 0 & 0 & 1
   & 0 & -1 & 0 & 0 & -1 & 0 & 1 & 0 & 0 & 0 & 0 & 0 & 0 & 0 & 0 &
   0 & 0 & 0 & 0 & 0 & 0 & 1 & -1 \\
 0 & 0 & 0 & 0 & 0 & 0 & 0 & 0 & 0 & 0 & 0 & 0 & 0 & 0 & 0 & 0 & 0
   & 0 & 0 & 0 & 0 & 0 & -1 & 0 & 0 & 0 & 0 & 0 & 0 & 0 & 0 & 0 & 0
   & 0 & 1 & 0 & 0 & 0 & -1 & 0 & 1 & 0 & 0 & 0 & 0 & 0 & 0 & 0 & 0
   & -1 & 0 & 0 & 0 & 0 & 0 & 1 \\
 0 & 0 & 0 & 0 & 0 & 0 & 0 & 0 & 0 & 0 & 0 & 0 & 0 & 0 & 0 & 0 & 0
   & 0 & 0 & 0 & 0 & 0 & 1 & 0 & -1 & 0 & 0 & 0 & 0 & 0 & 0 & 0 & 0
   & 0 & 0 & 0 & 0 & 0 & 0 & -1 & 0 & 1 & 0 & 0 & 0 & 0 & 0 & 0 & 0
   & 1 & -1 & 0 & 0 & 0 & 0 & 0 \\
 0 & 0 & 0 & 0 & 0 & 0 & 0 & 0 & 0 & 0 & 0 & 0 & 0 & 0 & 0 & 0 & 0
   & 0 & 0 & 0 & 0 & 0 & 0 & 0 & 1 & 0 & -1 & 0 & 0 & 0 & 0 & 0 & 0
   & 0 & 0 & 0 & 0 & 0 & 0 & 0 & -1 & 0 & 1 & 0 & 0 & 0 & 0 & 0 & 0
   & 0 & 1 & -1 & 0 & 0 & 0 & 0 \\
 0 & 0 & 0 & 0 & 0 & 0 & 0 & 0 & 0 & 0 & 0 & 0 & 0 & 0 & 0 & 0 & 0
   & 0 & 0 & 0 & 0 & 0 & 0 & 0 & 0 & 0 & 1 & 0 & -1 & 0 & 0 & 0 & 0
   & 0 & 0 & 1 & 0 & 0 & 0 & 0 & 0 & -1 & 0 & 0 & 0 & 0 & 0 & 0 & 0
   & 0 & 0 & 1 & -1 & 0 & 0 & 0 \\
 0 & 0 & 0 & 0 & 0 & 0 & 0 & 0 & 0 & 0 & 0 & 0 & 0 & 0 & 0 & 0 & 0
   & 0 & 0 & 0 & 0 & 0 & 0 & 0 & 0 & 0 & 0 & 0 & 0 & 1 & 0 & -1 & 0
   & 0 & 0 & 0 & -1 & 0 & 0 & 0 & 0 & 0 & 0 & 0 & 1 & 0 & 0 & 0 & 0
   & 0 & 0 & 0 & 0 & 1 & -1 & 0 \\
 0 & 0 & 0 & 0 & 0 & 0 & 0 & 0 & 0 & 0 & 0 & 0 & 0 & 0 & 0 & 0 & 0
   & 0 & 0 & 0 & 0 & 0 & 0 & 0 & 0 & 0 & 0 & 0 & 0 & 0 & 0 & 1 & 0
   & -1 & 0 & 0 & 0 & 0 & 0 & 0 & 0 & 0 & 0 & -1 & 0 & 1 & 0 & 0 &
   0 & 0 & 0 & 0 & 0 & 0 & 1 & -1 \\
 0 & 0 & 0 & 0 & 0 & 0 & 0 & 0 & 0 & 0 & 0 & 0 & 0 & 0 & 0 & 0 & 0
   & 0 & 0 & 0 & 0 & -1 & 0 & 0 & 0 & 0 & 0 & 0 & 0 & 0 & 0 & 0 & 0
   & 1 & 0 & 0 & 0 & 0 & 0 & 0 & 0 & 0 & 0 & 0 & -1 & 0 & 1 & 0 & 0
   & -1 & 0 & 0 & 0 & 0 & 0 & 1 \\
 0 & 0 & 0 & 0 & 0 & 0 & 0 & 0 & 0 & 0 & 0 & 0 & 0 & 0 & 0 & 0 & 0
   & 0 & 0 & 0 & 0 & 1 & 0 & -1 & 0 & 0 & 0 & 0 & 0 & 0 & 0 & 0 & 0
   & 0 & 0 & 0 & 0 & 0 & 0 & 0 & 0 & 0 & 0 & 0 & 0 & -1 & 0 & 1 & 0
   & 1 & -1 & 0 & 0 & 0 & 0 & 0 \\
 0 & 0 & 0 & 0 & 0 & 0 & 0 & 0 & 0 & 0 & 0 & 0 & 0 & 0 & 0 & 0 & 0
   & 0 & 0 & 0 & 0 & 0 & 0 & 1 & 0 & -1 & 0 & 0 & 0 & 0 & 0 & 0 & 0
   & 0 & 0 & 0 & 0 & 0 & 0 & 0 & 0 & 0 & 0 & 0 & 0 & 0 & -1 & 0 & 1
   & 0 & 1 & -1 & 0 & 0 & 0 & 0 \\
 0 & 0 & 0 & 0 & 0 & 0 & 0 & 0 & 0 & 0 & 0 & 0 & 0 & 0 & 0 & 0 & 0
   & 0 & 0 & 0 & 0 & 0 & 0 & 0 & 0 & 1 & 0 & -1 & 0 & 0 & 0 & 0 & 0
   & 0 & 0 & 0 & 1 & 0 & 0 & 0 & 0 & 0 & 0 & 0 & 0 & 0 & 0 & -1 & 0
   & 0 & 0 & 1 & -1 & 0 & 0 & 0 \\
 0 & 0 & 0 & 0 & 0 & 0 & 0 & 0 & 0 & 0 & 0 & 0 & 0 & 0 & 0 & 0 & 0
   & 1 & -1 & 0 & 0 & 0 & 0 & 0 & 0 & 0 & 0 & 0 & 0 & 0 & 0 & 0 & 0
   & 0 & 0 & 0 & 0 & 0 & 0 & 1 & -1 & 0 & 0 & 0 & 0 & 1 & -1 & 0 &
   0 & 0 & 1 & 0 & 0 & 0 & 0 & -1 \\
 0 & 0 & 0 & 0 & 0 & 0 & 0 & 0 & 0 & 0 & 0 & 0 & 0 & 0 & 0 & 0 & 0
   & 0 & 1 & -1 & 0 & 0 & 0 & 0 & 0 & 0 & 0 & 0 & 0 & 0 & 0 & 0 & 0
   & 0 & 0 & 0 & 0 & 0 & 0 & 0 & 1 & -1 & 0 & 0 & 0 & 0 & 1 & -1 &
   0 & -1 & 0 & 1 & 0 & 0 & 0 & 0 \\
 0 & 0 & 0 & 0 & 0 & 0 & 0 & 0 & 0 & 0 & 0 & 0 & 0 & 0 & 0 & 0 & 0
   & 0 & 0 & 1 & -1 & 0 & 0 & 0 & 0 & 0 & 0 & 0 & 0 & 0 & 0 & 0 & 0
   & 0 & 0 & 0 & 0 & 0 & 0 & 0 & 0 & 1 & -1 & 0 & 0 & 0 & 0 & 1 &
   -1 & 0 & -1 & 0 & 1 & 0 & 0 & 0 \\
 0 & 0 & 0 & 0 & 0 & 0 & 0 & 0 & 0 & 0 & 0 & 0 & 0 & 0 & -1 & 0 & 0
   & 0 & 0 & 0 & 1 & 0 & 0 & 0 & 0 & 0 & 0 & 0 & 0 & 0 & 0 & 0 & 0
   & 0 & 0 & -1 & -1 & 0 & 0 & 0 & 0 & 0 & 1 & 0 & 0 & 0 & 0 & 0 &
   1 & 0 & 0 & -1 & 0 & 1 & 0 & 0 \\
 0 & 0 & 0 & 0 & 0 & 0 & 0 & 0 & 0 & 0 & 0 & 0 & 0 & 0 & 1 & -1 & 0
   & 0 & 0 & 0 & 0 & 0 & 0 & 0 & 0 & 0 & 0 & 0 & 0 & 0 & 0 & 0 & 0
   & 0 & 0 & 1 & 1 & -1 & 0 & 0 & 0 & 0 & 0 & -1 & 0 & 0 & 0 & 0 &
   0 & 0 & 0 & 0 & -1 & 0 & 1 & 0 \\
 0 & 0 & 0 & 0 & 0 & 0 & 0 & 0 & 0 & 0 & 0 & 0 & 0 & 0 & 0 & 1 & -1
   & 0 & 0 & 0 & 0 & 0 & 0 & 0 & 0 & 0 & 0 & 0 & 0 & 0 & 0 & 0 & 0
   & 0 & 0 & 0 & 0 & 1 & -1 & 0 & 0 & 0 & 0 & 1 & -1 & 0 & 0 & 0 &
   0 & 0 & 0 & 0 & 0 & -1 & 0 & 1 \\
 0 & 0 & 0 & 0 & 0 & 0 & 0 & 0 & 0 & 0 & 0 & 0 & 0 & 0 & 0 & 0 & 1
   & -1 & 0 & 0 & 0 & 0 & 0 & 0 & 0 & 0 & 0 & 0 & 0 & 0 & 0 & 0 & 0
   & 0 & 0 & 0 & 0 & 0 & 1 & -1 & 0 & 0 & 0 & 0 & 1 & -1 & 0 & 0 &
   0 & 1 & 0 & 0 & 0 & 0 & -1 & 0 \\
\end{array}
\right)$}

\chapter{\MakeUppercase{Chapter 6 Appendix}}

\section{Symmetries of E-String and \texorpdfstring{\boldmath{$\mathcal{N} = (2,0)$}}{N = (2,0)} Theories} \label{app:JUSTHEFLUBRO}

In this Appendix, we examine the global symmetry structure of the E-string and $(2,0)$ theories, including the R-symmetry. As 
earlier, we leave implicit the action on the spacetime symmetries, as dictated by the group action on the supercharges of the theory.

At the level of the algebra, the symmetry of the rank $N$ E-string is $\mathfrak{e}_8 \oplus \mathfrak{su}(2)_L \oplus \mathfrak{su}(2)_R$, which reduces to $\mathfrak{e}_8  \oplus \mathfrak{su}(2)_R$ for $N=1$.
Starting from the latter case, which has the simple tensor branch configuration
\begin{align}
    [\mathfrak{e}_8] \, 1 \, ,
\end{align}
the global symmetry is encoded in the Green--Schwarz four-form:
\begin{align}\label{eq:GS-4-form_1-E-string}
    I = -c_2(F_{\mathfrak{e}_8}) + c_2(R) - \tfrac14 p_1(T) \, ,
\end{align}
where $-c_2(F_{\mathfrak{e}_8})$ is always integer since $E_8$ is simply-connected.
Following the discussion of Section \ref{sec:R-symmetry_twist}, we can consider a $\bbZ_2$-twist of the R-symmetry and tangent bundle, which indeed leads to an integer shift,
\begin{align}
    c_2(R) - \tfrac14 p_1(T) \equiv -\tfrac14 w_R^2 - \tfrac34 w_R^2 \equiv 0 \mod \bbZ \, .
\end{align}
So we conclude that the global symmetry group of the rank 1 E-string is
\begin{align}
    E_8 \times SO(3)_R \, .
\end{align}
Now, the additional $\mathfrak{su}(2)_L$ flavor symmetry of the rank $N>1$ E-string couples to all nodes of self-pairing 2 in the quiver,
\begin{align}
    [\mathfrak{e}_8] \, 1 \, \overbrace{\underbrace{\, 2 \, 2 \, \cdots \, 2 \,}_{N-1}}^{[\mathfrak{su}(2)_L]} \, ,
\end{align}
but does \emph{not} enter the topological coupling of the left-most tensor multiplet, which by itself would just be a rank one E-string.
Therefore, its topological coupling is formally identical to that in equation \eqref{eq:GS-4-form_1-E-string}, and allows a $\bbZ_2$ twisted $SU(2)_R$ bundle.
Meanwhile, the undecorated nodes of self-pairing 2 all have topological couplings of the same form \cite{Ohmori:2014kda},
\begin{align}\label{eq:GS-4-form_-2-curve}
    I^{i>1} = c_2(L) - c_2(R) \, ,
\end{align}
which, when $c_2(R)$ is fractional, also forces $c_2(L) \equiv c_2(F_{\mathfrak{su}(2)_L})$ to be fractional.
Hence, we conclude that the rank $N$ E-string has global symmetry group
\begin{align}
    E_8 \times [SU(2)_L \times SU(2)_R]/\bbZ_2 \cong E_8 \times SO(4) \, .
\end{align}

An ${\cal N}=(2,0)$ theory has a tensor branch quiver that takes the form of an ADE-type Dynkin diagram, with nodes being undecorated and having self-pairing 2.
While the SCFT has R-symmetry $\mathfrak{sp}_2 \cong \mathfrak{so}_5$, the tensor branch description sees only the $\mathfrak{so}_4 \cong \mathfrak{su}(2)_L \oplus \mathfrak{su}(2)_R$ subalgebra, where, from a $(1,0)$ perspective, the $\mathfrak{su}(2)_L$ appears as a flavor symmetry while the $\mathfrak{su}(2)_R$ is the $(1,0)$ R-symmetry.
This is analogous to the self-pairing-2 nodes of the rank $N$ E-string, including the form of the Green--Schwarz four-form in equation \eqref{eq:GS-4-form_-2-curve}, which does not couple to $p_1(T)$.
Therefore, we can naturally consider a diagonal $\bbZ_2 \subset Z(SU(2)_L \times SU(2)_R)$ twist with background field $w$, such that
\begin{align}
    c_2(L) - c_2(R) \equiv \tfrac14 w^2 - \tfrac14 w^2 \equiv 0 \mod \bbZ \, .
\end{align}
This would imply that the global symmetry of $(2,0)$ theories on the tensor branch is
\begin{align}
    SO(4) \cong (SU(2)_L \times SU(2)_R) / \bbZ_2 \, .
\end{align}
Since this is a subgroup of $SO(5)$, but not $Spin(5) \cong Sp(2)$, we predict that the $(2,0)$ SCFTs have $SO(5)$ R-symmetry group.

\section{SW-folds and Rank Two 4d \texorpdfstring{\boldmath{$\mathcal{N}=2$}}{N=2} SCFTs}\label{app:ranktwo}

In this Appendix we show that SW-folds can be used to construct nearly all of the known rank two 4d $\mathcal{N} = 2$ SCFTs.
In recent years, there has been much progress in the program of classifying low rank 4d $\mathcal{N}=2$ SCFTs by studying the complex geometry of the Coulomb branch \cite{Argyres:2015ffa,Argyres:2015gha,Argyres:2016xua,Argyres:2016xmc,Argyres:2016yzz,Caorsi:2018zsq,Caorsi:2019vex,Martone:2020nsy,Argyres:2020wmq,Argyres:2022kon}. An enumeration and analysis of the known rank two theories has appeared recently in \cite{Martone:2021ixp,Kaidi:2021tgr,Martone:2021drm,Bourget:2021csg, Kaidi:2022sng}. As has been emphasized in \cite{Martone:2021ixp}, this enumeration is by no means a classification, and there are many reasons to believe that there remain undiscovered rank two theories. Nevertheless, recent progress has unveiled an intricate structure to the rank two 4d $\mathcal{N}=2$ SCFT landscape.

Rank two SCFTs can be arranged into families that are connected via renormalization group flows. Each family possesses a collection of ``parent'' or ``top'' theories from which all other theories in the family can be obtained by mass deformation. We note that it is not necessary that each SCFT in the family comes from all top theories, only that it comes from a mass deformation of \emph{at least one} top theory. All of the ``top'' rank two theories of \cite{Martone:2021ixp} are given in Table \ref{tbl:toptwo}. In this section, we study which of the known rank two theories can be interpreted as arising from the Stiefel--Whitney twisted compactifications that have formed the topic of this chapter. It is necessary only to provide an origin for the top theories, as those theories obtained via 4d mass deformation follow from the addition of continuous Wilson lines, breaking the flavor symmetry, on the $T^2$.

\begin{table}[]
    \centering
    \renewcommand{\arraystretch}{1.5}
    \begin{threeparttable}
    \begin{tabular}{cccc}
    \toprule
         Flavor Algebra & $\left\{ \Delta_u, \Delta_v\right\}$ & $(24a, 12c)$
         &  SW-fold \\\midrule
         $\left(\mathfrak{e}_8\right)_{24} \oplus \left(\mathfrak{su}_2\right)_{13}$ & $\{6, 12\}$ & $(263, 161)$
         &  \multirow{2}{*}{``Trivial SW-folds''} \\
         $\left(\mathfrak{so}_{20}\right)_{16}$ & $\{6, 8\}$ & $(202, 124)$
         &   \\\midrule
         $\left(\mathfrak{usp}_{12}\right)_{8}$ & $\left\{4, 6\right\}$ & $(130, 76)$
         & $\mathcal{T}_2^{(1)}(0,0,0,0,2)$  \\
         $\left(\mathfrak{usp}_4\right)_{7} \oplus \left(\mathfrak{usp}_8\right)_{8}$& $\left\{4, 6\right\}$ & $(128, 74)$
         & I \\
         $\left(\mathfrak{su}_2\right)_{7}^2 \oplus \left(\mathfrak{f}_4\right)_{12}$ & $\left\{6, 6\right\}$ & $(156, 90)$
         &  $\mathcal{T}_2^{(1)}(1,0,0,0,0)$ \\\midrule
         $\left(\mathfrak{su}_6\right)_{16} \oplus \left(\mathfrak{su}_2\right)_{9}$ & $\{6, 8\}$ & $(179, 101)$
         & $\mathcal{S}_2^{(1)}(0,0,0,1,0)$  \\\midrule
         $\left(\mathfrak{usp}_{14}\right)_{9}$  & $\{6, 8\}$ & $(185, 107)$
         & $\mathcal{T}_2^{(1)}(0,0,0,0,3)$ \\\midrule
         $\left(\mathfrak{su}_{5}\right)_{16}$ & $\{6, 8\}$ & $(170, 92)$
         & $\mathcal{R}_3^{(1)}(0,0,2)$  \\\midrule
         $\left(\mathfrak{usp}_8\right)_{13} \oplus \left(\mathfrak{su}_2\right)_{26}$ & $\{6, 12\}$ & $(232, 130)$
         & $\mathcal{T}_2^{(2)}(0,0,0,0,1)$  \\
         $\left(\mathfrak{su}_2\right)_{2} \oplus \left(\mathfrak{su}_2\right)_{8}$ & $\left\{3, 6\right\}$ & $(102, 54)$
         & $\mathcal{T}_4^{(1)}(1,0)$ \\\midrule
         $\left(\mathfrak{g}_2\right)_{8} \oplus \left(\mathfrak{su}_2\right)_{14}$ & $\left\{4, 6\right\}$ & $(120, 66)$
         & $\mathcal{T}_3^{(1)}(1,0,0)$ \\\midrule
         $\left(\mathfrak{su}_3\right)_{26} \oplus \mathfrak{u}(1)$ & $\{6, 12\}$ & $(219, 117)$
         & $\mathcal{S}_3^{(2)}(0,0,1)$   \\\midrule
         $\left(\mathfrak{su}_2\right)_{16} \oplus \mathfrak{u}(1)$ & $\{6, 12\}$ & $(212, 110)$
         &  $\mathcal{S}_4^{(2)}(0,1)$ \\\midrule
         $\left(\mathfrak{usp}_4\right)_{14} \oplus \left(\mathfrak{su}_2\right)_{8}$ & $\left\{4, 6\right\}$ & $(118, 64)$
         & II \\\midrule
         $\left(\mathfrak{su}_{2}\right)_{14}$ & $\left\{ \frac{12}{5}, 6\right\}$ & $(\frac{456}{5}, \frac{234}{5})$
         & $\mathcal{T}_5^{(1)}(1)$ \\\midrule
         $\left(\mathfrak{su}_{2}\right)_{14}$ & $\left\{2, 6\right\}$ & $(84, 42)$
         & $\mathcal{T}_6^{(1)}(1)$  \\\midrule
         $\left(\mathfrak{usp}_{12}\right)_{11}$ & $\{4, 10\}$ & $(188, 110)$
         & III \\\midrule
         $\varnothing$ & $\{2, 4\}$ & $(58, 28)$
         & IV \\\bottomrule
    \end{tabular}
    \end{threeparttable}
    \caption{All the ``top'' theories from \cite{Martone:2021ixp}. We list their Coulomb branch operator dimensions and their 6d origin, if known. There are four theories, labelled by I--IV, for which no SW-fold description is known. The theories marked as ``Trivial SW-folds'' are those which are obtained via compactification of a 6d $(1,0)$ SCFT on $T^2$ without turning on a Stiefel--Whitney twist.}\label{tbl:toptwo}
\end{table}

When considering a Stiefel--Whitney twisted compactification, the Coulomb branch dimension of the resulting four-dimensional theory is always at least the number of tensor multiplets, i.e., compact curves in the geometric construction, of the parent 6d theory. As such, the 4d SCFTs of low rank can only come from a highly restrictive set of 6d SCFTs. There are two ways of engineering theories with a Coulomb branch of rank two. We can consider
\begin{equation}
    \overset{\mathfrak{g}}{1}/\mathbb{Z}_\ell \,,
\end{equation}
where the $\mathbb{Z}_\ell$ Stiefel--Whitney quotient breaks the gauge algebra to $\mathfrak{g}_\text{ub}$ which is either
\begin{equation}
    \mathfrak{so}_2 \quad \text{ or } \quad \mathfrak{su}_2 \,.
\end{equation}
In either case, the methodology of \cite{Ohmori:2018ona} allows us to determine that the Coulomb branch operators, $u$ and $v$, have dimensions\footnote{We do not need to worry about the subtleties with the prescription of \cite{Ohmori:2018ona} that were highlighted in Section \ref{sec:cbe6}, as the rank two requirement on the Coulomb branch ensures that the gauge algebra after Stiefel--Whitney twist is at most rank one.}
\begin{equation}
    \left\{ \Delta_u, \Delta_v \right\} = \begin{cases}
      \left\{6, \frac{6}{\ell} + 1 \right\} \quad &\text{if} \quad \mathfrak{g}_\text{ub} = \mathfrak{so}_2 \cr
      \left\{6, 8 \right\} \quad &\text{if} \quad \mathfrak{g}_\text{ub} = \mathfrak{su}_2 \,.
      \end{cases}
\end{equation}
On the other hand, we can have
\begin{equation}
    \overset{\mathfrak{g}}{1}\overset{\mathfrak{h}}{2}/\mathbb{Z}_\ell \,,
\end{equation}
where the quotient breaks the entirety of the gauge algebra. In this case, we allow $\mathfrak{g}$ and $\mathfrak{h}$ to be trivial, and we also allow $\ell = 1$.  We can determine that
\begin{equation}
    \left\{ \Delta_u, \Delta_v \right\} = \begin{cases}
      \left\{6, \frac{12}{\ell} \right\} \quad &\text{if} \quad \mathfrak{g} = \varnothing \cr
      \left\{6, 12 \right\} \quad &\text{if} \quad \mathfrak{g} \neq \varnothing \,.
      \end{cases}
\end{equation}
Putting this together, and using that fact that $\ell = 1, \cdots, 6$ are the only valid options, we find that the possible Coulomb branch operator spectra are highly constrained. Specifically, all rank two theories obtained by Stiefel--Whitney twisted compactification have a Coulomb branch operator with scaling dimension $\Delta = 6$.

In view of these restrictions on the scaling dimensions of the Coulomb branch operators, let us consider the theories I--IV from Table \ref{tbl:toptwo}; that is, those that do not have a known SW-fold description. Theories I and II have Coulomb branch operators of dimensions $\{4, 6\}$, which are consistent with them arising from a $\mathbb{Z}_3$ Stiefel--Whitney twisted compactification of 6d SCFTs of the form
\begin{equation}
    \overset{\mathfrak{g}}{1} \,, \qquad \text{ or } \qquad 1\overset{\mathfrak{g}}{2} \,,
\end{equation}
however, it is unclear that additional 6d SCFTs of this form admitting a $\mathbb{Z}_3$ center-flavor symmetry exist. The theory labelled III has Coulomb branch operators with dimensions $\{4, 10\}$, which does not include the requisite dimension six operator for it to be able to arise from a Stiefel--Whitney twisted compactification. There are two possibilities: either theory III is not a top theory, or else it is a theory that cannot be obtained from a Stiefel--Whitney twisted torus compactification. The analysis of \cite{Cecotti:2021ouq} appears to rule out any currently unknown theory with Coulomb branch dimensions $\{6,12\}$, which would be expected for a putative top theory that mass deforms to a theory with Coulomb branch operator spectrum $\{4,10\}$, and thus we conclude that this SCFT probably does not arise from a Stiefel--Whitney twisted compactification.\footnote{We thank M. Martone for discussions on this point.} Finally, we turn to theory IV. This is the Lagrangian theory with gauge algebra $\mathfrak{sp}_2$ and a single half-hypermultiplet in the $\bm{16}$ representation, and it is also the only theory in \cite{Martone:2021ixp} that does not have any known construction in string theory. As emphasized therein, one may speculate that this SCFT sits as a descendant inside of a currently unknown family of rank two SCFTs.

At rank two it appears that almost all of the 4d $\mathcal{N}=2$ SCFTs can be obtained from torus compactifications of 6d $(1,0)$ SCFTs. Of the sixty-nine 4d SCFTs listed in \cite{Martone:2021ixp}, there are only seven for which it is not known how to obtain them in this manner. It would be interesting to understand whether other ingredients can be included in the torus compactifications to generate the complete list of rank two theories, and to determine if the preponderance of Stiefel--Whitney twists persists to higher rank 4d SCFTs.

\section{Stiefel--Whitney Twists and the Literature}\label{app:lit}

In this Appendix we present a brief survey in table format of earlier work on Stiefel--Whitney compactifications.
4d $\mathcal{N}=2$ SCFTs have been constructed from Stiefel--Whitney twisted torus compactifications of very Higgsable 6d $(1,0)$ SCFTs in previous literature \cite{Ohmori:2018ona,Apruzzi:2020pmv,Giacomelli:2020jel,Heckman:2020svr,Giacomelli:2020gee,Bourget:2020mez}. These theories form a small subset of the landscape of Stiefel--Whitney twisted theories that we discuss in the present chapter, and we highlight for which values of the $E_8$-homomorphism parameters they have been studied. These particular theories are listed in Table \ref{tbl:lit}, together with the reference to where they were first explored.

\begin{table}[]
    \centering
    \small
    \begin{threeparttable}
    \begin{tabular}{ccccc}
        \toprule
        SW-fold & 6d Origin & $\mathbb{Z}_\ell$ & Alternate Name & Reference \\\midrule
        $\mathcal{T}_2^{(r)}(0,0,0,0,1)$ & $\overset{\mathfrak{su}_2}{1}\underbrace{\overset{\mathfrak{su}_2}{2}\cdots\overset{\mathfrak{su}_2}{2}}_{r-1}$ & $\mathbb{Z}_2$ & $\mathcal{S}_{E_6, 2}^{(r)}$ & \multirow{12}{*}{\cite{Giacomelli:2020jel}} \\
        $\mathcal{S}_3^{(r)}(0,0,1)$ & $\overset{\mathfrak{su}_3}{1}\underbrace{\overset{\mathfrak{su}_3}{2}\cdots\overset{\mathfrak{su}_3}{2}}_{r-1}$ & $\mathbb{Z}_3$ & $\mathcal{S}_{D_4, 3}^{(r)}$ & \\
        $\mathcal{S}_4^{(r)}(0,1)$ & $\overset{\mathfrak{su}_4}{1}\underbrace{\overset{\mathfrak{su}_4}{2}\cdots\overset{\mathfrak{su}_4}{2}}_{r-1}$ & $\mathbb{Z}_4$ & $\mathcal{S}_{A_2, 4}^{(r)}$ & \\
        $\mathcal{T}_2^{(r-1)}(1,0,0,0,0)$ & $1\underbrace{\overset{\mathfrak{su}_2}{2}\cdots\overset{\mathfrak{su}_2}{2}}_{r-1}$ & $\mathbb{Z}_2$ & $\mathcal{T}_{E_6, 2}^{(r)}$ & \\
        $\mathcal{T}_3^{(r-1)}(1,0,0)$ & $1\underbrace{\overset{\mathfrak{su}_3}{2}\cdots\overset{\mathfrak{su}_3}{2}}_{r-1}$ & $\mathbb{Z}_3$ & $\mathcal{T}_{D_4, 3}^{(r)} $ & \\
        $\mathcal{T}_4^{(r-1)}(1,0)$ & $1\underbrace{\overset{\mathfrak{su}_4}{2}\cdots\overset{\mathfrak{su}_4}{2}}_{r-1}$ & $\mathbb{Z}_4$ & $\mathcal{T}_{A_2, 4}^{(r)}$ & \\\midrule
        $\mathcal{T}_5^{(r-1)}(1)$ & $1\underbrace{\overset{\mathfrak{su}_5}{2}\cdots\overset{\mathfrak{su}_5}{2}}_{r-1}$ & $\mathbb{Z}_5$ & $\mathcal{T}_{\varnothing, 5}^{(r)}$ & \multirow{4}{*}{\cite{Giacomelli:2020gee}} \\
        $\mathcal{T}_6^{(r-1)}(1)$ & $1\underbrace{\overset{\mathfrak{su}_6}{2}\cdots\overset{\mathfrak{su}_6}{2}}_{r-1}$ & $\mathbb{Z}_6$ & $\mathcal{T}_{\varnothing, 6}^{(r)}$ & \\\midrule
        $\mathcal{S}_2^{(1)}(0,0,0,2r-1,n-2)$ & $\overset{\mathfrak{su}_{2n}}{1}\underbrace{\overset{\mathfrak{su}_{2n+8}}{2}\cdots\overset{\mathfrak{su}_{2n-8r-8}}{2}}_{r-1}$ & $\mathbb{Z}_2$ & --- & \multirow{18}{*}{\cite{Ohmori:2018ona}} \\
        $\mathcal{S}_3^{(1)}(0,0,3r-2)$ & $\overset{\mathfrak{su}_{3}}{1}\underbrace{\overset{\mathfrak{su}_{12}}{2}\cdots\overset{\mathfrak{su}_{9r-6}}{2}}_{r-1}$ & $\mathbb{Z}_3$ & --- & \\
        $\mathcal{R}_3^{(1)}(0,0,3r-1)$ & $\overset{\mathfrak{su}_{6}^\prime}{1}\underbrace{\overset{\mathfrak{su}_{15}}{2}\cdots\overset{\mathfrak{su}_{9r-3}}{2}}_{r-1}$ & $\mathbb{Z}_3$ & --- & \\
        $\mathcal{T}_3^{(1)}(0,0,3r-3)$ & $1\underbrace{\overset{\mathfrak{su}_{9}}{2}\cdots\overset{\mathfrak{su}_{9r-9}}{2}}_{r-1}$ & $\mathbb{Z}_3$ & --- & \\
        $\mathcal{S}_4^{(1)}(0,2r-1)$ & $\overset{\mathfrak{su}_{4}}{1}\underbrace{\overset{\mathfrak{su}_{12}}{2}\cdots\overset{\mathfrak{su}_{8r-4}}{2}}_{r-1}$ & $\mathbb{Z}_4$ & --- & \\
        $\mathcal{T}_4^{(1)}(0,2r-2)$ & $1\underbrace{\overset{\mathfrak{su}_{8}}{2}\cdots\overset{\mathfrak{su}_{8r-8}}{2}}_{r-1}$ & $\mathbb{Z}_4$ & --- & \\
        $\mathcal{T}_5^{(1)}(r-1)$ & $1\underbrace{\overset{\mathfrak{su}_{5}}{2}\cdots\overset{\mathfrak{su}_{5r-5}}{2}}_{r-1}$ & $\mathbb{Z}_5$ & --- & \\
        $\mathcal{T}_6^{(1)}(r-1)$ & $1\underbrace{\overset{\mathfrak{su}_{6}}{2}\cdots\overset{\mathfrak{su}_{6r-6}}{2}}_{r-1}$ & $\mathbb{Z}_6$ & --- & \\
         \bottomrule
    \end{tabular}
    \end{threeparttable}
    \caption{The Stiefel--Whitney twisted 4d $\mathcal{N}=2$ SCFTs that have appeared afore.}
    \label{tbl:lit}
\end{table}

\section{Nilpotent Orbits and Higgsing SW-folds}\label{app:nilp}

In this Appendix we track the structure of a particular class of Higgs branch flows in 6d, and their 4d descendants after a SW-twist.
Take the original 6d rank $N$ orbi-instanton SCFT of type $(\mathfrak{e}_8, \mathfrak{g)}$ which can be expressed on its partial tensor branch as
\begin{equation}
    [\mathfrak{e}_8] \overset{\mathfrak{g}}{1}  \overset{\mathfrak{g}}{2} \cdots \overset{\mathfrak{g}}{2}  [\mathfrak{g}] \,.
\end{equation}
Consider, as we did in Section \ref{sec:6d}, a Higgsing via an $E_8$-homomorphism $\Gamma_\mathfrak{g} \rightarrow E_8$ that leads to an SCFT with a non-trivial center-flavor symmetry $\mathbb{Z}_{\ell}$. Instead of immediately compactifying it on a $T^2$ with a $\bbZ_{\ell}$ Stiefel--Whitney twist, we first perform a nilpotent Higgsing in 6d of the $\mathfrak{g}$ flavor symmetry on the right of the tensor branch quiver:\footnote{Higgs branch renormalization group flows of 6d SCFTs triggered by nilpotent deformations have been studied in great detail in \cite{Heckman:2016ssk,Mekareeya:2016yal,Heckman:2018pqx,DeLuca:2018zbi,Hassler:2019eso,Baume:2021qho,Distler:2022yse}.}
\begin{equation}
    \mu_{\text{6d}}: \mathfrak{su}_2 \rightarrow \mathfrak{g} \,.
\end{equation}
Particular choices for the nilpotent orbit lead to 6d SCFT where the $\mathbb{Z}_\ell$ center-flavor symmetry is preserved. We can then take this Higgsed theory, and compactify it on a $T^2$ with a $\mathbb{Z}_\ell$ Stiefel--Whitney twist. In this way, we end up with a larger family of SW-fold theories in 4d, going beyond the scope of the theories listed in Table \ref{tbl:6dscfts}.

Nonetheless, we can still analyze these extra theories by computing their central charges and flavor central charges as in Section \ref{sec:sfolds}. We also point out that, these extra SW-fold theories can also be obtained by taking the SW-fold theories that we have obtained in the main text as in Table \ref{tbl:genSfolds} and performing the following ``induced'' nilpotent Higgsing in 4d:
\begin{equation}
    \mu_{4d}: \mathfrak{su}_2 \rightarrow \widetilde{\mathfrak{g}}
\end{equation}
where $\widetilde{\mathfrak{g}}$ is the flavor symmetry algebra in 4d that descended from the $\mathfrak{g}$ flavor symmetry on the right of the 6d SCFT. The nilpotent orbit $\mu_{4d}$ is then the nilpotent deformation in 4d which can be thought of as a ``folded'' version of the nilpotent deformation in 6d.

For the nilpotent Higgsings of the 6d SCFTs that were discussed in Section \ref{sec:e8CF}, we have $\mathfrak{g} = \mathfrak{su}_K$. Nilpotent orbits of $\mathfrak{su}_K$ are in one-to-one correspondence with integer partitions of $K$. We write
\begin{equation}\label{eqn:exppart}
    P = [1^{n_1}, 2^{n_2}, \cdots, K^{n_K}] \,,
\end{equation}
where $n_i \geq 0$ and
\begin{equation}
    \sum_{i=1}^K i n_i = K \,,
\end{equation}
to denote a partition of $K$. We are interested in the case where $K = \ell \widetilde{K}$ and where there exists a $\mathbb{Z}_\ell$ center-flavor symmetry. The RHS of the tensor branch configurations written in Table \ref{tbl:6dscfts} all have the form
\begin{equation}
    \cdots \overset{\mathfrak{su}_K}{2}\overset{\mathfrak{su}_K}{2}\cdots \overset{\mathfrak{su}_K}{2}\overset{\mathfrak{su}_K}{2} \,,
\end{equation}
and it is well-known how the tensor branch configuration is modified when Higgsing by a nilpotent orbit as in equation \eqref{eqn:exppart}. One finds\footnote{For ease of exposition, we do not focus here on the cases where the plateau is too short and the nilpotent Higgsing starts to correlate with the $E_8$-homomorphism Higgsing.}
\begin{equation}
    \cdots \underset{[\mathfrak{su}_{n_K}]}{\overset{\mathfrak{su}_{K_K}}{2}}\underset{[\mathfrak{su}_{n_{K-1}}]}{\overset{\mathfrak{su}_{K_{K-1}}}{2}}\cdots \underset{[\mathfrak{su}_{n_2}]}{\overset{\mathfrak{su}_{K_2}}{2}}\underset{[\mathfrak{su}_{n_1}]}{\overset{\mathfrak{su}_{K_1}}{2}} \,.
\end{equation}
The ranks of the flavor algebras are fixed by the exponents of the partition, and the $K_i$ are fixed from that data by anomaly cancellation. They are each required to satisfy
\begin{equation}
    2K_i - n_i - K_{i-1} - K_{i+1} = 0 \,,
\end{equation}
where we have defined $K_0 = 0$ and $K_{K+1} = K$. It is easy to see that the anomaly of the large gauge transformations of the two-forms fields, as discussed in Section \ref{sec:6d}, rules out a $\mathbb{Z}_\ell$ center-flavor symmetry unless
\begin{equation}
    n_i = \ell \widetilde{n}_i \,,
\end{equation}
for all $i$. The converse also can be shown. It is easy to see that a partition
\begin{equation}
    [1^{\ell\widetilde{n}_1}, 2^{\ell\widetilde{n}_2}, \cdots, (\ell\widetilde{K})^{\ell\widetilde{n}_{(\ell\widetilde{K})}}] \,,
\end{equation}
of $\ell \widetilde{K}$ can equivalently be written as a partition
\begin{equation}
    [1^{\widetilde{n}_1}, 2^{\widetilde{n}_2}, \cdots, \widetilde{K}^{\widetilde{n}_{\widetilde{K}}}] \,,
\end{equation}
of $\widetilde{K}$. More succinctly, $\mathbb{Z}_\ell$ center-flavor symmetry preserving nilpotent orbits of $\mathfrak{su}_K$ are in one-to-one correspondence with nilpotent orbits of $\mathfrak{su}_{\widetilde{K}}$. Physically, this reflects the fact that one can either first perform the nilpotent Higgsing by a ($\mathbb{Z}_\ell$-preserving) partition of $K$ in 6d and then compactify with Stiefel--Whitney twist to 4d, or else first perform the Stiefel--Whitney twisted compactification to 4d and then the nilpotent Higgsing by the associated partition of $\widetilde{K}$; either way, one ends up with same 4d $\mathcal{N}=2$ SCFT.

The family of theories obtained by nilpotent deformations forms a partially ordered set, capturing the network of renormalization group flows amongst the theories, that follows from the partial ordering of the nilpotent orbit inclusion: $\mu \prec \nu$ when $\text{Orbit}(\mu) \subset \overline{\text{Orbit}(\nu)}$. For $\mathfrak{su}_{K}$ nilpotent orbits, such partial ordering can be characterized by the ``dominant ordering'' of two partitions of $K$; let $\mu = [r_1, \cdots, r_{\ell_r}],\ \nu = [s_1, \cdots s_{\ell_s}]$ be weakly-decreasing partitions of $K$, then
\begin{equation}
   \mu \prec \nu\ \  \Leftrightarrow\ \  \sum_{i = 1}^j r_i \leq \sum_{i = 1}^j s_i, \quad \,\, 1 \leq j \leq \operatorname{max}(\ell_r, \ell_s) \,,
\end{equation}
where the partition with fewer elements is extended by zeroes until they are of equal length.

To be more specific, let us illustrate this construction by analyzing the network of theories obtained by starting from the SCFT in family $\mathcal{T}_3^{(N)}(p, s, 3q)$ with $p=s=q=1$. The 6d origin of this theory is given by the tensor branch description:
\begin{equation}\label{eq:6d_trivialOrbit}
   [\mathfrak{e}_8] 1 \underset{[\mathfrak{su}_3]}{\overset{\mathfrak{su}_{9}}{2}}
         \underset{[\mathfrak{su}_3]}{\overset{\mathfrak{su}_{15}}{2}}
         \underset{[\mathfrak{su}_3]}{\overset{\mathfrak{su}_{18}}{2}}
         \underbrace{\overset{\mathfrak{su}_{18}}{2}
         \cdots \overset{\mathfrak{su}_{18}}{2}}_{N-1} [\mathfrak{su}_{18}] \,.
\end{equation}
We label each theory in the nilpotent network via
\begin{equation}
    \mathcal{T}_3^{(N)}(p, s, 3q; \mu_{\text{4d}}) \,,
\end{equation}
where $\mu_{\text{4d}} = [1^{18}]$ corresponds to the $\mathcal{T}_3^{(N)}(p, s, 3q)$ SCFT discussed in Section \ref{sec:sfolds}.

We first discuss the nilpotent network formed by the $\mathbb{Z}_3$ center-flavor preserving nilpotent deformations of the $\mathfrak{su}_{18}$ flavor symmetry. As discussed, these are specified by partitions of eighteen such that each exponent is a multiple of three. Exhaustively, there are eleven such partitions:
\begin{equation}
    [1^{18}],\, [2^3, 1^{12}],\, [2^6, 1^6],\, [2^9],\, [3^3, 1^9],\, [3^3, 2^3, 1^3],\, [3^6],\, [4^3, 1^6],\, [4^3, 2^3],\, [5^3, 1^3],\, [6^3] \,.
\end{equation}
The nilpotent network/Higgs branch flows amongst the generated 6d SCFTs is depicted in Figure \ref{fig:sfold_nilpotent}. Similarly, one can consider the network formed by performing nilpotent Higgsing of the $\mathfrak{su}_6$ flavor algebra belonging to the 4d $\mathcal{T}_3^{(N)}(p=1, s=1, 3q=3)$ SCFT. These are described by partitions of six, and the generated nilpotent hierarchy of these 4d theories is shown in Figure \ref{fig:sfold_nilpotent_4d}. The nilpotent Higgsing in 6d and 4d commutes when combined with the $\mathbb{Z}_3$ Stiefel--Whitney twisted compactification of the theories in Figure \ref{fig:sfold_nilpotent} to the theories in Figure \ref{fig:sfold_nilpotent_4d}.

\begin{figure}
  \centering
\begin{tikzpicture}[node distance=1.8cm]

\node (0) [startstop, xshift=-1cm] {
$
\begin{gathered}
1\overset{\mathfrak{su}_{9}}{2}\overset{\mathfrak{su}_{15}}{2}\overset{\mathfrak{su}_{18}}{2}\cdots\overset{\mathfrak{su}_{18}}{2}\overset{\mathfrak{su}_{18}}{2}\overset{\mathfrak{su}_{18}}{2}\overset{\mathfrak{su}_{18}}{2}\overset{\mathfrak{su}_{18}}{2}\overset{\mathfrak{su}_{18}}{2}\end{gathered}:\ [1^{18}]
$};

\node (1) [startstop, below of=0] {
$
\begin{gathered}
1\overset{\mathfrak{su}_{9}}{2}\overset{\mathfrak{su}_{15}}{2}\overset{\mathfrak{su}_{18}}{2}\cdots\overset{\mathfrak{su}_{18}}{2}\overset{\mathfrak{su}_{18}}{2}\overset{\mathfrak{su}_{18}}{2}\overset{\mathfrak{su}_{18}}{2}\overset{\mathfrak{su}_{18}}{2}\overset{\mathfrak{su}_{15}}{2}\end{gathered}:\ [2^3, 1^{12}]
$};

\node (2) [startstop, below of=1] {
$
\begin{gathered}
1\overset{\mathfrak{su}_{9}}{2}\overset{\mathfrak{su}_{15}}{2}\overset{\mathfrak{su}_{18}}{2}\cdots\overset{\mathfrak{su}_{18}}{2}\overset{\mathfrak{su}_{18}}{2}\overset{\mathfrak{su}_{18}}{2}\overset{\mathfrak{su}_{18}}{2}\overset{\mathfrak{su}_{18}}{2}\overset{\mathfrak{su}_{12}}{2}\end{gathered}:\ [2^6, 1^6]
$};

\node (3) [startstop, below of=2, xshift=-5cm] {
$
\begin{gathered}
1\overset{\mathfrak{su}_{9}}{2}\overset{\mathfrak{su}_{15}}{2}\overset{\mathfrak{su}_{18}}{2}\cdots\overset{\mathfrak{su}_{18}}{2}\overset{\mathfrak{su}_{18}}{2}\overset{\mathfrak{su}_{18}}{2}\overset{\mathfrak{su}_{18}}{2}\overset{\mathfrak{su}_{18}}{2}\overset{\mathfrak{su}_{9}}{2}\end{gathered}:\ [2^9]
$};

\node (4) [startstop, below of=3, xshift=8cm] {
$
\begin{gathered}
1\overset{\mathfrak{su}_{9}}{2}\overset{\mathfrak{su}_{15}}{2}\overset{\mathfrak{su}_{18}}{2}\cdots\overset{\mathfrak{su}_{18}}{2}\overset{\mathfrak{su}_{18}}{2}\overset{\mathfrak{su}_{18}}{2}\overset{\mathfrak{su}_{18}}{2}\overset{\mathfrak{su}_{15}}{2}\overset{\mathfrak{su}_{12}}{2}\end{gathered}:\ [3^3, 1^9]
$};

\node (5) [startstop, below of=4, xshift=-5cm] {
$
\begin{gathered}
1\overset{\mathfrak{su}_{9}}{2}\overset{\mathfrak{su}_{15}}{2}\overset{\mathfrak{su}_{18}}{2}\cdots\overset{\mathfrak{su}_{18}}{2}\overset{\mathfrak{su}_{18}}{2}\overset{\mathfrak{su}_{18}}{2}\overset{\mathfrak{su}_{18}}{2}\overset{\mathfrak{su}_{15}}{2}\overset{\mathfrak{su}_{9}}{2}\end{gathered}:\ [3^3, 2^3, 1^3]
$};

\node (6) [startstop, below of=5, xshift=-3cm] {
$
\begin{gathered}
1\overset{\mathfrak{su}_{9}}{2}\overset{\mathfrak{su}_{15}}{2}\overset{\mathfrak{su}_{18}}{2}\cdots\overset{\mathfrak{su}_{18}}{2}\overset{\mathfrak{su}_{18}}{2}\overset{\mathfrak{su}_{18}}{2}\overset{\mathfrak{su}_{18}}{2}\overset{\mathfrak{su}_{12}}{2}\overset{\mathfrak{su}_{6}}{2}\end{gathered}:\ [3^6]
$};

\node (7) [startstop, below of=6, xshift=8cm] {
$
\begin{gathered}
1\overset{\mathfrak{su}_{9}}{2}\overset{\mathfrak{su}_{15}}{2}\overset{\mathfrak{su}_{18}}{2}\cdots\overset{\mathfrak{su}_{18}}{2}\overset{\mathfrak{su}_{18}}{2}\overset{\mathfrak{su}_{18}}{2}\overset{\mathfrak{su}_{15}}{2}\overset{\mathfrak{su}_{12}}{2}\overset{\mathfrak{su}_{9}}{2}\end{gathered}:\ [4^3, 1^6]
$};

\node (8) [startstop, below of=7, xshift=-5cm] {
$
\begin{gathered}
1\overset{\mathfrak{su}_{9}}{2}\overset{\mathfrak{su}_{15}}{2}\overset{\mathfrak{su}_{18}}{2}\cdots\overset{\mathfrak{su}_{18}}{2}\overset{\mathfrak{su}_{18}}{2}\overset{\mathfrak{su}_{18}}{2}\overset{\mathfrak{su}_{15}}{2}\overset{\mathfrak{su}_{12}}{2}\overset{\mathfrak{su}_{6}}{2}\end{gathered}:\ [4^3, 2^3]
$};

\node (9) [startstop, below of=8, xshift=2cm] {
$
\begin{gathered}
1\overset{\mathfrak{su}_{9}}{2}\overset{\mathfrak{su}_{15}}{2}\overset{\mathfrak{su}_{18}}{2}\cdots\overset{\mathfrak{su}_{18}}{2}\overset{\mathfrak{su}_{18}}{2}\overset{\mathfrak{su}_{15}}{2}\overset{\mathfrak{su}_{12}}{2}\overset{\mathfrak{su}_{9}}{2}\overset{\mathfrak{su}_{6}}{2}\end{gathered}:\ [5^3, 1^3]
$};

\node (10) [startstop, below of=9] {
$
\begin{gathered}
1\overset{\mathfrak{su}_{9}}{2}\overset{\mathfrak{su}_{15}}{2}\overset{\mathfrak{su}_{18}}{2}\cdots\overset{\mathfrak{su}_{18}}{2}\overset{\mathfrak{su}_{15}}{2}\overset{\mathfrak{su}_{12}}{2}\overset{\mathfrak{su}_{9}}{2}\overset{\mathfrak{su}_{6}}{2}\overset{\mathfrak{su}_{3}}{2}\end{gathered}:\ [6^3]
$};

\draw [arrow] (0) -- (1);
\draw [arrow] (1) -- (2);
\draw [arrow] (2) -- (3);
\draw [arrow] (2) -- (4);
\draw [arrow] (3) -- (5);
\draw [arrow] (4) -- (5);
\draw [arrow] (5) -- (6);
\draw [arrow] (5) -- (7);
\draw [arrow] (6) -- (8);
\draw [arrow] (7) -- (8);
\draw [arrow] (8) -- (9);
\draw [arrow] (9) -- (10);

\end{tikzpicture}
\caption{The subsector of the nilpotent hierarchy of the 6d SCFT in equation \eqref{eq:6d_trivialOrbit} in which each theory enjoys a $\mathbb{Z}_3$ center-flavor symmetry. We listed the quiver description of the tensor branch and the partition defining the nilpotent orbit in each case. The $\mathbb{Z}_3$ Stiefel--Whitney twisted torus compactification of the theories appearing here gives rise to the 4d $\mathcal{N}=2$ SCFTs whose nilpotent network is depicted in Figure \ref{fig:sfold_nilpotent_4d}.}
\label{fig:sfold_nilpotent}
\end{figure}

\begin{figure}
\centering
\begin{tikzpicture}[node distance=1.8cm]
\node (0) [startstop, xshift=-1cm] {
$
\mathcal{T}_3^{(N)}(1, 1, 3; \mu_{4d} = [1^6])
$};

\node (1) [startstop, below of=0] {
$
\mathcal{T}_3^{(N)}(1, 1, 3; \mu_{4d} = [2, 1^4])
$};

\node (2) [startstop, below of=1] {
$
\mathcal{T}_3^{(N)}(1, 1, 3; \mu_{4d} = [2^2, 1^2])
$};

\node (3) [startstop, below of=2, xshift=-3cm] {
$
\mathcal{T}_3^{(N)}(1, 1, 3; \mu_{4d} = [2^3])
$};

\node (4) [startstop, below of=3, xshift=6cm] {
$
\mathcal{T}_3^{(N)}(1, 1, 3; \mu_{4d} = [3, 1^3])
$};

\node (5) [startstop, below of=4, xshift=-3cm] {
$
\mathcal{T}_3^{(N)}(1, 1, 3; \mu_{4d} = [3, 2, 1])
$};

\node (6) [startstop, below of=5, xshift=-3cm] {
$
\mathcal{T}_3^{(N)}(1, 1, 3; \mu_{4d} = [3^2])
$};

\node (7) [startstop, below of=6, xshift=6cm] {
$
\mathcal{T}_3^{(N)}(1, 1, 3; \mu_{4d} = [4, 1, 1])
$};

\node (8) [startstop, below of=7, xshift=-3cm] {
$
\mathcal{T}_3^{(N)}(1, 1, 3; \mu_{4d} = [4, 2])
$};

\node (9) [startstop, below of=8, xshift=0cm] {
$
\mathcal{T}_3^{(N)}(1, 1, 3; \mu_{4d} = [5, 1])
$};

\node (10) [startstop, below of=9] {
$
\mathcal{T}_3^{(N)}(1, 1, 3; \mu_{4d} = [6])
$};

\draw [arrow] (0) -- (1);
\draw [arrow] (1) -- (2);
\draw [arrow] (2) -- (3);
\draw [arrow] (2) -- (4);
\draw [arrow] (3) -- (5);
\draw [arrow] (4) -- (5);
\draw [arrow] (5) -- (6);
\draw [arrow] (5) -- (7);
\draw [arrow] (6) -- (8);
\draw [arrow] (7) -- (8);
\draw [arrow] (8) -- (9);
\draw [arrow] (9) -- (10);

\end{tikzpicture}
\caption{The nilpotent hierarchy of 4d $\mathcal{N}=2$ SCFTs obtained from nilpotent deformations breaking the $\mathfrak{su}_6$ flavor symmetry of $\mathcal{T}_3^{(N)}(1,1,3)$. The structure of the network matches that of the 6d SCFTs in Figure \ref{fig:sfold_nilpotent}, and the $\mathbb{Z}_3$ Stiefel--Whitney twisted compactifications of each of those 6d SCFTs yields the associated 4d SCFT in this figure.}
\label{fig:sfold_nilpotent_4d}
\end{figure}

\subsection{Exceptional SW-folds and Nilpotent Higgsing}

In Section \ref{sec:Esfolds}, we considered a generalization of the SW-fold theories discussed in Section \ref{sec:sfolds} to those obtained from the rank $N$ orbi-instanton theory of type $(\mathfrak{e}_8, \mathfrak{g})$, where $\mathfrak{g}$ is an algebra of type DE. In particular, in Section \ref{sec:e6sfolds}, we showed that there exist seven homomorphisms $\Gamma_{\mathfrak{e}_6} \rightarrow E_8$ such that Higgsing the $\mathfrak{e}_8$ flavor symmetry of the orbi-instanton leads to a theory with $\mathbb{Z}_3$ center-flavor symmetry.  The 6d SCFTs obtained by such Higgsing retain the $\mathfrak{e}_6$ flavor symmetry on the right of the tensor branch quiver.

\begin{table}[t]
    \centering
    \begin{threeparttable}
    \begin{tabular}{cclcc}
        \toprule
        Bala--Carter Label & Weighted Dynkin Diagram & \multicolumn{1}{c}{6d Quiver} & $\mathfrak{f}$ & $\widetilde{\mathfrak{f}}$ \\\midrule
        $0$ & $\begin{gathered}00\overset{\displaystyle 0}{0}00\end{gathered}$ & $\begin{gathered} \cdots\overset{\mathfrak{e}_6}{6}1\overset{\mathfrak{su}_3}{3}1\overset{\mathfrak{e}_6}{6}1\overset{\mathfrak{su}_3}{3}1 \end{gathered}$ & $\mathfrak{e}_6$ & $\mathfrak{g}_2$ \\
        $A_1$ &  $\begin{gathered}00\overset{\displaystyle 1}{0}00\end{gathered}$ & $\begin{gathered} \cdots\overset{\mathfrak{e}_6}{6}1\overset{\mathfrak{su}_3}{3}1\overset{\mathfrak{e}_6}{6}1\overset{\mathfrak{su}_3}{2} \end{gathered}$ & $\mathfrak{su}_6$ & $\mathfrak{su}_2$ \\
        $3A_1$ &  $\begin{gathered}00\overset{\displaystyle 0}{1}00\end{gathered}$ & $\begin{gathered} \cdots\overset{\mathfrak{e}_6}{6}1\overset{\mathfrak{su}_3}{3}1\overset{\mathfrak{e}_6}{6}12 \end{gathered}$ & $\mathfrak{su}_3 \oplus \mathfrak{su}_2$ & $\mathfrak{su}_2$ \\
        $A_2$ & $\begin{gathered}00\overset{\displaystyle 2}{0}00\end{gathered}$ & $\begin{gathered} \cdots\overset{\mathfrak{e}_6}{6}1\overset{\mathfrak{su}_3}{3}1\underset{\displaystyle 1}{\overset{\mathfrak{e}_6}{6}}1 \end{gathered}$ &  $\mathfrak{su}_3 \oplus \mathfrak{su}_3$ & $\varnothing$ \\
        $D_4$ & $\begin{gathered}00\overset{\displaystyle 2}{2}00\end{gathered}$ & $\begin{gathered} \cdots\underset{\displaystyle 1}{\overset{\mathfrak{e}_6}{6}}1\overset{\mathfrak{su}_3}{3} \end{gathered}$ & $\mathfrak{su}_3$ & $\varnothing$ \\\bottomrule
    \end{tabular}
    \end{threeparttable}
    \caption{The nilpotent orbits of $\mathfrak{e}_6$ that are consistent with a $\mathbb{Z}_3$ center-flavor symmetry. The column labelled $\mathfrak{f}$, we write the subalgebra of the $\mathfrak{e}_6$ flavor symmetry that survives the nilpotent Higgsing, and in the $\widetilde{\mathfrak{f}}$ column, we write the remnant algebra after the $\mathbb{Z}_3$ Stiefel--Whitney twisted compactification down to 4d.}
    \label{tbl:e6nilp}
\end{table}

As in the spirit of this Appendix, the $\mathfrak{e}_6$ flavor symmetry can be Higgsed by a choice of nilpotent orbit of $\mathfrak{e}_6$, and if that nilpotent orbit is compatible then the $\mathbb{Z}_3$ center-flavor symmetry can be preserved. There are only five such nilpotent orbits, which we have listed in Table \ref{tbl:e6nilp}.\footnote{We label the exceptional nilpotent orbits using the Bala--Carter notation \cite{MR417306,MR417307}; see \cite{MR1251060} for the standard reference on nilpotent orbits, and \cite{Chacaltana:2012zy} for a useful summary for the exceptional Lie algebras from the perspective of nilpotent Higgsing.} Similarly to the case where $\mathfrak{g}$ is a special unitary algebra, the $\mathfrak{e}_6$ nilpotent orbits that are compatible with the $\mathbb{Z}_3$ center-flavor symmetry are in one-to-one correspondence with the nilpotent orbits of $\mathfrak{g}_2$. Again, the 4d $\mathcal{N}=2$ SCFT obtained by the operation of nilpotent Higgsing of the $\mathfrak{e}_6$ flavor symmetry and then compactifying with $\mathbb{Z}_3$ Stiefel--Whitney twist can alternatively be obtained by first performing the $\mathbb{Z}_3$ Stiefel--Whitney twisted compactification and then Higgsing the $\mathfrak{g}_2$ flavor symmetry by the appropriate nilpotent orbit.

\begin{table}[p!]
    \centering
    \renewcommand{\arraystretch}{1.6}
    \begin{threeparttable}
    \begin{tabular}{cclc}
        \toprule
        $\mathfrak{e}_7$ Orbit & $\mathfrak{f}_4$ Orbit & \multicolumn{1}{c}{6d Quiver} & $\mathfrak{f}$, $\widetilde{\mathfrak{f}}$ \\\midrule
        $0$ & $0$ &
        $\begin{gathered} \cdots\overset{\mathfrak{e}_7}{8}1\overset{\mathfrak{su}_2}{2}\overset{\mathfrak{so}_7}{3}\overset{\mathfrak{su}_2}{2}1\overset{\mathfrak{e}_7}{8}1\overset{\mathfrak{su}_2}{2}\overset{\mathfrak{so}_7}{3}\overset{\mathfrak{su}_2}{2}1\overset{\mathfrak{e}_7}{8}1\overset{\mathfrak{su}_2}{2}\overset{\mathfrak{so}_7}{3}\overset{\mathfrak{su}_2}{2}1 \end{gathered}$ &
        $\mathfrak{e}_7 \rightarrow \mathfrak{f}_4 \oplus \cancel{\mathfrak{su}_2}$ \\

        $A_1$ & $A_1$ &
        $\begin{gathered} \cdots\overset{\mathfrak{e}_7}{8}1\overset{\mathfrak{su}_2}{2}\overset{\mathfrak{so}_7}{3}\overset{\mathfrak{su}_2}{2}1\overset{\mathfrak{e}_7}{8}1\overset{\mathfrak{su}_2}{2}\overset{\mathfrak{so}_7}{3}\overset{\mathfrak{su}_2}{2}1\overset{\mathfrak{e}_7}{8}1\overset{\mathfrak{su}_2}{2}\overset{\mathfrak{so}_7}{3}\overset{\mathfrak{su}_2}{1} \end{gathered}$ &
        $\mathfrak{so}_{12} \rightarrow \mathfrak{sp}_3 \oplus \cancel{\mathfrak{su}_2}$ \\

        $2A_1$ & $\widetilde{A}_1$ &
        $\begin{gathered} \cdots\overset{\mathfrak{e}_7}{8}1\overset{\mathfrak{su}_2}{2}\overset{\mathfrak{so}_7}{3}\overset{\mathfrak{su}_2}{2}1\overset{\mathfrak{e}_7}{8}1\overset{\mathfrak{su}_2}{2}\overset{\mathfrak{so}_7}{3}\overset{\mathfrak{su}_2}{2}1\overset{\mathfrak{e}_7}{8}1\overset{\mathfrak{su}_2}{2}\overset{\mathfrak{so}_7}{3}1 \end{gathered}$ &
        $(\mathfrak{so}_9 \rightarrow \mathfrak{su}_4 \oplus \cancel{\mathfrak{su}_2}) \oplus \cancel{\mathfrak{su}_2}$  \\

        $3A_1^\prime$ & $A_1 + \widetilde{A}_1$ &
        $\begin{gathered} \cdots\overset{\mathfrak{e}_7}{8}1\overset{\mathfrak{su}_2}{2}\overset{\mathfrak{so}_7}{3}\overset{\mathfrak{su}_2}{2}1\overset{\mathfrak{e}_7}{8}1\overset{\mathfrak{su}_2}{2}\overset{\mathfrak{so}_7}{3}\overset{\mathfrak{su}_2}{2}1\overset{\mathfrak{e}_7}{8}1\overset{\mathfrak{su}_2}{2}\overset{\mathfrak{so}_7}{2} \end{gathered}$ &
        $(\mathfrak{sp}_3\rightarrow \mathfrak{su}_2 \oplus \cancel{\mathfrak{su}_2}) \oplus \mathfrak{su}_2$ \\

        {\color{red}$A_2$} & {\color{red}$A_2$ and $\widetilde{A}_2$} &
        {\color{red}$\begin{gathered} \cdots\overset{\mathfrak{e}_7}{8}1\overset{\mathfrak{su}_2}{2}\overset{\mathfrak{so}_7}{3}\overset{\mathfrak{su}_2}{2}1\overset{\mathfrak{e}_7}{8}1\overset{\mathfrak{su}_2}{2}\overset{\mathfrak{so}_7}{3}\overset{\mathfrak{su}_2}{2}1\overset{\mathfrak{e}_7}{8}1\overset{\mathfrak{su}_2}{2}\overset{\mathfrak{su}_4}{2} \end{gathered}$} &
        {\color{red}$\mathfrak{su}_6$} \\

        $A_2+2A_1$ & $A_2 + \widetilde{A}_1$ &
        $\begin{gathered} \cdots\overset{\mathfrak{e}_7}{8}1\overset{\mathfrak{su}_2}{2}\overset{\mathfrak{so}_7}{3}\overset{\mathfrak{su}_2}{2}1\overset{\mathfrak{e}_7}{8}1\overset{\mathfrak{su}_2}{2}\overset{\mathfrak{so}_7}{3}\overset{\mathfrak{su}_2}{2}1\overset{\mathfrak{e}_7}{8}1\overset{\mathfrak{su}_2}{2}\overset{\mathfrak{su}_2}{2} \end{gathered}$ &
        $\mathfrak{su}_2 \oplus \cancel{\mathfrak{su}_2} \oplus \cancel{\mathfrak{su}_2}$ \\

        $A_3$ & $B_2$ &
        $\begin{gathered} \cdots\overset{\mathfrak{e}_7}{8}1\overset{\mathfrak{su}_2}{2}\overset{\mathfrak{so}_7}{3}\overset{\mathfrak{su}_2}{2}1\overset{\mathfrak{e}_7}{8}1\overset{\mathfrak{su}_2}{2}\overset{\mathfrak{so}_7}{3}\overset{\mathfrak{su}_2}{2}1\underset{\displaystyle 1}{\overset{\mathfrak{e}_7}{8}}1\overset{\mathfrak{su}_2}{2} \end{gathered}$ &
        $(\mathfrak{so}_7\rightarrow \mathfrak{su}_2 \oplus \mathfrak{su}_2 \oplus \cancel{\mathfrak{su}_2}) \oplus \cancel{\mathfrak{su}_2}$ \\

        $2A_2+A_1$ & $\widetilde{A}_2 + A_1$ &
        $\begin{gathered} \cdots\overset{\mathfrak{e}_7}{8}1\overset{\mathfrak{su}_2}{2}\overset{\mathfrak{so}_7}{3}\overset{\mathfrak{su}_2}{2}1\overset{\mathfrak{e}_7}{8}1\overset{\mathfrak{su}_2}{2}\overset{\mathfrak{so}_7}{3}\overset{\mathfrak{su}_2}{2}1\overset{\mathfrak{e}_7}{8}122 \end{gathered}$ &
        $\mathfrak{su}_2 \oplus \cancel{\mathfrak{su}_2}$ \\

        $(A_3+A_1)^\prime$ & $C_3(a_1)$ &
        $\begin{gathered} \cdots\overset{\mathfrak{e}_7}{8}1\overset{\mathfrak{su}_2}{2}\overset{\mathfrak{so}_7}{3}\overset{\mathfrak{su}_2}{2}1\overset{\mathfrak{e}_7}{8}1\overset{\mathfrak{su}_2}{2}\overset{\mathfrak{so}_7}{3}\overset{\mathfrak{su}_2}{2}1\underset{\displaystyle 1}{\overset{\mathfrak{e}_7}{8}}12 \end{gathered}$ &
        $\mathfrak{su}_2 \oplus \cancel{\mathfrak{su}_2} \oplus \cancel{\mathfrak{su}_2}$ \\

        $D_4(a_1)$ & $F_4(a_3)$ &
        $\begin{gathered} \cdots\overset{\mathfrak{e}_7}{8}1\overset{\mathfrak{su}_2}{2}\overset{\mathfrak{so}_7}{3}\overset{\mathfrak{su}_2}{2}1\overset{\mathfrak{e}_7}{8}1\overset{\mathfrak{su}_2}{2}\overset{\mathfrak{so}_7}{3}\overset{\mathfrak{su}_2}{2}1\overset{\displaystyle 1}{\underset{\displaystyle 1}{\overset{\mathfrak{e}_7}{8}}}1 \end{gathered}$ &
        $\cancel{\mathfrak{su}_2} \oplus \cancel{\mathfrak{su}_2} \oplus \cancel{\mathfrak{su}_2}$ \\

        $A_5^\prime$ & $B_3$ &
        $\begin{gathered} \cdots\overset{\mathfrak{e}_7}{8}1\overset{\mathfrak{su}_2}{2}\overset{\mathfrak{so}_7}{3}\overset{\mathfrak{su}_2}{2}1\overset{\mathfrak{e}_7}{8}1\overset{\mathfrak{su}_2}{2}\overset{\mathfrak{so}_7}{3}1\overset{\mathfrak{so}_9}{4} \end{gathered}$ &
        $\mathfrak{su}_2 \oplus \cancel{\mathfrak{su}_2}$ \\

        $D_4$ & $C_3$ &
        $\begin{gathered} \cdots\overset{\mathfrak{e}_7}{8}1\overset{\mathfrak{su}_2}{2}\overset{\mathfrak{so}_7}{3}\overset{\mathfrak{su}_2}{2}1\overset{\mathfrak{e}_7}{8}1\overset{\mathfrak{su}_2}{2}\overset{\mathfrak{so}_7}{3}\overset{\mathfrak{su}_2}{1}\overset{\mathfrak{so}_{12}}{4} \end{gathered}$ &
        $\mathfrak{sp}_3 \rightarrow \mathfrak{su}_2 \oplus \cancel{\mathfrak{su}_2}$ \\

        $E_6(a_3)$ & $F_4(a_2)$ &
        $\begin{gathered} \cdots\overset{\mathfrak{e}_7}{8}1\overset{\mathfrak{su}_2}{2}\overset{\mathfrak{so}_7}{3}\overset{\mathfrak{su}_2}{2}1\overset{\mathfrak{e}_7}{8}1\overset{\mathfrak{su}_2}{2}\overset{\mathfrak{so}_7}{3}1\overset{\mathfrak{so}_8}{4} \end{gathered}$ &
        $\cancel{\mathfrak{su}_2}$ \\

        $D_5$ & $F_4(a_1)$ &
        $\begin{gathered} \cdots\overset{\mathfrak{e}_7}{8}1\overset{\mathfrak{su}_2}{2}\overset{\mathfrak{so}_7}{3}\overset{\mathfrak{su}_2}{2}1\underset{\displaystyle 1}{\overset{\mathfrak{e}_7}{8}}1\overset{\mathfrak{su}_2}{2}\overset{\mathfrak{so}_7}{3} \end{gathered}$ &
        $\cancel{\mathfrak{su}_2} \oplus \cancel{\mathfrak{su}_2}$ \\

        $E_6$ & $F_4$ &
        $\begin{gathered} \cdots\underset{\displaystyle 1}{\overset{\mathfrak{e}_7}{8}}1\overset{\mathfrak{su}_2}{2}\overset{\mathfrak{so}_7}{3}\overset{\mathfrak{su}_2}{2} \end{gathered}$ &
        $\cancel{\mathfrak{su}_2}$ \\\bottomrule
    \end{tabular}
    \end{threeparttable}
    \caption{The nilpotent orbits of $\mathfrak{e}_7$ that are consistent with a $\mathbb{Z}_2$ center-flavor symmetry. In the $\mathfrak{f}$, $\widetilde{\mathfrak{f}}$ column we write the flavor symmetry in 6d, and the remnant subalgebra in 4d after the SW-twisted compactification. Scored-out algebras are removed by the SW-twist. The remnant algebra, $\widetilde{\mathfrak{f}}$ matches the flavor symmetry associated to the $\mathfrak{f}_4$ nilpotent orbit. The {\color{red} red} line is exceptional, and is discussed in the text.}
    \label{tbl:e7nilp}
\end{table}

Further, we consider the case where $\mathfrak{g} = \mathfrak{e}_7$. There are fifteen nilpotent orbits of $\mathfrak{e}_7$ that are compatible with a $\mathbb{Z}_2$ center flavor symmetry, as depicted in Table \ref{tbl:e7nilp}. The sub-Hasse diagram formed by the subset of $\mathfrak{e}_7$ nilpotent orbits appears here almost matches the Hasse diagram for $\mathfrak{f}_4$ nilpotent orbits that appears in \cite{Hanany:2017ooe}. Similarly, the flavor symmetries surviving after the Stiefel--Whitney twist and those of the $\mathfrak{f}_4$ nilpotent orbits almost always match. The one subtlety is the line denoted in {\color{red}red} in Table \ref{tbl:e7nilp}; this appears to be associated to one $\mathfrak{e}_7$ nilpotent orbit, but \emph{two} $\mathfrak{f}_4$ nilpotent orbits. This case involves the $\mathbb{Z}_2$ Stiefel--Whitney twist of a 6d theory containing an $\overset{\mathfrak{su}_4}{2}$ factor. As we discussed at the conclusion of Section \ref{sec:classs}, this leads to curious features, similar to those that occur in six dimensions when one has $\overset{\mathfrak{su}_2}{2}$ \cite{Morrison:2016djb}. We expect that a deeper understanding of these highly special configurations will lead to the resolution of this subtlety in the $\mathfrak{e}_7$ and $\mathfrak{f}_4$ nilpotent orbits, however, we leave such a study for future work.

One can use the same methods from the 6d perspective to determine the central charges, flavor symmetries, Coulomb branch operator dimensions and so forth of the 4d $\mathcal{N}=2$ SCFTs obtained from the torus compactification. A similar analysis can be carried out when $\mathfrak{g} = \mathfrak{so}_{2k}$, however we leave such an enumeration to the reader.\footnote{For $\mathfrak{g} = \mathfrak{so}_{2k}$ there is a subtlety with the fact that two distinct SCFTs, obtained from very even nilpotent Higgsing, are associated to the same tensor branch description \cite{Distler:2022yse}.}

\section{Flavor Group for Conformal Matter \& Deformations}\label{app:CM}

In Section \ref{sec:6d}, we expanded upon the proposal to determine the global structure of the global symmetry group of a 6d $(1,0)$ SCFT that was put forth in \cite{Apruzzi:2020zot}. This proposal is based on the weakly-coupled spectrum of the effective theory that lives on the generic point of the tensor branch, together with the knowledge of the Green--Schwarz couplings. In Section \ref{sec:e8CF}, we applied this prescription to the 6d SCFTs obtained from the Higgsing of the $\mathfrak{e}_8$ flavor symmetry, by a choice of $E_8$-homomorphism $\rho: \mathbb{Z}_K \rightarrow E_8$, of the rank $N$ $(\mathfrak{e}_8, \mathfrak{su}_K)$ orbi-instanton; we found that the resulting global structure of the non-Abelian part of the flavor symmetry was encoded in a simple manner in the choice of $\rho$. We highlighted an extension of this analysis to the 6d $(1,0)$ SCFTs obtained via the $\mathfrak{e}_8$ Higgsing of the rank $N$ $(\mathfrak{e}_8, \mathfrak{g})$ orbi-instanton in Section \ref{sec:Esfolds}. Furthermore, in Appendix \ref{app:nilp}, we demonstrated that, when we consider the Higgsing of the orbi-instanton by both an $E_8$-homomorphism, $\rho$, and a nilpotent orbit of $\mathfrak{g}$, $\sigma$, there is a simple prescription for the non-Abelian center-flavor symmetry of the resulting SCFT in terms of $\rho$ and $\sigma$.

Throughout this chapter, we have focused on the 6d $(1,0)$ SCFTs known as the rank $N$ $(\mathfrak{e}_8, \mathfrak{g})$ orbi-instantons and the theories further obtained via Higgs branch renormalization group flows. Another broad class of 6d $(1,0)$ SCFTs are those commonly referred to as the ``Higgsable to $(2,0)$ of type $A_{N-1}$'' SCFTs \cite{Ohmori:2015pia}. These include the rank $N$ $(\mathfrak{g}, \mathfrak{g})$ conformal matter theories \cite{DelZotto:2014hpa}, corresponding to the worldvolume theory on a stack of $N$ M5-branes probing a $\mathbb{C}^2/\Gamma_\mathfrak{g}$ orbifold singularity, and the theories obtained via Higgsing of the $\mathfrak{g} \oplus \mathfrak{g}$ flavor symmetry of the conformal matter theory by a pair of nilpotent orbits $\sigma_L$ and $\sigma_R$ of $\mathfrak{g}$. The former have frequently been referred to in the literature as $\mathcal{T}_{\mathfrak{g},N}$, and the latter as $\mathcal{T}_{\mathfrak{g},N}(\sigma_L, \sigma_R)$.\footnote{We emphasize that these 6d SCFTs are distinct from the 4d SW-fold theories also labelled by $\mathcal{T}$ in Section \ref{sec:sfolds}.} In Section \ref{sec:cfanom}, we determined that the non-Abelian flavor group of rank $N$ $(\mathfrak{su}_K, \mathfrak{su}_K)$ conformal matter is
\begin{equation}
    (SU(K) \times SU(K))/\mathbb{Z}_K \,.
\end{equation}
It is straightforward to see, again from the methods presented in Section \ref{sec:6d}, that when $\mathfrak{su}_K$ is generalized to an arbitrary ADE Lie algebra $\mathfrak{g}$, the non-Abelian flavor group is
\begin{equation}
    (\widetilde{G} \times \widetilde{G})/Z(\widetilde{G}) \,,
\end{equation}
where $\widetilde{G}$ is the simply-connected group with Lie algebra $\mathfrak{g}$, and $Z(\widetilde{G})$ is the center of $\widetilde{G}$.

Now we turn to the determination of the global structure of the non-Abelian flavor symmetry after Higgsing via the pair of nilpotent orbits $(\sigma_L, \sigma_R)$. The nilpotent orbit $\sigma_L$ breaks the left $\mathfrak{g}$ flavor algebra to the semi-simple algebra $\mathfrak{h}_L$,\footnote{We ignore $\mathfrak{u}(1)$ factors in this Appendix.} and we let $\widetilde{H}_L$ denote the associated simply-connected Lie group; and similarly for the Higgsing of the right $\mathfrak{g}$ by $\sigma_R$. Both $\mathfrak{h}_L$ and $\mathfrak{h}_R$ can be read off directly from the nilpotent orbits \cite{Heckman:2016ssk}. Similarly, it is well-known how each nilpotent Higgsing modifies the tensor branch description, and thus one can use the analysis of Section \ref{sec:6d} to determine the subgroup of the center $Z(\widetilde{G})$ that is preserved after Higgsing; we refer to these subgroups as $Z_L(\widetilde{G})$ and $Z_R(\widetilde{G})$ for $\sigma_L$ and $\sigma_R$, respectively. For $\widetilde{G} = SU(K)$, $E_6$, and $E_7$ these subgroups have been discussed in Appendix \ref{app:nilp}.

Putting all this together, the global structure of the non-Abelian flavor group of the Higgsed conformal matter theory, $\mathcal{T}_{\mathfrak{g},N}(\sigma_L, \sigma_R)$, can be shown to be
\begin{equation}\label{eqn:NCMstr}
    (\widetilde{H}_L \times \widetilde{H}_R)/(Z_L(\widetilde{G})\cap Z_R(\widetilde{G})) \,.
\end{equation}
Here the quotient is by the common subgroup of $Z_L(\widetilde{G})$ and $Z_R(\widetilde{G})$ inside $Z(\widetilde{G})$. In particular, if we consider $\widetilde{G} \neq Spin(4K)$, then we have
\begin{equation}
    Z(\widetilde{G}) = \mathbb{Z}_K \,, \qquad Z_L(\widetilde{G}) = \mathbb{Z}_{K_L} \,, \qquad Z_R(\widetilde{G}) = \mathbb{Z}_{K_R} \,,
\end{equation}
for some $K$, and $K_L$, $K_R$ divisors of $K$. The quotient is then by
\begin{equation}
    (Z_L(\widetilde{G})\cap Z_R(\widetilde{G})) = \mathbb{Z}_{\gcd(K_L, K_R)} \,.
\end{equation}
For $\widetilde{G} = Spin(4K)$ it is a little more technical due to the product structure of the center.

We now make this explicit in one example. Consider the rank $N$ $(\mathfrak{su}_{18}, \mathfrak{su}_{18})$ conformal matter theory.\footnote{We assume that $N > 7$, as this guarantees that the specific nilpotent deformations that we turn on, on the left and on the right, do not start to cross-correlate.} As is by now familiar, the tensor branch description is 
\begin{equation}
    \overbrace{\,
    \underset{\displaystyle [\mathfrak{su}_{18}]}{\overset{\mathfrak{su}_{18}}{2}}
    \overset{\mathfrak{su}_{18}}{2}\,
    \cdots\,
    \overset{\mathfrak{su}_{18}}{2}
    \underset{\displaystyle [\mathfrak{su}_{18}]}{\overset{\mathfrak{su}_{18}}{2}}
    \,}^{N - 1 \text{ $(-2)$-curves}} \,.
\end{equation}
We consider the Higgs branch deformations triggered by turning on vacuum expectation values associated to the nilpotent orbits 
\begin{equation}
    \sigma_L = [1^6, 4^3] \,, \qquad \sigma_R = [1^6, 2^6] \,,
\end{equation}
of the $\mathfrak{su}_{18}$ flavor algebras on the left and right. It is clear from the analysis in Appendix \ref{app:nilp} that $\sigma_L$ preserves a $\mathbb{Z}_3$ center-flavor subgroup, as the exponents of the partition are all multiples of three, and similarly, $\sigma_R$ preserves a $\mathbb{Z}_6$ center-flavor subgroup:
\begin{equation}
    Z_L(SU(18)) = \mathbb{Z}_3 \,, \qquad Z_R(SU(18)) = \mathbb{Z}_6 \,.
\end{equation}
After performing the nilpotent Higgsing, the renormalization group flow ends at an interacting 6d $(1,0)$ SCFT with tensor branch description
\begin{equation}
    \overbrace{\,
    \underset{\displaystyle [\mathfrak{su}_6]}{\overset{\mathfrak{su}_9}{2}}\,
    \overset{\mathfrak{su}_{12}}{2}\,
    \overset{\mathfrak{su}_{15}}{2}\,
    \underset{\displaystyle [\mathfrak{su}_3]}{\overset{\mathfrak{su}_{18}}{2}}\,
    \overset{\mathfrak{su}_{18}}{2}\,
    \cdots\,
    \overset{\mathfrak{su}_{18}}{2}\,
    \underset{\displaystyle [\mathfrak{su}_6]}{\overset{\mathfrak{su}_{18}}{2}}\,
    \underset{\displaystyle [\mathfrak{su}_6]}{\overset{\mathfrak{su}_{12}}{2}}
    \,}^{N - 1 \text{ $(-2)$-curves}} \,.
\end{equation}
To determine the global structure of the non-Abelian flavor symmetry, we can apply the procedure described in Section \ref{sec:6d}. From that perspective, we determine that the non-Abelian flavor group is 
\begin{equation}
    (SU(6) \times SU(3) \times SU(6) \times SU(6))/\mathbb{Z}_3 \,,
\end{equation}
as expected from equation \eqref{eqn:NCMstr}.

\chapter{\MakeUppercase{Chapter 8 Appendix}}

\section{Other Realizations of Defects}\label{app:other}

In this chapter we have focused on one particular realization of 4D QFTs via D3-brane probes of 
a transverse geometry. This realization is especially helpful for studying duality defects since the 
corresponding defect is realized in terms of conventional bound states of $(p,q)$ 7-branes wrapped ``at infinity'' in the 
transverse geometry. On the other hand, some structures are more manifest in other duality frames. Additionally, 
there are other ways to engineer 4D QFTs via string constructions as opposed to considering worldvolumes of probe D3-branes.

With this in mind, in this Appendix we discuss how some of the structures considered in this chapter are represented in other top down constructions and how this can be used to obtain further generalizations. We begin by discussing how the defect group of $\mathcal{N} = 4 $ SYM is specified in different top down setups. After this, we discuss how duality defects are constructed in some of these alternative constructions.

\subsection{Defect Groups Revisited}

To begin, we recall that the ``defect group'' of a theory is obtained from the spectrum of heavy defects which are not screened by dynamical objects \cite{DelZotto:2015isa} (see also \cite{Tachikawa:2013hya, GarciaEtxebarria:2019caf, Albertini:2020mdx, Morrison:2020ool}). For example, in the context of a 6D SCFT engineered via F-theory on an elliptically fibered Calabi-Yau threefold $X \rightarrow B$ with base $B = \mathbb{C}^2 / \Gamma_{U(2)}$ a quotient by a finite subgroup of $U(2)$ (see \cite{Heckman:2013pva} and \cite{Heckman:2018jxk} for a review), we get extended surface defects (high tension effective strings) from D3-branes wrapped on non-compact 2-cycles of $\mathbb{C}^2 / \Gamma_{U(2)}$. The charge of these defects is screened by D3-branes wrapped on the collapsing cycles of the singularity. The corresponding quotient (as obtained from the relative homology exact sequence for $\mathbb{C}^2 / \Gamma_{U(2)}$ and its boundary geometry $S^3 / \Gamma_{U(2)}  = \partial B$) is:
\begin{equation}
0 \rightarrow H_{2}(B ) \rightarrow H_{2}(B , \partial{B}) \rightarrow H_{1}(\partial B) \rightarrow 0.
\end{equation}
In particular, the defect group for surface defects is simply given by $H_{1}(S^3 / \Gamma_{U(2)}) = \mathrm{Ab}[\Gamma]$, the abelianization of $\Gamma$.

Turning to the case of interest in this chapter, observe that for the $\mathcal{N} = (2,0)$ theory obtained from compactification of IIB on the A-type singularity $\mathbb{C}^2 / \mathbb{Z}_{N}$, this leads, upon further reduction on a $T^2$, to the 4D $\mathcal{N} =  4$ SYM theory with Lie algebra $\mathfrak{su}(N)$. The surface defects of the 6D theory can be wrapped on 1-cycles of the $T^2$, and this leads to line defects of the 4D theory. In this case, the defect group for lines is just $\mathbb{Z}_N^{\mathrm{(elec)}} \times \mathbb{Z}^{\mathrm{(mag)}}_N$, and a choice of polarization serves to specify the global form of the gauge group. Similar considerations hold for theories with reduced supersymmetry, as obtained from general 6D SCFTs compactified on Riemann surfaces (see \cite{DelZotto:2015isa} for further discussion).

Returning to the case of D3-branes probing a transverse singularity, this would appear to pose a bit of a puzzle for the ``defect group picture''. To see why, observe that for D3-branes probing $\mathbb{C}^3$, the heavy line defects are obtained from F1- and D1-strings which stretch from the D3-brane out to infinity. On the other hand, the boundary topology of the $\partial \mathbb{C}^3 = S^5$ has no torsional homology.\footnote{That being said, provided one just considers the dimensional reduction to the 5D Symmetry TFT, one can still readily detect the electric and magnetic 1-form symmetries (as we did in this chapter).} The physical puzzle, then, is to understand how the defect group is realized in this case.

The answer to this question relies on the fact that the precise notion of ``defect group'' is really specified by the Hilbert space of the string theory background, and quotienting heavy defects by dynamical states of the localized QFT. The general physical point is that in the boundary $S^5$, there is an asymptotic 5-form flux, and in the presence of this flux, the endpoints of the F1- and D1-strings will ``puff up'' to a finite size. The size is tracked by a single integer parameter, and the maximal value of this is precisely $N$.

To see how this flux picture works in more detail, it is helpful to work in a slightly different duality frame. Returning to the realization of $\mathcal{N} = 4$ SYM via type IIB on the background $\mathbb{R}^{3,1} \times T^2 \times \mathbb{C}^2 / \mathbb{Z}_N$, consider T-dualizing the circle fiber of the $\mathbb{C}^2 / \mathbb{Z}_N$ geometry. 
As explained in \cite{Ooguri:1995wj}, we then arrive in a type IIA background with $N$ NS5-branes filling $\mathbb{R}^{3,1} \times T^2$ and sitting at a point of the transverse $\mathbb{C}^2$. There is a dilaton gradient profile in the presence of the NS5-brane, but far away from it, the boundary geometry is simply an $S^3$ threaded by $N$ units of NSNS 3-form flux. In this realization of the 4D QFT, the line operators of interest are obtained from D2-branes which wrap a 1-cycle of the $T^2$ as well as the radial direction of the transverse geometry, ending at ``point'' of the boundary $S^3$ with flux. This appears to have the same puzzle already encountered in the case of the D3-brane realization of the QFT.

Again, the ordinary homology / K-theory for the $S^3$ is not torsional, but the \textit{twisted} K-theory is indeed torsional. Recall that the twisted K-theory for $S^3$ (see \cite{baraglia2015fourier}) involves the choice of a twist class $N \in H^3(S^3) \simeq \mathbb{Z}$, and for this choice of twist class, one gets:
\begin{equation}
K^{\ast}_{H}(S^3) \simeq \mathbb{Z}_{N}, 
\end{equation}
in the obvious notation. From a physical viewpoint, one can also see that the spectrum of ``point-like'' branes in this system is actually more involved. Indeed, the boundary $S^3$ is actually better described as an $SU(2)$ WZW model. Boundary states of the worldsheet CFT correspond to D-branes, and these are in turn characterized by fuzzy points of the geometry.\footnote{We are neglecting the additional boundary states provided by having a supersymmetric WZW model. This leads to additional extended objects / topological symmetry operators. For further discussion on these additional boundary states, see \cite{Maldacena:2001xj, Maldacena:2001ss}.} The interpretation of these boundary conditions can be visualized as ``fuzzy 2-spheres'' (see e.g., \cite{Alekseev:2000fd}), as specified by the non-commutative algebra:
\begin{equation}
[J^{a},J^{b} ] = i \varepsilon^{abc} J^c,
\end{equation}
for $a,b,c = 1,2,3$, namely a representation of $\mathfrak{su}(2)$. The size of the fuzzy 2-sphere is set by $J^{a} J^{a}$, the Casimir of the representation, and this leads to a finite list of admissible representations going from spin $j = 0,...,(N-1)/2$. Beyond this point, the stringy exclusion principle \cite{Maldacena:1998bw} is in operation, and cuts off the size of the fuzzy 2-sphere. The upshot of this is that the single point of ordinary boundary homology has now been supplemented by a whole collection of fuzzy 2-spheres, and these produce the required spectrum of heavy defects which cannot be screened by dynamical objects. Similar considerations clearly apply for topological symmetry operators generated by D4-branes wrapped on a 1-cycle of the $T^2$ and a fuzzy 2-sphere.

With this example in mind, we clearly see that similar considerations will apply in systems where the boundary geometry contains a non-trivial flux. In particular, in the D3-brane realization of $\mathcal{N} = 4$ SYM 
we can expect the F1- and D1-strings used to engineer heavy defects to also ``puff up'' to non-commutative cycles in the boundary $S^5$.

One point we wish to emphasize is that so long as we dimensionally reduce the boundary geometry to reach a lower-dimensional system (as we mainly did in this chapter), then the end result of the flux can also be detected directly in the resulting 5D bulk SymTFT.

\subsection{Duality Defects Revisited}

In the previous subsection we presented a general proposal for how to identify the defect group in 
duality frames where asymptotic flux is present. Now, one of the main reasons we chose to focus on the D3-brane 
realization of our QFTs is that the top down identification of duality defects is relatively straightforward (even if the defect group computation is more subtle). Turning the discussion around, one might also ask how our top down duality defects are realized in other string backgrounds which realize the same QFT. For related discussion on this point, see the recent reference \cite{Bashmakov:2022uek}.

To illustrate, consider the IIB background $\mathbb{R}^{3,1} \times T^2 \times \mathbb{C}^2 / \mathbb{Z}_{N}$, 
in which the 4D QFT is realized purely in terms of geometry. In this case, a duality of the 4D field theory will be specified by a large diffeomorphism of the $T^2$, namely as an $SL(2,\mathbb{Z})$ transformation of the complex structure of the $T^2$.

Because the duality symmetry is now encoded purely in the geometry, the ``brane at infinity'' which implements a topological defect / interface will necessarily be a variation in the asymptotic profile of the 10D metric far from the location of the QFT. Since, however, only the topology of the configuration actually matters, it will be enough to specify how this works at the level of a holomorphic Weierstrass model.

Along these lines, we single out one of the directions $x_{\bot}$ along the $\mathbb{R}^{3,1}$ such that the duality defect / interface will be localized at $x_{\bot} = 0$ in the 4D spacetime. Combining this with the radial direction of $\mathbb{C}^2 / \mathbb{Z}_N$, we get a pair of coordinates which locally fill out a patch of the complex line $\mathbb{C}$. It is helpful to introduce the complex combination:
\begin{equation}
z = x_{\bot} + \frac{i}{r},
\end{equation}
where $r = 0$ and $r = \infty$ respectively indicate the location of the QFT and the conformal boundary, where we reach the $S^3 / \mathbb{Z}_{N}$ lens space. In terms of this local coordinate, we can now introduce a Weierstrass model with the prescribed Kodaira fiber type at $z = 0$, namely $x_{\bot} = 0$ and $r = \infty$. 
For example, a type $IV^{\ast}$ and type $III^{\ast}$ fiber would respectively be written as:
\begin{align}
\mathrm{type} \,\, III^{\ast}: \, & y^2 = x^3 + x z^3 \\
\mathrm{type} \,\, IV^{\ast}: \, & y^2 = x^3 + z^4.
\end{align}
This sort of asymptotic profile geometrizes the duality / triality defects we considered.

\section{3D TFTs from 7-branes}\label{app:minimalTFT7branes}

In the main body of this chapter we showed how basic structure of 7-branes can account for duality / triality defects in 
QFTs engineered via D3-brane probes of a Calabi-Yau singularity $X$. In particular, we saw that anomaly inflow analyses constrain the resulting 3D TFT of the corresponding duality defect. Of course, given the fact that we are also claiming 
that these topological defects arise from 7-branes, it is natural to ask whether we can directly extract these terms from dimensional reduction of topological terms of the 7-brane. Our aim in this Appendix will be to show to what extent we can derive a 3D TFT living on the duality / triality defect whose 1-form anomalies match that of the appropriate minimal abelian 3D TFTs, $\mathcal{A}^{k,p}$. This is required due to in-flowing the mixed 't Hooft anomaly between the 0-form duality / triality symmetry and the 1-form symmetry of the 4D SCFT, as well as from the line operator linking arguments of Section \ref{ssec:linkingminimaltheory}. While we leave a proper match of these anomalies to future work, this appendix will highlight that, in general, that 3D TFTs on the 7-branes will differ from the minimal 3D TFTs due to the presence of a non-abelian gauge group. Additionally, we propose an 8D WZ term that allows us to determine the level of the 3D CS theory. 

We first review the case of a stack of $n$ D7-branes. The Wess-Zumino (WZ) terms are known from string perturbation theory to be \cite{Douglas:1995bn}\footnote{More generally $(p,q)$ 7-brane WZ topological actions can be inferred from $SL(2,\mathbb{Z})$ transformations of \eqref{eq:WZD7}.}:
\begin{equation}\label{eq:WZD7}
\begin{aligned}
    \mathcal{S}_{\text{WZ}} &= \int_{X_8} \left( \sum_{k}C_{2k} \mathrm{Tr} e^{ \mathcal{F}_2} \sqrt{\frac{\hat{\mathcal{A}}(R_T)}{\hat{\mathcal{A}}(R_N)}} \right)_{\text{8-form}}
\end{aligned}
\end{equation}
where 
\begin{equation}
\mathrm{Tr}\mathcal{F}_2=\mathrm{Tr}( F_2-i^{\ast}B_2)=nF^{U(1)}_2-n i^{\ast} B_2\,,
\end{equation}
with $i^{\ast}B_2$ denoting the pullback from the 10D bulk to the 8D worldvolume $X_8$ of the 7-brane, $F^{U(1)}_2$ is $U(1)$ the gauge curvature associated to factor in the numerator of the 7-brane gauge group $U(n)\simeq (U(1)\times SU(n))/\mathbb{Z}_n$. This precise combination is required because F1-strings can end on D7-branes. In particular, since F1-strings couple to the bulk 2-form, there is a gauge transformation 
$B_2 \rightarrow B_2 + d \lambda_1$ which is cancelled by introducing a compensating $U(1)$ curvature associated with the 
``center of mass'' of the 7-brane. $\mathcal{A}(R_T)$ and $\mathcal{A}(R_N)$ in \eqref{eq:WZD7} are the A-roof genera of the tangent and normal bundles which is given by the expansion
\begin{equation}
    \hat{\mathcal{A}}=1-\frac{1}{24}p_1+\frac{1}{5760}(7p^2_1-4p_2)+...
\end{equation}
where for completeness, we have included $p_i$, the $i^{\mathrm{th}}$ Pontryagin class of the tangent bundle / normal bundle. Since we are concerned with reducing the 7-brane on $S^5$, such contributions play little role in our analysis but could in principle play an important role in more intricate boundary geometries $\partial X$. Taking this into account, the only terms that concern us then are
\begin{equation}\label{eq:wzd7un}
    \int_{M_3\times S^5}\frac{1}{8\pi}C_4\wedge \mathrm{Tr}(F^2)+\frac{2\pi}{2}C_4\wedge \left(\mathrm{Tr}((F_2/2\pi)-B_2) \right)^2
\end{equation}
where we are now being careful with the overall factors of $2\pi$ and considering $\mathrm{Tr}(F)/2\pi$ as integrally quantized. Reducing \eqref{eq:wzd7un} on the $S^5$ surrounding $N$ D3s would then naively produce a level $N$ $U(n)$ 3D Chern-Simons theory living on $M_3$ with an additional coupling to the background $U(1)$ 1-form field $B_2$ that is proportional to $\int_{M_3}\mathrm{Tr}(A)\wedge B_2$. We say ``naive'' because one must first understand how the center-of-mass $U(1)$ of $n$ D7-branes is gapped out via the St\"uckelberg mechanism, i.e., how the gauge algebra reduces from $\mathfrak{u}(n)$ to $\mathfrak{su}(n)$. Indeed, observe that the coupling $C_6 \mathrm{Tr} F_2$ can gap out this $U(1)$ since integrating over $C_6$ produces the constraint\footnote{This is provided that we choose Neumann boundary conditions for $C_6$ along the D7-brane stack.}:
\begin{equation}\label{eq:Stukconstraint}
    \mathrm{Tr}(F_2)=nF^{U(1)}_2=0
\end{equation}
so $F^{U(1)}_2$ still survives as a $\mathbb{Z}_n$-valued 2-form field and is in fact equivalent to the generalized Stiefel-Whitney class \cite{Aharony:2013hda}
\begin{equation}
 F^{U(1)}_2  ~\rightarrow~ w_2\in H^2(X_8,\mathbb{Z}_n)\,.
\end{equation}
One sees this by supposing that there is a magnetic 4-brane monopole in the $U(n)$ gauge theory in the fundamental representation $\mathbf{n}_{+1}$. Then $\frac{1}{n}\int_{S^2}\mathrm{Tr}(F)$ measures a magnetic charge $+1$ with respect to the $U(1)$ in the numerator of $U(n) = (U(1) \times SU(n) ) / \mathbb{Z}_n$, where the $\mathbb{Z}_n$ embeds in the center of $SU(n)$ in the standard fashion. After gapping out the center of mass $U(1)$, we still measure a magnetic flux $1 \; \mathrm{mod}\; n$ around the monopole, which is a defining property of $w_2$. 

Similarly, when considering flat connections of the $U(n)$ gauge theory, we have that $\mathrm{Tr}(F_2)=d\mathrm{Tr}(A)$, and the constraint \eqref{eq:Stukconstraint} implies $n\mathrm{Tr}(A)=0$. Then, the integral $\int_{\gamma_1}\mathrm{Tr}A$ measures a $U(1)\subset U(n)$ monodromy around a 1-cycle $\gamma_1$ and becomes, after decoupling the center-of-mass $U(1)$, the value of $\int_{\gamma_1}w_1$ because this naturally measures the $\mathbb{Z}_n$ monodromy. A subtle distinction we should make is that while $w_2$ and $w_1$ are analogs of $\mathrm{Tr}(F)$ and $\mathrm{Tr}(A)$, the former are cohomology classes while the discrete remnants of the latter are discrete cocycles, i.e. members of $C^i(X_8,\mathbb{Z}_n)$. We therefore name these $\mathcal{J}_2$ and $a$, respectively, which satisfy $[\mathcal{J}_2]=w_2$ and $[a]=w_1$. Moreover, when $w_2=0$, we have that $\mathcal{J}_2=\delta a$ where $\delta$ is the coboundary operator. 

Generalizing to non-perturbative bound states of 7-branes,\footnote{Again, by this we mean 7-branes whose monodromy fixes $\tau$.} with some monodromy matrix $\rho$, we know from the main text that the analog of $B_2$ for the D7 case is generalized to $B^\rho_2$ which takes values in $\mathrm{ker}(\rho-1)$, it is natural then
ask whether there is an analogous discrete remnant of the ``center of mass $U(1)$'' for a general 7-brane 
of type $\mathfrak{F}$, namely the analog of the specific combination $\mathcal{F}_2 = \mathrm{Tr}(F_2 - B_2)$. From the discussion below \eqref{eq:Stukconstraint}, we can already make a reasonable guess that $\mathrm{Tr}(F_2)$ should be replaced by $\mathcal{J}_2\in C^2(M_3\times \partial X_6,\mathrm{ker}(\rho))$. In the case of perturbative IIB D7-branes, this involves the specific decomposition $U(n) = (SU(n) \times U(1)) / \mathbb{Z}_n$. In the case of constant axio-dilaton profiles, all of these cases can be obtained from the specific subgroups:\footnote{For example, we have the following subgroups of $E_8$: $(E_7 \times U(1)) / \mathbb{Z}_2$ and $(E_6 \times U(1)^2) / \mathbb{Z}_3$.} $E_8 \supset (G_{\mathfrak{F}} \times U(1)^m) / \mathbb{Z}_{k} $ with $n + m = 8$, where maximal torus of $G_{\mathfrak{F}}$ has dimension $n$, and $G_{\mathfrak{F}}$ has center $\mathbb{Z}_k$. We then see that all of the remarks below \eqref{eq:Stukconstraint} equally apply here if we start with an $E_8$ 7-brane and Higgs down to another 7-brane with constant axio-dilaton. Extending the treatment in \cite{Kapustin:2014gua} for A-type Lie groups, 
we introduce the gauge field:
\begin{equation}
    \mathbf{A}=A+\frac{1}{k}\widehat{a}
\end{equation}
and its field strength $\mathbf{F}=d\mathbf{A}+\mathbf{A}\wedge \mathbf{A}$ where the connections $A,\widehat{a}$ take values in the Lie algebras of $G_{\mathfrak{F}}$ and $U(1)^m$, respectively. As in our D7-brane discussion, the center of mass $U(1)\subset U(1)^m$ is gapped out up to a discrete $\mathbb{Z}_k$-valued gauge field $a$. The curvature on the 7-brane worldvolume therefore takes the form 
\begin{equation}
   \mathbf{F}=  F+\mathcal{J}_2
\end{equation} 
where $F$ is the traceless curvature of $A$ and $\mathcal{J}_2$ is the discrete remnant of the center of mass mode valued in $\Gamma=\mathbb{Z}_k$. Note that $\Gamma$ coincides with the center of the 7-brane gauge group (with electric polarization) which then nicely matches our guess that $\mathcal{J}_2$ should serve as our analog of $\mathrm{Tr}(F)$. Moreover, when we take electric polarization on the brane, $w_2$ is trivial for gauge bundles and therefore we have that $\mathcal{J}_2=\delta a$ with $a\in C^1(M_3\times \partial X_6, \mathrm{ker}(\rho))$

We now conjecture a non-perturbative generalization of the WZ terms in \eqref{eq:wzd7un} which applies to all types of 7-brane stacks (labelled by gauge algebra $\mathfrak{g}$) as listed in table \ref{tab:Fibs}:\footnote{A brief comment on the normalization of the instanton density: We have adopted a convention where $\mathrm{Tr} F^2 = \frac{1}{h_{G}^{\vee}} \mathrm{Tr}_{\mathrm{adj}} F^2$, with the latter a trace over the adjoint representation, and $h_{G}^{\vee}$ the dual Coxeter number of the Lie group $G$. Moreover, in our conventions, we have that for a single instanton on a compact four-manifold, $\frac{1}{4} \mathrm{Tr} F^2$ integrates to $1$. }
\begin{equation}\begin{aligned}\label{eq:WZtermsu3}
 \mathcal{S}_{WZ}^{\mathrm{(7)}} &\supset  \int_{M_3\times \partial X_6} \left(  C_4 \wedge \textnormal{tr}\exp \left(\mathbf{F}-B_2^\rho  \right)\right) \\
   &=  \int_{M_3\times \partial X_6} \left(  \frac{1}{8\pi} C_4 \wedge \mathrm{Tr}\,{F}^2 + C_4 \cup \frac{1}{2} (\mathcal{J}_2-B^\rho_2 )^2 \right)
\end{aligned}
\end{equation}
Here we have chosen normalizations in \eqref{eq:WZtermsu3} such that $\exp(i \mathcal{S}_{WZ})$ appears in the 7-brane path integral and $B_2^\rho, \mathcal{J}_2$ are $U(1)$ valued. The first term in \eqref{eq:WZtermsu3} is the D3-brane instanton density term. Namely, it can be obtained by considering a single D3-brane inside a 7-brane and viewing it as a charge-1 instanton which sets the normalization. The second term is a generalization of the term 
\begin{equation}
   \int_{D7}C_4\wedge n B_2\wedge \mathrm{Tr}F_2 = \int_{D7}C_4\wedge n B_2\wedge nF^{U(1)}_2
   \end{equation}
appearing in \eqref{eq:WZD7} by replacing $B_2$ with the more general $B_2^\rho$. The coefficient of $1$ for the second term of \eqref{eq:WZtermsu3} follows from the standard substitutions $B_2\rightarrow \frac{1}{n}B_2$ and $F^{U(1)}_2\rightarrow \frac{1}{n}F^{\mathbb{Z}_n}_2$ when one converts a $U(1)$-valued field to its  remnant field valued in $\mathbb{Z}_n\subset U(1)$.

After reducing on an $S^5$ with flux $\int F_5= N$, \eqref{eq:WZtermsu3} produces a 3D $(G_{\mathfrak{F}})_N$ CS theory\footnote{Subscript denotes the level.} along with a coupling to its electric 1-form symmetry background. In other words, our 3D action becomes:
\begin{equation}\label{eq:3dCS}
  \int_{M_3}   N \cdot CS(A) + 
\frac{N}{2 \pi}a \cup B_2^\rho+ N\cdot CS(a)\\
\end{equation}
where fields are treated as elements in $U(1)$ (suitably restricted): the background $B_2^\rho$ can be normalized to take values in $\mathbb{Z}_{\mathrm{gcd}(k,N)}$ (recall $k$ is the order of the monodromy matrix of the non-perturbative 7-brane), and similar considerations apply for $a$. Note also that a priori, the 3D TFT we get in this way need not match the minimal TFT of type $\mathcal{A}^{K,m}$, since anomaly inflow considerations do not fully fix the form of the 3D TFT. It would be interesting to carry out a complete match with the analysis presented in the main text, but we leave this for future work.

\section{D3-Brane Probe of \texorpdfstring{$\mathbb{C}^3 / \mathbb{Z}_3$}{C3/Gamma3}} \label{app:orbo}

In this Appendix we present further details on the case of a D3-brane probing the orbifold singularity $\mathbb{C}^3 / \mathbb{Z}_3$. The orbifold group action on $\mathbb{C}^3$ is defined by
\begin{equation}
    (z_1,z_2,z_3)\rightarrow (\zeta z_1, \zeta z_2, \zeta z_3),~  \zeta ^3=1.
\end{equation}
Following the procedure in \cite{Douglas:1996sw, Lawrence:1998ja, Kachru:1998ys}, 
the field content of the resulting 4D theory is given in Figure \ref{fig:quiverc3z3}. 
The superpotential of the theory is:
\begin{equation}
    W= \kappa \text{Tr}\big\{ X_{12}[Y_{23},Z_{31}]+X_{23}[Y_{31},Z_{13]}]+X_{31}[Y_{12},Z_{23}] \big\}.
\end{equation}
\begin{figure}[H]
    \centering
    \includegraphics[width=7cm]{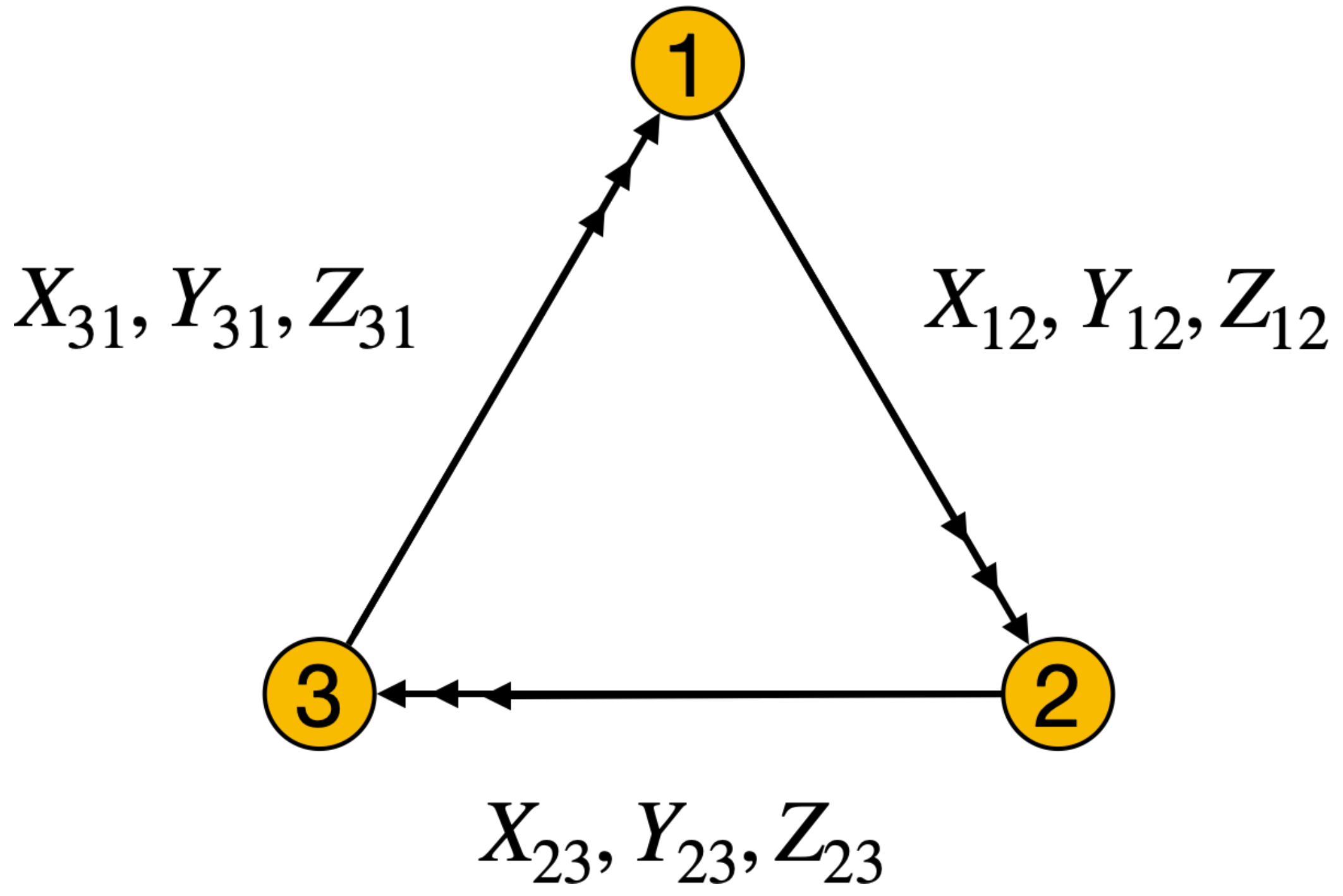}
    \caption{The quiver diagram of $\mathcal{N}=1$ theory on D3-branes probing $\mathbb{C}^3/\mathbb{Z}_3$.}
    \label{fig:quiverc3z3}
\end{figure}

The 5D boundary $\partial \mathbb{C}^3 / \mathbb{Z}_3 = S^5/\mathbb{Z}_3$ 
of the orbifold singularity has the cohomology classes:
\begin{equation}
    H^{\ast}(S^5/\mathbb{Z}_3)=\{ \mathbb{Z},0,\mathbb{Z}_3,0,\mathbb{Z}_3,\mathbb{Z}   \},
\end{equation}
with linking numbers: 
\begin{equation}
    \int_{S^5/\mathbb{Z}_3}t_2\star t_2 \star t_2=\int_{S^5/\mathbb{Z}_3}t_2\star t_4=-\frac{E_0^3}{3\cdot 3\cdot 3}=-\frac{1}{3}.
\end{equation}
$E_0^3$ is the triple self-intersection number of the compact divisor of $O(-3)\rightarrow \mathbb{P}^2$, which can be read from the toric data.\footnote{We refer the reader to \cite{hori2003mirror} for technical details on computing intersection numbers of toric varieties.}

The 5D TFT with the generic form (\ref{eq:TFT for generic N=1}) now reduces to 
\begin{equation}\label{eq:TFT for c3z3}
    \begin{split}
        S_{\text{symTFT}}&=-\int_{M_4\times \mathbb{R}_{\ge 0}} \bigg\{ N\breve{h}_3\star \breve{g}_3 - \frac{1}{3}\breve{E}_3\star \breve{B}_1\star \breve{C}_1-\frac{1}{3}\breve{E}_{1}\star \left( \breve{B}_1\star \breve{g}_3+\breve{h}_3\star \breve{C}_1 \right) \bigg\}.
    \end{split}
\end{equation}
Let us now identify the correspondence between the symmetries in the 4D SCFT and the 5D symmetry TFT fields. The first term in (\ref{eq:TFT for c3z3}) is obvious. It is just the differential cohomology version of the familiar $N\int B_2\wedge dC_2$ term which also appears in the $\mathcal{N}=4$ SYM case. In the $\mathcal{N}=1$ quiver gauge theory, $\breve{h}_3$ (resp. $\breve{g}_3$) corresponds to $B_2^{\text{diag}}$ (resp. $C_2^{\text{diag}}$) for the diagonal $\mathbb{Z}_N$ electric (resp. magnetic) 1-form symmetry of the $\mathfrak{su}(N)^3$ theory.

Based on our previous discussion, we know $E_1, B_1$ and $C_1$ are background gauge fields for $\mathbb{Z}_3$ 0-form symmetries. In fact, as explained in \cite{Gukov:1998kn}, there are indeed three candidate $\mathbb{Z}_3$ symmetries in the 4D SCFT which act on the fields of the quiver gauge theory as follows:
\begin{equation}
\begin{split}
    &\mathbf{B}:(X_{ij},Y_{ij},Z_{ij})\rightarrow (Y_{i+1,j+1},Z_{i+1,j+1},X_{i+1,j+1})\\
    &\mathbf{C}:(X_{ij},Y_{ij},Z_{ij})\rightarrow (\zeta X_{ij}, \zeta^2 Y_{ij},Z_{ij}),\\
    &\mathbf{E}:(X_{ij},Y_{ij},Z_{ij})\rightarrow (\zeta X_{ij},\zeta Y_{ij},\zeta Z_{ij}).
\end{split}
\end{equation}
where $i$ and $j$ are mod 3 numbers denoting gauge factors. These symmetry generators transform non-trivially under IIB dualities, and their transformations are the same as those already stated in section \ref{sec:N=1}.

\end{append}

\begin{bibliof}
\bibliography{main}
\end{bibliof}
\end{document}